# ESSAYS ON RESPONSIBLE AND SUSTAINABLE FINANCE

A Dissertation
Presented to
The Academic Faculty

By

Baridhi Malakar

In Partial Fulfillment
of the Requirements for the Degree
Doctor of Philosophy in the
Scheller College of Business

Georgia Institute of Technology

May 2023

**ESSAYS ON RESPONSIBLE AND SUSTAINABLE FINANCE**

Thesis committee:

Dr. Sudheer Chava, Advisor
Scheller College of Business
*Georgia Institute of Technology*

Dr. Rohan Ganduri
Goizueta Business School
*Emory University*

Dr. Manpreet Singh
Scheller College of Business
*Georgia Institute of Technology*

Dr. Alexander Oettl
Scheller College of Business
*Georgia Institute of Technology*

Dr. Manasa Gopal
Scheller College of Business
*Georgia Institute of Technology*

To my wife, Sukanya,

and my parents.

## ACKNOWLEDGMENTS

I am extremely fortunate and blessed to have Prof. Sudheer Chava as my advisor. I am deeply grateful for his continuous guidance, help, and support for my research. I want to extend my sincere gratitude, regard, and appreciation for Dr. Manpreet Singh for his invaluable support and guidance throughout my Ph.D.

I greatly appreciate Dr. Manasa Gopal, Dr. Rohan Ganduri, and Dr. Alex Oettl for their generosity with their time, and for providing constant advice and encouragement throughout the Ph.D. program. I also want to thank Prof. Narayanan Jayaraman, Dr. Nishant Dass, and Dr. Suzanne Lee for their advice and support during my initial years. I have looked up to Dr. Linghang Zeng and Dr. Nikhil Paradkar for guidance and encouragement during challenging times. I am thankful to all other faculty members for advice, comments, and suggestions in building a great research environment for PhD students. I thank Dr. Teng Zhang, Dr. Youngmin Choi, Dr. Minho Wang, Dr. Peter Simasek, Wendi Du, Yafei Zhang, Agam Shah, Richard Wang, Fred Hu, and other Ph.D. students for their help and interesting discussions. It is truly an honor and pleasure to have been working with all of you.

I would like to thank my parents, Mr. Banu Dhari and Mrs. Bithika, for their faith in me and their continuous support. I am thankful to my sister, Devopriya, my brother-in-law, Koushik, and my nephew (Mayukh) for their patience and support. My in-laws have been extremely encouraging and kept my spirits high during this process. Lastly, I express my deepest gratitude to my wife, Dr. Sukanya, for love and for accompanying me on this journey.

## TABLE OF CONTENTS

## LIST OF TABLES

# LIST OF FIGURES

# SUMMARY

The dissertation consists of three essays on responsible and sustainable finance. I show that local communities should be seen as stakeholders to decisions made by corporations. In the first essay, I examine whether the imposition of fiduciary duty on municipal advisors affects bond yields and advising fees. Using a difference-in-differences analysis, I show that bond yields reduce by 9% after the imposition of the SEC Municipal Advisor Rule. Larger municipalities are more likely to recruit advisors after the rule is effective and experience a greater reduction in yields. However, smaller issuers do not seem to significantly benefit from the SEC Rule in terms of offering yield. Instead, their borrowing cost increases if their primary advisor exits the market. Using novel hand-collected data, I find that the average advising fees paid by issuers does not increase after the regulation. Offering yields reduce due to lower markup at the time of underwriting, driven by issuers for whom advisors play a more significant ex-ante role in selecting underwriters. Overall, my results suggest that while fiduciary duty may mitigate the principal-agent problem between some issuers and advisors, it has a limited effect on small issuers.

In the second essay, we analyze the impact of $40 billion of corporate subsidies given by U.S. local governments on their borrowing costs. We find that winning counties experience a 15.2 bps increase in bond yield spread as compared to the losing counties. The increase in yields is higher (18 – 26 bps) when the subsidy deal is associated with a lower jobs multiplier or when the winning county has a lower debt capacity. However, a high jobs multiplier does not seem to alleviate the debt capacity constraints of local governments. Our results highlight the potential costs of corporate subsidies for local governments.

In the third essay, we provide new evidence that the bankruptcy filing of a locally-headquartered and publicly-listed manufacturing firm imposes externalities on the local governments. Compared to matched counties with similar economic trends, municipal bond yields for affected counties increase by 10 bps within a year of the firm's bankruptcy filing. Counties that are more economically dependent on the industry of the bankrupt firm are more affected and do not immediately recover from the negative impact of the corporate bankruptcy. Our results highlight that local communities are major stakeholders in public firms and how they are adversely affected by corporate financial distress.

The final essay examines whether managers walk the talk on the environmental and social discussion. We train a deep-learning model on various corporate sustainability frameworks to construct a comprehensive Environmental and Social (E&S) dictionary. Using this dictionary, we find that the discussion of environmental topics in the earnings conference calls of U.S. public firms is associated with higher pollution abatement and more future green patents. Similarly, the discussion of social topics is positively associated with improved employee ratings. The association with E&S performance is weaker for firms that give more non-answers and when the topic is immaterial to the industry. Overall, our results provide some evidence that firms do walk their talk on E&S issues.

# CHAPTER 1

# FIDUCIARY DUTY IN THE MUNICIPAL BONDS MARKET

## 1.1 Introduction

State and local governments in the United States finance various infrastructure and public utility projects through municipal bonds. As of 2022, $4 trillion of municipal bonds are outstanding, of which $457 billion were issued in 2021. Municipal issuers may not always have the necessary in-house expertise [1]. They typically hire advisors to help decide the method of sale, to structure the bonds and to develop and draft the offering statements [2]. However, due to misaligned incentives and decentralized information [3], advisors may not always act in the interest of bond issuers[1]. As a result of these potential principal-agent problems, the SEC imposed fiduciary duty on municipal advisors through the Municipal Advisor Rule (MA Rule) effective July 1, 2014. In this paper, I study how this imposition of fiduciary duty on the municipal advisors affects the offering yields and advising fees paid by the municipal issuers.

There are competing views on the implications of imposing a fiduciary duty [4]. Fiduciary duty may benefit issuers by alleviating the principal-agent problem and making it costly to offer poor advice (*advice channel*). Advisors may be held liable for adverse consequences to issuers during bond issuance if they fail to adhere to their fiduciary responsibility after the MA Rule[2]. Alternatively, fiduciary duty may impose an undue burden on municipal advisors by increasing their costs (*fixed costs channel*), leading to worse outcomes for the issuer. Therefore, it is unclear how the imposition of fiduciary duty on municipal advisors affects issuers.

[1]For instance, SEC Commissioner Kara M. Stein remarked in 2013, that the SEC had observed "numerous examples of bad behavior, including self-dealing and excessive fees". Also see comments by Municipal Securities Rulemaking Board (MSRB) and others.

[2]The fiduciary obligation under the MA Rule encompasses the twin requirements of the duty of care, i.e. advisors must exert effort on behalf of the issuer to make a recommendation; and the duty of loyalty, i.e. advisors must uphold the issuers' interests superior to their own. I provide more details in Section A1.1 in the Internet Appendix.

Using a canonical difference-in-differences research design, I compare the yield spreads for municipal bonds sold via negotiation to bonds sold via competitive bidding, before and after the SEC MA Rule. Bonds sold via negotiation are more likely to involve the underlying friction of misaligned incentives between the issuer and the advisor (treated group). This friction may be lower in bonds sold via competitive auction (control group), where price discovery happens without negotiation. I use this quasi-natural setting to understand the overall effect on bond yield spreads between the *advice channel* and the *fixed costs channel*.

The preferred specification indicates that offering yields decreased by 12.6 basis points (bps) for negotiated bonds relative to competitively bid bonds after the regulation. Given that municipalities may select into negotiating their bond sale, the baseline identification relies on within-issuer variation across advisors over time. Further, I absorb time-varying unobserved factors among advisors and among issuers based on the geographical state to which they belong. Finally, I include bond- and county-level controls to account for observable characteristics. Considering the average bond in the sample, this effect amounts to a nearly 9.7% reduction (=12.60/129) in yield spreads[3]. I conduct several robustness checks to show that the baseline specification is robust to alternative explanations.

The lower yields are driven by reduced markup of negotiated bonds at the time of issuance. With advisors' obligation to adhere to their fiduciary duty, they are more likely to overcome underpricing of bonds[4]. Using a measure of liquidity from post-issuance trades (see [5]), I find that the average price dispersion decreases for negotiated bonds after the MA Rule. Since markups on investor purchases increase with interdealer trading [6], my re-sults imply that negotiated bonds pass through fewer dealers before being held by investors. The lower underpricing (markup) and lower liquidity I document here likely support that the *advice channel* outweighs the *fixed costs channel* after the MA Rule.

[3]To offer a back-of-the-envelope calculation, I use a modified duration approach to estimate cost savings worth USD 2.1 million on the total issuance by the average county. This is nearly 133 times the amount spent on educating a pupil in a public school in the United States (see: https://nces.ed.gov/fastfacts/display.asp?id=66).

[4]As in [3], underwriter profit increases in yields in the primary market issuance because it pays a lower price to the municipal issuer.

The primary concern in this empirical setting is that of selection i.e., issuers' choice of the method of sale between negotiation versus competitive bidding is unlikely to be random. I provide evidence to mitigate concerns related to selection. Ex-ante, I show that observable factors do not explain the likelihood of selling bonds via negotiation *within* issuers. Further, a longitudinal analysis of the type of debt sold over the years reveals that nearly 40% of the bond volume is sold via negotiation, even around the SEC Municipal Advisor Rule. More formally, the likelihood of negotiation *within* issuers is stable around the timing of the SEC Rule. Taken together, these evidence suggest that issuers do not seem to select negotiated bonds around the regulation differentially. Finally, the distribution of important bond characteristics also tends to follow similar patterns between treated and control bonds, as seen in a quantile-quantile plot. Overall, I interpret my results as reflecting the change in yields for an issuer choosing to raise bonds via negotiation.

Identification in the difference-in-differences design assumes that outcomes for negotiated and competitively bid bonds would have trended in parallel without the regulatory change. Using a binned scatter plot of raw yield spreads around the MA Rule, I show that generally negotiated bond yield spreads are higher than competitively bid bonds[5]. Importantly, the fitted lines suggest that the negotiated and competitive yields were nearly parallel before the Rule. After July 2014, there seems to be a convergence in yields between the treated and control bonds. Beyond this eyeballing evidence, I formally assess the plausibility of the parallel trends assumption by testing the pre-trends *within* issuers. I plot the coefficients from a dynamic version of the difference-in-differences design using only issuer fixed effects. This analysis suggests no economic or statistical difference between the treated and control groups *within* issuers before the Rule. After the SEC Rule, negotiated

[5]This is consistent with previous evidence in the literature [7, 8, 9, 10, 11].

yield spreads tend to decrease compared to the control group.

I use the municipal bond transactions data from the Municipal Securities Rulemaking Board (MSRB) to explore why offering yields reduce for negotiated bonds at the time of issuance. In a negotiated sale, underwriters and issuers/advisors arrive at a price based on investor demand [12]. After the imposition of fiduciary duty on municipal advisors, issuers may be able to negotiate better with underwriters. I examine this by analyzing the average markups on new issuances using the baseline specification. Using transactions within the first month of trading, I construct the markup following [13]. The analysis reveals that underpricing of treated bonds reduced by nearly six basis points after the introduction of fiduciary duty on municipal advisors[6]. This evidence from post-issuance trades suggests that bonds are likely issued closer to the price investors pay in subsequent trading.

Advisors may also be involved in choosing an underwriting firm [14] based on their information. This becomes especially relevant for a negotiated sale because competitive bidding eliminates the pre-selection of the underwriting firm before the bond issuance. In this context, I present evidence using the heterogeneity among issuers for the role of advisors in choosing underwriters. Ideally, I would want to use precise information on advisors "selecting" underwriters. Without such data, I construct this measure by identifying whether an underwriter is introduced for the first time when a given issuer engages an advisor. Here, I show that the reduction in negotiated yield spreads is driven by issuers for whom advisors likely play a more significant ex-ante role in selecting the underwriter. This is consistent with fiduciary duty driving greater benefit where the principal-agent problem may be higher.

Another dimension to evaluate how the fiduciary duty on advisors affects issuers is the advisory fees paid by issuers. If the *fixed cost* channel dominates over the *advice channel*, then advising fees may increase after the MA Rule if advisors can pass on the cost to issuers. But if the *advice channel* outweighs it, regulatory oversight may discourage excessive fees. To the best of my knowledge, there is no standard database to obtain this information readily. As a result, I submit Freedom of Information Act (FOIA) requests to each of the states' bond issuance agencies to gather these data on advising fees. Overall, I find that issuers pay around 30 cents paid per USD 100 of municipal bonds raised during the sample period. This remains relatively stable around the MA Rule likely suggesting the importance of regulatory oversight[7].

[6]This result and magnitude are robust to using an alternative approach to compute the markup.

Taken together, the overall reduction in yields and stable advising fees for issuances may favor issuers in engaging advisors. Therefore, examining whether issuers are more likely to engage advisors after this regulation is reasonable. On an extensive margin, I find that the average likelihood of engaging an advisor increases by over 5% in the period after the SEC Municipal Advisor Rule. This estimation accounts for unobserved heterogeneity *within* issuers. [15] find that issuers may accept NPV losses for short term cash flow savings in the municipal bond market. While issuers seem to secure lower yields without paying higher advising fees, there may still be variation among issuers. To understand more, I examine the likelihood of engaging advisors based on the size of issuers. I find that large issuers are more likely to adopt advisors, with little change among small issuers.

The choice of engaging an advisor for a municipal bond issuance is endogenous to issuers. Issuers more likely to benefit from advised bonds may also be more likely to hire an advisor on the extensive margin. Therefore, I revisit the baseline analysis to distinguish between small and large issuers' overall effect on advised bonds. Using the average and median size of ex-ante bond issuances, this analysis shows that large issuers drive the reduction in yield spreads. This is consistent with the previous finding of large issuers being more likely to engage advisors after the SEC regulation. Similarly, I find that the evidence of reduced underpricing is also strongly driven by large issuers. Further, I show that more sophisticated issuers seem to drive the reduction in offering yields. I approximate sophistication among issuers based on the complexity of bonds issued, level of credit enhancement

[7]SEC Commissioner Kara Stein noted in 2013 that issuers may have faced excessive fees by municipal advisors.

purchased, wages paid to their finance staff, and fraction of advised bonds issued, respectively.

Some advisors may choose not to provide municipal advisory services after the SEC Municipal Advisor Rule [2]. Consistent with this, I find that there is a decline in the number of advisors after the regulation. The baseline reduction in yield spreads is driven by those issuers for whom the primary advisors do not exit after the MA Rule. In contrast, there is no effect on issuers linked to advisors that exit the market. Importantly, there is substantial heterogeneity among such issuers: while offering yield spreads reduce for large issuers, there is an increase in spreads for small issuers.

Finally, I evaluate whether lower yields drive changes in municipal bond issuance after the SEC Municipal Advisor Rule. I classify issuers with greater (above median) reliance on negotiated bonds before the regulation as the "treated" group. Compared to the one year before the Rule, treated issuers raise more municipal debt than control issuers. This is reasonable given that the ex-post reduction in yields may increase the debt capacity for these issuers. As before, this increased issuance is driven by large issuers who experience greater yield reduction. Meanwhile, there is a weak evidence for decrease in the amount of bonds raised by small issuers. Overall, issuers who benefit from lower yields after the imposition of fiduciary duty can raise more municipal debt.

This paper contributes to three strands of the literature in finance and economics. First, this paper is related to the economics of expert advice in financial decision-making. [16, 17] provide theoretical underpinnings to how competition and compensation structure are related to advice provided by financial intermediaries. Empirical work in this direction has shed light on advisors' commissions and other incentives in offering advice rather than clients' interests [18, 19, 20, 21, 22]. Recent work focuses on the geographic concentration of misconduct by financial advisors [23] and how the enforcement agency in charge may affect the quality [24]. This paper shows how enforcing discipline among municipal advisors through fiduciary duty may improve the average outcome for municipal borrowers.

Importantly, I introduce novel data on financial intermediaries' municipal advising fees to show that advising fees does not increase after the MA Rule.

Second, this paper follows a recent surge in academic work on the municipal bond market [25, 26, 27, 28, 29, 30, 31, 32]. Prior work leveraging the municipal bond market transactions has looked at liquidity [33, 6] and default [5] risk, as well as tax-effects [34, 35, 36, 37]. There is a growing literature documenting how local conditions affect municipal bond prices [38, 39, 40, 41, 42, 43]. In this context, this paper sheds new light on how federal regulation may reduce offering yields on average due to reduced underpricing at the time of bond issuance.

Finally, this paper contributes by showing the policy relevance of fiduciary duty. Recent work in this direction has estimated higher risk-adjusted returns by extending fiduciary duty [44]. [4] identify the effect of fiduciary duty in the reduced form by accounting for the entry margin using a structural model. Among municipal bonds, [3] provides empirical evidence on how reducing conflicts of interest for advisors may improve bond outcomes. I contribute to this literature by showing how regulation may mitigate the principal-agent problem between issuers and advisors but has a differential effect among issuers. While large issuers gain, small and newly advised issuers do not benefit after the fiduciary rule on municipal advisors. [3] focuses on how advisors may control the amount of information that is disclosed to competing underwriters. Regulatory change to increase this information provision by advisors leads to more efficient market outcomes. My paper addresses misaligned incentives and inadequate monitoring of advisors resulting in poor advice. I show that the MA Rule discourages unqualified advice by mitigating this principal-agent problem between some issuers and advisors.

## 1.2 Identification Challenges and Empirical Methodology

A direct analysis of offering yield spreads and the imposition of fiduciary duty on municipal advisors using Ordinary Least Squares (OLS) may be fraught with some problems. For example, unobserved issuer level characteristics may be driving the change in yields as well as the choice of advisors. Alternatively, municipal issuers that may be able to switch to more responsible advisors may have higher credit worthiness. This unobserved 'ability' may improve their borrowing costs and lead to inaccurate estimation of the causal effect of fiduciary duty. Overall, unobserved factors may impact observed yields, making issuers' decision to raise new municipal debt endogenous. In an ideal experiment, I may want to compare the yield spreads between two groups of otherwise identical bonds: those in which the SEC Rule was binding versus not. To this effect, I propose an approach that leverages a special aspect of the municipal bond market.

Municipal bonds are commonly sold via one of the two methods: competitive bidding or negotiated sale. Competitive bidding involves underwriters submitting their bids to buy the newly issued bonds. In a negotiated sale, a pre-selected underwriter works with the issuer to arrive at the terms of the sale. Municipal advisors play a crucial role in both methods of sale. They help the issuer in advertising the bond sale to potential underwriters in a competitive sale as well as in evaluating the final bids submitted. For negotiated sales, advisors may be involved from the time of selection of the underwriter to the final closure of the bond sale. Issuers typically rely on their advisors for bond structuring and pricing throughout their negotiation with the underwriter. In both negotiated and competitive sale, underwriters assume complete risk and responsibility for selling the bonds [12].

The primary friction that I propose to investigate arises from the principal-agent problem between the issuers and the advisors during the issuance of new municipal bonds. When hiring the more-informed advisors, municipal issuers may not be able to monitor advisors. By requiring municipal advisors to owe a fiduciary duty to their clients, the SEC Municipal Advisor Rule addresses this friction by mitigating the principal-agent problem. In this regard, this paper considers a quasi-natural experiment such that the underlying information asymmetry between the issuer and the advisor is potentially bid away in a competitive auction of bonds. This becomes especially relevant with regard to the choice of underwriter because competitive bidding eliminates the pre-selection of the underwriter. Thus, a bond issued via competitive bidding would constitute the control sample, with lower scope for misalignment of advisor incentives. Meanwhile, the treated group in this setting corresponds to bonds sold via negotiation. Prior to the Rule, issuers may have limited knowledge about the pricing of their bonds, especially in a negotiated sale. This is consistent with [11] who find that the use of negotiated sales increases yields by 15–17 bps. Under the fiduciary duty obligation, issuers may be better served by their and overcome some of the information asymmetry during the negotiation with the underwriter.

To formally characterize the baseline specification using a standard difference-in-differences [45] equation, as below:

$$\begin{aligned} y_{b,i,a,t} = \alpha + \beta_0 * Post_t \times Nego._{b,i,a} + \beta_1 * Post_t + \beta_2 * Nego._{b,i,a} + X_b + Z_{i,t} \\ + \gamma_i + \mu_{a,t} + \kappa_{s,t} + \epsilon_{b,i,a,t} \end{aligned} \tag{1.1}$$

here index $b$ refers to bond, $i$ refers to issuer, $a$ denotes the municipal adviser and $t$ indicates time. The main outcome variable in $y_{b,i,a,t}$ is the offering yield spread at which a bond is issued in the primary market. $X_b$ includes control variables at the bond level that influence its value. These include the coupon (%); log(amount issued in \$); dummies for callable bonds, bond insurance, general obligation bond, bank qualification, refunding, and credit enhancement; credit rating; remaining years to maturity; and inverse years to maturity. I provide the description of key variables in Table A1. $Z_{i,t}$ corresponds to local economic conditions of the issuer's county/state. Following [39], I use the lagged values for log(labor force) and unemployment rate, and the percentage change in unemployment rate and labor force, respectively. I aggregate county level metrics to arrive at the state level measures and use these for corresponding state level bonds. This specification includes three sets of fixed effects. First, $\gamma_i$ indicates issuer fixed effects, to make comparisons within a given issuer. I also include fixed effects at the advisor $\times$ year ($\mu_{a-t}$) level to account for unobserved time-varying changes among advisors. Finally, $\kappa_{s-t}$ corresponds to state-year fixed effects to account for unobserved time-varying changes across states in which the bonds are issued. I cluster standard errors by state due to the segmented nature of the municipal bond market [46][8].

### 1.2.1 Threats to Identification

An obvious concern with the identification strategy above relates to selection. Issuers that choose to raise municipal debt via negotiation may be different from those who do not. In order to test whether selection may likely bias the estimates of a difference-in-differences design, I evaluate the type of issuers who choose negotiation. I estimate a linear probability regression describing the choice of selling bonds via negotiation (indicator $\mathbb{1}\{Nego._{b,i,t}$) as,

$$\mathbb{1}\{Nego._{b,i,t} = 1\} = \kappa_t + \beta * X_{b,i,t} + \epsilon_{b,i,t} \tag{1.2}$$

where subscript $b$ indicates the bond issuance, $i$ indicates the issuer, and $t$ indicates the time. $X_{b,i,t}$ includes variables controlling for issue size, average bond size, coupon, years to maturity, callable status, credit enhancement, insurance status, bank qualification, number of bonds, type of security (general obligation vs revenue), type of issuer, frequency of issuer borrowing, and fixed effects for rating group as well as use of funds (bond purpose).

I estimate this regression over three approaches to demonstrate how various factors affect the choice of negotiation. I focus on the three years before the SEC Municipal Advisor Rule to capture the ex-ante snapshot. Figure 1.2 shows the estimated coefficients for each specification using bond issuances for the three years before the SEC Municipal Advisor Rule. I also provide the tabular results in Table A1.1. First, I show results from the *Overall* balance without using any geographic controls. The results suggest that there is some variation in terms of using negotiated sale for bonds that are insured as well as for state level bonds. Certain bonds that need outside insurance from a third-party may be well

[8]Other papers that follow similar clustering include [28, 27]. The results are robust to alternative dimensions of calculating standard errors including double clustering by state and year-month, as discussed in Section 1.4.3

suited for sale via negotiation. At the same time, state level issuers may be better suited for negotiation given their previous relationships [12]. When I include state fixed effects (*Within State*), I find similar factors driving the choice of negotiation.

However, importantly for identification here, these differences get wiped out when I consider the variation after including issuer fixed effects (*Within Issuer*). In this specification, I capture the likelihood of negotiation among issuers who borrow multiple times choosing negotiated or competitive sale for different issues. My strategy in Equation (4.1) looks to compare issuer-advisor pairs that sell bonds using negotiated versus competitive auction of bonds. By showing that issue characteristics do not seem to affect the choice of negotiation within issuers, I provide support against ex-ante selection concerns. Moreover, I also depict the portion of advised bonds sold via negotiation and competitive bidding in Figure A1.2. I do not find evidence for changing method of sale around the Municipal Advisor Rule. Nearly 40% of the bond volume is consistently sold via negotiation around 2014. I also show this more formally in Figure A1.3 by plotting the coefficients from regressing the likelihood of negotiation on half-year dummies after including issuer fixed effects. Benchmarked to the period before the Rule, I do not find any statistical difference around the SEC Rule. This evidence helps against concerns about selection into negotiation over time.

## 1.3 Data

This paper uses municipal bonds data are from FTSE Russell (formerly known as Mergent) Municipal Database and the Muncipal Securities Rulemaking Board (MSRB). Additionally, I hand-collect data on municipal advisor fees through Freedom Of Information Act (FOIA) requests made to state and local government bond issuers.

### 1.3.1 Municipal Bonds

Municipal bond characteristics are obtained from the Municipal Bonds dataset by FTSE Russell (formerly known as Mergent MBSD). I retrieve the key bond characteristics such as CUSIP, dated date, the amount issued, size of the issue, offering type (method of sale), state of the issuing authority, name of the issuer, yield to maturity, tax status, insurance status, pre-refunding status, coupon rate, and maturity date for bonds issued after 2004. The baseline sample consists of fixed rate, tax-exempt bonds issued during January 2010 to December 2019. This allows me to focus on a 4.5 year window around the SEC Municipal Advisor Rule. Issuers raised over USD 400 billion of municipal debt each year, mostly with advisors (Figure 1.1a). I also use the average credit ratings for these bonds within one year of issuance. These CUSIP-level ratings are provided by S&P, Moody's and Fitch. I encode character ratings into numerically equivalent values ranging from 28 for the highest quality to 1 for the lowest quality [25].

The FTSE Russell database also provides the names of municipal advisor and underwriters involved in the bond issuance. Most bond issuances have a single municipal advisor. For a few cases with two advisors, I assign the issuance to the advisor with a larger cumulative volume advised on a cumulative basis. Municipal bond issues are generally sold to a syndicate of underwriters with the lead manager identified in the data. I use the lead manager as the underwriter. In cases with multiple lead underwriters, I follow the same rule of assigning it to the larger player (based on volume underwritten on a cumulative basis). Since I observe some spelling errors and related incongruities in the string names provided by FTSE Russell, I manually check the names of advisors and underwriters in order to standardize them during the sample period. I provide more details in Section A1.2. [2] describe the evolution of the municipal advisor firm market. As shown in Figure 1.1b, there is an increase in the number of municipal advising firms during 2010 to 2013, followed by a decline. The number of withdrawals by advisor firms reaches a peak in 2014. In Panel A of Table 1.1, I summarize the top fifteen advisors in the sample along with their rela-

**Total Primary Market Issuance in Municipal Bonds**

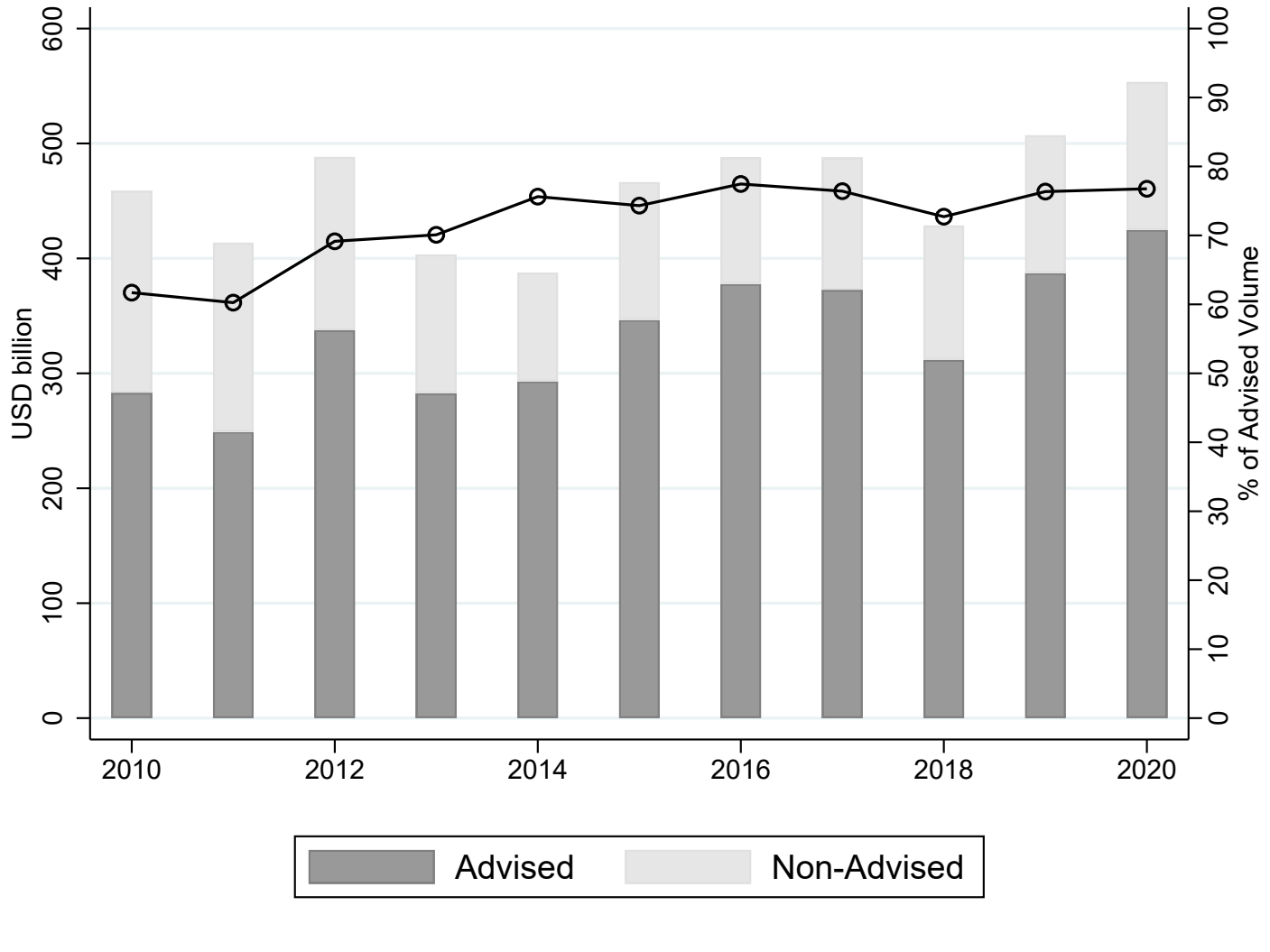

**(a)**

**Registration Activity of Municipal Advisor (MA) firms**

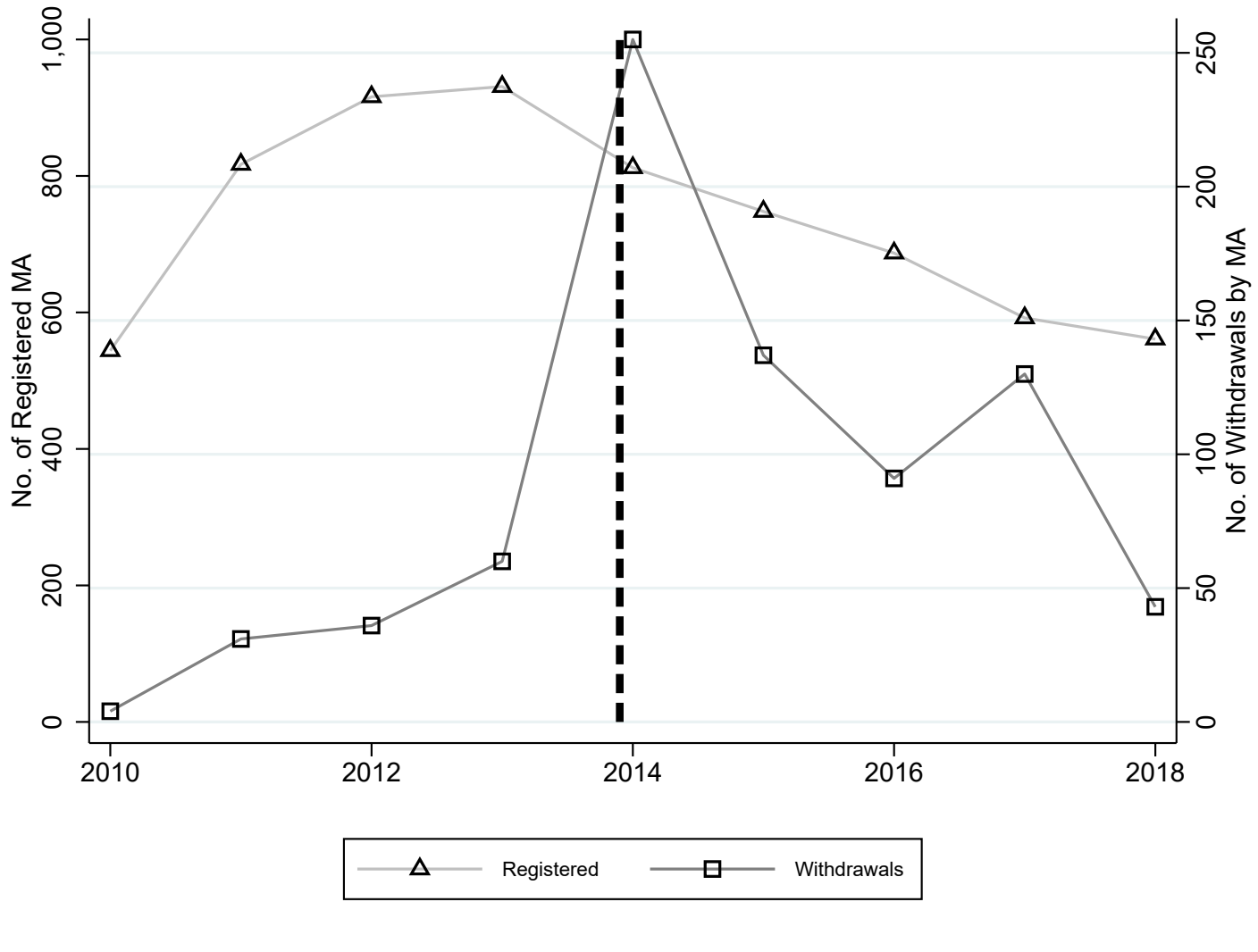

**(b)**

**Figure 1.1:** Municipal Market Issuance and Municipal Advisor Registration: In this figure, I show the issuance and registration activity in the municipal bond market. Panel (a) shows the total volume (in USD billion) of municipal debt issued during 2010-2019 in the primary market on the left axis. I also show the split based on advised versus non-advised bonds using the vertical bars. The line graph corresponds to the right axis, showing the percentage of advised bonds. Panel (b) shows the registration activity by municipal advisor (MA) firms during 2010-2018. This information is obtained from Table 1 in [2]. The left axis reports the number of firms registered as municipal advisors. The right axis provides the number of withdrawals filed by MA firms during this period.

tive share of volume advised. Since the FTSE Municipal Bonds dataset does not have the county name of each bond, I need to supplement this information from other sources like Bloomberg. However, in light of Bloomberg's download limit, it is not feasible to search for information on each CUSIP individually. Therefore, I first extract the first six digits of the CUSIP to arrive at the issuer's identity[9]. Out of 65,933 unique issuer identities (6-digit CUSIPs), Bloomberg provides me with county-state names on 60,884 issuers. For these issuers, I match the Federal Information Processing Standards (FIPS) code using [47]. I use the FIPS code to assign county level characteristics to bonds issued by local governments/issuers. Additionally, I define "issuers" based on the ultimate borrower identity from Bloomberg following [28]. The final sample comprises an average bond worth USD 2.7 million, issued at a coupon rate of 3.5% with a maturity of nearly ten years. Panel B of Table 1.1 provides more details on the distribution of bond characteristics.

**[Insert Table 1.1 here]**

I obtain information on the type of (issuer) government i.e., state, city, county or other, from the Electronic Municipal Market Access (EMMA) data provided by the Municipal Securities Rulemaking Board. This allows me to distinguish local government bonds from state-level bonds. I use the Municipal Securities Rulemaking Board (MSRB) database on secondary market transactions during 2005-2019. I follow the literature to perform standard data cleaning steps to drop observations as per [5]. Specifically, the main sample focuses on tax-exempt bonds while dropping variable coupon bonds. In matching the bond transactions from secondary market data to their respective issuance characteristics (from FTSE Russell), I rely on the CUSIP as the key identifier. I also use bond-level transactions during 2005-2020 from the Municipal Securities Rulemaking Board (MSRB) database to construct the measure for underpricing using post-issuance trades (see Section 1.4.2).

[9]The CUSIP consists of 9-digits. The first six characters represent the base that identifies the bond issuer. The seventh and eighth characters identify the type of the bond or the issue. The ninth digit is a check digit that is generated automatically.

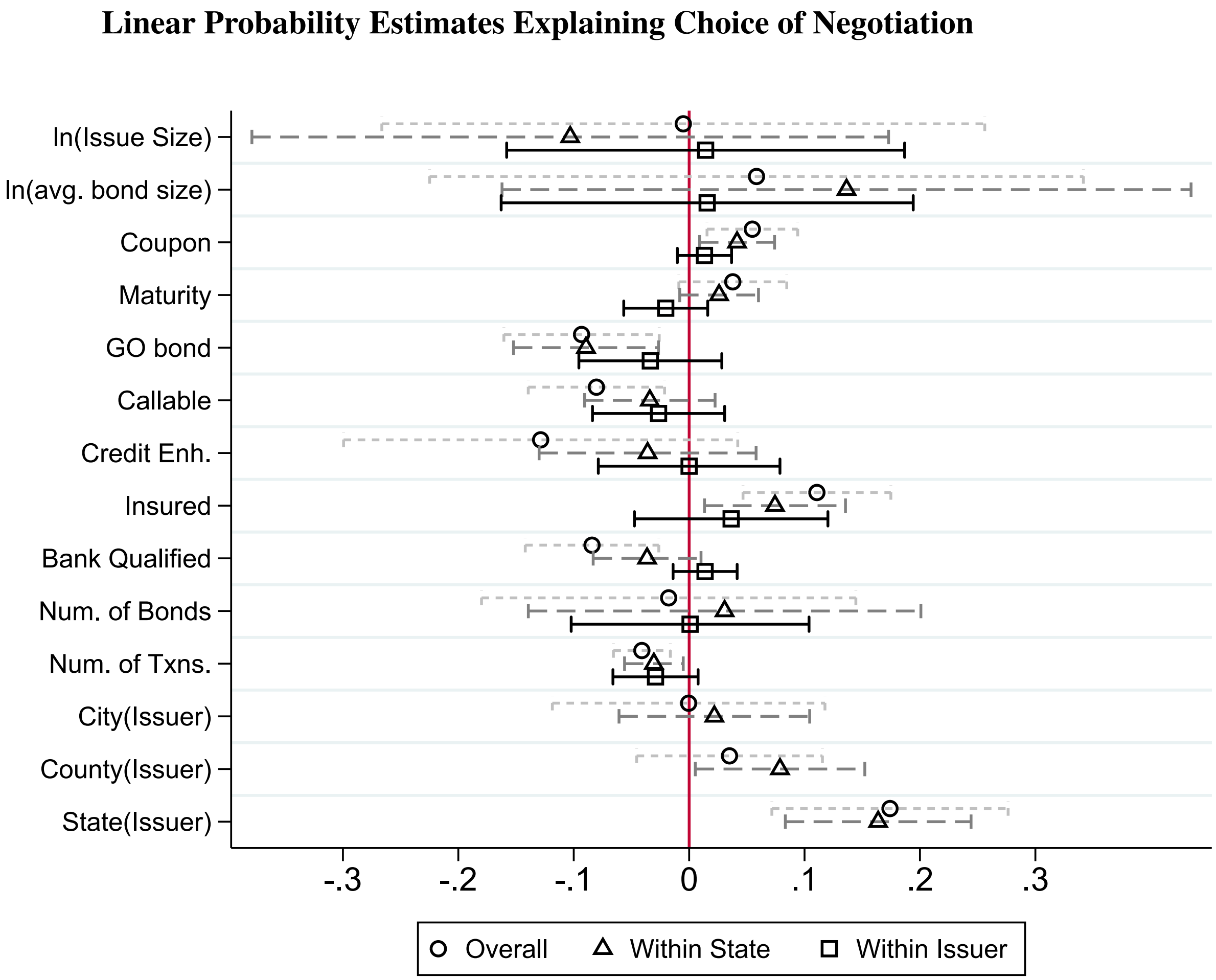

**Figure 1.2:** Linear Probability Estimates Explaining Choice of Negotiation:
The figure shows the point estimates and 95% confidence intervals using Equation (1.2) regressing the choice of negotiation on issuer and bond issue characteristics. The sample focuses on the three years before the SEC Municipal Advisor Rule to capture the ex-ante snapshot. Characteristics with continuous measurements are normalized to standard deviations. The tabular results for this analysis are shown in Table A1.1. *Overall* balance shows the estimates without including any geographic controls linked to the issuer. Next, *Within State* balance corresponds to the estimates after including state fixed effects for the issuer. Finally, *Within Issuer* shows results obtained from including issuer fixed effects.

The primary outcome variable used in Equation (4.1) is the offering yield spread at which a bond was issued. I calculate the bond's coupon-equivalent risk-free yield as in [39][10]. In robustness checks, I also show the results using tax adjusted yields where I follow [5] such that the marginal tax rate impounded in the tax-exempt bond yields is assumed to be the top statutory income tax rate in each state. This is consistent with the broad base of high net worth individuals and households who form a major section of investors in the US municipal bond market (often through mutual funds). A detailed study on tax segmentation across states by [46] shows significant costs on both issuers and investors in the form of higher yields. In Figure 1.3, I provide kernel densities to represent the distribution of the primary market bond features like amount issued, coupon, offering yield and maturity between the treated and control groups. I describe the key variables in Table A1. Importantly, I find that the two groups look similar in the pattern of their distributions. As further validation, I show the quantile-quantile plots between treated and control bonds for these characteristics in Figure A1.1. Most of the observations lie along the 45 degree slope, suggesting similarity between the two groups. I tabulate these characteristics between treated and control bonds in Table A1.2. The average bond in the sample is worth USD 2.7 million, with a maturity of 9.8 years and coupon of 3.5%.

### 1.3.2 Municipal Advisor Fees

The municipal bond market has been historically opaque and only recently been researched by academicians. It is not surprising that I do not find any commercial database that maintains a record of advisory fees charged by municipal advisors at the time of bond issuance. To overcome this hurdle, I hand-collect this information by requesting these data under the

[10]First, I calculate the present value of coupon payments and the face value of a municipal bond using the US treasury yield curve based on zero-coupon yields as given by [48]. Using this price of the coupon-equivalent risk-free bond, the coupon payments, and the face-value payment, I get the risk-free yield to maturity. Finally, the yield spread is calculated as the difference between the municipal bond yield observed in the trades and the risk-free yield to maturity calculated. This yield spread calculation is similar to [49].

**Distribution of Municipal Bond Characteristics (2010-2019)**

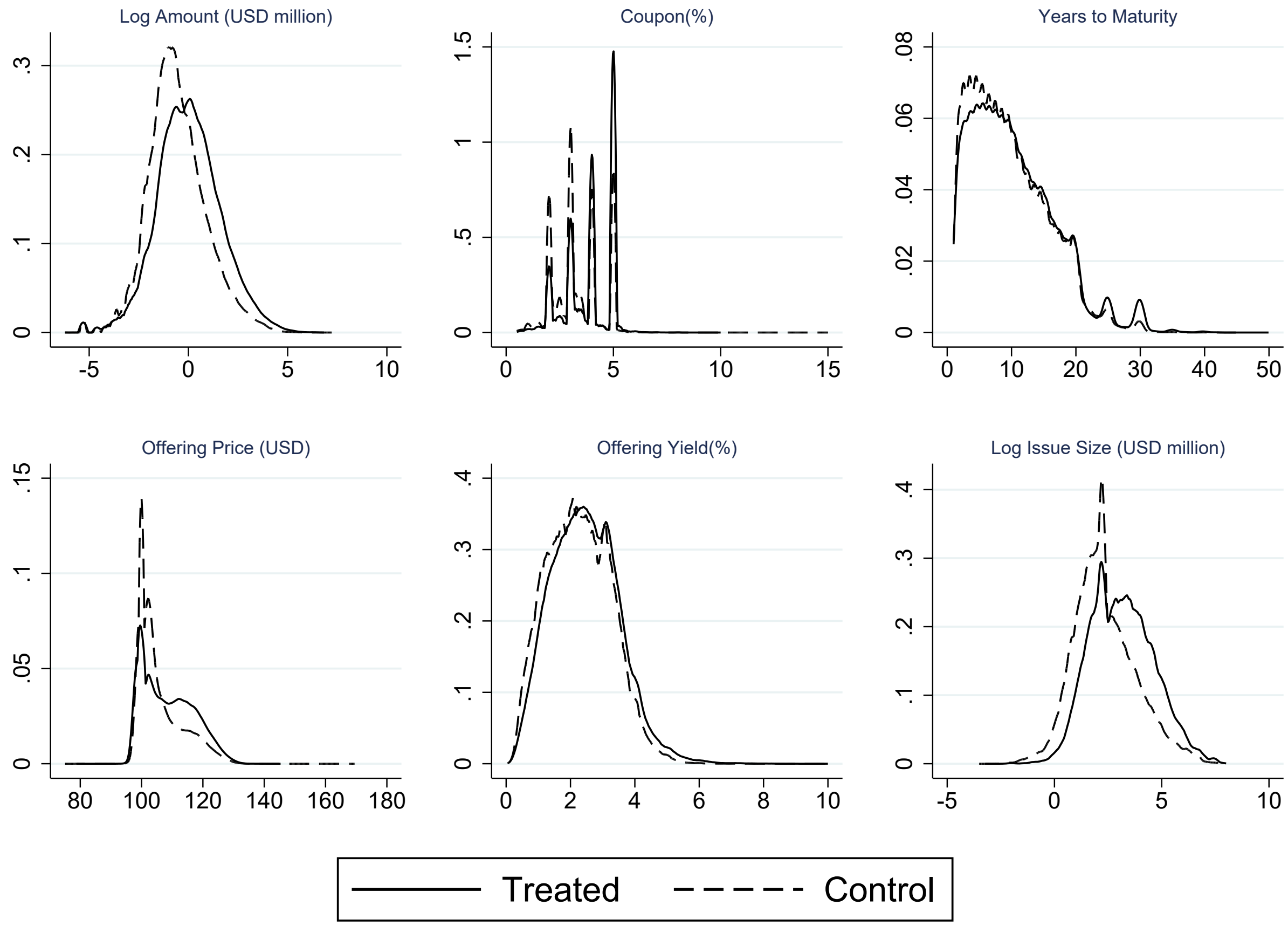

**Figure 1.3:** Municipal Bond Characteristics:
The figure shows the primary market characteristics of bonds issued with advisors at the time of issuance. Bonds sold via negotiated sale consist of the "treated" group, whereas competitively auctioned bonds comprise the "control" set. The sample focuses on fixed rate, tax-exempt bonds issued during 2010-2019. Figure A1.1 shows the quantile-quantile plot for these characteristics between treated and control bonds.

Freedom Of Information Act. States vary substantially in their handling and maintenance of these records. Some states like CA, TX and WA had detailed information on the break up of fees paid to various (financial and legal) agents in each bond issuance with in the state. This would include bonds issued by local governments as well as state agencies and authorities. In comparison, the state of New York was able to furnish information on the aggregate cost of issuance without providing a break-up to identify the fees paid to their municipal advisors. Few other states (like IL and PA) only had information on their state level general obligation bonds. They denied collecting similar information from the local governments within their jurisdiction and guided that the request be made to each issuer separately. The United States municipal bond market has more than 50,000 unique bond issuers because even at the county level, different agencies may be issuing bonds separately. This makes the pursuit of gathering information by requesting each local issuer painstakingly time-consuming and infeasible.

For three states (TN, VA and AR), the state public records agency is only responsible to entertaining queries from state residents only. This restriction is made to economically manage the limited resources available to the agency responding to FOIA requests. However, I was able to circumvent this limitation by directing this request through residents in TN and VA, respectively. Unfortunately, many states denied outright possessing this information. While that was disappointing, it also highlights the need to bring in greater transparency into the cost of intermediation in municipal bonds with respect to advising.

Overall, I was able to gather data from 11 states corresponding to nearly USD 100 billion of municipal issuance each year during 2010-2019. This represents nearly one-fourth of new municipal bond issuance volume in each year during this period. Specifically, I obtain these data from: CA, TX, WA, FL, MD, PA, NM, RI, VT, LA, NY. Figure 1.4 shows the trend in municipal advisor fees for every USD 100 of municipal debt raised. Interestingly, I find little change in average advising fees during the period. This may be partially attributed to regulatory oversight. For example, SEC Commissioner Kara Stein noted in

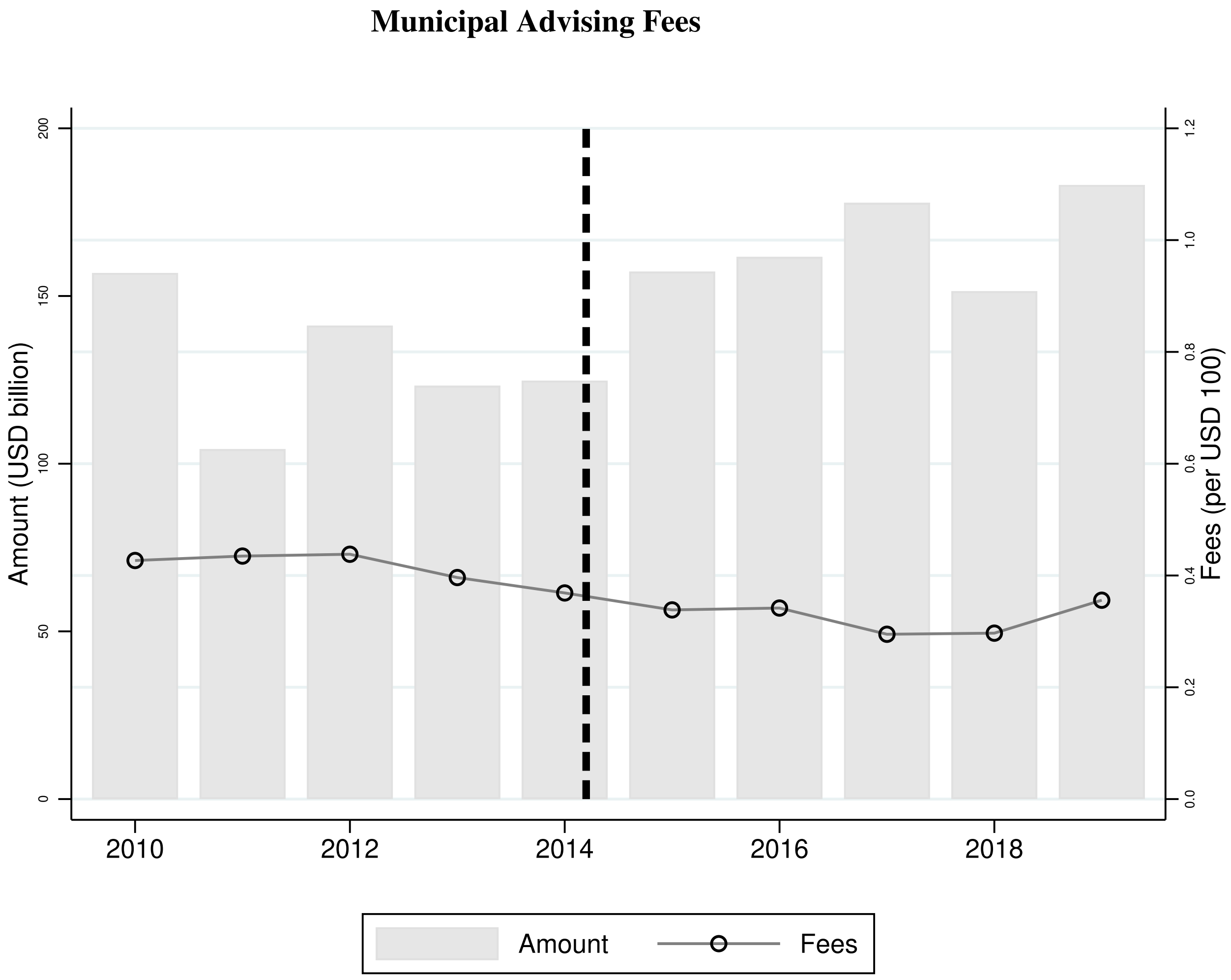

**Figure 1.4:** Municipal Advising Fees:
The figure shows the municipal advising fees paid by issuers alongside the corresponding amount of municipal debt raised. The vertical bars show the aggregated amount of municipal debt issued by state and local governments on the left axis for which I have the fees information during 2010-2019. The connected line depicts the fees paid to municipal advisors for every USD 100 of municipal debt raised on the right axis. These data were obtained under FOIA requests from 11 states: CA, TX, WA, FL, MD, PA, NM, RI, VT, LA, NY. See Section 1.3.2 for details.

September 2013 that issuers may have faced excessive fees by municipal advisors[11]. The SEC sought to address such problems by regulating the market for municipal advisors.

## 1.4 Results

I discuss the baseline results in Section 1.4.1 for Equation (4.1), including evidence from the dynamics using the raw data on municipal yield spreads and evidence on parallel pre-trends assumption. Section 3.4.2 focuses on the mechanism driving the main result. I show the impact on underpricing of new bonds and discuss the role of advisors in selecting underwriters. Section 1.4.3 shows robustness tests for the baseline specification. In Section 1.4.4, I discuss the heterogeneity by the size of issuers. The following analysis pertains to the sophistication of issuers (Section 1.4.5). I examine the exit of municipal advisors in Section 1.4.6. Finally, Section 1.4.7 discusses the implications for refunding bonds.

### 1.4.1 Impact on Offering Yield Spreads of Local Governments

In this Section, I begin by providing a graphical description of the raw yield spreads (Section 1.4.1). Following this visual summary, I provide evidence from a dynamic difference-in-differences regression estimation in Section 3.4.1. Finally, I discuss the baseline result with the full set of controls and fixed effects in Section 1.4.1.

*Raw Relationship in Offering Yield Spreads*

I start the analysis by a simple way of statistically summarizing the observed data: plotting the yield spreads and the corresponding fitted line, before and after the SEC Municipal Advisor Rule. Figure 1.5 shows the binscatter of negotiated/“treated” (circle) and competitive/“control” (diamond) yield spreads. I demarcate the promulgation of the Rule with a dashed vertical line. As shown, the yields tend to follow a downward trajectory until

[11] https://www.sec.gov/news/statement/2013-09-18-open-meeting-statement-kms

2012 before the Rule. Spreads decrease from about 300 basis points to 100 basis points. Thereafter, there is a slight increase in yield spreads leading up to the SEC Rule.

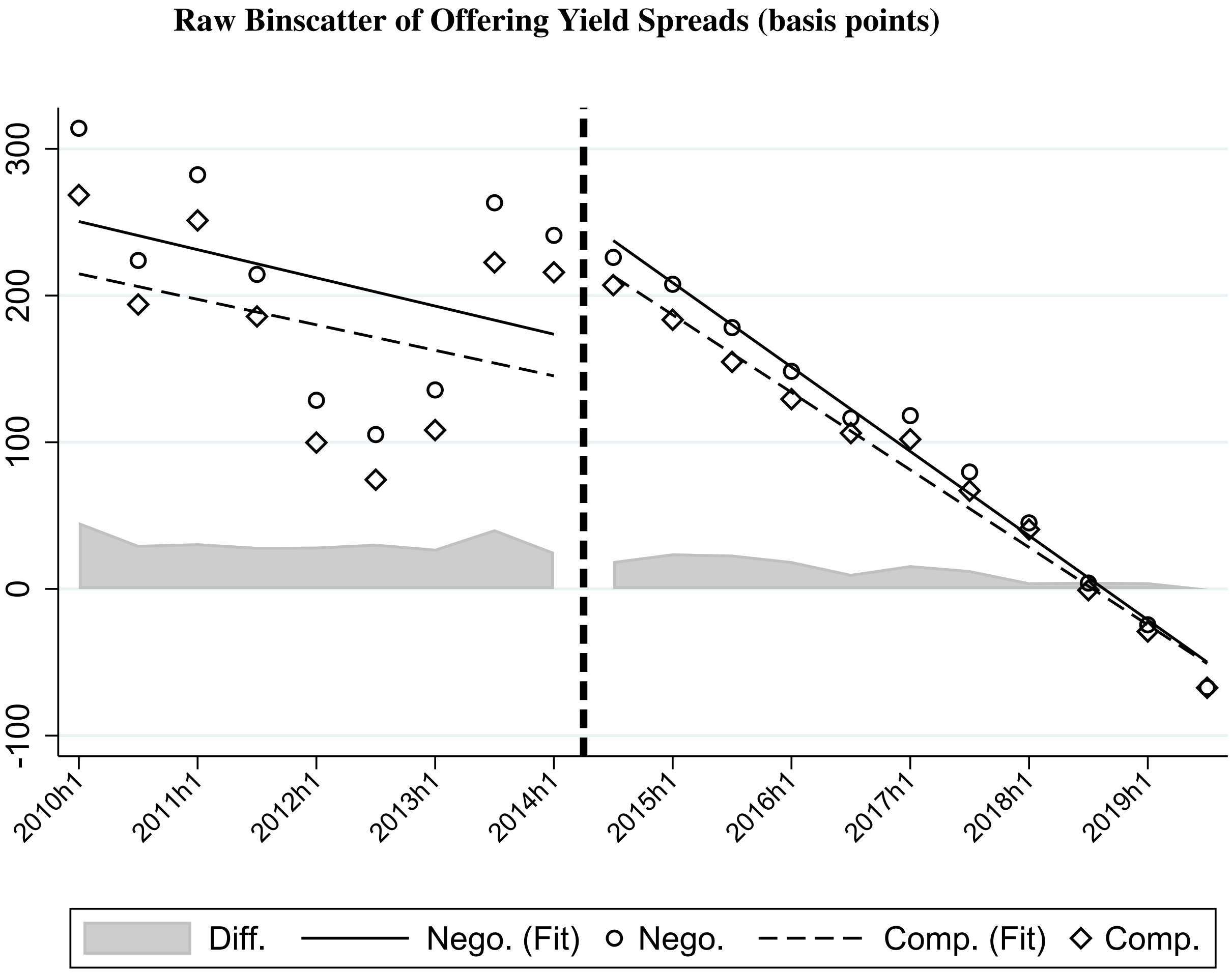

**Figure 1.5:** Binscatter of Offering Yield Spreads:
This figure reports the binscatter of offering yield spreads in basis points on the y-axis. Each dot is the average of the spreads calculated on every $5^{th}$ percentile of the sample for treated and control bonds, separately. The figure also shows the corresponding fitted lines for the negotiated and competitively bid bonds. The difference between the two groups is represented in the shaded portion. The observed pattern suggests a parallel trend until June 2014, followed by a downward trend of spreads in the treated bonds after the SEC Municipal Advisor Rule. This analysis should be interpreted in a non-causal way, as no fixed effects and controls are included.

Importantly, I observe nearly parallel trends in the fitted lines for the treated and control bonds before the Rule. After the Rule in 2014, I observe a downward trend for treated spreads resulting in a convergence of spreads. The figure suggests that the treated bonds depart from their original downward trend right after the SEC Municipal Advisor Rule.

Initially, there seems to be some gap between the treated and control bonds right after the Rule (marked by the dashed vertical line). Negotiated yields tend to be higher than competitive yields; this is consistent with the literature [7, 8, 9, 10, 11]. This difference reduces to nearly zero by the end of 2019. I plot this difference in the shaded area of the plot.

The binscatter plot does not account for unobserved time-unvarying factors pertaining to the adviser and the issuer. This analysis also does not control for calendar year effects. Moreover, a robust analysis of the standard errors and regression outcomes follows in the subsequent section. I also provide similar evidence using the bin-scatter plot of offering yields (unadjusted for spread to treasury yields) in Figure A1.4 in the Internet Appendix. Once again, I find a convergence between the treated and control yields only after the Municipal Advisor Rule, while observing nearly parallel trends between the fitted lines before 2014.

*Dynamics in Difference-in-Differences Design*

I begin the analysis by plotting the average yield spreads observed in the primary market between the treated and control bonds. The event window begins after the financial crisis in 2010 and concludes in 2019, before the Covid-19 pandemic. I show the regression coefficients obtained from the equation below:

$$y_{b,i,a,t} = \alpha + \delta_h * \sum_{h=2011H1}^{h=2019H1} Treated_{b,i,a} * Post_h + \beta_h * \sum_{h=2011H1}^{h=2019H1} Control_{b,i,a} * Post_h + \eta_i + \epsilon_{b,i,a,t} \tag{1.3}$$

where, $\eta_i$ represents issuer fixed effect and each coefficient $\delta_h$ corresponds to the twelve month period ending in June of that year. I estimate these time dummies for the treated and control bonds simultaneously, benchmarked to the twelve month period before the event window shown in Figure 1.6. Representing the yields from bond issuances on a twelve month scale ending June affords two advantages. First, I am clearly able to distinguish between the period before and after the SEC Municipal Advisor Rule, which became effective on July 1, 2014, and is depicted by the bold vertical line in the figure. Second, this frequency of representation is also consistent with the annual fiscal cycle of most local governments.

The coefficients in panel (a) of Figure 1.6 reveals a downward slope for yield spreads, in general. This is not surprising given the monetary policy environment leading to lower yields in the financial markets. Importantly, before the SEC Rule, I find that the treated and control groups tend to follow nearly parallel trends. This lends useful support to the main identification assumption that the treated group may have followed the control group in the absence of the regulation. At least for the period before the regulation, the treated and control groups follow similar trends on yields after I include the issuer fixed effects. After July 2014, I find that the yield spreads for the treated bonds tend to decrease in comparison to the control bonds.

In order to capture the impact on yields after the regulation in 2014, I plot the differences in coefficients over time along with their confidence intervals in panel (b) of Figure 1.6. The difference coefficients in the periods before the Rule (depicted by the bold vertical line) are nearly zero and statistically indistinguishable. On the other hand, I find that the coefficients after the SEC Rule are economically different from zero and statistically significant. Specifically, within the first 12 months after the Municipal Advisor Rule, the negotiated yields decrease by nearly 8 basis points and is statistically significant at the 5% level of significance. By the middle of 2019, this difference amounts to 30 basis points. I find a gradual increase in the magnitude given that municipal issuers are not frequent borrowers. It takes time for the effect to show up as more borrowers raise debt in the primary market after the SEC Rule.

Further, I replicate this approach using the raw offering yield (unadjusted for spreads to treasury) as the dependent variable. I present these findings in Figure A1.5 and find consistent results. Once again, I find evidence in support of the parallel trends assumption

**Yield Spreads**

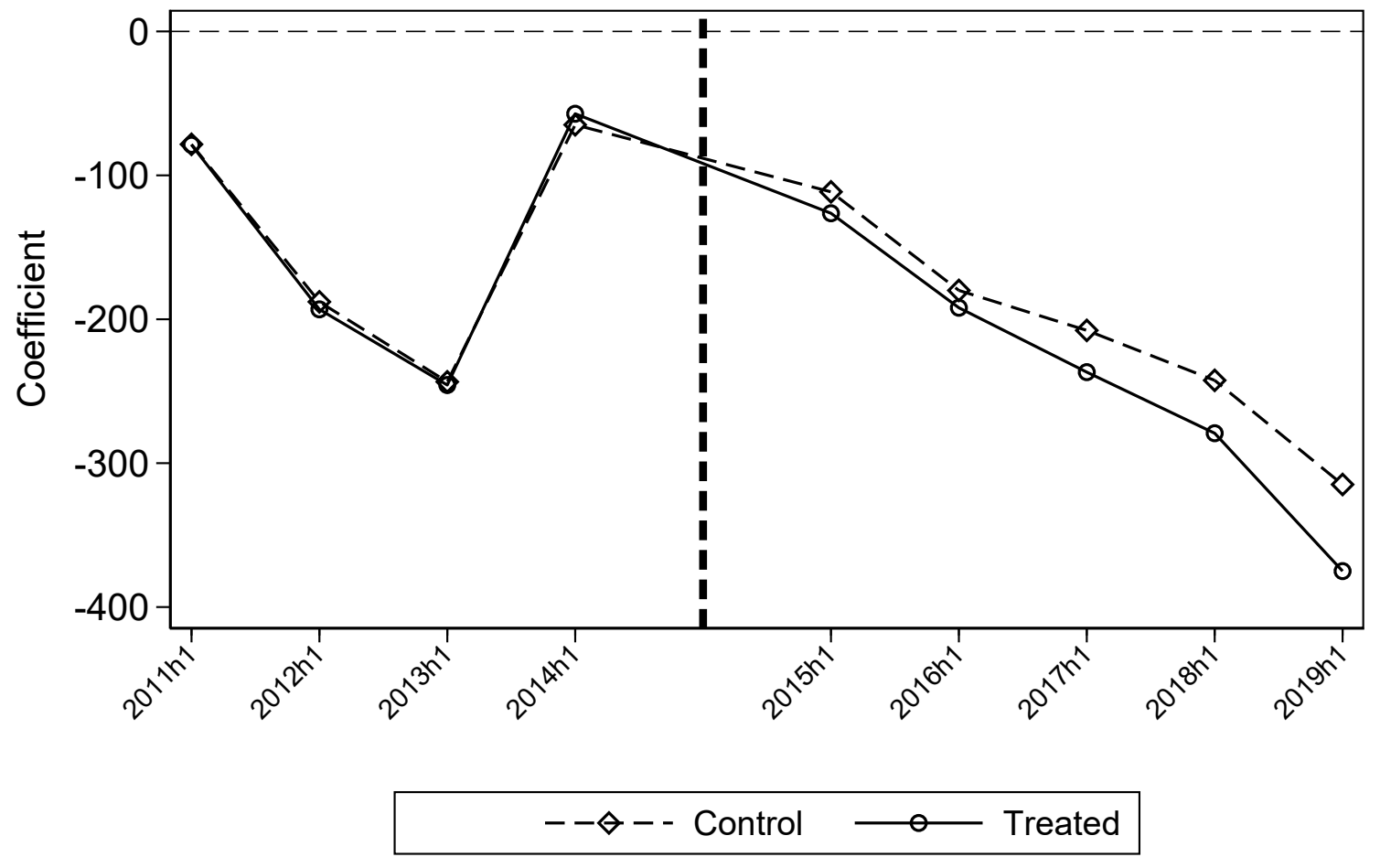

**(a)**

**Difference in Yield Spreads**

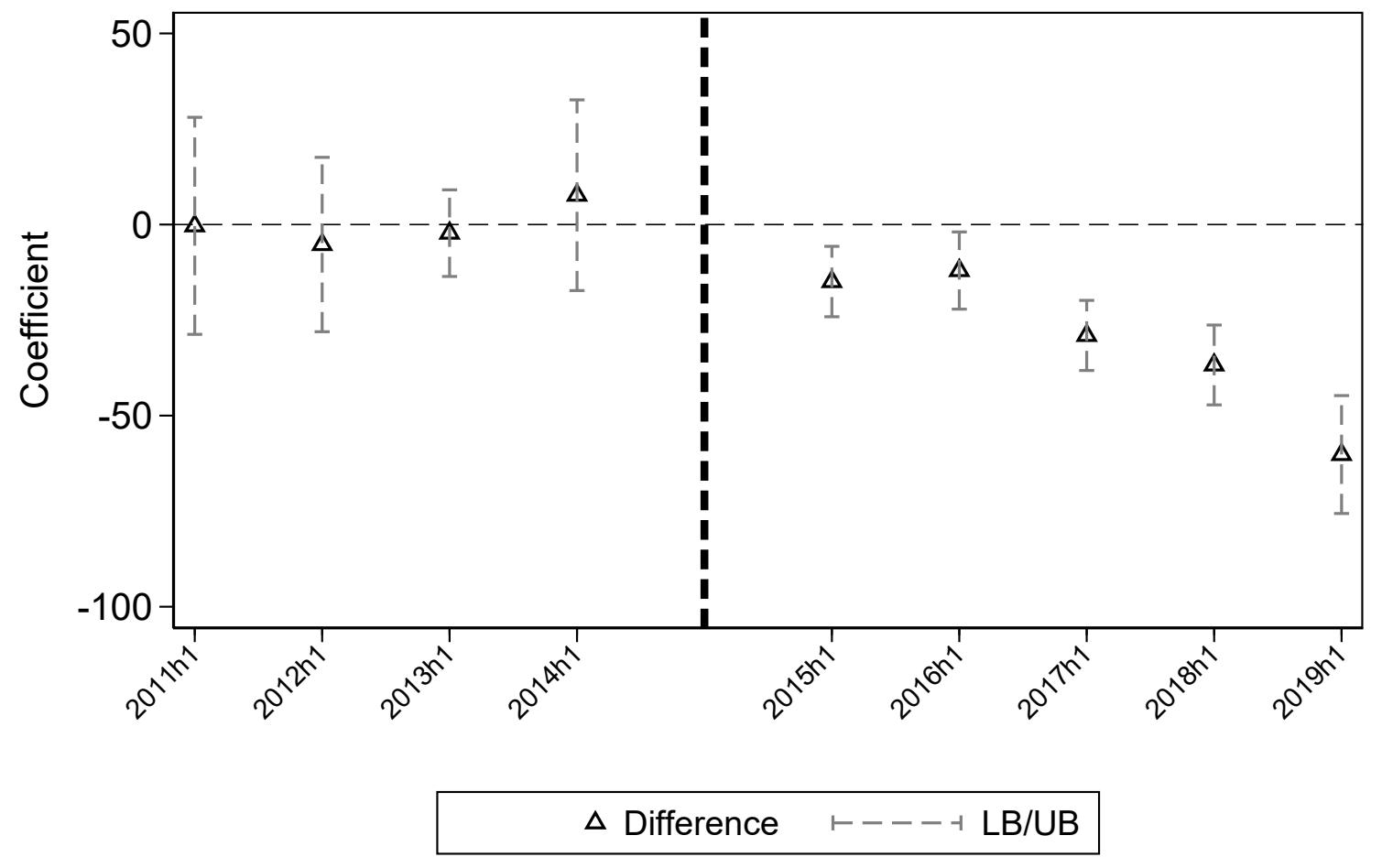

**(b)**

**Figure 1.6:** Baseline Result - Treated vs Control:
In this figure, I plot the average yield spread for municipal bonds issued based on Equation (1.3) in Panel (a). Panel (b) shows the differences between the spreads of treated and control bonds. See Table A1 for variables description. The coefficients are shown in basis points. Specifically, the coefficients are obtained from regressing the spreads on yearly interaction dummies for treated and control bonds using issuer fixed effects. These coefficients are depicted on a yearly scale on the x-axis, where the vertical line corresponds to the Municipal Advisor Rule. The omitted benchmark period is the twelve month period before the event window shown above. Standard errors are clustered by state. The dashed lines represent 95% confidence intervals.

between the treated and control bonds before the SEC Rule. After July 2014, I find a decline in negotiated offering yields when compared to bonds sold via competitive bidding.

*Baseline Difference-in-Differences*

So far I have shown the raw relationship in offering yield spreads suggesting an overall downward trend. I have also shown the nearly parallel trends between the treated and control bonds before the SEC Municipal Advisor Rule using a binned scatterplot (Figure 1.5) and regression within issuers (Figure 1.6). After providing these supporting evidence for the parallel trends assumption in the difference-in-differences design, I now turn to the baseline effect on yields to quantify the magnitude due to the imposition of fiduciary duty.

Table 1.2 reports the main result using Equation (4.1) to quantify the impact of advisors' fiduciary duty on yield spreads. The coefficient of interest ($\beta_0$) represents the interaction term *Treated* $\times$ *Post* corresponding to the difference in differences estimate. I begin the analysis in Column (1) without including any controls or fixed effects. I find that the treated yields decrease by 25.55 basis points. Hereafter, I incrementally introduce additional controls. First, in Column (2), I add the issuer fixed effects to find that the magnitude of $\beta_0$ changes to -26.56 bps. In Column (3), I control for unobserved time-varying factors among issuers based on the state in which they belong. Including this state $\times$ year fixed effect greatly diminishes the magnitude to -12.81 bps, suggesting the importance of these factors in determining yield spreads. Thereafter, including advisor fixed effect (Column (4)) to absorb unobserved heterogeneity among municipal advisors reduces the coefficient marginally (-12.43 bps).

**[Insert Table 1.2 here]**

Municipal bonds may differ along various observable characteristics and so I additionally control for them in Column (5). Specifically, I control for the coupon (%); log(amount issued in $); dummies for callable bonds, bond insurance, general obligation bond, bank

qualification, refunding status, and credit enhancement; credit rating; remaining years to maturity; and inverse years to maturity. I provide the description of key variables in Table A1. In this specification, the overall magnitude is -14.79 bps. In Column (6), I control for observable time-varying factors for the issuer based on the county-level economic conditions. Following [39], I use the lagged values for log(labor force) and unemployment rate, and the percentage change in unemployment rate and labor force, respectively. In this specification, I report a differential effect of -14.74 bps on yield spreads of the treated bonds after the regulation.

It is also important to account for unobserved time-varying heterogeneity among advisors. In Columns (7)-(8), I introduce advisor $\times$ year fixed effect. The baseline specification corresponds to Column (8), where I include county-level controls. I find that the differential effect on treated bonds after the SEC Municipal Advisor Rule (represented by $\beta_0$) amounts to -12.60 basis points. In other words, negotiated bonds are issued at yields that are lower by 12.60 basis points when compared to competitively issued bonds within the same issuer after the SEC Rule. This magnitude accounts for unobserved factors at the issuer level, as well as time-varying factors corresponding to the issuer's state. I also control for observable characteristics at the bond level and time-varying observed local economic conditions for the issuer based on their county. Further, I also absorb unobserved time-varying heterogeneity among advisors. For the average bond in the sample issued at a yield spread of 1.29%, this means a reduction in yields of about 9.7% (=12.60/129). This magnitude is comparable to the cost of switching to negotiated sale (15-17 bps), when they are allowed [11]. This is also close to the effect of Affordable Care Act (ACA) on hospital bonds yield spreads (-13.6 bps) reported in [28]. In Table A1.3 in the Internet Appendix, I show similar results using offering yields as the dependent variable.

To understand the magnitude more closely, I offer a back-of-the-envelope calculation on interest cost savings due to reduced yield spreads [28]. On average, issuers raised negotiated bonds worth USD 71 million during the sample period. With a reduction in yields of 12.6 basis points would amount to lower *annual* interest cost of USD 89,000. This is nearly six times the per pupil expenditure by the average public elementary school in the United States during 2018-19[12]. Given that the average negotiated bond in the sample has a maturity of 10 years, the realized benefit would be higher. An alternative method to quantify the impact relies on the average duration (about eight years) of bonds in the sample. Thus, a reduction of 12.6 bps reduces the cost for an average county by nearly USD 2.1 million[13]. In aggregate, nearly USD 700 billion was raised in municipal bonds via negotiated sale after fiduciary duty was imposed on municipal advisors. Using a similar approach, the expected benefit on the aggregate municipal bond issuance across issuers would be nearly USD 7 billion.

Overall, this section provides evidence supporting the identifying assumption in favor of the parallel pre-trends in yield spreads between the treated and control bonds. I show this effect using univariate distribution of yield spreads as well as from regression analysis. The difference-in-differences specification suggests that the yield spreads for negotiated bonds decreased by 12.60 basis points after the Rule. In the following section, I take a closer look at the pricing of securities using MSRB transactions to explain the main result.

### 1.4.2 Mechanism

In this Section, I first motivate the evidence from offering prices in Section 1.4.2 before showing the evidence on underpricing (Section 1.4.2). The corresponding impact on the liquidity of newly issued negotiated bonds is discussed in Section 1.4.2. Further, I explain my results in light of the role played by advisors in selecting underwriters in Section 1.4.2.

[12]https://nces.ed.gov/fastfacts/display.asp?id=66

[13]I obtain this estimate using a modified duration approach by multiplying the total issuance of $207 million by twelve basis points and then by the average duration of eight years, in a fashion similar to [39].

*Impact on Offering Price*

As further support for the main finding, I now provide evidence from offering prices as the dependent variable. Municipal bonds are usually priced on a face value of USD 100. In the absence of external monitoring, profit-maximizing underwriters may have incentives to price the municipal bonds below the market value. Specifically, underwriter profit increases in yields in the primary market issuance because it pays a lower price to the municipal issuer [3]. They may be able to take advantage of the limited information possessed by issuers with respect to investor demand on specific municipal bonds. The imposition of fiduciary duty on municipal advisors may more closely align the incentives of advisors with issuers. As a result, they may do a better job of monitoring underwriters at the time of issuing municipal bonds.

Another motivation for underwriters to distort the offering price of municipal bonds could come from the prospect of future business. It is easier to sell low-priced securities to clients and they may reward the underwriter with future business [50]. At the same time, lowering the offering price may also enable underwriters to generate higher profits in selling bonds to investors subsequently [51, 52]. In this regard, I evaluate the observed offering price of municipal bonds at the time of issuance. I use the baseline specification corresponding to Equation (4.1) and report the results in Table 1.3.

**[Insert Table 1.3 here]**

In Column (1), I show the results without any controls or fixed effects. The offering price increases by USD 1.94 (per USD 100 of face value of bond). As before, I introduce additional controls and fixed effects to take care of observed bond characteristics and unobserved factors across issuers and advisors. The final specification in Column (8) corresponds to the fully saturated difference-in-differences model. It shows that the offering price for treated bonds increases by USD 1.05 after the imposition of fiduciary duty on municipal advisors. For the median bond issued via negotiation before the regulation in the sample at a price of USD 104, this amounts to about 1% increase. This effect is statistically significant and is obtained after comparing treated and control bonds within the same issuer. The specification also accounts for unobserved time-varying factors for the issuer's state, as well as unobserved factors among advisors over time. Additionally, I control for local economic conditions for the issuer based on the county-level employment and labor-force. With a higher offering price on the face value, issuers may realize greater dollar proceeds from the bond sale.

*Impact on Markup (Underpricing) of New Bonds*

Several papers have examined the underpricing of securities [53, 54, 55, 56, 51, 52] with a long literature focusing on the dramatic underpricing of initial public offerings in the equities market. In the bond market, researchers have evaluated whether the opacity of the market facilitates underpricing by financial intermediaries. [6] shows that enhanced transparency with real-time reporting reduced price dispersion in the municipal bond market, but had little effect on average mark-ups to final investors. This paper evaluates the implications to municipal bond underpricing arising from the imposition of fiduciary duty on municipal advisors.

In a competitively sold bond, underwriters have to commit to the price submitted in the auction. But for negotiated sale, they arrive at the price based on discussions with the issuer. The pricing is not pre-determined and depends on demand from investors. I argue that this scenario affords possibility for price distortion by underwriters. With municipal advisors owing a fiduciary responsibility to their clients after the SEC regulation, issuers may be able to negotiate better with underwriters. To this effect, I begin the analysis based on average markups. The measure for underpricing follows [13] as the trade-weighted difference between the average price paid by customers and the offering price, as a percentage of the offering price. I use bond transactions within the first month of trading for a given a bond.

I follow the baseline specification in Equation (4.1). The hypothesis follows from the prospect of better informed issuers negotiating with underwriters due to the fiduciary duty on advisors after the SEC Municipal Advisor Rule. This might result in reduced underpricing of newly issued securities. The results are shown in Table 1.4. Specifically, Column (8) corresponds to the baseline specification with full set of fixed effects and controls. I find that the trade-weighted markup reduces by nearly 6 basis points for the treated bonds after the SEC Rule. In Column (9), I show similar evidence using an alternative measure by averaging (trade-weighted) the price paid by customers ($P$) over the offering price ($O$), as a percentage of offering price and shown in basis points. Once again, the markup reduces 5.5 basis points. These estimates are statistically significant at the conventional levels of significance. Finally, Column (10) shows the impact on interdealer markups by comparing the average price paid by customers ($P$) to the interdealer price ($V$), scaled by the interdealer price and shown in basis points. This measure shows a reduction of 11.12 basis points in the treated bonds after the regulation.

**[Insert Table 1.4 here]**

*Impact on Price Dispersion*

In the municipal bond market, markups on investor purchases increase with the amount of interdealer trading before the trade [6]. Therefore, lower underpricing (markup) among negotiated bonds may suggest that the bonds pass through fewer dealers before being held by investors. To test this hypothesis, I construct a measure of liquidity from post-issuance trades which is based on [5]. This measure is derived from [57] and captures the dispersion of traded prices around the market "consensus valuation". Bond-level estimates of the price dispersion are obtained by taking the average of daily estimates within the first month of bond trading.

**[Insert Table 1.5 here]**

Table 1.5 shows the results using the main specification in Equation (4.1). As expected, I find that the average price dispersion decreases for negotiated bonds in comparison to competitively bid bonds after the Rule. I show the results by incrementally introducing controls and fixed effects from Column (1) through (8). The coefficient estimates are fairly stable across each of these specifications, and remain statistically significant throughout. Using the baseline specification in Column (8), these results suggest that the price dispersion reduces by about USD 0.03 for treated bonds after the regulation. As an alterntive measure, I also compute these results using the Amihud measure of liquidity. Table A1.4 shows a similar effect of reduced liquidity for the treated bonds.

*Impact Due to Advisor's Ex-ante Role in "Selecting" Underwriter*

Advisors may also be entrusted with choosing an underwriting firm [14] based on their information. This becomes especially relevant for a negotiated sale because competitive bidding eliminates the requirement for pre-selection of the underwriting firm before the bond issuance. Previous literature has showsn that the market for underwriters in municipal bonds is segmented geographically, and may even lead to specialization based on the method of bond sale [58, 59]. This paper investigates the principal-agent problem arising from less-informed issuers engaging more informed advisors. In light of this, the decision to pre-select the underwriting firm in negotiated bonds becomes crucial.

In this context, I present evidence using the heterogeneity among issuers with respect to the role of advisors in choosing underwriters. Ideally, I would want to use the precise information on advisors "selecting" underwriters. In the absence of such data, I construct this measure by identifying whether an underwriter is introduced to the issuer for the first time when a given issuer engages an advisor. In Table 1.6, I analyze the ex-ante heterogeneity among issuers based on the average and weighted average metric. This measure is high for issuers among whom advisors likely played a greater role in introducing new underwriters.

**[Insert Table 1.6 here]**

Columns (3) and (6) correspond to the baseline specification with dummies interacted for the ex-ante classification of issuers. Additionally, this analysis also controls for group $\times$ year fixed effects. I find that the impact of fiduciary duty is driven by issuers where advisors play a greater (above median) role in selecting underwriters. The coefficient is -16.26 bps for the above median group and is statistically significant. Moreover, the difference in coefficients for the two groups is also significant. I find similar results using the weighted average measure in Column (6). The choice of underwriters is crucial with regard to the pricing of the bonds. Therefore, this evidence suggests that the imposition of fiduciary duty on advisors may drive greater yield reduction where advisors play a crucial role. I show similar results using offering yield and offering price as the dependent variables in Table A1.9.

Overall, I provide evidence in this section to explain the main finding of the paper. The reduction in borrowing cost for local governments is simultaneously associated with higher offering price and lower markups on newly issued bonds. These results point to the enhanced role of advisors in negotiated bonds at the time of bond issuance. Further, I show that the main result is driven by issuers among whom advisors likely play a greater role in selecting underwriters. In the following section, I consider several robustness checks to the baseline specification in Section 1.4.1.

### 1.4.3 Robustness

In this section, I test the robustness of the main result in Column (8) of Table 1.2 to various alternative econometric considerations. I present the results of these robustness checks in Table 1.7 ranging across different dependent variables, alternative specifications, tax and bond considerations, geographic considerations as well as alternative clustering of standard errors.

*Other Dependent Variables*

To show robustness of the main result to the choice of dependent variable, I show the results using after-tax yield (Column (1)) and after-tax yield spread (Column (2)). The construction of the tax-adjusted yield spread follows [5]. I describe these variables in Table A1. I find the main coefficient to be statistically significant with magnitudes of -18.13 bps and -19.69 bps, respectively. These higher magnitudes are not surprising because the tax adjustment tends to scale up the yields, as also shown in [28]. I also replicate the baseline table using offering yield as the dependent variable. Specifically, Column (8) in Table A1.3 corresponds to the main specification. I conclude that the main result is not sensitive to the specific choice of dependent variable.

**[Insert Table 1.7 here]**

There may be a concern that since municipal bonds are generally issued in a series [15], their pricing should be evaluated as a package. Ideally, the total interest cost may be a measure of such issue-level bond package. However, the FTSE Russell municipal bond database does not capture this variable. To overcome this limitation, I proceed by taking a weighted average of the yields across bonds in a given series/package. Similarly, I also aggregate the bond level characteristics across bonds in the same package to arrive at issue-level characteristics. Figure A1.6 shows the baseline result in a dynamic regression *within* issuers. I quantify the baseline magnitude using this definition of issue-level yields and issue-level yield spreads in Table A1.5 and Table A1.6, respectively. The preferred specification is shown in Column (8), where the magnitude looks similar to the main analysis and is statistically significant. Therefore, I conclude that the main specification is robust to this alternative consideration of measuring the main dependent variable.

*Alternative Specifications*

In Column (3), I show the baseline results by adding issuer-type $\times$ year fixed effects. By augmenting the model, I add flexible time trends for different types of issuers (city, county, state or other) interacted with year fixed effects. There may be a concern that municipal advisors specialize in the type of local governments they advise that may change around the same time as the SEC Municipal Advisor Rule. I attempt to absorb how such secular trends may unobservably affect different types of issuers. I find that the coefficient of interest is statistically significant and close to the baseline value.

Municipal bonds are issued with the objective of funding or financing specific types of projects/objectives. These could range from activities pertaining to infrastructure (like schools, highways, water-sewer), general purpose and public improvement, or servicing existing utilities like power generation. Each of these sub-markets may experience differences in underlying risks associated with bonds raised for that purpose. As a result, I show results after controlling for the underlying purpose of bonds in Column (4). I also control for such unobserved trends over time by including bond purpose $\times$ year fixed effects in Column (5). I find that the magnitude (-12.19 bps) is nearly similar to the main effect and is statistically significant. Thus, I rule out explanations about the main effect that may be linked to trends in the purpose of bonds which may change around the same time as the imposition of fiduciary duty.

[59] shows that underwriters tend to specialize in the method of sale in the municipal bond market. Moreover, [58] also argues that market concentration has increased in municipal bond underwriting, especially due to negotiated sales. In light of this, I address the possibility that unobserved changes among underwriters may simultaneously affect yields as the SEC Municipal Advisor Rule. First, in Column (6), I include underwriter fixed effects to the main specification to absorb unobserved characteristics across underwriters. I find the interaction coefficient of *Treated* $\times$ *Post* to be -12.92 bps and statistically significant. Further, I also absorb unobserved time-varying trends associated with underwriters

(underwriter $\times$ year fixed effects) and present the results in Column (7). The reported coefficient is -11.07 bps. Overall, I argue that the results are robust to unobserved time-unvarying and time-varying changes related to underwriters.

There may be a concern that issuers and advisors may rely on past relationships during new municipal bond issuances. This may allow advisors to gather more information about issuers with whom they work repeatedly. I account for this unobserved effect by including issuer-advisor pair fixed effect in Column (8), and find that the main result is robust to this consideration. Alternatively, one may argue that a similar alignment between issuers and underwriters may be driving the lower yields. Once again, I address this concern by introducing issuer-underwriter pair fixed effect to the main specification. Column (9) shows that the estimate is similar (-14.20 bps) after accounting for this. overall, I show that unobserved factors driving issuer-advisor and issuer-underwriter relationships do not drive the main result.

Besides the issuer-advisor relationship discussed above, advisors may be specializing over specific states. For example, Panel A Table 1.1 reports that most advisors among the top 15 operate over multiple states. However, over 50 percentile of the bond volume advised by them may come from less than ten states. In light of this, I examine the sensitivity of the baseline result to controlling for unobserved advisor-state pairing. Column (10) shows that the baseline effect is robust to this consideration. Finally, I also show results for a restrictive specification in Column (11), by including county $\times$ year fixed effect. Offering yield spreads may be unobservably driven by local economic conditions at the county level. This accounts for any unobserved time-varying heterogeneity among issuers for the county in which they belong. For state-issued bonds, this corresponds to the state $\times$ year fixed effect in the data. Even with this granular fixed effect, I find that the main coefficient is only slightly lower (-10.65 bps) and remains statistically significant.

*Additional Considerations on Taxability, Bond features and Geography*

An attractive feature of the municipal bond market is that interest income from most bonds is exempt from federal and state taxes [34, 5]. This greatly appeals to investors who fall in the high income tax bracket. [36] discusses the effects of tax distortions in interest income in municipal bond pricing. Given the heterogeneous effects due to tax considerations, I also show robustness of the main results to these aspects in Columns (12)-(14). First, in Column (12), I broaden the sample to bonds for which interest income is taxable under federal law. I find the effect to be -11.93 bps and statistically significant. Next, I drop bonds from states that do not provide income tax exemption for in-state or out-of-state municipal bonds (IL, IA, KS, OK, WI) [28]. Here, I report a baseline estimate of -11.08 bps in Column (13). Finally, I only focus on bonds that are exempt from federal as well as state level taxes in Column (14). The reported coefficient is -11.94 bps and is statistically significant. Thus, I conclude that the results are not sensitive to these tax considerations on interest income from municipal bonds.

For the baseline analysis, I present the results using a wide variety of municipal bonds. These may vary in terms of the purpose (and I show robustness in Section 1.4.3) or type of bond features [60]. These aspects may also drive and govern the advisors and underwriters that are associated with the bonds at the time of sale. For example, bonds that are advised and underwritten by the same financial agent may be unobservably different and may confound the estimates. However, in Column (15) of Table 1.7, I show the results by dropping a small number of such bonds during the sample period and find a similar effect. Another frequent feature in this market involves callability of the bonds. [61] show that municipalities lose money due to delays in refinancing their callable debt. To the extent that yields may vary unobservably differently for callable bonds around the same time as the SEC regulation, the estimates may be confounded. As a result, I show the baseline effect by dropping callable bonds in Column (16) and show that the findings are robust. Owing to similar considerations of unobserved heterogeneity, I also show results by dropping insured bonds (in Column (17)) and keeping only new money bonds (in Column (18)), respectively. I find that the baseline magnitude increases for these sub-samples and remains statistically significant.

Finally, I turn to geographic considerations that may confound with the identification strategy. In this regard, one may worry that there is substantial heterogeneity in respect of the size and type of issuers. While I also address some of these concerns in Section 1.4.3 by including issuer-type $\times$ year fixed effects, I revisit this aspect here. First, local bonds may be different from state level bonds/issuers. As a result, I show the result by keeping only local bonds in Column (19), followed by restricting the sample to state level bonds only in Column (20). While the magnitude reduces marginally to -10.81 bps in the former, I find a greater impact (-17.15 bps) for state level bonds. The greater effect on state bonds is consistent with the higher impact on more sophisticated issuers, discussed in Section 1.4.5. Next, I also address any concerns about the relative proportion of municipal bonds issued by different states. The states with the largest municipal debt issuance are California, New York and Texas. I show the robustness of the analysis by dropping observations from these states in Column (21) to show that the results are not driven by these states alone.

*Alternative Clustering of Standard Errors*

I follow a conservative approach in clustering standard errors by state in the baseline specification. However, one may worry that there might be correlation along other dimensions in municipal bond yields, for example, among advisors. In this regard, I consider alternative levels of clustering standard errors in Columns (22)-(32). First, I think of modifying the cross-sectional dimension of observations. In Column (22), I cluster standard errors by advisor and find the main result holds. The baseline effect is also robust to clustering standard errors by underwriters (Column (23)). Column (24) shows results by clustering standard errors by issuer. I also consider a weaker definition to identify issuers based on first six digits of the CUSIP and report the results in Column (25). The results also hold when I cluster by bond issue (Column (26)).

Next, I look to double cluster standard errors along two dimensions. If standard errors in yields are simultaneously correlated with state and advisor, I present results in Column (27) by double clustering errors. Likewise, I show results for clustering by advisor and issuer (Column (28)). Further, I shows robustness of the results to alternative specification involving double clustered standard errors along the cross-section and over time. Specifically, I double cluster by state and year in Column (29), and by advisor and year in Column (30). Finally, I modify the time dimension of clustering to year-month and report the results based on double clustering by state and year-month in Column (31), followed by advisor and year-month in Column (32). In all these specifications, the baseline result holds at the conventional levels of statistical significance, suggesting that the findings are robust to these alternative considerations of clustering standard errors.

Overall, I perform several robustness checks for the baseline specification and find that the main result is not driven by these alternative considerations/explanations. I show that the effect holds even after additional fixed effects, or stricter requirements involving smaller sub-samples or other dependent variables. Moreover, I find similar results by using alternative approaches to clustering standard errors. These evidence enhance the argument of a causal interpretation of the main result in the paper. In Table A1.7, I replicate these analyses using offering yields as the dependent variable. The results suggest that the main effect continues to be robust and significant across these alternative considerations. Finally, I perform similar evaluation using offering price as the dependent variable and report the results in Table A1.8.

### 1.4.4 Heterogeneity: Likelihood of Advising and Size of Issuers

So far, the evidence suggests that yield spreads reduce for negotiated bonds when compared to competitively bid bonds, after the SEC Rule. Hand-collected data from FOIA requests indicates that the advisory fees paid by issuers on average has not gone up. Given these favorable factors, I now examine whether issuers are more likely to engage advisors. It is reasonable to expect that the prospect of lower yields without incurring higher fees might encourage issuers to engage advisors.

As discussed in Section 3.2, issuers may have unobserved ability to engage advisors. For example, their past relationships with advisors may drive engagement of municipal advisors. To account for such factors, I evaluate the weighted average likelihood of recruiting an advisor *within* issuers, on the extensive margin. Figure 1.7a shows the results from a linear regression of the average likelihood over annual dummies with issuer fixed effect. I find that there is an increase in the likelihood of engaging an advisor after the SEC Municipal Advisor Rule. On average, there is an increase of over 5% in the likelihood of issuers to engage advisors. The omitted benchmark period corresponds to the half year at the start of the event window in 2010.

In the municipal bond market, [15] find that issuers may accept NPV losses for short term cash flow savings. While issuers seem to secure lower yields on average without paying higher advising fees, there may still be variation among issuers. To understand more, I examine the likelihood of engaging advisors based on the size of issuers. I present these results in Figure 1.7b by grouping issuers into small versus large, based on the size of their ex-ante municipal issuances. Surprisingly, the overall effect among all issuers is likely driven by large issuers only. In comparison, small issuers exhibit almost no change in their likelihood to engage advisors. I shed more light into this heterogeneity in subsequent analysis.

The choice of engaging an advisor for a municipal bond issuance is endogenous to issuers. Issuers who are more likely to benefit from advised bonds may also be more likely to engage an advisor on the extensive margin. I examine this by revisiting the baseline analysis on advised bonds by grouping issuers into small versus large issuers. First, in Panel A of Table 1.8, I show the heterogeneity with offering yield spread as the dependent variable. Using the ex-ante average and median size of issuance, I find that large (above

**All Issuers**

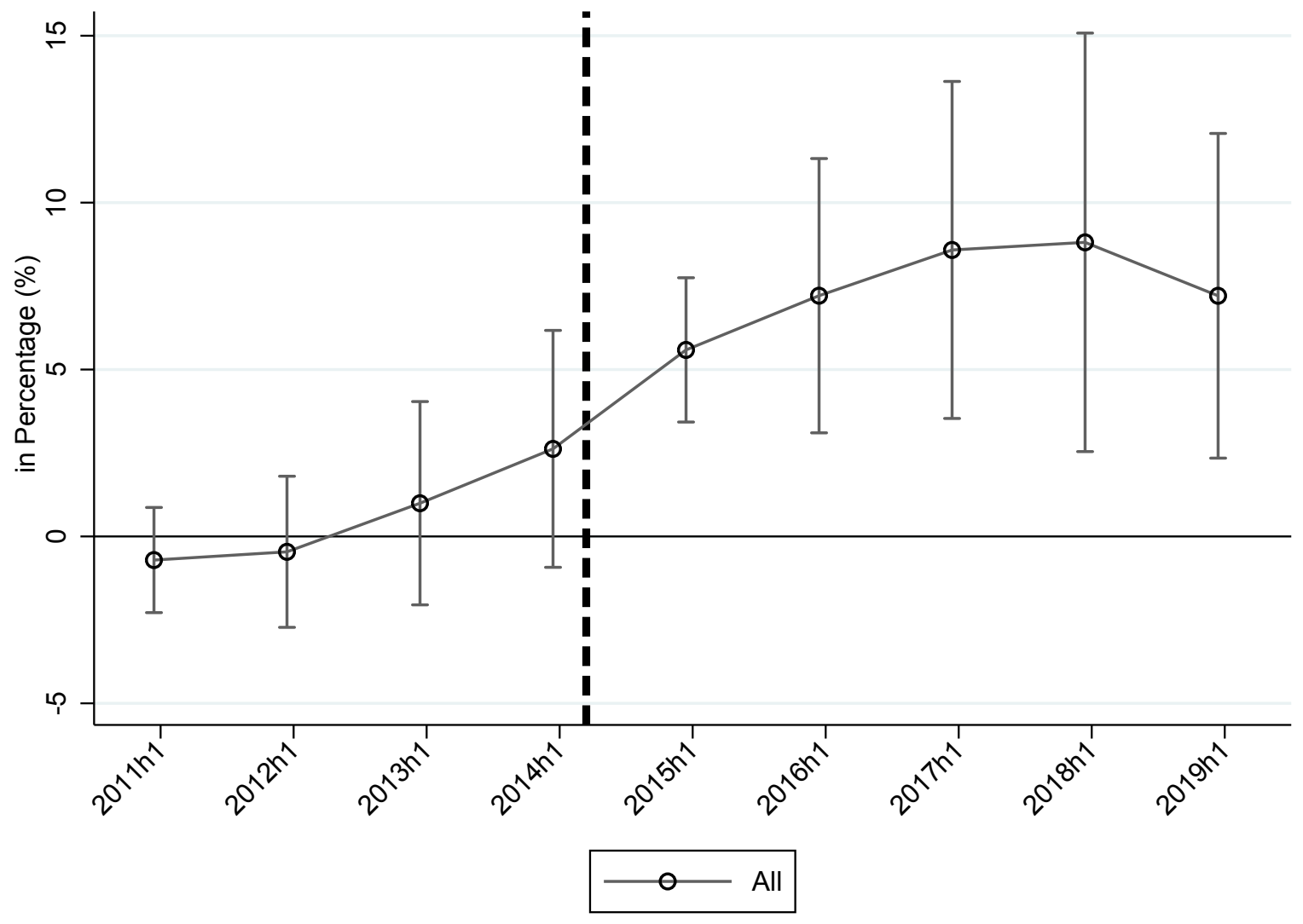

**(a)**

**Large vs Small Issuers**

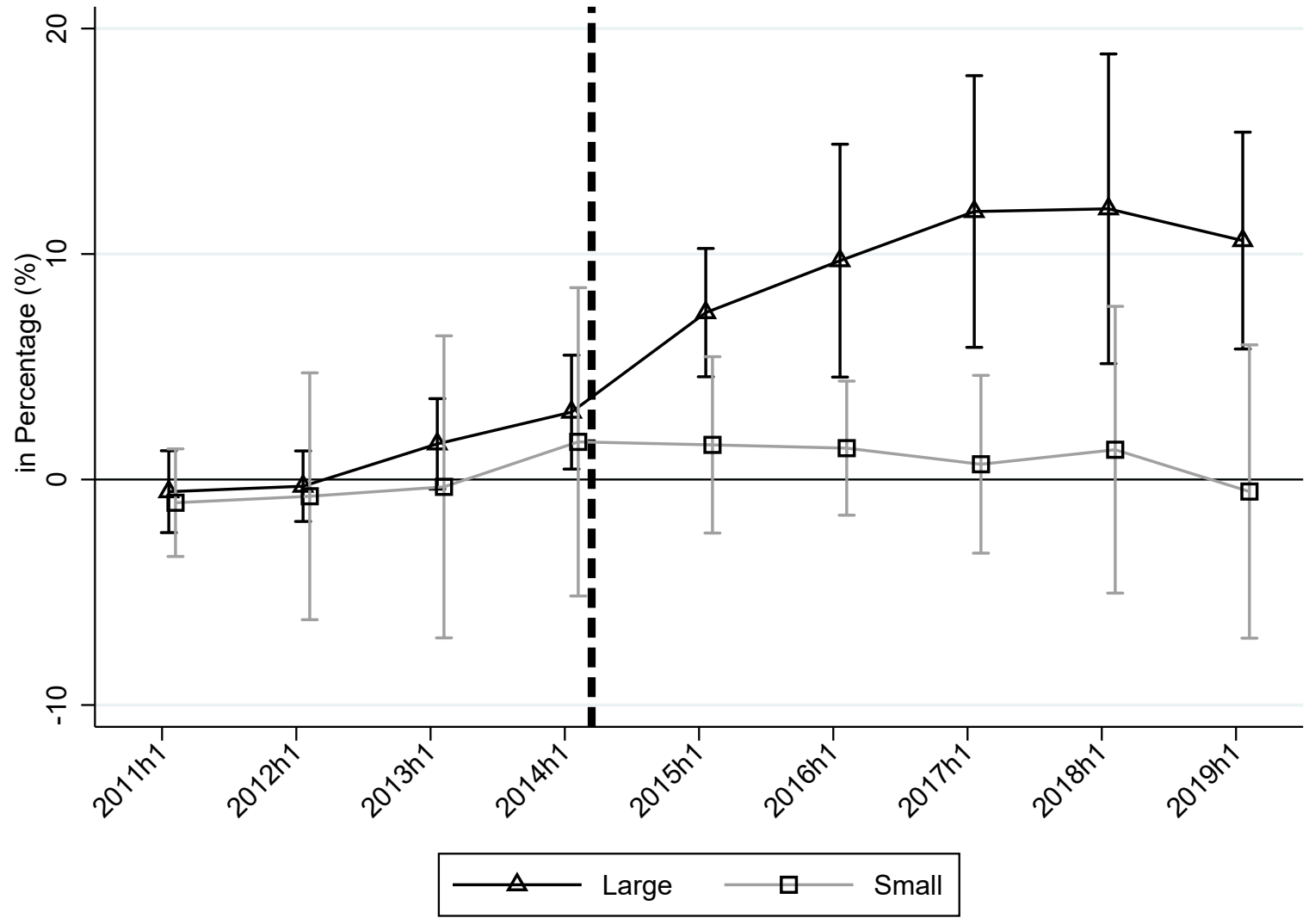

**(b)**

**Figure 1.7:** Likelihood of engaging advisor:
This figure reports the coefficients showing the weighted average likelihood of issuing advised bonds on the y-axis, *within* issuer. The vertical line corresponds to the SEC Municipal Advisor Rule. I represent these coefficients across all issuers in Panel (a), as well as the subsets of large versus small issuers in Panel (b). Standard errors are clustered at the state level. The dashed lines represent 95% confidence intervals.

median) issuers experience a greater reduction in yields for their treated bonds. Specifically, Columns (3) and (6) correspond to the baseline specification with full set of controls and fixed effects. Additionally, I control for group $\times$ year fixed effect to account for unobserved time-varying heterogeneity among small and large issuers. The impact on yield spreads is greater for large issuers by 12.67 bps (Column (3)) and 13.95 bps (Column (6)), respectively. These results suggest that conditional on engaging an advisor, large issuers benefit more from the reduction in yield spreads.

**[Insert Table 1.8 here]**

Following up on the analysis using offering yield spreads, I show results for the corresponding effect on offering price. In Panel B of Table 1.8, I find that the impact on offering prices at the time of issuance is higher for large issuers. The differences in Columns (3) and (6) suggest that the baseline effect of USD 1.05 is almost entirely driven by large issuers. Next, I examine the difference between large and small issuers in the context of underpricing. Taken together with the evidence in Section 1.4.2, large issuers should experience a greater impact on trade-weighted markups. Indeed the results in Panel C of Table 1.8 suggest that the overall effect on reduced underpricing is driven by large issuers. As before, I also control for group $\times$ year fixed effect in this analysis. The coefficient for small issuers is small and statistically insignificant. Large issuers show statistically significant reduction in underpricing by over 7 bps in each of the specifications. In Table A1.10, I show similar heterogeneity among issuers using offering yields as the dependent variable.

Finally, I extend these findings to learn whether lower yields drive changes in municipal bond issuance after the SEC Municipal Advisor Rule. [25] show that municipalities' financial constraints may impact the issuance of bonds. Similarly, [43] show that less wealthy school districts have difficulty obtaining municipal bond market funding. I classify issuers with greater (above median) reliance on negotiated bonds before the regulation as the "treated" group. The control group comprises issuers with below median reliance. Figure 1.9a shows the overall results. Compared to the one year before the Rule, treated issuers raise more municipal debt than control issuers. This is reasonable given that the ex-post reduction in yields may increase the debt capacity for these issuers. Once again, I explore this overall result based on the heterogeneity on the size of issuers. As shown in Figure 1.9b, this increased issuance is driven by large issuers who experience greater yield reduction. Meanwhile, there is weak evidence for decrease in the amount of bonds raised by small issuers.

Overall, the evidence in this Section shows that there is considerable difference in how issuers are affected due to the SEC Rule. I capture this heterogeneity based on the size of issuers. The results show that issuers who benefit from greater reduction in yields after the imposition of fiduciary duty under the SEC Municipal Advisor Rule are able to raise more municipal debt. Based on this heterogeneity, I also provide supporting evidence from the underpricing of bonds.

### 1.4.5 Heterogeneity: Sophistication of Issuers and Newly Advised Issuers

In this Section, I present more evidence to understand why the overall effect of fiduciary duty following the SEC Rule may vary among issuers. First, I show heterogeneity among issuers based on their level of ex-ante sophistication. Second, I consider another dimension of extensive margin for issuers that transition into advised bonds.

*Heterogeneity by Sophistication of Issuers*

Municipalities that rely more heavily on complex bond features may arguably represent a higher level of sophistication. I begin the analysis by identifying issuers that issued more complex bonds, ex-ante. I measure complexity of bonds following [33, 60] by aggregating over six bond features: callable bonds, sinking fund provision, special redemption/extraordinary call features, nonstandard interest payment frequency, nonstandard interest accrual basis, and credit enhancement. Based on the weighted average of ex-ante complexity among bonds for an issuer, I classify them into groups based on the median.

Thus, issuers with high complexity (above median) are likely to represent sophisticated issuers. I present the analysis by interacting the baseline Equation (4.1) with dummies corresponding to complexity in Column (1) of Table 1.9.

**[Insert Table 1.9 here]**

The results show that the differential impact on yield spreads of issuers with above median complexity is 14.68 bps higher than those below median. This difference is statistically significant and economically meaningful. As before, it accounts for the average effect among low versus high complexity issuers by including group $\times$ year fixed effect. Similarly, Column (2) shows the results by focusing on weighted average complexity of advised bonds only to classify issuers. I find similar results as in Column (1), suggesting higher effect on yield spreads of issuers with more complex bonds.

In Columns (3)-(5), I draw upon additional measures to quantify the ex-ante level of sophistication among issuers. First, in Column (3), I use the fraction of bonds with credit enhancement to represent the heterogeneity among issuers. Municipal issuers who are able to purchase more credit enhancement (usually include letters of credit and guarantees) are likely more sophisticated. Consistent with this, I find that yield spreads decrease by 15.76 bps for issuers with above median levels of credit enhancement. The difference between high and low groups of issuers is statistically significant, and similar to the previous case in Column (2). In Column (4), I use the ex-ante average wage paid to the finance staff of local governments. This data is obtained from the US Census Bureau's Annual Survey of of Public Employment & Payroll (ASPEP) for local governments. The results show that issuers with higher levels of wages for the finance staff benefit more in terms of yield reduction. Finally, Column (5) shows this result by using the ex-ante reliance on advised bonds. Issuers that are less reliant on advisors experience a greater reduction in yield spreads (-17.02 bps). Using different metrics, I show that the main result is driven by issuers who are likely more sophisticated.

*Evidence from Newly Advised Issuers*

In Section 1.4.4, I show results suggesting that issuers are more likely to engage advisors on the extensive margin, after the SEC Municipal Advisor Rule. Here, I evaluate how the imposition of fiduciary duty on municipal advisors may encourage issuers who previously did not (or could not) rely on advisors to engage them in raising future debt. Alternatively, the regulatory push may enhance the quality of advice offered while simultaneously raising the fees charged by municipal advisors. This may discourage such issuers from recruiting municipal advisors after the Rule. However, the evidence from the average fees paid to advisors (Figure 1.4) shows little support for this alternative case.

Drawing upon the evidence in Figure 1.7a, I examine "newly advised" issuers. These issuers engage advisers for the first time only after the SEC Rule. Table 1.10 shows the results using the baseline specification in Equation (4.1). In Columns (1)-(3), I show results using the yield spread as the dependent variable. For Columns (4)-(6), I use the offering yields as the dependent variable. Further, I distinguish between "complete" and "partial" transition of issuers. The former case applies to issuers who never issued bonds with advisors before the Rule but always do so afterward. On the other hand, issuers are classified as "partial" transitioned if they never issue bonds with advisors before but recruit advisors for some of the bonds ex-post. As a result, I do not use advisor $\times$ year fixed effect in this analysis. I do not find any significant effect on the treated bonds compared to the control group after the SEC Municipal Advisor Rule. Even though the sample size reduces substantially in this analysis, the absence of meaningful statistical significance and economic magnitude suggests that newly advised issuers do not experience a reduction in yield spreads.

**[Insert Table 1.10 here]**

Taken together, the evidence in this Section suggests that the baseline effect of reduced yields in negotiated bonds is driven by sophisticated issuers. Further, municipal issuers that recruit advisors for the first time after the imposition of fiduciary duty do not exhibit lower yields. This evidence supports the earlier finding showing that large issuers drive the main result because they may be more sophisticated.

### 1.4.6 Exit of Municipal Advising Firms

As discussed in Section 1.3.1, [2] suggest that some municipal advisors exit from the market after the SEC Municipal Advisor Rule. As in [4], this may be due to the additional cost of compliance with the new regulatory requirements, increased paperwork, increased overhead time required to deal with the regulation, or increased effort dedicated to oversight ex-post against the potential threat of being sued. I focus on the municipal advisors who advise on bonds issued during the sample period. In Figure 1.8, I depict the number of regular advisors operating in the municipal bond market. This corresponds to advisors in the sample who advise on at least one issuance in each calendar year until June 2014. Conditional on advising municipal bonds in the sample, the number of regular advisors decreases from 214 in 2010 to 136 in 2019. This implies that 78 municipal advising firms progressively exited the advisory market after the SEC Municipal Advisor Rule. Importantly, these municipal advisors worked on at least one issuance in each year before the regulation. Figure 1.8 also shows the share of municipal bonds advised by these regular advisors on the right-hand axis. Over 90% of the municipal bonds were advised by regular advisors before the regulation. After the MA Rule, their share declined to just over 80% in 2019. This suggests that focusing on these regular advisors with at least one issuance in the pre-period is representative of the overall market.

**[Insert Table 1.11 here]**

In this context, I analyze the impact of advisors exiting the market on municipal bond yield spreads in Table 1.11. In Column (1), I show results using Equation (4.1) interacted with dummies corresponding to whether the issuer primarily depended on an exiting advisor or not. I define issuers linked to advisors when more than 50% of their municipal

## Exit of Municipal Advising Firms

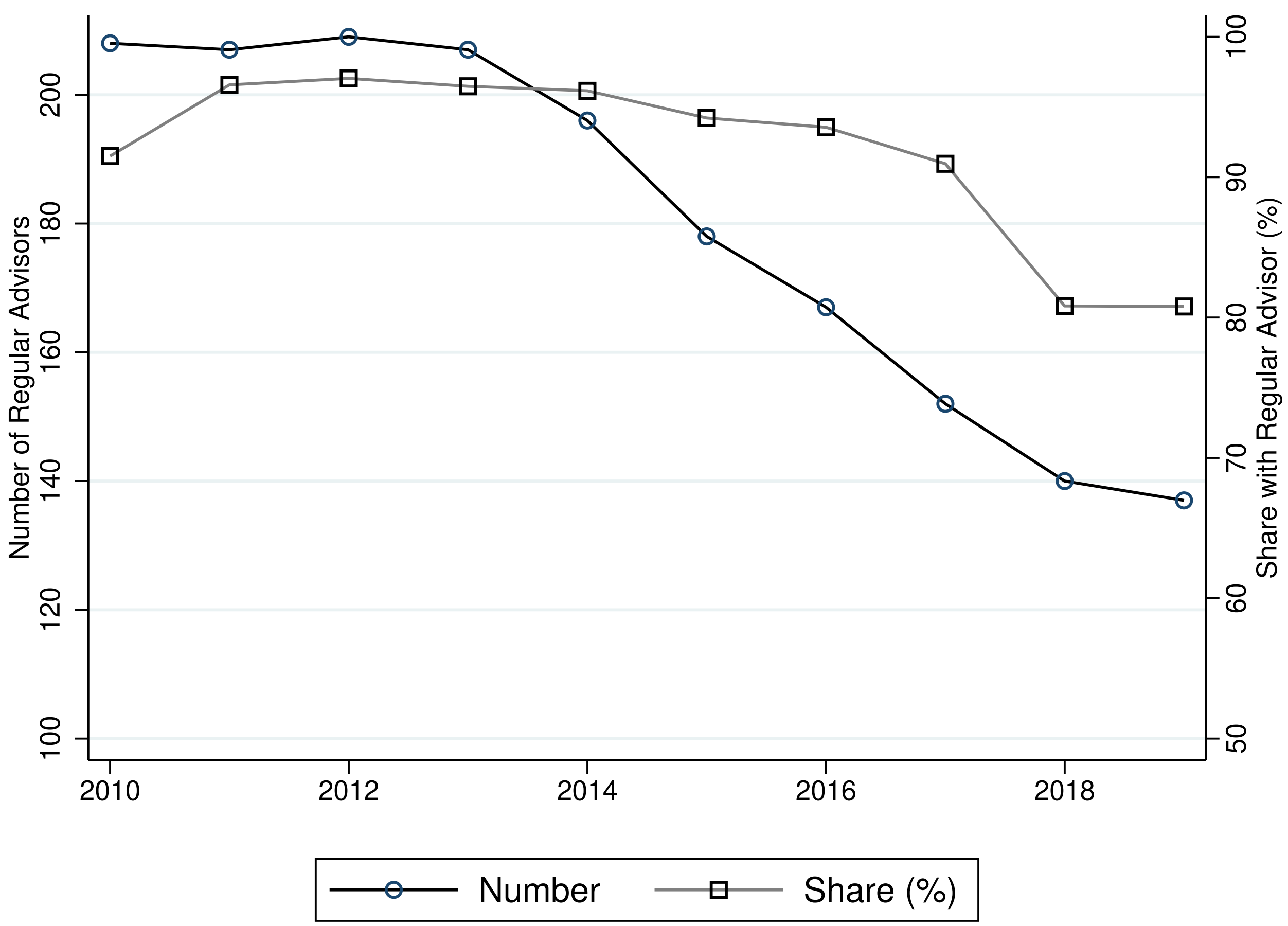

**Figure 1.8:** Number of Regular Advisors and Market Share:
The figure shows the number of municipal advising firms on the left axis that regularly advise on municipal bond issuance before the SEC Municipal Advisor Rule. This corresponds to advisors in the sample who advise on at least one issuance in each calendar year until June 2014. On the right-hand axis, I plot the market share of these regular advisors during the sample period of 2010-2019. This represents the proportion of municipal debt advised by these advisors to the total municipal debt issued during the year, expressed as a percentage.

debt issuance in the pre-period is advised by the exiting advisor. Alternatively, results are robust to using a lower threshold of 25% of municipal debt issuance by the exiting advisor, as shown in Table A1.11. The results suggest that the reduction in offering yield spreads (-16.61 bps) is driven by issuers that do not depend on an exiting advisor. This magnitude is slightly higher than the baseline effect reported in Table 1.2. Meanwhile, there is no significant impact on issuers that depend on exiting advisors. This is consistent with the analysis in Section 1.4.5 showing that the main effect is stronger for sophisticated issuers.

However, the muted effect on issuers dependent on exiting advisors masks the heterogeneity between small and large issuers. I present these results in Columns (2) and (3), by focusing on the sub-sample of issuers that depend on an exiting advisor, ex-ante. Column (2) suggests that the yield spreads increase by 15.89 bps for small issuers, whereas the large issuers continue to experience a reduction in yield spreads (8.59 bps) on their negotiated bonds after the MA Rule. The difference between the two groups is economically and statistically significant. It is possible that some issuers that depend on exiting advisors may raise debt without municipal advisors after the regulation. I account for this in Column (3) and show results after including non-advised bonds in the sample. The results look similar, suggesting an increase in yield spreads for small issuers that were ex-ante dependent on an exiting advisor. As before, I use the median size of issuance in the ex-ante period to distinguish small versus large issuers. Additionally, these results are robust to using the average size of ex-ante issuances (Table A1.12). Overall, the evidence in this section suggests that the regulatory burden may be associated with the exit of some municipal advisors.This may increase the borrowing cost of small issuers that were dependent on these exiting advisors.

#### 1.4.7 Evidence from Municipal Bond Refundings

Since the late 1990s, there has been an increasing trend of municipal bond issuance associated with refunding transactions [15]. More specifically, advance refundings allow municipalities to retire an existing callable bond to before its call date. Refunding bonds are

**New Municipal Bond Issuance**

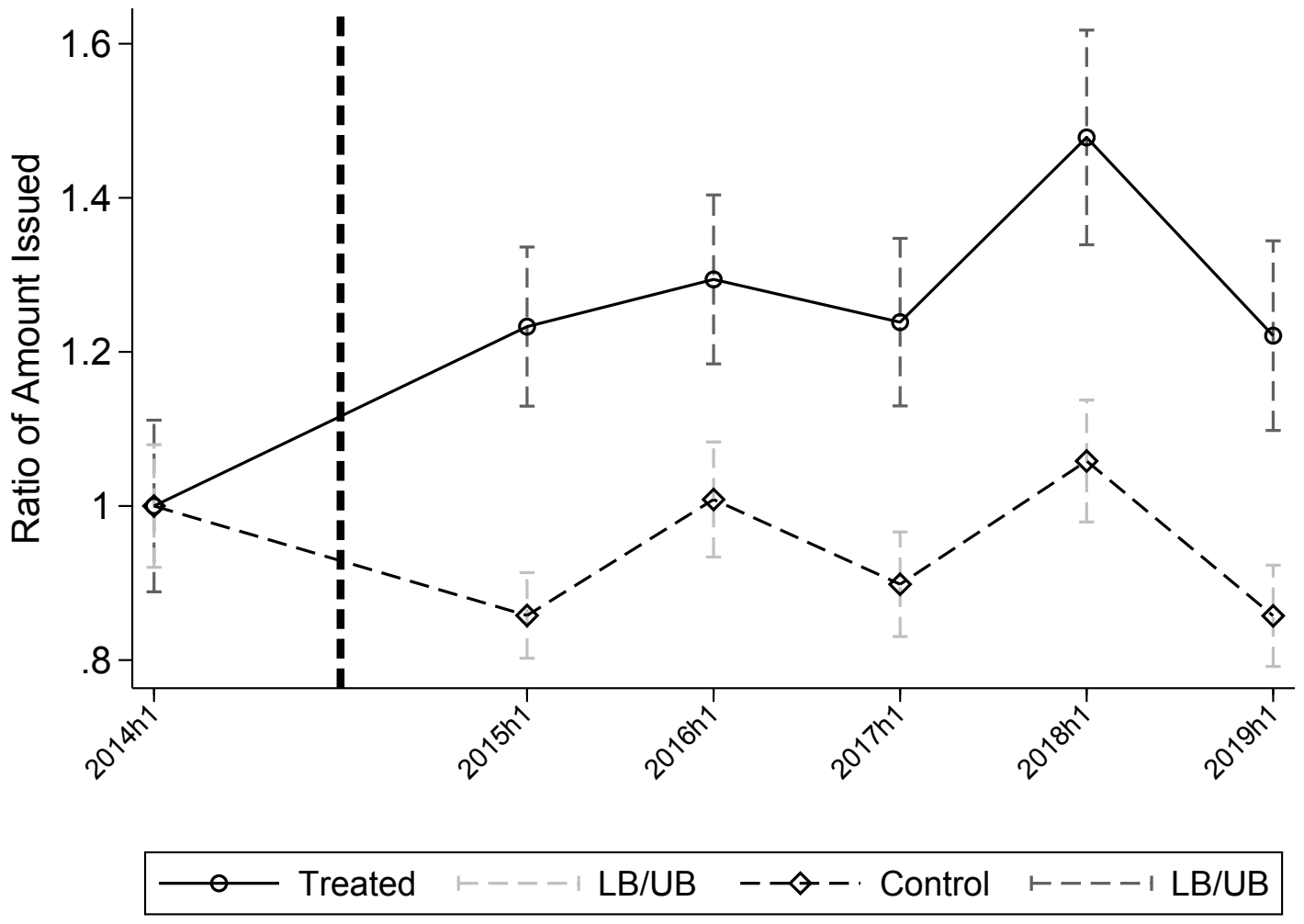

**(a)**

**New Municipal Bond Issuance by Size of Issuers**

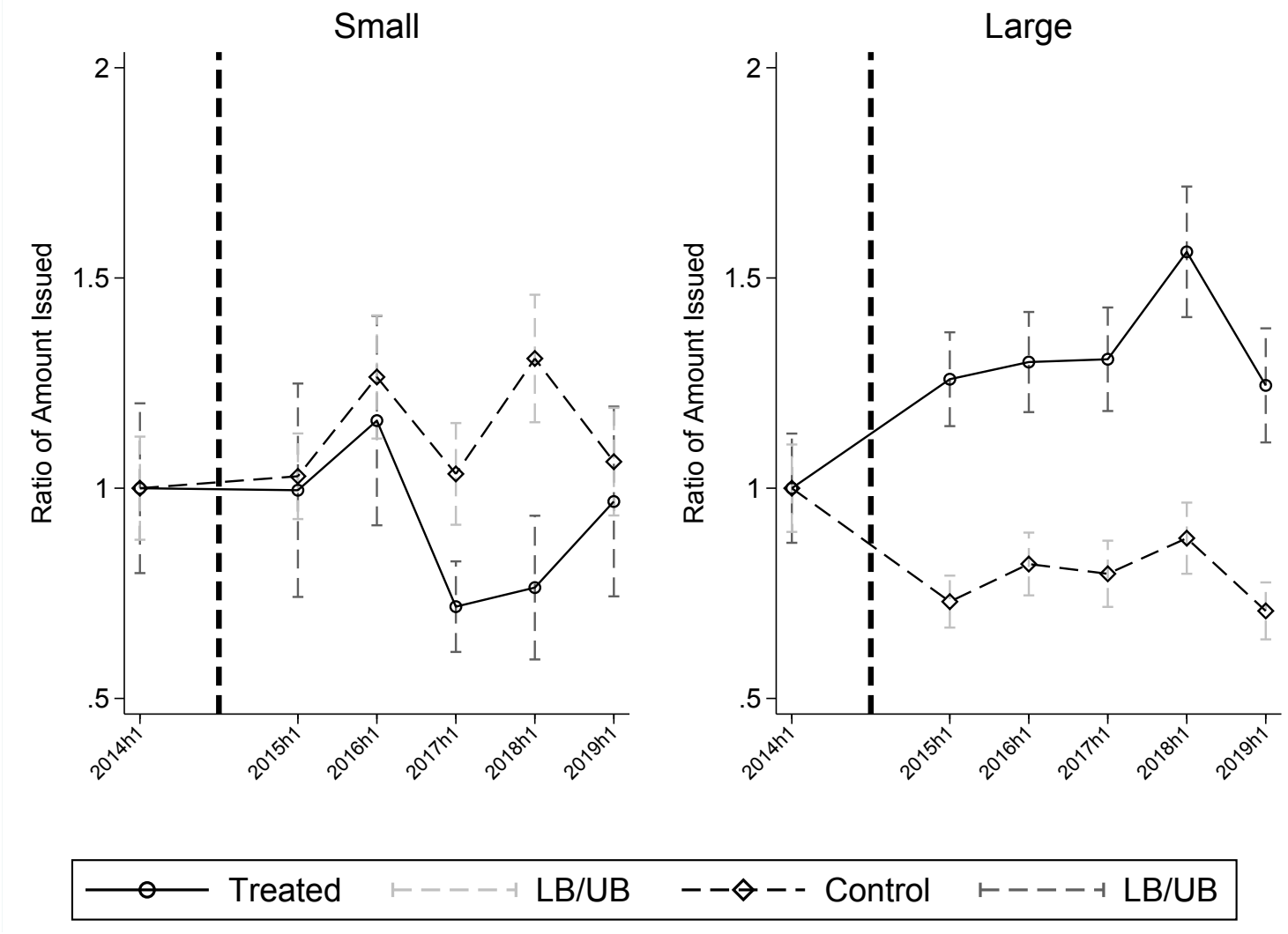

**(b)**

**Figure 1.9:** New Municipal Bond Issuance:

In this figure, I plot the amount of municipal debt issued with advisors in treated versus control issuers. The benchmark period is during the twelve months before the SEC Municipal Advisor Rule. Panel (a) shows results for all issuers. Panel (b) depicts the sub-samples corresponding to small versus large issuers. Standard errors are clustered by state. The dashed lines show the upper and lower limits based on the standard errors of the mean values.

used to refinance existing municipal debt. Previous evidence in the literature points toward compromised option value [62], and the timing of advance refundings [63]. In their analysis using an extensive data set, [15] find that a substantial number of advance refundings occur at a net present value loss. They explain their results based on financial constraints of issuers. Given that refundings constitute a large portion of debt in the municipal bond market, I shed light on these issuances in light of fiduciary duty on advisors.

I begin the analysis by first restricting the sample to bonds issued for the purpose of refundings. This allows me to isolate the impact of fiduciary duty by replicating the baseline Equation (4.1) for a relatively homogeneous group of bonds. Figure 1.10 shows the dynamic coefficients obtained from Equation (1.3) for the subset of refunding bonds. In Panel (a), I plot the coefficients on yield spreads over time and find justification for the parallel trends assumption in the period before the SEC Municipal Advisor Rule. Following the imposition of the fiduciary duty on advisors, I find that bond yields decrease for the treated bonds. Compared to the control group, this difference is statistically significant, as shown in Panel (b) of Figure 1.10. For the period before the Rule, I are able to verify that the differences are indistinguishable from zero and statistically insignificant.

**[Insert Table 1.12 here]**

To quantify the overall magnitude on yield spreads of bond refundings, I present the results in Table 1.12. Using the offering yields as the dependent variable in the full model specification of Equation (4.1) in Column (1), I find that yields decrease by 7.93 basis points for the treated bonds. In Column (2), I show the results using the after-tax yield as the dependent variable. As expected, the coefficient of interest is higher, given the tax-adjustments. The baseline specification corresponds to Column (3), where I show results for yield spreads as the dependent variable. I find that refunding yield spreads decrease by 7.54 bps for negotiated bonds after the SEC Municipal Advisor Rule, in comparison to competitively bid bonds. Finally, I also report the results using after-tax yield spreads as the dependent variable in Column (4). As in shown in Column (2), this effect tends

**Yield Spreads (Refunding Bonds)**

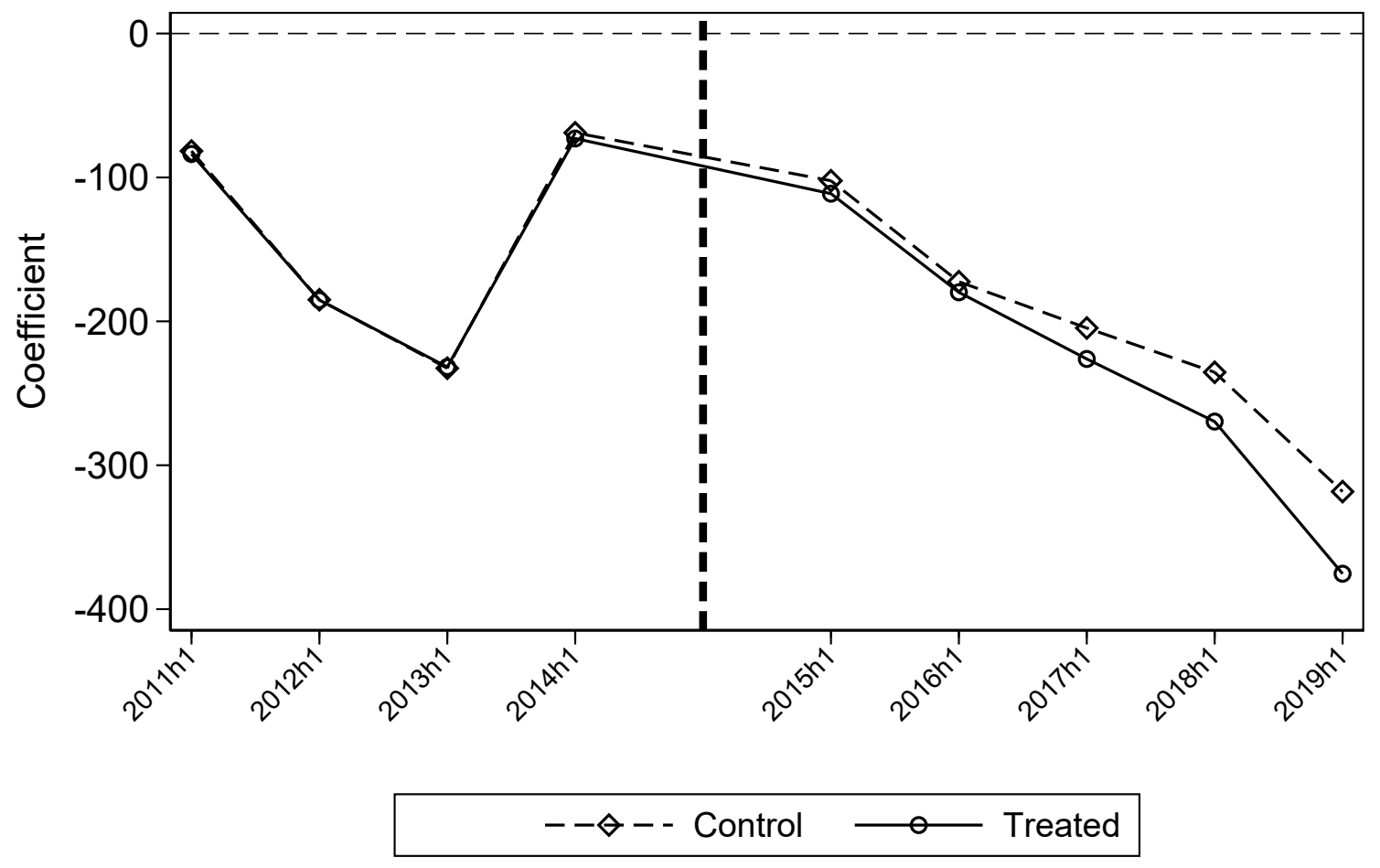

**(a)**

**Difference in Yield Spreads (Refunding Bonds)**

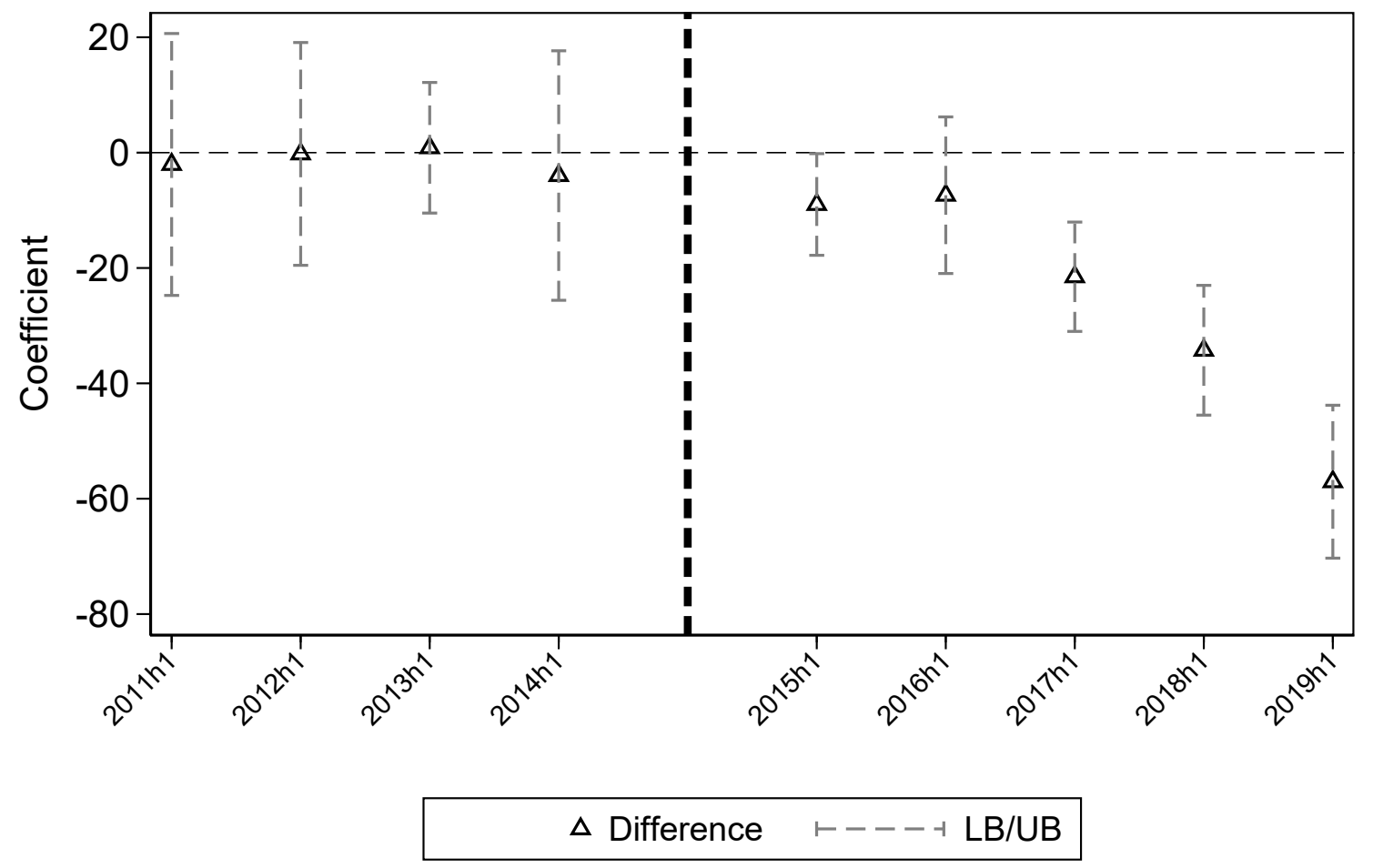

**(b)**

**Figure 1.10:** Refunding Bonds:

In this figure, I plot the average yield spread for refunding municipal bonds issued based on Equation (1.3) in Panel (a). Panel (b) shows the differences between the spreads of treated and control bonds. See Table A1 for variables description. The coefficients are shown in basis points. Specifically, the coefficients are obtained from regressing the spreads on yearly interaction dummies for treated and control bonds using issuer fixed effects. These coefficients are depicted on a yearly scale on the x-axis, where the vertical line corresponds to the Municipal Advisor Rule. The omitted benchmark period is the twelve month period before the event window shown above. Standard errors are clustered by state. The dashed lines represent 95% confidence intervals.

to be higher due to tax adjustments (see [28]). Overall, I find that the reduction in yields of treated bonds is lower for refunding bonds, when compared to the overall sample. This should be understood in light of the greater information that issuers have about these bonds. Since refundings are issued to refinance earlier bonds, issuers have more information about the underlying characteristics and finances of these projects/purposes. Further, issuers may learn about the demand for such bonds based on the secondary market trading of existing bonds to be refinanced. Thus, the principal-agent problem mitigated by the imposition of fiduciary duty is relatively lower in this setting.

Finally, I turn to another aspect of refunding bond issuances. [61] show that many local governments refinance their long-term debt with significant delays. This results in sizable losses for these issuers. Motivated by this evidence, I examine the timing of pre-refunding bonds in the sample. For these bonds, issuers raise the refunding bonds before the calling date of the bonds to be defeased. Ideally, issuers would benefit the most when such callable bonds are refinanced when they can borrow at the lowest yields. However, the future borrowing cost of issuers may not be precisely ascertained in advance.

The imposition of fiduciary duty on municipal advisors may hold them liable for adverse consequences to municipal issuers. Advisors may be more cautious in issuing refunding bonds in advance of the calling date. I examine this possibility using Equation (1.3) in Figure 1.11 with the gap between calling date and refunding date as the dependent variable. First, panel (a) shows that the treated and control bonds seem to follow nearly parallel trends *within* issuers. As before, I plot the difference between the treated and control bonds along with the confidence intervals in Figure 1.11b. These results show that the gap reduced significantly for the negotiated bonds in the period after the SEC Municipal Advisor Rule. This would suggest that issuers reduced earlier refunding of callable bonds after the imposition of fiduciary duty on their advisors.

I present this result more formally to quantify the magnitude in Table 1.13. The preferred specification corresponds to Column (5) where the dependent variable is the gap

**∆(Calling - Refunding) in days**

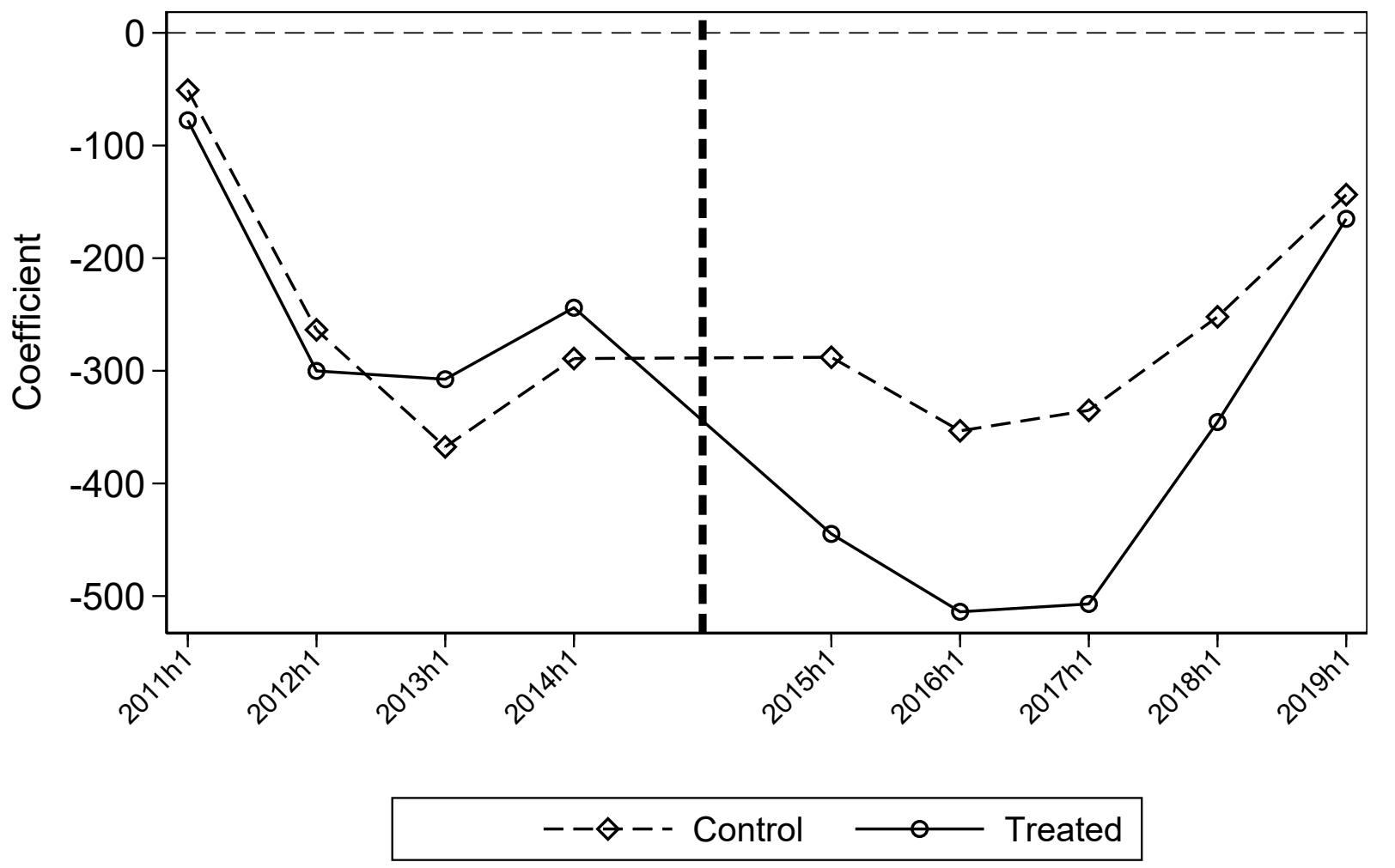

**(a)**

**Difference in ∆(Calling - Refunding) in days**

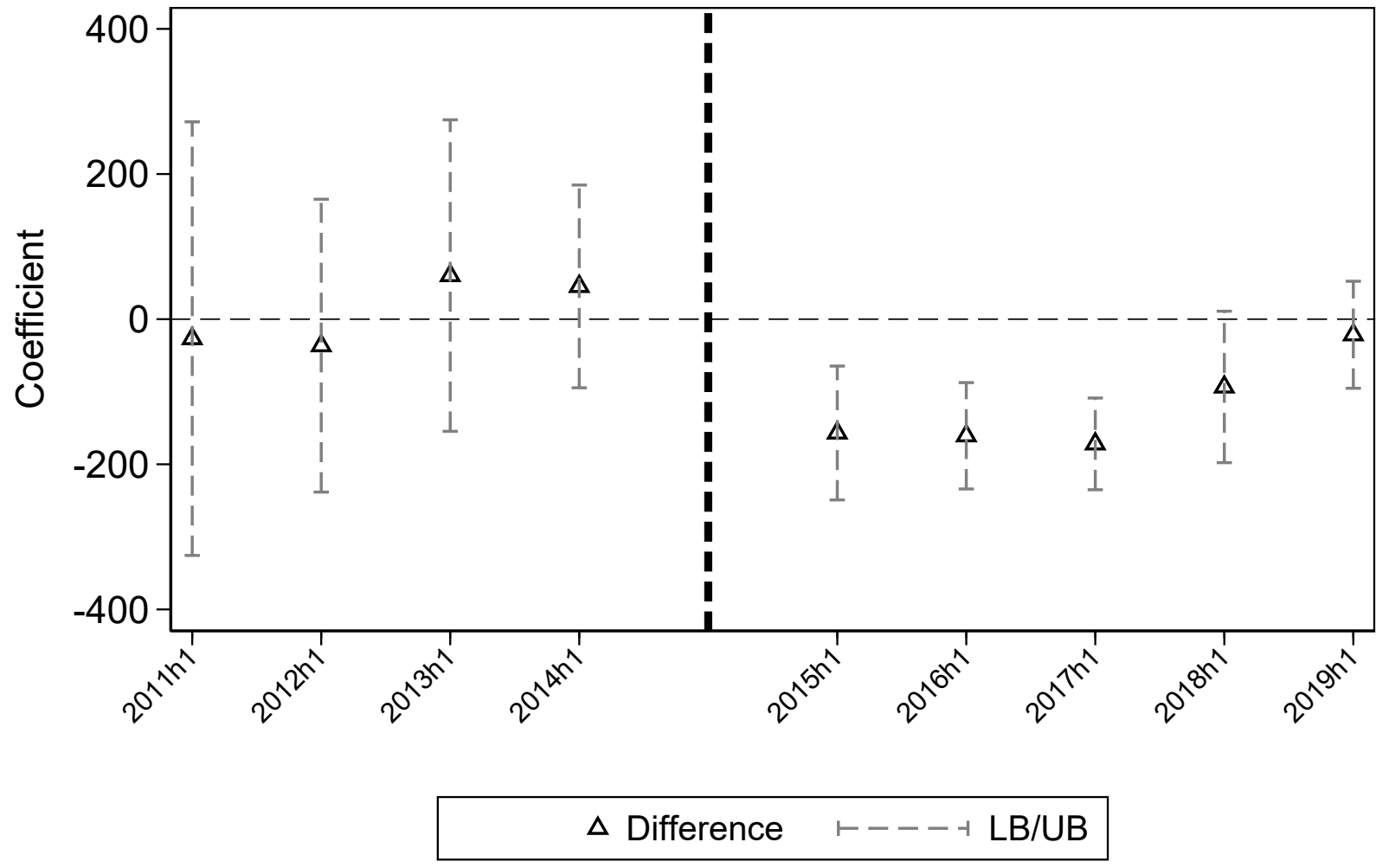

**(b)**

**Figure 1.11:** Timing of Pre-Refunding Bonds:
In this figure, I plot the average gap between calling and refunding dates (in days) for pre-refunding municipal bonds using Equation (1.3) in Panel (a). Panel (b) shows the differences between the gap for treated and control bonds. See Table A1 for variables description. The coefficients are shown in number of days. Specifically, I regress the gap between calling and refunding days on yearly interaction dummies for treated and control bonds using issuer fixed effects. I depict these coefficients on a yearly scale on the x-axis, where the thick vertical line corresponds to the effective date of the Municipal Advisor Rule. The omitted benchmark period is the twelve month period before the event window shown above. Standard errors are clustered by state. The dashed lines represent 95% confidence intervals.

between the calling and refunding dates. The coefficient suggests that this gap reduced by nearly 143 days, *within* issuers, after controlling for advisor fixed effect and state × year fixed effect. Due to the smaller sample in this setting, I do not use the more restrictive advisor × year fixed effect in this setting. Finally, Column (6) shows the result by using the log transformation of the gap in days, suggesting over 20% reduction for the treated bonds after the SEC Municipal Advisor Rule.

**[Insert Table 1.13 here]**

Overall, I show in this Section that there is a similar impact on yield spreads of treated bonds among refunding issuances. Focusing on these bonds also enables understanding more about how issuers reduce the extent of early refunding of callable bonds. These findings complement the main result showing that the imposition of fiduciary duty may benefit the average issuer by lowering their yield spreads at issuance.

## 1.5 Conclusion

I investigate how the imposition of fiduciary duty on municipal advisors affects municipal bond yields and advising fees. On the one hand, fiduciary duty may benefit municipal issuers by increasing the cost of poor advice by municipal advisors (*advice channel*). However, at the same time, the additional regulatory burden may increase the cost of doing business (*fixed cost channel*). It is unclear which of these two effects would dominate overall. In light of this, I evaluate the overall implication of these two competing channels in the context of municipal bonds.

By focusing on the SEC Municipal Advisor Rule of 2014, I provide the first evidence on how fiduciary duty on municipal advisors affects municipal issuers. The findings suggest that the offering yield spreads on negotiated bonds reduced by 12.6 bps after the SEC Municipal Advisor Rule. I explain this result based on reduced underpricing of bonds, and is driven by issuers for whom advisors likely played a more significant ex-ante role in se-lecting underwriters. However, this effect is not uniform. Further analysis shows that large and sophisticated issuers seem to experience a greater reduction in yield spreads. While issuers are more likely to engage advisors on the extensive margin after the regulation, this effect is driven by large issuers only.

Moreover, newly advised issuers do not benefit in terms of yields after the MA Rule. Using novel data, this paper also provides first evidence of little change in average advising fees around the SEC Municipal Advisor Rule. Overall, I find that the fiduciary duty of advisors helps to mitigate the principal-agent problem between issuers and advisors.

**Table A1:** Description of Key Variables

This table reports variable definitions. Data sources include the municipal bond transaction data from the Municipal Securities Rulemaking Board (MSRB), FTSE Russell's Municipal Bond Securities Database (FTSE, formerly known as Mergent MBSD), zero coupon yield provided by FEDS, and highest income tax bracket for the corresponding state of the bond issuer from the Federation of Tax Administrators (FTA).

| Variable | Description | Source |
|---|---|---|
| *Treated* | Dummy set to one for bonds sold via negotiation. This dummy equals zero for competitively bid bonds. | FTSE |
| *Post* | Dummy that is assigned a value of one for time after SEC Municipal Advisor Rule became effective on July 1, 2014 and zero otherwise. | SEC |
| *GO Bond Dummy* | Dummy variable for general obligation bond. A GO bond is a municipal bond backed by the credit and taxing power of the issuing jurisdiction rather than the revenue from a given project. | FTSE |
| *Log(Amount)* | Log transformation of the dollar amount of the individual bond's (9-digit CUSIP) original offering. | FTSE |
| *Callable Dummy* | Dummy variable that equals 1 if the issue is callable and is 0 otherwise. | FTSE |
| *Insured Dummy* | Dummy variable that equals 1 if the issue is insured and is 0 otherwise. | FTSE |
| *Remaining Maturity* | Individual bond maturity measured in years. | FTSE, MSRB |
| *Inverse Maturity* | Inverse of the value of *Remaining Maturity*; to account for non-linearity. | FTSE, MSRB |
| *Markup (underpricing)* | Trade-weighted difference between the average price paid by customers to buy the bonds versus the offering price of bonds, as a percentage of the offering price [13]. | FTSE, MSRB |

| Variable | Description | Source |
|---|---|---|
| *Average Yield* | Volume-weighted average yield for a CUSIP in a given month. Volume refers to the par value of the trade. | MSRB |
| *Yield Spread* | Calculated as the difference between the *Average Yield* and the coupon-equivalent risk free yield ($r_t$). The risk free yield is based on the present value of coupon payments and the face value of the municipal bond using the US treasury yield curve based on maturity-matched zero-coupon yields as given by [48]. This yield spread calculation is similar to [49]. | MSRB, FEDS |
| *After-tax Yield Spread* | Calculated as the difference between the tax-adjusted *Average Yield* and the coupon-equivalent risk free yield ($r_t$). The risk free yield is based on the present value of coupon payments and the face value of the municipal bond using the US treasury yield curve based on maturity-matched zero-coupon yields as given by [48]. This yield spread calculation is similar to [49]. We follow [5] in applying the tax adjustment. It is calculated as below: $$spread_{i,t} = \frac{y_{i,t}}{(1-\tau_t^{\text{fed}}) * (1-\tau_{s,t}^{\text{state}}))} - r_t$$ | MSRB, FEDS, FTA |

| Variable | Description | Source |
|---|---|---|
| *Price Dispersion* | Average dispersion using traded prices around the market "consensus valuation" using [5], based on [57]. Bond-level estimates of the price dispersion measure obtained by taking the average of daily estimates in the first month of trading. | FTSE, MSRB |
| *Rating* | Numeric value corresponding to the bond's credit rating from S&P, Moody's or Fitch. We use ratings within one year of bond issuance. Following [25], we map the ratings into numeric values where the lowest rating is assigned the value of one and the higher ratings are assigned higher numeric values, progressively. | FTSE |

**Table 1.1:** Summary Statistics: Municipal Advisors and Municipal Bonds

This table shows the summary statistics for municipal advisors (MA) and municipal bonds in the sample. Panel A reports the top 15 municipal advisors in the data during the sample period of 2010-2019, who advise on fixed rate, tax-exempt bonds sold by municipalities. The number of issues corresponds to the aggregate number of bond issuances advised by the MA. The number of states in which MA's advise is split by total states versus the number of states across which over 50% of the advisor's bond volume is spread. The share % indicates the relative percentage of the advisor's volume in the sample. Panel B provides the municipal bond level characteristics during 2010-2019 for the bonds in the sample. The summary statistics correspond to the new issuance of bonds in the primary market. The key variables are described in Table A1.

**Panel A: Municipal Advisors**

| | Number of Issues | Average No. of Bonds Per Issue | Number of States | | Total Volume (USD billion) | Share(%) |
|---|---|---|---|---|---|---|
| | | | Total | > 50%ile | | |
| THE PFM GROUP | 7,707 | 13 | 48 | 6 | 501.1 | 22.1 |
| PRAG | 1,058 | 16 | 27 | 2 | 286.2 | 12.6 |
| FIRST SOUTHWEST | 5,274 | 15 | 33 | 1 | 194.4 | 8.6 |
| ACACIA FINANCIAL | 787 | 14 | 13 | 3 | 56.0 | 2.5 |
| MONTAGUE DEROSE | 256 | 16 | 5 | 2 | 51.2 | 2.3 |
| HILLTOP SECURITIES | 1,056 | 16 | 25 | 2 | 50.0 | 2.2 |
| LAMONT FINANCIAL | 275 | 17 | 12 | 1 | 49.0 | 2.2 |
| RBC CAPITAL MARKETS | 1,484 | 14 | 12 | 1 | 47.3 | 2.1 |
| KAUFMAN AND HALL | 304 | 13 | 39 | 7 | 47.2 | 2.1 |
| KNN PUBLIC FINANCE | 522 | 15 | 1 | 1 | 42.4 | 1.9 |
| PONDER & COMPANY | 303 | 13 | 38 | 7 | 37.8 | 1.7 |
| PIPER JAFFRAY & CO. | 1,610 | 12 | 20 | 3 | 35.2 | 1.6 |
| ESTRADA HINOJOSA | 550 | 14 | 2 | 1 | 26.1 | 1.2 |
| DAVENPORT & CO. LLC | 573 | 17 | 10 | 2 | 23.8 | 1.0 |
| FIELDMAN ROLAPP | 630 | 17 | 4 | 1 | 22.0 | 1.0 |

**Panel B: Municipal Bonds**

| | Count | Mean | Std. Dev. | P25 | P50 | P75 |
|---|---|---|---|---|---|---|
| Amount (USD million) | 834,924 | 2.70 | 10.89 | 0.3 | 0.6 | 1.8 |
| Coupon(%) | 834,924 | 3.52 | 1.13 | 2.9 | 3.4 | 4.8 |
| Years to Maturity | 834,924 | 9.87 | 6.15 | 5.0 | 8.9 | 13.9 |
| Offering Price (USD) | 834,923 | 106.62 | 7.67 | 100.0 | 103.9 | 111.7 |
| Offering Yield(%) | 834,924 | 2.33 | 1.01 | 1.6 | 2.3 | 3.0 |
| Yield Spread(%) | 834,924 | 1.29 | 1.20 | 0.5 | 1.2 | 2.1 |
| Callable (Dummy) | 834,924 | 0.47 | 0.50 | 0.0 | 0.0 | 1.0 |
| General Obligation (Dummy) | 834,924 | 0.60 | 0.49 | 0.0 | 1.0 | 1.0 |
| Bank Qualified (Dummy) | 834,924 | 0.40 | 0.49 | 0.0 | 0.0 | 1.0 |
| Cred. Enh. (Dummy) | 834,924 | 0.21 | 0.41 | 0.0 | 0.0 | 0.0 |
| Insured (Dummy) | 834,924 | 0.16 | 0.37 | 0.0 | 0.0 | 0.0 |

**Table 1.2:** Impact on Offering Yield Spreads of Local Governments

This table reports the baseline results for the sample using Equation (4.1) estimating the differential effect on municipal bond yield spreads of treated and control bonds after the Municipal Advisor Rule of 2014. The primary coefficient of interest, $\beta_0$, is captured by the interaction term of *Treated* $\times$ *Post*. I show the results using offering yield spread as the dependent variable. Specifically, Column (1) reports the results without any controls or fixed effects. In Column (2), I first introduce issuer fixed effects. Column (3) shows results by introducing state $\times$ year fixed effects. In Column (4), I add advisor fixed effect. Column (5) reports the results by additionally including bond level controls consisting of coupon (%); log(amount issued in \$); dummies for callable bonds, bond insurance, general obligation bond, bank qualification, refunding and credit enhancement; credit rating; remaining years to maturity; and inverse years to maturity. I provide the description of key variables in Table A1. In Column (6), I control for the county-level economic conditions. I use the lagged values for log(labor force) and unemployment rate, and the percentage change in unemployment rate and labor force, respectively. Column (7) replaces advisor fixed effect with advisor $\times$ year fixed effect and does not include county-level controls. Finally, Column (8) shows results by also adding county-level controls. In Table A1.3, I show our results using offering yields as the dependent variable. T-statistics are reported in brackets and standard errors are clustered at the state level. $^{*}$ $p < 0.10$, $^{**}$ $p < 0.05$, $^{***}$ $p < 0.01$

| *Dependent Variable*: | Yield Spread (basis points) | | | | | | | |
|---|---|---|---|---|---|---|---|---|
| | (1) | (2) | (3) | (4) | (5) | (6) | (7) | (8) |
| Treated × Post | -25.55$^{***}$ | -26.56$^{***}$ | -12.81$^{***}$ | -12.43$^{***}$ | -14.79$^{***}$ | -14.74$^{***}$ | -12.63$^{***}$ | -12.60$^{***}$ |
| | [-3.45] | [-5.03] | [-3.12] | [-3.24] | [-5.04] | [-5.02] | [-4.27] | [-4.26] |
| Treated | 26.27$^{***}$ | -11.48 | 1.50 | 0.80 | 8.85$^{***}$ | 8.89$^{***}$ | 7.58$^{**}$ | 7.62$^{**}$ |
| | [2.79] | [-1.12] | [0.29] | [0.16] | [2.77] | [2.80] | [2.09] | [2.11] |
| Post | -81.12$^{***}$ | -81.97$^{***}$ | -4.83$^{**}$ | -5.60$^{***}$ | -3.74$^{*}$ | -3.62$^{*}$ | -4.07$^{**}$ | -3.96$^{**}$ |
| | [-32.73] | [-26.93] | [-2.48] | [-2.69] | [-1.89] | [-1.83] | [-2.15] | [-2.10] |
| Issuer FE | | ✓ | ✓ | ✓ | ✓ | ✓ | ✓ | ✓ |
| State-Year FE | | | ✓ | ✓ | ✓ | ✓ | ✓ | ✓ |
| Advisor FE | | | | ✓ | ✓ | ✓ | | |
| Bond Controls | | | | | ✓ | ✓ | ✓ | ✓ |
| County Controls | | | | | | ✓ | | ✓ |
| Advisor-Year FE | | | | | | | ✓ | ✓ |
| Adj.-$R^2$ | 0.135 | 0.345 | 0.635 | 0.638 | 0.847 | 0.847 | 0.855 | 0.855 |
| Obs. | 834,924 | 834,764 | 834,764 | 834,760 | 834,760 | 834,760 | 834,741 | 834,741 |

**Table 1.3:** Impact on Offering Price of New Bonds

This table reports the baseline results for the sample using Equation (4.1) estimating the differential effect on offering price of treated and control bonds after the Municipal Advisor Rule of 2014. The primary coefficient of interest, $\beta_0$, is captured by the interaction term of *Treated* $\times$ *Post*. I show our results using offering price as the dependent variable. Specifically, Column (1) reports the results without any controls or fixed effects. In Column (2), I first introduce issuer fixed effects. Column (3) shows results by introducing state $\times$ year fixed effects. In Column (4), I add advisor fixed effect. Column (5) reports the results by additionally including bond level controls consisting of coupon (%); log(amount issued in \$); dummies for callable bonds, bond insurance, general obligation bond, bank qualification, refunding and credit enhancement; credit rating; remaining years to maturity; and inverse years to maturity. I provide the description of key variables in Table A1. In Column (6), I control for the county-level economic conditions. I use the lagged values for log(labor force) and unemployment rate, and the percentage change in unemployment rate and labor force, respectively. Column (7) replaces advisor fixed effect with advisor $\times$ year fixed effect and does not include county-level controls. Finally, Column (8) shows results by also adding county-level controls. T-statistics are reported in brackets and standard errors are clustered at the state level. $^{*}$ $p < 0.10$, $^{**}$ $p < 0.05$, $^{***}$ $p < 0.01$

| *Dependent Variable*: | Offering Price (per USD 100) | | | | | | | |
|---|---|---|---|---|---|---|---|---|
| | (1) | (2) | (3) | (4) | (5) | (6) | (7) | (8) |
| Treated $\times$ Post | 1.94*** | 1.81*** | 1.32*** | 1.30*** | 1.25*** | 1.25*** | 1.05*** | 1.05*** |
| | [6.75] | [6.82] | [4.28] | [4.16] | [4.81] | [4.81] | [4.35] | [4.36] |
| Treated | 1.62*** | 0.57** | 0.63*** | 0.64** | -0.34 | -0.34 | -0.23 | -0.23 |
| | [3.84] | [2.03] | [2.84] | [2.57] | [-1.49] | [-1.49] | [-1.00] | [-1.00] |
| Post | 2.12*** | 2.37*** | 0.78*** | 0.77*** | 0.59*** | 0.58*** | 0.66*** | 0.65*** |
| | [9.22] | [17.13] | [4.29] | [4.55] | [3.84] | [3.82] | [4.54] | [4.51] |
| Issuer FE | | ✓ | ✓ | ✓ | ✓ | ✓ | ✓ | ✓ |
| State-Year FE | | | ✓ | ✓ | ✓ | ✓ | ✓ | ✓ |
| Advisor FE | | | | ✓ | ✓ | ✓ | | |
| Bond Controls | | | | | ✓ | ✓ | ✓ | ✓ |
| County Controls | | | | | | ✓ | | ✓ |
| Advisor-Year FE | | | | | | | ✓ | ✓ |
| Adj.-$R^2$ | 0.066 | 0.361 | 0.383 | 0.388 | 0.768 | 0.768 | 0.774 | 0.774 |
| Obs. | 834,924 | 834,763 | 834,763 | 834,759 | 834,759 | 834,759 | 834,740 | 834,740 |

**Table 1.4:** Impact on Markup (Underpricing) of New Municipal Bonds

This table reports the baseline results for the sample using Equation (4.1) estimating the differential effect on underpricing of treated and control bonds after the Municipal Advisor Rule of 2014. The primary coefficient of interest, $\beta_0$, is captured by the interaction term of *Treated* $\times$ *Post*. In Columns (1)-(8), the dependent variable is the trade-weighted mark-up paid by customers over offering price, as a percentage of the offering price (in basis points). I consider initial trades from the MSRB database within the first 30 days of trading. Specifically, Column (1) reports the results without any controls or fixed effects. In Column (2), I first introduce issuer fixed effects. Column (3) shows results by introducing state $\times$ year fixed effects. In Column (4), I add advisor fixed effect. Column (5) reports the results by additionally including bond level controls consisting of coupon (%); log(amount issued in $); dummies for callable bonds, bond insurance, general obligation bond, bank qualification, refunding and credit enhancement; credit rating; remaining years to maturity; and inverse years to maturity. I provide the description of key variables in Table A1. In Column (6), I control for the county-level economic conditions. I use the lagged values for log(labor force) and unemployment rate, and the percentage change in unemployment rate and labor force, respectively. Column (7) replaces advisor fixed effect with advisor $\times$ year fixed effect and does not include county-level controls. Finally, Column (8) shows results by also adding county-level controls. I replicate the results in Column (9) using the difference between the trade-weighted average price paid by customers and the offering price, scaled by the offering price. Column (10) shows the baseline results using the difference between the trade-weighted average price paid by customers and the trade-weighted interdealer price, scaled by the interdealer price. T-statistics are reported in brackets and standard errors are clustered at the state level. $^{*}$ $p < 0.10$, $^{**}$ $p < 0.05$, $^{***}$ $p < 0.01$

| *Dependent Variable*: | Trade-weighted Markup (basis points) | | | | | | | | $\bar{P}$-$\bar{O}$ (bps) | $\bar{P}$-$\bar{V}$ (bps) |
|---|---|---|---|---|---|---|---|---|---|---|
| | (1) | (2) | (3) | (4) | (5) | (6) | (7) | (8) | (9) | (10) |
| Treated $\times$ Post | -5.35$^{**}$ | -7.56$^{***}$ | -7.57$^{***}$ | -6.90$^{**}$ | -6.89$^{**}$ | -6.83$^{**}$ | -6.22$^{**}$ | -6.16$^{**}$ | -5.56$^{**}$ | -10.35$^{***}$ |
| | [-2.28] | [-4.04] | [-2.70] | [-2.40] | [-2.24] | [-2.24] | [-2.32] | [-2.31] | [-2.45] | [-7.50] |
| Treated | 16.29$^{***}$ | 12.29$^{***}$ | 12.60$^{***}$ | 11.49$^{***}$ | 14.49$^{***}$ | 14.47$^{***}$ | 14.38$^{***}$ | 14.35$^{***}$ | 13.10$^{***}$ | 2.49 |
| | [4.54] | [7.78] | [6.74] | [5.36] | [5.71] | [5.72] | [6.63] | [6.61] | [6.93] | [1.41] |
| Post | 4.49$^{***}$ | 3.74$^{***}$ | 2.03 | 2.40 | 2.76 | 2.76 | 2.63 | 2.63 | 2.45$^{*}$ | 1.72$^{*}$ |
| | [4.09] | [4.87] | [1.05] | [1.29] | [1.58] | [1.60] | [1.54] | [1.55] | [1.84] | [1.81] |
| Issuer FE | | ✓ | ✓ | ✓ | ✓ | ✓ | ✓ | ✓ | ✓ | ✓ |
| State-Year FE | | | ✓ | ✓ | ✓ | ✓ | ✓ | ✓ | ✓ | ✓ |
| Advisor FE | | | | ✓ | ✓ | ✓ | ✓ | | | |
| Bond Controls | | | | | ✓ | ✓ | ✓ | ✓ | ✓ | ✓ |
| County Controls | | | | | | ✓ | | ✓ | ✓ | ✓ |
| Advisor-Year FE | | | | | | | ✓ | ✓ | ✓ | ✓ |
| Adj.-$R^2$ | 0.007 | 0.110 | 0.117 | 0.126 | 0.185 | 0.186 | 0.190 | 0.190 | 0.223 | 0.244 |
| Obs. | 786,528 | 786,326 | 786,326 | 785,754 | 785,754 | 785,754 | 785,749 | 785,749 | 785,748 | 785,748 |

**Table 1.5:** Impact on Price Dispersion (Liquidity) of New Bonds

This table reports the baseline results for the sample using Equation (4.1) estimating the differential effect on price dispersion of treated and control bonds after the Municipal Advisor Rule of 2014. I use trades during the first month after issuance to construct this measure using [5], based on [57]. The primary coefficient of interest, $\beta_0$, is captured by the interaction term of *Treated* × *Post*. Specifically, Column (1) reports the results without any controls or fixed effects. In Column (2), I first introduce issuer fixed effects. Column (3) shows results by introducing state × year fixed effects. In Column (4), I add advisor fixed effect. Column (5) reports the results by additionally including bond level controls consisting of coupon (%); log(amount issued in $); dummies for callable bonds, bond insurance, general obligation bond, bank qualification, refunding and credit enhancement; credit rating; remaining years to maturity; and inverse years to maturity. I provide the description of key variables in Table A1. In Column (6), I control for the county-level economic conditions. I use the lagged values for log(labor force) and unemployment rate, and the percentage change in unemployment rate and labor force, respectively. Column (7) replaces advisor fixed effect with advisor × year fixed effect and does not include county-level controls. Finally, Column (8) shows results by also adding county-level controls. In Table A1.4, I show the results using the Amihud measure as an alternative way to capture liquidity. T-statistics are reported in brackets and standard errors are clustered at the state level. $^{*}$ $p < 0.10$, $^{**}$ $p < 0.05$, $^{***}$ $p < 0.01$

| *Dependent Variable*: | Price Dispersion (per USD 100) | | | | | | | |
|---|---|---|---|---|---|---|---|---|
| | (1) | (2) | (3) | (4) | (5) | (6) | (7) | (8) |
| Treated × Post | -0.05*** | -0.04*** | -0.03*** | -0.03*** | -0.03*** | -0.03*** | -0.03*** | -0.03*** |
| | [-8.68] | [-8.89] | [-6.47] | [-6.43] | [-6.80] | [-6.81] | [-7.02] | [-7.02] |
| Treated | 0.00 | -0.02*** | -0.03*** | -0.03*** | -0.01* | -0.01* | -0.01* | -0.01* |
| | [0.21] | [-4.91] | [-4.54] | [-5.18] | [-1.78] | [-1.78] | [-1.85] | [-1.86] |
| Post | 0.07*** | 0.06*** | 0.00 | 0.00 | 0.00 | 0.00 | 0.00 | 0.00 |
| | [13.49] | [10.35] | [0.24] | [0.23] | [1.15] | [1.13] | [0.65] | [0.64] |
| Issuer FE | | ✓ | ✓ | ✓ | ✓ | ✓ | ✓ | ✓ |
| State-Year FE | | | ✓ | ✓ | ✓ | ✓ | ✓ | ✓ |
| Advisor FE | | | | ✓ | ✓ | ✓ | | |
| Bond Controls | | | | | ✓ | ✓ | ✓ | ✓ |
| County Controls | | | | | | ✓ | | ✓ |
| Advisor-Year FE | | | | | | | ✓ | ✓ |
| Adj.-$R^2$ | 0.020 | 0.116 | 0.130 | 0.132 | 0.317 | 0.317 | 0.323 | 0.323 |
| Obs. | 786,528 | 786,326 | 786,326 | 786,322 | 786,322 | 786,322 | 786,303 | 786,303 |

**Table 1.6:** Impact Due to Advisor's Ex-ante Role in "Selecting" Underwriter

This table reports the results using Equation (4.1) to show the differential effect among issuers for whom advisors play a greater role in selecting underwriters. The dependent variable is offering yield spread. I use the average (Columns (1)-(3)) and weighted average (Columns (4)-(6)) of ex-ante likelihood of a new underwriter being introduced by an advisor for a given issuer, respectively. Specifically, I interact the equation with dummies corresponding to below and above median values for this measure among issuers. This analysis also includes group × year fixed effects. The baseline specification of Column (8) in Table 1.2 is shown in Columns (3) and (6). T-statistics are reported in brackets and standard errors are clustered at the state level. * $p < 0.10$, ** $p < 0.05$, *** $p < 0.01$

| *Dependent Variable*: | Yield Spread (basis points) | | | | | |
|---|---|---|---|---|---|---|
| *Based on Issuers'*: | Average | | | Weighted Average | | |
| Treated × Post | (1) | (2) | (3) | (4) | (5) | (6) |
| Below Median | -9.56* | -5.16 | -3.76 | -9.58* | -5.19 | -3.80 |
| | [-1.70] | [-1.03] | [-0.86] | [-1.71] | [-1.04] | [-0.86] |
| Above Median | -19.93*** | -18.56*** | -16.26*** | -19.91*** | -18.53*** | -16.23*** |
| | [-7.86] | [-8.21] | [-7.35] | [-7.86] | [-8.19] | [-7.32] |
| Difference | 10.37 | 13.40 | 12.50 | 10.33 | 13.35 | 12.43 |
| p-value | 0.05 | 0.01 | 0.01 | 0.05 | 0.01 | 0.01 |
| Issuer FE | ✓ | ✓ | ✓ | ✓ | ✓ | ✓ |
| Controls | ✓ | ✓ | ✓ | ✓ | ✓ | ✓ |
| Advisor FE | ✓ | ✓ | | ✓ | ✓ | |
| Group-Yr. FE | ✓ | ✓ | ✓ | ✓ | ✓ | ✓ |
| State-Yr. FE | | ✓ | ✓ | | ✓ | ✓ |
| Advisor-Yr. FE | | | ✓ | | | ✓ |
| Adj.-$R^2$ | 0.839 | 0.844 | 0.852 | 0.839 | 0.844 | 0.852 |
| Obs. | 776,304 | 776,304 | 776,286 | 776,304 | 776,304 | 776,286 |

## **Table 1.7:** Robustness Tests

In this table I report results for various robustness tests on the baseline specification, i.e., Column (8) of Table 1.2. Columns (1)-(2) show the baseline effect by changing the dependent variable to after-tax yield and after-tax yield spread, respectively. In Columns (3)-(11), I report results based on alternative econometric specifications. I introduce additional fixed effects to account for unobserved factors that may be varying over time. Specifically, Column (3) reports baseline results by adding issuer-type $\times$ year fixed effects. Column (4) shows results by adding bond purpose fixed effects. I add bond purpose $\times$ year fixed effect in Column (5). Columns (6) and (7) show results by adding underwriter fixed effect and underwriter $\times$ year fixed effect, respectively. In Columns (8) and (9), I control for unobserved pairing between issuers and advisors, as well as issuers and underwriters, separately, by adding issuer-advisor pair fixed effect and issuer-underwriter pair fixed effect, respectively. These specifications include time unvarying advisor and underwriter fixed effects, respectively. Column (10) shows results with advisor $\times$ state fixed effects. In Column (11), I also add county $\times$ year fixed effects. I consider additional tax considerations in Columns (12)-(14). First, I relax the sample of bonds to include taxable bonds in Columns (12)-(13). In Column (13), I further omit bonds from five states (IL, IA, KS, OK, WI) that tax interest income on municipal bonds issued in-state or out-of-state [28]. Column (14) shows results using bonds that are exempt from both state and federal income tax simultaneously. Columns (15)-(18) report results focusing on sub-samples of homogeneous bonds. Accordingly, in Column (15), I drop bonds in which the advisor and underwriter are same. Column (16) shows results by dropping callable bonds. I drop insured bonds in Column (17). Finally, I focus on only the new money bonds in Column (18). The results in Columns (19)-(21) focus on additional geographic considerations. I keep only local bonds (by dropping state level bonds) in Column (19). Conversely, I show results using only the state level bonds in Column (20). In Column (21), I report the baseline results by dropping issuances from the three largest municipal bond issuers, namely: California (CA), New York (NY) and Texas (TX). I consider alternative levels of clustering standard errors in Columns (22)-(32). In Columns (22)-(26), I cluster standard errors by advisor, underwriter, issuer, issuer(2), and bond issue, respectively. Issuer(2) in Column (25) refers to weakly identifying borrowers based on the first six-digits of the bond CUSIP [28]. Columns (27)-(28) double cluster standard errors along two dimensions in the geography of issuers: state-advisor, and advisor-issuer, respectively. Finally, in Columns (29)-(32), I double cluster standard errors over time and across bonds using: state and year, advisor and year, state and year-month, and advisor and year-month, respectively. In Table A1.7, I replicate these results using offering yield as the dependent variable. Additionally, I report robustness results for offering price as the dependent variable in Table A1.8. T-statistics are reported in brackets and standard errors are clustered at the state level, unless otherwise specified. * $p < 0.10$, ** $p < 0.05$, *** $p < 0.01$

| *Dependent Variable*: | Yield Spread (basis points) | | | | | | | | | | |
|---|---|---|---|---|---|---|---|---|---|---|---|
| | Other Dependent Variables | | Alternative Specifications | | | | | | | | |
| | After-tax Yield | After-tax Yield Spread | Add Issuer-Type Year FE | Add Purpose FE | Add Purpose Year FE | Add UW FE | Add UW-Yr. FE | Add Iss.-MA FE | Add Iss.-UW FE | Add MA-State FE | Add County-Yr. FE |
| | (1) | (2) | (3) | (4) | (5) | (6) | (7) | (8) | (9) | (10) | (11) |
| Treated × Post | -18.13*** | -19.69*** | -12.02*** | -12.63*** | -12.19*** | -12.92*** | -11.07*** | -14.13*** | -14.20** | -11.88*** | -10.65*** |
| | [-4.23] | [-4.33] | [-4.05] | [-4.19] | [-3.86] | [-4.45] | [-3.49] | [-4.25] | [-2.60] | [-3.86] | [-3.01] |
| Adj.-$R^2$ | 0.867 | 0.853 | 0.855 | 0.855 | 0.855 | 0.856 | 0.860 | 0.857 | 0.891 | 0.856 | 0.871 |
| Obs. | 833,343 | 833,343 | 833,667 | 833,348 | 833,348 | 834,730 | 834,707 | 834,670 | 834,297 | 834,731 | 834,629 |

| *Dependent Variable*: | Yield Spread (basis points) | | | | | | | | | |
|---|---|---|---|---|---|---|---|---|---|---|
| | Tax Considerations | | | Bond Considerations | | | | Geographic Considerations | | |
| | Include Taxable | | Exempt | Drop same | Drop | Drop | Only | Keep | Keep | Drop |
| | | (-Taxable States) | (Fed.+State) | Adv.-UW. bonds | Callable | Insured | New money | Local | State | CA,NY,TX |
| | (12) | (13) | (14) | (15) | (16) | (17) | (18) | (19) | (20) | (21) |
| Treated × Post | -11.93*** | -11.08*** | -11.94*** | -12.64*** | -11.44*** | -14.88*** | -22.90*** | -10.81*** | -17.15** | -15.94*** |
| | [-4.39] | [-4.23] | [-4.07] | [-4.32] | [-5.53] | [-5.92] | [-4.96] | [-3.50] | [-2.59] | [-5.77] |
| Adj.-$R^2$ | 0.836 | 0.834 | 0.854 | 0.854 | 0.876 | 0.856 | 0.880 | 0.861 | 0.836 | 0.859 |
| Obs. | 905,084 | 794,062 | 770,446 | 830,458 | 443,572 | 698,776 | 415,926 | 751,278 | 83,456 | 519,544 |

| *Dependent Variable*: | Yield Spread (basis points) | | | | | | | | | | |
|---|---|---|---|---|---|---|---|---|---|---|---|
| | Alternative Clustering | | | | | | | | | | |
| | Advisor | Underwriter | Issuer | Issuer(2) | Issue | State, Advisor | Advisor, Issuer | State, Year | Advisor, Year | State, YM | Advisor, YM |
| | (22) | (23) | (24) | (25) | (26) | (27) | (28) | (29) | (30) | (31) | (32) |
| Treated × Post | -12.60*** | -12.60*** | -12.60*** | -12.60*** | -12.60*** | -12.60*** | -12.60*** | -12.60*** | -12.60*** | -12.60*** | -12.60*** |
| | [-4.85] | [-6.42] | [-6.40] | [-8.40] | [-9.92] | [-3.99] | [-4.47] | [-3.49] | [-3.43] | [-4.12] | [-4.59] |
| Adj.-$R^2$ | 0.855 | 0.855 | 0.855 | 0.855 | 0.855 | 0.855 | 0.855 | 0.855 | 0.855 | 0.855 | 0.855 |
| Obs. | 834,741 | 834,741 | 834,741 | 834,741 | 834,741 | 834,741 | 834,741 | 834,741 | 834,741 | 834,741 | 834,741 |

**Table 1.8:** Heterogeneity by Size of Issuers

This table reports the results using Equation (4.1) to show the differential effect among issuers based on their ex-ante size. I use the average (Columns (1)-(3)) and median (Columns (4)-(6)) size of ex-ante issuances, respectively. Specifically, I interact the equation with dummies corresponding to small and large values for this measure among issuers. This analysis also includes group × year fixed effects. The baseline specification of Column (8) in Table 1.2 is shown in Columns (3) and (6). Panel A shows results using the offering yield spread at the time of issuance as the dependent variable. In Panel B, the dependent variable is the offering price at the time of issuance. Finally, Panel C corresponds to underpricing with trade-weighted mark-up being the dependent variable. T-statistics are reported in brackets and standard errors are clustered at the state level. * $p < 0.10$, ** $p < 0.05$, *** $p < 0.01$

## Panel A: Evidence from Yield Spreads

| *Dependent Variable*: | Yield Spread (basis points) | | | | | |
|---|---|---|---|---|---|---|
| *Based on Issuers'*: | Average size, ex-ante | | | Median size, ex-ante | | |
| Treated × Post | (1) | (2) | (3) | (4) | (5) | (6) |
| × Small | -5.62 | -1.08 | -1.66 | -4.45 | -0.47 | -1.13 |
| | [-0.71] | [-0.15] | [-0.24] | [-0.55] | [-0.06] | [-0.16] |
| × Large | -18.75*** | -16.75*** | -14.34*** | -19.50*** | -17.56*** | -15.08*** |
| | [-7.02] | [-7.85] | [-6.32] | [-7.23] | [-8.05] | [-6.72] |
| Difference | 13.13 | 15.67 | 12.67 | 15.05 | 17.09 | 13.95 |
| p-value | 0.08 | 0.02 | 0.06 | 0.07 | 0.02 | 0.06 |
| Issuer FE | ✓ | ✓ | ✓ | ✓ | ✓ | ✓ |
| Controls | ✓ | ✓ | ✓ | ✓ | ✓ | ✓ |
| Advisor FE | ✓ | ✓ | | ✓ | ✓ | |
| Group-Yr. FE | ✓ | ✓ | ✓ | ✓ | ✓ | ✓ |
| State-Yr. FE | | ✓ | ✓ | | ✓ | ✓ |
| Advisor-Yr. FE | | | ✓ | | | ✓ |
| Adj.-$R^2$ | 0.843 | 0.847 | 0.855 | 0.843 | 0.847 | 0.855 |
| Obs. | 834,760 | 834,760 | 834,741 | 834,760 | 834,760 | 834,741 |

### Panel B: Evidence from Offering Price

| *Dependent Variable*: | Offering Price (per USD 100) | | | | | |
|---|---|---|---|---|---|---|
| *Based on Issuers'*: | Average size, ex-ante | | | Median size, ex-ante | | |
| Treated × Post | (1) | (2) | (3) | (4) | (5) | (6) |
| × Small | 0.55 | 0.15 | 0.13 | 0.53 | 0.18 | 0.17 |
| | [1.14] | [0.33] | [0.29] | [1.09] | [0.40] | [0.38] |
| × Large | 1.57*** | 1.38*** | 1.18*** | 1.59*** | 1.41*** | 1.21*** |
| | [6.79] | [7.81] | [7.18] | [7.41] | [8.53] | [7.88] |
| Difference | 1.03 | 1.24 | 1.05 | 1.06 | 1.23 | 1.04 |
| p-value | 0.00 | 0.00 | 0.01 | 0.00 | 0.00 | 0.01 |
| Issuer FE | ✓ | ✓ | ✓ | ✓ | ✓ | ✓ |
| Controls | ✓ | ✓ | ✓ | ✓ | ✓ | ✓ |
| Advisor FE | ✓ | ✓ | | ✓ | ✓ | |
| Group-Yr. FE | ✓ | ✓ | ✓ | ✓ | ✓ | ✓ |
| State-Yr. FE | | ✓ | ✓ | | ✓ | ✓ |
| Advisor-Yr. FE | | | ✓ | | | ✓ |
| Adj.-$R^2$ | 0.765 | 0.769 | 0.774 | 0.765 | 0.769 | 0.774 |
| Obs. | 834,759 | 834,759 | 834,740 | 834,759 | 834,759 | 834,740 |

### Panel C: Evidence from Underpricing

| *Dependent Variable*: | Trade-weighted Mark-up (basis points) | | | | | |
|---|---|---|---|---|---|---|
| *Based on Issuers'*: | Average size, ex-ante | | | Median size, ex-ante | | |
| Treated × Post | (1) | (2) | (3) | (4) | (5) | (6) |
| × Small | -2.97 | -2.80 | -4.56 | -2.44 | -2.15 | -3.95 |
| | [-1.01] | [-0.91] | [-1.50] | [-0.57] | [-0.47] | [-0.90] |
| × Large | -7.64*** | -7.70** | -7.46*** | -7.63*** | -7.88*** | -7.60*** |
| | [-3.12] | [-2.64] | [-2.81] | [-3.43] | [-3.11] | [-3.32] |
| Difference | 4.67 | 4.90 | 2.90 | 5.19 | 5.73 | 3.65 |
| p-value | 0.03 | 0.02 | 0.23 | 0.08 | 0.04 | 0.22 |
| Issuer FE | ✓ | ✓ | ✓ | ✓ | ✓ | ✓ |
| Controls | ✓ | ✓ | ✓ | ✓ | ✓ | ✓ |
| Advisor FE | ✓ | ✓ | | ✓ | ✓ | |
| Group-Yr. FE | ✓ | ✓ | ✓ | ✓ | ✓ | ✓ |
| State-Yr. FE | | ✓ | ✓ | | ✓ | ✓ |
| Advisor-Yr. FE | | | ✓ | | | ✓ |
| Adj.-$R^2$ | 0.177 | 0.180 | 0.190 | 0.177 | 0.180 | 0.190 |
| Obs. | 786,322 | 786,322 | 786,303 | 786,322 | 786,322 | 786,303 |

**Table 1.9:** Heterogeneity by Sophistication of Issuers

This table reports the results using Equation (4.1) to show the differential effect among issuers based on their ex-ante size. The dependent variable is offering yield spread. I use the average (Columns (1)-(3)) and median (Columns (4)-(6)) size of ex-ante issuances, respectively. Specifically, I interact the equation with dummies corresponding to small and large values for this measure among issuers. This analysis also includes group $\times$ year fixed effects. The baseline specification of Column (8) in Table 1.2 is shown in Columns (3) and (6). T-statistics are reported in brackets and standard errors are clustered at the state level. $^{*}$ $p < 0.10$, $^{**}$ $p < 0.05$, $^{***}$ $p < 0.01$

| *Dependent Variable*: | Yield Spread (basis points) | | | | |
|---|---|---|---|---|---|
| *Based on Issuers'*: | Complexity of Bonds (ex-ante) | | Credit | Average Wages | Fraction |
| | All | Advised | Enhancement | of Finance Staff | advised |
| Treated × Post | (1) | (2) | (3) | (4) | (5) |
| Below Median | -4.22 | -7.71$^{***}$ | -8.73$^{**}$ | -5.97$^{*}$ | |
| | [-1.40] | [-3.33] | [-2.04] | [-1.87] | |
| Above Median | -18.89$^{***}$ | -14.71$^{***}$ | -15.76$^{***}$ | -12.63$^{***}$ | |
| | [-5.81] | [-3.33] | [-5.86] | [-4.84] | |
| High | | | | | -9.85$^{**}$ |
| | | | | | [-2.62] |
| Low | | | | | -17.02$^{***}$ |
| | | | | | [-6.18] |
| Difference | 14.68 | 6.99 | 7.02 | 6.66 | 7.18 |
| p-value | 0.00 | 0.08 | 0.07 | 0.02 | 0.09 |
| Issuer FE | ✓ | ✓ | ✓ | ✓ | ✓ |
| Advisor-Year FE | ✓ | ✓ | ✓ | ✓ | ✓ |
| State-Year FE | ✓ | ✓ | ✓ | ✓ | ✓ |
| Group-Year FE | ✓ | ✓ | ✓ | ✓ | ✓ |
| Adj.-$R^2$ | 0.853 | 0.851 | 0.853 | 0.854 | 0.853 |
| Obs. | 712,422 | 685,294 | 709,792 | 610,873 | 709,785 |

**Table 1.10:** Evidence from Newly Advised Issuers

This table reports the results using Equation (4.1) estimating the differential effect on yield spreads of treated and control bonds after the Municipal Advisor Rule of 2014 for newly advised issuers. In Columns (1)-(3), I show our results using yield spreads as the dependent variable. I restrict the sample to issuers who did not use advisors before the SEC Rule but always recruited them afterward in Columns (1)-(2). In Column (3), I also include some issuers who engage advisors for only some of their bonds after the Rule. The primary coefficient of interest, $\beta_0$, is captured by the interaction term of *Treated* $\times$ *Post*. Specifically, Column (1) reports the results without any controls. Thereafter, I introduce bond and county level controls, consisting of coupon (%); log(amount issued in $); dummies for callable bonds, bond insurance, general obligation bond, refunding, bank qualification, refunding, and credit enhancement; credit rating; remaining years to maturity; and inverse years to maturity; lagged values for log(labor force) and unemployment rate, and the percentage change in unemployment rate and labor force, respectively. I provide the description of key variables in Table A1. I follow a similar scheme for Columns (4)-(6) with offering yield as the dependent variable. T-statistics are reported in brackets and standard errors are clustered at the state level. * $p < 0.10$, ** $p < 0.05$, *** $p < 0.01$

| *Dependent Variable*: | Yield Spread (basis points) | | | Offering Yield (basis points) | | |
|---|---|---|---|---|---|---|
| *Extent of Transition*: | Complete | | Partial | Complete | | Partial |
| | (1) | (2) | (3) | (4) | (5) | (6) |
| Treated × Post | 5.01 | 2.75 | 5.58 | -2.46 | -1.34 | 4.35 |
| | [0.82] | [0.48] | [1.00] | [-0.30] | [-0.26] | [1.10] |
| Issuer FE | ✓ | ✓ | ✓ | ✓ | ✓ | ✓ |
| State-Yr. FE | ✓ | ✓ | ✓ | ✓ | ✓ | ✓ |
| Bond Controls | | ✓ | ✓ | | ✓ | ✓ |
| County Controls | | ✓ | ✓ | | ✓ | ✓ |
| Adj.-$R^2$ | 0.725 | 0.913 | 0.885 | 0.293 | 0.900 | 0.864 |
| Obs. | 24,242 | 24,242 | 34,078 | 24,242 | 24,242 | 34,078 |

**Table 1.11:** Evidence from the Exit of Municipal Advisors (MA)

This table reports the results using Equation (4.1) with interactions to show the differential effect among issuers based on the exit of municipal advisors (MA). This analysis also includes group $\times$ year fixed effects. Column (1) shows results among all issuers with interactions corresponding to whether the issuer primarily depended on an exiting advisor or not. I define issuers linked to advisors when more than 50% of their municipal debt issuance in the pre-period is advised by the exiting advisor. Alternatively, results are robust to using a lower threshold of 25% of municipal debt issuance by the exiting advisor, as shown in Table A1.11. I focus on the exit of regular advisors. These represent municipal advisors with at lease one issuance in each calendar year before the SEC Municipal Advisor Rule in the sample. Columns (2) and (3) show results for issuers that depend on exiting advisors. For these issuers, I show the heterogeneity between small and large issuers based on the median size of ex-ante issuances. Results are robust to using the average size of ex-ante issuances (Table A1.12). Column (2) shows results for advised bonds only. Column (3) also includes bonds issued without any advisors, and the analysis does not include advisor $\times$ year fixed effects. T-statistics are reported in brackets and standard errors are clustered at the state level. $^{*}$ $p < 0.10$, $^{**}$ $p < 0.05$, $^{***}$ $p < 0.01$

| *Dependent Variable*: | Yield Spread (basis points) | | |
|---|---|---|---|
| *Sample of Issuers*: | All | Depending on Exiting MA | |
| Treated $\times$ Post | (1) | (2) | (3) |
| $\times$ Other | -16.61$^{***}$ | | |
| | [-5.60] | | |
| $\times$ Depended on Exiting MA | -0.73 | | |
| | [-0.22] | | |
| $\times$ Small | | 15.89$^{***}$ | 14.08$^{*}$ |
| | | [4.18] | [1.94] |
| $\times$ Large | | -8.59$^{**}$ | -8.20$^{***}$ |
| | | [-2.43] | [-2.68] |
| Difference | -15.87 | 24.47 | 22.28 |
| p-value | 0.00 | 0.00 | 0.01 |
| Issuer FE | ✓ | ✓ | ✓ |
| State-Yr. FE | ✓ | ✓ | ✓ |
| Controls | ✓ | ✓ | ✓ |
| Group-Yr. FE | ✓ | ✓ | ✓ |
| Advisor-Yr. FE | ✓ | ✓ | |
| Adj.-$R^2$ | 0.855 | 0.861 | 0.854 |
| Obs. | 834,741 | 234,679 | 275,381 |

**Table 1.12:** Evidence from Refunding Bonds

This table reports the baseline results for the sample of refunding bonds using Equation (4.1) estimating the differential effect on yield spreads of treated and control bonds after the Municipal Advisor Rule of 2014. The primary coefficient of interest, $\beta_0$, is captured by the interaction term of *Treated* $\times$ *Post*. I include the full set of controls and fixed effects in each Column corresponding to the baseline specification in Column (8) of Table 1.2, showing results for different dependent variables. Specifically, Column (1) reports the results using offering yields as the outcome. I use the after-tax yields as the dependent variable in Column (2). For Column (3), I use the preferred dependent variable in the form of yield spread. Finally, Column (4) shows the effect on after-tax yield spreads. I provide the description of key variables in Table A1. T-statistics are reported in brackets and standard errors are clustered at the state level. $^{*}$ $p < 0.10$, $^{**}$ $p < 0.05$, $^{***}$ $p < 0.01$

| *Dependent Variable*: | Offering Yield (bps) (1) | After-tax Yield (bps) (2) | Yield Spread (bps) (3) | After-tax Yield Spread (bps) (4) |
|---|---|---|---|---|
| Treated × Post | -7.93*** | -11.93*** | -7.54*** | -11.75*** |
| | [-2.95] | [-2.82] | [-3.07] | [-2.97] |
| Treated | 7.10** | 11.13** | 6.39*** | 10.37*** |
| | [2.59] | [2.52] | [2.77] | [2.69] |
| Post | -8.19*** | -14.90*** | -6.14*** | -12.78*** |
| | [-5.62] | [-5.76] | [-2.92] | [-4.09] |
| Issuer FE | ✓ | ✓ | ✓ | ✓ |
| Bond Controls | ✓ | ✓ | ✓ | ✓ |
| County Controls | ✓ | ✓ | ✓ | ✓ |
| State-Yr. FE | ✓ | ✓ | ✓ | ✓ |
| Advisor-Yr. FE | ✓ | ✓ | ✓ | ✓ |
| Adj.-$R^2$ | 0.880 | 0.880 | 0.864 | 0.865 |
| Obs. | 418,989 | 418,318 | 418,623 | 417,952 |

**Table 1.13:** Evidence from Timing of pre-refundings

This table reports the baseline results for the sample using Equation (4.1) estimating the differential effect on the timing of pre-refunding bonds for the treated and control bonds after the Municipal Advisor Rule of 2014. I restrict the sample to bonds that are pre-refunded and are callable. In Columns (1)-(6), I focus on all bonds with call option unlocking during the sample period. The primary coefficient of interest, $\beta_0$, is captured by the interaction term of *Treated* $\times$ *Post*. Our dependent variable is the gap between calling and refunding dates (in days) for bonds in Columns (1)-(5). I use the logged transformation of gap days in Columns (6) as the dependent variable. Specifically, Column (1) reports the results with issuer fixed effects. Column (2) shows results by additionally including state $\times$ year fixed effects. In Column (3), I include advisor fixed effects. Column (4) shows results with bond level controls consisting of coupon (%); log(amount issued in $); dummies for callable bonds, bond insurance, general obligation bond, bank qualification, refunding, and credit enhancement; remaining years to maturity; and inverse years to maturity. I provide the description of key variables in Table A1. In Column (5), I control for the county-level economic conditions. I use the lagged values for log(labor force) and unemployment rate, and the percentage change in unemployment rate and labor force, respectively. Column (6) shows the results with logged value of gap days as the dependent variable. T-statistics are reported in brackets and standard errors are clustered at the state level. $^{*}$ $p < 0.10$, $^{**}$ $p < 0.05$, $^{***}$ $p < 0.01$

| *Dependent Variable*: | Δ(Calling - Refunding) in days | | | | | Log(Δ) |
|---|---|---|---|---|---|---|
| | (1) | (2) | (3) | (4) | (5) | (6) |
| Treated × Post | -147.39** | -144.21** | -139.16** | -143.48** | -143.39** | -0.23* |
| | [-2.07] | [-2.26] | [-2.14] | [-2.13] | [-2.17] | [-1.70] |
| Treated | 42.60 | 51.16 | 44.44 | 21.20 | 16.83 | -0.14* |
| | [0.78] | [0.99] | [0.88] | [0.47] | [0.38] | [-1.72] |
| Post | -6.63 | 68.57 | 71.66* | 78.22* | 77.27* | 0.02 |
| | [-0.30] | [1.61] | [1.75] | [1.77] | [1.71] | [0.21] |
| Issuer FE | ✓ | ✓ | ✓ | ✓ | ✓ | ✓ |
| State-Yr. FE | | ✓ | ✓ | ✓ | ✓ | ✓ |
| Advisor FE | | | ✓ | ✓ | ✓ | ✓ |
| Bond Controls | | | | ✓ | ✓ | ✓ |
| County Controls | | | | | ✓ | ✓ |
| Adj.-$R^2$ | 0.491 | 0.561 | 0.583 | 0.611 | 0.612 | 0.654 |
| Obs. | 131,841 | 131,840 | 131,838 | 131,838 | 131,838 | 130,572 |

# CHAPTER 2

# IMPACT OF CORPORATE SUBSIDIES ON BORROWING COSTS OF LOCAL GOVERNMENTS: EVIDENCE FROM MUNICIPAL BONDS

## 2.1 Introduction

State and local governments in the United States compete intensely, by offering subsidies in the form of tax abatement and grants, to attract firms to their regions.[1] Targeted business incentives may help job creation, firm growth, and investment due to local agglomeration economies [64]. However, the forgone tax revenue and the additional demand for public services may require local governments, especially those that are financially constrained, to raise additional resources either by increasing taxes or by issuing additional debt, or both. Alternatively, local governments may cut spending on public services [65]. In this paper, we analyze the financial and real impact of winning large corporate investment deals on local communities.

A direct assessment of the economic impact of corporate subsidies on the local community is challenging given the significant uncertainty about the level and timing of the proposed investment, the number and type of jobs created[2], wages offered, and potential multiplier effects of these jobs [66]. Moreover, confounding events during the long gestation period complicates the measurement of multiplier effects and the associated costs of these corporate subsidy deals. In this paper, we shed light on the potential economic impact of large corporate subsidy deals by documenting their effects on the borrowing costs of local governments.

The $4 trillion municipal bond market is a significant source of financing for local

[1]238 cities made bids for Amazon HQ2 that promised $5 billion investment. The winners, New York City and Northern Virginia, offered tax rebates and other incentives totaling $5.5 billion.

[2]For example, in 2017, Wisconsin announced over $4 billion in subsidies to Foxconn. However, there was still significant uncertainty about the actual investment and the number of jobs it would create.

governments. They use these funds to build roads, and highways, waterworks, schools, hospitals, and other related infrastructure. For example, healthcare municipal bonds are one of the main sources of financing for non-profit hospitals, representing about 70% of all hospitals in the US [28]. Despite a low (0.08%) default rate, default risk accounts for at least 74% of the average municipal bond spread [5]. We test how municipal bond yields change with the announcement of corporate subsidy deals as they may affect both the revenue and expenditures of the local governments and hence any changes to their default risk.

Using hand-collected county-level data on the winners and runner-up bidders for large corporate investment deals during 2005–2018, we find that the bond yields of winning counties increase by 7.2 basis points (bps) in the secondary market within one year after the announcement of the subsidy. However, the losing counties do not experience a significant change. Within thirty-six months after the deal, we find an increase of 15.25 bps in after-tax yield spreads. This is equivalent to a 12.1% increase in credit spread or a reduction in bondholders' wealth of about $7.6 billion[3] in the secondary market.

Local governments are likely to benefit from subsidy deals that bring in greater economic activity including new jobs. However, local governments may need to raise capital through new municipal debt to provide for additional infrastructure, and many local governments are already highly indebted. For example, New York ranks the highest among states with a per capita debt of $18,000, representing 20% of the state GDP [67]. New York City itself has a per capita debt of nearly $14,997. A subsidy deal with high multiplier benefits may alleviate these debt capacity constraints. We find that our main result is mostly driven by subsidy deals with low (anticipated) jobs multiplier. Consistent with the cost-benefit trade-offs that counties face, we find that winning counties with a lower debt capacity experience higher borrowing costs. Finally, we find that even a high jobs multiplier does not seem to alleviate the binding debt capacity constraints of local governments.

[3] The average after-tax credit spread between A- and AAA-rated municipal bonds of winning counties in the sample before the deal equals 126 basis points. The outstanding municipal debt in the deal year was $\sim$ $631 billion for all winning counties. The average duration of bonds for winning counties in the year before the deal was 8.04 years, with an average yield of 2.89%. We use the modified duration approach to compute this impact on the bondholders' value for a yield increase of 15.25 bps three years after the deal on a semi-annual basis as $7.6 billion (=$631 $\times$ 8.04 $\times$ 0.0015/(1+2.89%/2) billion).

Identifying the causal impact of the corporate subsidy events on the borrowing costs of local governments is challenging since we cannot observe what would have happened if the winning county did not win the bid. So, we follow [68] and [69] and consider the closest runner-up bidder for the project (the 'losing county') as the counterfactual county. We manually parse through local print media/newspapers to find out the losing county (and state) and the earliest date of the announcement for the subsidy (see Section B2.2 of Internet Appendix for details). Typically, local governments and economic development board officials maintain secrecy about subsidy offers to avoid other competitors. For example, when Missouri bid for Freightquote's facility in 2012, the project was encoded as "Apple". Even after the project investment is announced, competing local governments may be bound by non-disclosure requirements from releasing details about the subsidy. Given such constraints, it was difficult to collect data on the subsidy packages offered by losing counties.[4] In the absence of such data, we first report evidence supporting the identifying assumption, i.e., the winning county and losing county follow similar economic trends before the deal. We find an upward (downward) trend for aggregate employment (unemployment rate) for both winning and losing counties after the subsidy announcement. Similarly, trends for county ratings and the underlying county risk using local betas [70] for the winning-losing county pairs are statistically similar. For instance, if a bidding county is too aggressive in the hopes of reversing its fortunes, we expect to find differential pre-trends for economic indicators like the unemployment rate, county ratings, etc., before the deal. However, the absence of such trends suggests that the losing county is an appropriate counterfactual for our analysis.

We use secondary market trades for 163,771 municipal bonds of the winning-losing county pairs for the 199 deals. We focus on secondary market trades to avoid any con-founding endogeneity due to market-timing in the issuance market of new municipal bonds. We use tax-adjusted yield spreads as the main dependent variable. We estimate event-study style difference-in-differences regression with winning-losing county-pair fixed effects, county fixed effects, and calendar year-month fixed effects, and include bond-specific controls and county-specific controls[5]. We find that the bond yields of the winning counties increase by approximately $15$ bps as compared to the losing counties within 36 months after the deal.

[4]There may be a related concern about state governments backing the subsidy deal. For 73 deal pairs in our sample, we are unable to find evidence of the subsidy package being sponsored by the state. Our baseline coefficient estimate is 16.29 bps and is statistically significant for this subset of deals.

Corporations do not randomly choose between the bidding counties [68]. In order to address this concern, we estimate a predictive regression of the winner dummy using various county-level ex-ante observable characteristics such as the level and changes in the unemployment rate, level and changes in the labor force, house price index, and income per capita. We do not find any of these observable county-level characteristics systematically predict the probability of winning the deal. These results show that a declining trend in economic conditions of the winning county is unlikely to be driving the higher bond yields after winning the deal[6].

To shed light on the economic mechanism underlying the higher bond yields of the winning counties, we analyze the winning counties based on the expected benefits and costs of corporate subsidies. To proxy for expected benefits, we use two measures of the expected jobs multiplier for a deal. First, we construct the measure of anticipated jobs multiplier by summing up the proportion of value-added in the upstream and downstream segments of a given industry, weighted by the corresponding county's share of wages. We find that the difference between winners and losers is $21$ bps with a low multiplier effect. However, for high multiplier deals, the difference between winners and losers is much lower. In our second measure, we focus on knowledge spillovers using the economic importance of the aggregate value of prior patents granted to the firm winning the subsidy deal [71]. We find that deals involving firms with low-value patents result in 21–23 bps higher bond yield spreads for the winning counties. These results suggest that deals with a lower expected jobs multiplier lead to a greater increase in bond yields for winners.

[5]We include coupon rate, size of issuance, remaining maturity, callability, bond insurance, and type of security based on bond repayment source (tax revenues for general obligation bonds and project-specific revenues for revenue bonds). For county-specific controls, we include lagged level and changes in the unemployment rate and labor force to control for local economic conditions.

[6]We also conduct a falsification test, wherein we consider the impact on the bonds of the winning county that have negligible credit risk (i.e., insured bonds which are pre-refunded with escrow accounts in state and local government securities). We find that the announcement of the subsidy deal does not have an impact on bonds of winning counties with minimal credit risk, further reinforcing our main results.

Following the subsidy deal, apart from the forgone tax revenues, the increased demand for public services may require the winning county to raise more municipal debt. We consider the ex-ante debt capacity of local governments in two ways: a) interest expenditure, and b) county credit ratings. We find that counties with higher interest expenditure (scaled by revenue) show a higher impact on yields (15–27 bps). We also find a higher probability of bond rating downgrades for winning counties with lower debt capacity (or higher interest expenditure). We also show that counties with lower ex-ante credit rating experience higher yield spread on their existing debt after the subsidy announcement. This effect is similar across general obligation and revenue bonds, separately. These results suggest that winning counties with a lower debt capacity experience a greater increase in bond yields after the subsidy deal.

Next, we test whether a high jobs multiplier can alleviate the debt capacity constraints of the local governments. We divide the winning counties into low and high groups based on the anticipated jobs multiplier and interact with our baseline coefficient. We find that the differential impact due to high-interest expenditure is similar in magnitude across both groups of anticipated jobs multiplier. The results suggest that a high multiplier does not alleviate the debt capacity constraints of local governments.[7]

One possible reason for the increase in municipal bond yields after winning the subsidy deal could be the aggressive bidding by local governments. There is heterogeneity in municipal governance as proxied by the number of corruption-related convictions among public officials [73]. We find that the impact on municipal bond yields is greater when the winning county belongs to a state with a higher level of corruption among public officials.

[7] Our results on bargaining power show how the relative size between the firm and the county also affects the distribution of gains from corporate subsidy deals [72].

We also find that compared to a year before the deal, the new municipal issuance for the winning counties increases about three times in the year after the deal. This is driven by counties with more debt capacity, i.e., those with lower interest expenditure (scaled by revenue). However, this increase is only about 1.5 times for the losing counties. Compared to the losing counties, there is an increase of about 5 bps in new issuance offering yields after the deal for the winners with low debt capacity. This is consistent with our results in the secondary market of municipal bond yields.

To understand the real impact of corporate subsidies on the local economy, we also examine employment at the county level around the deal. Our results suggest no meaningful change in employment growth, unemployment rate, annual payroll growth, and the number of establishments. However, we find that there is a significant decrease in the overall expenditure on public services at the county level. We measure these across elementary education, police and protection, and other services. For winning counties with lower debt capacity, the per capita expenditure on elementary education, hospitals, police and protection, and other expenditures decrease. Overall, debt-constrained counties appear to reduce their per capita expenditure on some local services without making meaningful gains in employment.

We contribute to the literature on the economics of location-based tax incentives [74, 75]. Our paper builds on the literature[8] about the overall implications of subsidy-based location economics by studying their impact on the yields of municipal bonds, a critical source of financing for local governments. Prior literature has focused on employment-related outcomes and spillovers of subsidies [69, 77]. To the best of our knowledge, our paper is the first attempt to use the municipal bond market as a lens to evaluate the impact of corporate subsidies on local communities. In this regard, our approach sheds new light on how policymakers' decisions may affect the wealth of local bondholders and the default risk of local bonds [5]. We also contribute to the recent literature documenting how local conditions affect municipal bond prices [38, 39, 40, 78]. Further, our results on the debt capacity constraints of borrowing counties contribute to the literature on the county's debt capacity [25].

[8] [68] document a 12% increase in total factor productivity (TFP) in incumbents of the winning county five years after the opening of a large plant, suggesting agglomeration gain to the county. [76] uses the state-level bidding process to show that the firms capture the welfare gains in subsidy competition.

Our hand-collected records of winning and losing counties for subsidy deals in the United States may also be useful for future studies. Our results suggest that some winning counties, especially those with a lower debt capacity, experience an increase in their bond yields after winning the subsidy deal.

The rest of the paper proceeds as follows. We discuss our empirical methodology and identification concerns in Section 3.2. Section 3.3 describes our data and provides summary statistics. We present empirical results in Section 4.4, and conclude in Section 4.5.

## 2.2 Identification Challenges and Methodology

In this section, we first discuss the challenges in identifying the impact of corporate subsidies and then describe our empirical specification.

### 2.2.1 Identification Challenges

The first econometric challenge is that the targeted subsidies are not random. Large corporations usually invite bids on subsidy packages from various counties that wish to attract investment in their jurisdiction. However, if a specific location is endowed with natural resources [74] or other strategic advantages pertinent to a specific kind of firms, they are more likely to get repeated investments in that sector or industry. Therefore, the assignment of the *winner* of a corporate subsidy deal may depend on multiple local factors. (See Section B2.1 of the Internet Appendix for details on corporate subsidies in the U.S.)

[79] argue that firms' decisions are governed by the expected future supply of inputs and the magnitude of subsidy offered by the county. This results in a two-way matching

between government decision-makers and corporate agents to determine the ‘winner’ between the bidding counties. To the extent that local officials cannot fully determine their chances of winning the plant by merely offering the higher subsidy, the assignment of ‘winner’ is closer to being random. The uncertainty in the final treatment assignment after the subsidy bids provides some support to the causal effect.

The next challenge is to identify the control group. Following [68], we denote a ‘winner’ as the bidding county that was chosen by the firm to locate their project and use the closest runner-up bidder, the ‘losing’ county, as a counterfactual. In an ideal experiment, we would like to have the same incentive package offered by the competing locations. However, it is difficult to obtain the data of subsidy offer made by the losing county because of the inherent secrecy maintained by local governments (see Figure B2.1 of the Internet Appendix). Regardless, there is some anecdotal evidence in support of a bidding process involving competitive subsidy bids offered by both the bidders[9].

Finally, another potential threat to our identification stems from the local economic conditions resulting in a negative selection. The underlying assumption in our identification strategy requires that the winning and corresponding losing county follow similar economic trends before the subsidy deal announcement. If the winning county is in worse economic shape, then its bond yields should be higher, which implies that our main effect is over-estimated. We plot the trends for bond yields, county-level aggregate employment, unemployment rate, bond rating, and local beta around the subsidy announcement. We do not find supporting evidence for negative selection (see Section 3.4.1 for details.)

### 2.2.2 Methodology

Our baseline event study focuses on the impact of corporate subsidies on the borrowing costs of local governments. Consistent with [68], we rely on the matching between firms and counties to identify the closest bidder as the counterfactual. This approach has the advantage of not introducing any researcher-specific biases in choosing the counterfactual. We carefully read newspaper articles to identify 199 winner-loser deal pairs at the county level spanning 42 states during 2005-2018 (See Section 2.3.1 for details). We use a three-year window before and after the subsidy announcements.[10] We use secondary market trades as the baseline case because these bonds are already trading in the winner-loser county pairs at the time of the deal announcement (and mitigate any concerns about bond issuance related to subsidy deals driving our results).

[9]For example, Kansas and Missouri arrived at a subsidy armistice only in August 2019 after a history of shuffling jobs across the border: https://www.wsj.com/articles/the-kansas-missouri-subsidy-armistice-11565824671

We use a standard difference-in-differences approach between the treatment and control counties' bond yields in the secondary market for municipal bonds. This results in the baseline specification as below:

$$
\begin{aligned}
y_{i,c,p,t} = {} & \alpha + \beta_0 * Winner_{i,c,p} * Post_{i,c,t} + \beta_1 * Winner_{i,c,p} + \beta_2 * Post_{i,c,t} \\
& + BondControls + CountyControls + \eta_p + \gamma_c + \kappa_t + \epsilon_{i,c,p,t}
\end{aligned}
\tag{2.1}
$$

where index $i$ refers to bond, $c$ refers to county, $p$ denotes the (winner-loser) county pair and $t$ indicates the year-month. After-tax yield spread is the main dependent variable in $y_{i,c,p,t}$ obtained from secondary market trades in local municipal bonds (described in Section 2.3.2). We also use the raw average yield, after-tax yield, and yield spread as dependent variables in additional tests. *Winner* corresponds to a dummy set to one for a county that ultimately wins the subsidy deal. This dummy equals zero for the runner-up county in that subsidy deal. *Post* represents a dummy that is assigned a value of one for *event* months after the deal is announced and zero otherwise. The main coefficient of interest is $\beta_0$ which comes from the interaction term, *Winner* $\times$ *Post*. The baseline specification also includes three sets of fixed effects: $\eta_p$, denoting county pair fixed effects to ensure that the comparisons are within bonds mapped to a winner-loser pair; $\gamma_c$, denoting county fixed effects to absorb unobserved heterogeneity at the county-level; and, $\kappa_t$, denoting *calendar* year-month fixed effects to control for time trends. The main coefficient of interaction ($\beta_0$) captures the difference in the outcome variables within a deal between winner-loser county pairs, after controlling for the average effect across counties and time. We follow [80, 39] to include amount issued, coupon rate, dummy for status of insurance and dummy based on general obligation versus revenue bond security type, collectively represented as *BondControls*. *CountyControls* refers to a vector of county level measures to control for local economic conditions. It includes the lagged value of log of labor force in the county, lagged county unemployment rate, the percentage change in the annual labor force level, and the percentage change in the annual unemployment rate. In all our specifications we double cluster standard errors at the county-specific bond issuer and year-month level, unless specified otherwise.

[10]Our results are robust to using other windows, as shown in Panel B of Table 2.3

Our difference-in-differences approach following [68] affords us some advantages over previously used methods in the literature. First, we do not compare the winning counties with all other counties in the US. Such a regression is likely to lead to biased estimates due to unobserved heterogeneity between the two sets of counties. Counties that offer large subsidies could be fundamentally different from the rest of the counties within the US. Plausibly, a county that is likely to gain substantially from a particular firm locating within it is more likely to attract the project with greater incentives. Simultaneously, a county with a greater need to increase jobs is likely to offer an aggressive incentives package. By doing so, it could try to overcome its inherent disadvantages and influence the firms' location decisions. These omitted factors may also be correlated with the bond yields of the respective local issuers. By restricting the sample to only those that were also involved in bidding for the same corporation at the same time, we reduce the bias from such unobserved heterogeneity.

## 2.3 Data

In this section, we provide details about the data used in this paper. First, in Section 2.3.1, we describe our data on corporate subsidies. In Section 2.3.2, we discuss the data used from the municipal bond market. Finally, we describe some other variables used in this study in Section 3.3.3.

### 2.3.1 Corporate Subsidies

The *Good Jobs First Subsidy Tracker* [81] provides a starting point with its compilation on establishment-level spending data. As shown in Figure 2.1, states and the federal government spent more than USD 10 billion every year in corporate subsidies after the financial crisis of 2009. Further, there has been an increase in the portion of subsidies offered by state governments during the sample period of 2005-2018. States differ in the amount of subsidy they have offered in the past, with New York, Louisiana, and Michigan ranking among the top three (see Figure 2.2 for a ranking among states). On a per-capita basis, Washington, Oregon, and Louisiana spent over USD 1,500 during this period. (Figure B2.2 of the Internet Appendix depicts the subsidy value per capita using a choropleth map with five breaks shown in the legend.)

One of the challenges that previous studies faced in evaluating the impact of corporate subsidies was the lack of comprehensive data at the county-level. (See literature on location-based incentives in Section B2.1 of the Internet Appendix.) The identification used in this paper relies on close-bidding auctions where two cities compete against each other to attract a firm. Their respective states may back local governments in sponsoring the subsidy. However, there is no published data source documenting such competing bids based on subsidy. One contribution of our paper is to provide the first records of winning and losing counties for large subsidy (defined as those exceeding USD 1 million) deals in the United States. We detail the construction of the data in Section B2.2 of the In-

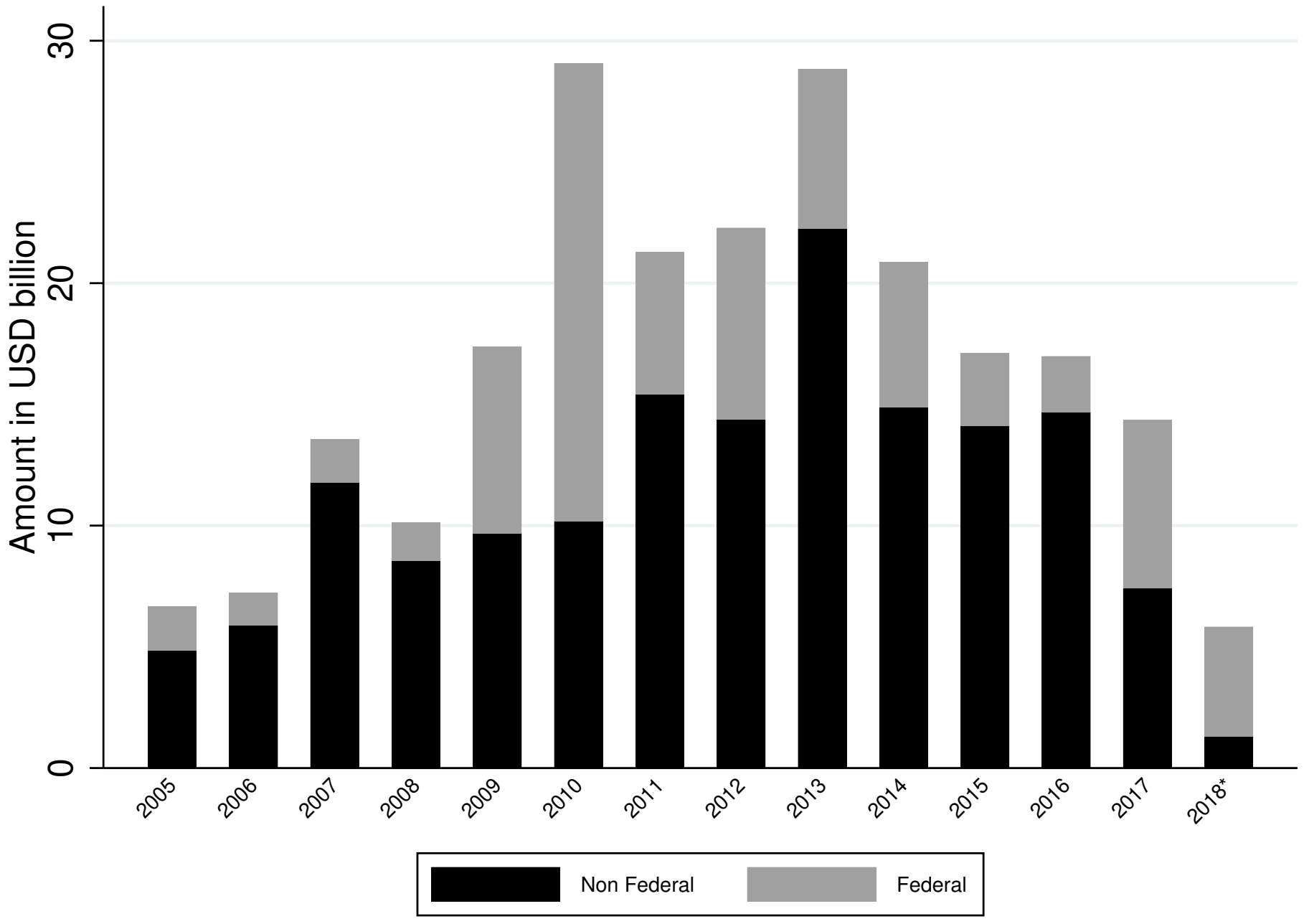

**Figure 2.1:** Total Subsidy:
The vertical bars show the aggregated value of total subsidy offered by federal and non-federal (state and local) governments for each year during 2005-2018. This does not include federal loans. Calculated based on Source: Good Jobs First, Subsidy Tracker. *Denotes incomplete data for the year, until June 2018 ternet Appendix. In Table B2.1 of the Internet Appendix, we show a comparison of the original data set from *Good Jobs First Subsidy Tracker* versus that constructed after the hand-collection of relevant variables. Hand-collection was especially difficult due to inherent secrecy maintained by the bidding local governments. For example, when Missouri bid for Freightquote's facility in 2012, the project was encoded as "Apple" (see Internet Appendix, Table B2.2). Further, it is difficult to obtain all the bidders in a given subsidy project due to multiple stages involved in the negotiations.

We identify winner-loser deal pairs at the county level. We provide a summary of the subsidy deals in our final sample in Table 2.1. Panel A shows the distribution across all deals. The mean subsidy amount in the deals is USD 202 million, whereas the median amount is USD 81 million. Comparing this to the proposed investment, we find that the median deal gets a 22% subsidy as a proportion of investment. The median deal involved 800 jobs promised by the firm. The average subsidy per job promised by the firm amounts

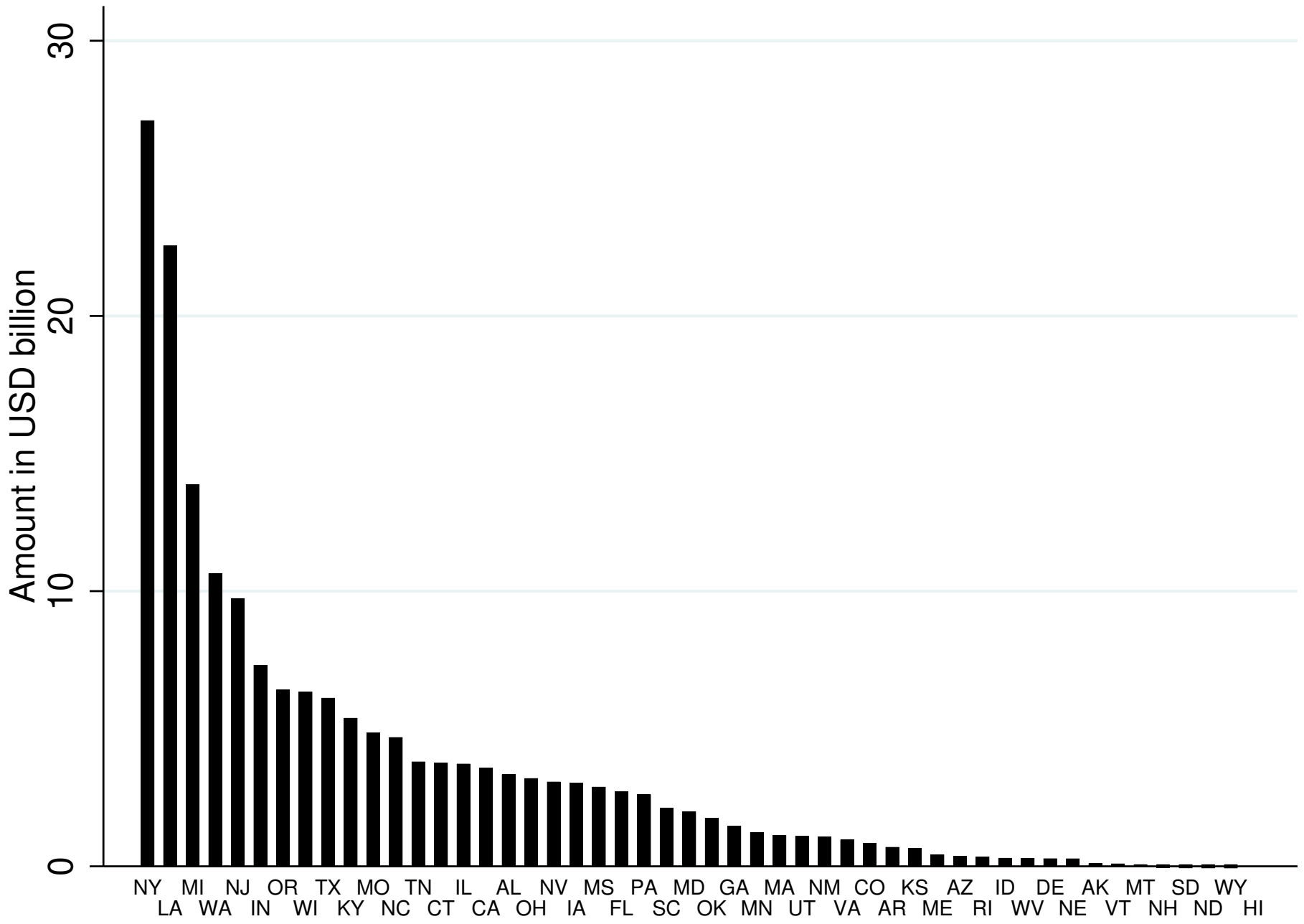

**Figure 2.2:** Total Subsidy by States:
The figure shows ranking among US states based on total non-federal subsidy offered during 2005-2018. Calculated based on Source: Good Jobs First, Subsidy Tracker to nearly USD 324,000. Panel B shows that most of these deals are for new/expansion projects with about half the deals in manufacturing (Panel C)[11]. In Table B2.3 of the Internet Appendix, we evaluate probable metrics in the data, which may help predict the level of subsidy offered by the winning counties. We find the amount of investment and jobs promised to be strongly correlated. Figure B2.3 of the Internet Appendix provides a distribution of the subsidy amounts over different buckets. Each bin between USD 5 million to USD 500 million has at least 20 deals in it.

**[Insert Table 2.1 here]**

[11]For some subsidy deals, we are not able to identify the runner-up counties. However, we argue that the deals included in the sample are not affected by this sample selection issue. In Figure B2.4 of the Internet Appendix, we provide a comparison of counties in the sample (for which we could find the loser) and those excluded from the sample (for which we could not find the losing county) using the interest expense on general debt. We scale the interest expenditure by various county-level measures such as total debt, total revenue, and total expenditure. The distribution for the two groups looks similar which provides comfort against sample selection issues.

### 2.3.2 Municipal Bonds

Municipal bond characteristics are obtained from the Municipal Bonds dataset by FTSE Russell (formerly known as Mergent MBSD). We retrieve the key bond characteristics such as CUSIP, dated date, amount issued, size of the issue, state of the issuing authority, name of the issuer, offering yield, status of tax exemption, insurance status, pre-refunding status, type of bid, coupon rate, and maturity date for the bonds. We also use S&P credit ratings for these bonds by reconstructing the time-series of the most recent ratings from the history of CUSIP-level rating changes. We encode character ratings into numerically equivalent values ranging from 28 for the highest quality to 1 for the lowest.

An important step in our data construction is to link the bonds issued at the local level to the counties that make the subsidy bids. This geographic mapping allows us to study the implications on other economic variables using data on demographics and county-level financial metrics. Since the FTSE Municipal Bonds dataset does not have the county name for each bond, we need to supplement this information from other sources, such as Bloomberg. However, in light of Bloomberg's download limit, it is not feasible to search for information on each CUSIP individually. Therefore, we first extract the first six digits of the CUSIP to arrive at the issuer's identity[12]. Out of 63,754 unique issuer identities (6-digit CUSIPs), Bloomberg provides us with county-state names on 59,901 issuers. For these issuers, we match the Federal Information Processing Standards (FIPS) code. The FIPS code is then used as the matching key between bonds and bidding counties involved in offering corporate subsidies. We also match the names of issuers to the type of (issuer) government (state, city, county, other) on Electronic Municipal Market Access (EMMA) provided by Municipal Securities Rulemaking Board. We use this information to distinguish local bonds from state-level bonds because we are interested in the non-state bonds.

We use the Municipal Securities Rulemaking Board (MSRB) database on secondary

[12]The 9-digit CUSIP consists of the first six characters representing the base that identifies the bond issuer. The seventh and eighth digits identify the type of the bond or the issue. The ninth digit is a check digit that is generated automatically.

market transactions during 2005-2019. Our paper closely follows [39] in aggregating the volume-weighted trades to a monthly level. Following [82, 38], we only use customer buy trades to eliminate the possibility of bid-ask bounce effects. Table B2.4 of the Internet Appendix summarizes each step of the sample construction [5]. Given our primary focus on the borrowing cost from secondary market yields, our sample is derived from the joint overlap between the bond characteristics and bond trades at the CUSIP level. In matching the bond transactions from secondary market data to their respective issuance characteristics (from FTSE Russell), we rely on the CUSIP as the key identifier. In Table 2.2, we provide descriptive statistics on bond features pertaining to the primary market and secondary market. The average bond in the sample has a weighted average yield of 2.8% in the secondary market, with a remaining maturity of 10.7 years and 11.2 years for the winning and losing counties, respectively. We describe the key variables in Table B1.

**[Insert Table 2.2 here]**

The primary outcome variable used in Equation (4.1) is the tax-adjusted spread over the risk-free rate. We calculate the bond's coupon-equivalent risk-free yield as in [39][13]. Tax adjustment follows [5] wherein the marginal tax rate impounded in the tax-exempt bond yields is assumed to be the top statutory income tax rate in each state. This is consistent with the broad base of high net worth individuals and households who form a major section of investors in the US municipal bond market (often through mutual funds). A detailed study on tax segmentation across states by [46] shows significant costs on both issuers and investors in the form of higher yields. In particular, we use:

$$1 - \tau_{s,t} = (1 - \tau_t^{\text{fed}}) * (1 - \tau_{s,t}^{\text{state}})$$

[13]First, we calculate the present value of coupon payments and the face value of a municipal bond using the US treasury yield curve based on zero-coupon yields as given by [48]. Using this price of the coupon-equivalent risk-free bond, the coupon payments, and the face-value payment, we get the risk-free yield to maturity. Finally, the yield spread is calculated as the difference between the municipal bond yield observed in the trades and the risk-free yield to maturity calculated. This yield spread calculation is similar to [49].

To compute the tax-adjusted spread on secondary market yields:

$$spread_{i,t} = \frac{y_{i,t}}{(1 - \tau_{s,t})} - r_t,$$

where $r_t$ corresponds to the maturity–matched coupon-equivalent risk-free yield for a bond traded at time t. Similar to [5], we use the top federal income tax rate as 35% from 2005 to 2012, 39.6% from 2013 to 2017, and 37% from 2018 to 2019. We also consider tax-exemption at county-level and discuss this in Section 3.4.4.

### 2.3.3 Other Variables

We use data on county finances from the Census Bureau Annual Survey of Local Government Finances to get details on revenue, property tax, expenditures, and indebtedness of the local bodies. This gives us detailed constituents of revenue and tax components at the local level, which we use in additional tests to examine the implications for our main results. Our data on county-level household income is from the Internal Revenue Service (IRS) and is used as the total personal income at the county level. Our unemployment data comes from the Bureau of Labor Statistics. We use input-output tables from the Bureau of Labor Statistics (BLS). For the county-level population, we use data from the Surveillance, Epidemiology, and End Results (SEER) Program under the National Cancer Institute. We obtain county-level data on the number of establishments and annual payroll growth from County Business Pattern (CBP). As a proxy for the risk-free rate, we use zero-coupon yield provided by FEDS, which provides continuously compounded yields for maturities up to 30 years. To get tax-adjusted yield spreads, we use the highest income tax bracket for the corresponding state of the bond issuer from the Federation of Tax Administrators.

## 2.4 Results

We discuss our baseline results (Section 3.4.1) for Equation (4.1), including evidence from the dynamics using the raw data on secondary market municipal yields and evidence on the parallel pre-trends assumption. Section 3.4.4 shows robustness tests for our baseline specification. We propose the potential mechanism to explain our results in Section 3.4.2. Finally, we discuss the impact on the primary market of municipal bonds (Section 3.4.5) and then on the local economy, property taxes, and public expenditure (Section 2.4.5).

### 2.4.1 Impact on Borrowing Costs of Local Governments

*Dynamics and Baseline Results*

We begin our analysis by plotting the yield spreads observed in the secondary market between the winning and losing counties. Our event window comprises three years before and three years after the subsidy deal announcement. We use the quarter before the event window (T=-37 to T=-39 months) as the benchmark period to evaluate the pre-trends between the treatment and control groups. We depict the observations aggregated to a quarterly scale to mitigate the inherent limitations of liquidity in the municipal bond market. We plot the yield spreads based on Equation (3.4) below:

$$y_{i,c,p,t} = \alpha + \beta_q * \sum_{q=-12}^{q=12} Winner_{i,c,q} * Post_{i,c,q} + \delta_q * \sum_{q=-12}^{q=12} Loser_{i,c,q} * Post_{i,c,q} \quad (2.2)$$
$$+ \eta_p + \kappa_t + t.\gamma_c + \epsilon_{i,c,p,t}$$

where index $i$ refers to bond, $c$ refers to county, $p$ denotes the (winner-loser) county pair, $t$ indicates the event month, and $q$ refers to the quarter corresponding to the event month $t$. Yield spreads and after-tax yield spread are the dependent variables in $y_{i,c,p,t}$ obtained from secondary market trades in local municipal bonds. $\eta_p$ represents the (winner-loser) county pair fixed effects and $\kappa_t$ represents the calendar year-month fixed effects. $t.\gamma_c$ represents linear time (year-month) trends interacted with county dummies to control for time trends for a given county in this dynamic analysis. The coefficients $\beta_q$ and $\delta_q$ represent the average change in yield spreads with respect to the benchmark period for the winning and losing counties, respectively. We depict the coefficient estimates from the regression in Figure 2.3.

**Yield Spread (bp)**

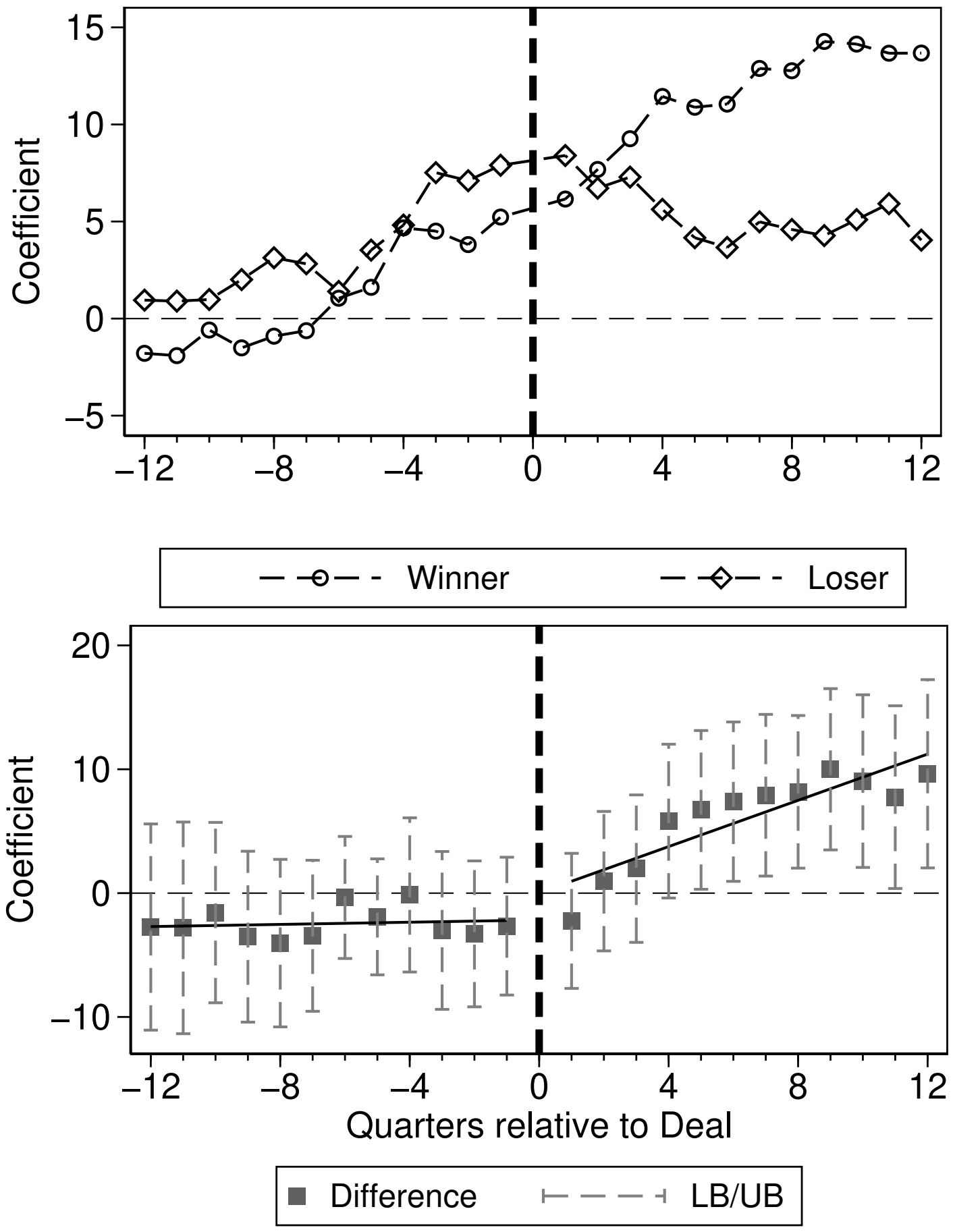

**Figure 2.3:** Baseline Result - Winner vs Loser:
In this figure, we plot the yield spread for municipal bonds traded using Equation (3.4). We also show the differences between the yields of winning and losing counties. See Table B1 for variables description. The coefficients are shown in basis points. We regress the yield spreads on monthly interaction dummies for winner and loser using county-pair (winner-loser pair) fixed effects, county fixed effects, and year-month fixed effects. We also add county specific year-month trends. We depict the coefficients on a quarterly scale on the x-axis, where 0 corresponds to the month of the subsidy deal announcement. The first quarter dummy after announcement subsumes the month of announcement. The omitted benchmark period is a quarter before the event window around the deal, i.e., (-12,12) quarters. Standard errors are double clustered by county bond issuer and year-month. The dashed lines represent 95% confidence intervals.

In Figure 2.3, the dashed line with circles plots the yield spreads over the 3-year window for winning counties depicted quarterly. The losing counties are depicted using a dashed line with diamonds. We plot the corresponding differences with 95% confidence intervals in the figure. We notice that in the pre-period, the difference between the yield spreads for winners and losers is negligible. Based on the fitted line in the pre-period, we find supporting evidence for the parallel pre-trends assumption between the winning and losing counties (See Section 3.4.1 for additional evidence on parallel trends assumption). Interestingly, we notice that for both winning and losing counties, the yield spreads increase similarly one year before the announcement of the subsidy deal. However, the difference between the two groups remains statistically insignificant. Second, the secondary market yield spreads for the winning counties appear to be higher than those of the losing counties in the first year after the deal. This difference increases to 5.8 bps by the fourth quarter after the subsidy deal and is statistically significant. The fitted line for the post-period suggests an increase in the difference between the winning and losing counties after the subsidy announcement.[14]

Note that the above results only represent the raw difference in yield spreads between the two groups by stacking the 199 deal-pairs in our sample into an aggregated set. These findings do not control for differences in bond characteristics and local economic conditions over time. Next, we estimate our difference-in-differences using our baseline Equation (4.1). Here, the coefficient $\beta_0$ of the interaction term, $Winner \times Post$, identifies the differential effect after the subsidy deal announcement on after-tax yield spreads. We compare the winning counties to the losing counties while controlling for observable characteristics. To revisit our identifying assumption: the losing county serves as an adequate counterfactual to map how the winner's yield spreads would have changed in the absence of the deal announcement. The county–pair fixed effects ensure estimation from within each deal pair. Using county fixed effects helps us absorb any unobserved variation due to the bidding county itself. The calendar year-month fixed effects control for declining yield spreads in the overall municipal bond market during our sample period, over and above the spread adjustment for coupon-equivalent risk-free yield spreads.

[14]We extend the post-window up to five years and find an increasing effect in Panel B of Table 2.3.

**[Insert Table 2.3 here]**

Table 2.3, Panel A reports the effect of winning a subsidy deal on the municipal bond yield spreads using Equation (4.1). In Column (1) - Column (3), we estimate the regression equation using the yield spread as the dependent variable. Specifically, Column (1) denotes the estimates without using any controls. We use bond level controls in Column (2), which consist of the coupon (%); log(amount issued in USD); dummies for callable bonds, bond insurance, general obligation bond, and competitively issued bonds; remaining years to maturity; and inverse years to maturity. We describe the key variables in Table B1. In Column (3), we use county controls for levels and trends in the local economy. We use the lagged values (to the year of deal announcement) for log(labor force) and unemployment rate, and the percentage change in the unemployment rate and labor force, respectively. Since subsidies are often motivated by job creation, we use these measures at the county level consistent with the previous literature.

We follow the same scheme and show our results using after-tax yield spread as a dependent variable in Column (4) - Column (6). Consistent with [5], we adjust the yield spread for taxes because most municipal bonds in our sample are tax-exempt securities. For robustness, we report our results using unadjusted raw average yield and yield spread as dependent variables in Section B2.3.2 of the Internet Appendix.

Using Column (6) of after-tax yield spread as our baseline case implies that the yield spread for winning counties increases by 15.25 bps after the subsidy announcement, in comparison to the losing counties. The 15.25 bps is equivalent to a reduction in bond-holders' wealth amounting to 19.07% (=7.63/40) of the total subsidy ($40 billion) offered during the sample period. To arrive at this magnitude, we start with the outstanding municipal debt of the winning counties in our sample. We find that this amount is $\sim$ $631 billion in the deal year. The average duration of bonds for winning counties in the year before the deal was 8.04 years, with an average yield of 2.89%. Using a modified duration approach, we compute the aggregate impact for three years after the deal on a semi-annual basis as \$631 $\times$ 8.04 $\times$ 0.001525/(1+2.89%/2) billion = \$7.63 billion.

Next, in Panel B, we show the baseline result of Column (6) using different forward windows, keeping the pre-event window the same as three years. We find that the magnitude of the differential impact increases from seven bps within the first twelve months after the event (Column (1)) to about 21 bps in 5 years (Column(7)). There seems to be a gradual increase in magnitude, which likely persists beyond the immediate near-term. To evaluate the sensitivity of our results against the choice of the window used, we discuss robustness of our main result in Section 3.4.4.[15] In the next sub-section, we provide more evidence on the parallel trends assumption.

*Do bond yield spreads respond to underlying local economic differences?*

Our baseline comparison between winning and losing counties' yield spreads assumes similar local economic conditions between the treatment and control groups during the event window around the deal. The results in Section 3.4.1 suggest that winning and losing counties exhibit parallel pre-trends in their bond yield spreads. However, as discussed before, the decision by local governments to engage in the bidding process to attract firms may not be random. The local administration may be attempting to create new jobs or to retain existing ones by offering incentives. It could be the case that bondholders from these counties are responding to underlying differences between the winning and losing counties. We test such underlying economic differences based on some relevant observable economic indicators. We present the comparison of the average trends at the county-level in a) aggregate employment, b) unemployment rate, c) county-level municipal bond rating, and d) local beta between winners and losers in Figure 2.4. In each of these subplots, we use the annualized version of Equation (3.4).

[15]We find some evidence for an increase in volume traded for both customers buy and customers sell trades (see Internet Appendix, Table B2.5).

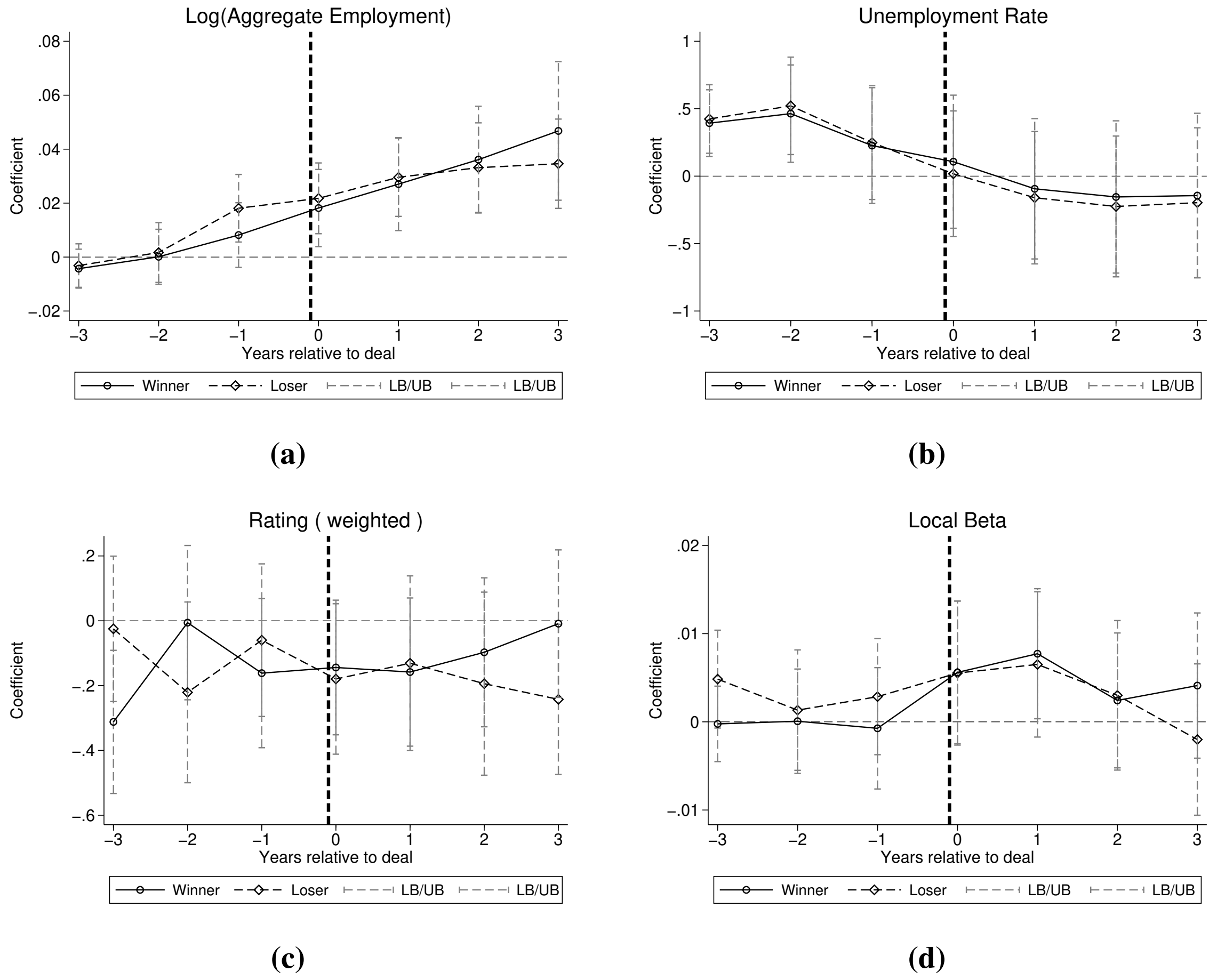

**Figure 2.4:** Identifying Assumption - Local Economy:
The figure shows the local economic conditions at the county level between the bidding counties, around the event of subsidy announcement. We use the annualized version of Equation (3.4). Here, we cluster standard errors at the deal level. The benchmark period is the year before the window (-3,3) years. The dashed lines represent 95% confidence intervals.

In Figures 2.4a and 2.4b, we find that aggregate employment shows an upward trend while the unemployment rate decreases. Both winning and losing counties seem to follow a similar trajectory with no statistical difference between them. This supports our parallel trends assumption on these key metrics related to employment. However, it is worth noting that after the subsidy deal, the increase in employment in the winning county is similar to that in the losing county.

Further, Figures 2.4c and 2.4d provide a comparison of the county level credit-worthiness and riskiness, respectively. We use credit ratings from general obligation bonds aggregated up to the county-year to get the county level ratings. As shown in the figure, the two groups do not show any difference in trends. Since the rating of the winners is not worse than that of the losers, this also helps against the concern of negative selection. Finally, the local beta is a measure defined in [70]. Using this as a proxy for the underlying riskiness of the counties, we find that both winning and losing counties had similar local beta during the event window. Overall, the results suggest that winning and losing counties look similar based on local economic conditions during the event period[16].

Next, we estimate a multivariate linear probability model to understand if the local economic factors jointly determine the probability of winning a deal by the county. We use the local conditions during the three years before the deal as the regressors. In addition to using the four control variables in our baseline specification on unemployment and labor force, we further introduce income per capita and house price index. Table B2.6 of the Internet Appendix shows the regression results where we introduce each regressor successively. We plot the coefficients from Column (6) in Figure B2.5 of the Internet Appendix and show the confidence intervals at the 95% level. For each metric on the y-axis, we show the explanatory power in determining the 'winner' dummy. We find that the coefficient for none of these key local metrics significantly differs from zero[17].

In our baseline specification, we use county-level controls to absorb variations in key economic metrics that may be relevant to the subsidy offer. However, different issuers in the same county may have different risks and credit ratings (and thus yield spreads), and may respond differently to investment by a subsidized firm. To address this heterogeneity among municipal bonds, we include bond purpose fixed effects and bond purpose $\times$ year-month fixed effects in additional robustness checks to mitigate these concerns (see Section B2.3.2,

[16] We also show the parallel trends along these dimensions by dividing the winners based on their debt capacity using $\textit{Interest/Revenue}_1$ ratio (see Figure B2.6, Internet Appendix).

[17] Additionally, in Figure B2.7 and Figure B2.8 of the Internet Appendix, we also verify against improper pairing between winners and losers based on local economic conditions. By dividing the winning counties into low and high groups based on the unemployment rate and household income, we show that there is no statistical difference in the unemployment rates between correspondingly matched winners and losers. Further, even the matching based on county-level credit rating and local beta suggests against any evidence of mismatching good counties with bad ones.

Internet Appendix). Our results remain similar and statistically significant. Compared to the baseline magnitude of 15.25 bps, our coefficients in this analysis range between 13-15 bps.

### 2.4.2 Robustness Tests

In this section, we test the robustness of our main result in Column (6) of Table 2.3 (Panel A) to various alternative specifications. We present the results of these robustness checks in Table 2.4.

**[Insert Table 2.4 here]**

*Controlling for Unobservables*

Our baseline specification controls for relevant time-varying county-level observables. We now consider whether our results are robust to a host of unobserved factors at the county level, such as the economic environment, which cannot be controlled for using observables. Panel A of Table 2.4 shows results after controlling for unobservable factors. First, in Column (1) we use deal $\times$ county fixed effects to account for unobserved variation at the county level within deals, where deals represent winner-loser county pairs. The resulting magnitude for the baseline effect is 15.29 bps. There could be time-specific county-level unobserved factors that may determine the winner-loser status for a given corporate subsidy deal. In Column (2), we address this issue by using a stricter specification where we include deal $\times$ county $\times$ year fixed effects. This absorbs unobserved variation over the years across counties within subsidy deals. Thus we are left with the monthly variation in yield spreads across bonds within deal-county-year. We find that the baseline effect reduces to 5.73 bps and remains statistically significant. Our results in Column (3) use an alternative way to absorb unobserved heterogeneity at the county level over year-months. We use indicator dummies to control for monthly county trends and find that after-tax yield spreads increase by 9.10 bps for the winning counties. Next, to control for issuer-level character-istics among the bidding counties, we show our results with issuer-fixed effects introduced to the baseline specification in Column (4). This may be relevant because municipal bonds are issued in a series corresponding to the given issuer's bond sale program. The increase in bond yield spreads is about 13.41 bps under this specification.

*Additional Tax Considerations and Bond Duration*

The change in yields may be affected by differences in the tax treatment of municipal bonds. Municipal bonds are exempt from federal and state-level income taxes in most states, especially for bonds issued within the state itself. However, there are cases where some bonds issued by local governments may not qualify for tax exemption. Additionally, there are four states which do not offer an exemption on state-level taxes for municipal bonds issued by them, namely: Illinois, Iowa, Oklahoma and Wisconsin. We account for these considerations in Columns (1) and (2) of Table 2.4. We provide evidence from additional tax considerations and duration in Panel B of Table 2.4. First, we drop bonds that do not qualify for exemption from state-level taxes in any state. Our main result in this case amounts to 16.25 bps in Column (1). Second, we drop subsidy deals involving the four states mentioned above (as either of the bidders). For this consideration, we report our results in Column (2) as 15.83 bps. Since both of these magnitudes are higher than the baseline effect, we argue that accounting for these additional tax considerations does not weaken our baseline result.

Additionally, there may be a concern that local residents may price in expectations of higher local individual income tax rates after the subsidy announcement. Municipal bond interest income may often be exempt from local taxes as well, leading to triple tax exemption. In this case, a higher future expectation of local tax rates may decrease bond yield spreads in winning counties. On the other hand, if the bondholders expect that the local administration may reduce local tax rates to attract more businesses after the subsidy, the bond yield spreads may go up. To understand more about this ambiguity, we next consider the individual income tax rates in Column (3). In the United States, most cities and counties do not impose a local income tax. Only 4,943 jurisdictions in 17 states impose local income taxes, as per Tax Foundation[18]. We obtain local individual income tax rates for 2011 for counties in our sample from the Tax Foundation website. We assume these values to hold for other years. We apply the local tax rate adjustment to bond yield spreads, over and above the state tax rate. In Column (3), the main coefficient of interest is estimated as 15.34 bps, which is very similar to our baseline result. Based on these results, we argue that our results are not explained by the bondholders' expectations for a change in local individual income tax rates.

Finally, we consider the non-linearity in bond level payoffs by accounting for duration effects in the baseline specification. In this regard, we modify the baseline specification in Columns (4)-(5). First, in Column (4), we show our main effect by replacing years to maturity and inverse years to maturity at the bond level by the corresponding duration using (pre-tax) average yield for the bond-month observation. This results in a higher impact of 16.05 bps. We show the same result by re-calculating duration based on after-tax yields in Column (5). This tax adjustment further increases the impact to 16.42 bps.

*Alternative Event Window and Sample*

We use a three-year window in our baseline specification. There might be a concern about how sensitive our results are to the choice of the event window. Specifically, our results in Panel C bring out the robustness of our main effect to the choice of the event window. First, in Column (1), we use a shorter window than the baseline. We find that the bond yield spreads increase by 10.53 bps when using a 24-month event window. In Columns (2) and (3), we show our main result using longer windows of 48 months and 60 months, respectively. We find the magnitude to be higher than our baseline effect and statistically significant. In Column (4), we show the baseline result by dropping deals involving big cities. We identify these based on the number of bond-month observations for each deal. We drop the top 10 deals. In summary, these alternative definitions of the event window and sample show robustness to our main specification in this regard.

[18]https://taxfoundation.org/local-income-taxes-city-and-county-level-income-and-wage-taxes-continue-wane/

Overall, we provide results in this section for the robustness of our baseline specification to observable and unobservable factors, additional tax considerations and duration, and choice of the event window. Separately, we conduct a falsification test, wherein we consider the impact on the bonds of the winning county that have negligible credit risk (i.e., insured bonds which are pre-refunded with escrow accounts in state and local government securities). We find that the announcement of the subsidy deal does not have an impact on bonds of winning counties with minimal credit risk, further reinforcing our main results (see Section B2.3.1, Internet Appendix for further details). We provide further robustness checks on the size of the trade, subsidy deals related to the financial crisis of 2009, recently issued bonds, additional county-level controls, heterogeneity of bond purpose, clustering standard errors, and the choice of other dependent variables in Section B2.3.2 of the Internet Appendix. We also show that our results are robust to dropping deals involving subsidies backed by both state and local governments.

#### 2.4.3 Mechanism

As we discussed before, the local governments face a trade-off while using targeted business incentives, i.e., anticipated jobs multiplier benefit (see [68]) versus forgoing future tax revenue. Our results in the previous section suggest that bondholders' value reduces on average after the subsidy announcement. However, it is unclear whether this reduction is due to lower anticipated benefits (jobs) from the project or the higher cost of financing the deal with forgone revenue. This section sheds light on how our results vary with the anticipated jobs multiplier and the underlying debt capacity of the winning counties.

*Anticipated Jobs Multiplier Effects*

First, we evaluate the heterogeneity of our results based on potential benefits after the subsidy deal. As noted before, a direct assessment of the future economic impact of the subsidies on the local community is challenging. Most of these projects have a long gestation period and the benefits get realized over a longer horizon. We hypothesize that the municipal bond prices reflect the expected local benefit from the subsidy deal. We measure the anticipated multiplier effects using two proxies: a) anticipated jobs multiplier using input-output tables and b) knowledge spillover using firm patents.

We construct the measure of the anticipated jobs multiplier effect by summing up the proportion of value-added in the upstream and downstream segments of the firm's industry, weighted by the corresponding county's share of wages and share of employment, respectively. This ex-ante measure is constructed three years before the subsidy deal announcement. We obtain sector-level data on upstream and downstream value-added fractions from real-valued input-output tables provided by the Bureau of Labor Statistics (BLS). Our sector-level NAICS mapping comes from the BLS crosswalk. See Table B1 for variables description.

**[Insert Table 2.5 here]**

For all the triple-interaction analysis henceforth, we interact *Winner* × *Post*, *Winner*, and *Post* dummies in our baseline Equation (4.1) with the corresponding groups. In this case, we interact with dummies corresponding to high and low values of anticipated jobs multiplier effect, based on the median value. We additionally control for the average impact within a particular group for that year-month by adding group × year-month fixed effects. We provide our results for the coefficient estimates in Table 2.5. Using the upstream and downstream segments of the proportion of value-added weighted by the county-level share of wages, our results in Columns (1)-(2) show that the bond yield spreads increase by 21-24 bps for deals involving a low anticipated jobs multiplier. The difference between the winning counties with low and high values is statistically significant. We find qualitatively similar results in Columns (3) and (4). Here, we use the county-level share of employment to value-weight the proportion of value-added.

**After-tax Yield Spreads(bp) by Ex-Ante Firm Value of Patents**

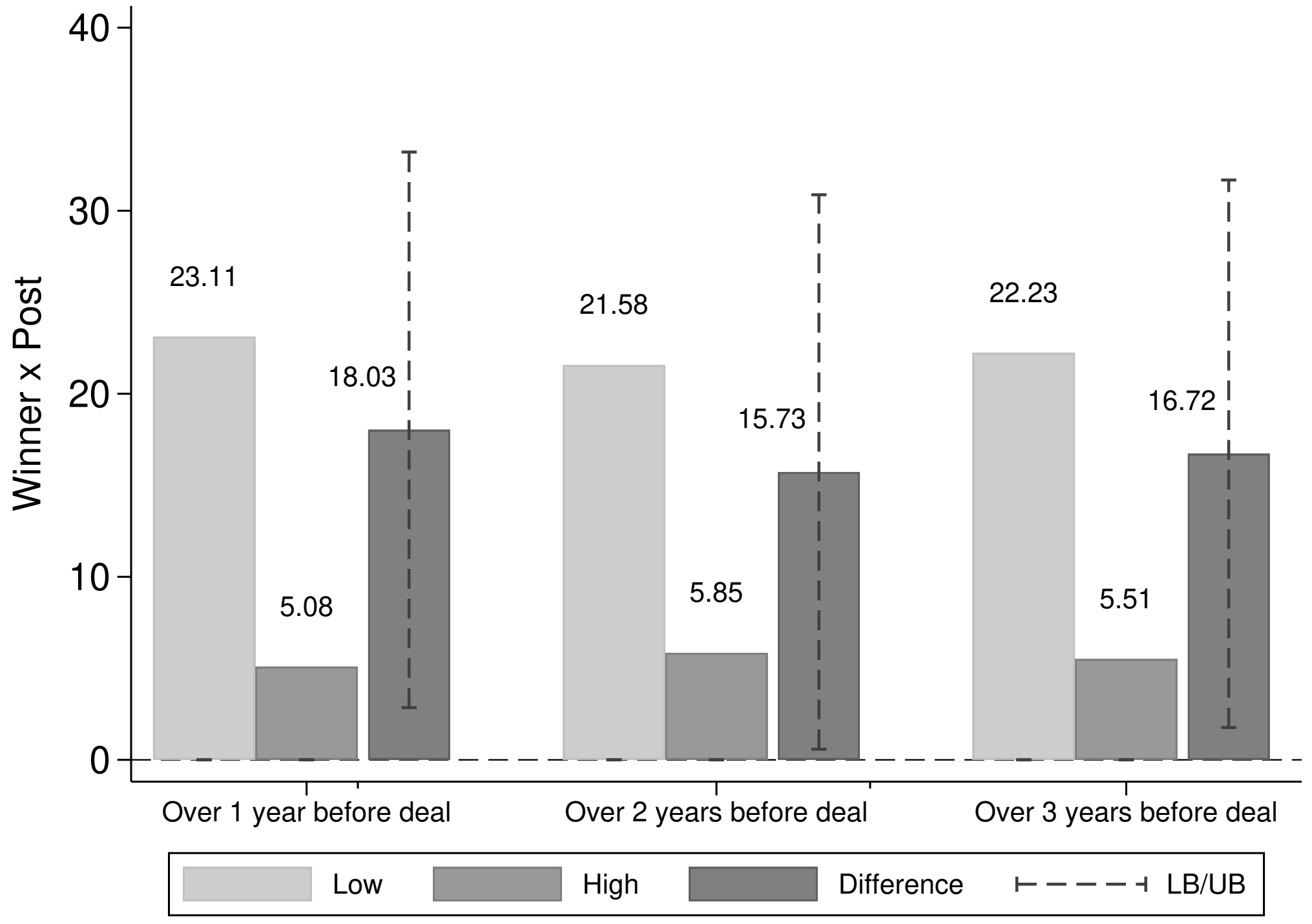

**Figure 2.5:** Effect of Anticipated Jobs Multiplier:
The figure shows results for our main interaction term, $\beta_0$, from Equation (4.1). We modify the baseline equation to interact with dummies corresponding to terciles of ex-ante firm value of patents. We classify the lowest tercile in the *Low* group. The remaining subsidy deal events are classified as *High*. We additionally control for group-month fixed effects in the regression. We show results by aggregating the value of patents at the firm level over 1, 2 and 3 years before the deal, respectively. Standard errors are double clustered by county bond issuer and year-month. The dashed lines represent 95% confidence intervals.

Next, to measure knowledge spillover, we follow [71] to quantify the economic importance of patents originating from firms receiving subsidies. Specifically, we use the aggregated dollar value of innovation for patents granted over one, two, and three years before the deal, respectively[19]. We use our baseline Equation (4.1) interacted with dummies corresponding to values above and below the median for the winning counties. We additionally control for group $\times$ year-month fixed effects. We show our results in Figure 2.5. Our results show that the difference in winning and losing counties' municipal bond yield spreads after the deal announcement increases by 15-18 basis points more for winner-loser pairs with low patent value in comparison to winner-loser pairs with high patent value.

[19]We are able to match 94 deal-pairs in our subsidy database to the patents granted. Following [71], we use the issuing year of patents. For deals that can be linked to the patent-CRSP (firm) database but do not have any patents associated with them, we assign their value of innovation as zero.

Overall, we highlight the mechanism based on expected benefits from the anticipated jobs multiplier. We show that the impact on bond yields is higher and statistically different for winning counties that attract/retain firms in industries with lower anticipated jobs multiplier or lower-valued innovation. These results are consistent with municipal bond investors incorporating the expected benefits of the deal in their valuation. Next, we discuss the impact of a county's debt capacity on the cost of borrowing.

*County Debt Capacity*

We consider the ex-ante debt capacity of local governments using three proxies: a) interest expenditure, b) net debt, and c) county credit ratings. First, we expect the secondary market impact to be higher for winning counties with large ex-ante interest expenditures. To this end, we use the ex-ante interest on general debt scaled by three measures of revenue, and one measure of total debt, namely: (i) $\text{Revenue}_1$, (ii) $\text{Revenue}_2$, (iii) $\text{Revenue}_3$, and (iv) total long term debt outstanding. We define the three approaches of calculating revenue in Table B1. We use interest expenditure in the year preceding the deal, scaled by the corresponding fiscal metric two years before the deal. A high value of the interest expenditure measure corresponds to a low debt capacity.

**[Insert Table 2.6 here]**

We divide the winning counties into two bins based on the median of the interest expenditure measures (as defined above). Using our baseline Equation (4.1) with interactions for the bins, we estimate the differential impact on high versus low debt capacity counties. We also include group-month fixed effects in this regression. We present our results in Table 2.6, which shows the coefficient of the interaction term for each group. First, Column (1) shows that for counties with a high $Interest/Revenue_1$ ratio, the bond yield spreads increase by 26.69 bps. This corresponds to the higher debt burden these counties face since more of their revenue is devoted to meeting general debt interest costs. Similarly, Column (2) suggests that the borrowing cost for counties with higher $Interest/Revenue_2$ ratio goes up by 26.12 bps. We find a similar result in Column (3) using $Revenue_3$. Column (4) shows the impact on bond yield spreads due to differential interest to debt ratio. A higher value of the measure results in an increase in borrowing cost by 27.11 bps. The difference between the two groups is economically meaningful and statistically significant when using the scaled measure of interest expenditure in each approach.

Next, we use the county's ex-ante measure of net debt outstanding to uncover differences in debt capacity. We define the net debt outstanding as the difference between the total debt outstanding and the amount of debt retired. In Column (5) of Table 2.6, we show that counties with higher ex-ante net debt experience an increase in bond yield spreads of 15.75 bps. A higher net debt burden implies a lower debt capacity. As before, the difference between the two groups is statistically significant.

**[Insert Table 2.7 here]**

Finally, we consider our third proxy for debt capacity based on county credit ratings. Table 2.7 shows the results for our analysis. We interact our baseline Equation (4.1) with dummy variables corresponding to ex-ante high credit rating versus low credit rating winning counties[20]. We divide the winning counties based on the median value of the credit rating numeral. As before, we control for the average effect within a group in a given month using the group-month fixed effects. We find that a lower credit rating is associated with a higher increase in the winning county's municipal bond yield spreads. In Column (1), using all bonds in our sample, we find that yield spreads increase by 22.00 bps for counties with a rating below the median. We identify the two major classes of bond repayment in the sample. General obligation bonds correspond to the underlying taxing power of the issuer. Revenue bonds generally have their cash flows linked to the specific project backing the bonds. Proceeding similarly, we show results using the sub-sample of general obligation bonds (Column (2)) and revenue bonds (Column (3)). As before, we find that there is a greater increase in after-tax yield spreads for winning counties with below the median credit rating. This suggests that we find similar results for both types of municipal bonds.

[20]The county-level credit rating is based on the average S&P municipal bond rating obtained from the FTSE Russell Municipal Bonds dataset. To use a clean period before the event, we focus on ratings of corresponding bonds issued from 12 to 24 months before the deal month. We use the numeric equivalent for the bond ratings with AAA representing the highest value.

To summarize, our results in this subsection provide evidence suggesting the ex-ante debt capacity of winners as the underlying channel. We show that winning counties with a higher interest expenditure are associated with up to 26 bps as additional borrowing cost. Similarly, winning counties with high net debt (low debt capacity) experience a greater increase in their bond yield spreads. Further, low rated winning counties also pay more for their debt after the deal.

*Does a high multiplier alleviate debt capacity constraint?*

Our results in Sections 2.4.3 and 2.4.3 provide evidence based on the anticipated jobs multiplier and county-level debt capacity. These results motivate our evaluation of the interaction effect between debt capacity and anticipated jobs multiplier to address: Does a high multiplier alleviate debt capacity constraints of local governments? We divide the winning counties into two groups based on the median value of the anticipated jobs multiplier. We modify the baseline Equation (4.1) to interact with dummies corresponding to the ex-ante county-level debt capacity based on $\textit{Interest/Revenue}_1$. We additionally control for group-month fixed effects in the regression. Figure 2.6 shows our results from this analysis. For subsidy deals with low anticipated multiplier, we find that bond yield spread increase by 25.31 bps when the $\textit{Interest/Revenue}_1$ ratio is above the median. The corresponding im-pact on the low *Interest/Revenue*$_1$ ratio group is 14.06 bps. Meanwhile, for subsidy deals involving a high anticipated jobs multiplier, the bond yield spreads increase by 26.91 bps when the interest expenditure is high, but there is no significant impact when the value is low. The difference between these two groups is economically meaningful and statistically significant.

**After-tax Yield Spreads(bp) by Interest/Revenue$_1$**

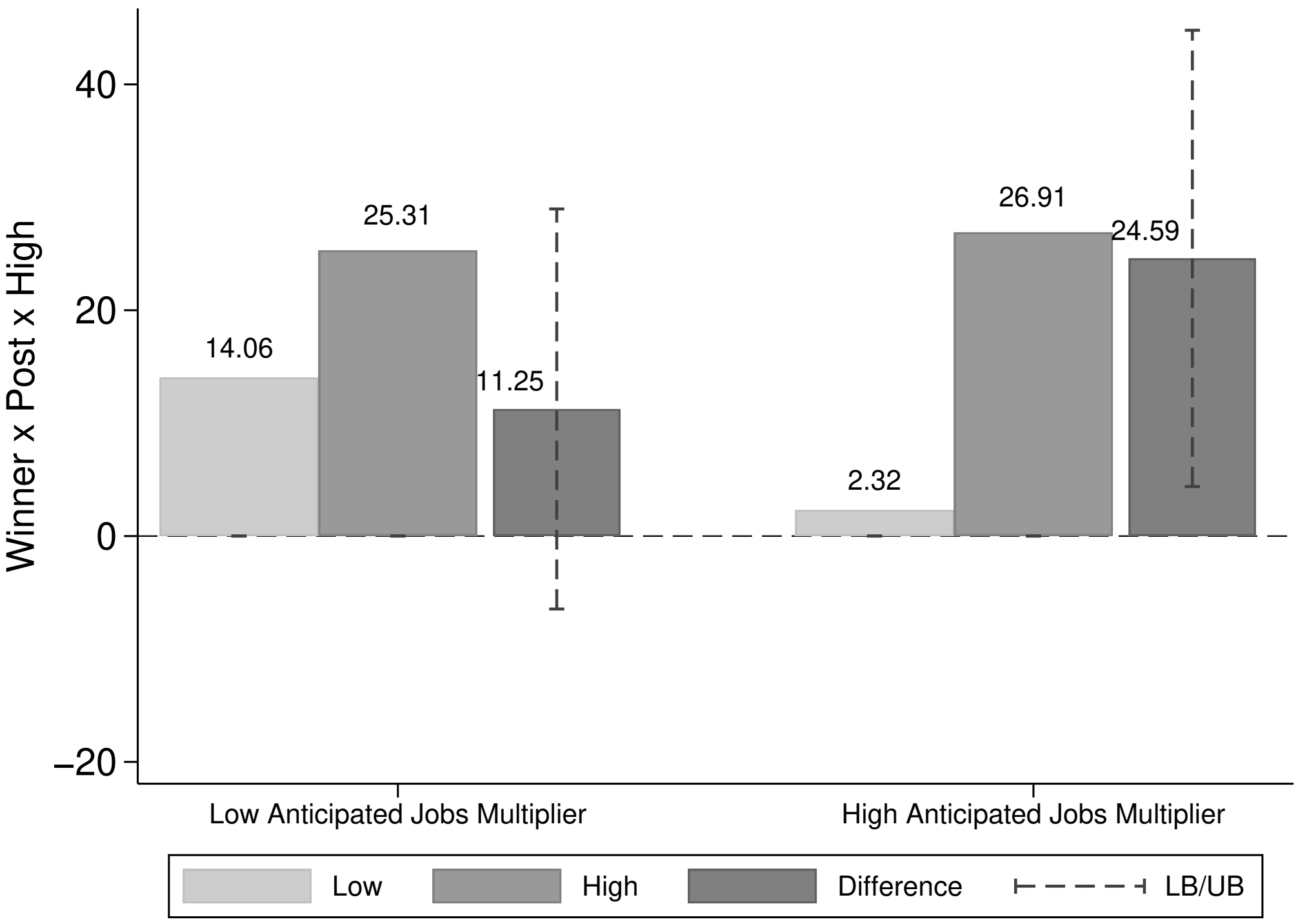

**Figure 2.6:** Effect of Anticipated Jobs Multiplier and Debt Capacity:
The figure shows results for our main interaction term, $\beta_0$, from Equation (4.1). We modify the baseline equation to interact with dummies corresponding to the ex-ante county-level debt capacity based on *Interest/Revenue*$_1$. We additionally control for group-month fixed effects in the regression. We show results by using sub-samples across subsidy deals involving low and high anticipated jobs multiplier effect, respectively. See Table B1 for details on the construction of anticipated jobs multiplier effect. Standard errors are double clustered by county bond issuer and year-month. The dashed lines represent 95% confidence intervals.

Regardless of whether the anticipated jobs multiplier is low or high, our analysis shows that the differential impact due to a high *Interest/Revenue*$_1$ ratio is 11–24 bps. This evidence seems to suggest that even a high jobs multiplier effect does not seem to alleviate the debt capacity constraints of local governments.

In addition to debt capacity of the county and anticipated jobs multiplier effects, the amount of subsidy given to attract or retain firms to the county relative to the projected benefits is likely to influence the response of municipal bond investors after the deal. The amount of subsidy offered likely could depend on the relative bargaining power between the counties and the firm involved in the deal. We argue that while firms may hire site consultants to conduct their search through a bidding mechanism[21], local governments may not have access to such sophisticated resources. To assess the relative bargaining power between the county and the firm, we use the following: a) *Proposed Value*, b) ratio of investment to state revenue, c) intensity of bidding competition, and d) county's unemployment rate. A lower bargaining power causes a greater increase in yields (between 14-20 bps) (see Section B2.3.4, Internet Appendix for details).

*How does the corruption of public officials affect municipal bond yields?*

One possible reason for the increase in municipal bond yields after winning the subsidy deal could be the aggressive bidding by local governments. We formulate a test to examine this by incorporating heterogeneity in municipal governance. We follow [73] to create a state-level measure of corruption among public officials. We use the ratio of the number of corruption-related convictions to state population in millions for each state-year to represent ex-ante level of corruption one year before the subsidy deal. Our analysis aims to capture instances when local governments bid too aggressively to attract corporations.

We present our results in Figure 2.7. We modify the baseline equation to interact with dummies corresponding to the ex-ante public officials' corruption in the year before the subsidy deal. We additionally control for group-month fixed effects in the regression. We show results by classifying subsidy deal events with above median and top tercile levels as *High* corruption events, alternatively. The remaining events correspond to the *Low* group.

[21] For instance, The Wall Street Journal reported on a cadre of consultants who help companies decide the location of their projects: https://www.wsj.com/articles/meet-the-fixers-pitting-states-against-each-other-to-win-tax-breaks-for-new-factories-11558152005

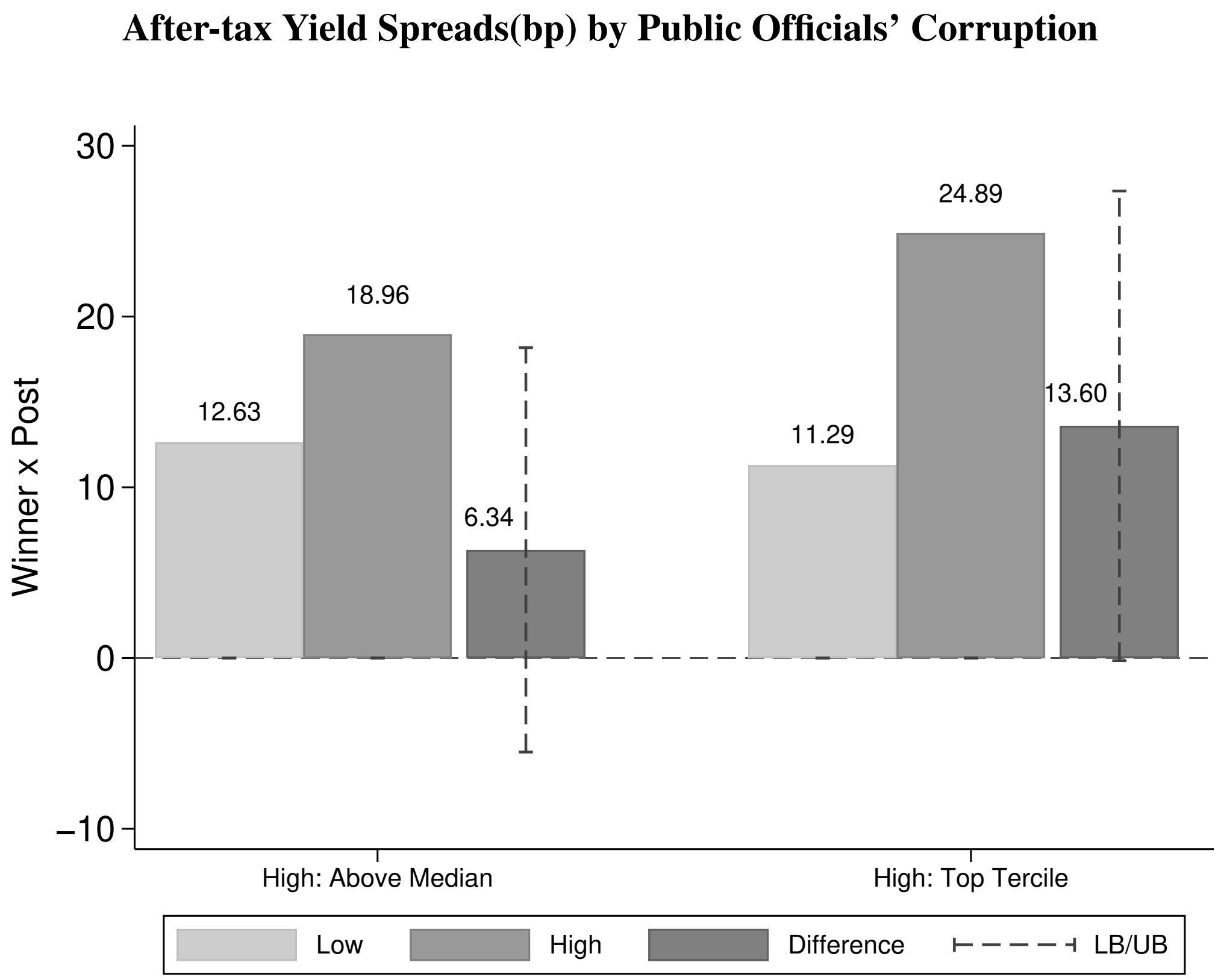

**Figure 2.7:** Effect of Public Officials' Corruption:
The figure shows results for our main interaction term, $\beta_0$, from Equation (4.1). We modify the baseline equation to interact with dummies corresponding to the ex-ante state-level corruption in the year before the subsidy deal. We additionally control for group-month fixed effects in the regression. We show results by classifying subsidy deal events with above median and top tercile levels as *High* corruption events, alternatively. The remaining events correspond to the *Low* group. We obtain the state-level corruption from [73] and extend the data to 2020 from the website of the Department of Justice. Our ex-ante measure corresponds to the ratio of the number of corruption-related convictions to state population in millions for each state-year. Standard errors are double clustered by county bond issuer and year-month. The dashed lines represent 95% confidence intervals.

We find that in deals where corruption in the winning county is above the median, the municipal bond yield spreads increase by 18.96 bps. The effect is lower for deals with lower corruption. Similarly, for deals involving the highest tercile of public officials' corruption, we find that yields increase by 24.89 bps. The difference over the low group in this setting is weakly statistically significant and economically meaningful at 13.60 bps. We find that the impact on municipal bond yields is greater when the winning county belongs to a state with a higher level of corruption among public officials.

Taken together, our results in Section 3.4.2 suggest that a lower anticipated jobs multiplier and lower debt capacity (higher interest expenditure or higher net debt) increase the impact on borrowing cost. The combined effect is dominated by the debt capacity, with the jobs multiplier not attenuating the effect. We find consistent evidence in terms of the probability of bond rating downgrades for the winning counties using our measures of debt capacity (see Section B2.3.3, Internet Appendix). We also show that the impact on municipal bond yields is higher when the winning state has higher ex-ante corruption among its public officials. Further, we find that impact on municipal bond yields is higher when there are closely contested or upcoming local elections (see Section B2.3.5, Internet Appendix). We also analyze the heterogeneity among subsidy deals based on the subsidy-to-jobs ratio in Section B2.3.6 of the Internet Appendix. We find that a higher value of the subsidy-to-jobs ratio results in a greater impact on bond yield spreads. In the next section, we consider the impact on the issuance of new municipal bonds.

#### 2.4.4 Impact on New Issuance of Municipal Bonds

First, we consider the volume of municipal debt issued in the form of bonds. Given that some of the additional economic activity/expansion would have to be financed through borrowings, we expect the winning counties to issue more debt. This especially could be the case when the winners need to create the infrastructure required to support the large plant. Instead of diverting cash from regular sources of revenue (which may have been already earmarked for dedicated uses), borrowing in the public market could be a feasible option. In this light, we present our results in Figure 2.8a where we compare the volume of bond issuance at the county level between the winning and losing counties after the deal announcement.

For each county, we calculate the total par value of bonds issued during the six months before the corresponding deal event window (comprising T=-13 to T=-18 months). We normalize this value to one and compute the total par value of new issuances relative to this amount in subsequent half years. The ratio represents the relative growth in issuance among winners, compared to the corresponding growth of losers. The vertical bars in the figure show the upper and lower limits based on the standard error of the mean values. It is possible that winning counties may prepare for the deal before the actual announcement and start to raise funds before the announcement date. We see some evidence of an increase in municipal bond issuance for winning counties during the six months before the subsidy deal in Figure 2.8a. However, this impact is not very strong. We further study the issuance based on the debt capacity of winning counties. In Figure 2.8b, we find evidence consistent with our proposed mechanism. Interestingly, we find that the counties with a low $\textit{Interest/Revenue}_1$ ratio (higher debt capacity) are able to issue more debt than the corresponding losing counties after the deal announcement. Meanwhile, counties with a low debt capacity (high $\textit{Interest/Revenue}_1$ ratio) may issue bonds before the announcement. One possible reason is that these counties may strategically raise municipal debt before the deal announcement to avoid higher borrowing costs. We know from Table 2.6 that the effect on secondary market yields is stronger for counties with higher $\textit{Interest/Revenue}_1$ ratio after the deal announcement.

**[Insert Table 2.8 here]**

As a final step in our analysis of the primary market of new municipal bonds, we evaluate the impact of the subsidy announcement on offering yields. Our coefficient of interest is the interaction term in the difference-in-differences estimate using Equation (4.1).

**New Municipal Bond Issuance**

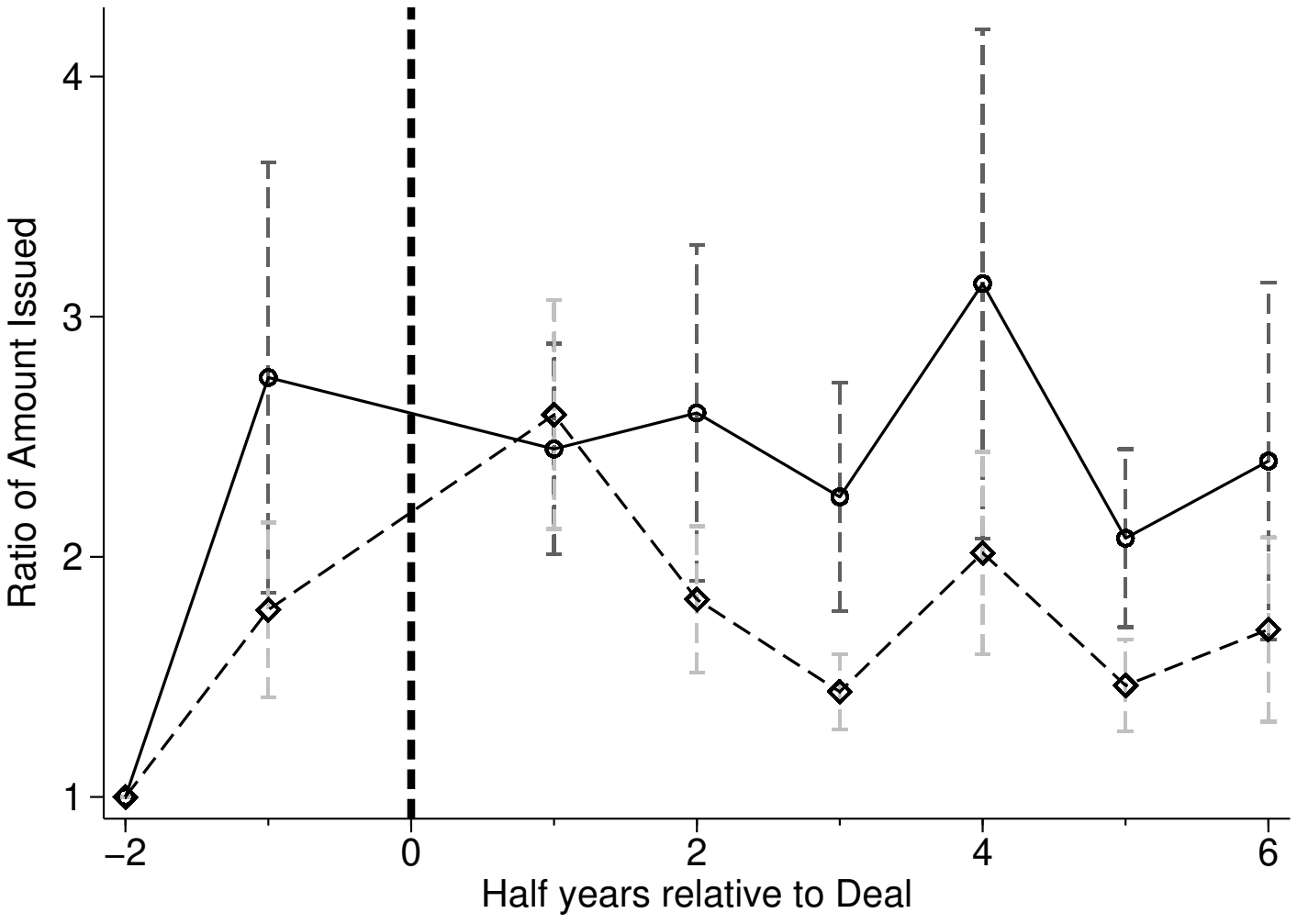

**(a)**

**By Interest/Revenue$_1$:**

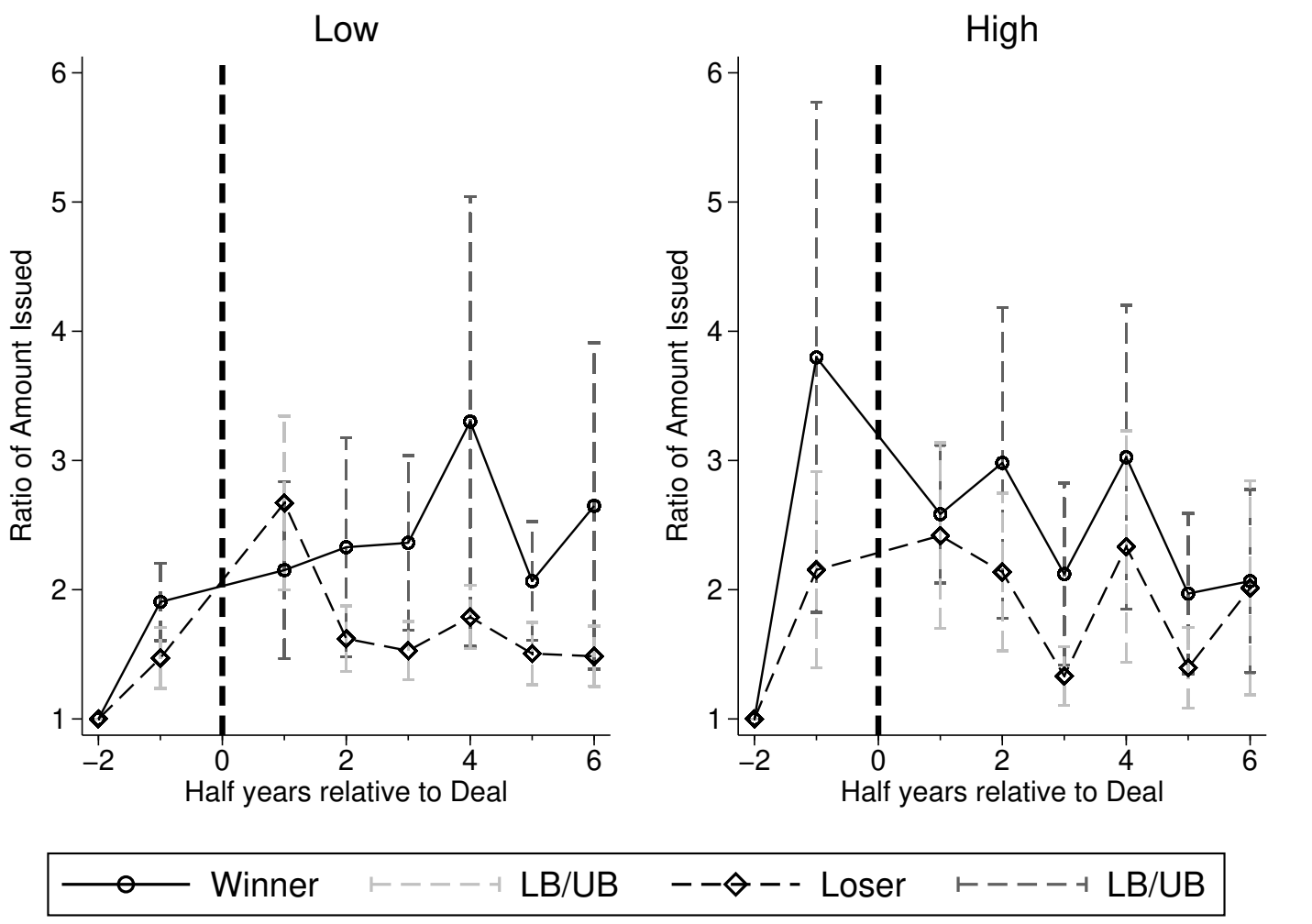

**(b)**

**Figure 2.8:** New Municipal Bond Issuance:

The figure shows the county level aggregate volume of bond issuance for winners and losers after the deal announcement. For each county, we calculate the total par value of bonds issued during the six months before the corresponding deal event window. We normalize this value to one and compute the total par value of new issues relative to this amount in the half years after the announcement. The ratio represents the relative growth in issuance among winners, compared to the corresponding growth of issuance among losers. In sub-figure (a), we show the total issuance and in sub-figure (b), we split the sample based on the *Interest/Revenue*$_1$ ratio (defined in Table B1). A low value of *Interest/Revenue*$_1$ ratio suggests a high debt capacity for the county, and vice-versa. The vertical bars show the upper and lower limits based on the standard errors of the mean values.

However, we additionally introduce issuer fixed effects to absorb unobserved heterogeneity among different issuers within a county. We also control for bond ratings at the time of issuance. We double cluster standard errors at the county-specific bond issuer and dated month level. We show our results in Table 2.8. In Column (1), we estimate the difference-in-differences coefficient from within the same county-pair, absorbing for the county fixed effect and issuer fixed effect. We report no significant effect on the offering yields, overall. Next, we show our results in Columns (2)-(4) using the debt capacity measures discussed in Section 2.4.3. For example, in Column (2), we find that offering yields increase by 5.53 bps in winning counties with a high $\textit{Interest/Revenue}_1$ ratio. On the other hand, there is weak evidence to suggest that offering yields go down after the subsidy deal in winning counties with low $\textit{Interest/Revenue}_1$ ratio. We show qualitatively similar results in Columns (2) and (3). This suggests that offering yields of new municipal bonds increase when the debt capacity is low (interest expenditure and net debt are relatively high).

Overall, our results suggest an increase in offering yields of new municipal issuance by about 5 bps after the subsidy announcement for counties with low debt capacity. However, one caveat to these results is that counties may rationally expect a higher borrowing cost following the deal announcement and may try to time the market in raising new debt. That is one of the reasons why we focus on the secondary market trades of existing bonds in our baseline analysis to evaluate the impact of the corporate subsidy deal on borrowing costs.

### 2.4.5 Impact on Local Economy and Public Expenditure

So far, we find evidence suggesting an increase in the borrowing cost of local governments after the deal both in the primary and secondary bond markets. We also show that the new municipal bond issuance increases for winning counties (compared to the losing counties), but only for counties with low-interest expenditure, i.e., high debt capacity. Keeping these results in mind, we now investigate the impact on the local economy and local public expenditure similar to [25].

**[Insert Table 2.9 here]**

Given that the reported motivation behind offering corporate subsidies is to promote county-level economic growth, we first consider the impact on the local economy and local public expenditure over a five-year window. First, in Panel A of Table 2.9, we show the implications for county-level employment growth from the Quarterly Census of Employment and Wages (QCEW) and the unemployment rate from the Bureau of Labor Statistics. We use the annualized version of Equation (4.1). Here, we do not include county controls because of limitations in using the lagged county-level variable as controls in county-level regressions [83]. Column (1) and (4) suggest that there is no meaningful overall effect on the employment growth and the unemployment rate after a subsidy deal, respectively. Further, we interact the main equation with dummies corresponding to above-median (high) and below-median (low) values of the anticipated jobs multiplier effect (Column (2) and Column (5)) and $\textit{Interest/Revenue}_1$ ratio (Column (3) and Column (6)) among winning counties. We do not find any evidence to suggest a differential impact among winning counties with high anticipated jobs multiplier subsidy deals or high value of interest expenditure.

We also study the impact on the local economy based on entrepreneurship and payroll growth. We present our analysis in Panel B of Table 2.9. In Column (1), we report the aggregate effect on all winners when compared to the losing counties in the three years after the subsidy deal. Our dependent variable is the logged number of establishments at the county-level obtained from the County Business Pattern. We do not find a meaningful change in entrepreneurship, as measured from the number of establishments. In Column (2), we find that winning counties with a high multiplier see an increase in the number of establishments by nearly 1%. However, this effect is only weakly statistically significant. Meanwhile, the difference between the high versus low multiplier groups is 2% and is statistically significant. There is no differential impact between winning counties based on the debt capacity measure (Column (3)). Further, we use the annual payroll growth (%) as the dependent variable for our analysis in Columns (4)-(6). Column (4) suggests that overall there is no effect on the annual payroll growth for winning counties when compared to the losers. We do not find statistical difference between winning counties of low and high values of anticipated jobs multiplier and debt capacity (Columns (5)-(6)).

**[Insert Table 2.10 here]**

Finally, we show the impact on county-level expenditures around the corporate subsidy deal announcement in Table 2.10. We use the annualized version of Equation (4.1) as the primary specification for this table. First, in Columns (1)-(5), we show the aggregate effect for the coefficient of interest from our difference-in-difference setting. We scale all the dependent variables by the county population of the lagged year to get the per capita impact. Our measure of *Elementary Education* consists of per capita expenditure on elementary education. For expenditure on *Hospitals* in Columns (3) and (8), we consider per capita expenditure on hospitals. *Police and Protection Expenditure* in Columns (4) and (9) consists of per capita expenditure on police protection. Finally, *Other* in Columns (5) and (10) captures per capita expenditure on all other expenses. In Columns (1)-(5), we find a reduction in total expenditure, elementary education expenditure, and other expenditures but no significant change in hospitals expenditure or police and protection expenditure. For the average county in the sample with a total expenditure of $5,272 per capita during the period, the reduction in Column (1) amounts to nearly 6%. Next, we show the interaction effects in Columns (5)-(10) by using our proxy for debt capacity in the form of the $\textit{Interest/Revenue}_1$ ratio. In each of these columns, we find a significant decline in expenditure for winning counties with high $\textit{Interest/Revenue}_1$ ratio.

Overall, we find evidence suggesting that there is an increase in borrowing costs for the winning county after the subsidy announcement. This is driven by counties with low anticipated jobs multiplier effect and low debt capacity. As a result, only counties with high debt capacity are able to issue new debt after the deal. We find no evidence of employment growth but a significant reduction in total public expenditure on local public services after the deal, especially hospitals. We also consider the impact on local house prices and property tax revenue around the announcement of subsidy deals. We find an average increase in the house price index for both winning and losing counties, after the deal announcement. However, the differential effect for winning counties is negative and is driven by counties with a higher ex-ante interest expenditure. Finally, we do not observe any differential in property tax revenue after the subsidy deal announcement. This suggests that winning counties may be able to retain the same level of property tax revenue even with lower house price gain compared to losing counties. We report these results in Section B2.3.7 of the Internet Appendix.

## 2.5 Conclusion

Corporate subsidies have recently attracted much attention in the United States. Some policymakers favoring corporate incentives highlight the importance of creating more jobs, while others worry about the costs of financing the incentives and the additional burden on public services. In light of this divergence, our paper evaluates how subsidy deals impact the borrowing costs of local governments and their expenditure on civic services. Counties face a trade-off while using targeted business incentives for economic development, i.e., forgoing future tax revenue versus anticipated jobs multiplier gains. If the additional civic burden requires local governments to raise more debt, the underlying debt capacity may impact the borrowing cost. On the other hand, the overall benefit from the expected jobs multiplier may attenuate the impact of debt capacity.

Using detailed hand-collected data on corporate subsidy deals worth USD 40 billion during 2005-2018, we provide new evidence through the lens of municipal bond yields. We find that the cost of municipal debt in the secondary market increases for the winning counties compared to the losing counties. This amounts to a reduction in bondholders' wealth of about $7.6 billion, which is 19.07% of the total subsidy offered. We propose a mechanism based on the ex-ante debt capacity of the winning counties and the anticipated

jobs multiplier effects. We find a more significant increase in yield spreads after the deal for counties with lower debt capacity and a lower anticipated jobs multiplier. We also document additional debt issuance by counties that have a high debt capacity.

To the best of our knowledge, our paper is the first to document the impact of corporate subsidies on the borrowing cost of local governments. Our results highlight that the costs of some of the corporate subsidy deals to some of the counties may be more than the benefits from attracting or retaining firms through the subsidy deal.

**Table B1:** Description of Key Variables

This table reports variable definitions. Data sources include the municipal bond transaction data from the Municipal Securities Rulemaking Board (MSRB), FTSE Russell's Municipal Bond Securities Database (FTSE, formerly known as Mergent MBSD), zero coupon yield provided by FEDS, highest income tax bracket for the corresponding state of the bond issuer from the Federation of Tax Administrators (FTA), Census data from the Census Bureau Annual Survey of Local Government Finances (CLGF), input-output tables from Bureau of Labor Statistics (BLS), wages and employment data from Quarterly Census of Employment and Wages (QCEW), and subsidy data from Subsidy Tracker which was enhanced through hand-collection (ST-HC).

| Variable | Description | Source |
|---|---|---|
| *Winner* | Dummy set to one for a county that ultimately wins the subsidy deal. This dummy equals zero for the runner-up county in that subsidy deal. | ST-HC |
| *Post* | Dummy that is assigned a value of one for *event* months after the deal is announced and zero otherwise. | ST-HC, MSRB |
| *Average Yield* | Volume-weighted average yield for a CUSIP in a given month. Volume refers to the par value of the trade. | MSRB |
| *Yield Spread* | Calculated as the difference between the *Average Yield* and the coupon-equivalent risk free yield. The risk free yield is based on the present value of coupon payments and the face value of the municipal bond using the US treasury yield curve based on maturity-matched zero-coupon yields as given by [48]. This yield spread calculation is similar to [49]. | MSRB, FEDS |

| Variable | Description | Source |
|---|---|---|
| *After-tax Yield* | The tax-adjusted *Average Yield* calculated as below: $$after-taxYield_{i,t} = \frac{y_{i,t}}{(1-\tau_t^{\text{fed}}) * (1-\tau_{s,t}^{\text{state}}))}$$ | MSRB, FTA |
| *After-tax Yield Spread* | Calculated as the difference between the tax-adjusted *Average Yield* and the coupon-equivalent risk free yield. The risk free yield is based on the present value of coupon payments and the face value of the municipal bond using the US treasury yield curve based on maturity-matched zero-coupon yields as given by [48]. This yield spread calculation is similar to [49]. We follow [5] in applying the tax adjustment. It is calculated as below: $$spread_{i,t} = \frac{y_{i,t}}{(1-\tau_t^{\text{fed}}) * (1-\tau_{s,t}^{\text{state}}))} - r_t$$ | MSRB, FEDS, FTA |
| *Competitive Bond Dummy* | Dummy variable that equals 1 if the issue is sold to underwriters on a competitive basis and is 0 otherwise | FTSE |

| Variable | Description | Source |
|---|---|---|
| *GO Bond Dummy* | Dummy variable for general obligation bond. A GO bond is a municipal bond backed by the credit and taxing power of the issuing jurisdiction rather than the revenue from a given project. | FTSE |
| *Log(Amount)* | Log transformation of the dollar amount of the individual bond's (9-digit CUSIP) original offering. | FTSE |
| *Callable Dummy* | Dummy variable that equals 1 if the issue is callable and is 0 otherwise. | FTSE |
| *Insured Dummy* | Dummy variable that equals 1 if the issue is insured and is 0 otherwise. | FTSE |
| *Remaining Maturity* | Individual bond maturity measured in years. | FTSE, MSRB |
| *Inverse Maturity* | Inverse of the value of *Remaining Maturity*; to account for non-linearity. | FTSE, MSRB |

| Variable | Description | Source |
| --- | --- | --- |
| *Anticipated Jobs Multiplier* | This metric represent the county's exposure in the industry ($j$) using the upstream (or downstream) sector ($s$) based on the fraction of total wages in that sector ($\eta_{s,t}^{county}$), derived from QCEW. The input-out tables from BLS also provide us with a share of value added by upstream sectors in a given industry ($w_{s,t}^{j}$). To arrive at the county's sector level exposure for a given industry ($j$) ($e_{s,j,t}^{county}$) by summing up the upstream sector linkages, we follow: $$e_{s,j,t}^{county} = \sum_{s} w_{s,t}^{j} * \eta_{s,t}^{county}$$ | BLS, QCEW |
| *Revenue*$_1$ | (Total Revenue - State Inter-Governmental Transfers) - (Total Expenditure - Interest on Total Debt) | CLGF |
| *Revenue*$_2$ | (Total Revenue - State Inter-Governmental Transfers) - (Total Expenditure - Interest on General Debt) | CLGF |
| *Revenue*$_3$ | (Total Revenue - State Inter-Governmental Transfers) - ( - Interest on General Debt) | CLGF |
| *Interest/Revenue*$_1$ | Interest/Revenue$_1$=$\frac{\text{Interest on general debt}}{\text{Revenue}_1}$ | CLGF |

| Variable | Description | Source |
|---|---|---|
| *Interest/Revenue$_2$* | Interest/Revenue$_2$=$\frac{\text{Interest on general debt}}{\text{Revenue}_2}$ | CLGF |
| *Interest/Revenue$_3$* | Interest/Revenue$_3$=$\frac{\text{Interest on total debt}}{\text{Revenue}_3}$ | CLGF |
| *Interest to debt* | Ratio of interest on general debt to total long term debt outstanding for the county. | CLGF |
| *Net debt* | Difference between total debt outstanding and total debt retired. | CLGF |

## **Table 2.1:** Summary Statistics: Subsidy Deals

This table summarizes the deal level characteristics on subsidy in our sample during 2005-2018. In Panel A, we provide summary statistics on all deals together. In Panel B, the deals are sub-divided based on the purpose for which the subsidy was offered. Panel C shows the subsidy amount across various industry groups of the subsidy firms, based on NAICS classification, which are recombined further. Non-tradeables include wholesale trade and retail trade. Data centers include information, finance, and insurance sectors, professional and scientific, and management and administrative services. Transportation includes transportation and warehousing, and real estate sectors. Mining is a combination of mining and energy, utilities, and construction sectors.

**Panel A: All Deals**

| | Count | Mean | Median | Std. Dev. |
|---|---|---|---|---|
| Subsidy (USD million) | 199 | 202.3 | 81.8 | 471.1 |
| Investment (USD million) | 192 | 805.3 | 308.5 | 1,503.1 |
| Subsidy/Investment(%) | 192 | 82.2 | 22.3 | 352.8 |
| Jobs promised | 199 | 1,490.5 | 800.0 | 2,414.5 |
| Subsidy (USD) per job | 199 | 324,548.5 | 105,488.4 | 882,497.5 |

**Panel B: By Purpose of Subsidy**

| | | Subsidy ($ million) | | |
|---|---|---|---|---|
| | Count | Mean | Median | Std. Dev. |
| New/Expansion | 145 | 241.9 | 84.1 | 544.9 |
| Relocation | 34 | 74.3 | 73.5 | 61.6 |
| Retention | 20 | 132.6 | 95.9 | 96.6 |

**Panel C: By Industry of Firms**

| | | Subsidy ($ million) | | |
|---|---|---|---|---|
| | Count | Mean | Median | Std. Dev. |
| Manufacturing | 93 | 240.0 | 91.8 | 592.0 |
| Data Centers | 35 | 120.5 | 86.2 | 119.6 |
| Non-Tradables | 25 | 357.7 | 90.0 | 541.6 |
| Mining-Energy-Utilities | 15 | 190.5 | 81.9 | 409.3 |
| Transportation | 12 | 45.5 | 26.5 | 49.6 |
| All Other | 19 | 72.3 | 35.0 | 106.1 |

**Table 2.2:** Summary Statistics: Municipal Bonds

This table summarizes the municipal bond level characteristics during 2005-2019 for our sample of bonds linked to corporate subsidies. Panel A reports the secondary market attributes. Panel B reports the primary market features. The key variables are described in Table B1.

**Panel A: Secondary market**

| | Count | Mean | Median | Std. Dev. |
|---|---|---|---|---|
| **Winner** | | | | |
| Wtd. Avg. Yield(%) | 1,433,430 | 2.8 | 2.8 | 1.5 |
| After-tax Yield (%) | 1,433,430 | 4.6 | 4.7 | 2.4 |
| After-tax Yield Spread (%) | 1,433,430 | 3.4 | 3.0 | 2.5 |
| Remaining Maturity (years) | 1,433,430 | 10.7 | 9.3 | 7.0 |
| **Loser** | | | | |
| Wtd. Avg. Yield(%) | 2,347,209 | 2.8 | 2.9 | 1.5 |
| After-tax Yield (%) | 2,347,209 | 4.7 | 4.9 | 2.4 |
| After-tax Yield Spread (%) | 2,347,209 | 3.5 | 3.1 | 2.6 |
| Remaining Maturity (years) | 2,347,209 | 11.2 | 9.8 | 7.3 |
| *Observations* | 3,780,639 | | | |

**Panel B: Primary market**

| | Count | Mean | Median | Std. Dev. |
|---|---|---|---|---|
| **Winner** | | | | |
| Offering Yield(%) | 205,159 | 2.9 | 2.9 | 1.3 |
| Offering Price (USD) | 205,159 | 103.9 | 101.8 | 7.6 |
| Coupon(%) | 205,159 | 3.6 | 4.0 | 1.2 |
| Years to Maturity | 205,159 | 9.6 | 8.6 | 6.4 |
| Years to Call | 87,645 | 8.9 | 9.7 | 1.8 |
| Amount (USD million) | 205,159 | 3.0 | 0.7 | 15.1 |
| Issue Size (USD million) | 205,159 | 43.5 | 12.2 | 108.3 |
| **Loser** | | | | |
| Offering Yield(%) | 281,874 | 3.0 | 3.0 | 1.4 |
| Offering Price (USD) | 281,859 | 103.5 | 101.9 | 10.4 |
| Coupon(%) | 281,874 | 3.7 | 4.0 | 1.3 |
| Years to Maturity | 281,874 | 10.1 | 9.0 | 6.7 |
| Years to Call | 125,583 | 8.8 | 9.7 | 2.1 |
| Amount (USD million) | 281,874 | 4.6 | 0.8 | 20.4 |
| Issue Size (USD million) | 281,874 | 75.2 | 15.0 | 174.3 |
| *Observations* | 487,033 | | | |

**Table 2.3:** Impact on Borrowing Costs of Local Governments: Evidence from Municipal Bonds Secondary Market

This table reports the baseline results for our sample using Equation (4.1) estimating the differential effect on municipal bond yields of winning and losing counties after the subsidy announcement. The primary coefficient of interest, $\beta_0$, is captured by the interaction term of Winner × Post. Panel A compares winners and losers in the secondary market around an equal window of three years of the event. Columns (1)-(3) show the results for monthly yield spread as the dependent variable. Specifically, Column (1) reports the effect using county-pair fixed effects, county fixed effects and year-month fixed effects. In Column (2), we also introduce bond level controls consisting of coupon (%); log(amount issued in $); dummies for callable bonds, bond insurance, general obligation bond and competitively issued bonds; remaining years to maturity; and inverse years to maturity. We provide the description of key variables in Table B1. In Column (3), we additionally control for the county-level variation in unemployment rate and labor force. We use the lagged values (to the year of deal announcement) for log(labor force) and unemployment rate, and the percentage change in unemployment rate and labor force, respectively. We use a similar scheme for the remaining columns. In Columns (4)-(6), the dependent variable is after-tax yield spread (see Section 2.3.2 for details). Our baseline specification comes from Column (6) in Panel A. In Panel B, we report the baseline specification with incremental duration after the subsidy, holding the pre-event window constant at 36 months before the subsidy announcement. T-statistics are reported in brackets and standard errors are double clustered at county bond issuer and year-month level, unless otherwise specified. * $p < 0.10$, ** $p < 0.05$, *** $p < 0.01$

**Panel A: Three-year Window**

| *Dependent Variable*: | Yield Spread | | | After-tax Yield Spread | | |
|---|---|---|---|---|---|---|
| | (1) | (2) | (3) | (4) | (5) | (6) |
| Winner × Post | 8.20*** | 8.37*** | 9.20*** | 13.72*** | 13.59*** | 15.25*** |
| | [3.58] | [3.78] | [4.16] | [3.65] | [3.94] | [4.40] |
| County-pair FE | ✓ | ✓ | ✓ | ✓ | ✓ | ✓ |
| Year-month FE | ✓ | ✓ | ✓ | ✓ | ✓ | ✓ |
| County FE | ✓ | ✓ | ✓ | ✓ | ✓ | ✓ |
| Bond Controls | | ✓ | ✓ | | ✓ | ✓ |
| County Controls | | | ✓ | | | ✓ |
| Adj.-$R^2$ | 0.733 | 0.755 | 0.755 | 0.529 | 0.595 | 0.595 |
| Obs. | 3,780,639 | 3,780,639 | 3,780,639 | 3,780,639 | 3,780,639 | 3,780,639 |

## Panel B: Different Forward Windows (in months)

| *Dependent Variable*: | After-tax Yield Spread | | | | | | |
|---|---|---|---|---|---|---|---|
| *Window (months)*: | [-36,+12] | [-36,+18] | [-36,+24] | [-36,+30] | [-36,+36] | [-36,+48] | [-36,+60] |
| | (1) | (2) | (3) | (4) | (5) | (6) | (7) |
| Winner × Post | 7.27** | 9.58*** | 11.77*** | 13.73*** | 15.25*** | 19.63*** | 21.60*** |
| | [2.45] | [2.97] | [3.66] | [4.14] | [4.40] | [5.46] | [5.70] |
| County-pair FE | ✓ | ✓ | ✓ | ✓ | ✓ | ✓ | ✓ |
| Year-month FE | ✓ | ✓ | ✓ | ✓ | ✓ | ✓ | ✓ |
| County FE | ✓ | ✓ | ✓ | ✓ | ✓ | ✓ | ✓ |
| Bond Controls | ✓ | ✓ | ✓ | ✓ | ✓ | ✓ | ✓ |
| County Controls | ✓ | ✓ | ✓ | ✓ | ✓ | ✓ | ✓ |
| Adj.-$R^2$ | 0.625 | 0.619 | 0.610 | 0.603 | 0.595 | 0.591 | 0.582 |
| Obs. | 2,381,818 | 2,732,469 | 3,082,486 | 3,430,109 | 3,780,639 | 4,478,343 | 5,167,919 |

**Table 2.4:** Robustness Tests

In this table we report results for various robustness tests on our baseline specification, i.e., Column (6) of Panel A in Table 2.3. In Panel A, we present evidence to control for other observables and unobservables. First, Columns (1)-(3) report results controlling for unobserved factors at the county level. Specifically, in Column (1), we use deal $\times$ county fixed effect in the baseline, where the deal represents winner-loser county pair. We show our strictest specification in Column (2) by including deal $\times$ county year fixed effects. In Column (3), we control for county specific year-month trends. Column (4) shows results with issuer fixed effect added to the baseline for the specific borrower in the county. Panel B corresponds to further tax considerations and duration. First, in Column (1), we show the effect of using the sub-sample of municipal bonds in the sample that are tax-exempt at the state level. Next, in Column (2), we show results by dropping deals involving Illinois, Iowa, Oklahoma and Wisconsin. These states do not offer tax exemption on municipal bonds issued even by the respective states. Column (3) shows the result using yield spreads adjusted for local (county) level individual income tax rates, over and above the state tax rate. We use individual income tax rates from 2011 for deals where we could find the data, while assuming zero local individual income tax rates for the remaining deals. Finally, we show results after controlling for duration to account for non-linear effects in remaining maturity. In Column (4), we use duration in the controls by replacing years to maturity and inverse of years to maturity. Thereafter, in Column (5), we use tax-adjusted duration to replace years to maturity and inverse of years to maturity. We show robustness to the choice of event window in Panel C, by using alternative windows. First, in Column (1), we use a shorter window of 24 months. In Columns (2)-(3) we use a longer event window of 48 months and 60 months around the subsidy announcement, respectively. In Column (4), we show the baseline result by dropping deals involving big cities. We identify these based on the number of bond-month observations for each deal. We drop the top 10 deals. T-statistics are reported in brackets and standard errors are double clustered at county bond issuer and year-month level, unless otherwise specified. $^{*}\ p < 0.10$, $^{**}\ p < 0.05$, $^{***}\ p < 0.01$

**Panel A: Other Observables and Unobservables**

| *Dependent Variable*: | After-tax Yield Spread | | | |
|---|---|---|---|---|
| | Other Unobservables | | | |
| | Deal × county FE | Deal × county × year FE | Control for monthly county trends | Add Issuer FE |
| | (1) | (2) | (3) | (4) |
| Winner × Post | 15.29*** | 5.73*** | 9.10** | 13.41*** |
| | [4.35] | [13.29] | [2.36] | [3.84] |
| Adj.-$R^2$ | 0.595 | 0.603 | 0.599 | 0.680 |
| Obs. | 3,780,639 | 3,780,620 | 3,780,639 | 3,780,514 |

**Panel B: Additional Tax Considerations and Duration**

| *Dependent Variable*: | After-tax Yield Spread | | | | |
|---|---|---|---|---|---|
| | | | | Controlling for Duration | |
| | Tax Exempt Bonds Only | Drop States w/o Tax Exemption | Adjusting Spreads for Local Taxes | Use Duration | Use After-tax Duration |
| | (1) | (2) | (3) | (4) | (5) |
| Winner × Post | 16.25*** | 15.83*** | 15.34*** | 16.05*** | 16.42*** |
| | [4.79] | [4.32] | [4.42] | [4.46] | [4.43] |
| Adj.-$R^2$ | 0.603 | 0.606 | 0.595 | 0.580 | 0.566 |
| Obs. | 3,513,025 | 3,132,137 | 3,780,639 | 3,756,901 | 3,756,901 |

**Panel C: Alternative Window and Sample**

| *Dependent Variable*: | After-tax Yield Spread | | | |
|---|---|---|---|---|
| | [-24,+24 months] | [-48,+48 months] | [-60,+60 months] | Drop big cities |
| | (1) | (2) | (3) | (4) |
| Winner × Post | 10.53*** | 19.46*** | 20.55*** | 14.70*** |
| | [3.80] | [5.54] | [5.60] | [4.45] |
| Adj.-$R^2$ | 0.597 | 0.597 | 0.598 | 0.614 |
| Obs. | 2,600,773 | 4,915,490 | 5,991,406 | 2,649,800 |

**Table 2.5:** Anticipated Jobs Multiplier: Evidence Based on Input-Output Tables

This table shows the evidence based on ex-ante county level expected jobs multiplier effect among winning counties, using the baseline Equation (4.1). We construct the measure of anticipated jobs multiplier effect by summing up the proportion of value-added in the upstream and downstream segments of a given industry three years before the deal announcement, weighted by the corresponding county's share of wages or share of employment. See Table B1 for variables description. We additionally control for the average impact within a particular group for that month by adding group-month fixed effects. Specifically, we show results by constructing this measure using the county's share of wages in Columns (1)-(2). In Columns (3)-(4), we show results by constructing this using the county's share of employment. Columns (1) and (3) show results without including the group-month fixed effects. T-statistics are reported in brackets and standard errors are double clustered at county bond issuer and year month level. * $p < 0.10$, ** $p < 0.05$, *** $p < 0.01$

| *Dependent Variable*: | After-tax Yield Spread | | | |
|---|---|---|---|---|
| *Weighted by County's:*: | Share of Wages | | Share of Employment | |
| Winner × Post | (1) | (2) | (3) | (4) |
| × High dummy | 11.10*** | 11.45*** | 10.84*** | 11.79*** |
| | [3.05] | [3.28] | [2.82] | [3.19] |
| × Low dummy | 24.28*** | 21.17*** | 22.83*** | 18.91*** |
| | [4.30] | [3.95] | [4.16] | [3.54] |
| Difference | 13.18 | 9.72 | 11.99 | 7.13 |
| p-value | 0.02 | 0.08 | 0.04 | 0.22 |
| County-pair FE | ✓ | ✓ | ✓ | ✓ |
| Year-month FE | ✓ | ✓ | ✓ | ✓ |
| County FE | ✓ | ✓ | ✓ | ✓ |
| Controls | ✓ | ✓ | ✓ | ✓ |
| Group-Month FE | | ✓ | | ✓ |
| Adj.-$R^2$ | 0.597 | 0.597 | 0.597 | 0.597 |
| Obs. | 3,739,687 | 3,739,687 | 3,739,687 | 3,739,687 |

**Table 2.6:** County Debt Capacity: Evidence based on Interest Expenditure and Local Government Debt

This table shows the evidence based on interest expenditure and outstanding local government debt, using the baseline Equation (4.1). We modify the equation to interact with dummies for high and low values of ex-ante county level measures of debt capacity using interest expenditure and debt. Specifically, we use the *Interest/Revenue*$_1$ ratio in Column (1). In Column (2), we use the *Interest/Revenue*$_2$ ratio, followed by the *Interest/Revenue*$_3$ ratio in Column (3). We show results using interest to debt in Column (4). Finally, in Column (5), we use the net debt. See Table B1 for a description of these variables. We additionally control for group-month fixed effects in these regressions. T-statistics are reported in brackets and standard errors are double clustered by county bond issuer and year-month. $^{*}$ $p < 0.10$, $^{**}$ $p < 0.05$, $^{***}$ $p < 0.01$

| *Dependent Variable*: | After-tax Yield Spread | | | | |
|---|---|---|---|---|---|
| *Interaction Variable*: | Interest<br>Revenue$_1$ | Interest<br>Revenue$_2$ | Interest<br>Revenue$_3$ | Interest<br>Debt | Net Debt |
| Winner × Post | (1) | (2) | (3) | (4) | (5) |
| × Low dummy | 7.77 | 8.23$^{*}$ | 9.63$^{*}$ | 1.76 | 4.10 |
| | [1.60] | [1.69] | [1.90] | [0.41] | [0.74] |
| × High dummy | 26.69$^{***}$ | 26.12$^{***}$ | 19.58$^{***}$ | 27.11$^{***}$ | 15.75$^{***}$ |
| | [4.45] | [4.38] | [4.63] | [5.46] | [3.80] |
| Difference | 18.92 | 17.89 | 9.95 | 25.35 | 11.65 |
| p-val | 0.02 | 0.03 | 0.10 | 0.00 | 0.09 |
| County-pair FE | ✓ | ✓ | ✓ | ✓ | ✓ |
| Year-month FE | ✓ | ✓ | ✓ | ✓ | ✓ |
| County FE | ✓ | ✓ | ✓ | ✓ | ✓ |
| Controls | ✓ | ✓ | ✓ | ✓ | ✓ |
| Group-Month FE | ✓ | ✓ | ✓ | ✓ | ✓ |
| Adj.-$R^2$ | 0.595 | 0.595 | 0.595 | 0.596 | 0.595 |
| Obs. | 3,765,485 | 3,765,485 | 3,765,485 | 3,765,485 | 3,765,485 |

**Table 2.7:** County Debt Capacity: Evidence based on County Credit Ratings

This table shows the evidence based on ex-ante county level credit ratings among winning counties, using the baseline Equation (4.1). We interact the main equation with dummies corresponding to the ex-ante average S&P municipal bond rating group of the county. We use municipal bonds issued before the subsidy announcement to determine the county level credit rating. The rating group *Above Median* corresponds to higher credit rating quality, while *Below Median* represents lower credit quality. We additionally control for the average effect within a particular group for that month by adding group-month fixed effects. First, in Column (1), we show the impact on the full sample of bonds. Second, in Column (2), we show the results using a sub-sample of general obligation (GO) bonds only. Finally, Column (3) shows the impact on revenue (RV) bonds alone. T-statistics are reported in brackets and standard errors are double clustered at county bond issuer and year-month level. $^{*}$ $p < 0.10$, $^{**}$ $p < 0.05$, $^{***}$ $p < 0.01$

| *Dependent Variable*: | After-tax Yield Spread | | |
|---|---|---|---|
| *Type of Bonds*: | All Bonds | GO Bonds | RV Bonds |
| Winner × Post | (1) | (2) | (3) |
| × Above Median | 5.58 | -6.46 | 9.75 |
| Rating dummy | [0.86] | [-0.66] | [1.24] |
| × Below Median | 22.00$^{***}$ | 21.07$^{***}$ | 19.10$^{***}$ |
| Rating dummy | [5.21] | [4.38] | [3.55] |
| Difference | 16.42 | 27.53 | 9.35 |
| p-val | 0.05 | 0.03 | 0.33 |
| County-pair FE | ✓ | ✓ | ✓ |
| Year-month FE | ✓ | ✓ | ✓ |
| County FE | ✓ | ✓ | ✓ |
| Controls | ✓ | ✓ | ✓ |
| Group-Month FE | ✓ | ✓ | ✓ |
| Adj.-$R^2$ | 0.596 | 0.602 | 0.608 |
| Obs. | 3,738,567 | 1,357,552 | 2,381,013 |

**Table 2.8:** Impact on Offering Yields of Municipal Bonds

This table shows the effect of subsidy announcement on offering yields of new bond issuances using a difference-in-differences estimate based on Equation (4.1). Here, we also introduce the issuer fixed effects in each of the specifications. In Column (1), we show the result for the baseline equation. Next, we show evidence based on interest expenditure and outstanding local government debt, using the baseline Equation (4.1). We modify the equation to interact with dummies for high and low values of ex-ante county level measures of debt capacity using interest expenditure and debt. Specifically, we use the *Interest/Revenue*$_1$ ratio in Column (2). In Column (3), we use the *Interest/Revenue*$_2$ ratio to capture the debt capacity of the winning county. Finally, in Column (4), we use the net debt as the interaction variable. See Table B1 for a description of these variables. We additionally control for group fixed effects in these regressions. T-statistics are reported in brackets and standard errors are double clustered at county bond issuer and dated month level. $^{*}$ $p < 0.10$, $^{**}$ $p < 0.05$, $^{***}$ $p < 0.01$

| *Dependent Variable*: | Offering Yield | | | |
|---|---|---|---|---|
| *Interaction Variable*: | | $\frac{\text{Interest}}{\text{Revenue}_1}$ | $\frac{\text{Interest}}{\text{Revenue}_2}$ | Net Debt |
| Winner × Post | (1) | (2) | (3) | (4) |
| Overall | -0.13 | | | |
| | [-0.08] | | | |
| × Low dummy | | -4.17$^{*}$ | -3.89$^{*}$ | -7.51$^{*}$ |
| | | [-1.93] | [-1.80] | [-1.96] |
| × High dummy | | 5.53$^{**}$ | 5.01$^{*}$ | 4.35$^{**}$ |
| | | [1.98] | [1.80] | [2.40] |
| High vs Low | | 9.71 | 8.91 | 11.86 |
| p-val | | 0.01 | 0.02 | 0.01 |
| County-pair FE | ✓ | ✓ | ✓ | ✓ |
| County FE | ✓ | ✓ | ✓ | ✓ |
| Issuer FE | ✓ | ✓ | ✓ | ✓ |
| Controls | ✓ | ✓ | ✓ | ✓ |
| Group FE | | ✓ | ✓ | ✓ |
| Adj.-$R^2$ | 0.821 | 0.822 | 0.822 | 0.822 |
| Obs. | 487,033 | 484,254 | 484,254 | 484,254 |

**Table 2.9:** Impact on Local Economy

This table shows the impact of subsidy on employment growth (%) from QCEW and unemployment rate (%) from the BLS at the county level in Panel A. In Panel B, we show results using the number of establishments and annual payroll growth (%) from the County Business Pattern. We use the annualized version of Equation (4.1) without county controls over a five year window as the primary specification for this table. In each panel, Columns (1) and (4) report the overall effect. Columns (2) and (5) show the results by interacting the equation with dummies corresponding to ex-ante county level anticipated jobs multiplier effect among winning counties. We construct this measure by summing up the proportion of value-added in the upstream and downstream segments of a given industry, weighted by the corresponding county's share of wages. See Table B1 for variables description. Finally, Columns (3) and (6) report the results by interacting Equation (4.1) with a dummy variable (*Low Int./Rev.*$_1$ and *High Int./Rev.*$_1$) based on the median value of the *Interest/Revenue*$_1$ ratio among winning counties. In these interacted specifications, we replace the event-year fixed effects with group-event year fixed effects. T-statistics are reported in brackets and standard errors are clustered at the deal level. $^{*}$ $p < 0.10$, $^{**}$ $p < 0.05$, $^{***}$ $p < 0.01$

**Panel A**

| *Dependent Variable*: | Employment Growth (%) | | | Unemployment Rate (%) | | |
|---|---|---|---|---|---|---|
| | (1) | (2) | (3) | (4) | (5) | (6) |
| Winner × Post | 0.29 | | | -0.02 | | |
| | [0.88] | | | [-0.18] | | |
| Low Multiplier × Winner × Post | | -0.18 | | | 0.01 | |
| | | [-0.33] | | | [0.07] | |
| High Multiplier × Winner × Post | | 0.65 | | | -0.13 | |
| | | [1.40] | | | [-0.84] | |
| Low Int./Rev$_1$ × Winner × Post | | | 0.60 | | | -0.08 |
| | | | [1.43] | | | [-0.60] |
| High Int./Rev$_1$ × Winner × Post | | | 0.12 | | | 0.05 |
| | | | [0.23] | | | [0.31] |
| Difference | | 0.83 | -0.48 | | -0.14 | 0.13 |
| P-value | | 0.25 | 0.46 | | 0.52 | 0.53 |
| County-pair FE | ✓ | ✓ | ✓ | ✓ | ✓ | ✓ |
| County FE | ✓ | ✓ | ✓ | ✓ | ✓ | ✓ |
| Year FE | ✓ | | | ✓ | | |
| Group-Yr. FE | | ✓ | ✓ | | ✓ | ✓ |
| $R^2$ | 0.350 | 0.351 | 0.358 | 0.890 | 0.890 | 0.893 |
| Obs. | 3,859 | 3,453 | 3,828 | 3,993 | 3,575 | 3,962 |

**Panel B**

| *Dependent Variable*: | Log(Number of establishments) | | | Annual Payroll Growth (%) | | |
|---|---|---|---|---|---|---|
| | (1) | (2) | (3) | (4) | (5) | (6) |
| Winner × Post | -0.00 | | | 0.61 | | |
| | [-0.23] | | | [1.26] | | |
| Low Multiplier × Winner × Post | | -0.01 | | | 1.24 | |
| | | [-1.53] | | | [1.54] | |
| High Multiplier × Winner × Post | | 0.01* | | | -0.30 | |
| | | [1.71] | | | [-0.54] | |
| Low Int./$Rev_1$ × Winner × Post | | | 0.01 | | | 0.10 |
| | | | [0.86] | | | [0.15] |
| High Int./$Rev_1$ × Winner × Post | | | -0.01 | | | 1.11 |
| | | | [-0.99] | | | [1.54] |
| Difference | | 0.02 | -0.02 | | -1.54 | 1.01 |
| P-value | | 0.02 | 0.19 | | 0.12 | 0.30 |
| County-pair FE | ✓ | ✓ | ✓ | ✓ | ✓ | ✓ |
| County FE | ✓ | ✓ | ✓ | ✓ | ✓ | ✓ |
| Year FE | ✓ | | | ✓ | | |
| Group-Yr. FE | | ✓ | ✓ | | ✓ | ✓ |
| $R^2$ | 1.000 | 0.999 | 1.000 | 0.232 | 0.253 | 0.237 |
| Obs. | 4,150 | 3,700 | 4,106 | 4,150 | 3,700 | 4,106 |

# **Table 2.10:** Impact on Local Public Expenditure

This table shows the impact of subsidy on local government expenditure per capita. We use the annualized version of Equation (4.1) as the primary specification for this table over a five year window. Columns (1)-(5) show the aggregate impact between winners and losers, while Columns (6)-(10) present the results based on sub-groups of the *Interest/Revenue*$_1$ ratio. Columns (1) and (6) show the impact on total expenditure at the local level. *Elementary Education* in Columns (2) and (7) consists of per capita expenditure on elementary education. *Hospitals* in Columns (3) and (8) consists of per capita expenditure on hospitals. *Police and Protection Expenditure* in Columns (4) and (9) consists of per capita expenditure on police protection. *Other* in Columns (5) and (10) consists of per capita expenditure on all other expenses. Specifically, we show the interacted form of the difference-in-differences estimate using a dummy variable (*Low Int./Rev.*$_1$ and *High Int./Rev.*$_1$) based on the median value of the *Interest/Revenue*$_1$ ratio among winning counties. In these interacted specifications, we replace the event-year fixed effects with group-event year fixed effects. T-statistics are reported in brackets and standard errors are clustered at the deal level. $^{*}$ $p < 0.10$, $^{**}$ $p < 0.05$, $^{***}$ $p < 0.01$

| *Dependent Variable*: | Total | Elementary Education | Hospitals | Police and Protection | Other | Total | Elementary Education | Hospitals | Police and Protection | Other |
|---|---|---|---|---|---|---|---|---|---|---|
| | (1) | (2) | (3) | (4) | (5) | (6) | (7) | (8) | (9) | (10) |
| Winner × Post | -311.72$^{*}$ | -89.05$^{*}$ | -5.72 | -7.44 | -209.51$^{*}$ | | | | | |
| | [-1.79] | [-1.96] | [-0.17] | [-1.17] | [-1.86] | | | | | |
| Winner × Post | | | | | | -40.94 | -25.64 | 61.15 | 3.83 | -80.28 |
| ×Low Int./Rev.$_1$ | | | | | | [-0.18] | [-0.58] | [1.20] | [0.54] | [-0.49] |
| Winner × Post | | | | | | -588.08$^{**}$ | -152.70$^{*}$ | -79.95$^{**}$ | -18.93$^{*}$ | -336.50$^{**}$ |
| ×High Int./Rev.$_1$ | | | | | | [-2.16] | [-1.82] | [-2.07] | [-1.72] | [-2.05] |
| Difference | | | | | | -547.14 | -127.06 | -141.10 | -22.76 | -256.22 |
| P-value | | | | | | 0.13 | 0.19 | 0.03 | 0.09 | 0.27 |
| County-pair FE | ✓ | ✓ | ✓ | ✓ | ✓ | ✓ | ✓ | ✓ | ✓ | ✓ |
| County FE | ✓ | ✓ | ✓ | ✓ | ✓ | ✓ | ✓ | ✓ | ✓ | ✓ |
| County Controls | ✓ | ✓ | ✓ | ✓ | ✓ | ✓ | ✓ | ✓ | ✓ | ✓ |
| Year FE | ✓ | ✓ | ✓ | ✓ | ✓ | | | | | |
| Group-Year FE | | | | | | ✓ | ✓ | ✓ | ✓ | ✓ |
| Adj.-$R^2$ | 0.979 | 0.957 | 0.958 | 0.985 | 0.980 | 0.980 | 0.958 | 0.958 | 0.986 | 0.981 |
| Obs. | 3,885 | 3,885 | 3,885 | 3,885 | 3,885 | 3,864 | 3,864 | 3,864 | 3,864 | 3,864 |

# CHAPTER 3
# COMMUNITIES AS STAKEHOLDERS: IMPACT OF CORPORATE BANKRUPTCIES ON LOCAL GOVERNMENTS

## 3.1 Introduction

While extensive literature in finance has studied how corporate decisions impact various stakeholders of firms like customers, workers, and suppliers [84], less attention has been paid to the impact on local governments. Local communities are essential stakeholders of firms because they provide factors of production [85, 86] and firms require their cooperation to create and capture value [87, 64]. In fact, business taxes account for more than one-third of revenues of local governments (Census Bureau Annual Survey of Local Government Finance, 2019). In this paper, we analyze the impact of the bankruptcy filing of a locally headquartered public manufacturing firm on the bond yields of the affected counties to shed light on the externalities imposed on the local governments.

It is not clear how a local firm's bankruptcy affects the local economies and local government finances. When growth options partially reside in workers' human capital, firm financial distress may boost entrepreneurship [88, 89, 90]. Employees may draw upon valuable ideas or projects from financially distressed firms to launch their enterprise. For example, [91] find that labor market declines can lead to more firms and better firms. If the human capital of the bankrupt firm is redeployed productively, the area may attract new residents and businesses. This "creative destruction" may result in higher tax revenue for the affected counties and lower credit risk. Alternatively, as part of the bankruptcy process, firms typically implement many cost-saving policies, including laying off workers. These worker layoffs may reduce local demand and future cash flows to the county, resulting in an increase in local government's bond yields [92][1].

[1] For example, the decline of Detroit with the fall of auto manufacturing highlights the hardships commu-

We find that bankruptcy filing of a publicly listed manufacturing firm leads to a 10% increase in secondary market yields for the headquarter counties, compared to a matched county with similar ex-ante economic trends. We find similar results using new issuance yields from the primary market. This negative impact is amplified when the county is more dependent (based on the share of wages and employment) on the industry of the bankrupt firm. Local gross domestic product (GDP) growth declines more for the affected counties. Using various proxies for a city's appeal to workers [94], entrepreneurship, start-up formation rate, patents, and mobility across counties, we find limited evidence supporting the creative destruction argument.

Identifying the causal impact of the announcement of a corporate bankruptcy filing on local governments' borrowing costs is challenging since we cannot observe the counterfactual. Additionally, it is possible that a deterioration in local economic conditions and local demand may have been responsible for the firm filing for a bankruptcy. We attempt to ameliorate these concerns by focusing on the impact of a manufacturing firm's bankruptcy on the headquarter (HQ) county and other locations where the firm has a significant presence. Publicly listed manufacturing firm bankruptcies are less likely to stem from local economic conditions, as they are less dependent on the local demand for their revenue in the HQ county or other plant locations[2]. Second, we use nearest neighbor method to match each HQ county where manufacturing firm filed for bankruptcy to another observably similar county based on the level and changes in the unemployment rate, labor force, and ex-ante average municipal bond yields as the counterfactual county. We provide evidence for the validity of the identifying assumption, i.e., the treated county and the matched control county follow similar economic trends before the shock.

We use firm bankruptcy data during 2006-2016 using [95] which was extended in [96].

---

nities face. In contrast, Rochester recovered from Kodak's financial distress [93]

[2]Additionally, the role manufacturing plays in the US economy remains an important public policy issue. According to the Bureau of Economic Analysis (BEA), in 1998, the manufacturing sector accounted for 16% of the GDP by employing 17.6 million workers. In 2019, the manufacturing sector's GDP share reduced to 11% with the employment of 12.8 million workers.

In our primary analysis, we use secondary market trades for 162,740 municipal bonds of the counties with HQ of bankrupt manufacturing firms. We estimate an event-study style difference-in-differences regression with event fixed effects (i.e., treatment-control county-pair fixed effects), county fixed effects, and calendar-time month fixed effects (to control for declining trends in the yields during our sample period). First, we confirm that the yields of municipal bonds of treated and control counties follow similar trends before the shock. We find that the difference between the two groups is statistically insignificant before the announcement of a bankruptcy filing by a publicly-listed manufacturing firm (hereafter "firm bankruptcy"). Next, we find that within a quarter after the announcement of firm bankruptcy, there is an upward trend in the yields for bonds of treated counties, but there is no change in the yields for the bonds of control counties. The tax-adjusted yield spreads for treated counties increased by 10.01 basis points (bps) compared to the control counties within 36 months after the firm bankruptcy announcement. The median credit spread between AA- and AAA-rated municipal bonds in the sample is 93 basis points. This suggests that the average yield increase after the bankruptcy filing represents $\sim$ 10% (=10.01/93) of the credit spread.

In robustness checks, we identify the control counties using additional considerations such as county-level debt capacity measures and primary market bond characteristics like the average amount issued and average maturity of bonds to control for underlying differences in treated and control counties. We find consistent results when we impose a geographic restriction of identifying the control counties from within the same region in the US and when we consider matching to three nearest neighbors. We also test if declining local consumer demand derives our results. A similar pre-trend for treatment and control counties in consumer bankruptcies and a significant increase in consumer bankruptcies after firm bankruptcy suggests a decline in local demand may not have been responsible for the firm filing for bankruptcy.

In other robustness specifications, we show results after accounting for the average per capita county income and house prices to control for the local economic conditions. We also control for bond fixed effects to absorb bond-specific unobservables in a restrictive specification. To control for county and region-specific unobservables, we include the county and region/region-year fixed effects in some specifications. We also find similar results when we expand our sample of bankrupt firms beyond the manufacturing sector by including firms in the tradeable sector. Also, our results are robust if we only consider trades after the financial crisis of 2008-2009.

In addition to HQ counties, we also analyze the impact on other plant locations of the bankrupt firms. We utilize the plant location data from the Environmental Protection Agency (EPA) and Mergent Intellect. We find that most of the impact is limited to the location of the firm's headquarters. Next, we sort firm locations based on the number of employees at the site. We find a more significant increase in municipal bond yields for the counties with 25% or more operations (measured using the number of employees at the site). We focus on secondary market trades to avoid any confounding endogeneity due to market timing in the new municipal bond issuance market. Our baseline results include multiple controls specific to the bond. For example, we control for coupon rate, issuance size, remaining maturity, callability, bond insurance, and type of security based on bond repayment source (tax sources for general obligation bonds and specific revenue stream for revenue bonds). Further, we utilize county-specific controls, including lagged levels and changes in the unemployment rate and labor force.

The impact of a large firm's bankruptcy may spill over to competitors or suppliers. For example, [97] shows intra-industry negative stock reactions to a competitor's bankruptcy filing. Using the fraction of county-level employment and county-level wages in the industry of the bankrupt firm, we capture the county's exposure to the bankrupt firm's industry. Our results on municipal bond yield spreads suggest that most treated counties with high (above median) exposure drive the results. Finally, we use an alternative way to understand the firm's impact on the county. To this end, we construct three measures to proxy for the relative size of the bankrupt firm vis-à-vis the HQ county. Here again, we find that relatively larger firms that may be more economically significant to the county drive the results.

We leverage county-level exposure analysis to show that revenue bonds backed by project-specific revenues drive the overall effect. In contrast, general obligation bonds backed by the overall tax revenues of the counties are not affected because of higher revenue diversification. This is consistent with the cash flows based impact of natural disasters in [92]. To further examine this result, we also evaluate the heterogeneity in secondary market yield spreads due to the type of project financed by the bond. We find significant increase in yields for improvement-and-development related bonds used to finance infrastructure projects in the treated county. This finding further supports our evidence linked to project-specific revenue bonds.

Finally, we test the implication of higher borrowing costs reflected by the secondary market yields. First, we find that compared to 6 months before a firm bankruptcy, the new municipal bond issuance for the treated counties increases by about 1.5 times in the year after the firm bankruptcy. For the control counties, however, this increase is about two times. We find that the treatment counties' offering yields (after-tax) on new bonds issued increase by about six bps compared to the control group. Overall, we find an increase in the unemployment rate and a decline in county GDP growth for the local economy. This effect is more for treated counties with high (above median) exposure to the industry of bankrupt firms. Importantly, we do not find evidence for "creative destruction" after a manufacturing firm's bankruptcy. Using metrics for county-level attractiveness based on education, weather [94], entrepreneurship, and innovation, we do not find support for recovery after firm bankruptcy.

Our paper relates to the literature discussing shareholder vs. stakeholder theories [98, 99, 100].[3] Moreover, [87] and [64] document the role of firms in urban vibrancy, while our paper highlights the importance of firms for the local area. We show the adverse impact of bankruptcy filings by public firms on municipal bond yields. Our paper also relates to the recent academic literature documenting how firms that file for bankruptcies impact competitors, industry peers [102], and local employment [103]. However, the evidence on the impact of firm bankruptcies on local governments' borrowing costs and municipal bond issuance is limited. In this paper, we identify a new channel by which bankrupt firms impose negative externalities on local communities through their impact on the yields of municipal bonds issued by local governments, school districts, hospitals, and public service agencies. To our knowledge, this is the first paper that documents how large firms' financial distress adversely impacts financing cost for local communities. In this regard, we also contribute to the recent literature on municipal bonds [25, 5, 39, 38, 104, 105, 92, 106, 107].

[3] [98] argued that the social responsibility of business is to increase its profits and maximizing shareholder value benefits other stakeholders of the firm. Consistent with this view, the Business Roundtable, an associ-

The rest of the paper proceeds as follows. We discuss our empirical methodology and identification concerns in Section 3.2. Section 3.3 provides details about our data and relevant summary statistics. Our main empirical results are presented in Section 4.4, followed by our conclusion in Section 4.5.

## 3.2 Identification Challenges and Empirical Methodology

In this section, we first discuss the challenges in identifying the impact of firm bankruptcies on local communities' municipal borrowing costs and then describe our empirical specification.

### 3.2.1 Identification Challenges

The first econometric challenge in identifying firm bankruptcy's impact on local governments is whether the underlying local economic conditions drive firms to file for bankruptcy.

ation of chief executive officers of America's leading companies, has emphasized shareholder primacy since 1997. However, policymakers and academics have revisited the idea of maximizing shareholder value [101]. Accordingly, Business Roundtable updated its view that redefined the purpose of corporations as benefiting all their stakeholders.

In that case, certain omitted variables (e.g., reduced local demand) could impact the firm and the corresponding county's municipal bond yields. To overcome this threat, we specifically focus on bankruptcies in the manufacturing sector, arguably relatively exogenous to demand in the local area. In the US, manufacturing firms, especially large ones, are less likely to depend on local demand to sell their finished products. Further, we also drop counties that experience bankruptcies filed by multiple publicly listed firms in the same calendar year in any sector. We argue that these potentially are linked to unobserved county-level unobservable trends and, thus, exclude them from our sample.

Since we do not observe the treated counties' counterfactual scenario, we use a matching strategy to identify a suitable control group. To this end, we first show the kernel density plot for the matching variables between the treated and control counties. The patterns look very similar, with no appreciable difference between the two groups. This affords us considerable comfort in matching the counties. Further, we evaluate the robustness of our matching considerations by including additional variables based on debt capacity, primary market issuance, and geographic considerations. We find consistent results using all these approaches (see Section C3.1.1). We also verify the kernel density plots between the treated and control groups for these additional matching strategies.

We also provide evidence from consumer bankruptcies suggesting that deteriorating local economic conditions do not drive our main result. In short, we find no statistical difference in the number of consumer bankruptcies between the treated and control counties before the firm bankruptcy event. The result suggests that local economic conditions were not deteriorating before the bankruptcy filing (see Section 3.4.1).

Finally, there may be a concern for our primary dependent variable–bond yield spreads. Unobserved factors at the county level other than the bankruptcy filing may drive the impact on yields. To address this, we evaluate and verify the pre-trends in the bond yield spreads between the treatment and control counties before the bankruptcy filing. We show that there is no statistical difference during the quarters before the bankruptcy filing. Since we do not find a significant difference in the bond yields before the event dates, we derive more comfort about our identifying assumption. The control groups do indeed represent a suitable counterfactual to the treated counties. Also, the bond yield spreads do not demonstrate any deviation before the filing dates (see Section 3.4.1).

### 3.2.2 Methodology

Our baseline event study focuses on the impact of firm bankruptcies on local governments' borrowing costs. Identifying the causal impact in this setting is challenging since we cannot observe what would have happened to the county's municipal bond yields if the firm had not filed for bankruptcy. To overcome the lack of a ready counterfactual available to us, we use the nearest neighbor in Euclidian distance as the control county based on the level of and changes in the unemployment rate and labor force, along with ex-ante average municipal bond yields. We discuss robustness to additional matching considerations to our baseline in Section C3.1.1. We identify 128 treatment-control event pairs of bankruptcy filings during 2006-2016, spanning 43 states in the US. We restrict our sample period based on the availability of ex-ante municipal bond yields in the secondary market at the beginning and county-level information on ex-ante unemployment and labor for the ending year. We use a three-year window before and after the bankruptcy filings. We use secondary market trades as the baseline case because these bonds are already trading in the event county pairs at the time of the bankruptcy filing. This also mitigates any concerns with filing-related bond issuance driving our results.

Using a standard difference-in-differences approach between the treatment and control counties' bond yields in the secondary market for municipal bonds results in the baseline specification as below:

$$
\begin{aligned}
y_{i,c,e,t} = {} & \alpha + \beta_0 * Treated_{i,c,e} * Post_{i,c,e,t} + \beta_1 * Treated_{i,c,e} + \beta_2 * Post_{i,c,e,t} \\
& + BondControls + CountyControls + \eta_e + \delta_c + \gamma_t + \epsilon_{i,c,e,t}
\end{aligned}
\tag{3.1}
$$

where index $i$ refers to bond, $c$ refers to county, $e$ denotes the county event pair, and $t$ indicates the year-month. The dependent variable $y_{i,c,e,t}$ is after-tax yield spread and is obtained from secondary market trades in local municipal bonds (described in Section 3.3). *Treated* is a dummy equal to one for a county where the headquarter of the firm filing bankruptcy is located. This dummy equals zero for the control county in that event pair. Our baseline specification uses a three-year event window around the firm bankruptcy filing. *Post* represents a dummy that is assigned a value of one for months after the bankruptcy is filed and zero otherwise. The coefficient of interest is $\beta_0$. The baseline specification also includes three sets of fixed effects: county event pair fixed effects ($\eta_e$), so the comparisons are within bonds mapped to a treated-control pair; $\delta_c$, corresponding to county fixed effects; and $\gamma_t$, denoting year-month fixed effects to control for time trends. We follow [80, 39] to include the amount issued, coupon rate, a dummy for the status of bond insurance, a dummy for the type of bid (competitive versus negotiated), and a dummy based on general obligation versus revenue bond security type, collectively represented as *BondControls*. *CountyControls* refers to a vector of county-level measures to control local economic conditions. It includes a log of the lagged value of the labor force in the county, lagged county unemployment rate, the percentage change in the annual labor force level, and the percentage change in the annual unemployment rate. All our specifications are similar to [39] in double clustering standard errors at the county bond issuer and year-month level unless specified otherwise.

### 3.3 Data

We use data on firm-level corporate bankruptcies matched to municipal bonds corresponding to the locations of firms. Our firm locations come from Compustat for the headquarters. In addition, we use data from Mergent Intellect and the Environmental Protection Agency (EPA) to identify other firm facilities. Our municipal bonds data is based on FTSE Russell (formerly known as Mergent) and the Municipal Securities Rulemaking Board (MSRB).

### 3.3.1 Corporate Bankruptcies

Our data on corporate bankruptcies come from the data collected by [95] and [96]. We supplement the firm-level bankruptcy announcement dates with headquarters locations from Compustat. To avoid potential endogeneity problems in our research design, we drop headquarters counties with multiple bankruptcies filed in the same calendar year. Further, our focus on the manufacturing sector (using two-digit NAICS codes 31-33) during 2006-2016 resulted in a sample of 128 firm bankruptcies. We present summary statistics for our firms in the sample from Compustat in Panel A of Table 3.1. We use the most recent information available before the bankruptcy date. The sample cut-off years are based on the nearest-neighbor matching strategy to identify the control group. The median firm employs 167 personnel and has total assets of $43 million with total revenue of $31 million.

**[Insert Table 3.1 here]**

To identify counties similar to the treated headquarter locations, we use five county-level variables for the nearest neighbor matching strategy: unemployment rate, change in unemployment rate, log(labor force), change in the labor force, and average yield in the year before the bankruptcy. We provide the kernel density plot of these matching variables between the treated and control groups in Figure C3.1. The two groups look similar in terms of these matching characteristics. In Panel B of Table 3.1, we tabulate the difference between additional county-level economic variables. On average, the treated counties are larger than the control counties in terms of population, average municipal bond trading volume, revenues, and expenditures. A closer look at the distribution of these metrics in Panel C of Table 3.1 provides greater comfort in our matching.

### 3.3.2 Municipal Bonds

We obtain Municipal bond characteristics from the Municipal Bonds dataset by FTSE Russell (formerly known as Mergent MBSD). We retrieve the key bond characteristics such as CUSIP, dated date, the amount issued, size of the issue, state of the issuing authority, name of the issuer, yield to maturity, tax status, insurance status, pre-refunding status, type of bid, coupon rate, and maturity date for bonds issued after 1990. We also use S&P credit ratings for these bonds by reconstructing the time series of the most recent ratings from the history of CUSIP-level rating changes. Finally, we encode character ratings into numerically equivalent values ranging from 28 for the highest quality to 1 for the lowest quality.

An important step in our data construction is linking the bonds issued at the local level to the counties, which form the treatment and control pairs. This geographic mapping allows us to study the implications of other economic variables using data on demographics and county-level financial metrics. Since the FTSE Municipal Bonds dataset does not have the county name of each bond, we need to supplement this information from other sources like Bloomberg. However, in light of Bloomberg's download limit, it is not feasible to search for information on each CUSIP individually. Therefore, we first extract the first six digits of the CUSIP to arrive at the issuer's identity[4]. Out of 63,754 unique issuer identities (6-digit CUSIPs), Bloomberg provides us with county-state names on 59,901 issuers. Next, we match these issuers' Federal Information Processing Standards (FIPS) code. The FIPS is then used as the matching key between bonds and treatment/control counties. We also match the names of issuers to the type of (issuer) government (state, city, county, other) on Electronic Municipal Market Access (EMMA) data provided by the Municipal Securities Rulemaking Board. We use this information to distinguish local bonds from state-level bonds because we are interested in non-state bonds.

We use the Municipal Securities Rulemaking Board (MSRB) database on secondary

[4]The CUSIP consists of 9-digits. The first six characters represent the base that identifies the bond issuer. The seventh and eighth characters identify the type of the bond or the issue. The ninth digit is a check digit that is generated automatically.

market transactions during 2005-2019. Our paper closely follows [39] in aggregating the volume-weighted trades to a monthly level. Following [82, 38], we only use customer buy trades to eliminate the possibility of bid-ask bounce effects. Given our primary focus on borrowing costs from secondary market yields, our sample is derived from the joint overlap between the bond characteristics and bond trades at the CUSIP level. In matching the bond transactions from secondary market data to their respective issuance characteristics (FTSE Russell), we rely on the CUSIP as the key identifier. In Figure C3.2, we provide kernel densities for secondary market bond features like after-tax yield spreads and remaining maturity between the treatment and control groups. Figure C3.3 represents corresponding municipal bond characteristics from the primary market. We describe the key variables in Table C1. Importantly, we find that the two groups look similar in the pattern of their distributions. We also tabulate these characteristics in Table C3.1.

The primary outcome variable used in Equation 3.1 is the tax-adjusted spread over the risk-free rate. First, we calculate the bond's coupon-equivalent risk-free yield as in [39][5]. Tax adjustment follows [5] wherein the marginal tax rate impounded in the tax-exempt bond yields is assumed to be the top statutory income tax rate in each state. This is consistent with the broad base of high-net-worth individuals and households who form a major section of investors in the US municipal bond market (often through mutual funds). A detailed study on tax segmentation across states by [46] shows significant costs for both issuers and investors in the form of higher yields. In particular, we use:

$$1 - \tau_{s,t} = (1 - \tau_t^{\mathrm{fed}}) * (1 - \tau_{s,t}^{\mathrm{state}}) \tag{3.2}$$

[5]First, we calculate the present value of coupon payments and the face value of a municipal bond using the US treasury yield curve based on zero-coupon yields as given by [48]. We get the risk-free yield to maturity using this price of the coupon-equivalent risk-free bond, the coupon payments, and the face-value payment. Finally, the yield spread is calculated as the difference between the municipal bond yield observed in the trades and the risk-free yield to maturity calculated. This yield spread calculation is similar to [49].

To compute the tax-adjusted spread on secondary market yields, we use:

$$spread_{i,t} = \frac{y_{i,t}}{(1 - \tau_{s,t})} - r_t, \tag{3.3}$$

where $r_t$ corresponds to the coupon-equivalent risk-free yield for a bond traded at time $t$. From [5], we use the top federal income tax rate as 35% from 2005 to 2012, 39.6% from 2013 to 2017, and 37% from 2018 to 2019.

### 3.3.3 Other Variables

We use Census data from the Census Bureau Annual Survey of Local Government Finances to get details on revenue, property tax, expenditures, and indebtedness of the local bodies. This gives us detailed constituents of revenue and tax components at the local level, which we use in additional tests to examine the implications for our main results. We obtain the gross domestic product (GDP) measures by county from the Bureau of Economic Analysis (BEA). Our data on county-level household income is from the Internal Revenue Service (IRS) and is used as the total personal income at the county level. Our county-level establishments, employment, and wages data come from the Quarterly Census of Employment and Wages (QCEW) by the Bureau of Labor Statistics (BLS). We use unemployment data from the Bureau of Labor Statistics. We use data from the Surveillance, Epidemiology, and End Results (SEER) Program under the National Cancer Institute for the county-level population. As a proxy for the risk-free rate, we use the zero-coupon yield provided by FEDS, which provides continuously compounded yields for maturities up to 30 years. To get tax-adjusted yield spreads, we use the highest income tax bracket for the corresponding state of the bond issuer from the Federation of Tax Administrators.

## 3.4 Results

We discuss our baseline results for Equation (3.1) documenting the impact on the borrowing costs of local governments in Section 3.4.1. Next, we propose the potential mechanism to explain our results in Section 3.4.2. In Section 3.4.3, we analyze the heterogeneity in our main result. Finally, we present robustness tests in Section 3.4.4, before discussing the impact on the primary market for municipal bonds (Section 3.4.5) and the local economy (Section 3.4.6).

### 3.4.1 Impact on Borrowing Costs of Local Governments

In this section, we start with dynamic evidence from the raw data on yields (Section 3.4.1), followed by evidence against underlying economic differences driving our results in Section 3.4.1. Finally, we present results on headquarters versus other locations of the firm filing for bankruptcy in Section 3.4.1.

*Dynamics and Baseline Results*

We begin our analysis by depicting the after-tax yields aggregated to a quarterly scale in order to overcome the inherent limitations of liquidity in the municipal bond market. We use Equation (3.4) below to represent our approach in comparing the raw difference between the treatment and control groups:

$$y_{i,c,e,t} = \beta_q * \sum_{n=-12}^{n=12} Treated_{i,c,q} * Post_{i,c,q} + \delta_q * \sum_{n=-12}^{n=12} Control_{i,c,q} * Post_{i,c,q} \quad (3.4)$$
$$+ \eta_e + \gamma_c + \kappa_t + \epsilon_{i,c,e,t}$$

where index $i$ refers to bond, $c$ refers to county, $e$ denotes the event pair, $t$ indicates the event year-month, and $q$ refers to the quarter corresponding to the event month $t$. The dependent variable $y_{i,c,d,t}$ is obtained from secondary market trades in local municipal bonds.

$\eta_e$ represents event pair fixed effects, $\gamma_c$ corresponds to county fixed effects, and $\kappa_t$ denotes year-month fixed effects. We cluster standard errors by the county bond issuer and year-month.

First, we evaluate the difference between the treated and control groups around bankruptcy filing. In Figure 3.1, the solid line with circles plots the after-tax yields over the 3-year window for the treated group on average. The control group is depicted using a dashed line. First, the figure reveals no statistical difference between the two groups during the eight quarters before the bankruptcy announcement; the treatment and control groups trend parallel. Second, on average, the treatment group's yields become higher than those of the control group just one quarter before the announcement. However, this effect is statistically indistinguishable from zero. Third, the effect increases and becomes statistically significant in subsequent quarters. Finally, the difference between the two groups persists until the $12^{th}$ quarter, when it shrinks marginally. We test these observations in our baseline regressions.

Note that the above results represent the raw difference in yield spreads between the two groups by stacking the 128 event pairs in our sample into an aggregated set. These findings do not control for differences in bond characteristics and local economic conditions over time. Next, we estimate our difference-in-differences using our baseline Equation (3.1). Here, the coefficient $\beta_0$ of the interaction term $Treated_{i,e} * Post_{i,t}$ identifies the differential effect after the bankruptcy announcement on average yields of treated counties in comparison to the control groups by additionally accounting for observable characteristics. To revisit our identifying assumption: the control county serves as an adequate counterfactual to map how the treated county's yields would have changed in the absence of the bankruptcy filing. We discuss robustness to additional matching considerations in Section C3.1.1. The event fixed effects ensure estimation from within each event pair. We absorb the unobserved county-level variation using the county-fixed effects. The year-month fixed effects control for declining yields in the overall municipal bond market during our sample

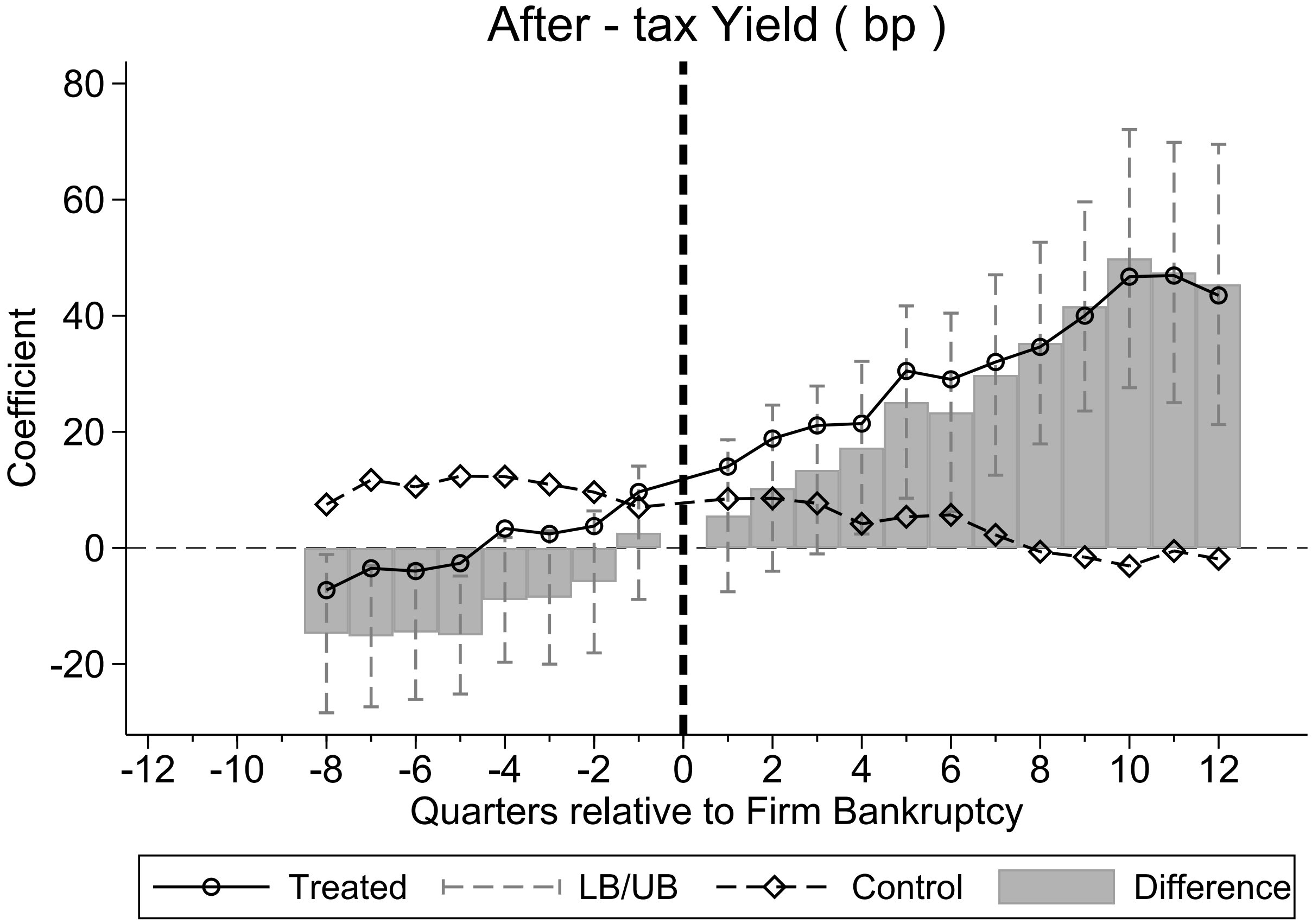

**Figure 3.1:** Municipal Bonds Secondary Yields: Treatment vs Matched Control:
The figure shows the impact on secondary market bond yield spreads between the treated and control counties. We report the coefficients from Equation 3.4. We cluster standard errors at the county bond issuer and year-month level. The control county was matched based on five variables in the year before firm bankruptcy: unemployment rate, change in unemployment rate, log(labor force), change in labor force, and average yield of the county in that year.

period, over and above the treasury adjustment for spreads.

**[Insert Table 3.2 here]**

Table 3.2, reports the effect of firm bankruptcy on the municipal bond yields using Equation (3.1). In Column (1) - Column (3), we estimate the regression equation using the after-tax yield as the dependent variable. Specifically, Column (1) denotes the estimates without using any controls. We use bond level controls in Column (2), which consist of coupon (%); log(amount issued in USD); dummies for callable bonds, bond insurance, general obligation bond, and competitively issued bonds; remaining years to maturity; and inverse years to maturity. We describe key variables in Table C1. In Column (3), we control for the county-level variation in the unemployment rate and labor force. We use the lagged values (to the year of bankruptcy filing) for log(labor force) and unemployment rate and the percentage change in the unemployment rate and labor force, respectively.

We show results with after-tax yield spread as the dependent variable in Columns (4)-(8) by following a similar progression across columns. Using Column (6) with after-tax yield spread as our baseline specification implies that the yield spread for treated counties increases by 10.01 bps after the bankruptcy filing compared to the control counties. The effect is statistically significant and economically meaningful. To understand the magnitude in context better, consider the average credit spread between AA- and AAA-rated municipal bonds, which amounts to 93 bps for our sample of municipal bonds in the treated group before the firm bankruptcy. This implies that the average increase in yield spreads after the bankruptcy represents $\sim$ 10% ($= 10.01/93$) of the credit spread. Further, we introduce more granular fixed effects in Columns (7) and (8). In Column (7), we also introduce the event pair $\times$ county fixed effect. The baseline magnitude increases to 10.15 bps, with similar statistical significance. As expected, we find that the "treated" dummy gets absorbed in this estimation. Finally, we introduce the event pair $\times$ year-month fixed effect in Column (8). Both "treated" and "post" dummies get absorbed, but the overall effect remains similar to the baseline specification in Column (6). In the next sub-section, we provide evidence against underlying local economic differences driving our main result.

*Do bond yields respond to underlying local economic differences?*

We have already shown that the secondary market and primary market bond features look almost identical in their density plots. We provide evidence from consumer bankruptcies to further alleviate concerns about underlying economic differences between the treated and control counties. If it is the case that the increase in bond yield spreads is due to worse local economic conditions in the treated counties, similar evidence should show up at the consumer level. However, in Figure 3.2 we do not find such evidence. This plot shows the regression coefficients corresponding to the treated and control groups similar to Equation (3.4). The dependent variable is the log transformation of the number of consumer bankruptcies filed in the county in a given quarter. Our evidence shows no statistical difference between the number of consumer bankruptcies filed between the two groups in the quarters before the firm bankruptcy filing. The solid line representing the treated group shows an increase in consumer bankruptcies after the penultimate quarter of firm bankruptcy filing. The difference becomes statistically significant beyond the second quarter and persists until the end of the event window (three years after). This evidence suggests that it is unlikely that the deterioration in ex-ante local economic conditions drives the results in bond yield spreads. We discuss the importance of the headquarter versus other firm locations in the next sub-section.

*Headquarters vs Other Firm Locations*

[94] document that value creation has become heavily concentrated in a few headquarters cities over the last 20 years as per stock market indicators. One may argue that while the location of a firm's headquarters may represent a hub of economic activity, it may not be the only location. Manufacturing firms may have operations in other locations besides their

**Log(Number of Consumer Bankruptcies)**

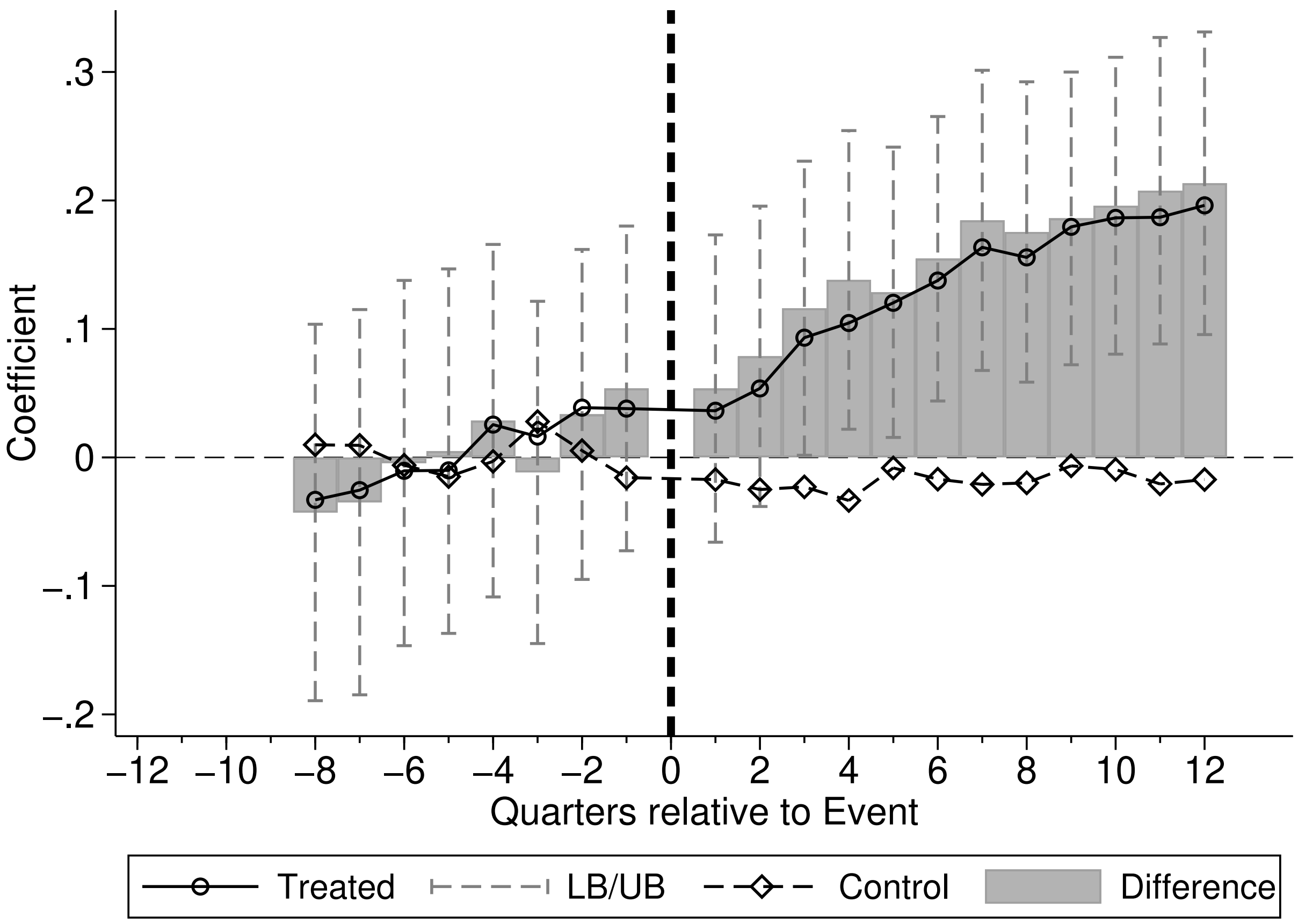

**Figure 3.2:** Consumer Bankruptcies : Treatment vs Matched Control:
The figure shows the coefficients based on Equation 3.4. We regress the logged value of the number of consumer bankruptcies as the dependent variable. We cluster standard errors at the event pair level. The control county was matched based on five variables in the year before firm bankruptcy: unemployment rate, change in unemployment rate, log(labor force), change in labor force, and average yield of the county in that year.

headquarters. To this end, we analyze the impact on additional facilities for the firms in our sample. We report our results in Table 3.3. We obtain data on other locations using two sources: Environmental Protection Agency and Mergent Intellect. Our data from Mergent Intellect is hand-collected and allows us to observe each firm's facility level DUNS number and employment information. In Column (1), we show the baseline result by including non-headquarter locations from the EPA and find the impact as 4.02 bps. However, Column (2) reveals that most of this effect comes from the headquarter locations (12.74 bps). We find similar results using non-headquarter facilities from Mergent Intellect in Columns (3)-(4). Column (3) shows that the aggregate effect across all facilities amounts to an increase in bond yield spreads of 6.34 bps. Once again, Column (4) reveals that the headquarter locations account for 10.82 bps. The effect is insignificant and too small for other locations.

**[Insert Table 3.3 here]**

Further, we also investigate the importance of facilities based on employment at the site. We show these results in Columns (5)-(7), where we consider all facilities from Mergent Intellect, which may or may not overlap with the headquarter site. However, this sample does not include all headquarters from the baseline. First, in Column (5), we find an overall increase in bond yield spreads of 3.83 bps among these facilities listed in Mergent Intellect. However, this effect only seems to be weakly significant. Column (6) shows that most of this effect belongs to multi-facility firm bankruptcies, which might include larger firms with a more significant economic footprint. We find this magnitude to be 6.63 bps. The effect is statistically significant and about half the baseline magnitude. On the other hand, the impact on single facility locations seems to be insignificant. This result is consistent with our mechanism discussing the county's dependence on the firm filing for bankruptcy (See Section 3.4.2). To further consolidate our findings on the relative economic importance of facilities, we rank them based on the fraction of employees. Sites with $\geq 25\%$ of total employees are ranked 1, followed by rank 2 for sites with $> 5\%$ of total employees. Rank 3 is assigned to the remaining locations. Finally, we report our results in Column (7), where we find the greatest impact (of 14.77 bps) due to facilities that are ranked highest (for counties with the highest employment share in the firm). The effect on counties with lower-ranked facilities is insignificant.

Our results in this section show that the increase in bond yield spreads after the firm filing for bankruptcy is mainly due to the headquarters locations. Counties with non-headquarter facilities are not affected as much. Additionally, counties with a higher fraction of employment show a more significant impact on bond yield spreads. We argue that this is because the employment share may represent the site's economic importance.

### 3.4.2 Mechanism

To explain our results, we propose a mechanism based on the exposure of treated (HQ) counties to the bankrupt firm's industry.

*Source of Repayment and County's exposure to the bankrupt firm's industry*

Our analysis in this section begins with the type of repayment among municipal bonds. We identify the two major classes of bond repayment in the sample. General obligation bonds correspond to the underlying taxing power of the issuer. Revenue bonds generally have their cash flows linked to the specific project backing the bonds. We show results for our baseline specification in Equation (3.1) in Panel A of Table 3.4.

**[Insert Table 3.4 here]**

In Column (1), we report the baseline effect of 10.01 bps across all bonds. Column (2) shows that the effect on revenue bonds is higher, amounting to 12.56 bps. This effect is primarily driven by the sub-sample of uninsured revenue bonds (Column (3)). Insured revenue bonds show a weak and statistically insignificant effect in Column (4). Finally, we do not find any significant effect on general obligation bonds in Column (5). The fact that we do not find any effect on general obligation bonds suggests that the underlying mechanism of firm bankruptcies affecting local bonds runs through cash flows linked to project revenues [92] of revenue bonds. General obligation bonds are backed by the taxing authority of local institutions and, therefore, relatively immune to such shocks based on the local economy's exposure. We build on this argument in subsequent analysis using the county-level exposure described below.

We use the 3-digit NAICS code to identify the industry of a bankrupt firm. We expect the baseline effect to be higher for counties that rely heavily on the bankrupt firm's industry. As a direct measure of the county's dependence on the industry, we calculate the county's ex-ante share of wages and employment from the NAICS 3-digit industry of the bankrupt firm in the year before firm bankruptcy. A higher proportion would reflect a county's greater reliance on that sector.

We divide the treated counties into two groups based on the median values of the share of wages and share of employment in the industry. Our specification is suitably modified to additionally include group-month fixed effects. We report our results in Panel B of Table 3.4. We split the treated counties based on the share of aggregate industry wages in Columns (1)-(3). Using all bonds in Column (1), we show that a higher dependence based on wages results in an increase in yield spreads of 15.01 bps. This effect is negligibly small and statistically insignificant for treated counties with low exposure (below median). The difference between the high and low groups is statistically significant. Thereafter, we consider the sub-samples of revenue bonds and uninsured revenue bonds in Columns (2) and (3). As before, bonds associated with treated counties of high exposure are more strongly affected.

Next, we consider the heterogeneity among treated counties based on their share of aggregate employment in the industry of the bankrupt firm. We present these results in Columns (4)-(6). Once again, we examine the impact among all municipal bonds in the sample. Column (4) shows that the after-tax yield spreads increase in treated counties with high exposure to firm bankruptcy by 13.73 bps. The corresponding effect on counties with below median exposure is only 3.88 bps and is statistically insignificant. Our results show that the difference between these two groups is statistically significant. We perform a similar analysis using the sub-samples of revenue bonds (Column (5)) and uninsured revenue bonds (Column (6)). Once again, the heterogeneity seems to be driven by counties with high exposure to the industry of bankrupt firm.

To demonstrate further support for our mechanism, we compare the county-level unemployment rate between the treated and control counties. Since the unemployment rate is reported annually, we use the annualized version of Equation (3.4) at the county level. We replace month-fixed effects with year-fixed effects and cluster standard errors at the event pair level. We show our results from the regression coefficients in Figure 3.3a. We find that in the three years before a firm's bankruptcy filing, the unemployment rates in the treated and control counties trend similarly, with negligible differences between them. The difference in the unemployment rate begins to pick up in the year of the bankruptcy filing by the firm and becomes statistically significant in the first year after the filing. Subsequently, the difference between the two groups tends to increase later. We substantiate this evidence further in light of the county-level dependence using the share of employment in the industry of firm bankruptcy. Specifically, in Figure 3.3b, we show that treated counties with higher dependence seem to drive the overall unemployment rate. However, the differential effect is relatively muted and statistically insignificant for counties with low exposure to the industry of bankrupt firms. We provide similar evidence using county exposure based on the county-level share of wages in the industry of firm bankruptcy in Figure C3.11. Once again, we find that treated counties with a higher dependence on the industry of the firm filing for bankruptcy are more severely affected. Separately, we also discuss our results in light of state-specific budgetary regulations and incentive policies in Section C3.1.2.

To summarize, our results in this subsection provide evidence in favor of the county's dependence on the industry for firm bankruptcy. We show that treated counties with a higher dependence and a greater exposure to the industry of firm bankruptcy experience a

**Panel A: All Counties**

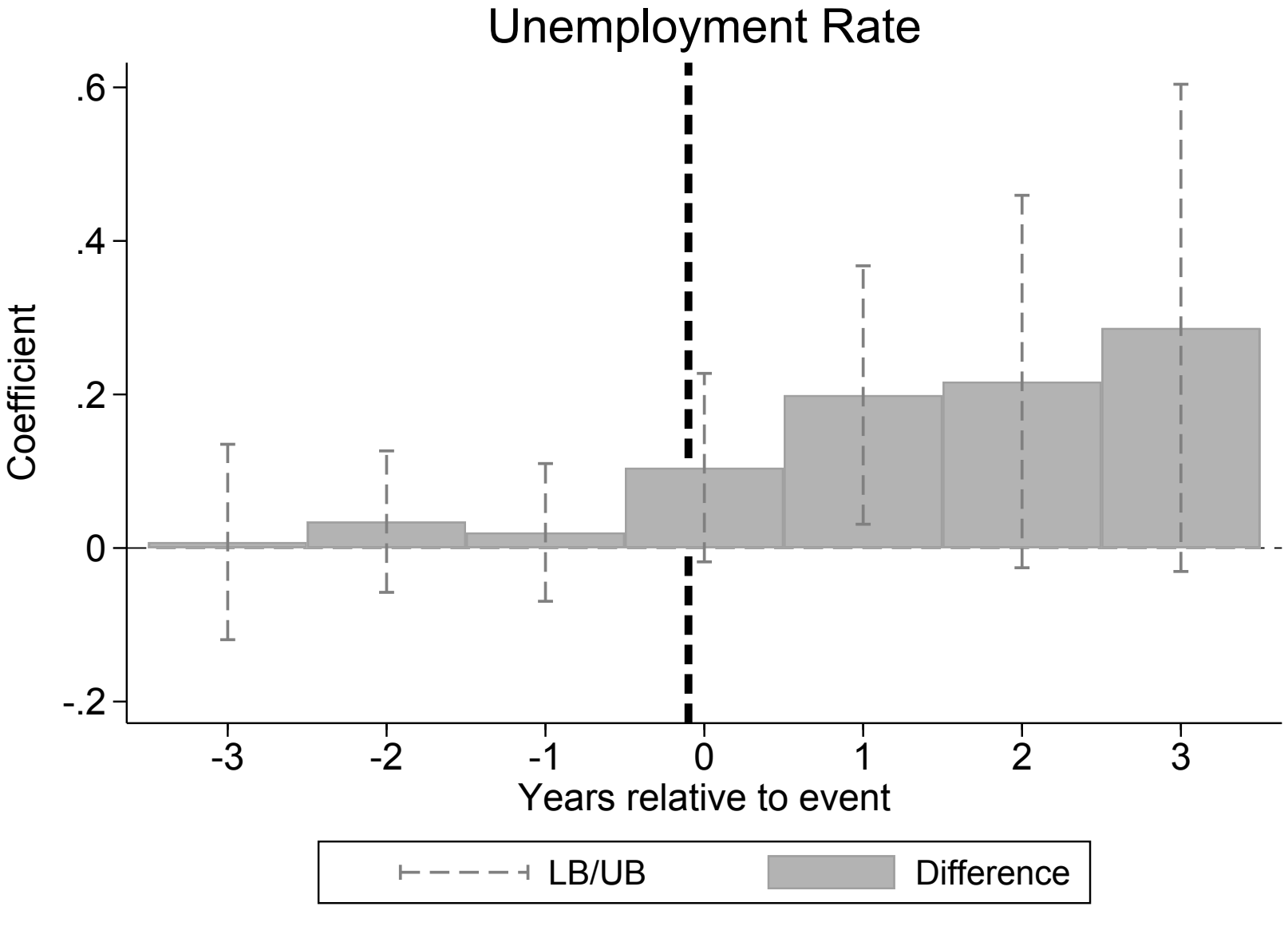

**(a)**

**Panel B: Based on County Dependence (Employment)**

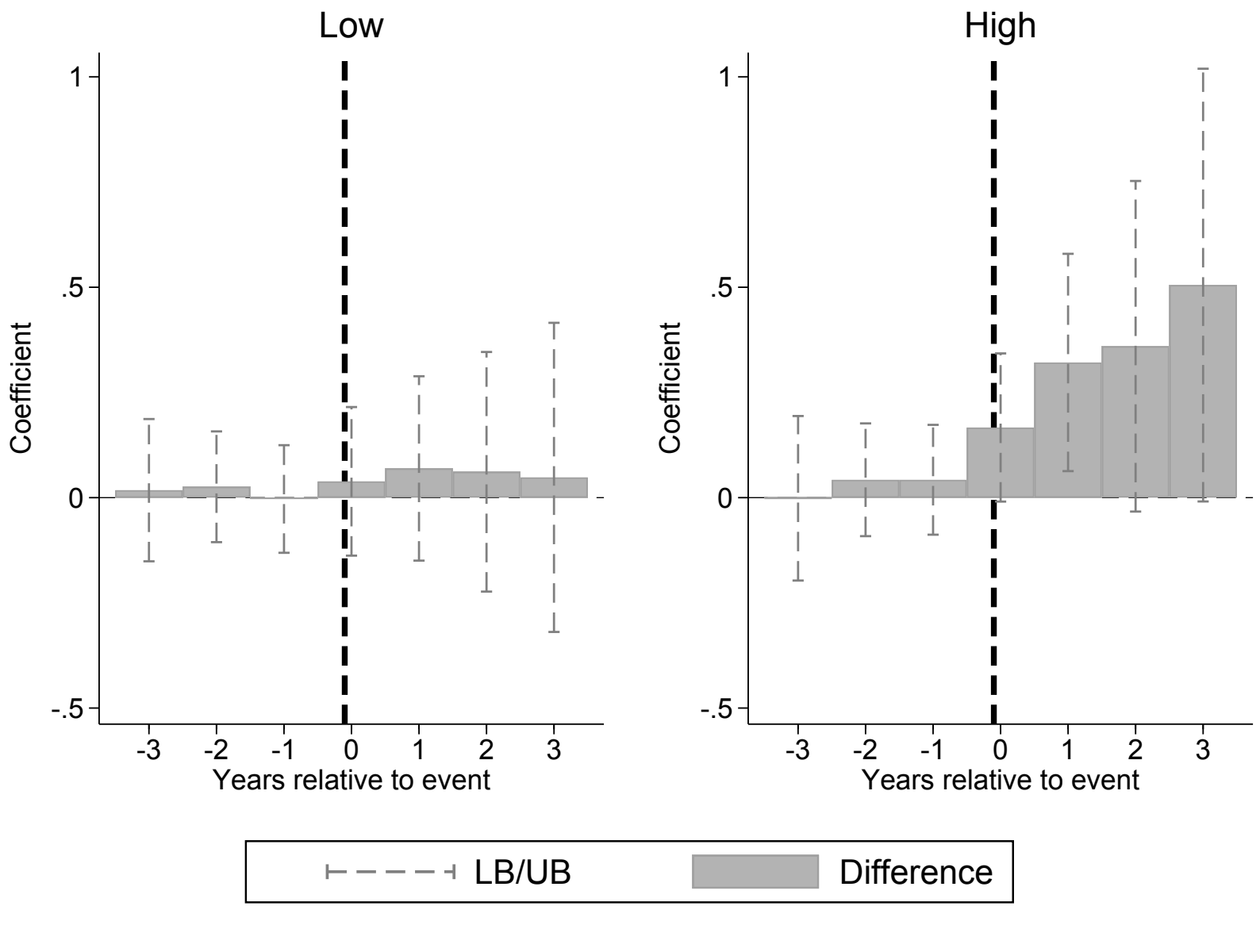

**(b)**

**Figure 3.3:** Unemployment Rate: Treatment vs Matched Control:
The figure shows the impact on unemployment rates between the treated and control counties. We report coefficients from the annualized version of Equation 3.4 using unemployment rate as the dependent variable. We cluster standard errors at the event pair level. In Panel A, we show all treated counties versus their matching control counties. In Panel B, we show the impact on sub-samples of low versus high level of county-level dependence on the industry of bankrupt firm among treated counties. We measure dependence based on the share of aggregate employment in the county. The control county was matched based on five variables in the year before firm bankruptcy: unemployment rate, change in unemployment rate, log(labor force), change in labor force, and average yield of the county in that year.

higher increase (13-15 bps) in bond yield spreads. These counties also seem to drive the result showing an aggregate increase in the unemployment rate.

### 3.4.3 Heterogeneity

In this sub-section, we test the heterogeneity of our results. We study how factors such as the relative importance of the firm filing for bankruptcy and bond-specific features may affect our results. We report our results in Table 3.5 and in Figure 3.4, respectively. In these analyses, we suitably modify our baseline Equation (3.1) to include group-year month fixed effects.

*Relative Importance of Firm*

To understand how the relative importance of a firm may determine its impact on the local economy, we construct three measures in Columns (1)-(3) of Table 3.5. First, we use a ratio of the firm's operating income (EBIT) to the county's revenue. We divide the bankruptcy events into two groups based on the median value of this ratio. We find that a high value of the ratio results in an increase in bond yield spreads of 14.98 bps. The effect on the low group is negligible and insignificant. This result suggests that firms with a relatively large size compared to the HQ county have a greater impact on the borrowing cost of the treated county. We replicate this approach using two other measures of a firm's importance to the county. Since we do not observe the number of employees on the firm's payroll for each location, we proxy for this by using selling, general, and administrative (SG&A) expenses. In Column (2), we use the ratio of SG&A expenses of the firm to the county's revenue. We find that a higher ratio of expenses to county revenue increases bond yields by 20.84 bps. The differential effect in the high group counties is statistically different from the treated counties with low values. Finally, in Column (3), we evaluate the relative importance of the firm by comparing the firm's plant, property, and equipment to the county's property tax revenue. Since property tax consists of a large portion of a county's income, we argue that in the absence of precise data on commercial property tax revenue from the bankrupt firm, this measure would capture the economic significance of the firm. Consistent with our previous results, we find that counties with a high value of this ratio experience a yield spread increase of 11.96 bps.

**[Insert Table 3.5 here]**

*Purpose of Bonds*

Next, we evaluate the differential impact among different types of municipal bonds. Specifically, we classify them based on the use of proceeds associated with municipal bonds. We report our results in Figure 3.4. The highest impact (32.55 bps) occurs with bonds associated with improvement and development. This is understandable because many of these bonds are associated with constructing new amenities and features in the locality with the prospect of providing infrastructure for private sector firms. For example, industrial revenue bonds may be for building or acquiring factories or heavy equipment and tools. We do not find a large impact on bonds that are linked to services and fee-based utilities.

Overall, our results in this section provide evidence of the heterogeneity of the impact of firm bankruptcies. We show that the impact is higher if the relative importance of the firm in the county is higher. Further, the increase in bond yields comes from specific types of bonds that are likely affected due to the firm bankruptcy.

### 3.4.4 Robustness Tests

In this section, we test the robustness of our main result in Column (6) of Panel A in Table 3.2 to various potential confounding factors. We present the results of these robustness checks in Table 3.6.

**[Insert Table 3.6 here]**

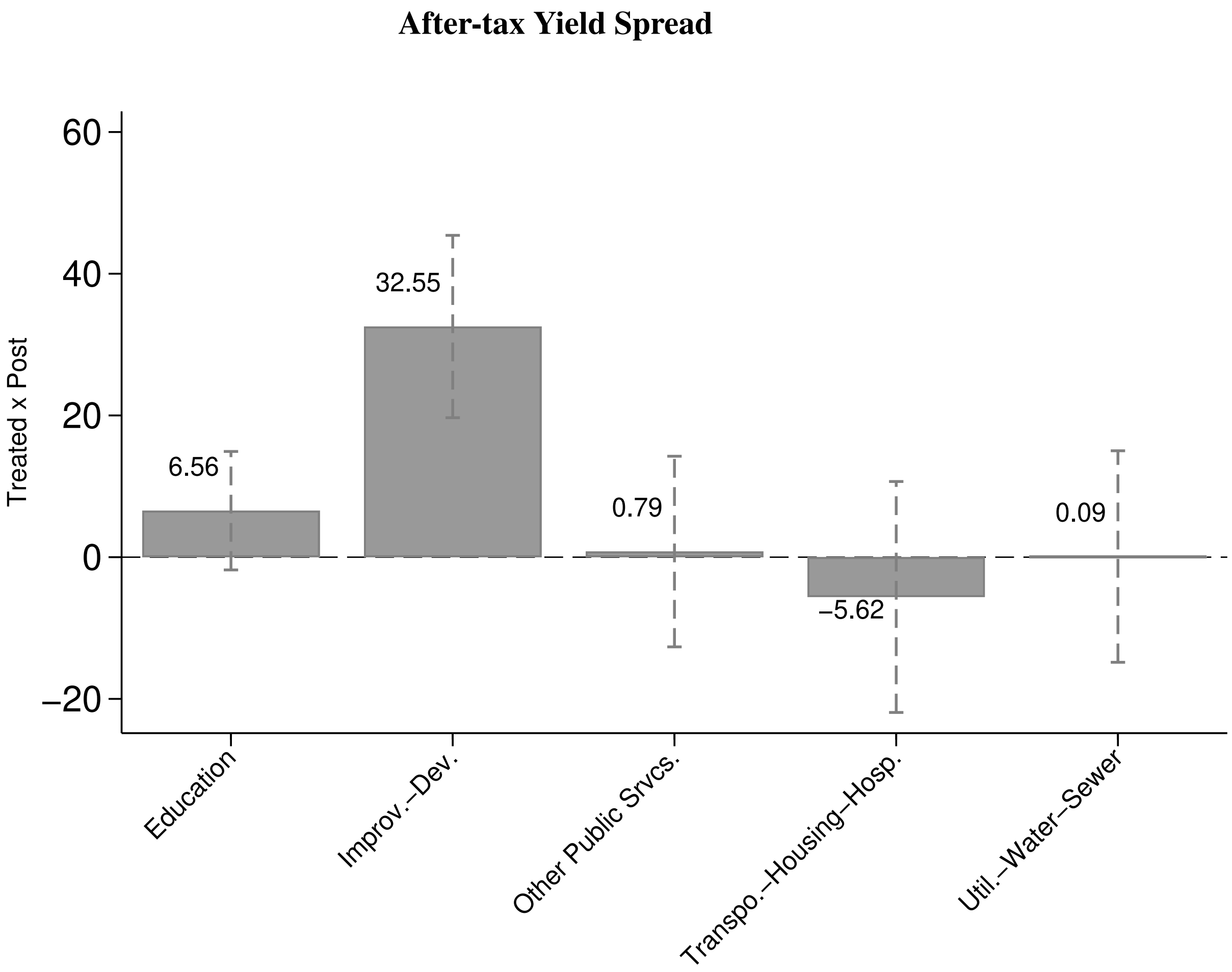

**Figure 3.4:** Heterogeneity based on Purpose of Bonds:
The figure shows results for our main interaction term, $\beta_0$, from Equation (3.1). We modify the baseline equation to interact with indicators corresponding to the different types of municipal bonds, based on the use of proceeds. We additionally control for group-month fixed effects in the regression. Standard errors are double clustered by county bond issuer and year-month. The dashed lines represent 95% confidence intervals.

*Other unobservables*

We consider whether our results are robust to various unobserved factors at the bond, county, and geographic region levels. First, we absorb all bond level, time-invariant variation to control for unobserved bond features in Column (1). This yields a baseline effect of 8.23 bps. To account for unobserved regional variation in the US, we show our results in Columns (2)-(4). First, in Column (2), we show the result after introducing region-fixed effects to the baseline specification. Our magnitude changes to 10.01 bps. Further, we impose region $\times$ year fixed effects to account for time-varying unobserved factors at the regional level. The main result in Column (3) is 10.42 bps. Finally, in Column (4), we show our results by controlling for a more granular time-varying component by introducing region $\times$ year-month fixed effects. Even with this restrictive specification, we report an increase in yield spreads of 10.66 bps. Moreover, if there is a concern that the municipal bond market yields were different for general obligation versus revenue bonds in a given year-month, we control for year-month $\times$ GO bond indicator fixed effects in Column (5). We find that the bond yield spreads increase by 8.74 bps. Likewise, in Column (6), we control for year-month $\times$ insured bond status fixed effects to control for unobserved market-wide factors for insured bonds. Using bond purpose $\times$ year-month fixed effects, we report the baseline effects as 7.88 bps in Column (7). This specification controls for any time-varying unobservables that may drive our result in yields for a given use of proceeds raised in the municipal bond market. Our analysis suggests that our main effect is robust to unobserved factors at the bond, county, region, and municipal bond market levels.

*Additional controls*

Following unobserved factors, we now control for additional observables in Panel B. In Column (1), we present our results by introducing some more county-level time-varying covariates. We introduce the lagged values of log(personal income) and log(house price index) to account for any changes in these metrics that may be simultaneously changing at the time of firm bankruptcy filing. In this case, the bond yield spreads increase by 9.60 bps. Our next check on the robustness of our main results involves using only those bonds for which the most recent bond ratings are available from Standard & Poor's credit ratings in the FTSE Russell municipal bonds database. Since the restriction of requiring bond ratings reduces the sample, we do not impose the rating requirement on the baseline specification. We show our results by controlling for bond ratings in Column (2) of Table 3.6. We find that the bond yield spreads increase by 8.29 bps. This test mitigates any concern about unrated bonds solely driving our results.

To verify our results' robustness to the bonds' duration effects, we modify the baseline specification in Columns (3)-(4). First, Column (3) shows our main effect by replacing years to maturity and inverse years to maturity at the bond level with the corresponding duration for the bond-month observation. We report a higher impact of 10.60 bps. Next, in Column (4), we show the same result by calculating duration based on after-tax yields. This tax adjustment also increases the impact to 10.61 bps. Our main result is robust to altering the baseline specification to include bond level duration instead of years to maturity and inverse years to maturity among the bond level controls.

*Is the effect driven by recently issued bonds?*

[6] documents that price dispersion in municipal bond trades in the days after issuance has declined dramatically since the introduction of 15-minute trade reporting by the MSRB in 2005. While our municipal bond trades started in 2005, we further exercise caution in this regard by dropping bonds issued within a few months of the bankruptcy filing. We show these results in Columns (1)-(4) in Panel C. For example, in Column (1), we report the baseline result by dropping bonds dated within 6 months of the bankruptcy filing and find that the increase in bond yield spreads is 10.01 bps. In each sub-sample, the coefficient of interest is higher in magnitude than in the baseline result. Based on this evidence, we argue that trades from recently issued bonds do not drive our results.

*Alternative specifications*

We also consider some modifications to the baseline equation to evaluate potential concerns about alternative specifications that might be used in this setting. We report our results in Panel D. First, we consider our choice of clustering standard errors in the baseline specification in Equation (3.1), which is similar to the existing literature [39, 38]. In Columns (1)-(3), we show our main result using alternative definitions for clustering. First, we consider the possibility that the standard errors may be correlated across different bond issues over the calendar months (see Column (1)). It could also be that the error term in our main specification is correlated with specific bond issuers. We report our results for this in Column (2). Finally, we also consider clustering standard errors at the county bond issue level in Column (3). In all these specifications, we find results similar to our baseline specification. Thus, our results are robust to these choices of clustering standard errors.

Next, we add event-month fixed effects to the baseline. If there is a concern that the impact on bond yields may be driven by specific timing effects on the event scale, this modification should account for that. In other words, if the relative age of the bond trades with respect to the bankruptcy filing may affect our main result, then these fixed effects should absorb that variation. In Column (4), the baseline effect is reported as 10.03 bps. Further, if one believes that controlling for event-month fixed effects is more important than absorbing calendar year-month fixed effects, we address this in Column (5). We replace the year-month fixed effects with event-month fixed effects. The overall magnitude goes up to 23.27 bps and is statistically significant. Finally, we show robustness in replacing year-month fixed effects with year-fixed effects in Column (6). The increase in bond yields is reported as 10.35 bps. Overall, we show robustness to alternative specifications to our baseline Equation (3.1).

*Is the effect driven by size of trades?*

One potential concern with interpreting the results presented so far is that they may be driven by different customer groups in the municipal bond market. [51] show that municipal bond dealers earn lower average markups on larger trades even though they bear a higher risk of losses on such transactions. In this market, retail customers dominate the holdings [5] through direct ownership or via investment vehicles like bond mutual funds. However, there has been a recent rise in holdings by institutional investors as well[6]. Accounting for such differences, we dissect our results into sub-samples of trades constituting various buckets and show our results in Panel E. Columns (1)-(4) depict the main effect derived from trade sizes with threshold cut-offs worth $25,000 or $50,000. The increase in borrowing cost is around 8 bps or more in each of these specifications except in Column (2), which has substantially fewer trades due to the sample restriction. The results are comparable to the baseline effect regarding statistical significance and economic meaning. This suggests that any specific group of clientele does not seem to drive our main result.

*Additional Considerations*

We focus on some additional considerations regarding our sample and specification in Panel F. First, we address the concern that our results may be capturing the differences in taxation of income from municipal bonds across states in Column (1). By focusing only on tax-exempt bonds in the sample, we try to minimize the differential impact due to taxation and report our main result as 11.02 bps.

Another potential worry stems from the fact that the sample period spans the financial crisis of 2008-2009. This was a period of major unrest in the financial markets across asset classes, and municipal bonds were not immune to this unrest. As a result, we report our findings by excluding this period from our data. Column (2) shows the baseline results from Equation (3.1) by excluding all trades up to 2010. The increase in bond yield spreads among the treated counties is 11.04 bps higher than those of the control counties. The sample size is reduced by almost half, but the main effect is still statistically significant and economically meaningful with this reduced sample. This suggests that our result is not primarily driven by the financial crisis of 2008-2009.

[6]http://www.msrb.org/msrb1/pdfs/MSRB-Brief-Trends-Bond-Ownership.pdf

Next, we test our main result from Equation (3.1) by changing the dependent variable. Column (3) uses the monthly average yield observed from MSRB as the outcome variable. We find that the average increase in bond yields is 7.19 bps. The effect is statistically significant and economically meaningful. Similarly, by using the yield spread as the dependent variable in Column (4), we find that the baseline effect is 6.51 bps. These outcome variables do not account for the state-level tax exemption on income from municipal bonds. Overall, our results suggest that our baseline effect is robust to the choice of the dependent variable used in the analysis.

It is possible that our matching control counties may also face the repercussions of recent firm bankruptcy. If so, this might underestimate our main result. We address this concern in Column (5) by restricting the control counties to include only those that do not also see a firm bankruptcy from our sample within two years of the treated county's firm bankruptcy filing. We find that the main result shows up as 8.69 bps in this revised sample of restricted counties for the control set. We argue that this restriction could likely reduce the availability of matching counties based on size. As a result, the matched counties may likely be smaller than the treated group, leading to higher yields.

Finally, to evaluate if our results are sensitive to the choice of using manufacturing firm bankruptcies, we broaden the set of firms to include all tradeables in Column (6). Our classification of industries is based on [103]. We exclude all service industries as per the Census [7] and classify the nontradable sector as retail trade (NAICS 44-45) and accommodation and food services (NAICS 72). All remaining industries are considered under tradeables, which are mostly manufacturing (NAICS 31-33). In Column (6), we report our baseline effect as 8.27 bps by using firm bankruptcies in tradeables.

[7] www.census.gov/econ/services.html

Overall, we provide evidence to show that our results are not sensitive to redefining the control group or to the choice of focusing on manufacturing firm bankruptcies. Our results are also robust to dropping taxable bonds from the sample.

*Sensitivity to Length of Event Window*

To evaluate the sensitivity of our baseline results against the choice of event window used, we show our main result in Panel G using different lengths of the event window. We modify the baseline choice of a 36-month event window to range between 12 and 60-month event windows. In Column (1), we find that even over a short window of one year, the magnitude of 3.86 bps is statistically significant. The increase in yield spreads ranges between 5 to 12 bps when we consider alternative windows, as shown in Columns (2)-(6). We argue that a longer period is needed in the preceding and succeeding periods around the announcement of firm bankruptcy to arrive at sharper estimates of the effect, especially given the limited trading in the municipal bond market.

*Sensitivity to Start/End of Event Window*

Next, in Panel H, we show the baseline result of Column (6) in Table 3.2 by changing either the starting or the ending point in the event window. Specifically, in Columns (1)-(3), we use shorter pre-event periods ranging from 6 to 24 months before the bankruptcy filing. We find that the magnitude of the baseline impact ranges between 6 to 10 bps. In Columns (4)-(6), we keep the pre-event window similar to the baseline specification (at 36 months) and modify the post-event window. Our results remain statistically significant and economically meaningful, ranging between 4 and 8 bps.

Overall, this section demonstrates robustness to our baseline specification using alternative considerations. We consider additional matching strategies to identify the counterfactual counties for our setting in Section C3.1.1. In the following sections, we present results from new municipal bonds and the local economy.

### 3.4.5 Impact on Primary Market of Municipal Bonds

As a related implication of our first result on increasing yields for the treated county in the secondary market, we evaluate the event's impact on the primary market using new bond issuance data. The average new issuance in the sample is $53 million (median is $15 million) for the treated counties. Therefore, it becomes important to understand the borrowing cost implications of raising new money. To this end, we use a similar estimate as the difference-in-differences test of the secondary market, with suitable modifications to Equation (3.1). The equation is presented below:

$$y_{i,c,e,t} = \beta_0 * Treated_{i,c,e} * Post_{i,c,t} + \beta_1 * Treated_{i,c,e} + \beta_2 * Post_{i,c,t} \quad (3.5)$$
$$+ BondControls + CountyControls + \eta_e + \delta_c + \gamma_j + \kappa_t + \epsilon_{i,c,e,t}$$

where index $i$ refers to the bond, $c$ refers to the county, $e$ denotes the event pair, and $t$ indicates the time (in months). We use tax-adjusted yields at the time of bond issuance obtained from the primary market as the dependent variable in $y_{i,c,e,t}$. Since municipal bonds are tax-exempt at the state level, we argue that adjusting for variation in tax effects across states is relevant in our sample. The coefficient of interest is $\beta_0$. We use the same items for *BondControls* and *CountyControls* as in Equation (3.1). $\eta_e$ refers to event pair fixed effects, $\delta_c$ corresponds to county fixed effects, and $\gamma_j$ represents issuer fixed effect for the bond $i$ in county $c$. The issuer fixed effects help us to control for characteristics specific to the issuer, especially in the context of the purpose for which the bond may be issued. We also control for the month in which the bond was issued by including year-month fixed effects in the form of $\kappa_t$. We cluster standard errors at the bond issue level.

**[Insert Table 3.7 here]**

Table 3.7 shows the results for the specification above. Column (1) estimates the difference-in-differences coefficient from within the same event-pair, absorbing for issuer fixed effects. We report an increase of 13.69 bps in the after-tax offering yields. However, this does not account for the sample's bond level and county-level heterogeneity. In Column (2), we introduce bond level controls to show an increase in offering yields of new issuances by 10.49 bps. Next, we introduce county-level time-varying characteristics along with county-fixed effects in Column (3) to get a 5.79 bps increase in offering yields for the treated counties compared to the control group.

Overall, our results suggest that the primary market for bond issuance is associated with an increase in tax-adjusted yields of about 6 bps after a firm bankruptcy filing. However, we admit a major caveat. If counties and local governments rationally expect a higher borrowing cost in anticipation of the bankruptcy of a distressed firm, they may try to time the market and raise money well before or after the event. This inherent endogeneity problem limits our ability to test our hypothesis in this market. Existing bonds that are already trading in the secondary market are not riddled with this limitation and, hence are used in our baseline analysis to evaluate the impact on borrowing cost. Since the information on bond ratings reduces our sample size, we do not consider this in the main analysis above. However, we show similar results after controlling for bond ratings, resulting in a smaller sample shown in Table C3.3.

In Figure 3.5, we also show evidence indicating lower municipal bond issuance among the treated counties when compared to their matching control counties in the three years after the firm bankruptcy. This is consistent with the increase in offering yields of new bond issuances. We substantiate the evidence on the underlying mechanism discussed in Section 3.4.2 by showing these results based on the treated county's exposure. We find that the reduction in issuance is driven by counties with a high dependence on the industry of the bankrupt firm. As before, we measure dependence based on the fraction of aggregate employment of the county in the industry of a firm filing for bankruptcy. Alternatively, we find similar results when we measure the dependence based on the aggregate share of

**Panel A: Overall**

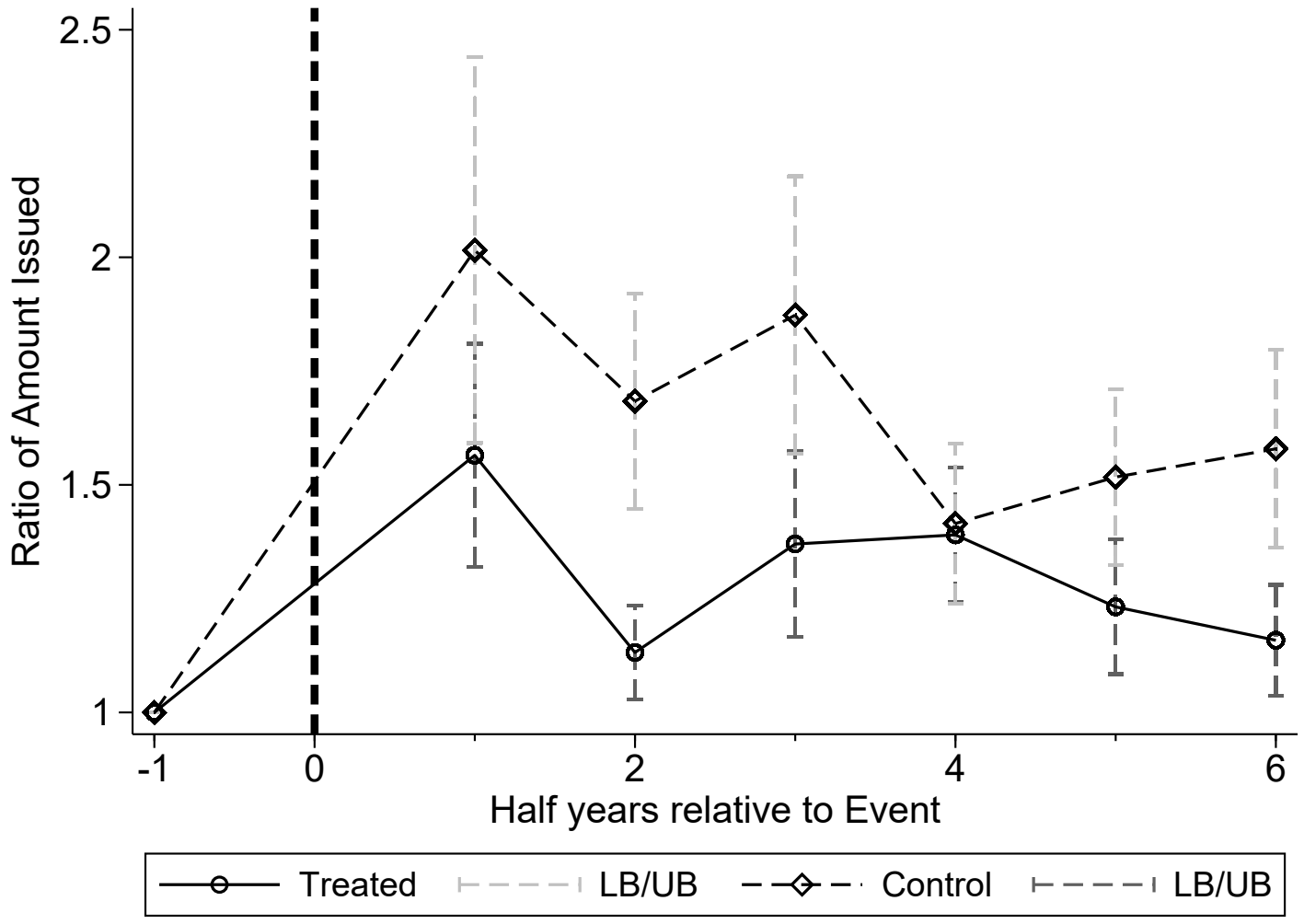

**(a)**

**Panel B: Based on County Dependence (Employment)**

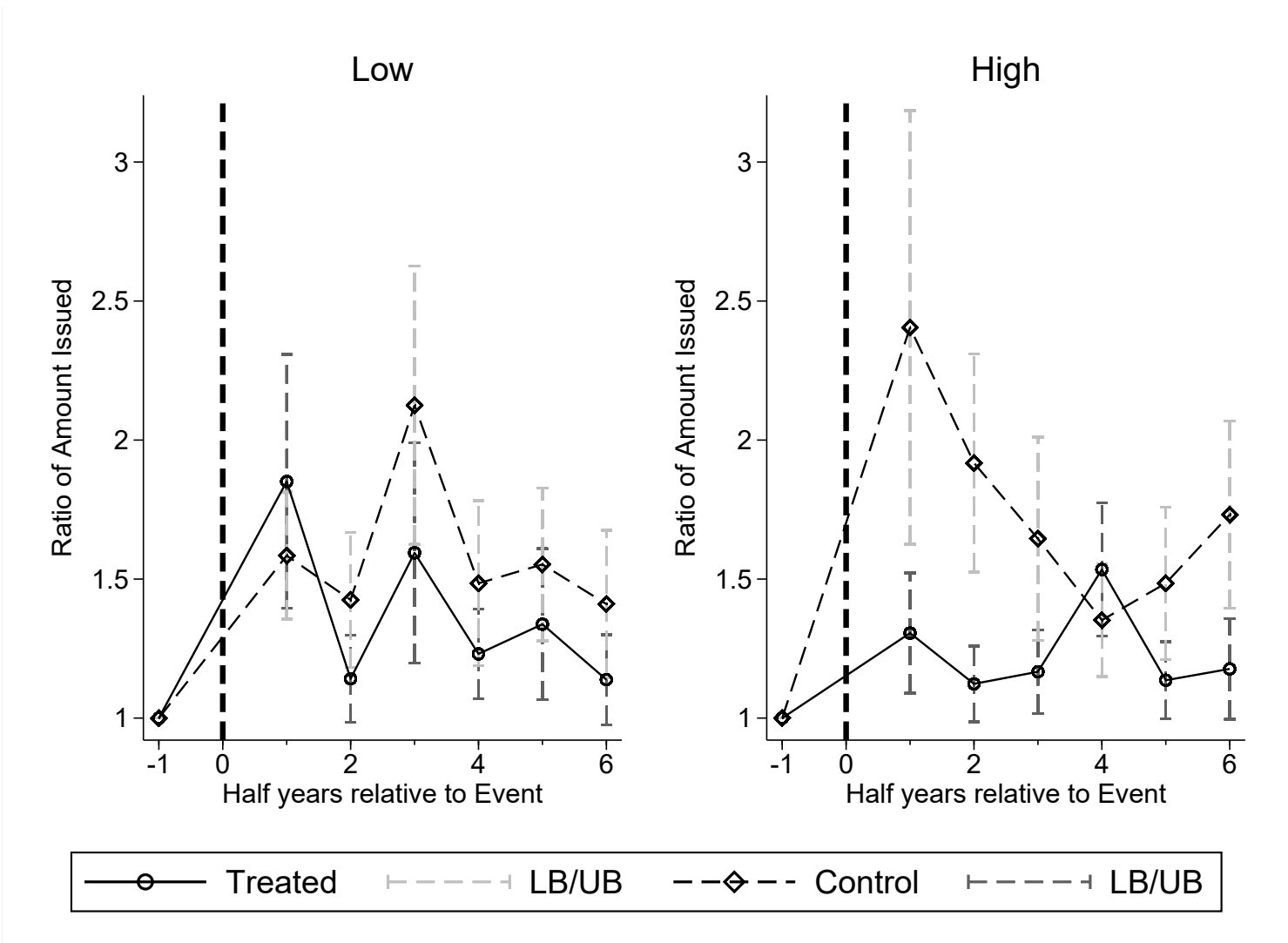

**(b)**

**Figure 3.5:** Primary Market Bond Issuance:
The figure shows the county level aggregate volume of bond issuance for treated and control counties after the firm bankruptcy filing. For each county, we calculate the total par value of bonds issued in the six month rolling window before the corresponding bankruptcy event. We normalize this value to one and compute total par value of new issues relative to this amount in the half years after the announcement. The ratio represents the relative growth in issuance among treated counties, compared to the corresponding growth of control counties' issuance. The vertical bars show the upper and lower limits based on the standard errors of the mean values. In Panel A, we show all treated counties versus their matching control counties. In Panel B, we show the impact on sub-samples of low versus high level of county-level dependence on the industry of bankrupt firm among treated counties. We measure dependence based on the share of aggregate employment in the county. The control group counties are matched using five variables in the year before firm bankruptcy: unemployment rate, change in unemployment rate, log(labor force), change in labor force, and average yield of the county in that year.

wages belonging to the industry of a bankrupt firm (Figure C3.12).

### 3.4.6 Impact on County's Local Economy

Having shown thus far that a county's borrowing cost is affected following the bankruptcy of a listed manufacturing firm, we now turn to some real economy effects after this event.

Business taxes account for more than one-third of local government revenue. Since the bankruptcy filing of a listed manufacturing firm can be interpreted as a major shock to expected future revenues, we investigate the implications on the local economy in this section. We examine the impact on the county's GDP growth (%) and growth in county-level establishments. We present our results in Table 3.8.

**[Insert Table 3.8 here]**

To study the county-level metrics reported on an annual basis, we modify our baseline Equation (3.1) as below:

$$y_{c,e,t} = \beta_0 * Treated_{c,e} * Post_{c,t} + \beta_1 * Treated_{c,e} + \beta_2 * Post_{c,t} + CountyControls + \eta_e + \gamma_c + \epsilon_{c,e,t} \quad (3.6)$$

where index $c$ refers to the county, $e$ denotes the event pair, and $t$ indicates the year. The dependent variable $y_{c,e,t}$ represents the county level outcome variables. The coefficient of interest is $\beta_0$, representing the differential impact on treated counties compared to the control group after the bankruptcy filing by the firm. The above specification also includes two sets of fixed effects: event pair fixed effects ($\eta_e$), so the comparisons are within counties mapped to an event pair; and county fixed effects ($\gamma_c$) to account for unobserved variation across counties. Here, we cluster standard errors at the event level.

First, we show our results using the baseline difference-in-differences strategy in Panel A. We start by analyzing the overall GDP growth (%) in the county in Column (1) for the baseline event window of three years around the firm bankruptcy. Our results show a 0.67% decline in GDP growth among headquarters counties in the three years after the listed manufacturing firm filed for bankruptcy. This effect persists over five years (Column (2)) and seven years (Column (3)) after the event. However, the magnitude of the impact declines marginally. Next, we follow a similar estimation for the growth in county-level establishments in Columns (3)-(6). Again, we find weak evidence for a decline in the coefficient of interest, but the magnitude is much smaller and statistically insignificant.

Next, we examine these county-level dependent variables per our proposed county-level exposure mechanism. We show these results in Panel B, by splitting the treated counties into two groups based on the median value of county-level employment dependence. Using county GDP growth as the dependent variable in Columns (1)-(3), we find evidence consistent with our mechanism. The decline in GDP growth is driven by counties with above median (high) dependence on the industry of bankrupt firm. The difference between the high and low groups is statistically significant during the three, five, and seven years after the firm's bankruptcy. The evidence of county-level growth in establishments is not clear.

#### 3.4.7 Recovery of Treated Counties

Following the evidence about the local economy, we now turn to the long-term implications of large firm bankruptcy. We consider conventional proxies for a city's appeal to workers [94] like education rate and weather to understand the probable recovery over time. We extend this argument to other factors like entrepreneurship, start-up formation rate, patent formation, race, and mobility across counties. Using the county-level measures of growth in GDP and the growth in establishments discussed in Section 3.4.6, we show results based on the heterogeneity among treated counties.

Table 3.9 shows these results using the corresponding interactions for Equation (3.6). First, based on the county-level rate of bachelor's education (Panel A), we find no difference between high and low groups of treated counties for growth in GDP. This analysis represents the top quartile of treated counties in the "high" group. The remaining counties are classified as the low group. As before, we perform this analysis for the baseline event window of three years around bankruptcy and extended periods of five and seven years after the event, respectively. However, Columns (4)-(6) show some evidence for higher county-level growth in establishments for treated counties with high levels of bachelor's education.

**[Insert Table 3.9 here]**

In Panel B, we use the number of pleasant days to group the treated counties into low versus high categories. Here, we find that growth in county GDP declines more for counties with a high number of pleasant days. This difference is statistically significant and economically large. However, the corresponding evidence from the growth in establishments in Columns (3)-(6) is inconclusive. We report similar mixed evidence in Panels C to H. Thus, along these observable dimensions, we find limited evidence supporting the recovery of treated counties after bankruptcy.

## 3.5 Conclusion

Using bankruptcies of publicly listed manufacturing firms during 2006-2016, we assess the impact on local government finances at headquarters counties and other locations. We focus on listed manufacturing firms since they are less likely to depend on local demand. This allows us to identify the impact of firm bankruptcy on HQ county and other locations. We find that the cost of municipal debt increases by 10 bps in the secondary market three years after the bankruptcy filing by a local manufacturing firm. We approximate the treated county's exposure to the firm bankruptcy using aggregate industry-county-level employment and wage data. This negative impact is higher when the county is more dependent on the industry of the bankrupt firm.

We also explore the heterogeneity based on the economic exposure of the county to the bankrupt firm and the bankrupt firm's industry. When the firm is relatively more im-portant to the county, the impact on the bond yield spreads is higher. Consistent with the increase in the secondary market yields, we document an increase in yields for new municipal bond issuances also. Counties experience a decline in GDP growth in the three years after the firm bankruptcy compared to their control group. Overall, our results highlight local communities are an important stakeholder in corporations and how they get affected by corporate bankruptcies.

**Table C1:** Description of Key Variables

This table reports variable definitions. Data sources include municipal bond transaction data from the Municipal Security Rulemaking Board (MSRB), FTSE Russell's Municipal Bond Securities Database (FTSE, formerly known as Mergent MBSD), zero coupon yield provided by FEDS, highest income tax bracket for the corresponding state of the bond issuer from the Federation of Tax Administrators (FTA), and Census data from the Census Bureau Annual Survey of Local Government Finances (CLGF). Our data on firm bankruptcy filings comes from [95] and [96] (CST-ACK).

| Variable | Description | Source |
|---|---|---|
| *Treated* | Dummy set to one for a county that has a firm bankruptcy filing. This dummy equals zero for the control group county in that event pair. | CST-ACK |
| *Post* | Dummy that is assigned a value of one for months after the bankruptcy filing and zero otherwise. | CST-ACK, MSRB |
| *Average Yield* | Volume-weighted average yield for a CUSIP in a given month. Volume refers to the par value of the trade. | MSRB |
| *Yield Spread* | Calculated as the difference between the *Average Yield* and the coupon-equivalent risk free yield. The risk free yield is based on the present value of coupon payments and the face value of the municipal bond using the US treasury yield curve based on maturity-matched zero-coupon yields as given by [48]. This yield spread calculation is similar to [49]. | MSRB, FEDS |

| Variable | Description | Source |
| --- | --- | --- |
| *After-tax Yield Spread* | Calculated as the difference between the tax-adjusted *Average Yield* and the coupon-equivalent risk free yield. The risk free yield is based on the present value of coupon payments and the face value of the municipal bond using the US treasury yield curve based on maturity-matched zero-coupon yields as given by [48]. This yield spread calculation is similar to [49]. We follow [5] in applying the tax adjustment. It is calculated as below: $spread_{i,t} = \frac{y_{i,t}}{(1-\tau_t^{\text{fed}}) * (1-\tau_{s,t}^{\text{state}}))} - r_t$ | MSRB, FEDS, FTA |
| *Competitive Bond Dummy* | Dummy variable that equals 1 if the issue is sold to underwriters on a competitive basis, and 0 otherwise. | FTSE |
| *GO Bond Dummy* | Dummy variable for general obligation bond. A GO bond is a municipal bond backed by the credit and taxing power of the issuing jurisdiction rather than the revenue from a given project. | FTSE |
| *Log(Amount)* | Log transformation of the dollar amount of the individual bond's (9-digit CUSIP) original offering. | FTSE |
| *Callable Dummy* | Dummy variable that equals 1 if the issue is callable, and 0 otherwise. | FTSE |

| Variable | Description | Source |
|---|---|---|
| *Insured Dummy* | Dummy variable that equals 1 if the issue is insured, and 0 otherwise. | FTSE |
| *Remaining Maturity* | Individual bond maturity measured in years. | FTSE, MSRB |
| *Inverse Maturity* | Inverse of the value of *Remaining Maturity*; to account for non-linearity. | FTSE, MSRB |

## **Table 3.1:** Summary Statistics

This table summarizes the ex-ante financial characteristics for our sample of bankruptcy firms during 2006-2016 and the corresponding headquarter counties. All figures in Panel A are in USD Million, except the number of employees. In Panel B, we report the difference between the average values of the ex-ante characteristics between the treatment (HQ) and control counties. Panel C provides more details about the distribution of those variables at the county level.

**Panel A: Firm Bankruptcies in Manufacturing**

| | Mean | Median | Std. Dev. |
|---|---|---|---|
| Total Revenue | 578.5 | 31.1 | 1,518.3 |
| EBITDA | 14.6 | -4.7 | 101.5 |
| Net Income | -75.0 | -13.2 | 206.1 |
| Total Assets | 461.6 | 43.7 | 1,142.8 |
| LT Debt - Total | 105.0 | 1.6 | 315.9 |
| Equity | -27.5 | 0.1 | 337.6 |
| Employees | 2,168.9 | 167.5 | 7,171.5 |

**Panel B: Difference between Treated and Control Counties**

| | Mean (Treated) | Mean (Control) | Difference | T-Stat |
|---|---|---|---|---|
| Annual Trading Volume (USD mio) | 761.9 | 589.2 | 172.74 | 1.23 |
| Annual Average Yield (%) | 3.5 | 3.5 | -0.05 | -0.39 |
| Population | 884,806.9 | 707,202.2 | 177,604.66 | 1.71 |
| Unemployment Rate (%) | 0.1 | 0.1 | 0.00 | 0.10 |
| Laborforce | 458,578.2 | 371,886.1 | 86,692.05 | 1.67 |
| Log(Laborforce) | 12.6 | 12.5 | 0.13 | 1.00 |
| Revenue (USD million) | 4.6 | 3.4 | 1.22 | 1.92 |
| Expenditure (USD million) | 4.6 | 3.4 | 1.13 | 1.82 |
| Surplus (USD million) | 0.0 | -0.1 | 0.09 | 1.58 |
| Zillow HPI | 300,816.0 | 270,647.2 | 30,168.80 | 1.04 |

**Panel C: Distribution between Treated and Control Counties**

| | Mean | p25 | p50 | p75 | Std. Dev. |
|---|---|---|---|---|---|
| **Treated** | | | | | |
| Annual Trading Volume (USD mio) | 761.9 | 85.0 | 242.3 | 880.7 | 1,176.2 |
| Annual Average Yield (%) | 3.5 | 2.9 | 3.8 | 4.1 | 1.0 |
| Population | 884,806.9 | 361,000.0 | 686,600.0 | 1,023,859.0 | 1,033,083.0 |
| Unemployment Rate (%) | 0.1 | 0.0 | 0.1 | 0.1 | 0.0 |
| Log(Laborforce) | 12.6 | 12.1 | 12.8 | 13.3 | 1.0 |
| Revenue (USD million) | 4.6 | 1.3 | 2.7 | 5.9 | 6.5 |
| Expenditure (USD million) | 4.6 | 1.3 | 2.8 | 6.0 | 6.3 |
| Zillow HPI | 300,816.0 | 138,200.0 | 237,100.0 | 379,700.0 | 245,562.8 |
| **Control** | | | | | |
| Annual Trading Volume (USD mio) | 589.2 | 87.0 | 213.9 | 601.9 | 1,059.5 |
| Annual Average Yield (%) | 3.5 | 3.1 | 3.7 | 4.1 | 1.0 |
| Population | 707,202.2 | 350,858.0 | 647,187.0 | 911,626.0 | 553,934.2 |
| Unemployment Rate (%) | 0.1 | 0.0 | 0.1 | 0.1 | 0.0 |
| Log(Laborforce) | 12.5 | 12.1 | 12.7 | 13.1 | 1.0 |
| Revenue (USD million) | 3.4 | 1.2 | 2.3 | 4.8 | 3.1 |
| Expenditure (USD million) | 3.4 | 1.2 | 2.3 | 4.9 | 3.2 |
| Zillow HPI | 270,647.2 | 148,300.0 | 206,000.0 | 323,600.0 | 211,200.1 |

**Table 3.2:** Impact of Corporate Bankruptcies on Borrowing Costs of Local Governments

This table reports the baseline results for our sample using Equation (3.1) estimating the differential effect on municipal bond yields of treated versus control counties after a firm bankruptcy filing. The primary coefficient of interest, $\beta_0$, is captured by the interaction term of Treated × Post. Panel A compares treatment and control bonds in the secondary market around an equal window of 3 years of the event. Columns (1)-(3) show the results for monthly after-tax average yield as the dependent variable. Specifically, Column (1) reports the effect using event pair fixed effects and year month fixed effects. In Column (2), we also introduce bond level controls consisting of coupon (%); log(amount issued in \$); dummies for callable bonds, bond insurance, general obligation bonds and competitively issued bonds; remaining years to maturity; and inverse years to maturity. We provide descriptions of key variables in Table C1. In Column (3), we additionally control for the county-level variation in unemployment rate and labor force. We use the lagged values (to the year of bankruptcy filing) for log(labor force) and unemployment rate, and the percentage change in unemployment rate and labor force, respectively. We use a similar scheme for the remaining columns. In Columns (4)-(8), the dependent variable is after-tax yield spread, which is calculated using Equations (3.2) and (3.3). Our baseline specification comes from Column (6). Column (7) shows results by introducing event × county fixed effect. In Column (8), we also add event × year-month fixed effect. T-statistics are reported in brackets and standard errors are double clustered at county bond issuer and year month level, unless otherwise specified. $^{*}\ p < 0.10$, $^{**}\ p < 0.05$, $^{***}\ p < 0.01$

| *Dependent Variable*: | After-tax yield | | | After-tax yield spread | | | | |
|---|---|---|---|---|---|---|---|---|
| | (1) | (2) | (3) | (4) | (5) | (6) | (7) | (8) |
| Post × Treated | 16.56$^{***}$ | 16.53$^{***}$ | 11.35$^{***}$ | 15.40$^{***}$ | 15.01$^{***}$ | 10.01$^{***}$ | 10.15$^{***}$ | 10.47$^{***}$ |
| | [4.38] | [4.44] | [3.59] | [4.23] | [4.12] | [3.23] | [3.29] | [3.48] |
| Post | -7.99$^{***}$ | -7.35$^{***}$ | -3.56$^{*}$ | -7.19$^{***}$ | -6.67$^{***}$ | -3.13 | -3.21 | 0.00 |
| | [-3.22] | [-2.92] | [-1.67] | [-2.89] | [-2.64] | [-1.46] | [-1.50] | [0.00] |
| Treated | -8.86$^{**}$ | -4.65 | -1.39 | -7.29$^{*}$ | -4.69 | -1.62 | 0.00 | 0.00 |
| | [-1.99] | [-1.04] | [-0.34] | [-1.85] | [-1.14] | [-0.44] | [0.00] | [0.00] |
| Event FE | ✓ | ✓ | ✓ | ✓ | ✓ | ✓ | ✓ | ✓ |
| County FE | ✓ | ✓ | ✓ | ✓ | ✓ | ✓ | ✓ | ✓ |
| Year-month FE | ✓ | ✓ | ✓ | ✓ | ✓ | ✓ | ✓ | ✓ |
| Bond Controls | | ✓ | ✓ | | ✓ | ✓ | ✓ | ✓ |
| County Controls | | | ✓ | | | ✓ | ✓ | ✓ |
| Event × County FE | | | | | | | ✓ | ✓ |
| Event × Year-month FE | | | | | | | | ✓ |
| Adj.-$R^2$ | 0.313 | 0.554 | 0.559 | 0.594 | 0.645 | 0.648 | 0.648 | 0.655 |
| Obs. | 2,703,342 | 2,703,342 | 2,703,342 | 2,703,342 | 2,703,342 | 2,703,342 | 2,703,342 | 2,703,273 |
| HQ Location | ✓ | ✓ | ✓ | ✓ | ✓ | ✓ | ✓ | ✓ |
| Other Locations | × | × | × | × | × | × | × | × |

## **Table 3.3:** Why do we focus on Headquarters?

This table reports the results from our analysis on the intensive margin. We use the baseline Equation (3.1), with suitable modifications, where necessary. In Columns (1)-(4), we use data from the EPA and Mergent Intellect on other non-HQ locations for firms in the sample. Columns (1) and (3) show the baseline results with aggregated locations across HQ and non-HQ counties from the baseline Equation 3.1. In Columns (2) and (4), we show the interaction effect by adding group-month fixed effects. Columns (5)-(7) show results from using all facilities from the Mergent Intellect database only. This may or may not include all the HQ locations. We report the results using all such locations in Column (5) based on Equation 3.1. We show the interaction effect from multi-facility locations in Column (6), while additionally controlling for group-month fixed effects. Finally, Column (7) shows results from ranking only the multi-facility locations in Column (6). We use the fraction of employment to assign the ranks. Sites with $\geq 25\%$ firm employees are ranked 1, followed by rank 2 for sites with $> 5\%$ of the firm employment. Rank 3 is assigned to the remaining locations. T-statistics are reported in brackets and standard errors are double clustered at county bond issuer and year month level, unless otherwise specified. $^{*}$ $p < 0.10$, $^{**}$ $p < 0.05$, $^{***}$ $p < 0.01$

| *Dependent Variable*: | After-tax yield spread | | | | | | |
|---|---|---|---|---|---|---|---|
| | Compustat HQ vs Other Locations | | | | All Facilities from Mergent | | |
| *Other Locations:* | From EPA | | From Mergent | | | | |
| | (1) | (2) | (3) | (4) | (5) | (6) | (7) |
| Post × Treated | 4.02$^{**}$ | | 6.34$^{***}$ | | 3.83$^{*}$ | | |
| | [2.18] | | [3.16] | | [1.66] | | |
| Post × Treated × HQ County | | 12.74$^{***}$ | | 10.82$^{***}$ | | | |
| | | [3.96] | | [3.45] | | | |
| Post × Treated × Facility County | | -2.83 | | 2.92 | | | |
| | | [-1.17] | | [1.24] | | | |
| Post × Treated × Multi-Facility | | | | | | 6.63$^{**}$ | |
| | | | | | | [2.54] | |
| Post × Treated × Single-Facility | | | | | | -2.86 | |
| | | | | | | [-0.59] | |
| Post × Treated ×Multi-facility Rank=1 | | | | | | | 14.77$^{**}$ |
| | | | | | | | [2.43] |
| Post × Treated ×Multi-facility Rank=2 | | | | | | | 4.76 |
| | | | | | | | [0.56] |
| Post × Treated ×Multi-facility Rank=3 | | | | | | | 0.46 |
| | | | | | | | [0.20] |
| Difference | | 16.89 | | 8.75 | | 9.38 | |
| P-value | | 0.00 | | 0.02 | | 0.10 | |
| Event FE | ✓ | ✓ | ✓ | ✓ | ✓ | ✓ | ✓ |
| County FE | ✓ | ✓ | ✓ | ✓ | ✓ | ✓ | ✓ |
| Year-month FE | ✓ | ✓ | ✓ | ✓ | ✓ | ✓ | ✓ |
| Controls | ✓ | ✓ | ✓ | ✓ | ✓ | ✓ | ✓ |
| Adj.-$R^2$ | 0.620 | 0.620 | 0.693 | 0.693 | 0.711 | 0.711 | 0.742 |
| Obs. | 6,369,196 | 6,369,196 | 6,675,920 | 6,675,920 | 4,347,334 | 4,347,334 | 3,146,863 |
| HQ Location | ✓ | ✓ | ✓ | ✓ | ✓ | ✓ | ✓ |
| Other Locations | ✓ | ✓ | ✓ | ✓ | ✓ | ✓ | ✓ |

**Table 3.4:** Mechanism: Source of Repayment and County's exposure to the bankrupt firm's industry

This table reports the results for the proposed mechanism at work. Panel A reports results from the baseline specification of (3.1) using subsamples of bonds, as shown. In Panel B, we report results for our baseline specification from Equation (3.1), which is interacted with dummies corresponding to county dependence (or county exposure). We additionally include group-month fixed effects in the modified baseline equation. We define county dependence based on the ex-ante median values of the respective treated county's share of employment (or wages) in the NAICS-3 industry of firm bankruptcy. Columns (1)-(3) show the results using *Dependence* based on aggregate share of wages. We use the aggregate share of employment in Columns (4)-(6). Further, we show results among all bonds in Column (1), followed by the sub-sample of revenue (RV) bonds in Column (2) and uninsured revenue (RV) bonds in Column (3). We follow a similar approach in Columns (4)-(6). T-statistics are reported in brackets and standard errors are double clustered at county bond issuer and year month level, unless otherwise specified. * $p < 0.10$, ** $p < 0.05$, *** $p < 0.01$

## Panel A: By Source of Repayment

| *Dependent Variable*: | After-tax Yield Spread | | | | |
|---|---|---|---|---|---|
| *Sample*: | All bonds | RV bonds | Uninsured RV | Insured RV | GO bonds |
| | (1) | (2) | (3) | (4) | (5) |
| Post × Treated | 10.01*** | 12.56*** | 13.70** | 9.67 | 3.59 |
| | [3.23] | [2.74] | [2.26] | [1.62] | [1.11] |
| Event FE | ✓ | ✓ | ✓ | ✓ | ✓ |
| County FE | ✓ | ✓ | ✓ | ✓ | ✓ |
| Year-month FE | ✓ | ✓ | ✓ | ✓ | ✓ |
| Controls | ✓ | ✓ | ✓ | ✓ | ✓ |
| Adj.-$R^2$ | 0.648 | 0.644 | 0.576 | 0.701 | 0.678 |
| Obs. | 2,703,342 | 1,544,732 | 694,988 | 849,744 | 1,158,610 |

## Panel B: By County Exposure using Dependence

| *Dependent Variable*: | After-tax yield spread | | | | | |
|---|---|---|---|---|---|---|
| *Type of Bonds*: | All | RV | Uninsured RV | All | RV | Uninsured RV |
| *Split by*: | Share of Agg. Industry Wages | | | Share of Agg. Industry Employment | | |
| | (1) | (2) | (3) | (4) | (5) | (6) |
| Post × Treated × | | | | | | |
| Above Median | 15.01*** | 16.45*** | 17.84** | 13.73*** | 16.46*** | 18.11** |
| | [3.69] | [2.83] | [2.14] | [3.34] | [2.78] | [2.08] |
| Below Median | 1.90 | 5.33 | 9.18 | 3.88 | 5.96 | 9.66 |
| | [0.41] | [0.84] | [1.11] | [0.87] | [0.96] | [1.23] |
| Difference | 13.11 | 11.12 | 8.66 | 9.85 | 10.50 | 8.45 |
| p-val | 0.03 | 0.17 | 0.45 | 0.10 | 0.20 | 0.47 |
| Event FE | ✓ | ✓ | ✓ | ✓ | ✓ | ✓ |
| County FE | ✓ | ✓ | ✓ | ✓ | ✓ | ✓ |
| Group Month FE | ✓ | ✓ | ✓ | ✓ | ✓ | ✓ |
| Controls | ✓ | ✓ | ✓ | ✓ | ✓ | ✓ |
| Adj.-$R^2$ | 0.649 | 0.645 | 0.577 | 0.649 | 0.645 | 0.577 |
| Obs. | 2,703,342 | 1,544,732 | 694,988 | 2,703,342 | 1,544,732 | 694,988 |

**Table 3.5:** Heterogeneity: Importance of the Bankrupt Firm for the HQ County

This table reports the results showing the heterogeneity in our main effect. We report results for our baseline specification from Equation (3.1), which is interacted with dummies corresponding to sub-groups indicated in each column. We additionally include group-month fixed effects in the modified baseline equation. Specifically, Columns (1)-(3) show the relative importance of the firm filing for bankruptcy. We divide the treated counties into two groups based on the median value of the ratios defined hereafter. These ratios are calculated using the corresponding values in the year before the bankruptcy filing. In Column (1), we use the ratio of the firm's operating income (EBIT) to the county's revenue. Column (2) uses the ratio of the firms' selling, general and administrative (SGA) expenses to the county's revenue. In Column (3), we use the ratio of the firm's plant, property, and equipment (PPE) to the property tax revenue of the county. T-statistics are reported in brackets and standard errors are double clustered at county bond issuer and year month level, unless otherwise specified. $^{*}\ p < 0.10$, $^{**}\ p < 0.05$, $^{***}\ p < 0.01$

| *Dependent Variable*: | After-tax yield spread | | |
|---|---|---|---|
| *Split by:* | $\frac{FirmEBIT}{CountyRevenue}$ (1) | $\frac{FirmSG\&A}{CountyRevenue}$ (2) | $\frac{FirmPP\&E}{PropertyTax}$ (3) |
| Post × Treated | | | |
| High | 14.98*** | 20.84*** | 11.96** |
| | [3.41] | [3.20] | [2.26] |
| Low | 3.49 | 5.15 | 7.80** |
| | [0.82] | [1.20] | [1.99] |
| Difference | 11.49 | 15.69 | 4.16 |
| P-value | 0.06 | 0.06 | 0.53 |
| Event FE | ✓ | ✓ | ✓ |
| County FE | ✓ | ✓ | ✓ |
| Group Month FE | ✓ | ✓ | ✓ |
| Controls | ✓ | ✓ | ✓ |
| Adj.-$R^2$ | 0.651 | 0.660 | 0.651 |
| Obs. | 2,555,676 | 2,210,711 | 2,555,676 |

## **Table 3.6:** Robustness Tests

In this table we report results for various robustness tests on our baseline specification, i.e., Column (6) of Table 3.2. In Panel A, results controlling for unobserved factors at the bond, county, region, and municipal bond market levels. Specifically, in Column (1), we introduce bond fixed effect to the baseline. In Column (2), we impose region fixed effects. Columns (3)-(4) show the results after adding time-varying region-year and region-year month fixed effects, respectively. Column (5) shows the baseline results with year-month $\times$ GO bond indicator fixed effects. Thereafter, in Column (6), we use year-month $\times$ Insured bond status fixed effects. Finally, Column (21) shows the baseline results after including bond purpose $\times$ year-month fixed effects. Panel B corresponds to robustness checks using additional controls. In Column (1), we introduce additional county level time-varying covariates using lagged values of log(personal income) and log(house price index). In Column (2), we show the results for bonds with non-missing S&P credit ratings. We use the most recent ratings for a given CUSIP transaction. In Columns (3)-(4), we control for duration. First, in Column (3), we use duration in the controls by replacing years to maturity and inverse of years to maturity. Second, in Column (4), we use tax-adjusted duration to replace years to maturity and inverse of years to maturity. We drop bonds recently issued with respect to firm bankruptcy in Panel C. Columns (1)-(4) report regression results where we drop bonds that are dated within 6 months, 12 months, 24 months, and 36 months of the bankruptcy filing, respectively. We consider alternative econometric specifications in Panel D. Specifically, Columns (1)-(3) report results with alternative choices for clustering the standard errors, namely: issue and year-month, county bond issuer, and county bond issue, respectively. Column (4) shows the results by introducing event-month fixed effects to the baseline. In Column (5), we drop the year month fixed effect and replace it with the event-month fixed effect in the baseline. Finally, Column (6) shows the result by changing year-month fixed effect to calendar year fixed effects. We focus on the sensitivity to trade size in Panel E. In Columns (1)-(4), we report results using only customer-buy trades with transaction size $>$ \$25,000, $>$ \$50,000, $\leq$25,000, and $\leq$ \$50,000, respectively. Panel F shows robustness in results to additional considerations. In Column (1), we restrict our sample of bonds to tax-exempt municipal bonds only. In Column (2), we only use transactions after the financial crisis of 2010. Columns (3)-(4) show the baseline effect by changing the dependent variable to average yield and average yield spread, respectively. In Column (5), we restrict the choice of control group among counties which do not also observe a publicly listed manufacturing firm bankruptcy within two years of a given treated county. This exclusion criterion is imposed based on the publicly listed manufacturing bankruptcy events in our main sample set. Column (6) shows results by expanding the original sample of events to the broader set of bankruptcy filings from tradeable sector industries. In Panel G, we report the baseline results by altering the length of the event window duration. Finally, Panel H reports the baseline specification by changing the beginning or ending point of the event scale, holding the other period constant at 36 months. T-statistics are reported in brackets and standard errors are double clustered at county bond issuer and year month level, unless otherwise specified. $^{*}$ $p < 0.10$, $^{**}$ $p < 0.05$, $^{***}$ $p < 0.01$

## Panel A: Other Unobservables

| *Dependent Variable*: | After-tax yield spread | | | | | | |
|---|---|---|---|---|---|---|---|
| *Add:* | Bond FE | Region FE | Region × Year FE | Region × YM FE | GO bond × YM FE | Insured × YM FE | Purpose × YM FE |
| | (1) | (2) | (3) | (4) | (5) | (6) | (7) |
| Post × Treated | 8.23** | 10.01*** | 10.42*** | 10.66*** | 8.74*** | 7.74*** | 7.88*** |
| | [2.57] | [3.23] | [3.38] | [3.45] | [2.95] | [2.62] | [3.02] |
| Adj.-$R^2$ | 0.833 | 0.648 | 0.650 | 0.651 | 0.655 | 0.652 | 0.677 |
| Obs. | 2,697,225 | 2,703,342 | 2,703,342 | 2,703,342 | 2,703,342 | 2,703,342 | 2,703,113 |

## Panel B: Additional Controls

| *Dependent Variable*: | After-tax yield spread | | | |
|---|---|---|---|---|
| *Add:* | More Controls | Rating | Duration | After-tax Duration |
| | (1) | (2) | (3) | (4) |
| Post × Treated | 9.60*** | 8.29** | 10.60*** | 10.61*** |
| | [2.95] | [2.33] | [3.35] | [3.32] |
| Adj.-$R^2$ | 0.651 | 0.663 | 0.637 | 0.626 |
| Obs. | 2,504,271 | 2,016,734 | 2,688,584 | 2,688,584 |

## Panel C: Drop Recent Bonds

| *Dependent Variable*: | After-tax yield spread | | | |
|---|---|---|---|---|
| *Dated Within:* | 6 months | 12 months | 24 months | 36 months |
| | (1) | (2) | (3) | (4) |
| Post × Treated | 10.01*** | 10.42*** | 11.81*** | 11.70*** |
| | [3.14] | [3.17] | [3.24] | [2.91] |
| Adj.-$R^2$ | 0.641 | 0.632 | 0.615 | 0.595 |
| Obs. | 2,518,961 | 2,348,294 | 2,024,963 | 1,713,545 |

## Panel D: Alternative Specifications

| *Dependent Variable*: | After-tax yield spread | | | | | |
|---|---|---|---|---|---|---|
| | Clustering | | | Specification | | |
| *Modification:* | Issue and YM | Issuer | Issue | Add event-month FE | Change YM FE to event-month FE | Change YM FE to and year FE |
| | (1) | (2) | (3) | (4) | (5) | (6) |
| Post × Treated | 10.01*** | 10.01*** | 10.01*** | 10.03*** | 23.27*** | 10.35*** |
| | [4.44] | [3.30] | [4.81] | [3.23] | [3.69] | [3.38] |
| Adj.-$R^2$ | 0.648 | 0.648 | 0.648 | 0.648 | 0.416 | 0.512 |
| Obs. | 2,703,342 | 2,703,342 | 2,703,342 | 2,703,342 | 2,703,342 | 2,703,342 |

**Panel E: Sensitivity to Trade Size**

| *Dependent Variable*: | After-tax yield spread | | | |
|---|---|---|---|---|
| *in USD:* | >25,000 | >50,000 | ≤25,000 | ≤50,000 |
| | (1) | (2) | (3) | (4) |
| Post × Treated | 7.84** | 5.83* | 9.43*** | 9.86*** |
| | [2.38] | [1.68] | [2.91] | [3.13] |
| Adj.-$R^2$ | 0.646 | 0.627 | 0.667 | 0.661 |
| Obs. | 1,482,804 | 872,465 | 1,915,893 | 2,323,079 |

**Panel F: Additional Considerations**

| *Dependent Variable*: | After-tax yield spread | | | | | |
|---|---|---|---|---|---|---|
| | Tax-exempt Bonds | Financial Crisis | Other DV | | Restricted Control Counties | Bankruptcies in Tradeables |
| | | | Yield | Spread | | |
| | (1) | (2) | (3) | (4) | (5) | (6) |
| Post × Treated | 11.02*** | 11.04** | 7.19*** | 6.51*** | 8.69** | 8.27*** |
| | [3.55] | [2.38] | [3.91] | [3.39] | [2.45] | [3.05] |
| Adj.-$R^2$ | 0.662 | 0.438 | 0.571 | 0.801 | 0.648 | 0.634 |
| Obs. | 2,453,646 | 1,272,300 | 2,835,820 | 2,703,829 | 2,567,695 | 3,731,872 |

**Panel G: Sensitivity to Length of Event Window**

| *Dependent Variable*: | After-tax yield spread | | | | | |
|---|---|---|---|---|---|---|
| *Window (months)*: | [-12,+12] | [-18,+18] | [-24,+24] | [-30,+30] | [-48,+48] | [-60,+60] |
| | (1) | (2) | (3) | (4) | (5) | (6) |
| Post x Treated | 3.86** | 5.17** | 7.42*** | 9.28*** | 10.67*** | 11.98*** |
| | [2.34] | [2.40] | [2.93] | [3.21] | [3.29] | [3.77] |
| Adj.-$R^2$ | 0.673 | 0.662 | 0.651 | 0.651 | 0.638 | 0.628 |
| Obs. | 971,594 | 1,427,506 | 1,861,432 | 2,288,998 | 3,471,961 | 4,159,651 |

**Panel H: Sensitivity to Start/End of Event Window**

| *Dependent Variable*: | After-tax yield spread | | | | | |
|---|---|---|---|---|---|---|
| *Window (months)*: | [-6,+36] | [-12,+36] | [-24,+36] | [-36,+6] | [-36,+12] | [-36,+24] |
| | (1) | (2) | (3) | (4) | (5) | (6) |
| Post x Treated | 6.64*** | 8.73*** | 9.56*** | 4.41* | 5.62** | 7.89*** |
| | [2.68] | [3.49] | [3.39] | [1.78] | [2.15] | [2.71] |
| Adj.-$R^2$ | 0.632 | 0.636 | 0.642 | 0.679 | 0.675 | 0.658 |
| Obs. | 1,751,064 | 1,977,783 | 2,373,529 | 1,457,062 | 1,697,153 | 2,191,245 |

**Table 3.7:** Impact on Primary Market of Municipal Bonds

This table shows the effect of bankruptcy filing on new bond issuances using a difference-in-differences estimate similar to the baseline specification. It is based on primary market bonds in Equation (3.5) for offering yields. The dependent variable in Columns (1)-(3) is the after-tax offering yield. In Column (1), we show the result by using only event-pair fixed effects and issuer fixed effects in the baseline equation. Next, in Column (2), we introduce bond level controls. Column (3) shows the results with county controls and county fixed effects. T-statistics are reported in brackets and standard errors are clustered at issue level. $^{*}$ $p < 0.10$, $^{**}$ $p < 0.05$, $^{***}$ $p < 0.01$

| *Dependent Variable*: | After-tax Offering Yield | | |
|---|---|---|---|
| | (1) | (2) | (3) |
| Post × Treated | 13.69*** | 10.49*** | 5.79*** |
| | [0.00] | [0.00] | [0.00] |
| Post | -9.41*** | -8.19*** | -6.02*** |
| | [0.00] | [0.00] | [0.00] |
| Treated | -3.09 | -1.71 | 1.41 |
| | [0.11] | [0.22] | [0.30] |
| Event FE | ✓ | ✓ | ✓ |
| Issuer FE | ✓ | ✓ | ✓ |
| Bond Controls | | ✓ | ✓ |
| County Controls | | | ✓ |
| Adj.-$R^2$ | 0.474 | 0.843 | 0.845 |
| Obs. | 424,655 | 424,655 | 424,655 |

## **Table 3.8:** Impact on the Local Economy of the County

This table shows the impact of a firm bankruptcy on the local economy. We use the annualized version of our baseline equation as the primary specification for this table, shown in Equation (3.6). Panel A corresponds to the overall effect on county-level GDP growth and growth in establishments. In Panel B, we show results based on the county-level dependence. We define county dependence using the median values of the respective treated county's share of employment in the NAICS-3 industry of firm bankruptcy. T-statistics are reported in brackets and standard errors are clustered at the event-pair level. * $p < 0.10$, ** $p < 0.05$, *** $p < 0.01$

**Panel A: Overall**

| *Dependent Variable*: | County GDP growth (%) | | | County Establishments growth (%) | | |
|---|---|---|---|---|---|---|
| *Window (years)*: | [-3, +3] | [-3, +5] | [-3, +7] | [-3, +3] | [-3, +5] | [-3, +7] |
| | (1) | (2) | (3) | (4) | (5) | (6) |
| Treated × Post | -0.67** | -0.53* | -0.52* | -0.13 | -0.24 | -0.28 |
| | [0.04] | [0.08] | [0.08] | [0.74] | [0.52] | [0.44] |
| Event FE | ✓ | ✓ | ✓ | ✓ | ✓ | ✓ |
| County FE | ✓ | ✓ | ✓ | ✓ | ✓ | ✓ |
| County Controls | ✓ | ✓ | ✓ | ✓ | ✓ | ✓ |
| Adj.-$R^2$ | 0.411 | 0.377 | 0.355 | 0.120 | 0.130 | 0.139 |
| Obs. | 1,704 | 2,094 | 2,423 | 1,732 | 2,128 | 2,462 |

**Panel B: By County Exposure using Dependence (Employment)**

| *Dependent Variable*: | County GDP growth (%) | | | County Establishments growth (%) | | |
|---|---|---|---|---|---|---|
| *Window (years)*: | [-3, +3] | [-3, +5] | [-3, +7] | [-3, +3] | [-3, +5] | [-3, +7] |
| Treated × Post × | (1) | (2) | (3) | (4) | (5) | (6) |
| Above Median | -1.38*** | -1.23*** | -1.22*** | -0.08 | -0.18 | -0.28 |
| | [0.00] | [0.00] | [0.00] | [0.83] | [0.60] | [0.38] |
| Below Median | 0.11 | 0.26 | 0.27 | -0.18 | -0.31 | -0.28 |
| | [0.81] | [0.52] | [0.50] | [0.80] | [0.66] | [0.68] |
| Difference | -1.48 | -1.48 | -1.48 | 0.10 | 0.13 | 0.00 |
| p-val | 0.02 | 0.01 | 0.01 | 0.90 | 0.88 | 1.00 |
| Event FE | ✓ | ✓ | ✓ | ✓ | ✓ | ✓ |
| County FE | ✓ | ✓ | ✓ | ✓ | ✓ | ✓ |
| Group FE | ✓ | ✓ | ✓ | ✓ | ✓ | ✓ |
| County Controls | ✓ | ✓ | ✓ | ✓ | ✓ | ✓ |
| Adj.-$R^2$ | 0.412 | 0.378 | 0.356 | 0.119 | 0.130 | 0.140 |
| Obs. | 1,704 | 2,094 | 2,423 | 1,732 | 2,128 | 2,462 |

## **Table 3.9:** Evidence on Mild Recovery of the County

This table shows the impact of a firm bankruptcy on the local economy. We use the annualized version of our baseline equation as the primary specification for this table, shown in Equation (3.6). The dependent variables are county-level GDP growth and growth in establishments. In each of the panels, we split the treated counties along the metric mentioned in the panel. We classify the counties in the top quartile as high, while the remaining are grouped as low. T-statistics are reported in brackets and standard errors are clustered at the event-pair level. * $p < 0.10$, ** $p < 0.05$, *** $p < 0.01$

**Panel A: Recovery (By County Bachelor Education)**

| *Dependent Variable*: | County GDP growth (%) | | | County Establishments growth (%) | | |
|---|---|---|---|---|---|---|
| *Window (years)*: | [-3, +3] | [-3, +5] | [-3, +7] | [-3, +3] | [-3, +5] | [-3, +7] |
| Treated × Post × | (1) | (2) | (3) | (4) | (5) | (6) |
| High | -0.34 | -0.09 | 0.06 | 0.85** | 0.62 | 0.54 |
| | [0.58] | [0.87] | [0.92] | [0.05] | [0.10] | [0.14] |
| Low | -0.78** | -0.67* | -0.71** | -0.47 | -0.54 | -0.57 |
| | [0.05] | [0.06] | [0.04] | [0.35] | [0.27] | [0.23] |
| High vs Low | 0.44 | 0.58 | 0.77 | 1.31 | 1.16 | 1.10 |
| p-val | 0.54 | 0.39 | 0.27 | 0.05 | 0.07 | 0.07 |
| Event FE | ✓ | ✓ | ✓ | ✓ | ✓ | ✓ |
| County FE | ✓ | ✓ | ✓ | ✓ | ✓ | ✓ |
| Group FE | ✓ | ✓ | ✓ | ✓ | ✓ | ✓ |
| County Controls | ✓ | ✓ | ✓ | ✓ | ✓ | ✓ |
| Adj.-$R^2$ | 0.411 | 0.378 | 0.356 | 0.121 | 0.131 | 0.140 |
| Obs. | 1,704 | 2,094 | 2,423 | 1,732 | 2,128 | 2,462 |

## Panel B: Recovery (By Number of Pleasant Days)

| *Dependent Variable*: | County GDP growth (%) | | | County Establishments growth (%) | | |
|---|---|---|---|---|---|---|
| *Window (years)*: | [-3, +3] | [-3, +5] | [-3, +7] | [-3, +3] | [-3, +5] | [-3, +7] |
| Treated × Post × | (1) | (2) | (3) | (4) | (5) | (6) |
| High | -1.66*** | -1.45*** | -1.42*** | 0.53 | 0.29 | 0.14 |
| | [0.00] | [0.01] | [0.00] | [0.27] | [0.52] | [0.74] |
| Low | -0.37 | -0.25 | -0.24 | -0.34 | -0.43 | -0.44 |
| | [0.35] | [0.49] | [0.50] | [0.49] | [0.37] | [0.36] |
| High vs Low | -1.30 | -1.21 | -1.18 | 0.86 | 0.72 | 0.58 |
| p-val | 0.06 | 0.06 | 0.05 | 0.23 | 0.29 | 0.37 |
| Event FE | ✓ | ✓ | ✓ | ✓ | ✓ | ✓ |
| County FE | ✓ | ✓ | ✓ | ✓ | ✓ | ✓ |
| Group FE | ✓ | ✓ | ✓ | ✓ | ✓ | ✓ |
| County Controls | ✓ | ✓ | ✓ | ✓ | ✓ | ✓ |
| Adj.-$R^2$ | 0.412 | 0.378 | 0.356 | 0.119 | 0.130 | 0.139 |
| Obs. | 1,704 | 2,094 | 2,423 | 1,732 | 2,128 | 2,462 |

## Panel C: Recovery (By County-level Entrepreneurship Quotient)

| *Dependent Variable*: | County GDP growth (%) | | | County Establishments growth (%) | | |
|---|---|---|---|---|---|---|
| *Window (years)*: | [-3, +3] | [-3, +5] | [-3, +7] | [-3, +3] | [-3, +5] | [-3, +7] |
| Treated × Post × | (1) | (2) | (3) | (4) | (5) | (6) |
| High | -1.48*** | -1.36*** | -1.30*** | 0.44 | 0.31 | 0.09 |
| | [0.00] | [0.01] | [0.01] | [0.33] | [0.46] | [0.81] |
| Low | -0.40 | -0.24 | -0.25 | -0.31 | -0.43 | -0.41 |
| | [0.32] | [0.50] | [0.49] | [0.53] | [0.39] | [0.39] |
| High vs Low | -1.07 | -1.12 | -1.05 | 0.75 | 0.74 | 0.51 |
| p-val | 0.11 | 0.08 | 0.09 | 0.29 | 0.28 | 0.43 |
| Event FE | ✓ | ✓ | ✓ | ✓ | ✓ | ✓ |
| County FE | ✓ | ✓ | ✓ | ✓ | ✓ | ✓ |
| Group FE | ✓ | ✓ | ✓ | ✓ | ✓ | ✓ |
| County Controls | ✓ | ✓ | ✓ | ✓ | ✓ | ✓ |
| Adj.-$R^2$ | 0.411 | 0.378 | 0.355 | 0.122 | 0.133 | 0.142 |
| Obs. | 1,704 | 2,094 | 2,423 | 1,732 | 2,128 | 2,462 |

## Panel D: Recovery (By County-level Start-up Formation Rate)

| *Dependent Variable*: | County GDP growth (%) | | | County Establishments growth (%) | | |
|---|---|---|---|---|---|---|
| *Window (years)*: | [-3, +3] | [-3, +5] | [-3, +7] | [-3, +3] | [-3, +5] | [-3, +7] |
| Treated × Post × | (1) | (2) | (3) | (4) | (5) | (6) |
| High | -1.23*** | -1.05*** | -1.01** | -0.15 | -0.42 | -0.49 |
| | [0.01] | [0.01] | [0.01] | [0.75] | [0.38] | [0.26] |
| Low | -0.49 | -0.36 | -0.36 | -0.13 | -0.20 | -0.23 |
| | [0.23] | [0.33] | [0.32] | [0.80] | [0.67] | [0.62] |
| High vs Low | -0.75 | -0.69 | -0.65 | -0.03 | -0.22 | -0.26 |
| p-val | 0.23 | 0.21 | 0.23 | 0.97 | 0.75 | 0.69 |
| Event FE | ✓ | ✓ | ✓ | ✓ | ✓ | ✓ |
| County FE | ✓ | ✓ | ✓ | ✓ | ✓ | ✓ |
| Group FE | ✓ | ✓ | ✓ | ✓ | ✓ | ✓ |
| County Controls | ✓ | ✓ | ✓ | ✓ | ✓ | ✓ |
| Adj.-$R^2$ | 0.413 | 0.378 | 0.355 | 0.120 | 0.130 | 0.139 |
| Obs. | 1,704 | 2,094 | 2,423 | 1,732 | 2,128 | 2,462 |

## Panel E: Recovery (By County-level Regional Entrepreneurship Acceleration)

| *Dependent Variable*: | County GDP growth (%) | | | County Establishments growth (%) | | |
|---|---|---|---|---|---|---|
| *Window (years)*: | [-3, +3] | [-3, +5] | [-3, +7] | [-3, +3] | [-3, +5] | [-3, +7] |
| Treated × Post × | (1) | (2) | (3) | (4) | (5) | (6) |
| High | 0.11 | 0.20 | 0.59 | 0.54 | 0.51 | 0.45 |
| | [0.87] | [0.75] | [0.37] | [0.10] | [0.16] | [0.21] |
| Low | -0.87** | -0.72** | -0.81** | -0.30 | -0.44 | -0.48 |
| | [0.02] | [0.03] | [0.01] | [0.52] | [0.34] | [0.29] |
| High vs Low | 0.99 | 0.91 | 1.40 | 0.84 | 0.96 | 0.93 |
| p-val | 0.20 | 0.20 | 0.06 | 0.14 | 0.10 | 0.10 |
| Event FE | ✓ | ✓ | ✓ | ✓ | ✓ | ✓ |
| County FE | ✓ | ✓ | ✓ | ✓ | ✓ | ✓ |
| Group FE | ✓ | ✓ | ✓ | ✓ | ✓ | ✓ |
| County Controls | ✓ | ✓ | ✓ | ✓ | ✓ | ✓ |
| Adj.-$R^2$ | 0.410 | 0.376 | 0.355 | 0.118 | 0.128 | 0.138 |
| Obs. | 1,704 | 2,094 | 2,423 | 1,732 | 2,128 | 2,462 |

## Panel F: Recovery (By County-level Number of Patents)

| *Dependent Variable*: | County GDP growth (%) | | | County Establishments growth (%) | | |
|---|---|---|---|---|---|---|
| *Window (years)*: | [-3, +3] | [-3, +5] | [-3, +7] | [-3, +3] | [-3, +5] | [-3, +7] |
| Treated × Post × | (1) | (2) | (3) | (4) | (5) | (6) |
| High | -1.67*** | -1.50** | -1.45** | 0.38 | 0.22 | 0.17 |
| | [0.01] | [0.01] | [0.01] | [0.48] | [0.67] | [0.73] |
| Low | -0.36 | -0.23 | -0.24 | -0.30 | -0.41 | -0.44 |
| | [0.35] | [0.50] | [0.48] | [0.53] | [0.38] | [0.33] |
| High vs Low | -1.31 | -1.26 | -1.21 | 0.68 | 0.64 | 0.61 |
| p-val | 0.08 | 0.07 | 0.07 | 0.35 | 0.37 | 0.36 |
| Event FE | ✓ | ✓ | ✓ | ✓ | ✓ | ✓ |
| County FE | ✓ | ✓ | ✓ | ✓ | ✓ | ✓ |
| Group FE | ✓ | ✓ | ✓ | ✓ | ✓ | ✓ |
| County Controls | ✓ | ✓ | ✓ | ✓ | ✓ | ✓ |
| Adj.-$R^2$ | 0.411 | 0.378 | 0.356 | 0.120 | 0.131 | 0.140 |
| Obs. | 1,704 | 2,094 | 2,423 | 1,732 | 2,128 | 2,462 |

## Panel G: Recovery (By County-level Percentage White)

| *Dependent Variable*: | County GDP growth (%) | | | County Establishments growth (%) | | |
|---|---|---|---|---|---|---|
| *Window (years)*: | [-3, +3] | [-3, +5] | [-3, +7] | [-3, +3] | [-3, +5] | [-3, +7] |
| Treated × Post × | (1) | (2) | (3) | (4) | (5) | (6) |
| High | 0.04 | -0.07 | -0.13 | -0.07 | -0.21 | -0.25 |
| | [0.95] | [0.92] | [0.84] | [0.90] | [0.69] | [0.61] |
| Low | -0.90** | -0.68** | -0.65* | -0.15 | -0.26 | -0.30 |
| | [0.01] | [0.04] | [0.05] | [0.75] | [0.58] | [0.52] |
| High vs Low | 0.94 | 0.61 | 0.51 | 0.07 | 0.04 | 0.05 |
| p-val | 0.26 | 0.42 | 0.49 | 0.93 | 0.95 | 0.94 |
| Event FE | ✓ | ✓ | ✓ | ✓ | ✓ | ✓ |
| County FE | ✓ | ✓ | ✓ | ✓ | ✓ | ✓ |
| Group FE | ✓ | ✓ | ✓ | ✓ | ✓ | ✓ |
| County Controls | ✓ | ✓ | ✓ | ✓ | ✓ | ✓ |
| Adj.-$R^2$ | 0.410 | 0.376 | 0.354 | 0.117 | 0.128 | 0.138 |
| Obs. | 1,704 | 2,094 | 2,423 | 1,732 | 2,128 | 2,462 |

## Panel H: Recovery (By County-level Percentage Moved from a different county)

| *Dependent Variable*: | County GDP growth (%) | | | County Establishments growth (%) | | |
|---|---|---|---|---|---|---|
| *Window (years)*: | [-3, +3] | [-3, +5] | [-3, +7] | [-3, +3] | [-3, +5] | [-3, +7] |
| Treated × Post × | (1) | (2) | (3) | (4) | (5) | (6) |
| High | -1.20* | -1.04* | -1.04* | 0.36 | 0.17 | -0.01 |
| | [0.08] | [0.07] | [0.06] | [0.43] | [0.69] | [0.99] |
| Low | -0.51 | -0.38 | -0.36 | -0.28 | -0.37 | -0.37 |
| | [0.18] | [0.29] | [0.30] | [0.57] | [0.44] | [0.43] |
| High vs Low | -0.69 | -0.66 | -0.68 | 0.64 | 0.54 | 0.37 |
| p-val | 0.39 | 0.32 | 0.30 | 0.35 | 0.41 | 0.56 |
| Event FE | ✓ | ✓ | ✓ | ✓ | ✓ | ✓ |
| County FE | ✓ | ✓ | ✓ | ✓ | ✓ | ✓ |
| Group FE | ✓ | ✓ | ✓ | ✓ | ✓ | ✓ |
| County Controls | ✓ | ✓ | ✓ | ✓ | ✓ | ✓ |
| Adj.-$R^2$ | 0.411 | 0.378 | 0.356 | 0.118 | 0.128 | 0.138 |
| Obs. | 1,704 | 2,094 | 2,423 | 1,732 | 2,128 | 2,462 |

# CHAPTER 4

# DO MANAGERS WALK THE TALK ON ENVIRONMENTAL AND SOCIAL ISSUES?

## 4.1 Introduction

From 2018-2020, the total assets under management for sustainable investment grew by 42%, from $12 trillion to $17 trillion in the U.S. Commensurate with the increase in investor interest, around 90% of S&P 500 firms voluntarily published a sustainability report in 2020 compared to just 20% in 2011. Many firms have also joined various sustainability initiatives such as UN Global Compact and Carbon Disclosure Project. At the same time, there are significant concerns that both corporations and institutional investors may be greenwashing[1]. In this paper, we study whether managers walk their talk on the E&S issues by analyzing the firm-level E&S discussion in earnings calls and firms' E&S performance.

We develop a comprehensive dictionary of E&S phrases from multiple sustainability frameworks using deep learning techniques. Our methodology can distinguish whether E&S phrases are material to the specific industry. We use this dictionary to analyze E&S discussions in firm's earnings conference calls. E&S ratings can diverge significantly across various sustainability rating agencies (see [109]). Hence, we analyze firm's environmental performance based on publicly observable sustainability outcomes such as pollution abatement activities, emissions and green patents. We use anonymous employee ratings to measure social performance. Our evidence suggests that on average, the E&S discussion in earnings conference calls is credible and is associated with improved performance on the E&S dimensions. However, this association is weaker for firms that have more non-answers in their earnings calls and when they discuss E&S topics that are not material to that industry. Our results highlight when firms walk their E&S talk.

[1]2021 EU Sustainable Finance Disclosure Regulation (SFDR) highlighted some of the concerns about *Greenwashing*, i.e. making people believe that the company is doing more in sustainability dimensions than it really is. For example, recently, Business Roundtable (BRT), an association of chief executive officers of America's leading companies, redefined the purpose of corporations as benefiting all their stakeholders rather than just stockholders. However, [108] find no evidence that BRT members have engaged in stakeholder-centric practices.

Identifying E&S discussion in earnings conference calls is challenging for multiple reasons. The first challenge is about capturing the heterogeneous environmental and social topics comprehensively. E&S issues can vary widely, for example, from water pollution to minimum wage. Second, there is no widely accepted definition on the environmental and social topics. Relying on one particular sustainability framework or rating agency may lead to an incomplete list of E&S related topics. Third, even if there is a consensus on the E&S topics, the same terminologies may not necessarily be used verbatim by firms' executives in the business context. They may instead rely on short-hand synonyms or "nicknames" that refer to these topics. Finally, the same topic may not be material for all the industries. For instance, air quality is important and costly for the transportation firms to make improvements, but not for financial services industry. Thus, it is necessary to distinguish the material E&S discussion from the immaterial topics for each firm.

We mitigate these concerns by using a state-of-the-art deep learning model, Robustly optimized BERT approach (RoBERTa, [110]), to construct a comprehensive E&S dictionary[2]. Intuitively, we treat the task of constructing a dictionary of E&S phrases as a text classification problem, where a powerful model automatically assigns "environmental", "social", or other labels to a large number of phrases. This task involves three main steps. First, we create a high-quality training sample, within which each textual sequence is labeled as "environmental", "social", or other. Second, we use this training sample to train the deep-learning model, so that it acquires knowledge about E&S issues. Third, we use this fine-tuned model to make predictions on candidate phrases and validate the selected

[2]While traditional NLP techniques like word2vec can identify words with similar meanings, they lack the ability to process sentence-level information from the sustainability documents and perform classification in both the E&S and materiality dimensions. In section 4.2.1, we present more details on why a deep learning model is required and why keyword based dictionaries and word2vec may not work as well in our setting.

ones manually.

We construct a self-labeled E&S training sample by aggregating the documents from nine leading sustainability standards and rating agencies. The resulting training sample has two main advantages. First, combining domain knowledge from multiple organizations allows us to incorporate the broad language used by the business community to discuss these narrow topics and helps us incorporate the diversity and disagreement about the E&S topics. Second, the sequences in the training sample are self-labeled based on the conceptual frameworks of the sustainability organizations, which requires less effort, domain expertise, and time commitment. For example, the Global Reporting Initiative (GRI) organizes its environment-related documents in Series 300, while those related to social issues in Series 400 [111]. Thus, it is straightforward to label sequences in those documents as per the corresponding dimension. Based on this method, we construct a high-quality E&S training sample with five labels as shown in appendix Table D4.1.

We use the constructed E&S training sample to further train the deep–learning model, RoBERTa. RoBERTa is pre-trained on a large corpus and can comprehend human language, even for the words that are not included in our E&S training sample. Through further training with our E&S labeled data, RoBERTa model can acquire the sustainability domain knowledge (e.g., aspects related to climate change). In this step, RoBERTa learns which parts of an input sequence need more attention to make an accurate prediction in the E&S context. Moreover, during the training process, RoBERTa absorbs the contextual information by analyzing sequences of multiple lengths with the word order information, which is helpful in a better understanding of the E&S language[3].

We use the fine-tuned RoBERTa model to make predictions on a large number of phrases extracted from the sustainability documents (45,595) and Wikipedia (approximately 3.8 mn). We keep the phrases predicted as “environmental” or “social” with high probability. To validate whether the selected phrases are actually used in the corporate dis-closure, we extract text from 415 corporate social responsibility (CSR) reports posted on SASB website during 2017 to 2020 and filter out the phrases which never occur in this CSR corpus. After extensive human checking, the resulting dictionary consists of 614 "environmental" phrases and 697 "social" phrases. Since materiality is important for sustainability analysis [112], we further train RoBERTa to classify the resulting phrases into "material" and "immaterial" for each industry under a specific E&S topic.

[3]In addition, RoBERTa’s 12-layer architecture allows for a cascading of non-linear transformations to define the hyperspace of textual information.

We use our E&S dictionary to parse the earnings conference calls by firms in the United States during 2007 to 2019. On average, managers refer to E&S topics about four times during an earnings conference call. Specifically, these discussions are dominated by environmental references with a gradual increase over the years in terms of the proportion of earnings calls with such discussions. Meanwhile, the social discussion are relatively stable between 40–50% in these documents. In addition, there is significant variation in E&S discussion in the cross section. As expected, firms in utilities, chemicals, and energy industry are more likely to discuss E&S topics in their earnings calls than finance and healthcare firms.

We begin our analysis by examining whether the firms' E&S discussions in the earnings calls translate into environmental actions with respect to pollution abatement. To this end, we rely on pollution prevention (P2) data from the Environmental Protection Agency (EPA). Our analysis focuses on the two most common forms of abatement: changes to operating practices and production improvements. We find that the discussions of E&S topics is positively associated with the investment in pollution abatement. Firms with more environmental discussion in their earnings calls appear to increase operations-related pollution abatement initiatives.[4] The effect is stronger when the environmental discussion is material to the industry. We find that a 1% change in environmental discussion in the earnings call is associated with a 9.81% increase in operations-related abatement activities. We do not find significant improvements in production-related processes on aggregate; however, there is a significant positive association when the environmental discussion is material to that industry.

[4]We acknowledge that this association may represent either past actions or planned future actions i.e. firms that have already invested more or plan to invest more in abatement initiatives may discuss more of these activities in their earnings calls. However, it still suggests that the discussion is credible and accompanied by real corporate actions and can be considered a further validation of our E&S dictionary.

We also test the credibility of firms' E&S discussion by examining the firm's pollution emissions and green patenting activities. We find that a 10% increase in discussion of E&S topics in the earnings call is associated with a 0.99% decrease in the firm's total pollution. For the average firm in our sample, this amounts to 25,807 pounds. We find similar results by focusing on environmental discussion alone. Notably, this significant negative association persists for the next 1-2 years, suggesting the managerial E&S talk in earnings calls is associated with firms' pollution emissions in the future. We also find a positive association between the managers' discussion about E&S topics and the number of green patents granted to their firms. This relationship holds over the one to three years after the earnings call.

For the social dimension, we examine the linkage between firms' E&S talk and their anonymous employee ratings from Glassdoor. We average the employee ratings at the firm-year-quarter level, and find that a 1% increase in the discussion of E&S topics is associated with a 0.24-notch improvement in the overall rating from employees. This represents a 7.62% increase relative to the average overall rating of 3.15 in our sample. We also find similar results by focusing on discussions of social topics alone. These effects are mainly driven by the individual components of compensation and benefits, work-life balance, and organizational culture. As further evidence, we show that the positive associations in work-life balance and organizational culture ratings are more strongly associated with the discussion of social topics that are material to that industry.

Overall, our findings provide some evidence that on average, firms do walk their talk on E&S issues. However, it is possible that some firms engage in cheap E&S talk that is not associated with real actions. We consider two situations where a firm's E&S talk may be cheap. First, managers who discuss E&S topics immaterial to the industry may be less willing to take concrete actions on E&S issues. We test this hypothesis by distinguishing the material E&S discussions from the immaterial ones and examining their relationship with the firms' total pollution. We find that the negative association between the pollution and E&S talk are mainly driven by the discussions on the material E&S topics. Specifically, the coefficient magnitude of material discussions on environmental topics is nearly twice that of the immaterial ones. This evidence suggests that discussions of immaterial E&S topics in earnings calls are less meaningful about the firms' commitment on the E&S issues.

Second, we consider the case when there are many non-answers in the earnings conference call. The concurrence of managers' E&S talk and their high unwillingness to answer analysts' questions may signal the intention of greenwashing or distracting audience's attention from critical business issues to E&S topics. The significant positive estimate of the interaction term between the level of E&S discussions and the high non-answer level suggests that given the same level of E&S talk, firms with high level of unwillingness to answer questions tend to have higher pollution. Our results highlight some cases where firms' E&S talk may not be credible.

In a recent paper, [113] (SVVZ henceforth) construct firm-level measures of earning call participants' topical attention to climate change exposures. They also show that these measures are useful in predicting real outcomes related to the net-zero transition such as job creation in disruptive green technologies and green patenting. Our work complements the results in their paper, but also differs from their paper in the following ways. First, we focus on broader environmental and social issues rather than just climate change. Second, while SVVZ use 50 pre-specified bigrams to initiate a keyword discovery algorithm that finally identify over 7,000 bigrams related to climate change from earnings calls, our E&S dictionary is constructed by learning domain knowledge from a group of leading sustainability frameworks. It allows us to not only better capture flexible language (unigrams, bigrams, and trigrams) on E&S topics, but also can be easily applied to analyzing discussion in other environments (e.g., news articles, SEC filings, CSR reports). In addition, our method deploys the pre-trained deep learning model, which is superior in understanding human language to the traditional machine learning models. Finally, our measure is able to distinguish the material and immaterial E&S discussions in earnings calls. The climate-change measure in SVVZ positively correlates with our measure of E&S discussions. However, this correlation is mostly driven by the immaterial discussion rather than discussion material to the given industry. Understanding how the materiality of sustainability information varies across industries is an important aspect in firms' disclosure and for guiding investors' asset allocation [114].

We contribute to the literature on ESG measurement and ratings by constructing a firm-level textual measure of E&S profile of a firm. The major sustainability rating agencies differ significantly in the E&S issues they measure, how they measure, and the relative importance of attributes. As a result, there is a significant divergence in their ratings [109]. Our E&S measure can complement the existing E&S ratings as our E&S dictionary can be used to quickly and transparently score the firm's E&S profile. We focus on publicly available E&S related outcomes to transparently analyze whether firm's discussion of E&S issues is simply greenwashing. We interpret the positive association that we find between firm's E&S discussion and the E&S outcomes as a further validation of our firm-level E&S measure. However, one shortcoming of our current measure is that it doesn't differentiate between discussion of past E&S actions and planned E&S actions. There can also be feedback effect wherein managers may change their language that are unfavorably perceived by algorithms [115] as E&S profile of firm is related to its cost of capital [116].

We also contribute to the emerging literature that uses machine learning techniques in analyzing financial documents. To our knowledge, we are the first to apply a RoBERTa approach to analyze financial documents in the context of sustainability. [117] predict the probability of a firm being financially constrained based on the textual analysis of firms' annual reports. A recent study by [118] applies the BERT model to study the cross-cultural differences in sentiment towards finance across eight countries and [119] analyze the re-sponse of this finance sentiment to natural disasters and its impact on economic outcomes. [120] use BERT to analyze whether the discussion of emerging technologies in earnings calls is just hype or whether it conveys credible information to investors. In contrast, we train the RoBERTa model with sustainability documents, to first generate a dictionary of E&S-related phrases. We then apply this dictionary to parse out references to E&S topics in earnings conference calls and also incorporate materiality. Our E&S dictionary can be easily applied by future researchers to capture the sustainability-related linguistic information.

## 4.2 Constructing a dictionary of E&S phrases

Our main goal in this paper is to examine the credibility of a firms' discussion of environmental and social issues in their earnings conference calls. A key challenge to this exercise is how to capture firm-level environmental and social discussions over time. We address this challenge by using a deep-learning model to construct a dictionary of environmental and social phrases. We treat the construction of an E&S dictionary as a text classification task to automatically classify phrases into environmental and social categories. Our methodology involves three broad steps, as also shown in Figure 4.1: (1) constructing an E&S training sample; (2) training the deep learning model; (3) applying the trained model to select the E&S phrases and to validate the selected phrases. We further identify the E&S phrases that are material to each industry. We first describe why we use deep-learning techniques to construct an E&S dictionary. Then we provide details about each step below.

### 4.2.1 Do we need deep learning classification for E&S discussion?

As discussed, the key challenge in this paper is to measure how E&S topics are discussed (or language pattern) in firms' earnings calls. For example, the vice president of Exxon Mobil (NYSE: XOM) in the Q3 earnings conference call of 2019 was discussing the E&S issues as follows, with the underlined phrases highlighting the environmental and social

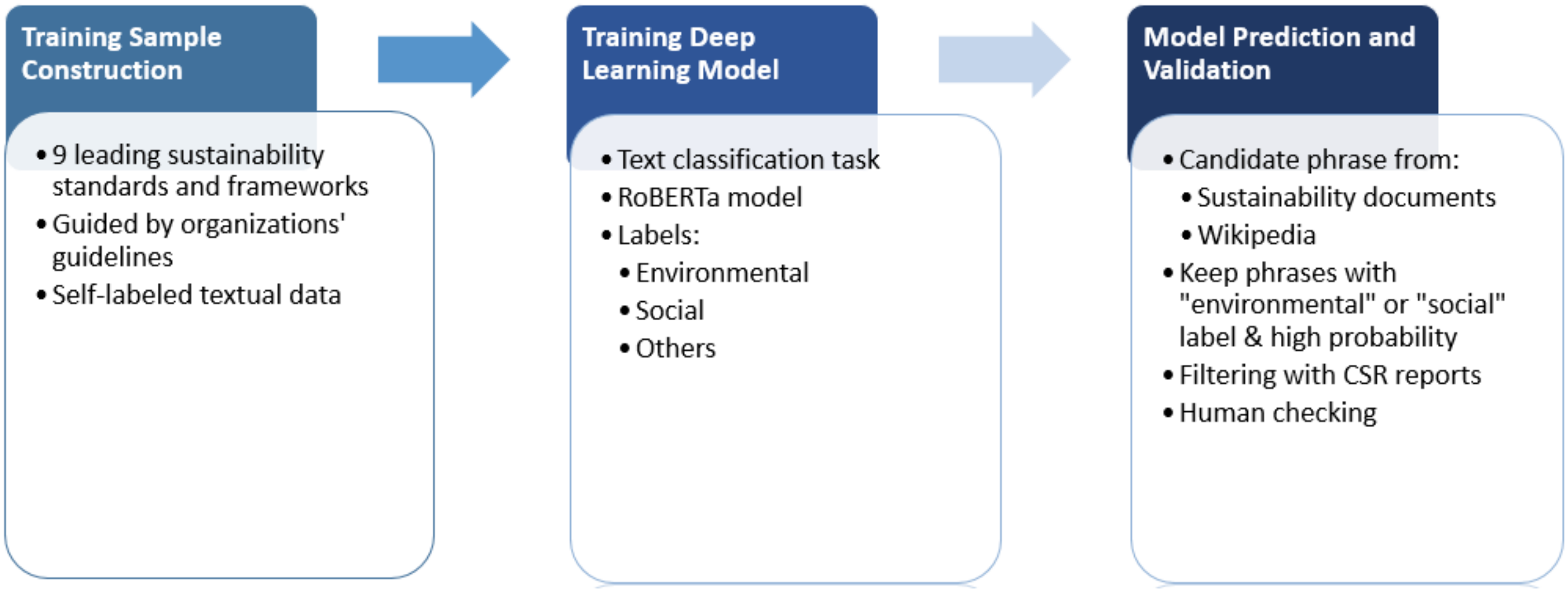

**Figure 4.1:** Work Flow of Dictionary Construction:
This figure provides a high-level overview of the steps we take to construct our dictionary of environmental and social (E&S) topics. We provide detailed descriptions of these aspects in Section 4.2.

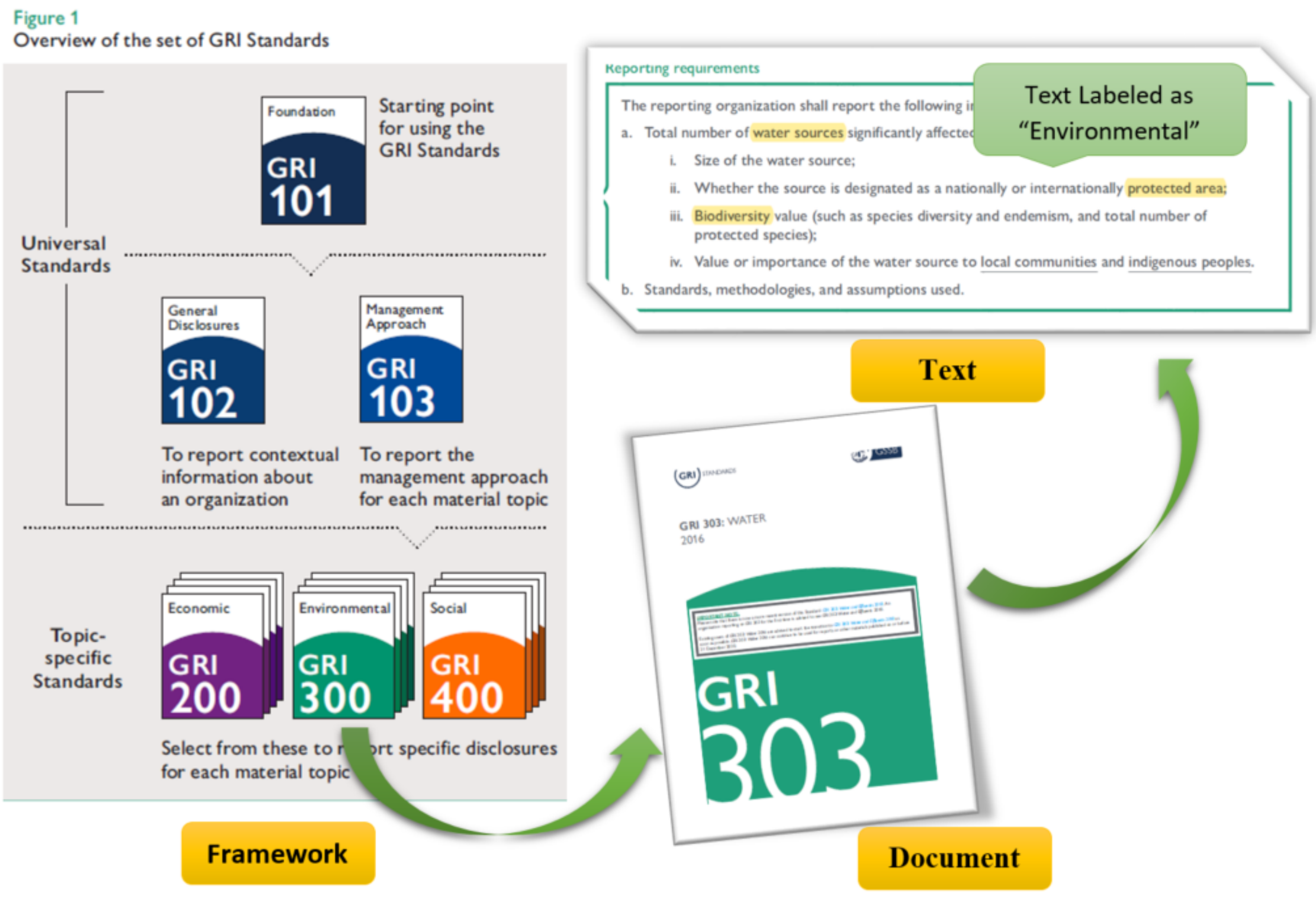

**Figure 4.2:** Constructing Training Sample–GRI Example:
This figure shows an example from GRI, one of the sustainability frameworks used in constructing our E&S training sample. With the overview of GRI Standards, we are able to identify which dimension, like environmental or social, each document and their textual data belong to.

content:

> *Finally, but importantly, we are building on our extensive network of partnerships to develop new technologies to address the dual challenge of providing reliable and* ***affordable*** *energy, while mitigating impacts to the environment, including the risk of* ***climate change****... if the underlying concern is about risk of* ***climate change*** *and* ***emissions reduction****, we certainly share similar concerns.*

We construct a dictionary of E&S phrases to capture such discussion as task-specific word lists are extensively used in the literature [121, 122, 123, 124] to extract information from unstructured textual data. Also, as the above example shows, the E&S issues are often discussed with certain key words (e.g., climate change). It is possible that relying on a word list and ignoring context information may result in Type I error, i.e. some discussion may be misidentified as E&S-related. However, given that the topics discussed in earnings calls are concentrated around business-related issues that are of interest to the firms' investors, we believe a dictionary of E&S phrases can capture such information. Further, our E&S dictionary can be easily and efficiently adopted by researchers in the future without incurring our carbon footprint.

However, sustainability covers a wide spectrum of topics, from global warming to minimum wage. Thus, a comprehensive E&S dictionary is necessary to capture the sustainability-related discussion. Second, in the context of firms' earnings conference calls, managers and analysts may use short-hand synonyms to discuss the E&S topics. Moreover, identifying the E&S phrases is challenging due to the lack of agreement on E&S topics viewed as important by the sustainability raters and their definitions. Relying on one sustainability framework versus another may lead to missing a big part of E&S-related content [109]. Thus, it is important to learn the E&S topics from a wide range of sustainability bodies in the marketplace. While traditional NLP techniques like word2vec can identify words with similar meanings, they lack the ability to process sentence-level information from the sustainability documents and perform classification on both the E&S and materiality dimensions.

In this paper, we tackle this task by aggregating E&S documents from nine leading sustainability frameworks and deploying RoBERTa [110], a state-of-the-art deep learning model for these kind of tasks. The first advantage of RoBERTa is its superiority in understanding human language, including the words not shown in the E&S training sample. RoBERTa is adept at understanding the general semantic and syntactic knowledge as it is pre-trained on a large corpus of over 160GB uncompressed text, consisting of BookCorpus + Wikipedia (16GB), CC-News (76GB), OpenWebText (38GB), and Stories (31GB). It can capture the nuances and complexities of the English language as phrases used in similar contexts tend to have similar meanings [125], as discussed in the *Distributional Hypothesis* [126]. Specifically, RoBERTa can understand the meaning of candidate phrases and relate them to the E&S terminologies in the sustainability documents with similar meanings.

In addition, RoBERTa is able to learn sustainability-related knowledge to make predictions in E&S contexts. There is no need to build an E&S-related language model from scratch as the pre-trained RoBERTa model, through *transfer learning*, can easily absorb the domain knowledge of sustainability topics through further training with the E&S labeled sample. Second, RoBERTa is able to analyze sequences of multiple lengths with word order information, which is helpful in understanding the E&S language. Why is word order information helpful? If we ignore it and look at some single words in isolation in a business context, for example, the word "diversity" could mean a company's product diversity, and the word "inclusion" could be used to describe the inclusion of new product lines. However, if they are put in a linear order like "diversity and inclusion", we know this phrase represents the goal to respect and value individuals from different backgrounds along the social dimension. Thus, reading sequences of various lengths enables RoBERTa to get additional knowledge about E&S topics.

Moreover, RoBERTa learns which part of an input sequence is important to make ac-curate E&S predictions during the training process, based on its attention mechanism (see [127] for details.). For example, the input sequence could be "we have to take actions to address the risk of climate change". During the training process, RoBERTa learns to put more attention to the phrases like "climate change" than the phrases like "take actions" to minimize noise. So, limited domain knowledge or feature design is needed in the pre-processing of training data in contrast to traditional ML models that split the sequence into fixed-length features in isolation. It also ensures an effective understanding of the language used in the E&S context. Some of the other advantages of RoBERTa model for our task are: RoBERTa can extract a word's contextual information bidirectionally, which is an upgrade over the unidirectional deep learning models like OpenAI GPT [128]. These would process words sequentially. Further, based on the contextual information, RoBERTa is able to understand multiple connotations of the *same* phrase depending on how the phrase is used. This is in contrast to models such as Word2Vec and Glove [129], which are only capable of understanding one meaning for each unique phrase.

#### 4.2.2 E&S labeled training data

Constructing a high-quality E&S training sample is the fundamental step to training a model that is learning the domain knowledge of sustainability. We tackle this task by focusing on the sustainability documents from nine leading sustainability standards and frameworks, namely: Sustainability Accounting Standards Board (SASB), Global Reporting Initiative (GRI), Task Force on Climate-related Financial Disclosures (TCFD), Asset4, Dow Jones Sustainability Indices (DJSI), FTSE4Good, Innovest, MSCI KLD, and Sustainalytics.

The collected sustainability documents contain the textual information of the E&S topics' definitions and the corresponding metrics. Guided by the conceptual framework provided by each organization, we label the sequences in the documents as per the corresponding dimension, i.e. "environmental", "social" or other. For example, as shown in Figure 4.2, the Global Reporting Initiative (GRI) provides an overview of its standards. For any GRI document in Series 300, like Document 303, its content is related to environmental issues. Thus, we are able to identify and label the sequences from Document 303 as "environmental", even without the corresponding domain knowledge We rely on the expert judgment of sustainability organizations for the E&S labels, this minimizing researcher bias. Based on this method, we construct a high-quality E&S training sample with five labels as shown in appendix Table D4.1.[5]

### 4.2.3 Model predictions and validation

After training with the E&S labeled sample, RoBERTa learns the inherent association between the input sequences and their E&S labels. We then use this RoBERTa model to make out-of-sample predictions on a large number of candidate phrases, and focus on those which are labeled as "environmental" and "social".

To collect the potentially related candidate phrases, we extract 45,595 key phrases from the sustainability documents and augment them by collecting 3,798,530 topics from Wikipedia under the categories related to sustainability. For each candidate phrase, the RoBERTa model not only labels it into one dimension (e.g., "environmental" or "social"), but also reports how confident it is about this prediction.

We keep the 998,089 phrases predicted as "environmental" or "social" with a probability higher than or equal to 80%.[6] In order to collect an exhaustive list of keyword phrases with limited redundant strings, we extract all unigrams (one-), bigrams (two-), and trigrams (three-word combinations) from each phrase. For instance, *LGBT rights in the Cook Islands* would yield *LGBT*, *LGBT rights*, *the Cook Islands*, etc. as the shorter phrases. This process generates over 2 million such phrases in our sample. We then supply these shorter phrases into RoBERTa, and retain those with the same threshold as above. Next, we get

[5]The labels other than "environmental" and "social" are the ones generated by sustainability organizations. Keeping them in the training sample helps to teach the model to identify the information not related to the E&S topics.

[6]This threshold is also used in [113].

792,453 unique phrases after cleaning those with stop words at the head or tail.

Having thus identified the phrases, it remains unclear whether they are actually used by the businesses to discuss issues on sustainability. To address this concern, we downloaded 415 CSR reports for 2017–2020 from SASB website as of Nov 26, 2020. We drop the phrases which never occur in this CSR corpus that includes 10,151,937 words . Thereafter, we manually check 14,037 phrases with a frequency larger than five in the CSR corpus. After three rounds of manual checking, we arrive at our dictionary of E&S topics with 614 "environmental" phrases and 697 "social" phrases.

#### 4.2.4 Material environmental and social phrases

Some E&S topics maybe more important for some industries than others. For example, climate change is material for the firms in the oil and gas industry, but not as important for those operating in financial services. Using industry-specific guidance on material sustainability issues from SASB (Figure 4.3), [112] find that firms with good ratings on material sustainability issues significantly outperform those with poor ratings on these issues. In earnings calls, some firms may focus on the sustainability issues which are material for their business. Other firms may prefer to discuss immaterial issues while avoiding important and potentially costly issues. To analyze this aspect of E&S discussion, we follow the guidelines of SASB and identify the E&S phrases referring to the sustainability issues material to each industry.

For the E&S dictionary we constructed, we use RoBERTa model to further classify each phrase as "material" or "immaterial" for each industry. For each industry, we generate the training data by labeling the sequences under the SASB definition as topics "material" to this industry, and as "immaterial" otherwise. We then train the RoBERTa model to make predictions on the E&S phrases. Based on the E&S phrases which are material for each industry, we can further split the E&S discussion into material and immaterial ones.

| | | Consumer Goods | Extractives & Minerals Processing | Financials |
|---|---|---|---|---|
| Dimension | General Issue Category | Click to expand | Click to expand | Click to expand |
| Environment | GHG Emissions | | | |
| | Air Quality | | | |
| | Energy Management | | | |
| | Water & Wastewater Management | | | |
| | Waste & Hazardous Materials Management | | | |
| | Ecological Impacts | | | |
| Social Capital | Human Rights & Community Relations | | | |
| | Customer Privacy | | | |
| | Data Security | | | |
| | Access & Affordability | | | |
| | Product Quality & Safety | | | |
| | Customer Welfare | | | |
| | Selling Practices & Product Labeling | | | |

**Figure 4.3:** Sustainability Framework Example – SASB:
This figure shows an example of SASB Materiality Map. For each industry, SASB provides detailed descriptions about which topics are material in the environmental and social dimensions. We extract industry-specific materiality information based on the SASB Materiality Map.

## 4.3 Data and methodology

### 4.3.1 Firm-level E&S Discussion

Our main task is to identify the firm-level environmental and social discussion over time. With the dictionary of the E&S phrases we constructed in Section 4.2, we capture the sustainability-related discussion in firms' earnings conference calls incorporating industry-level materiality.

*Earnings conference calls*

Earnings conference calls provide a good setting to analyze the discussion of E&S topics by firms' management for multiple reasons. First, earnings conference calls convey critical corporate information to the market [130]. Second, managers' discussion at earnings conference calls is less constrained than in the regulatory filings to the Securities and Exchange Commission (SEC) and managers are able to interact with participants during the Q&A session in a more conversational format. Third, earnings call could be a fruitful source for corporate communication on E&S issues. In fact, Nasdaq reports that around 65% of Russell 3000 firms discuss sustainability issues in their earnings calls [131], suggesting the importance of earnings calls in E&S discussion.

We collect data on 159,138 earnings conference call transcripts of U.S. public firms from 2007 to 2019 from SeekingAlpha, focusing on the period before Covid-19 pandemic. Using the firm name, the stock ticker, and the earnings conference call date, we find that 143,473 transcripts in our data can be merged to CRSP data. Next, we follow the approach in [130] and only retain the 132,295 transcripts with more than one thousand words[7]

[7] We retain the longest earnings call transcript for a given firm-earnings call date if there are multiple filings for that date. Some of the shorter files correspond to the previous version of the earnings call transcript.

*Measuring E&S discussion in earnings calls*

Based on the dictionary constructed (see Section 4.2) to identify the level of E&S discussion, we aggregate the frequency of E&S phrases mentioned in a given firm's earnings calls during a period of time, and scale it by the total number of words in the corresponding earnings call transcripts as follows:

$$Level(E+S)_{i,t} = \frac{\textit{Environmental Frequency}_{i,t} + \textit{Social Frequency}_{i,t}}{\textit{Number of Words in Transcript}_{i,t}}$$

The higher value of $Level(E+S)$ represents the greater emphasis placed on the E&S issues during the firm's earnings conference calls. Moreover, we generate the measure $Level(E)$ ($Level(S)$) by dividing the frequency of only environmental (social) phrases by the total number of words in earnings call transcripts. Importantly, we incorporate materiality by constructing variable $Level(E - Material)$ if the discussed environmental issues are material for the firm's industry, and $Level(E - Immaterial)$ if not. A similar method is used to construct $Level(S - Material)$ and $Level(S - Immaterial)$ on the social dimension. We report the transcript-level summary statistics for the frequency and fraction of environmental and social words of the 132,295 earnings call transcripts in Panel A of Table 4.1. On average, the E&S phrases are mentioned around 3.7 times during each earnings conference call. The average frequency of environmental phrases is more than twice as large as that of social phrases. There are around five environmental and social words per 10,000 words in the earnings call transcripts. When we split these into their constituent groups, we find an average of 3.4 environmental words and 1.6 social words for every 10,000 words in the conference call transcripts.

**[Insert Table 4.1 here]**

*Validation of the E&S dictionary*

We validate our E&S dictionary from several aspects. First, we list the top 20 environmental and social phrases in Table 4.2 in terms of their frequency shown in the earnings call sample of 132,295 transcripts. In the environmental dimension, “sustainable”, “environmental”, “ecosystem”, “renewable”, and “$CO_2$” are the most often discussed phrases in earnings conference calls, while the top social phrases are “affordable”, “minority”, “demographic(s)”, “affordability”, and “HIV”.

**[Insert Table 4.2 here]**

In addition, Table D4.2 shows several earnings call snippets for the top 10 firms with the most E&S discussions. Overall, it seems that our E&S dictionary captures the keywords used in the earnings calls. For example, Jaguar Animal Health (now Jaguar Health) talked about their large advertising campaign related to “HIV” and “LGBT” in 2017, and environmental phrases like “industrial waste”, and “wastewater” were discussed in the earnings call of Tri-Tech Holding in 2012. Not surprisingly, many of these firms operate businesses focused on environmentally friendly products or services, like clean energy, water/waste treatment, etc.

Next, there is significant industry-level variation in the discussion of E&S topics captured by our dictionary. Figure 4.4 shows the proportion of earnings calls mentioning E&S phrases across years and Fama-French 12 industries. It appears that companies in the utilities, chemicals, and energy industries are more likely to discuss E&S topics in their earnings conference calls than firms in financial services, business equipment, and healthcare businesses.

Further, in time series, we present a yearly variation of the proportion of transcripts mentioning E&S topics during the earnings calls in Figure 4.5. Over 40% of the transcripts consistently include some discussion on social topics throughout the sample period. On the other hand, there is a gradual increase in the discussion of environmental topics during

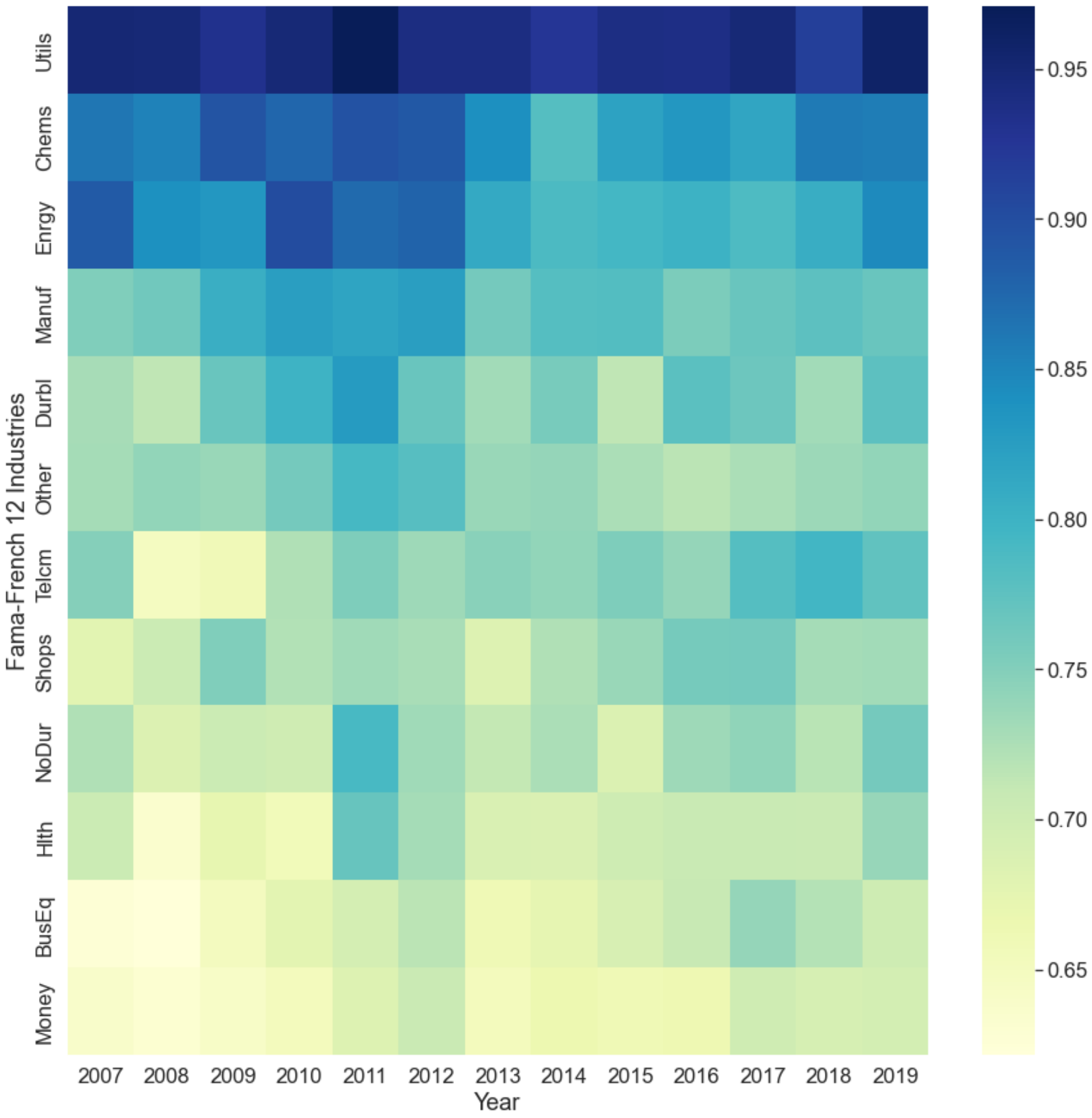

**Figure 4.4:** Distribution of Transcripts with E&S Phrases by Industry and Year:
This figure shows the percentage of earnings calls mentioning E&S phrases across industries and years. The *y*-axis shows the abbreviations of Fama-French 12 industries sorted by the cumulative percentage of sample mentioning E&S phrases. We show the industry distribution as per Fama-French 12 industries classification due to space constraints. The *x*-axis represents different years of the sample. The density bar on the right of the graph displays the percentage of earnings calls mentioning E&S phrases for each industry between 2007 and 2019.

2007-2011. Thereafter, this metric remains fairly uniform between 50-60%.

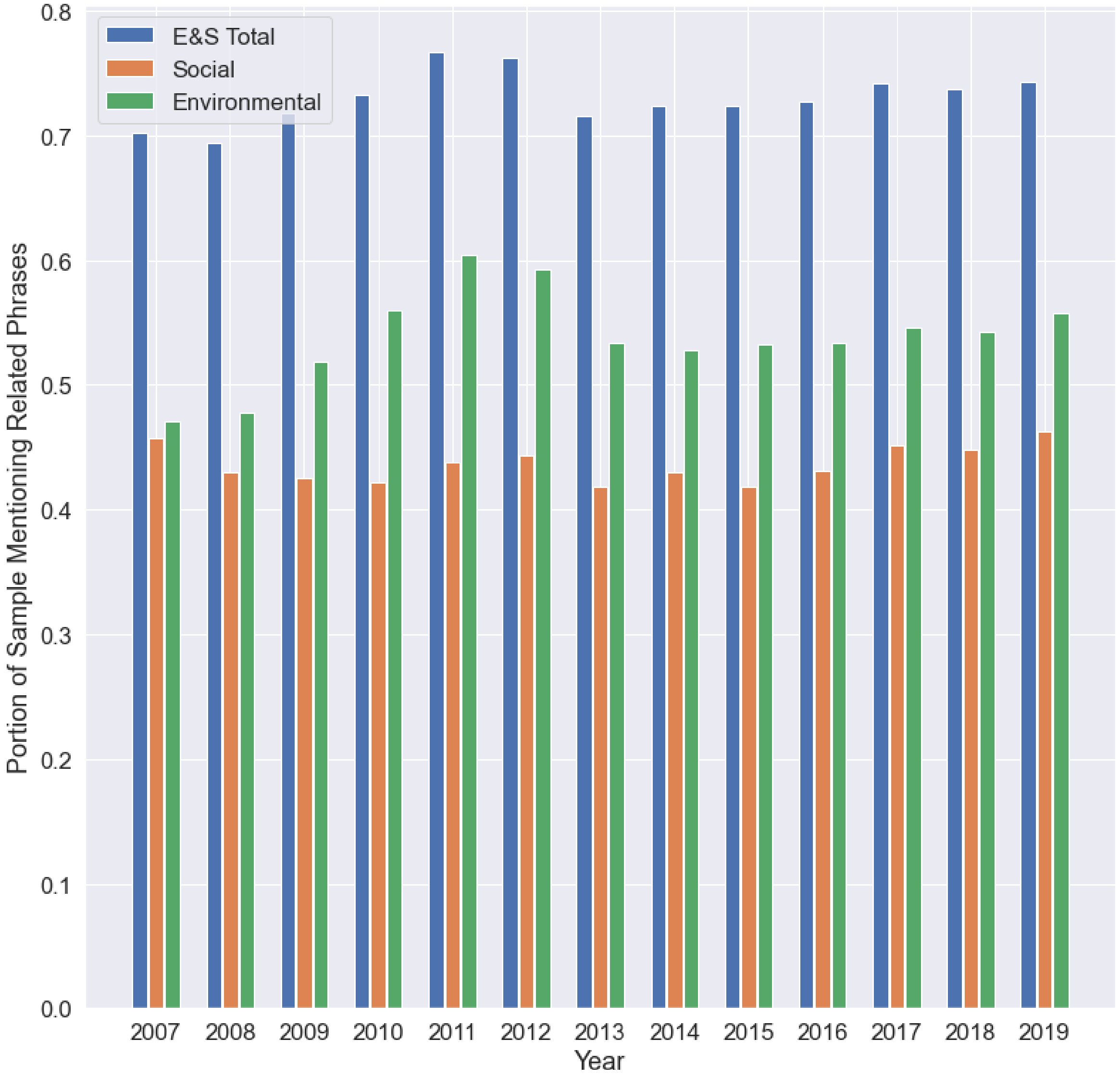

**Figure 4.5:** Distribution of Transcripts with E&S Phrases by Year:
This figure shows the annual percentage of earnings calls mentioning E&S phrases during 2007-2019. The figure also separately documents the annual percentage of earnings calls mentioning environmental and social phrases separately.

Finally, we assess the dynamics of our measurement of environmental discussion in earnings calls after a major environmental event, the Paris Agreement. Adopted on December 12, 2015, this legally binding international treaty on climate change implies a tightening regulatory environment for companies on climate change and other environmental topics [132]. Thus, we expect a higher level of environmental discussion in firms' earnings calls to address the increasing concern on these issues. We regress the *Level(E)* (in %) on annual dummies corresponding to calendar years and use the years prior to 2015 in our sample as the omitted benchmark[8]. Panel A in Figure 4.6 shows the estimates of the regression coefficients along with the 95% confidence intervals. Our results suggest that there is an increase in environmental discussion in earnings calls after the Paris Agreement, with the effect nearly doubling in 2019. We find similar results when we use the *Level(E-Material)* (in %) as the dependent variable, as shown in Panel B of Figure 4.6. We notice that the estimate slightly decreases in 2017, which might be driven by the U.S. announcement to withdraw from the Paris Agreement. Next, we explore whether the tone in which environmental topics are discussed in earnings calls changes over time[9]. As shown in Panel C and Panel D, we find the positive sentiment has increased from 2016 onward, while there was a dip in the negative sentiment around 2018.

Taken together, these results suggest our E&S dictionary can capture sustainability-related content in firms' earnings conference calls.

### 4.3.2 Firm-level E&S Performance

To examine the credibility of firms' E&S discussion in earnings calls, we evaluate the firm-level E&S performance on three dimensions. We provide details on each one of them and the data sources in this section.

*Environmental pollution and abatement activities*

We use plant-level data on toxic chemical releases reported under the Toxic Release Inventory (TRI) program from the Environmental Protection Agency (EPA) during 2007-2019[10].

[8]We control for firm and industry fixed effects and double cluster at firm and year-quarter levels.

[9]For each earnings call transcript in our data, we conduct sentiment analysis on sentences containing environmental phrases. We use the RoBERTa model trained on the Stanford Sentiment Treebank data (SST-2, [133]) to classify sentences containing environmental phrases into those with "positive" or "negative" sentiment.

[10]While the EPA has collected this information since 1987, we restrict our sample to overlap with the availability of earnings conference call transcripts from Seeking Alpha (Section 4.3.1).

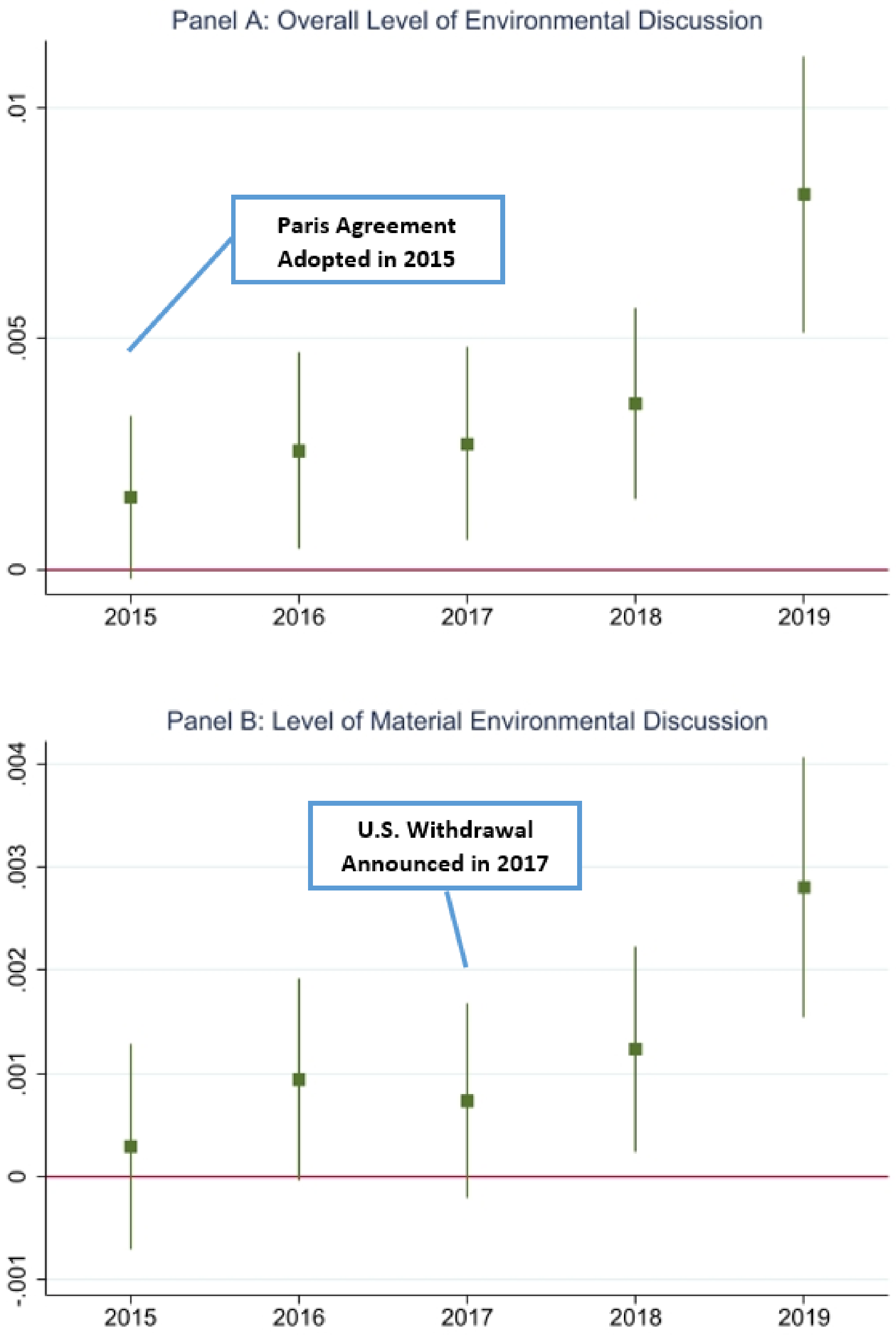

**Figure 4.6:** Paris Agreement and Environmental Talk in Earnings Calls:
This figure shows the dynamic trends of the level of environmental discussion in Panel A after the Paris Agreement on December 12, 2015. Panels B, C and D show the coefficients using material environmental discussion, and the discussion with positive or negative sentiment as the dependent variables. The y-axis represents the estimated coefficients obtained from regressing the level of various kinds of environmental discussion (%) in the earnings conference call transcripts over yearly dummies, after including firm and industry fixed effects. The benchmark period comprises the years before 2015 in our sample. Standard errors are double clustered by firm and year-quarter. The verticals lines represent 95% confidence intervals.

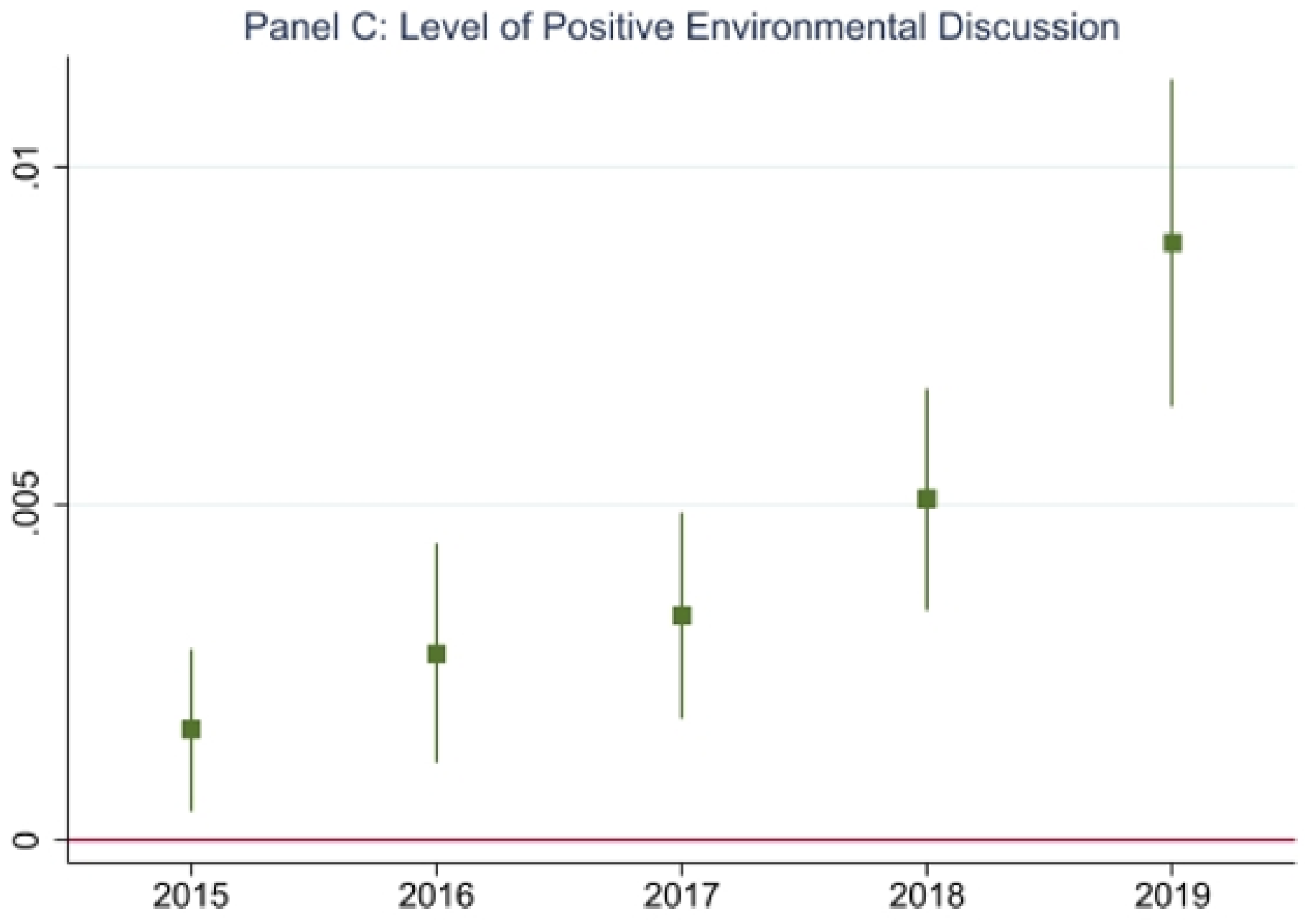

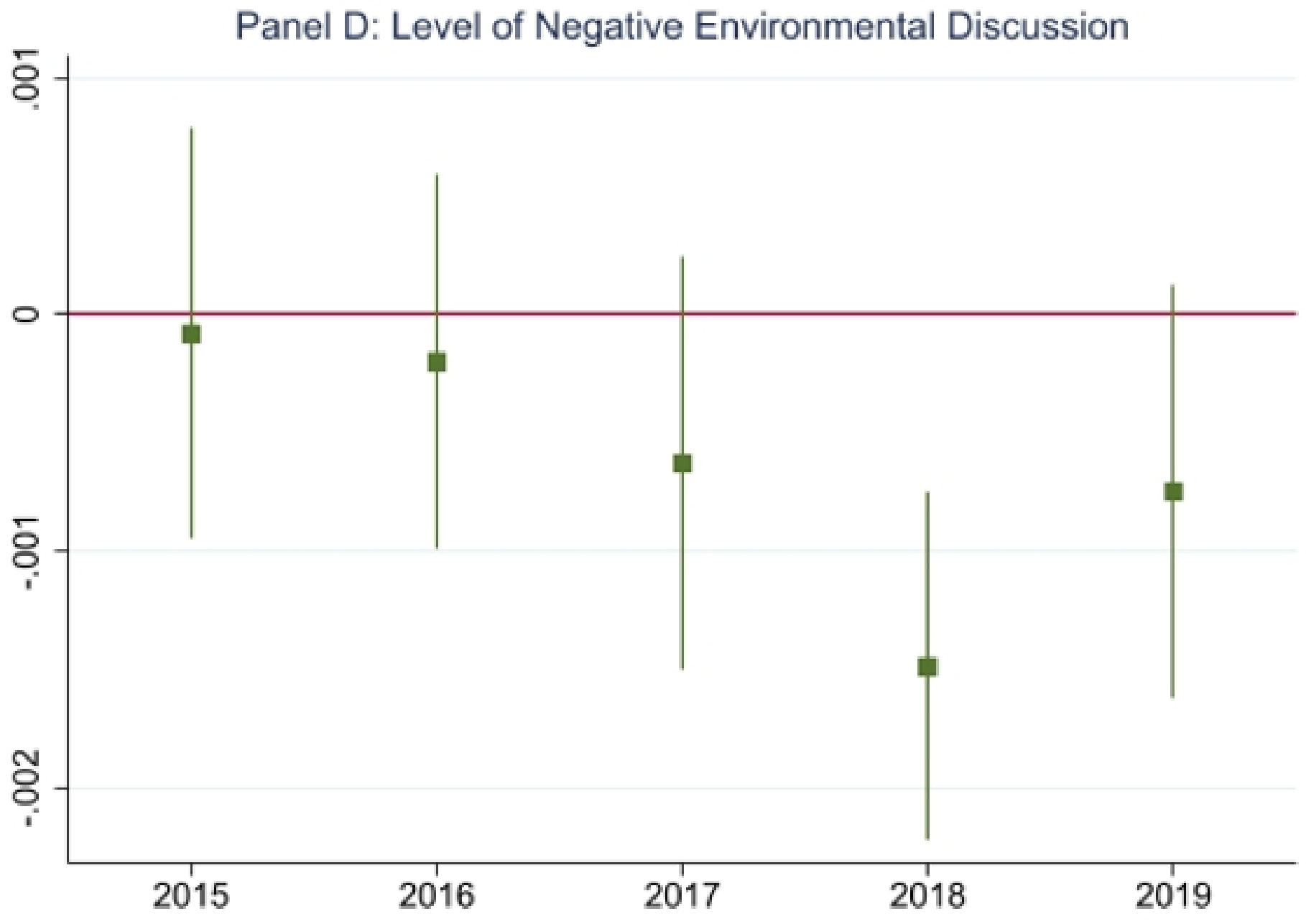

The EPA reports chemical-level emissions data in TRI for plants with a facility of over 10 full-time employees, operating in one of the 400 industries defined at the six-digit NAICS level, and using one of nearly 600 chemicals. The TRI data are self-reported by firms and the EPA conducts audits to investigate anomalies. Misreporting could lead to penalties of a civil or criminal nature ([134]). [135] find limited evidence in favor of over- or under-estimation of toxic releases in the EPA-TRI data. For each firm in a given year of our sample, we aggregate the number of pounds released into the ground, air, and water from all its plants. We require each firm to be present throughout all the years of our sample period. Next, we collect data on abatement activities from EPA's Pollution Prevention (P2) database. Under this program, plants included in the EPA-TRI database are required to document source reduction activities at the chemical level to restrict the amount of hazardous substances released into the environment.

We use the firm names from various datasets by the EPA to create a crosswalk between EPA and firm names in the earnings conference calls data. We implement a combination of such matches across various files in the EPA to capture permutations of facility spellings and firm names used in the EPA.[11]. Our matched dataset contains 1,067 unique firms for which we observe E&S discussions in the earnings call transcripts and EPA-TRI pollution. We provide the coverage of unique firms over the years in Figure D4.1. This remains broadly uniform during the sample period with a slight decline toward 2019. The associated coverage of the plant-level facilities from EPA-TRI is shown in Figure D4.2.

The average firm in our sample emits total emissions of about 2.58 million pounds, with ground emissions forming the largest component amounting to 1.50 million pounds. Figure D4.3 shows the time series of aggregate emissions for the three categories of pollution over the sample period. The overall trend is consistent with previous findings that report a decline in emissions, primarily driven by air pollution [136, 137]. On average, there are 24% of firms in the sample that adopt some measure of pollution abatement based on operating procedures. The corresponding fraction for process improvement methods is 23%. We provide details on the distribution of pollution in Panel B of Table 4.1.

[11]Unfortunately, these names are not absolutely consistent over the years. To account for possible differences in names, we standardize the strings by removing common suffixes like "Corp.", "LLC", "Inc." etc.

*Green patents*

We study firms' green patenting activities to analyze their performance on green innovation. The data on firms' granted patents are provided by [71]. The primary challenge is to identify whether a patent is related to the environmental-related technology or not. The Organization for Economic Co-operation and Development (OECD) reports a list of Cooperative Patent Classification (CPC) codes [138] to identify patents that contribute to solving environmental problems, like climate change, air pollution, water scarcity, etc.[12] [139] make use of this framework to analyze green innovation and ESG scores across industries. Based on the CPC data obtained from USPTO, we identify the set of green patents granted to U.S. public firms.

For the examination of the link between the E&S discussion and green innovation, we look at how many green patents are granted in the future, given the level of E&S discussion during one earnings call. In this analysis, our sample construction follows the literature. Specifically, we require that the earnings call transcripts are associated with firms with positive total assets, with non-missing earnings announcement dates in COMPUSTAT, with stock prices greater than $1 at the end of the fiscal quarter, with market values of equity larger than $5 million at the end of the fiscal quarter, and with non-missing dependent and control variables. Following the approach in [140], we further restrict the sample to only include potentially innovative firms that have filed at least a single patent from 2000–2019. We use Poisson regression for our analysis of the number of green patents. After accommodating these considerations, we are left with 32,074 earnings call transcripts during the one-year window analysis. We provide the summary statistics for this sample in Panel C of Table 4.1. The average firm in the sample has 2 green patents granted. Of these, nearly 1.85 are classified as related to climate change mitigation by the OECD. In addition, we report the distribution of the two-year and three-year samples in Panel C of Table D4.3.

[12] https://www.oecd.org/environment/consumption-innovation/ENV-tech%20search%20strategies,%20version%20for%20OECDstat%20(2016).pdf

*Employee ratings*

Our employee ratings come from Glassdoor reviews for the corresponding firm during 2008-2019[13]. The Glassdoor website was launched in 2008 to host anonymously written voluntary reviews from current and former employees. We use the overall firm rating and its components provided by employees on a scale of one to five. Glassdoor requires email verification from an active email address or a valid social networking account to prevent self-promotion by the company. Further, the site also moderates content using its two-step approach consisting of algorithmic detection and human checking to eliminate fraudulent reviews.

We consider the average employee ratings in a given calendar year-quarter for each firm in the sample. Our final sample consists of 2,125 unique firms matched to their earnings conference call transcripts. We report the summary statistics for the distribution of overall employee reviews and its components in Panel D of Table 4.1. The mean value of the overall ratings is 3.15 which is similar to the average rating for compensation and benefits. The standard deviation across all the ratings is over 0.70 notches. Given that this distribution is not very skewed, we do not perform the logarithmic transformation of the raw values thus obtained. Our distribution of the ratings looks similar to other papers in the literature [141, 142].

### 4.3.3 Empirical specification

We use a multivariate panel regression approach to measure the association between firm-level outcomes and discussion of environmental and social issues in earnings conference call transcripts. Using the dictionary of E&S phrases, we quantify the fraction of environ-mental and social topics in earnings call transcripts during a period of time as detailed in Section 4.3.1. Our baseline empirical specification is given below:

$$Y_{i,f,t} = \alpha + \beta_0 * Level(E+S)_{i,f,t} + Controls_{i,t} + \delta_{f,t} + \phi_i + \epsilon_{i,f,t} \tag{4.1}$$

where $Y_{i,f,t}$ represents the outcome variable corresponding to firm $i$ (operating in industry $f$) in time $t$. The key independent variable, $Level(E+S)_{i,f,t}$, represents the fraction of words devoted to the discussion of environmental and social topics, in firm $i$'s earnings conference calls during time $t$, belonging to industry $f$. $Controls_{i,t}$ represents a vector of firm-level time-varying observable characteristics which might be correlated with the outcome variables. We obtain firm-level financial information from Compustat, IBES, and CRSP. Detailed definitions of variables can be found in Table D1. The baseline specification includes firm fixed effects ($\phi_i$) and industry $\times$ time fixed effects ($\delta_{f,t}$). The inclusion of firm fixed effects in the specification helps account for any firm-specific, time-invariant unobserved characteristics, such as a firm's general proclivity to always or never include any E&S discussion in its conference calls. Moreover, controlling for the industry $\times$ time fixed effects helps us account for any time-varying trends within industries that are potentially correlated with the E&S concerns relevant to specific industries. We double cluster standard errors at the firm and time levels.

[13]We are thankful to Reza Ferhadi for sharing their data with us.

We suitably modify the equation when we apply it to outcome variables that are available at a quarterly or annual frequency. For example, while studying the association between a firm's E&S discussion and its annual pollution performance (see Section 4.4.1), we sum up the frequency of E&S phrases appearing in this firm's earnings call transcripts for each year, scaled by the total number of words in these transcripts during the year, and then test its relationship with the firm's pollution performance on an annual basis.

## 4.4 Results

We begin our analysis by showing results for the association of firm-level pollution abatement with the discussion of environmental words in Section 4.4.1. In this regard, we also evaluate the firm-level emissions. Next, we examine firms' innovation activities using green patents (Section 4.4.2). In Section 4.4.3, we provide evidence from employee ratings and discussion of social topics in earnings conference calls. Finally, we consider the stock market's immediate reaction to the discussion of environmental and social topics in earnings conference calls (Section 4.4.5).

### 4.4.1 Firm pollution and E&S talk

We first examine whether firms with E&S discussions in their earnings conference calls reduce their overall emission of pollution. We study the pollution emissions aggregated up to the firm level using Equation (4.1) at the firm-year level. To this end, we consider the air, water, and ground emissions from the Toxic Release Inventory database provided by the EPA. Table 4.3 shows the results with standard errors double clustered at the firm and year levels. We normalize emissions by the firm-level cost of goods sold in our baseline approach. Our dependent variable is the natural logarithm of one plus the normalized toxic chemical release, $\log(1 + \frac{Release_t}{COGS_{t-1}})$. We show robustness to our choice of scaling the pollution emissions in Section 4.4.5.

**[Insert Table 4.3 here]**

In Column (1) of Panel A, we show the unconditional correlation of total pollution with the proportion of environmental and social words used in the earnings conference call discussions with firm and year fixed effects. In Column (2), we show results by introducing firm-level controls consisting of leverage, logged assets, and capital intensity. We further consider the unobserved heterogeneity across industries in Column (3) with the firm and industry × year fixed effects. We use Fama-French 30 industry classification. The coefficient of -68.59 suggests that a 10% increase in discussion of E&S topics is associated with 0.998% decrease in total pollution. For the average firm in our sample, this amounts to 25,807 pounds. This effect is statistically significant and economically meaningful. We find similar results by using only the level of environmental discussion in Panel B, as shown in Columns (1)-(3).

Next, we exploit the novelty of our dictionary data to understand whether our results are driven by the topics that are material to a given industry or not[14]. Section 4.2.4 provides details on how we classify environmental and social discussions into those that are material to the firm's industry or otherwise. Based on this classification, we report our results from the baseline Equation (4.1) in Column (4) of Table 4.3. We find that most of the correlation between total pollution and E&S discussion is driven by topics that are material to the industry. A 10% increase in those discussions is associated with almost 1% reduction in total pollution. This nearly accounts for the overall effect reported earlier in Column (3).

We find similar results when using only the environmental discussion in Column (4) of Panel B. Importantly the magnitude of the coefficient associated with material discussion on environmental topics is nearly twice that of the coefficient for immaterial discussion. At the same time, the statistical significance is more prominent for material discussion of environmental aspects. These evidence suggest that the material topics relevant to a given industry are more meaningful in the context of pollution reduction. Further, we draw support that our dictionary is able to capture these nuanced aspects of materiality across industries reasonably well.

We also evaluate the association of future pollution and E&S discussion. In Panel C, we show our results by changing the dependent variable to total pollution in the future. Column (1) suggests a similar negative association implying a reduction in total pollution even one year after the discussion of environmental and social topics. This effect reduces gradually in the longer horizon of two and three years ahead, as shown in Columns (2) and (3), respectively. Further, in Panel D we replicate our analysis using forward years corresponding to the discussion of environmental topics only. As before, Columns (1)-(3) suggest a gradually weakening association with environmental discussion for a reduction in firm pollution during the future years. Overall, the evidence in this section suggests an association between reduction in firm pollution and discussion of environmental topics in managers' earnings conference calls. This effect is stronger when the topics are material to the industry of the firm. Moreover, the effect tends to reduce and become insignificant in the three years after.

[14]While it is appealing to consider whether the impact is stronger for consumer-facing industries, we are restricted by our sample. Only a small fraction of the firm-year observations belong to the retail industry as per two-digit SIC classification.

Next, firms that are concerned about the environment may likely invest in pollution abatement. Such investments may reduce overall firm emissions provided firms do not increase production in order to capitalize on the emission reduction technologies. We measure abatement activities at the firm level using EPA's P2 database. Specifically, we focus on the two most common abatement categories: modifications to operating practices and process improvements. While good operating practices are related to improving maintenance or quality control, process improvements may be associated with improving chemical reactions or enhancing process controls. We follow [136] to classify abatement activities into these two types.

**[Insert Table 4.4 here]**

In Table 4.4, we test whether discussion on environmental and social issues is associated with changes in annual pollution abatement by firms using Equation (4.1). We double cluster standard errors at the firm and year levels. The dependent variable in Columns (1)-(3) is an indicator for abatement related to operating practices, and the dependent variable in Columns (4)-(6) is an indicator for abatement related to process improvements. We count the total number of abatement activities of each type across all plants for a given firm in a year. We denote the abatement indicator as one if the firm undertakes at least one such activity for the year. Our results suggest that firms increase abatement related to operating practices when they discuss environmental issues. The association with process improvements is relatively weaker.

Column (1) shows that when the level of environmental and social discussion increases by 1%, the associated abatement in operations increases by 8.83%. Similarly, in Column (2) we find evidence for a 9.81% increase in operations associated with a 1% change in environmental discussions alone. In Column (3), we provide results by splitting the environmental topics into those that are material or immaterial to the firm's industry. We find that the impact is mostly driven by words material to the industry. The economic magnitude of this association is sizable given that the unconditional average likelihood of operating process abatement is 24%. For abatement related to the production process, we find similar evidence in Column (6) for association with environmental topics material to the industry. However, the statistical significance is relatively weaker.

#### 4.4.2 Future green innovation and E&S discussion

In this section, we examine whether firms follow up the discussion of E&S phrases in their earnings calls with higher environmental-related innovation later. For each earnings call, we identify the green patents granted to the firm in the three-year window after the earnings call date.

In Table 4.5, we examine how the level of E&S discussion in an earnings conference call is related to the number of green patents granted ex-post. We use a modified version of Equation (4.1), by omitting the firm fixed effects while ensuring comparison across firms in the same industry over time. The specifications are estimated using a Poisson regression model with fixed effects based on industry $\times$ year-quarter level and standard errors double clustered at the firm and year-quarter levels. In Columns (1)-(3) of Panel A, the dependent variable is the number of green patents granted in the one-year horizon after the earnings call. We find that firms with a higher level of E&S discussion in their earnings calls have a higher number of green patents granted in the one-year window after the earnings calls. In addition, this effect is mainly driven by the level of environmental discussion in the earnings calls. For the average firm in the sample, one standard deviation increase (0.08%) in the level of environmental discussion is associated with around 16.77% ($e^{193.84*0.08\%} - 1$) increase in the number of green patents granted in the next one year (Column (2)). In Column (3), we further examine how the environmental discussions about material and immaterial topics relate to future green innovation. We find that both types of environmental discussions predict a higher number of green patents in the future, with the coefficient estimate of the material discussions larger than that of the immaterial ones. This inference remains unchanged when we examine the two-year window (Panel B) and three-year window (Panel C) after the earnings calls.

**[Insert Table 4.5 here]**

Beyond the evidence on the overall green innovation provided above, we also separately examine the green patents of different classifications after the earnings calls. Based on OECD's guidelines, we further consider the green patents which are related to environmental management and water-related adaptation technologies ($Env\&Water$, Columns (4)-(6)), and those which are related to climate change mitigation ($Climate$, Columns (7)-(9)). These groups are not mutually exclusive. As before, we find that the level of E&S discussion is positively associated with the number of green patents of each group. Moreover, this association is stronger if we focus on the discussions of environmental topics, especially the material ones. Overall, our results in this section suggest that there is a positive relationship between the discussion of E&S topics with the firms' green patenting activity.

### 4.4.3 Employee ratings and E&S content

In this Section, we present our analysis of the correlation between firm outcomes that might reflect the discussion on social issues in earnings conference calls. Specifically, we eval-uate the association of average firm-level employee ratings aggregated up to the quarterly frequency from Glassdoor.

**[Insert Table 4.6 here]**

First, in Panel A of Table 4.6, we report the results from our baseline Equation (4.1) with average employee overall ratings from Glassdoor as the outcome variable. We double cluster standard errors at the firm and year-quarter levels. Our firm level controls consist of the logged value of assets and market-to-book ratio, leverage, return on assets, and stock returns. Column (1) shows the correlation with just the firm and year-quarter fixed effects. We find a statistically strong positive correlation amounting to 25.26 units. In Column (2), we show our results by adding firm-quarter level controls. Our preferred specification comes from Column (3) where we introduce industry $\times$ year-quarter fixed effects to further absorb unobserved variation across industries over time. The coefficient of 24.60 is statistically significant and suggests that a 1% increase in the discussion of environmental and social topics is associated with a 0.24-notch improvement in overall rating from employees in that quarter[15]. This represents an improvement of 7.62% on the average overall rating of 3.15 in our sample. Further, in Column (4), we report our results by changing the dependent variable to the overall rating observed in the subsequent quarter. The impact is economically weaker, although still statistically significant. Columns (2) and (3) show results based on the overall ratings observed two and three quarters ahead. We find that the statistical relationship does not hold in these cases.

We follow a similar analysis in Panel B using only the discussion of social topics. As before, our preferred baseline specification comes from Column (3) where we control for firm and industry-quarter fixed effects, in addition to firm-quarter controls. Our results show a magnitude of 29.73 although the statistical significance is weaker. A 1% increase in discussion of social topics is associated with a 0.29-notch improvement in overall ratings from employees in that quarter. When we project this relationship onto the future quarters in Columns (4)-(6), we do not find this relationship to hold. Overall, we find that there is a strong positive association between overall ratings from employees and the discussion of environmental and social topics. The relationship seems to become weaker gradually over the subsequent quarters. Interestingly, our results in Panel C highlight that employee ratings are also positively associated with the level of environmental discussion alone by managers. This effect is somewhat stronger than in Panel B and even shows up one quarter after the conference call.

[15]It is not surprising to find that employees seem to care about environmental topics as well when submitting employer reviews. For example, survey evidence suggests that 72% of respondents among UK office workers were concerned about environmental ethics. A focus on sustainability may also help firms in retaining their employees and in engaging with them more beneficially.

Next, we examine the different components of the Glassdoor employee ratings that constitute the overall rating. We show our results in Panel D of Table 4.6 using the fraction of social topics as the key explanatory variable. We start with our baseline specification in Column (1) with the overall rating as the dependent variable. In the subsequent columns, we show the association corresponding to each component of the overall rating as the dependent variable. Column (3) suggests that a 1% increase in discussion of social topics is associated with a 0.24-notch improvement in ratings corresponding to compensation and benefits in that quarter. Likewise, we find that a higher discussion on social topics is also correlated with better ratings on work-life balance (Column (5)) and culture (Column (6)). These associations are statistically significant and the economic magnitudes are comparable to the impact on overall rating discussed before. In summary, the improvement in overall ratings seems to be driven by ratings on compensation and benefit, work-life balance, and culture.

In order to understand the drivers of these results, we divide the fraction of social topics into those that are material to the given firm's industry versus otherwise. We report our results in Panel E using the preferred baseline specification. In Column (1), we find that there is no statistically significant correlation between overall rating and the fraction of industry-material social topics. However, Columns (3) and (5) suggest that the impact on ratings for compensation and benefits, and work-life balance respectively, is indeed driven by social topics material to that industry. The coefficients suggest an improvement of 0.34-notch and 0.39-notch in compensation and benefits and work-life balance ratings, respectively. These correspond to 10.76% and 12.04% of the average values of the corresponding ratings in the sample.

Taken together, our results in this section provide some evidence of marginal improvement in average employee ratings, especially following the discussion of industry-specific material words. The effect is driven by ratings corresponding to compensation and benefits, and work-life balance. We do not find evidence for future improvements in ratings.

#### 4.4.4 Which firms do not walk the talk?

The empirical results so far imply that in general, firms walk the talk on environmental and social issues. However, it is possible that not all firms walk their E&S talk. In this section, we conduct cross-sectional analysis to understand the characteristics of firms where E&S discussion in earnings conference calls may not be associated with E&S outcomes.

We first define one *interaction* in the Q&A session of an earnings call as the conversations between one analyst and executives until the next analyst shows up in the transcript, to address the issue that there might exist some back-and-forth between this analyst and the management team as discussed by [143]. Next, we classify the sentences of the managers' answers in one interaction into "non-answer" or not, and then identify this *interaction* as "non-answer" if any of these sentences is classified as "non-answer". Finally, we calculate the call-level non-answer ratio as the number of non-answer interactions divided by the total number of interactions in the Q&A session of the earnings call.

**[Insert Table 4.7 here]**

We then conduct the cross-sectional analysis with the firms' yearly pollution from the EPA Toxic Release Inventory (TRI) as the dependent variable during 2007-2019. For each year, we sort firms into 10 groups based on the average non-answer ratio of their earnings calls during that year. The dummy variable, $NoAnswer$, equals to one if the firm belongs to the top 10% in terms of the non-answer ratio for that year, and 0 otherwise. The Column (1)-(2) of Table 4.7 shows the empirical results. As shown in Column (1), the level of E&S discussions still has a significantly negative association with the firms' pollution, but the coefficient estimate to the interaction term of the E&S discussion and $NoAnswer$ is significantly positive. It means that even though a higher fraction of E&S discussion is generally associated with a decrease in firms' pollution, the firms with high level of unwillingness to disclose information in response to analysts' questions in earnings calls tend to have more pollution given their E&S discussion. This implies that they likely do not walk their talk on E&S matters. This inference remains unchanged when we focus on the discussions of environmental topics (Column (2)).

Next, we explicitly combine the E&S talk with the interactions within one earnings call when the firm's managers refuse or are unable to disclose information asked by analysts. We distinguish the level of the E&S discussions shown in the non-answered interactions ($Level(E+S) - -NoAnswer$) from the E&S talk in the answered interactions ($Level(E+S) - -Answer$) within one earnings call. The results in Column (3) show that only the E&S talk in the answered interaction is significantly associated with a decrease in firms' pollution. There is no statistically significant relation between the firms' polluting actions and the managerial E&S talk when the managers at the same time refuse or are unable to answer the questions from analysts. This inference holds when we only look at the environmental discussions and distinguish them based on whether they occur in the answered or non-answered interactions (Column (4)).

### 4.4.5 Supplemental Tests

*The Paris Agreement and environmental discussion*

The Paris Agreement is a legally binding international treaty on climate change that was adopted on December 12, 2015. Its primary goal is to restrict the rise in global temperatures to 1.5 degrees Celsius during the twenty-first century. The Agreement calls upon signatory nations to submit their plans aimed at reducing greenhouse gas emissions. In light of the national plans, firms would likely have to comply with more stringent environmental regulations to reduce pollution and mitigate climate change.

In Section 4.3.1, we assess how environmental discussion in the earnings conference calls changed around the Paris Agreement. Our results suggest that managers seem to discuss environmental issues more prominently after the Paris Agreement. However, it is unclear whether this increasing "talk" is followed by more "walk" in the environmental dimension. On one hand, firms may be motivated to acknowledge the importance of environmental and climate-related issues due to greater emphasis on the news media and related coverage. On the other hand, firms may indeed be taking actions to align their carbon footprint and other environmental performance in line with the expected regulatory changes after the Paris Agreement.

In order to understand how firms respond to the Agreement, we use a modified version of our baseline Equation (4.1). Since the United States announced to cease all participation in the Paris Agreement on June 1, 2017, we consider three periods around the Agreement. We interact the equation with dummies corresponding to before, during, and after withdrawal from the Paris Agreement. We additionally control for group-year fixed effects in this analysis.

**[Insert Table 4.8 here]**

We report our regression results in Table 4.8. First, in Column (1), we consider the periods in our sample that correspond to before and during the Paris Agreement. We find that the negative association between total pollution and the discussion of environmental topics becomes stronger after the Paris Agreement. However, the difference between the two periods is not statistically significant. Second, in Column (2), we show our results focusing on the periods during and after the withdrawal from the Accord only. Here, we do not find any meaningful association between the environmental discussion and firm pollution.

Finally, Column (3) shows our results by interacting with each period before, during, and after the Paris Agreement. We find that the negative association between pollution and environmental discussion tends to increase linearly. By comparing the difference in coefficients across the three periods, we conclude that firms did reduce their pollution after the Paris Accord. This difference was statistically significant. Importantly, the negative correlation increased after the US announced its withdrawal from the climate accord. This is not surprising in light of some anecdotal evidence to suggest that CEOs and business leaders had been supportive[16] of the US maintaining its commitment to the Paris Agreement. In particular, the inverse association between pollution and *Level(E)* is statistically more pronounced during the withdrawal period than during the Agreement. Overall, our results suggest that firms seem to be reducing emissions when their managers discuss environmental topics even after the US withdrew from the international treaty on climate change.

*Robustness to log one plus transformation*

As discussed in [144], the existing literature takes two different approaches in measuring pollution. On one hand, [136, 145] use the logarithmic transformation of emissions without any scaling. On the other hand, [146, 147] rescale the emissions in their outcome variables. Our baseline specification on pollution uses a scaled version of the dependent variable as discussed in Section 4.4.1. We argue that this is relevant because the amount of pollution is a function of actual production by the firm.

[16] https://www.reutersevents.com/sustainability/75-ceos-call-us-stay-paris-agreement-emissions-continue-rise
https://hbr.org/2017/05/u-s-business-leaders-want-to-stay-in-the-paris-climate-accord

Accordingly, [148] recommend that one way of working with skewed outcomes limited to non-negative values could be to scale them to produce a rate. The cost of goods sold is a suitable scaling variable that captures the exposure of a firm to pollution. [148] find that rate regressions produce estimates with a simple interpretation and show that their efficiency is similar to Poisson regression estimates. This is likely because the relevant scaling would substantially deskew the outcome variable.

**[Insert Table 4.9 here]**

We address this concern in Table 4.9. First, in Panel A, we demonstrate that our regression results are robust to using an alternative variable in scaling the pollution outcome. For example, our coefficient estimate of -68.82 reported in Column (2) is very similar to our baseline estimate of -68.59 (reported in Column (3) of Panel A in Table 4.3) and is statistically significant. Here, we scale the total pollution by the firm's revenue in the previous year. We show similar estimates by alternatively scaling the pollution using the contemporaneous cost of goods sold (Column (1)), lagged sales (Column (3)), and lagged total assets (Column (4)). Our results are statistically significant in each of these alternative approaches to scale the outcome variable.

In addition, we also report results from alternative specifications shown in Panel B of Table 4.9. Specifically, in Column (1), we consider the unscaled version of pollution in our log transformation, similar to [136]. Next, Column (2) shows results using observations with positive levels of pollution. We use inverse hyperbolic sine (IHS) transformation for the dependent variable in Column (3). Finally, Columns (4) and (5) show our results using a Poisson regression specification using the unscaled and scaled versions of the dependent variable. We find that our results remain similar across all these additional specifications. Overall, we conclude that our results are not sensitive to the log transformation of the dependent variable.

*Stock price response to E&S discussion*

We also explore the immediate price response to the use of E&S phrases in earnings conference calls. The dependent variable in the analysis is the seven-day cumulative abnormal return (CAR[$-3$,$+3$]), calculated using the market model. Our independent variables of interest are the level of environmental and social discussion ($Level(E + S)$). We also consider them separately ($Level(E)$, $Level(S)$) for our analysis. The construction of these variables is described in Section 4.3. Our findings are presented in Table 4.10.

**[Insert Table 4.10 here]**

In Column (1), we study the impact of the overall E&S discussion on the immediate stock price response to earnings conference calls. We control for time-varying firm characteristics at the time of the earnings conference call, firm fixed effects, and industry $\times$ year-quarter fixed effects. We find that the coefficient estimate on $Level(E + S)$ is positive and significant at the 10% level, which suggests that earnings calls with a greater discussion of E&S issues generate a positive immediate price reaction. In Column (4), we separately examine the impact of environmental and social discussion on the immediate stock price response to the earnings conference calls. We find that the market only positively reacts to the discussion of environmental topics in the earnings call.

## 4.5 Conclusion

There has been an increasing focus on firms' performance on environmental and social dimensions. However, recent evidence suggests that there is much disagreement among ESG ratings provided by different agencies and there are significant concerns about greenwashing. We construct a novel dictionary on environmental and social topics using a deep learning–based NLP model trained on comprehensive sustainability frameworks. Our approach also incorporates industry-specific materiality of E&S topics. We use this dictionary to parse the earnings conference calls of U.S. public firms. We show that firms discussing more about the environment also invest more in pollution abatement initiatives, especially when such topics are material to that industry. Consistent with these results, we find a negative association between E&S discussion and the firm's overall emissions. We also find a positive association between the green patenting activities and managers' talk about the environmental topics. On the social dimension, we find a positive relationship between the discussion of environmental and social topics, and the overall employee rating, suggesting that employees care about the E&S profile of the firm. However, the association with E&S performance is weaker for firms that give more non-answers and when the topic is immaterial to the industry. Overall, our evidence suggests that managers do walk some of their talk on the E&S issues.

**Table D1:** Description of Key Variables

Data sources in the above table include the Environmental Protection Agency's Toxic Releases Inventory (EPA-TRI), Environmental Protection Agency's Pollution Prevent (EPA-P2) database, and Glassdoor reviews (GD), Centre for Research in Security Prices (CRSP), Transcripts: Seeking Alpha (TSA), [71] (KPSS).

| Variable | Description | Source |
|---|---|---|
| *Dependent variables* | | |
| 1*(Abatement-Operating)* | Dummy variable indicating one if the firm undertakes operations-related abatement activities across any of its plants in the year, and zero otherwise | EPA-P2 |
| 1*(Abatement-Process)* | Dummy variable indicating one if the firm undertakes process-related abatement activities across any of its plants in the year, and zero otherwise | EPA-P2 |
| Pollution | Firm-level pollution obtained by aggregating emissions corresponding to various chemicals across all plants of a firm in a given year. In our analysis, we aggregate the pollution across all media: air, water, and ground, to compute the total pollution. | EPA-TRI |
| #Patents[$t$,$t+x$] | The number of green patents granted to the firm in the [$t$,$t+x$] window (measured in years) after the earnings call occurring on date $t$. Using information of the Cooperative Patent Classification from the USPTO, we identify the green patents following the guidelines from OECD. | KPSS |
| Overall Rating | The average employee overall ratings in a given calendar year-quarter. | GD |

| Variable | Description | Source |
|---|---|---|
| CAR[−3,+3] | Seven-day cumulative abnormal return centered on the earnings conference call date, calculated using the market model. | CRSP |
| *Key independent variables* | | |
| Level(E+S) | The total frequency of environmental and social phrases in the earnings conference call scaled by the total number of words in the earnings call. | TSA |
| Level(E) | The total frequency of environmental phrases in the earnings conference call scaled by the total number of words in the earnings call. | TSA |
| Level(S) | The total frequency of social phrases in the earnings conference call scaled by the total number of words in the earnings call. | TSA |
| Level(E_Material) | The total frequency of environmental phrases which are material for the firm's industry in its earnings conference call scaled by the total number of words in the earnings call. | TSA |
| Level(E_Immaterial) | The total frequency of environmental phrases which are immaterial for the firm's industry in its earnings conference call scaled by the total number of words in the earnings call. | TSA |

| Variable | Description | Source |
|---|---|---|
| Level(S_Material) | The total frequency of social phrases which are material for the firm's industry in its earnings conference call scaled by the total number of words in the earnings call. | TSA |
| Level(S_Immaterial) | The total frequency of social phrases which are immaterial for the firm's industry in its earnings conference call scaled by the total number of words in the earnings call. | TSA |
| *Firm-level control variables* | | |
| Earnings surprise | Actual earnings per share (EPS) from IBES minus the consensus (median) of EPS forecasts issued or reviewed in 90 days before the earnings announcement date. The difference is scaled by stock price at the end of quarter. | IBES |
| HighUE | Dummy variable which equals to 1 if the earnings surprise is in the highest decile in a given quarter, 0 otherwise. | IBES |
| LowUE | Dummy variable which equals to 1 if the earnings surprise is in the lowest decile in a given quarter, 0 otherwise. | IBES |
| Negative EPS | Dummy variable which equals 1 if announced earnings per share (EPS) is negative, 0 otherwise. | IBES |

| Variable | Description | Source |
|---|---|---|
| Pre-event return | Average stock return in window [−71,−11] in terms of trading days relative to the earnings conference call date. | CRSP |
| Pre-event volume | Average trading volume in window [−71,−11] in terms of trading days relative to the earnings conference call date. | CRSP |
| ROA | Earnings before extraordinary items divided by total assets. | COMPUSTAT |
| Accruals | Earnings minus cash flows from operations scaled by total assets. | COMPUSTAT |
| Size | Natural logarithm of the market cap at the end of quarter. | COMPUSTAT |
| MTB | Market cap plus book value of liabilities scaled by total assets at the end of quarter. | COMPUSTAT |
| Leverage | Long-term debt divided by total assets. | COMPUSTAT |
| NetPPEA | Net property, plant and equipment scaled by total assets | COMPUSTAT |
| Earnings volatility | Standard deviation of earnings-to-assets ratio in the past 20 quarters, with a minimum of 8 quarters required. | COMPUSTAT |
| Return volatility | Standard deviation of monthly stock returns in the last 12 months, with a minimum of 6 months required. | CRSP |

| Variable | Description | Source |
|---|---|---|
| # Analysts (log) | Natural logarithm of the number of analysts whose forecasts were issued or reviewed in the 90 days before the earnings announcement date. | IBES |
| Firm age (log) | Natural logarithm of the number of years since the stock's first date in CRSP. | CRSP |
| PreROA | Earnings before extraordinary items change of current quarter ($q$) over one year ago ($q-4$), scaled by total assets of one year ago ($q-4$). | COMPUSTAT |
| PreSale | Sales change of current quarter ($q$) over one year ago ($q-4$), scaled by total assets of one year ago ($q-4$). | COMPUSTAT |
| *Call Transcript Controls* | | |
| Uncertainty | Percentage of uncertain words in earnings call transcript based on the dictionary and code from Loughran-McDonald website. | TSA |
| Sentiment | Percentage of positive words minus percentage of negative words in earnings call transcript based on the dictionary and code from Loughran-McDonald website. | TSA |

## **Table 4.1:** Summary Statistics

This table reports the summary statistics for key variables in our analysis during the sample period. Panel A shows the transcript-level frequency and fraction of E&S phrases in earnings call transcripts during 2007-2019. In Panel B, we report the distribution of annual firm pollution from the Toxic Releases Inventory database of the EPA during 2007-2018. Panel C shows the number of green patents granted in the 1-year window after the earnings call date in the sample used in green innovation analysis. Panel D provides the statistics for overall employee rating, as obtained from Glassdoor reviews for the sample period of 2008-2019.

**Panel A: Transcript-Level E&S Discussion**

| | Count | Mean | Median | Std. Dev. |
|---|---|---|---|---|
| *Frequency of E&S Phrases* | | | | |
| Freq(E+S) | 132,295 | 3.662 | 2.000 | 6.599 |
| Freq(E) | 132,295 | 2.444 | 1.000 | 5.860 |
| Freq(S) | 132,295 | 1.218 | 0.000 | 2.852 |
| Freq(E-Material) | 132,295 | 0.814 | 0.000 | 2.643 |
| Freq(E-Immaterial) | 132,295 | 1.630 | 0.000 | 4.549 |
| Freq(S-Material) | 132,295 | 0.566 | 0.000 | 1.849 |
| Freq(S-Immaterial) | 132,295 | 0.653 | 0.000 | 1.929 |
| *Fraction of E&S Discussion (in %)* | | | | |
| Level(E+S) | 132,295 | 0.051 | 0.024 | 0.090 |
| Level(E) | 132,295 | 0.034 | 0.010 | 0.082 |
| Level(S) | 132,295 | 0.016 | 0.000 | 0.037 |
| Level(E-Material) | 132,295 | 0.011 | 0.000 | 0.038 |
| Level(E-Immaterial) | 132,295 | 0.023 | 0.000 | 0.063 |
| Level(S-Material) | 132,295 | 0.008 | 0.000 | 0.024 |
| Level(S-Immaterial) | 132,295 | 0.009 | 0.000 | 0.026 |

**Panel B: Annual Firm Pollution (in pounds)**

| | Count | Mean | Median | Std. Dev. |
|---|---|---|---|---|
| Total Pollution | 9,103 | 2,583,485 | 30,140 | 11,636,114 |
| 1*(Abatement-Operating)* | 9,081 | 0.24 | 0 | 0.42 |
| 1*(Abatement-Process)* | 9,081 | 0.23 | 0 | 0.42 |

**Panel C: Number of Green Patents Granted**

| | Count | Mean | Median | Std. Dev. |
|---|---|---|---|---|
| Overall | 32,074 | 2.236 | 0.000 | 8.597 |
| Env&Water | 32,074 | 0.465 | 0.000 | 2.159 |
| Climate | 32,074 | 1.850 | 0.000 | 7.073 |

**Panel D: Quarterly Glassdoor Employee Reviews**

| | Count | Mean | Median | Std. Dev. |
|---|---|---|---|---|
| Overall Rating | 41,387 | 3.15 | 3.17 | 0.75 |
| Career Opp. | 41,307 | 3.03 | 3.00 | 0.71 |
| Comp.+Benefits | 41,305 | 3.16 | 3.18 | 0.76 |
| Senior Mgmt. | 32,548 | 2.78 | 2.79 | 0.76 |
| Work-life Balance | 41,317 | 3.24 | 3.25 | 0.75 |
| Culture | 41,311 | 3.08 | 3.07 | 0.80 |

**Table 4.2:** Top E&S Phrases In Earnings Conference Calls

This table reports the Top 20 environmental and social phrases based on their frequency shown in the earnings call transcripts during the sample period. Panel A shows the top environmental phrases with their frequency and the number of earnings call transcripts with their occurrence. Panel B reports the top phrases in social dimension.

| | Frequency | # of Transcripts | % of Transcripts |
|---|---|---|---|
| ***Panel A: Top Environmental Phrases*** | | | |
| Sustainable | 66,570 | 37,798 | 28.57% |
| Environmental | 30,269 | 12,878 | 9.73% |
| Ecosystem | 17,889 | 8,004 | 6.05% |
| Renewable | 16,164 | 5,289 | 4.00% |
| $CO_2$ | 7,481 | 1,780 | 1.35% |
| Power generation | 6,846 | 3,194 | 2.41% |
| EPA | 6,665 | 2,537 | 1.92% |
| Oil field | 6,275 | 2,566 | 1.94% |
| Asphalt | 6,154 | 953 | 0.72% |
| Renewables | 5,874 | 2,001 | 1.51% |
| Fuel cell | 4,363 | 422 | 0.32% |
| Our water | 4,258 | 2,239 | 1.69% |
| Flooding | 4,231 | 2,519 | 1.90% |
| Landfill | 4,095 | 733 | 0.55% |
| Recycle | 3,638 | 2,400 | 1.81% |
| Natural resource | 3,465 | 1,463 | 1.11% |
| Clean energy | 3,250 | 1,107 | 0.84% |
| Water treatment | 3,227 | 1,353 | 1.02% |
| Hydrocarbon | 2,419 | 1,359 | 1.03% |
| Gas plant | 2,275 | 1,295 | 0.98% |
| ***Panel B: Top Social Phrases*** | | | |
| Affordable | 14,180 | 8,259 | 6.24% |
| Minority | 10,340 | 6,479 | 4.90% |
| Demographic | 7,061 | 5,056 | 3.82% |
| Demographics | 6,652 | 5,149 | 3.89% |
| Affordability | 6,492 | 3,108 | 2.35% |
| HIV | 5,979 | 852 | 0.64% |
| Accountability | 5,040 | 3,652 | 2.76% |
| Our community | 4,502 | 2,566 | 1.94% |
| Our communities | 4,370 | 2,431 | 1.84% |
| Employee benefit | 4,267 | 2,728 | 2.06% |
| Public service | 4,124 | 1,498 | 1.13% |
| Disability | 3,749 | 1,155 | 0.87% |
| Minimum wage | 3,288 | 1,533 | 1.16% |
| Cyber security | 2,834 | 1,448 | 1.09% |
| Food safety | 2,707 | 848 | 0.64% |
| Infectious disease | 2,674 | 1,216 | 0.92% |
| Health and wellness | 2,175 | 1,195 | 0.90% |
| Hazard | 1,896 | 1,205 | 0.91% |
| CDC | 1,782 | 856 | 0.65% |
| Fairness | 1,778 | 1,549 | 1.17% |

**Table 4.3:** Firm Pollution and Discussion on Environmental and Social Issues

This table reports the results from Equation (4.1) using yearly firm pollution from the EPA Toxic Release Inventory (TRI) as the dependent variable during 2007-2019. *Level(E+S)* corresponds to the fraction of words devoted to discussion of environmental and social topics in the earnings conference calls during the year. *Level(E)* denotes the fraction of words devoted to discussion of environmental topics in the earnings conference calls during the year. In Panel A, we show regression results using *Level(E+S)* as the explanatory variable. Panel B shows results using *Level(E)* as the independent variable. In Panel C, we show results using forward years of total pollution as the dependent variable. Panel D reports results using forward years of air pollution as the dependent variable. Industry definition is based on Fama-French 30 classification. The T-statistics (in brackets) are based on double clustering standard errors at the firm and year levels, unless otherwise specified. * $p < 0.10$, ** $p < 0.05$, *** $p < 0.01$

**Panel A: Using Environmental & Social Discussion**

| *Dependent Variable*: | Log (1+Pollution/COGS$_{t-1}$) | | | |
|---|---|---|---|---|
| | (1) | (2) | (3) | (4) |
| Level(E+S) | -81.02*** | -75.13*** | -68.59*** | |
| | [-3.63] | [-3.35] | [-3.46] | |
| Level(E+S-Material) | | | | -96.80** |
| | | | | [-2.34] |
| Level(E+S-Immaterial) | | | | -58.36** |
| | | | | [-2.22] |
| Leverage $_{t-1}$ | | 0.24 | 0.19 | 0.19 |
| | | [1.54] | [1.28] | [1.28] |
| Log(Assets) $_{t-1}$ | | -0.39*** | -0.40*** | -0.40*** |
| | | [-5.44] | [-5.68] | [-5.69] |
| NetPPEA $_{t-1}$ | | 0.16 | 0.16 | 0.16 |
| | | [0.52] | [0.51] | [0.50] |
| Firm FE | ✓ | ✓ | ✓ | ✓ |
| Year FE | ✓ | ✓ | | |
| Controls | | ✓ | ✓ | ✓ |
| Industry-Year FE | | | ✓ | ✓ |
| Adj.-$R^2$ | 0.936 | 0.938 | 0.940 | 0.940 |
| Obs. | 9,001 | 9,001 | 8,967 | 8,967 |

**Panel B: Using Only Environmental Discussion**

| *Dependent Variable*: | Log (1+Pollution/COGS$_{t-1}$) | | | |
|---|---|---|---|---|
| | (1) | (2) | (3) | (4) |
| Level(E) | -76.48*** | -70.06** | -65.31** | |
| | [-3.11] | [-2.81] | [-2.91] | |
| Level(E-Material) | | | | -104.32** |
| | | | | [-2.33] |
| Level(E-Immaterial) | | | | -52.34* |
| | | | | [-1.85] |
| Firm FE | ✓ | ✓ | ✓ | ✓ |
| Year FE | ✓ | ✓ | | |
| Controls | | ✓ | ✓ | ✓ |
| Industry-Year FE | | | ✓ | ✓ |
| Adj.-$R^2$ | 0.936 | 0.938 | 0.940 | 0.940 |
| Obs. | 9,001 | 9,001 | 8,967 | 8,967 |

**Panel C: Using Forward Years with Environmental & Social Discussion**

| *Dependent Variable*: | Log (1+Poll.$_{t+1}$/COGS$_{t-1}$) | Log (1+Poll.$_{t+2}$/COGS$_{t-1}$) | Log (1+Poll.$_{t+3}$/COGS$_{t-1}$) |
|---|---|---|---|
| | (1) | (2) | (3) |
| Level(E+S) | -72.71*** | -55.19** | -27.16 |
| | [-3.77] | [-2.77] | [-1.36] |
| Firm FE | ✓ | ✓ | ✓ |
| Industry-Year FE | ✓ | ✓ | ✓ |
| Controls | ✓ | ✓ | ✓ |
| Adj.-$R^2$ | 0.943 | 0.947 | 0.949 |
| Obs. | 8,343 | 7,696 | 7,035 |

**Panel D: Using Forward Years with Environmental Discussion**

| *Dependent Variable*: | Log (1+Poll.$_{t+1}$/COGS$_{t-1}$) | Log (1+Poll.$_{t+2}$/COGS$_{t-1}$) | Log (1+Poll.$_{t+3}$/COGS$_{t-1}$) |
|---|---|---|---|
| | (1) | (2) | (3) |
| Level(E) | -68.21*** | -51.76** | -25.72 |
| | [-3.27] | [-2.46] | [-1.28] |
| Firm FE | ✓ | ✓ | ✓ |
| Industry-Year FE | ✓ | ✓ | ✓ |
| Controls | ✓ | ✓ | ✓ |
| Adj.-$R^2$ | 0.943 | 0.947 | 0.949 |
| Obs. | 8,343 | 7,696 | 7,035 |

**Table 4.4:** Abatement of Pollution

This table reports the results from Equation (4.1) using EPA Pollution Prevention data on firm pollution abatement during 2007-2019. The dependent variables are indicators for whether a firm undertakes operations- or process-related abatement activities across any of its plants in the year. *Level(E+S)* corresponds to the fraction of words devoted to discussion of environmental and social topics in the earnings conference calls during the year. Whereas, *Level(E)* denotes the fraction of words devoted to discussion of environmental topics in the earnings conference calls during the year. *Level(E-Material)* denotes the fraction of words devoted to the discussion of environmental topics material to the industry in the earnings conference calls during the year. Similarly, *Level(E-Immaterial)* corresponds to the discussion of remaining environmental topics. Industry definition is based on Fama-French 30 classification. The T-statistics (in brackets) are based on double clustering standard errors at the firm and year levels, unless otherwise specified. $^{*}$ $p < 0.10$, $^{**}$ $p < 0.05$, $^{***}$ $p < 0.01$

| *Dependent Variable*: | 1*(Abatement-Operating)* | | | 1*(Abatement-Process)* | | |
|---|---|---|---|---|---|---|
| | (1) | (2) | (3) | (4) | (5) | (6) |
| Level(E+S) | 8.83$^{**}$ | | | 8.24 | | |
| | [2.50] | | | [1.24] | | |
| Level(E) | | 9.81$^{**}$ | | | 7.59 | |
| | | [2.47] | | | [1.05] | |
| Level(E-Material) | | | 31.90$^{**}$ | | | 30.26$^{*}$ |
| | | | [2.43] | | | [1.93] |
| Level(E-Immaterial) | | | 2.32 | | | -0.09 |
| | | | [0.66] | | | [-0.01] |
| Leverage $_{t-1}$ | -0.01 | -0.01 | -0.01 | -0.10$^{**}$ | -0.10$^{**}$ | -0.10$^{**}$ |
| | [-0.20] | [-0.20] | [-0.17] | [-2.76] | [-2.76] | [-2.75] |
| Log(Assets) $_{t-1}$ | -0.01 | -0.01 | -0.01 | 0.01 | 0.01 | 0.01 |
| | [-0.70] | [-0.70] | [-0.74] | [0.87] | [0.87] | [0.83] |
| NetPPEA $_{t-1}$ | 0.09 | 0.09 | 0.09 | 0.05 | 0.05 | 0.05 |
| | [1.13] | [1.13] | [1.14] | [0.64] | [0.63] | [0.65] |
| Firm FE | ✓ | ✓ | ✓ | ✓ | ✓ | ✓ |
| Industry-Year FE | ✓ | ✓ | ✓ | ✓ | ✓ | ✓ |
| Controls | ✓ | ✓ | ✓ | ✓ | ✓ | ✓ |
| Adj.-$R^2$ | 0.452 | 0.452 | 0.452 | 0.387 | 0.387 | 0.387 |
| Obs. | 9,035 | 9,035 | 9,035 | 9,035 | 9,035 | 9,035 |

**Table 4.5:** Discussion of E&S Topics and Future Green Patenting Activity

This table presents results exploring whether the greater discussion of environmental and social topics in earnings calls is associated with the ex post granting of a higher number of green patents. The regressions are estimated using a Poisson model. The dependent variable in Columns (1)-(3) is the number of overall green patents granted after the earnings calls ($Overall$). The dependent variable in Columns (4)-(6) is the the number of green patents which are related to environmental management and water-related adaptation technologies ($Env\&Water$). The dependent variable in Columns (7)-(9) is the the number of green patents about climate change mitigation ($Climate$). In Panel A, we use the green patents which are granted during the one-year window after the earnings call date. Similarly, Panel B (Panel C) shows the number of green patents which are granted during the two-year (three-year) window after the earnings call date. The number of green patents and control variables are winsorized at 1% and 99%. All variables are defined in Table A1. The Z-statistics (in brackets) are based on double clustering standard errors at the firm and year-quarter levels, unless otherwise specified. *, **, and *** indicate significance at the 10%, 5%, and 1% levels, respectively.

**Panel A: 1-Year Window**

| *Dependent Variable*: | Overall | | | Env&Water | | | Climate | | |
|---|---|---|---|---|---|---|---|---|---|
| | (1) | (2) | (3) | (4) | (5) | (6) | (7) | (8) | (9) |
| Level(E+S) | 168.03*** | | | 319.59*** | | | 143.43** | | |
| | (2.82) | | | (3.89) | | | (2.45) | | |
| Level(E) | | 193.84*** | | | 366.85*** | | | 168.96*** | |
| | | (3.05) | | | (4.52) | | | (2.74) | |
| Level(E-Material) | | | 338.03** | | | 700.65*** | | | 293.43* |
| | | | (2.08) | | | (4.18) | | | (1.84) |
| Level(E-Immaterial) | | | 132.62** | | | 212.27** | | | 117.95* |
| | | | (2.29) | | | (2.44) | | | (1.93) |
| Ind. × Year-Qtr. FE | ✓ | ✓ | ✓ | ✓ | ✓ | ✓ | ✓ | ✓ | ✓ |
| Controls | ✓ | ✓ | ✓ | ✓ | ✓ | ✓ | ✓ | ✓ | ✓ |
| Pseudo $R^2$ | 0.587 | 0.588 | 0.589 | 0.531 | 0.534 | 0.537 | 0.571 | 0.572 | 0.572 |
| Obs. | 32,074 | 32,074 | 32,074 | 29,310 | 29,310 | 29,310 | 31,884 | 31,884 | 31,884 |

**Panel B: 2-Year Window**

| *Dependent Variable*: | Overall | | | Env&Water | | | Climate | | |
|---|---|---|---|---|---|---|---|---|---|
| | (1) | (2) | (3) | (4) | (5) | (6) | (7) | (8) | (9) |
| Level(E+S) | 170.79*** | | | 319.23*** | | | 150.13** | | |
| | (2.81) | | | (3.82) | | | (2.49) | | |
| Level(E) | | 195.80*** | | | 367.64*** | | | 173.90*** | |
| | | (3.04) | | | (4.47) | | | (2.76) | |
| Level(E-Material) | | | 341.30** | | | 689.66*** | | | 302.56* |
| | | | (2.05) | | | (4.17) | | | (1.83) |
| Level(E-Immaterial) | | | 132.62** | | | 213.10** | | | 119.75* |
| | | | (2.27) | | | (2.46) | | | (1.93) |
| Ind. × Year-Qtr. FE | ✓ | ✓ | ✓ | ✓ | ✓ | ✓ | ✓ | ✓ | ✓ |
| Controls | ✓ | ✓ | ✓ | ✓ | ✓ | ✓ | ✓ | ✓ | ✓ |
| Pseudo $R^2$ | 0.619 | 0.619 | 0.620 | 0.565 | 0.568 | 0.571 | 0.609 | 0.610 | 0.610 |
| Obs. | 32,593 | 32,593 | 32,593 | 30,642 | 30,642 | 30,642 | 32,517 | 32,517 | 32,517 |

**Panel C: 3-Year Window**

| *Dependent Variable*: | Overall | | | Env&Water | | | Climate | | |
|---|---|---|---|---|---|---|---|---|---|
| | (1) | (2) | (3) | (4) | (5) | (6) | (7) | (8) | (9) |
| Level(E+S) | 169.28*** | | | 313.05*** | | | 151.48** | | |
| | (2.71) | | | (3.63) | | | (2.43) | | |
| Level(E) | | 194.63*** | | | 362.26*** | | | 175.11*** | |
| | | (2.95) | | | (4.28) | | | (2.69) | |
| Level(E-Material) | | | 337.46** | | | 682.19*** | | | 303.56* |
| | | | (1.99) | | | (4.08) | | | (1.81) |
| Level(E-Immaterial) | | | 131.14** | | | 203.27** | | | 119.81* |
| | | | (2.16) | | | (2.30) | | | (1.86) |
| Ind. × Year-Qtr. FE | ✓ | ✓ | ✓ | ✓ | ✓ | ✓ | ✓ | ✓ | ✓ |
| Controls | ✓ | ✓ | ✓ | ✓ | ✓ | ✓ | ✓ | ✓ | ✓ |
| Pseudo $R^2$ | 0.628 | 0.629 | 0.630 | 0.584 | 0.586 | 0.589 | 0.619 | 0.620 | 0.620 |
| Obs. | 32,686 | 32,686 | 32,686 | 31,505 | 31,505 | 31,505 | 32,644 | 32,644 | 32,644 |

**Table 4.6:** Employee Ratings and Discussion on Environmental and Social Topics

This table reports the results from Equation (4.1) using overall quarterly ratings from Glassdoor as the dependent variable during 2007-2019. *Level(E+S)* corresponds to the fraction of words devoted to discussion of environmental and social topics in the earnings conference calls during the year-quarter. Whereas, *Level(S)* denotes the fraction of words devoted to discussion of social topics in the earnings conference calls during the year-quarter. *Level(S-Material)* denotes the fraction of words devoted to the discussion of social topics material to the industry in the earnings conference calls during the year-quarter. Similarly, *Level(S-Immaterial)* corresponds to the discussion of remaining social topics. $\text{Rating}_{t+i}$ denotes the overall rating in the $i^{th}$ year-quarter ahead. In Panel A, we show regression results using *Level(E+S)* as the main explanatory variable. Panel B shows results using *Level(S)* as the main independent variable. In Panel C, we analyze different components of the overall rating. Panel D provides regression results based on the industry-specific material discussion across different components of overall rating. Industry definition is based on Fama-French 30 classification. The T-statistics (in brackets) are based on double clustering standard errors at the firm and year-quarter levels, unless otherwise specified. * $p < 0.10$, ** $p < 0.05$, *** $p < 0.01$

## Panel A: Using Environmental & Social Discussion

| *Dependent Variable*: | Overall Rating | | | | | |
|---|---|---|---|---|---|---|
| | $\text{Rating}_t$ | | | $\text{Rating}_{t+1}$ | $\text{Rating}_{t+2}$ | $\text{Rating}_{t+3}$ |
| | (1) | (2) | (3) | (4) | (5) | (6) |
| Level(E+S) | 25.26*** | 24.76*** | 24.60*** | 17.66** | 6.28 | -3.11 |
| | [2.84] | [2.80] | [2.88] | [2.05] | [0.82] | [-0.33] |
| $\text{Log(Assets}_{t-1})$ | | 0.03 | 0.01 | 0.01 | 0.00 | 0.01 |
| | | [1.44] | [0.78] | [0.72] | [0.24] | [0.54] |
| Log(MTB Ratio) | | 0.10*** | 0.10*** | 0.10*** | 0.07*** | 0.06*** |
| | | [4.87] | [4.58] | [4.49] | [3.98] | [2.69] |
| Leverage | | -0.05 | -0.05 | -0.09** | -0.07 | -0.01 |
| | | [-1.41] | [-1.36] | [-2.12] | [-1.63] | [-0.38] |
| Return on Assets | | 0.47** | 0.42 | 0.81*** | 0.46** | 0.24 |
| | | [2.06] | [1.66] | [3.61] | [2.07] | [1.09] |
| $\text{Return}_{t-1}$ | | 0.03 | 0.03 | 0.09*** | 0.04* | 0.01 |
| | | [1.34] | [1.56] | [4.05] | [2.00] | [0.63] |
| Firm FE | ✓ | ✓ | ✓ | ✓ | ✓ | ✓ |
| Year-Qtr. FE | ✓ | ✓ | | | | |
| Controls | | ✓ | ✓ | ✓ | ✓ | ✓ |
| Industry × Year-Qtr. FE | | | ✓ | ✓ | ✓ | ✓ |
| Adj.-$R^2$ | 0.276 | 0.277 | 0.283 | 0.285 | 0.282 | 0.287 |
| Obs. | 41,267 | 41,267 | 41,199 | 41,139 | 40,884 | 40,401 |

**Panel B: Using Only Social Discussion**

| *Dependent Variable*: | Overall Rating | | | | | |
|---|---|---|---|---|---|---|
| | $\text{Rating}_t$ | | | $\text{Rating}_{t+1}$ | $\text{Rating}_{t+2}$ | $\text{Rating}_{t+3}$ |
| | (1) | (2) | (3) | (4) | (5) | (6) |
| Level(S) | 28.18* | 27.87* | 29.73* | 15.79 | 9.89 | -0.23 |
| | [1.83] | [1.82] | [1.99] | [1.03] | [0.69] | [-0.02] |
| Firm FE | ✓ | ✓ | ✓ | ✓ | ✓ | ✓ |
| Year-Qtr. FE | ✓ | ✓ | | | | |
| Controls | | ✓ | ✓ | ✓ | ✓ | ✓ |
| Industry × Year-Qtr. FE | | | ✓ | ✓ | ✓ | ✓ |
| Adj.-$R^2$ | 0.276 | 0.277 | 0.283 | 0.285 | 0.282 | 0.287 |
| Obs. | 41,267 | 41,267 | 41,199 | 41,139 | 40,884 | 40,401 |

**Panel C: Using Only Environmental Discussion**

| *Dependent Variable*: | Overall Rating | | | | | |
|---|---|---|---|---|---|---|
| | $\text{Rating}_t$ | | | $\text{Rating}_{t+1}$ | $\text{Rating}_{t+2}$ | $\text{Rating}_{t+3}$ |
| | (1) | (2) | (3) | (4) | (5) | (6) |
| Level(E) | 25.29** | 24.66** | 23.13* | 19.46* | 4.66 | -4.81 |
| | [2.12] | [2.07] | [1.97] | [1.86] | [0.50] | [-0.39] |
| Firm FE | ✓ | ✓ | ✓ | ✓ | ✓ | ✓ |
| Qtr. FE | ✓ | ✓ | | | | |
| Industry-Qtr. FE | | | ✓ | ✓ | ✓ | ✓ |
| Controls | | ✓ | ✓ | ✓ | ✓ | ✓ |
| Adj.-$R^2$ | 0.276 | 0.277 | 0.283 | 0.285 | 0.282 | 0.287 |
| Obs. | 41,267 | 41,267 | 41,199 | 41,139 | 40,884 | 40,401 |

**Panel D: Components of Rating**

| *Dependent Variable*: | Overall Rating | Career Opp. | Comp.+Benefits | Senior Mgmt. | Work-life Ba |
|---|---|---|---|---|---|
| | (1) | (2) | (3) | (4) | (5) |
| Level(S) | 29.73* | 22.43 | 24.56* | 29.85 | 26.12** |
| | [1.99] | [1.45] | [1.95] | [1.67] | [2.14] |
| Firm FE | ✓ | ✓ | ✓ | ✓ | ✓ |
| Controls | ✓ | ✓ | ✓ | ✓ | ✓ |
| Industry × Year-Qtr. FE | ✓ | ✓ | ✓ | ✓ | ✓ |
| Adj.-$R^2$ | 0.283 | 0.276 | 0.378 | 0.277 | 0.307 |
| Obs. | 41,199 | 41,118 | 41,116 | 32,403 | 41,128 |

**Panel E: Based on Industry-Specific Material Discussion**

| *Dependent Variable*: | Overall Rating | Career Opp. | Comp.+Benefits | Senior Mgmt. | Work-life Ba |
|---|---|---|---|---|---|
| | (1) | (2) | (3) | (4) | (5) |
| Level(S-Material) | 28.29 | 25.31 | 34.33** | 37.85 | 39.29** |
| | [1.30] | [1.09] | [2.23] | [1.66] | [2.73] |
| Level(S-Immaterial) | 31.61 | 18.68 | 11.84 | 18.80 | 8.96 |
| | [1.41] | [0.90] | [0.59] | [0.72] | [0.45] |
| Firm FE | ✓ | ✓ | ✓ | ✓ | ✓ |
| Controls | ✓ | ✓ | ✓ | ✓ | ✓ |
| Industry × Year-Qtr. FE | ✓ | ✓ | ✓ | ✓ | ✓ |
| Adj.-$R^2$ | 0.283 | 0.276 | 0.378 | 0.277 | 0.307 |
| Obs. | 41,199 | 41,118 | 41,116 | 32,403 | 41,128 |

**Table 4.7:** Firms That Do Not "Walk the Talk" on E&S Issues

This table reports the results from Equation (4.1) using yearly firm pollution from the EPA Toxic Release Inventory (TRI) as the dependent variable during 2007-2019. *Level(E+S)* (*Level(E)*) corresponds to the fraction of words devoted to discussion of environmental and social (only environmental) topics in the earnings conference calls during the year. For each year, we sort firms into 10 groups with the average non-answer ratio of their earnings calls happening in that year. The dummy variable, $NoAnswer$, equals to one if the firm belongs to the top 10% in terms of the non-answer ratio for that year. In Column (3), we distinguish the E&S discussions discussions shown in the non-answer interactions ($Level(E+S)--NoAnswer$) from the sustainability talk in the answered interactions ($Level(E+S)--Answer$) of the earnings calls. Similar test is in Column (4) when we focus on the environmental discussions. Industry definition is based on Fama-French 30 classification. The T-statistics (in brackets) are based on double clustering standard errors at the firm and year levels, unless otherwise specified. $^{*}$ $p<0.10$, $^{**}$ $p<0.05$, $^{***}$ $p<0.01$

| **Dependent Variable**: | Total Pollution | | | |
|---|---|---|---|---|
| | (1) | (2) | (3) | (4) |
| NoAnswer = 1 | -0.00 | 0.01 | | |
| | (-0.04) | (0.18) | | |
| Level(E+S) | -66.41** | | | |
| | (-2.86) | | | |
| Level(E+S) × NoAnswer | 36.27** | | | |
| | (2.90) | | | |
| Level(E) | | -61.01** | | |
| | | (-2.36) | | |
| Level(E) × NoAnswer | | 31.15** | | |
| | | (2.43) | | |
| Level(E+S)–Answer | | | -57.32** | |
| | | | (-2.39) | |
| Level(E+S)–NoAnswer | | | -106.22 | |
| | | | (-1.71) | |
| Level(E)–Answer | | | | -59.16** |
| | | | | (-2.48) |
| Level(E)–NoAnswer | | | | -111.20 |
| | | | | (-1.49) |
| Controls | ✓ | ✓ | ✓ | ✓ |
| Firm FE | ✓ | ✓ | ✓ | ✓ |
| Industry-Year FE | ✓ | ✓ | ✓ | ✓ |
| Adj.-$R^2$ | 0.945 | 0.945 | 0.945 | 0.940 |
| Obs. | 7,318 | 7,318 | 7,318 | 8,967 |

**Table 4.8:** Pollution and Paris Agreement

This table reports the results based on a modified version of Equation (4.1) using yearly firm pollution from the EPA TRI as the dependent variable during 2007-2019. We interact the equation with dummies corresponding to periods around the inclusion of the US into the Paris Agreement. We additionally control for group-year fixed effects. *Level(E)* denotes the fraction of words devoted to the discussion of environmental topics. The dummy *Before Paris* corresponds to the years before the Paris Agreement up to 2015. *During Paris* refers to years during which the United States was part of the Paris Agreement in 2016 and 2017. Finally, *Paris Withdrawal* indicates years after the withdrawal from the Paris Agreement during 2018 and 2019. Industry definition is based on Fama-French 30 classification. The T-statistics (in brackets) are based on double clustering standard errors at the firm and year levels, unless otherwise specified. * $p < 0.10$, ** $p < 0.05$, *** $p < 0.01$

| *Dependent Variable*: | Log (1+Pollution/COGS$_{t-1}$) | | |
|---|---|---|---|
| | (1) | (2) | (3) |
| Before Paris × Level(E) | -36.32* | | -32.88* |
| | [-2.10] | | [-1.90] |
| During Paris × Level(E) | -71.69** | 4.42 | -63.81* |
| | [-2.25] | [0.13] | [-2.06] |
| Paris Withdrawal × Level(E) | | -28.03 | -96.92** |
| | | [-0.66] | [-2.84] |
| Diff.(*Before-During*) | 35.37 | | 30.93 |
| P-value | 0.25 | | 0.31 |
| Diff(*Before-Withdrawal*) | | | 64.05 |
| P-value | | | 0.06 |
| Diff(*During-Withdrawal*) | | 32.45 | 33.11 |
| P-value | | 0.36 | 0.05 |
| Firm FE | ✓ | ✓ | ✓ |
| Industry-Year FE | ✓ | ✓ | ✓ |
| Controls | ✓ | ✓ | ✓ |
| Group-Year FE | ✓ | ✓ | ✓ |
| Adj.-$R^2$ | 0.946 | 0.963 | 0.942 |
| Obs. | 7,711 | 2,574 | 8,967 |

**Table 4.9:** Robustness to Log(1+Pollution/COGS) as dependent variable

This table reports the results from Equation (4.1) using yearly firm pollution from the EPA Toxic Release Inventory (TRI) as the dependent variable during 2007-2019. In Panel A, we show robustness to alternative scaling of the pollution. First, in Column (1), we report results from scaling pollution by the same year's cost of goods sold. In Column (2), we us the previous year's revenue as the scaling variable. We use the lagged value of sale of goods to rescale pollution. Finally, Column (4) shows results after rescaling by total assets in the previous year. Panel B shows results using alternative specifications. We do not scale the dependent variable in Column (1). In Column (2), we omit the one plus transformation to report results from non-zero pollution observations. We use inverse hyperbolic sine (IHS) transformation for the dependent variable in Column (3). Columns (4)-(5) report results from Poisson regressions using the dependent variables as shown. The T-statistics (in brackets) are based on double clustering standard errors at the firm and year levels, unless otherwise specified. * $p < 0.10$, ** $p < 0.05$, *** $p < 0.01$

**Panel A: Using Alternative Scaling**

| *Dependent Variable*: | Log (1+Total Pollution/Scaling Variable) | | | |
|---|---|---|---|---|
| *Scaling Variable*: | $COGS_t$ | $REV_{t-1}$ | $SALE_{t-1}$ | $TA_{t-1}$ |
| | (1) | (2) | (3) | (4) |
| Level(E+S) | -63.51*** | -68.82*** | -68.25*** | -54.75*** |
| | [-3.29] | [-3.85] | [-3.79] | [-3.76] |
| Firm FE | ✓ | ✓ | ✓ | ✓ |
| Industry-Year FE | ✓ | ✓ | ✓ | ✓ |
| Controls | ✓ | ✓ | ✓ | ✓ |
| Adj.-$R^2$ | 0.941 | 0.947 | 0.946 | 0.938 |
| Obs. | 8,975 | 9,046 | 8,968 | 9,059 |

**Panel B: Using Alternative Specifications**

| *Dependent Variable*: | Log (1+Total Pollution) | Log (Total Pollution) | IHS (Total Pollution) | Log (1+ Pollution) | Log (1+Pollution/ $COGS_{t-1}$) |
|---|---|---|---|---|---|
| | (1) | (2) | (3) | (4) | (5) |
| Level(E+S) | -38.59* | -48.62** | -37.75* | -3.42* | -8.42*** |
| | [-1.98] | [-2.53] | [-1.89] | [-1.87] | [-2.68] |
| Firm FE | ✓ | ✓ | ✓ | ✓ | ✓ |
| Industry-Year FE | ✓ | ✓ | ✓ | ✓ | ✓ |
| Controls | ✓ | ✓ | ✓ | ✓ | ✓ |
| Adj.-$R^2$ | 0.936 | 0.933 | 0.934 | | |
| Obs. | 8,967 | 8,342 | 8,967 | 8,687 | 8,687 |

**Table 4.10:** Stock Price Response to E&S Discussion

This table presents results for the immediate stock price response to the E&S discussion in earnings calls. The dependent variable across all columns is CAR[$-3$,+3], calculated using the market model. The key independent variables are $Level(E+S)$, $Level(E)$ and $Level(S)$, which are computed as the frequency of environmental and social phrases, environmental phrases, and social phrases in an earnings call scaled by the total number of words in the earnings call. The market reaction variable and control variables are winsorized at 1% and 99%. All control variables are described in Table A1. The T-statistics (in brackets) are based on double clustering standard errors at the firm and year-quarter levels, unless otherwise specified. *, **, and *** indicate significance at the 10%, 5%, and 1% levels, respectively.

| *Dependent Variable*: | CAR[$-3$,+3] | | | |
|---|---|---|---|---|
| | (1) | (2) | (3) | (4) |
| Level(E+S) | 1.10* | | | |
| | [1.72] | | | |
| Level(E) | | 1.56** | | 1.56** |
| | | [2.35] | | [2.37] |
| Level(S) | | | -0.27 | -0.34 |
| | | | [-0.21] | [-0.26] |
| Controls | ✓ | ✓ | ✓ | ✓ |
| Firm FE | ✓ | ✓ | ✓ | ✓ |
| Industry × Year-Qtr. FE | ✓ | ✓ | ✓ | ✓ |
| Adj.-$R^2$ | 0.143 | 0.143 | 0.143 | 0.143 |
| Obs. | 81,662 | 81,662 | 81,662 | 81,662 |

# Appendices

# APPENDIX A

# MISCELLANEOUS SECTION FOR CHAPTER 1

## A1.1 SEC Municipal Advisor Rule

As shown in Figure 1.1a, municipalities issue over USD 400 billion of municipal bonds each year to finance various infrastructure and public utility projects. However, these municipalities may often lack the financial sophistication to navigate the issuance process [3]. Under the Congressional mandate of June 1975, the MSRB has been charged with protecting investors to prevent fraud and financial irregularities in the municpial bond market[1]. Following the financial crisis of 2007-09, Congress passed the Dodd-Frank Wall Street Reform and Consumer Protection Act in 2010. This Act introduced major changes to the regulatory framework and operations of the financial service industry. It also included several provisions concerning SEC rulemaking.

Under this framework, the SEC drew up the Municipal Advisor Rule which became effective on July 1, 2014. Specifically, it became unlawful for a municipal advisor (MA) to provide advice to or on behalf of a municipal entity or obligated person with respect to municipal financial products or the issuance of municipal securities unless the MA is registered with the SEC. Prior to this reform, only broker-dealers and banks were subject to federal regulatory requirements. The SEC sought to mitigate some of the problems observed in the conduct of municipal advisors in the municipal bond market. This included the municipal advisor's failure to place the duty of loyalty to their clients ahead of their own interests [149]. Additionally, the SEC Commissioner noted in a statement in 2013[2]:

> Our dedicated public servants were relying on municipal advisors whose advisory activities generally did not require them to register with the Com-

[1] https://www.msrb.org/About-MSRB/About-the-MSRB/Creation-of-the-MSRB.aspx

[2] https://www.sec.gov/news/statement/2013-09-18-open-meeting-statement-kms

> mission, or any other federal, state, or self-regulatory entity. And a lack of meaningful regulation over these advisors created confusion, and in some instances, horrific abuses. Sadly, the shortcomings of this hands-off regulatory regime became glaringly apparent during the last several years as we learned about numerous examples of bad behavior, including self-dealing and excessive fees.

Importantly, advisors owe fiduciary responsibility to municipal clients under the MA Rule and cannot act in ways that my be unfavorable to their clients. This would encompass the twin requirements of duty of care, and duty of loyalty. According to the former, advisors must exert effort on behalf of the issuer in order to make a recommendation. The duty of loyalty requires advisors to uphold the interests of the issuer superior to their own. Failing to adhere to their fiduciary responsibility, advisors may be held liable for adverse outcomes to issuers during municipal bond issuance. The SEC MA Rule also clarified what constitutes municipal advice and therefore under what circumstances the registration requirements would be applicable. Broadly, the SEC documented "advice" as any recommendation particularized to specific needs, objectives, and circumstances of a municipal entity. Overall, the SEC Municipal Advisor Rule introduced a new set of standards applicable to municipal advisors, with an aim to help municipal issuers.

### A1.2 Advisor Names

There is substantial variation in the names used by FTSE Russell municipal bond database to record municipal advisors involved in the issuance of bonds. Often these variations are due to spellings like "BACKSTORM MCCARLEY BERRY & COMPANY LLC" vs "BACKSTROM MCCARLEY BERRY & COMPANY". I investigate the bond offer statement associated with the CUSIP to verify the corresponding entity and update the standardized name. Besides the jumbled spelling errors, I also come across typos and mistakes due to omitted letters. For example, "BERKSHIRE BANK" and "BERSHIRE BANK" correspond to the same municipal advisor. Further, I also account for alternative company name extensions, such as, “LLC”, “INC”, “Co.” or “CO.”. These extensions are recorded differently over time for the same company. Occasionally, the names would omit portions of the names altogether: “SUDSINA & ASSOCIATES” was also reported as “SUDSINA”. To verify against mismatching firm names, I rely on logos printed on the bond’s offering statement in addition to the official address provided.

Other instances of spelling differences involve special characters like “.” or “,” and “&”. Clearly, these are easy to handle and to resolve. I also need to update names for subsidiaries and affiliated company names. These would also often involve mergers and acquisitions. Following [59], I retroactively replace the names of the merged entities with the name of the new company. To an extent, this assumes that client relationships prevailed even after the M&A activity. Indeed I do find evidence where both sets of names are found in the offering statements. This could also be due to delay in sale of bonds, especially via negotiation. Thus, offer statements may reflect contractual engagements before or after the merger activity. As a result, I identify these entities under a common name: for example, “MERRILL LYNCH & COMPANY” and “BANC OF AMERICA SECURITIES” as “BANC OF AMERICA SECURITIES”. I also find limited anecdotal evidence suggesting that acquisitions involve retention of officials in the new company.

Still other challenges include names where firms operate using different brand names. For example, “SUSAN D. MUSSELMAN INC”, “SDM ADVISORS INC” and “DASHEN-MUSSELMAN INC” are captured differently. However, I represent them as “A DASHEN & ASSOCIATES” by checking the names of principals and office locations recorded under each company. I believe that it is very difficult to identify abbreviated names corresponding to the same umbrella company without verifying each name separately. In some cases, I need to ascertain the office locations from internet search versus that reported in the bond offering statement. But I also come across simpler associations where alternative trading names used by companies are seemingly related, such as, “RV NORENE & ASSOCIATE

INC" and "CROWE NORENE".

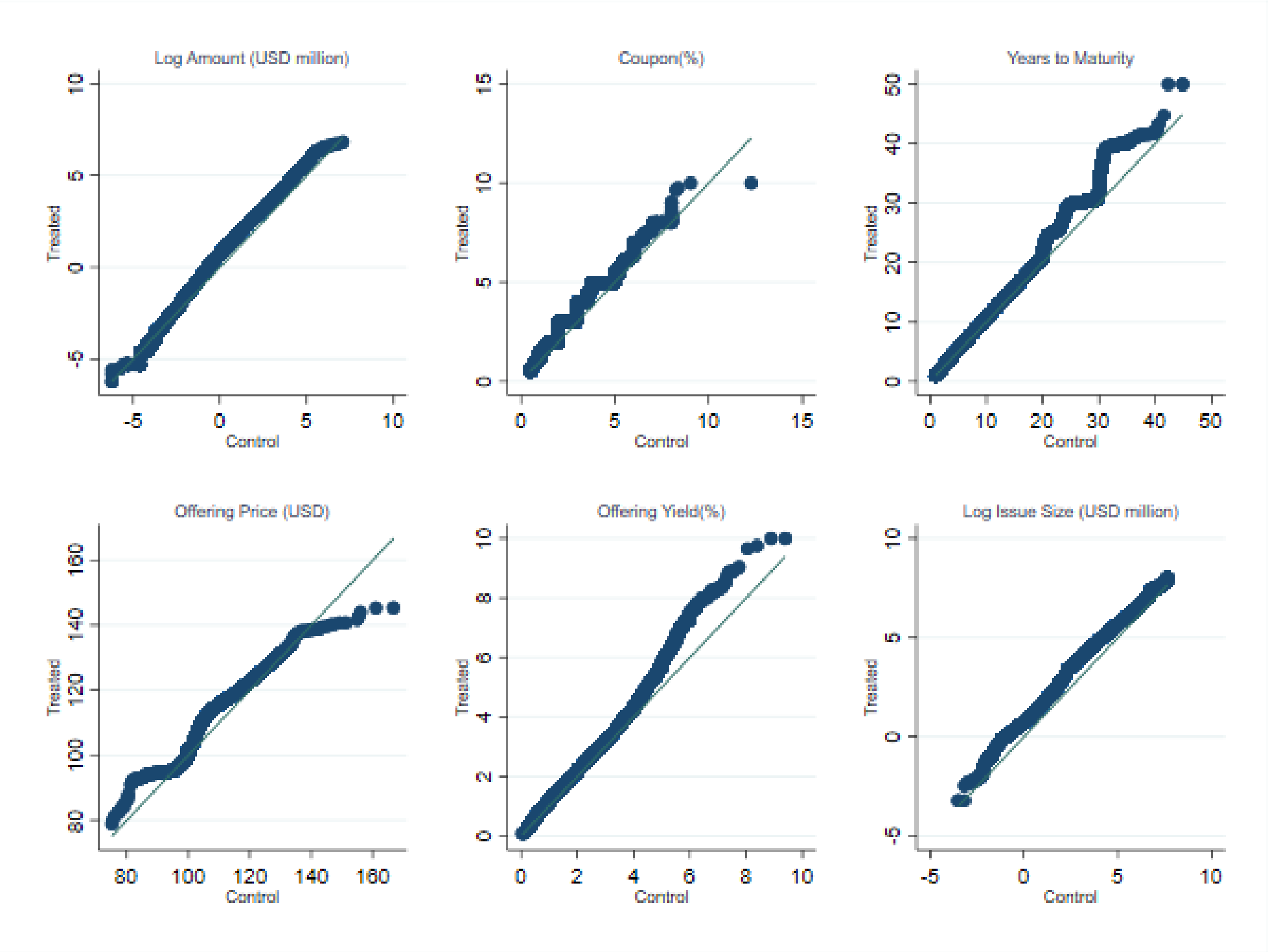

**Figure A1.1:** Quantile-quantile plot of bond characteristics:
The figure shows the primary market characteristics of bonds with advisors at the time of issuance, using a quantile-quantile plot. In my setting, bonds sold via negotiation consist of the "treated" group, whereas competitively auctioned bonds comprise the "control" group. I focus on fixed rate, tax-exempt bonds issued during 2010-2019.

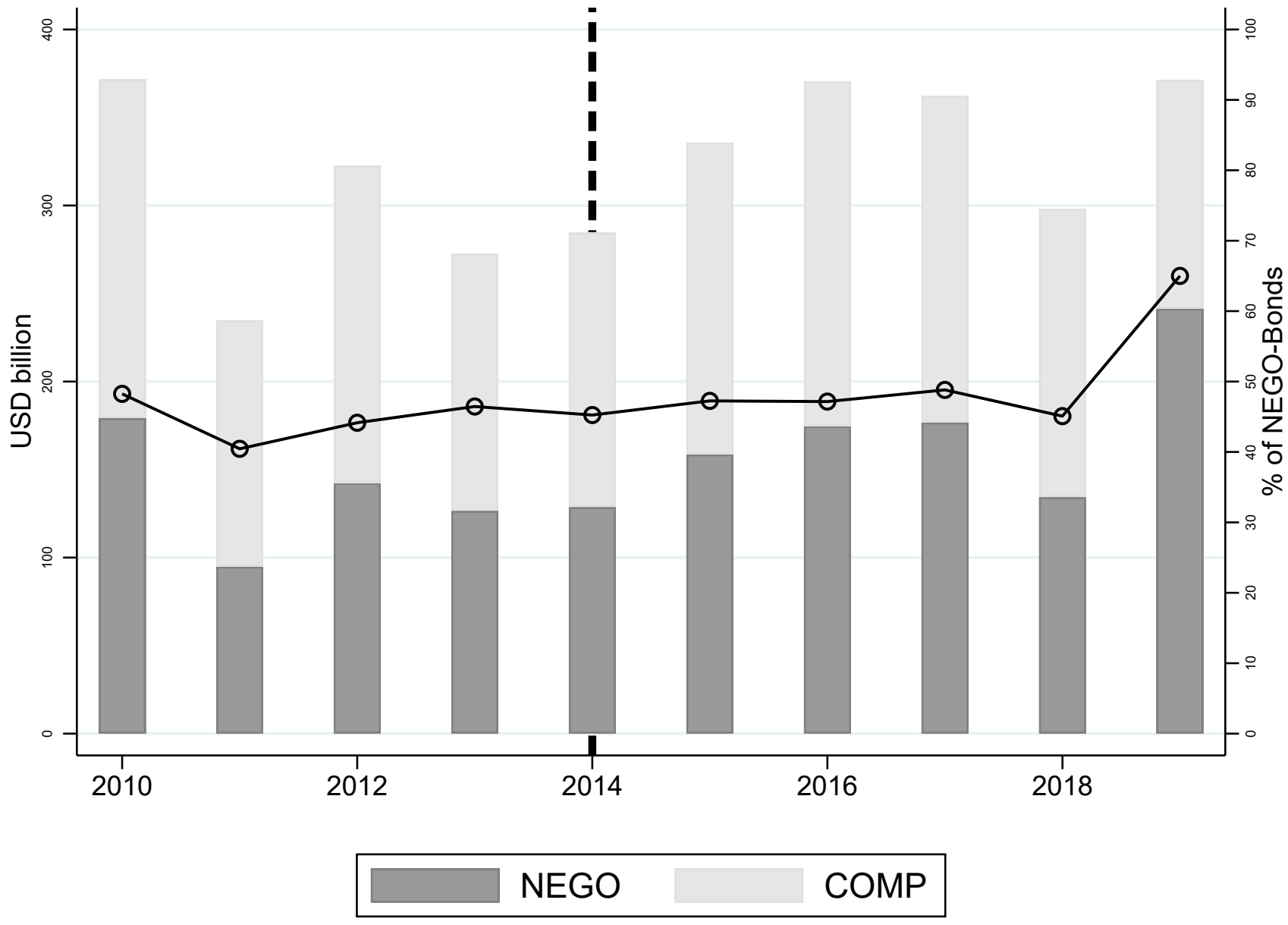

**Figure A1.2:** Total Issuance: Negotiated vs Competitive:
This figure shows the annual amount of advised bonds sold via negotiation. The left hand axis corresponds to the vertical bars denoting the amounts in USD billion. The line graph represents the percentage of dollar value sold via negotiation, shown on the right axis.

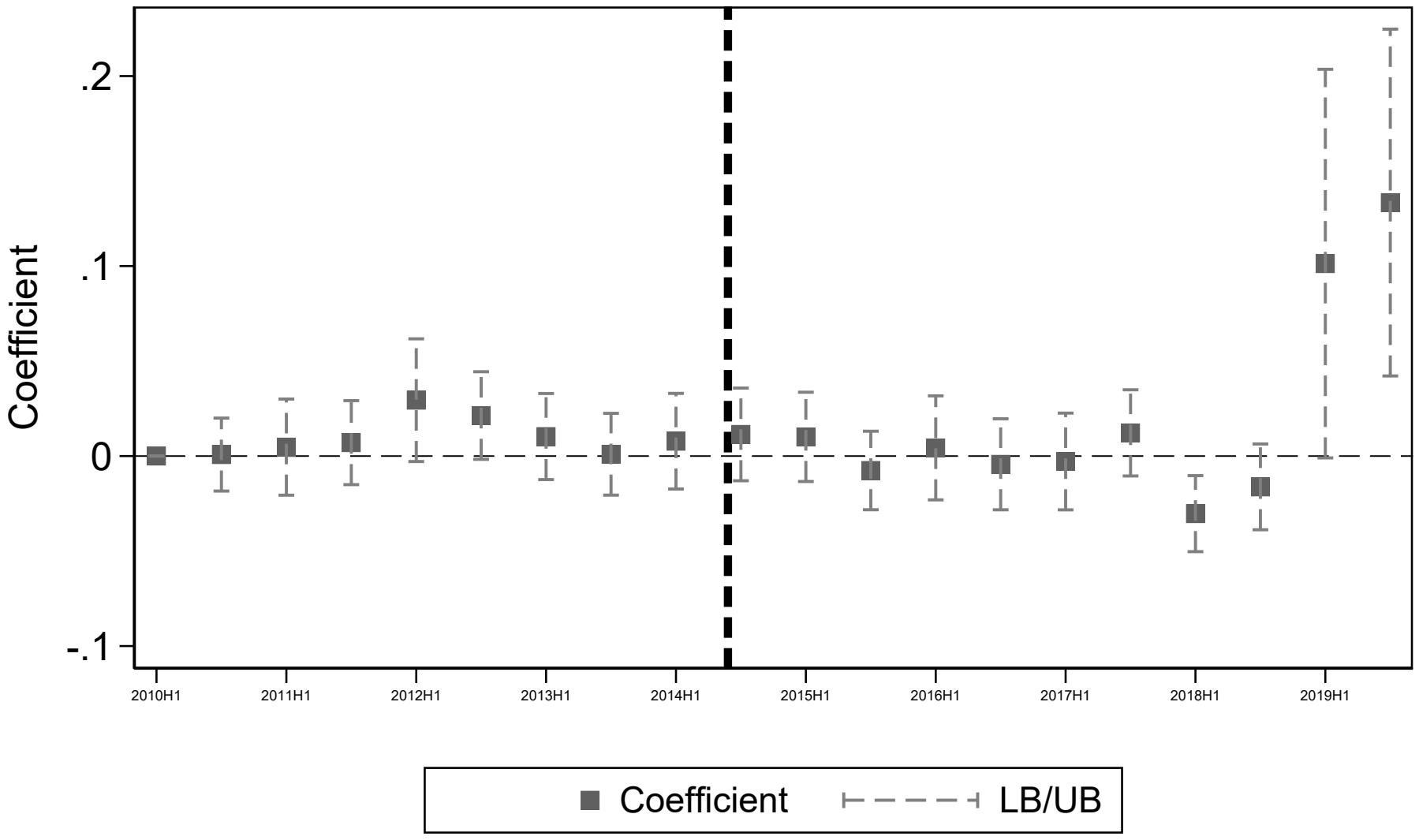

**Figure A1.3:** Likelihood of Negotiated Sale:
This figure reports the likelihood of issuing a bond via negotiation, *within* issuer. The coefficients are shown relative to a benchmark period of half year period at the start of the event window. Standard errors are clustered at the state level.

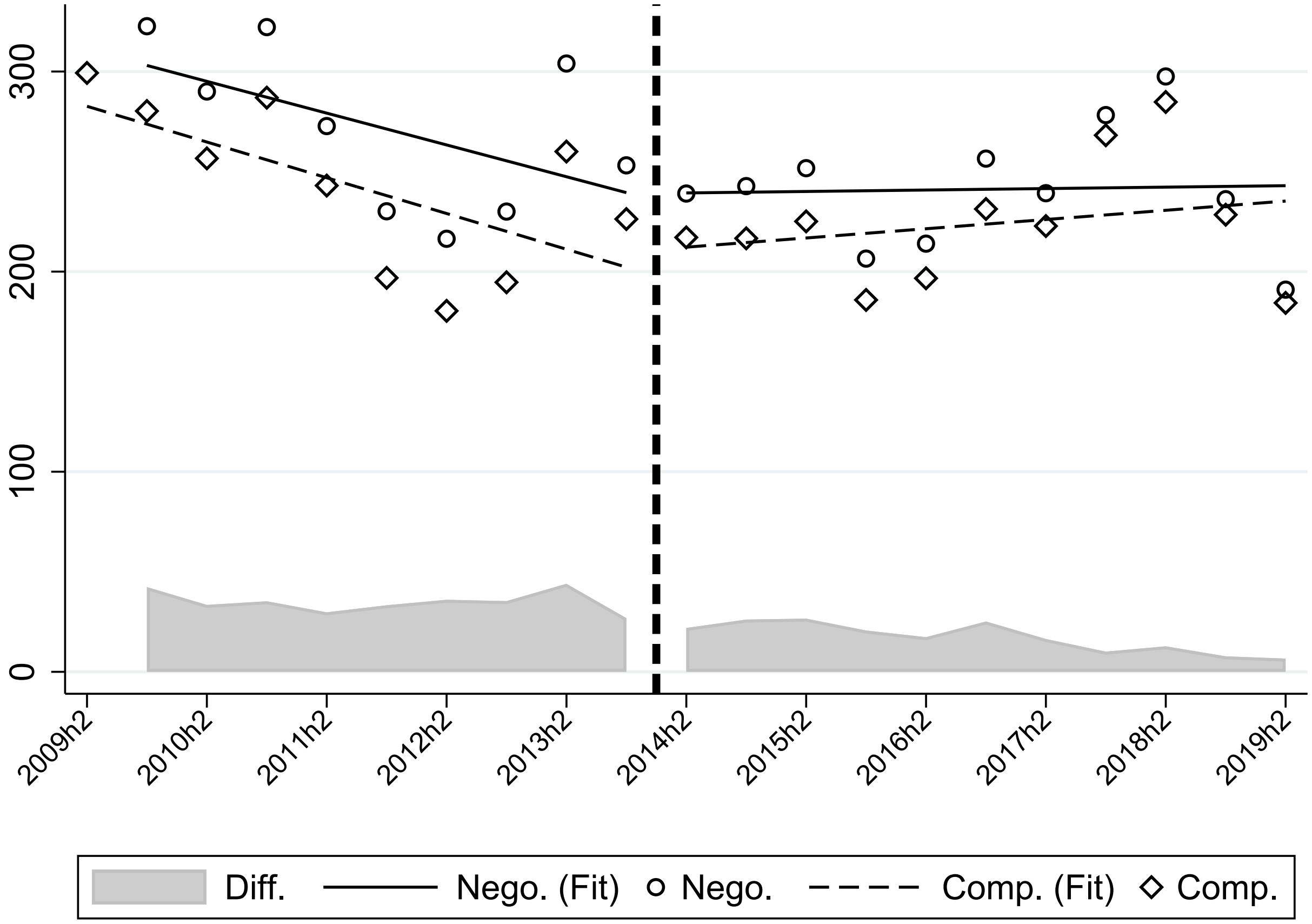

**Figure A1.4:** Raw Binscatter of Offering Yields:
This figure reports the binscatter of offering yields in basis points on the y-axis. Each dot is the average of the offering yields calculated on every $5^{th}$ percentile of the sample for treated and control bonds, separately. The figure also shows the corresponding fitted lines for the negotiated and competitively bid bonds. The difference between the two groups is represented in the shaded portion. The observed pattern suggests a convergence of yields between the treated and control bonds after the SEC Municipal Advisor Rule. This analysis should be interpreted in a non-causal way, as no fixed effect and controls are included.

**Offering Yields**

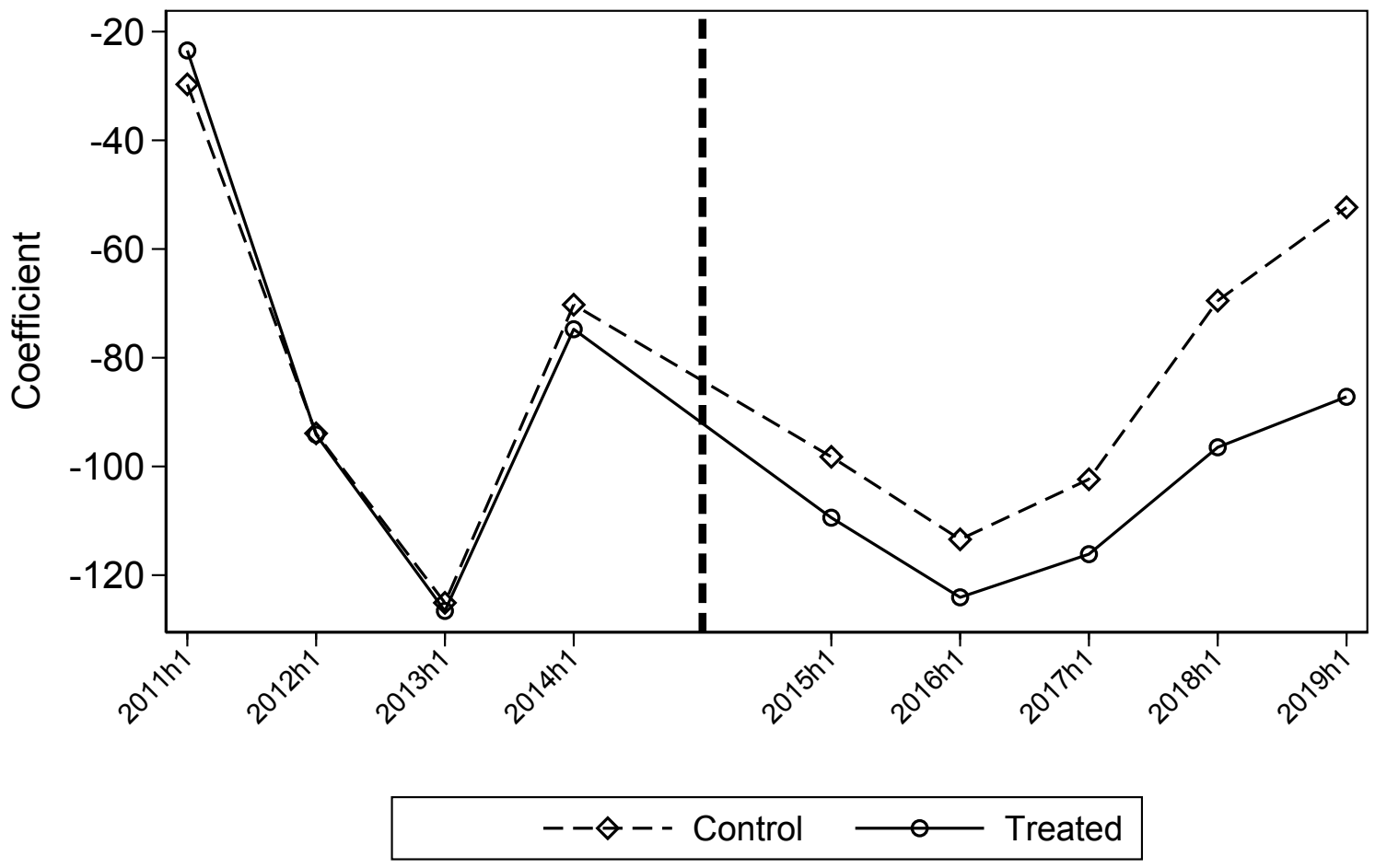

**(a)**

**Difference in Offering Yields**

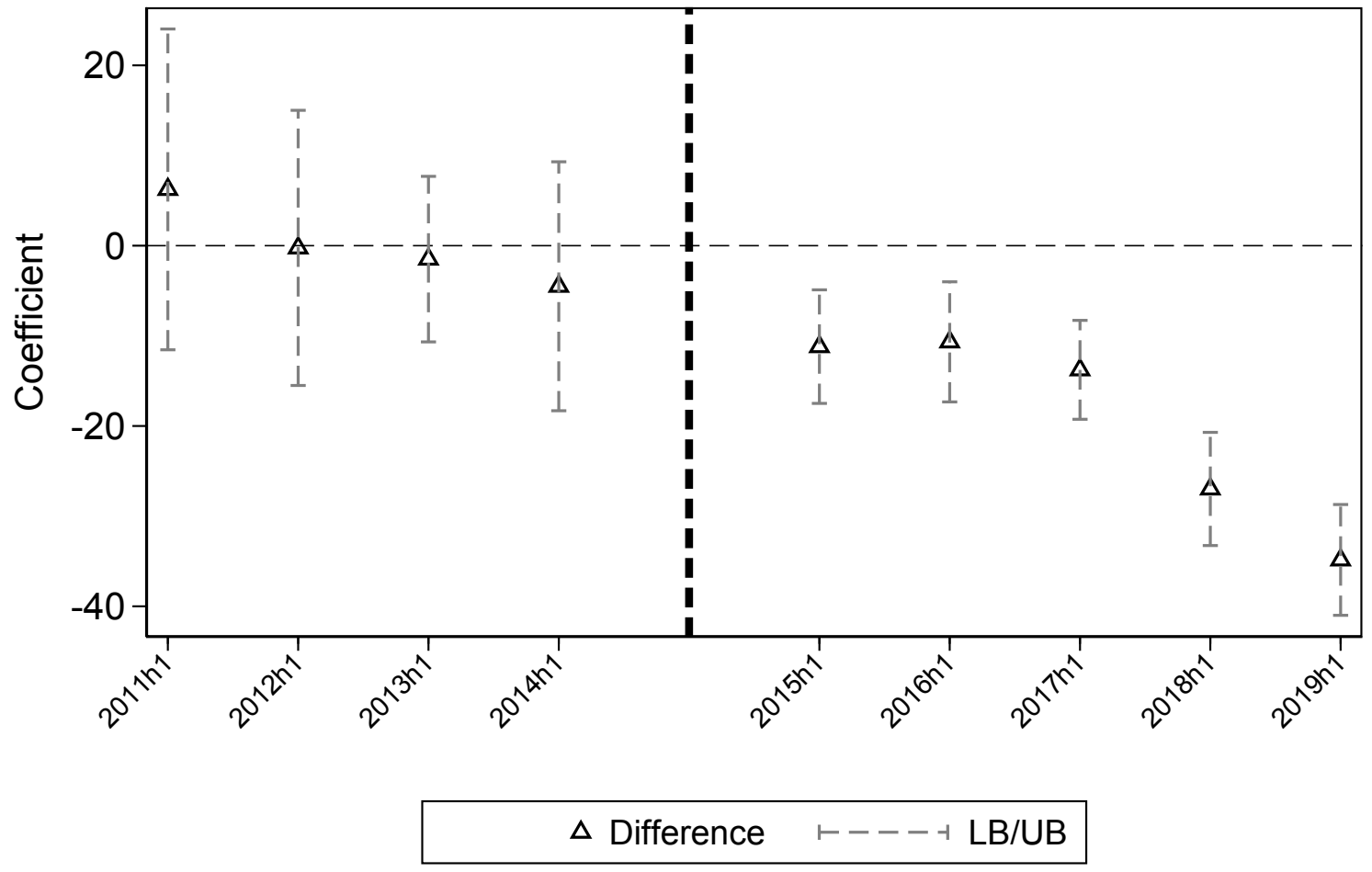

**(b)**

**Figure A1.5:** Baseline Result - offering yields:

In this figure, I plot the offering yield for new municipal bonds using Equation (1.3) in Panel (a). Panel (b) shows the differences between the yields of treated and control bonds. See Table A1 for variables description. The coefficients are shown in basis points. Specifically, I regress the offering yields on yearly interaction dummies for treated and control bonds using issuer fixed effects. These coefficients are depicted on a yearly scale on the x-axis, where the vertical line corresponds to the Municipal Advisor Rule. The omitted benchmark period is the twelve month period before the event window shown above. Standard errors are clustered by state. The dashed lines represent 95% confidence intervals.

## Issue Yields

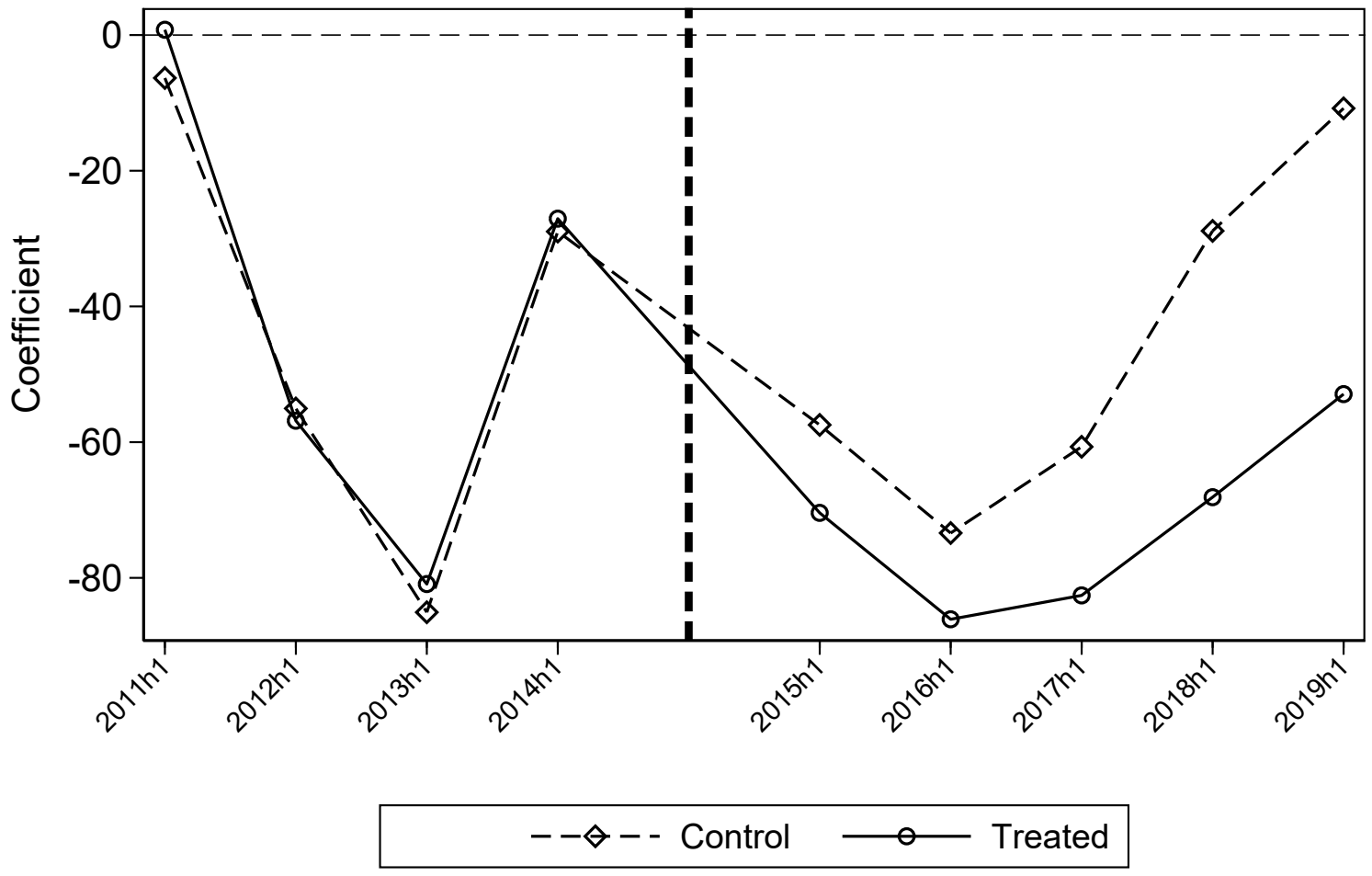

**(a)**

## Difference in Issue Yields

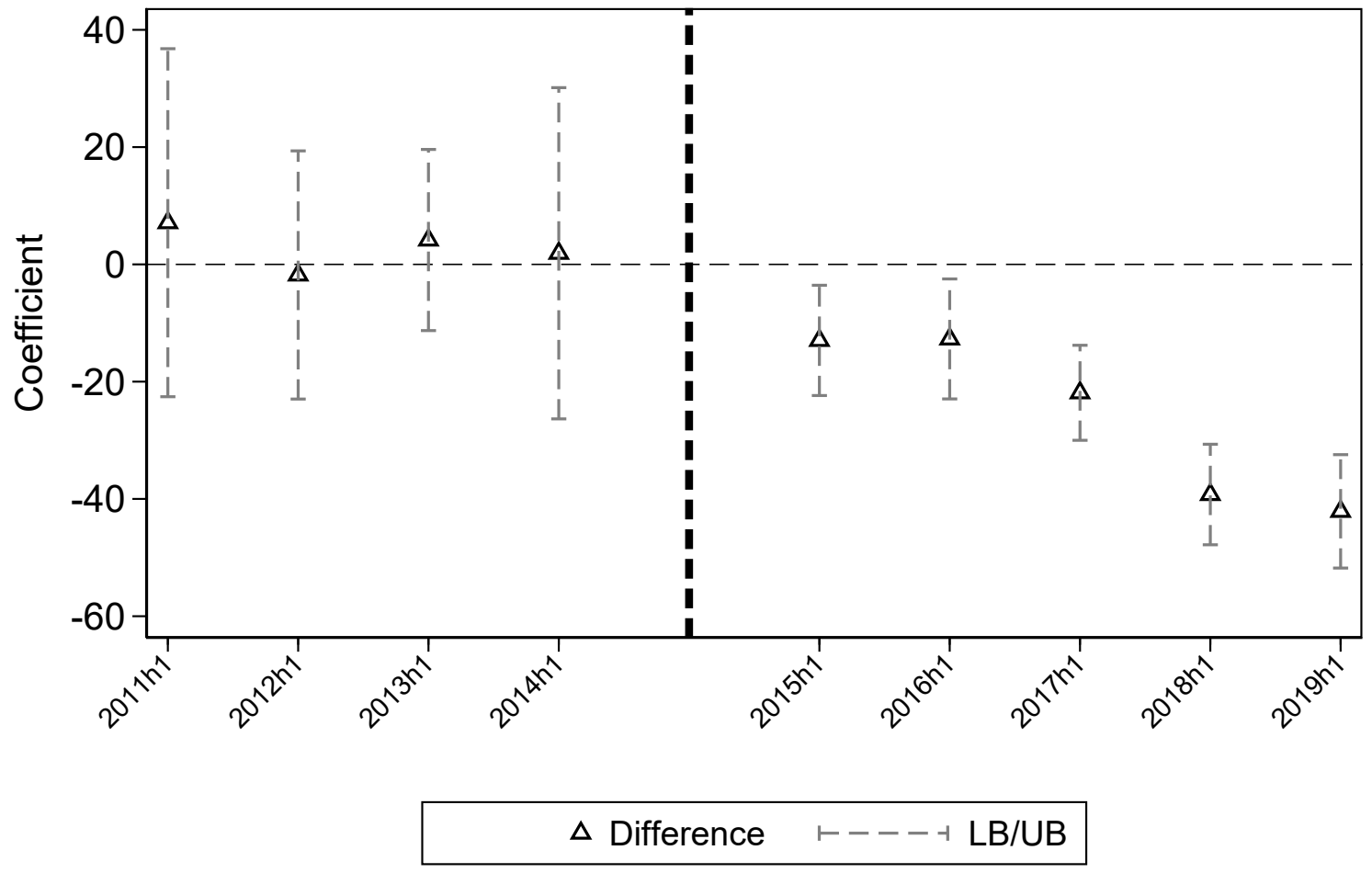

**(b)**

**Figure A1.6:** Baseline Result - Issue Yields:
In this figure, I plot the average offering yield for new municipal bond issues using Equation (1.3) in Panel (a). Panel (b) shows the differences between the yields of treated and control bonds. See Table A1 for variables description. The coefficients are shown in basis points. Specifically, I regress the average issue yields on yearly interaction dummies for treated and control bonds using issuer fixed effects. These coefficients are depicted on a yearly scale on the x-axis, where the vertical line corresponds to the Municipal Advisor Rule. The omitted benchmark period is the twelve month period before the event window shown above. Standard errors are clustered by state. The dashed lines represent 95% confidence intervals.

**Table A1.1:** Probability of Choosing Negotiated Sale Conditional on Ex-ante Observables

This table shows the estimates from a linear probability regression of choice of negotiated sale on issue characteristics. See Section 1.2.1 for details. T-statistics are reported in brackets and standard errors are clustered at the state level. * $p < 0.10$, ** $p < 0.05$, *** $p < 0.01$

| *Dependent Variable*: | Likelihood of Negotiation | | | | | |
|---|---|---|---|---|---|---|
| | (1) | (2) | (3) | (4) | (5) | (6) |
| ln(Issue Size) | -0.07 | -0.06 | -0.14 | -0.13 | 0.02 | 0.01 |
| | [-0.55] | [-0.50] | [-1.00] | [-0.97] | [0.19] | [0.17] |
| ln(avg. bond size) | 0.13 | 0.12 | 0.17 | 0.17 | 0.01 | 0.02 |
| | [0.92] | [0.84] | [1.18] | [1.13] | [0.16] | [0.18] |
| Coupon | 0.05** | 0.04** | 0.03** | 0.03** | 0.01 | 0.01 |
| | [2.34] | [2.34] | [2.04] | [2.03] | [1.11] | [1.11] |
| Maturity | 0.03 | 0.03 | 0.02 | 0.02 | -0.02 | -0.02 |
| | [1.44] | [1.29] | [1.24] | [1.10] | [-1.18] | [-1.17] |
| Callable | -0.06* | -0.06** | -0.02 | -0.02 | -0.02 | -0.02 |
| | [-1.98] | [-2.01] | [-0.71] | [-0.76] | [-0.86] | [-0.87] |
| Credit Enh. | -0.14* | -0.14* | -0.04 | -0.04 | -0.00 | -0.00 |
| | [-1.71] | [-1.68] | [-0.76] | [-0.75] | [-0.02] | [-0.01] |
| Insured | 0.10*** | 0.12*** | 0.07** | 0.08** | 0.04 | 0.04 |
| | [2.88] | [3.26] | [2.35] | [2.62] | [0.88] | [0.89] |
| Bank Qualified | -0.10*** | -0.09*** | -0.05** | -0.04* | 0.01 | 0.01 |
| | [-3.42] | [-3.19] | [-2.10] | [-1.80] | [0.92] | [1.03] |
| Num. of Bonds | 0.02 | 0.01 | 0.05 | 0.05 | -0.00 | -0.00 |
| | [0.19] | [0.17] | [0.62] | [0.60] | [-0.03] | [-0.00] |
| GO bond | | -0.08** | | -0.08** | | -0.03 |
| | | [-2.31] | | [-2.29] | | [-0.97] |
| Num. of Txns. | -0.04*** | -0.04*** | -0.03** | -0.03** | | |
| | [-3.09] | [-2.78] | [-2.39] | [-2.25] | | |
| City(Issuer) | 0.01 | 0.01 | 0.04 | 0.04 | | |
| | [0.17] | [0.16] | [0.95] | [0.83] | | |
| County(Issuer) | 0.05 | 0.05 | 0.08** | 0.09** | | |
| | [1.09] | [1.14] | [2.08] | [2.25] | | |
| State(Issuer) | 0.20*** | 0.18*** | 0.18*** | 0.16*** | | |
| | [4.26] | [3.57] | [4.58] | [3.90] | | |
| Year FE | ✓ | ✓ | ✓ | ✓ | ✓ | ✓ |
| Purpose FE | ✓ | ✓ | ✓ | ✓ | ✓ | ✓ |
| Rating FE | ✓ | ✓ | ✓ | ✓ | ✓ | ✓ |
| State FE | | | ✓ | ✓ | | |
| Issuer FE | | | | | ✓ | ✓ |
| Adj.-$R^2$ | 0.174 | 0.178 | 0.307 | 0.311 | 0.686 | 0.686 |
| Obs. | 15,578 | 15,578 | 15,578 | 15,578 | 15,578 | 15,578 |

**Table A1.2:** Summary Statistics: Municipal Bonds

This table summarizes the municipal bond level characteristics during 2010-2019 for the bonds in the sample. The two panels correspond to the treated (negotiated sale) versus control (competitively bid) bonds. The key variables are described in Table A1.

| | Count | Mean | Std. Dev. | P25 | P50 | P75 |
|---|---|---|---|---|---|---|
| **Treated - Negotiated Sale** | | | | | | |
| Amount (USD million) | 270,504 | 4.08 | 14.83 | 0.4 | 1.0 | 2.8 |
| Coupon(%) | 270,504 | 3.90 | 1.10 | 3.0 | 4.0 | 5.0 |
| Years to Maturity | 270,504 | 10.34 | 6.50 | 5.2 | 9.2 | 14.4 |
| Offering Price (USD) | 270,504 | 108.62 | 8.30 | 101.1 | 107.4 | 115.1 |
| Offering Yield(%) | 270,504 | 2.47 | 1.04 | 1.7 | 2.4 | 3.2 |
| Yield Spread(%) | 270,504 | 1.32 | 1.31 | 0.4 | 1.3 | 2.2 |
| Callable (Dummy) | 270,504 | 0.47 | 0.50 | 0.0 | 0.0 | 1.0 |
| General Obligation (Dummy) | 270,504 | 0.42 | 0.49 | 0.0 | 0.0 | 1.0 |
| Bank Qualified (Dummy) | 270,504 | 0.23 | 0.42 | 0.0 | 0.0 | 0.0 |
| Cred. Enh. (Dummy) | 270,504 | 0.13 | 0.34 | 0.0 | 0.0 | 0.0 |
| Insured (Dummy) | 270,504 | 0.19 | 0.40 | 0.0 | 0.0 | 0.0 |
| **Control - Competitively Bid** | | | | | | |
| Amount (USD million) | 564,420 | 2.04 | 8.29 | 0.2 | 0.5 | 1.3 |
| Coupon(%) | 564,420 | 3.33 | 1.10 | 2.5 | 3.0 | 4.0 |
| Years to Maturity | 564,420 | 9.65 | 5.96 | 4.9 | 8.7 | 13.7 |
| Offering Price (USD) | 564,419 | 105.65 | 7.16 | 100.0 | 103.1 | 109.2 |
| Offering Yield(%) | 564,420 | 2.27 | 0.99 | 1.5 | 2.3 | 3.0 |
| Yield Spread(%) | 564,420 | 1.27 | 1.14 | 0.5 | 1.2 | 2.0 |
| Callable (Dummy) | 564,420 | 0.47 | 0.50 | 0.0 | 0.0 | 1.0 |
| General Obligation (Dummy) | 564,420 | 0.68 | 0.47 | 0.0 | 1.0 | 1.0 |
| Bank Qualified (Dummy) | 564,420 | 0.48 | 0.50 | 0.0 | 0.0 | 1.0 |
| Cred. Enh. (Dummy) | 564,420 | 0.24 | 0.43 | 0.0 | 0.0 | 0.0 |
| Insured (Dummy) | 564,420 | 0.15 | 0.35 | 0.0 | 0.0 | 0.0 |

**Table A1.3:** Impact on Offering Yields of Local Governments: Evidence from Municipal Bonds Primary Market

This table reports the baseline results for my sample using Equation (4.1) estimating the differential effect on municipal bond yields of treated and control bonds after the Municipal Advisor Rule of 2014. The primary coefficient of interest, $\beta_0$, is captured by the interaction term of *Treated* × *Post*. I show the results using offering yields as the dependent variable. Specifically, Column (1) reports the results without any controls or fixed effects. In Column (2), I first introduce issuer fixed effects. Column (3) shows results by introducing state × year fixed effects. In Column (4), I add advisor fixed effect. Column (5) reports the results by additionally including bond level controls consisting of coupon (%); log(amount issued in $); dummies for callable bonds, bond insurance, general obligation bond, bank qualification, refunding and credit enhancement; credit rating; remaining years to maturity; and inverse years to maturity. I provide the description of key variables in Table A1. In Column (6), I control for the county-level economic conditions. I use the lagged values for log(labor force) and unemployment rate, and the percentage change in unemployment rate and labor force, respectively. Column (7) replaces advisor fixed effect with advisor × year fixed effect and does not include county-level controls. Finally, Column (8) shows results by also adding county-level controls. T-statistics are reported in brackets and standard errors are clustered at the state level. $^{*}$ $p < 0.10$, $^{**}$ $p < 0.05$, $^{***}$ $p < 0.01$

| *Dependent Variable*: | Offering Yield (basis points) | | | | | | | |
|---|---|---|---|---|---|---|---|---|
| | (1) | (2) | (3) | (4) | (5) | (6) | (7) | (8) |
| Treated × Post | -17.62$^{***}$ | -20.84$^{***}$ | -11.10$^{***}$ | -10.60$^{***}$ | -13.96$^{***}$ | -13.92$^{***}$ | -11.10$^{***}$ | -11.08$^{***}$ |
| | [-2.78] | [-2.72] | [-2.75] | [-2.75] | [-4.79] | [-4.78] | [-4.14] | [-4.14] |
| Treated | 31.55$^{***}$ | -2.18 | -1.93 | -2.80 | 8.98$^{***}$ | 8.98$^{***}$ | 7.44$^{**}$ | 7.45$^{**}$ |
| | [2.91] | [-0.23] | [-0.32] | [-0.46] | [2.70] | [2.71] | [2.11] | [2.11] |
| Post | -11.60$^{***}$ | -14.28$^{***}$ | -10.23$^{***}$ | -11.05$^{***}$ | -8.40$^{***}$ | -8.34$^{***}$ | -9.19$^{***}$ | -9.14$^{***}$ |
| | [-3.64] | [-2.93] | [-4.78] | [-5.20] | [-4.92] | [-4.88] | [-5.55] | [-5.50] |
| Issuer FE | | ✓ | ✓ | ✓ | ✓ | ✓ | ✓ | ✓ |
| State-Year FE | | | ✓ | ✓ | ✓ | ✓ | ✓ | ✓ |
| Advisor FE | | | | ✓ | ✓ | ✓ | | |
| Bond Controls | | | | | ✓ | ✓ | ✓ | ✓ |
| County Controls | | | | | | ✓ | | ✓ |
| Advisor-Year FE | | | | | | | ✓ | ✓ |
| Adj.-$R^2$ | 0.017 | 0.197 | 0.263 | 0.267 | 0.862 | 0.862 | 0.868 | 0.868 |
| Obs. | 834,925 | 834,764 | 834,764 | 834,760 | 834,760 | 834,760 | 834,741 | 834,741 |

## **Table A1.4:** Impact on Amihud Measure (Liquidity) of New Bonds

This table reports the baseline results for the sample using Equation (4.1) estimating the differential effect on Amihud measure of liquidity between treated and control bonds after the Municipal Advisor Rule of 2014. I use trades during the first month after issuance to construct this measure using [5], based on [150]. The primary coefficient of interest, $\beta_0$, is captured by the interaction term of *Treated* × *Post*. Specifically, Column (1) reports the results without any controls or fixed effects. In Column (2), I first introduce issuer fixed effects. Column (3) shows results by introducing state × year fixed effects. In Column (4), I add advisor fixed effect. Column (5) reports the results by additionally including bond level controls consisting of coupon (%); log(amount issued in $); dummies for callable bonds, bond insurance, general obligation bond, bank qualification, refunding and credit enhancement; credit rating; remaining years to maturity; and inverse years to maturity. I provide the description of key variables in Table A1. In Column (6), I control for the county-level economic conditions. I use the lagged values for log(labor force) and unemployment rate, and the percentage change in unemployment rate and labor force, respectively. Column (7) replaces advisor fixed effect with advisor × year fixed effect and does not include county-level controls. Finally, Column (8) shows results by also adding county-level controls. T-statistics are reported in brackets and standard errors are clustered at the state level. $^{*}\ p < 0.10$, $^{**}\ p < 0.05$, $^{***}\ p < 0.01$

| *Dependent Variable*: | Amihud measure | | | | | | | |
|---|---|---|---|---|---|---|---|---|
| | (1) | (2) | (3) | (4) | (5) | (6) | (7) | (8) |
| Treated × Post | -2.53*** | -2.19*** | -1.67*** | -1.58*** | -1.67*** | -1.66*** | -1.47*** | -1.46*** |
| | [-5.68] | [-6.37] | [-4.82] | [-4.67] | [-8.37] | [-8.30] | [-7.65] | [-7.62] |
| Treated | -1.87*** | -2.93*** | -3.21*** | -3.30*** | -1.29*** | -1.29*** | -1.40*** | -1.40*** |
| | [-4.69] | [-8.83] | [-9.23] | [-9.34] | [-7.18] | [-7.20] | [-8.04] | [-8.07] |
| Post | 2.41*** | 1.86*** | 0.47 | 0.43 | 0.72*** | 0.72*** | 0.56** | 0.56** |
| | [8.21] | [5.72] | [1.54] | [1.44] | [2.75] | [2.76] | [2.03] | [2.02] |
| Issuer FE | | ✓ | ✓ | ✓ | ✓ | ✓ | ✓ | ✓ |
| State-Year FE | | | ✓ | ✓ | ✓ | ✓ | ✓ | ✓ |
| Advisor FE | | | | ✓ | ✓ | ✓ | | |
| Bond Controls | | | | | ✓ | ✓ | ✓ | ✓ |
| County Controls | | | | | | ✓ | | ✓ |
| Advisor-Year FE | | | | | | | ✓ | ✓ |
| Adj.-$R^2$ | 0.014 | 0.107 | 0.112 | 0.113 | 0.277 | 0.277 | 0.281 | 0.281 |
| Obs. | 526,910 | 526,285 | 526,285 | 526,278 | 526,278 | 526,278 | 526,245 | 526,245 |

**Table A1.5:** Impact on Offering Yields of Local Governments using Issue Yields

This table reports the baseline results for the sample using Equation (4.1) estimating the differential effect on municipal issue yields of treated and control bonds at the issue level, after the Municipal Advisor Rule of 2014. The primary coefficient of interest, $\beta_0$, is captured by the interaction term of *Treated* $\times$ *Post*. Specifically, Column (1) reports the results without any controls or fixed effects. In Column (2), I first introduce issuer fixed effects. Column (3) shows results by introducing state $\times$ year fixed effects. In Column (4), I add advisor fixed effect. Column (5) reports the results by additionally including bond level controls consisting of coupon (%); log(amount issued in \$); dummies for callable bonds, bond insurance, general obligation bond, bank qualification, refunding and credit enhancement; credit rating; remaining years to maturity; and inverse years to maturity. I provide the description of key variables in Table A1. In Column (6), I control for the county-level economic conditions. I use the lagged values for log(labor force) and unemployment rate, and the percentage change in unemployment rate and labor force, respectively. Column (7) replaces advisor fixed effect with advisor $\times$ year fixed effect and does not include county-level controls. Finally, Column (8) shows results by also adding county-level controls. T-statistics are reported in brackets and standard errors are clustered at the state level. $^{*}$ $p < 0.10$, $^{**}$ $p < 0.05$, $^{***}$ $p < 0.01$

| *Dependent Variable*: | Issue Level Offering Yield (basis points) | | | | | | | |
|---|---|---|---|---|---|---|---|---|
| | (1) | (2) | (3) | (4) | (5) | (6) | (7) | (8) |
| Treated $\times$ Post | -27.38$^{***}$ | -29.74$^{***}$ | -15.67$^{**}$ | -14.50$^{**}$ | -17.60$^{***}$ | -17.55$^{***}$ | -14.10$^{***}$ | -14.08$^{***}$ |
| | [-3.30] | [-2.92] | [-2.28] | [-2.20] | [-4.34] | [-4.34] | [-3.78] | [-3.78] |
| Treated | 53.90$^{***}$ | 1.97 | -0.54 | -2.28 | 8.29$^{**}$ | 8.28$^{**}$ | 6.63$^{*}$ | 6.64$^{*}$ |
| | [3.86] | [0.16] | [-0.06] | [-0.24] | [2.55] | [2.56] | [1.94] | [1.95] |
| Post | -7.61$^{**}$ | -10.96$^{*}$ | -8.45$^{**}$ | -9.38$^{***}$ | -8.80$^{***}$ | -8.74$^{***}$ | -9.92$^{***}$ | -9.88$^{***}$ |
| | [-2.10] | [-1.77] | [-2.62] | [-2.79] | [-3.43] | [-3.39] | [-3.89] | [-3.86] |
| Issuer FE | | ✓ | ✓ | ✓ | ✓ | ✓ | ✓ | ✓ |
| State-Year FE | | | ✓ | ✓ | ✓ | ✓ | ✓ | ✓ |
| Advisor FE | | | | ✓ | ✓ | ✓ | ✓ | ✓ |
| Bond Controls | | | | | ✓ | ✓ | ✓ | ✓ |
| County Controls | | | | | | ✓ | | ✓ |
| Advisor-Year FE | | | | | | | ✓ | ✓ |
| Adj.-$R^2$ | 0.052 | 0.337 | 0.454 | 0.461 | 0.856 | 0.857 | 0.863 | 0.863 |
| Obs. | 56,793 | 51,321 | 51,321 | 51,220 | 51,220 | 51,220 | 50,668 | 50,668 |

**Table A1.6:** Impact on Offering Yield Spreads of Local Governments using Issue Yield Spreads

This table reports the baseline results for my sample using Equation (4.1) estimating the differential effect on municipal issue yield spreads of treated and control bonds at the issue level, after the Municipal Advisor Rule of 2014. The primary coefficient of interest, $\beta_0$, is captured by the interaction term of *Treated* $\times$ *Post*. Specifically, Column (1) reports the results without any controls or fixed effects. In Column (2), I first introduce issuer fixed effects. Column (3) shows results by introducing state $\times$ year fixed effects. In Column (4), I add advisor fixed effect. Column (5) reports the results by additionally including bond level controls consisting of coupon (%); log(amount issued in \$); dummies for callable bonds, bond insurance, general obligation bond, bank qualification, refunding and credit enhancement; credit rating; remaining years to maturity; and inverse years to maturity. I provide the description of key variables in Table A1. In Column (6), I control for the county-level economic conditions. I use the lagged values for log(labor force) and unemployment rate, and the percentage change in unemployment rate and labor force, respectively. Column (7) replaces advisor fixed effect with advisor $\times$ year fixed effect and does not include county-level controls. Finally, Column (8) shows results by also adding county-level controls. T-statistics are reported in brackets and standard errors are clustered at the state level. $^{*}$ $p < 0.10$, $^{**}$ $p < 0.05$, $^{***}$ $p < 0.01$

| *Dependent Variable*: | Issue Level Yield Spread (basis points) | | | | | | | |
|---|---|---|---|---|---|---|---|---|
| | (1) | (2) | (3) | (4) | (5) | (6) | (7) | (8) |
| Treated $\times$ Post | -34.30$^{***}$ | -33.57$^{***}$ | -14.00$^{***}$ | -12.90$^{**}$ | -14.81$^{***}$ | -14.74$^{***}$ | -14.16$^{***}$ | -14.13$^{***}$ |
| | [-4.21] | [-4.52] | [-2.70] | [-2.68] | [-4.95] | [-4.92] | [-4.71] | [-4.71] |
| Treated | 40.00$^{***}$ | -7.23 | 2.35 | 1.03 | 6.74$^{***}$ | 6.74$^{***}$ | 6.49$^{**}$ | 6.51$^{**}$ |
| | [4.08] | [-0.66] | [0.39] | [0.17] | [3.05] | [3.08] | [2.49] | [2.52] |
| Post | -72.93$^{***}$ | -74.34$^{***}$ | -4.50 | -5.20 | -5.16$^{*}$ | -5.05$^{*}$ | -4.92$^{*}$ | -4.81$^{*}$ |
| | [-26.09] | [-20.66] | [-1.46] | [-1.59] | [-1.87] | [-1.83] | [-1.78] | [-1.74] |
| Issuer FE | | ✓ | ✓ | ✓ | ✓ | ✓ | ✓ | ✓ |
| State-Year FE | | | ✓ | ✓ | ✓ | ✓ | ✓ | ✓ |
| Advisor FE | | | | ✓ | ✓ | ✓ | ✓ | ✓ |
| Bond Controls | | | | | ✓ | ✓ | ✓ | ✓ |
| County Controls | | | | | | ✓ | | ✓ |
| Advisor-Year FE | | | | | | | ✓ | ✓ |
| Adj.-$R^2$ | 0.145 | 0.251 | 0.699 | 0.702 | 0.816 | 0.816 | 0.824 | 0.824 |
| Obs. | 56,793 | 51,321 | 51,321 | 51,220 | 51,220 | 51,220 | 50,668 | 50,668 |

## **Table A1.7:** Robustness Tests using Offering Yields

In this table I report results for various robustness tests on the baseline specification, i.e., Column (8) of Table 3.2 with offering yield as the dependent variable. Columns (1)-(2) show the baseline effect by changing the dependent variable to after-tax yield and after-tax yield spread, respectively. In Columns (3)-(11), I report results based on alternative econometric specifications. I introduce additional fixed effects to account for unobserved factors that may be varying over time. Specifically, Column (3) reports baseline results by adding issuer-type × year fixed effects. Column (4) shows results by adding bond purpose fixed effects. I add bond purpose × year fixed effect in Column (5). Columns (6) and (7) show results by adding underwriter fixed effect and underwriter × year fixed effect, respectively. In Columns (8) and (9), I control for unobserved pairing between issuers and advisors, as well as issuers and underwriters, separately, by adding issuer-advisor pair fixed effect and issuer-underwriter pair fixed effect, respectively. These specifications include time unvarying advisor and underwriter fixed effects, respectively. Column (10) shows results with advisor × state fixed effects. In Column (11), I also add county × year fixed effects. I consider additional tax considerations in Columns (12)-(14). First, I relax the sample of bonds to include taxable bonds in Columns (12)-(13). In Column (13), I further omit bonds from five states (IL, IA, KS, OK, WI) that tax interest income on municipal bonds issued in-state or out-of-state [28]. Column (14) shows results using bonds that are exempt from both state and federal income tax simultaneously. Columns (15)-(18) report results focusing on sub-samples of homogeneous bonds. Accordingly, in Column (15), I drop bonds in which the advisor and underwriter are same. Column (16) shows results by dropping callable bonds. I drop insured bonds in Column (17). Finally, I focus on only the new money bonds in Column (18). The results in Columns (19)-(21) focus on additional geographic considerations. I keep only local bonds (by dropping state level bonds) in Column (19). Conversely, I show results using only the state level bonds in Column (20). In Column (21), I report the baseline results by dropping issuances from the three largest municipal bond issuers, namely: California (CA), New York (NY) and Texas (TX). I consider alternative levels of clustering standard errors in Columns (22)-(32). In Columns (22)-(26), I cluster standard errors by advisor, underwriter, issuer, issuer(2), and bond issue, respectively. Issuer(2) in Column (25) refers to weakly identifying borrowers based on the first six-digits of the bond CUSIP [28]. Columns (27)-(28) double cluster standard errors along two dimensions in the geography of issuers: state-advisor, and advisor-issuer, respectively. Finally, in Columns (29)-(32), I double cluster standard errors over time and across bonds using: state and year, advisor and year, state and year-month, and advisor and year-month, respectively. T-statistics are reported in brackets and standard errors are clustered at the state level, unless otherwise specified. * $p < 0.10$, ** $p < 0.05$, *** $p < 0.01$

| *Dependent Variable*: | Offering Yield (basis points) | | | | | | | | | | |
|---|---|---|---|---|---|---|---|---|---|---|---|
| | Other Dependent Variables | | Alternative Specifications | | | | | | | | |
| | After-tax Yield (1) | After-tax Yield Spread (2) | Add Issuer-Type Year FE (3) | Add Purpose FE (4) | Add Purpose Year FE (5) | Add UW FE (6) | Add UW-Yr. FE (7) | Add Iss.-MA FE (8) | Add Iss.-UW FE (9) | Add MA-State FE (10) | Add County-Yr. FE (11) |
| Treated × Post | -18.13*** | -19.69*** | -10.51*** | -11.08*** | -10.61*** | -11.00*** | -9.29*** | -12.61*** | -11.43** | -10.25*** | -8.57*** |
| | [-4.23] | [-4.33] | [-4.02] | [-4.06] | [-3.89] | [-4.18] | [-3.14] | [-4.28] | [-2.44] | [-3.81] | [-2.87] |
| Adj.-$R^2$ | 0.867 | 0.853 | 0.868 | 0.868 | 0.868 | 0.869 | 0.873 | 0.870 | 0.893 | 0.869 | 0.878 |
| Obs. | 833,343 | 833,343 | 833,667 | 833,348 | 833,348 | 834,730 | 834,707 | 834,670 | 834,297 | 834,731 | 834,629 |

| *Dependent Variable*: | Offering Yield (basis points) | | | | | | | | | |
|---|---|---|---|---|---|---|---|---|---|---|
| | Tax Considerations | | | Bond Considerations | | | | Geographic Considerations | | |
| | Include Taxable | | Exempt | Drop same | Drop | Drop | Only | Keep | Keep | Drop |
| | | (-Taxable States) | (Fed.+State) | Adv.-UW. bonds | Callable | Insured | New money | Local | State | CA,NY,TX |
| | (12) | (13) | (14) | (15) | (16) | (17) | (18) | (19) | (20) | (21) |
| Treated × Post | -11.08*** | -9.83*** | -10.83*** | -11.17*** | -10.75*** | -13.16*** | -20.92*** | -9.42*** | -12.66** | -15.35*** |
| | [-4.14] | [-3.88] | [-3.95] | [-4.26] | [-6.21] | [-6.33] | [-6.21] | [-3.42] | [-2.39] | [-7.58] |
| Adj.-$R^2$ | 0.868 | 0.865 | 0.866 | 0.867 | 0.828 | 0.870 | 0.879 | 0.875 | 0.838 | 0.871 |
| Obs. | 834,741 | 737,424 | 770,446 | 830,458 | 443,572 | 698,776 | 415,926 | 751,278 | 83,456 | 519,544 |

| *Dependent Variable*: | Offering Yield (basis points) | | | | | | | | | | |
|---|---|---|---|---|---|---|---|---|---|---|---|
| | Alternative Clustering | | | | | | | | | | |
| | Advisor (22) | Underwriter (23) | Issuer (24) | Issuer(2) (25) | Issue (26) | State, Advisor (27) | Advisor, Issuer (28) | State, Year (29) | Advisor, Year (30) | State, YM (31) | Advisor, YM (32) |
| Treated × Post | -11.08*** | -11.08*** | -11.08*** | -11.08*** | -11.08*** | -11.08*** | -11.08*** | -11.08*** | -11.08*** | -11.08*** | -11.08*** |
| | [-5.11] | [-7.25] | [-7.24] | [-9.91] | [-12.47] | [-3.92] | [-4.75] | [-3.96] | [-4.09] | [-4.13] | [-5.02] |
| Adj.-$R^2$ | 0.868 | 0.868 | 0.868 | 0.868 | 0.868 | 0.868 | 0.868 | 0.868 | 0.868 | 0.868 | 0.868 |
| Obs. | 834,741 | 834,741 | 834,741 | 834,741 | 834,741 | 834,741 | 834,741 | 834,741 | 834,741 | 834,741 | 834,741 |

**Table A1.8:** Robustness Tests using Offering Price

In this table I report results for various robustness tests on the baseline specification, i.e., Column (8) of Table 1.3 with offering price as the dependent variable. In Columns (1)-(9), I report results based on alternative econometric specifications. I introduce additional fixed effects to account for unobserved factors that may be varying over time. Specifically, Column (1) reports baseline results by adding issuer-type $\times$ year fixed effects. Column (2) shows results by adding bond purpose fixed effects. I add bond purpose $\times$ year fixed effect in Column (3). Columns (4) and (5) show results by adding underwriter fixed effect and underwriter $\times$ year fixed effect, respectively. In Columns (6) and (7), I control for unobserved pairing between issuers and advisors, as well as issuers and underwriters, separately, by adding issuer-advisor pair fixed effect and issuer-underwriter pair fixed effect, respectively. These specifications include time unvarying advisor and underwriter fixed effects, respectively. Column (8) shows results with advisor $\times$ state fixed effects. In Column (9), I also add county $\times$ year fixed effects. I consider additional tax considerations in Columns (10)-(12). First, I relax the sample of bonds to include taxable bonds in Columns (10)-(11). In Column (11), I further omit bonds from five states (IL, IA, KS, OK, WI) that tax interest income on municipal bonds issued in-state or out-of-state [28]. Column (12) shows results using bonds that are exempt from both state and federal income tax simultaneously. Columns (13)-(16) report results focusing on sub-samples of homogeneous bonds. Accordingly, in Column (13), I drop bonds in which the advisor and underwriter are same. Column (14) shows results by dropping callable bonds. I drop insured bonds in Column (15). Finally, I focus on only the new money bonds in Column (16). The results in Columns (17)-(19) focus on additional geographic considerations. I keep only local bonds (by dropping state level bonds) in Column (17). Conversely, I show results using only the state level bonds in Column (18). In Column (19), I report the baseline results by dropping issuances from the three largest municipal bond issuers, namely: California (CA), New York (NY) and Texas (TX). I consider alternative levels of clustering standard errors in Columns (20)-(30). In Columns (20)-(24), I cluster standard errors by advisor, underwriter, issuer, issuer(2), and bond issue, respectively. Issuer(2) in Column (23) refers to weakly identifying borrowers based on the first six-digits of the bond CUSIP [28]. Columns (25)-(26) double cluster standard errors along two dimensions in the geography of issuers: state-advisor, and advisor-issuer, respectively. Finally, in Columns (27)-(30), I double cluster standard errors over time and across bonds using: state and year, advisor and year, state and year-month, and advisor and year-month, respectively. T-statistics are reported in brackets and standard errors are clustered at the state level, unless otherwise specified. $^{*}$ $p < 0.10$, $^{**}$ $p < 0.05$, $^{***}$ $p < 0.01$

| *Dependent Variable*: | Offering Price (per USD 100) | | | | | | | | |
|---|---|---|---|---|---|---|---|---|---|
| | Alternative Specifications | | | | | | | | |
| | Add Issuer-Type Year FE | Add Purpose FE | Add Purpose Year FE | Add UW FE | Add UW-Yr. FE | Add Iss.-MA FE | Add Iss.-UW FE | Add MA-State FE | Add County-Yr. FE |
| | (1) | (2) | (3) | (4) | (5) | (6) | (7) | (8) | (9) |
| Treated $\times$ Post | $1.00^{***}$ | $1.05^{***}$ | $1.02^{***}$ | $1.03^{***}$ | $0.92^{***}$ | $1.11^{***}$ | $0.92^{**}$ | $0.97^{***}$ | $0.84^{***}$ |
| | [4.42] | [4.31] | [4.33] | [4.22] | [3.82] | [4.44] | [2.28] | [4.11] | [3.51] |
| Adj.-$R^2$ | 0.774 | 0.774 | 0.774 | 0.775 | 0.777 | 0.776 | 0.796 | 0.775 | 0.783 |
| Obs. | 833,666 | 833,347 | 833,347 | 834,729 | 834,706 | 834,669 | 834,296 | 834,730 | 834,628 |

| *Dependent Variable*: | Offering Price (per USD 100) | | | | | | | | | |
|---|---|---|---|---|---|---|---|---|---|---|
| | Tax Considerations | | | Bond Considerations | | | | Geographic Considerations | | |
| | Include Taxable | | Exempt | Drop same | Drop | Drop | Only | Keep | Keep | Drop |
| | | (-Taxable States) | (Fed.+State) | Adv.-UW. bonds | Callable | Insured | New money | Local | State | CA,NY,TX |
| | (10) | (11) | (12) | (13) | (14) | (15) | (16) | (17) | (18) | (19) |
| Treated $\times$ Post | $1.05^{***}$ | $0.94^{***}$ | $1.01^{***}$ | $1.05^{***}$ | $0.72^{***}$ | $1.24^{***}$ | $1.62^{***}$ | $0.90^{***}$ | $1.02^{**}$ | $1.45^{***}$ |
| | [4.36] | [3.99] | [4.15] | [4.43] | [4.81] | [6.82] | [6.99] | [3.79] | [2.54] | [9.28] |
| Adj.-$R^2$ | 0.774 | 0.774 | 0.774 | 0.774 | 0.806 | 0.773 | 0.776 | 0.776 | 0.755 | 0.777 |
| Obs. | 834,740 | 737,423 | 770,445 | 830,457 | 443,571 | 698,775 | 415,926 | 751,277 | 83,456 | 519,544 |

| *Dependent Variable*: | Offering Price (per USD 100) | | | | | | | | | | |
|---|---|---|---|---|---|---|---|---|---|---|---|
| | Alternative Clustering | | | | | | | | | | |
| | Advisor | Underwriter | Issuer | Issuer(2) | Issue | State, Advisor | Advisor, Issuer | State, Year | Advisor, Year | State, YM | Advisor, YM |
| | (20) | (21) | (22) | (23) | (24) | (25) | (26) | (27) | (28) | (29) | (30) |
| Treated $\times$ Post | $1.05^{***}$ | $1.05^{***}$ | $1.05^{***}$ | $1.05^{***}$ | $1.05^{***}$ | $1.05^{***}$ | $1.05^{***}$ | $1.05^{***}$ | $1.05^{***}$ | $1.05^{***}$ | $1.05^{***}$ |
| | [5.14] | [7.64] | [8.71] | [12.49] | [15.68] | [3.89] | [4.89] | [3.90] | [4.11] | [4.26] | [4.96] |
| Adj.-$R^2$ | 0.774 | 0.774 | 0.774 | 0.774 | 0.774 | 0.774 | 0.774 | 0.774 | 0.774 | 0.774 | 0.774 |
| Obs. | 834,740 | 834,740 | 834,740 | 834,740 | 834,740 | 834,740 | 834,740 | 834,740 | 834,740 | 834,740 | 834,740 |

**Table A1.9:** Impact Due to Advisor's Ex-ante Role in "Selecting" Underwriter

This table reports the results using Equation (4.1) to show the differential effect among issuers for whom advisors play a greater role in selecting underwriters. The dependent variable is offering yield spread. I use the average (Columns (1)-(3)) and weighted average (Columns (4)-(6)) of ex-ante likelihood of a new underwriter being introduced by an advisor for a given issuer, respectively. Specifically, I interact the equation with dummies corresponding to below and above median values for this measure among issuers. This analysis also includes group × year fixed effects. The baseline specification of Column (8) in Table 3.2 is shown in Columns (3) and (6). T-statistics are reported in brackets and standard errors are clustered at the state level. * $p < 0.10$, ** $p < 0.05$, *** $p < 0.01$

## Panel A: Evidence from Offering Yields

| *Dependent Variable*: | Offering Yield (basis points) | | | | | |
|---|---|---|---|---|---|---|
| *Based on Issuers'*: | Average | | | Weighted Average | | |
| Treated × Post | (1) | (2) | (3) | (4) | (5) | (6) |
| Below Median | -12.60*** | -6.82* | -4.88 | -12.61*** | -6.84* | -4.90 |
| | [-2.76] | [-1.80] | [-1.34] | [-2.77] | [-1.80] | [-1.34] |
| Above Median | -16.97*** | -15.21*** | -12.45*** | -16.95*** | -15.19*** | -12.43*** |
| | [-5.69] | [-7.16] | [-6.66] | [-5.69] | [-7.15] | [-6.65] |
| Difference | 4.37 | 8.39 | 7.57 | 4.34 | 8.36 | 7.52 |
| p-value | 0.19 | 0.02 | 0.02 | 0.19 | 0.02 | 0.02 |
| Issuer FE | ✓ | ✓ | ✓ | ✓ | ✓ | ✓ |
| Controls | ✓ | ✓ | ✓ | ✓ | ✓ | ✓ |
| Advisor FE | ✓ | ✓ | | ✓ | ✓ | |
| Group-Yr. FE | ✓ | ✓ | ✓ | ✓ | ✓ | ✓ |
| State-Yr. FE | | ✓ | ✓ | | ✓ | ✓ |
| Advisor-Yr. FE | | | ✓ | | | ✓ |
| Adj.-$R^2$ | 0.858 | 0.863 | 0.868 | 0.858 | 0.863 | 0.868 |
| Obs. | 776,304 | 776,304 | 776,286 | 776,304 | 776,304 | 776,286 |

**Panel B: Evidence from Offering Price**

| *Dependent Variable*: | Offering Price (per USD 100) | | | | | |
|---|---|---|---|---|---|---|
| *Issuers' reliance*: | Average | | | Wtd. average | | |
| Treated × Post | (1) | (2) | (3) | (4) | (5) | (6) |
| Below Median | 0.98** | 0.52 | 0.40 | 0.98** | 0.52 | 0.40 |
| | [2.43] | [1.50] | [1.26] | [2.43] | [1.50] | [1.26] |
| Above Median | 1.47*** | 1.36*** | 1.20*** | 1.47*** | 1.36*** | 1.20*** |
| | [5.66] | [7.27] | [6.76] | [5.66] | [7.27] | [6.77] |
| Difference | -0.50 | -0.84 | -0.81 | -0.49 | -0.84 | -0.80 |
| p-value | 0.06 | 0.00 | 0.00 | 0.06 | 0.00 | 0.00 |
| Issuer FE | ✓ | ✓ | ✓ | ✓ | ✓ | ✓ |
| Controls | ✓ | ✓ | ✓ | ✓ | ✓ | ✓ |
| Advisor FE | ✓ | ✓ | | ✓ | ✓ | |
| Group-Yr. FE | ✓ | ✓ | ✓ | ✓ | ✓ | ✓ |
| State-Yr. FE | | ✓ | ✓ | | ✓ | ✓ |
| Advisor-Yr. FE | | | ✓ | | | ✓ |
| Adj.-$R^2$ | 0.762 | 0.766 | 0.772 | 0.762 | 0.766 | 0.772 |
| Obs. | 776,303 | 776,303 | 776,285 | 776,303 | 776,303 | 776,285 |

**Table A1.10:** Heterogeneity by Size of Issuers

This table reports the results using Equation (4.1) to show the differential effect among issuers based on their ex-ante size. The dependent variable is offering yields. I use the average (Columns (1)-(3)) and median (Columns (4)-(6)) size of ex-ante issuances, respectively. Specifically, I interact the equation with dummies corresponding to small and large values for this measure among issuers. This analysis also includes group $\times$ year fixed effects. The baseline specification of Column (8) in Table 3.2 is shown in Columns (3) and (6). T-statistics are reported in brackets and standard errors are clustered at the state level. $^{*}$ $p < 0.10$, $^{**}$ $p < 0.05$, $^{***}$ $p < 0.01$

| *Dependent Variable*: | Offering Yields (basis points) | | | | | |
|---|---|---|---|---|---|---|
| *Based on Issuers'*: | Average Size | | | Median Size | | |
| Treated $\times$ Post | (1) | (2) | (3) | (4) | (5) | (6) |
| $\times$ Small | -9.29 | -3.34 | -2.49 | -8.14 | -2.84 | -2.05 |
| | [-1.41] | [-0.55] | [-0.40] | [-1.26] | [-0.47] | [-0.34] |
| $\times$ Large | -17.84$^{***}$ | -15.17$^{***}$ | -12.19$^{***}$ | -18.23$^{***}$ | -15.72$^{***}$ | -12.68$^{***}$ |
| | [-6.32] | [-7.68] | [-6.67] | [-6.80] | [-8.33] | [-7.42] |
| Difference | 8.55 | 11.83 | 9.70 | 10.09 | 12.88 | 10.63 |
| p-value | 0.11 | 0.02 | 0.08 | 0.07 | 0.02 | 0.07 |
| Issuer FE | ✓ | ✓ | ✓ | ✓ | ✓ | ✓ |
| Controls | ✓ | ✓ | ✓ | ✓ | ✓ | ✓ |
| Advisor FE | ✓ | ✓ | ✓ | ✓ | ✓ | ✓ |
| Group-Yr. FE | ✓ | ✓ | ✓ | ✓ | ✓ | ✓ |
| State-Yr. FE | | ✓ | ✓ | | ✓ | ✓ |
| Advisor-Yr. FE | | | ✓ | | | ✓ |
| Adj.-$R^2$ | 0.858 | 0.863 | 0.868 | 0.858 | 0.863 | 0.868 |
| Obs. | 834,760 | 834,760 | 834,741 | 834,760 | 834,760 | 834,741 |

**Table A1.11:** Evidence from the Exit of Municipal Advisors (MA) - lower dependence

This table reports the results using Equation (4.1) with interactions to show the differential effect among issuers based on the exit of municipal advisors (MA). This analysis also includes group × year fixed effects. Column (1) shows results among all issuers with interactions corresponding to whether the issuer primarily depended on an exiting advisor or not. I define issuers linked to advisors when more than 25% of their municipal debt issuance in the pre-period is advised by the exiting advisor. I focus on the exit of regular advisors. These represent municipal advisors with at lease one issuance in each calendar year before the SEC Municipal Advisor Rule in the sample. Columns (2) and (3) show results for issuers that depend on exiting advisors. For these issuers, I show the heterogeneity between small and large issuers based on the median size of ex-ante issuances. Column (2) shows results for advised bonds only. Column (3) also includes bonds issued without any advisors, and the analysis does not include advisor × year fixed effects. T-statistics are reported in brackets and standard errors are clustered at the state level. * $p < 0.10$, ** $p < 0.05$, *** $p < 0.01$

| *Dependent Variable*: | Yield Spread (basis points) | | |
|---|---|---|---|
| *Sample of Issuers*: | All | Depending on Exiting MA | |
| Treated × Post | (1) | (2) | (3) |
| × Other | -16.93*** | | |
| | [-5.58] | | |
| × Depend on Exiting MA | -1.80 | | |
| | [-0.49] | | |
| × Small | | 14.69*** | 12.46 |
| | | [3.74] | [1.56] |
| × Large | | -8.85*** | -8.53*** |
| | | [-2.70] | [-2.85] |
| Difference | -15.13 | 23.54 | 20.98 |
| p-value | 0.00 | 0.00 | 0.02 |
| Issuer FE | ✓ | ✓ | ✓ |
| State-Yr. FE | ✓ | ✓ | ✓ |
| Controls | ✓ | ✓ | ✓ |
| Group-Yr. FE | ✓ | ✓ | ✓ |
| Advisor-Yr. FE | ✓ | ✓ | |
| Adj.-$R^2$ | 0.855 | 0.860 | 0.850 |
| Obs. | 834,741 | 250,538 | 294,592 |

**Table A1.12:** Evidence from the Exit of Municipal Advisors (MA) - by size of average issuance

This table reports the results using Equation (4.1) with interactions to show the differential effect among issuers based on the exit of municipal advisors (MA). This analysis also includes group × year fixed effects. Column (1) shows results among all issuers with interactions corresponding to whether the issuer primarily depended on an exiting advisor or not. I define issuers linked to advisors when more than 25% of their municipal debt issuance in the pre-period is advised by the exiting advisor. I focus on the exit of regular advisors. These represent municipal advisors with at lease one issuance in each calendar year before the SEC Municipal Advisor Rule in the sample. Columns (2) and (3) show results for issuers that depend on exiting advisors. For these issuers, I show the heterogeneity between small and large issuers based on the average size of ex-ante issuances. Column (2) shows results for advised bonds only. Column (3) also includes bonds issued without any advisors, and the analysis does not include advisor × year fixed effects. T-statistics are reported in brackets and standard errors are clustered at the state level. * $p < 0.10$, ** $p < 0.05$, *** $p < 0.01$

| *Dependent Variable*: | Yield Spread (basis points) | | |
|---|---|---|---|
| *Sample of Issuers*: | All | Depending on Exiting MA | |
| Treated × Post | (1) | (2) | (3) |
| × Other | -16.61*** | | |
| | [-5.60] | | |
| × Depend on Exiting MA | -0.73 | | |
| | [-0.22] | | |
| × Small | | 15.30*** | 14.96** |
| | | [5.66] | [2.52] |
| × Large | | -5.62 | -6.17* |
| | | [-1.49] | [-1.94] |
| Difference | -15.87 | 20.92 | 21.13 |
| p-value | 0.00 | 0.00 | 0.00 |
| Issuer FE | ✓ | ✓ | ✓ |
| State-Yr. FE | ✓ | ✓ | ✓ |
| Controls | ✓ | ✓ | ✓ |
| Group-Yr. FE | ✓ | ✓ | ✓ |
| Advisor-Yr. FE | ✓ | ✓ | |
| Adj.-$R^2$ | 0.855 | 0.861 | 0.854 |
| Obs. | 834,741 | 234,679 | 275,381 |

# APPENDIX B
# MISCELLANEOUS SECTION FOR CHAPTER 2

## B2.1 Corporate Subsidies in the U.S.

Since colonial times, businesses have been offered tax-related incentives [151, 152]. [153] provides some interesting details about the history of subsidy competition. As early as 1800, states financed infrastructure and offered capital to businesses. For example, Pennsylvania had invested USD 100 million in more than 150 corporations and placed directors on their boards by 1844. While intense rivalry between Pittsburgh and Philadelphia led to substantial investment in public infrastructure, widespread corruption also ensued. As a result, constitutional amendments outlawed some of these practices [154]. Relevant to many deals in our setting, Mississippi pioneered tax-exempt municipal bonds to attract industries in 1936. Subsequently, by 1959, 21 states had established state-level business development corporations. Much of this new economic development was often financed through debt. In the latter part of the $20^{th}$ century, the unemployment crises of the 1970s and the recessions in the early 1980s resulted in an aggressive war between states to win/retain jobs.

[155] document 15 of the most common business tax incentives ranging from corporate and personal tax exemption to various forms of tax credits related to job creation or research and development. [156] provides some broad conclusions about the usefulness of different types of tax incentives. Property taxes and tax concessions are fully capitalized into property values. As a result, tax increment financing (TIF) is not an effective economic redevelopment tool. On the other hand, increasing the corporate tax rate reduces employment and decreases business entry. Even so, there has been justification for such tax incentives with various motivations: protecting (retaining) businesses from being lost to other states, shielding businesses from competition, revitalizing failing firms [157, 158, 159] or attracting new firms from outside. When most states offer such subsidy bids and incentives, other states also make room for such developmental tools [160]. There is also an argument made in favor of subsidies since they are revenues forgone but not actual cash paid out. Another justification comes from [161]: if society has underemployed resources, then said resources could be used more productively through corporate incentive programs. The primary difficulty in understanding the overall impact of using subsidies for local economic development stems from the endogeneity: policy changes/moves are directly correlated with outcomes of interest [156]. In this regard, [162] provide a detailed toolkit of methods and best practices in evaluating spatially targeted urban redevelopment incentives. We try to incorporate some of those recommendations in our methodology and identification.

### B2.2 Data on Corporate Subsidies in the U.S.

#### B2.2.1 Subsidy Deals

The *Good Jobs First Subsidy Tracker* [81] provides a starting point with its compilation on establishment-level spending data. It sources these data from the state level dossiers on revenue forgone/credit offered in the Tax Expenditure Reports. Further, states also report incentives allocated through various programs provided by their respective economic development offices. Such disclosures are usually cited in the annual (or biennial) state-level budgets. The state Department of Revenue or Budget Office may be responsible for updating and maintaining such (web) archives. States that do not report establishment-level monetary spending through subsidies in their financial data also are present in the Subsidy Tracker dataset. News articles, press releases, and Freedom of Information Act (FOIA) requests are used/cited in the dataset for these additional deals. However, *Good Jobs First* does not contain an exhaustive list of all the subsidy programs launched and run by various states. At best, it may be most relevant for the larger set of discretionary subsidies floated by states and local governments. Overall, the existing dataset reports state-year level observations. For our purposes, the dataset needs to be enhanced with key variables that are not already recorded.

As of June 2018, the Subsidy Tracker files contained 606,899 records of subsidy items listed in their full dataset. We focus on records after 1990, wherein the year of subsidy is not missing. Also, our setting requires bidding competition between non-federal governments. Hence, we omit deals where the money is sponsored by the federal government of the United States. (Figure 2.1 brings out the proportion of federal versus non-federal incentives.) As a further check to verify for deals containing federal sponsorship, we check the raw data included in our sample for loans granted by the US government. While some of the composite subsidy packages may contain components offered by the US government, the total subsidy listed under state-level deals excludes these federal loans. In order to focus on large, economically meaningful deals for the local governments, we restrict our sample to subsidies with values exceeding USD 1 million. After dropping records below this threshold, we manually parse these records because they include repetitions at the firm or parent company level. Due to a lack of consistent nomenclature of firms, we parse this information through careful reading. Specifically, a “Megadeal” may include various incentives stitched together from money/tax abatement offered by the city, town, county and state governments. The Subsidy Tracker data may or may not include overlapping items at the state level. For instance, in 2006 the state of Florida offered USD 310 million as subsidy to Burnham Institute for Medical Research to locate their medical research facility in Orlando (Orange County), which included USD 155.3 million from the Innovation Incentive Fund. Given the existing overlap in the raw Subsidy Tracker data, both these observations show up after the above filters. Florida statutes list an Innovation Incentive Program[1] which is intended ‘to respond expeditiously to extraordinary economic opportunities and to compete effectively for high-value research and development, innovation business, and alternative and renewal energy projects.’ In archival reports, grants approved under the scheme date back to 1995–96.

We focus our analysis on 2005–2018 for our sample period. This imposition of calendar years chosen is based on the availability of secondary market transactions in municipal bonds, described in Section 2.3.2. From the variables listed by *Good Jobs First*, we are primarily interested in the company name, parent firm, firm location, year, subsidy amount, subsidy adjusted, level of subsidy (based on the government level), city/county of the facility, number of jobs promised and total investment. There are cases of missing information. Additional data is gathered on the FIPS code for the county, NAICS code for the proposed facility or firm, and the purpose of the subsidy: new plant/expansion, retention, or relocation. To distinguish between retention versus expansions, we rely on documented evidence in newspaper articles. A retention must be for a facility already operating in a location, while expansions may be a new unit/assembly line. Understandably, retentions are often without any fresh investments made by the firms. However, significant effort is devoted to comprehensively parse through local print media/newspapers to find out the losing county (and state) and earliest date of announcement for the subsidy/plant. These two variables were the most painstaking aspects of the data collection procedure. In this context, our dataset construction is more granular and focused than [76], who uses state-level bidding competition. The Subsidy Tracker dataset never provides information on the losing county nor the precise announcement date. In Table B2.1, we show a comparison of the original dataset versus the one constructed after hand-collection of relevant variables.

Through a careful manual reading of newspaper articles, we finally identify 199 winner-loser deal pairs at the county level. Where it was not possible to reasonably align a winning/losing city to a single county, all counties were included. Since losing county information in these articles is worded differently, it is challenging to automate the process through a programmed algorithm. We argue that given the incongruity between the size of the state and the subsidy offered, the local governments’ lens would be more relevant as a setting, in terms of proportion. Motivated by reasons similar to those cited in [163] on using the exact dates of law passage for announcement effects of anti-takeover provisions, we rely on the earliest available dates for a given deal. Specifically, if the date of incentive approval/announcement is before the plant announcement date, we use this date because the market already learns of the potential subsidy offer. However, occasionally the facility announcement predates the disclosures of all incentives that may have been offered to attract the firm.

There is inherent secrecy maintained by local governments and economic development board officials about such subsidy offers. The underlying assumption is that disclosures would invite other competitors; alternatively, it could invite moral hazard problems for counties that may be desperate to win new jobs. For instance, consider the case of Burnham Institute for Medical Research choosing Florida in 2006. Local officials refused to share details of the economic incentives bid in the public media for fear of instigating more competition from other locations/bidders (See Figure B2.1). A snapshot of some project names attributed to subsidy deals is provided in Table B2.2 of the Appendix.

[1] https://tinyurl.com/y2jze7ys

#### B2.2.2 Caveats

State economic development boards often revamp their (web) archives when the officer/Governor in charge loses power. This also becomes a hurdle in collecting information. As indicated before, we do not claim to have collected the full universe of the subsidies offered to corporations. In fact, doing so may complicate the task of identifying the impact of corporate subsidies on the local governments, based on insignificant amounts waived off in abatement. Also, there is no way to ascertain what subsidy bid was offered by the runner-up county/location. Only in some cases do newspaper stories carry information about the competing bid offered. Largely, this remains unobserved in the current setting - and we acknowledge this as a major limitation in the data. This is especially true in cases where more than one city is known to have competed. For deals with multiple losers, there is no direct way to ascertain which of the losers was the closest to getting the deal. We base our judgment on a subjective assessment of grammatical hints available in the article documenting the story. Therefore, this is not a robust way to identify the closest runner-up location. To replicate the interstate competition, wherever possible, priority is offered to a location outside the winning state in assigning the runner-up county (for multiple runners-up).

### B2.3 Additional Results

In Section B2.3.1, we provide results of falsification tests. Next, in Section B2.3.2, we provide robustness of our baseline specifications. Our results in Section B2.3.3 show the impact on the probability of rating downgrades. In Section B2.3.4, we analyze the heterogeneity in relative bargaining power between the county and the firm involved in the subsidy deal.

#### B2.3.1 Falsification Tests: Pre-Refunded Bonds

It is typical for municipal bond issuers to pre-refund bonds before the call date by issuing new debt and holding the proceeds in a trust to fund remaining payments until the call date. This would effectively render the pre-refunded bonds nearly risk free [164, 165, 5]. Local governments may choose to pre-refund their bonds, thereby offering a clean change of said bonds from risky to risk-free. We exploit this argument to claim that bonds which have been thus "insured" would not see any significant change in their yield spreads in our setting of Equation (4.1).

To construct the sample of pre-refunded bonds for this test, we follow [15] and [5]. We apply the following filters to our sample: keep only the pre-refunded bonds, excluding bonds that are not exempt from federal and within-state income taxes; and exclude pre-refunded bonds that are not escrowed by Treasury securities, State and Local Government Series (SLGS), or cash. Finally, we exclude bonds that are pre-refunded within 90 days of the call date since the Internal Revenue Service treats these transactions as different from pre-refunding, instead classifying them as current refunding. Our analysis focuses on insured bonds only. Table B2.7 shows the results of the falsification test. In Column (1), we find that the average yield for winners goes up by a small magnitude of about 2.17 bps, but the measure is not statistically significant. Likewise, in Column (2), we do not find any significant change to the yield spread as the outcome variable among the pre-refunded bonds. Finally, Columns (3)-(5) report the effect on after-tax yield spreads without and with county controls, respectively. The magnitude is similar but again the estimates are statistically insignificant. Thus, we do not find any impact on these pre-refunded bonds as they have been secured against the escrow of funds earmarked for their outstanding payments. The absence of any marginal impact in the subset of pre-refunded bonds suggests that our main effect is not driven by overall market conditions in the US municipal bond market.

#### B2.3.2 Further Robustness Tests

In Table B2.8, we report results from additional considerations of robustness to our baseline specification using Equation (4.1) as reported in Column (6) of Panel A in Table 3.2.

*Is the effect driven by the size of trades?*

In 2018, about USD 0.96 trillion out of USD 3.25 trillion of the municipal bond holdings was managed by money market mutual funds and exchange-traded funds. One potential concern is that few large institutional trades may be driving our main result. We separate our results into sub-samples of trades constituting various buckets in Panel A. Retail-sized transactions usually correspond to $100,000 or less [2]. Columns (1)-(3) depict the main effect from Equation (4.1), as derived from trade sizes worth $\leq$ $25,000, $\leq$ $50,000, and $\leq$ $100,000. The increase in borrowing cost is over 15 bps in each of these sub-samples, which is higher than our baseline estimate. This suggests that our main result is also present in smaller transactions.

[2] http://www.msrb.org/~/media/Files/Resources/Mark-Up-Disclosure-and-Trading.ashx?

On the other hand, the lack of information among retail investors may be driving our results. To address this, we now use bond-month observations based on trade sizes worth > \$25,000, > \$50,000, and > \$100,000 in Columns (4)-(6). As before, we still report an increase in bond yield spreads of over 15 bps in each of these sub-samples. Based on this evidence, we argue against the size of trades explaining our main result.

*Are results driven by newly issued bonds?*

Even though our data from MSRB on secondary market bond yield spreads is cleaned for primary-market transactions recorded therein, we assume further precaution in favor of seasoned bonds. We show these results in Panel B. In Columns (1)-(5), we report our baseline results by dropping bonds that were recently issued with respect to the deal i.e., within 6, 12, 18, 24, and 36 months of the subsidy announcement date, respectively. By doing so, we remove bonds from the sample that have been newly issued before or after the subsidy and thus may demonstrate unusual trading in the initial phases. Our main effect still shows up as nearly 10–14 bps in each of these columns, even as the sample size reduces in Column (5). This shows that our results are not solely driven by trading activity in newly issued bonds around the subsidy announcement dates. Moreover, in Columns (6)-(8), we use the complementary sub-samples by only keeping bonds that were recently issued around the subsidy announcement dates. Understandably, the sample size shrinks substantially in these analyses focusing on bonds issued within 18, 24, and 36 months of the subsidy dates. Even so, our results show that we still find the baseline effect to be over 9 bps. This provides further evidence that our results are not affected by including or dropping newly issued bonds.

*Additional County and Bond Level Considerations*

In Panel C, we show our baseline results for robustness against additional considerations, as detailed below. A potential worry is that the sample period spans the financial crisis of 2009. Understandably, this was a period of major volatility in the financial markets across asset classes, and municipal bonds were no exception. As a result, we report our findings by showing our results for periods before and after 2009. In Columns (1)-(2), we report the main coefficient of interest, $\beta_0$, by interacting the baseline Equation (4.1) with dummies corresponding to events before and after 2009, respectively. We find that the increase in yield spreads is 17.61 bps for subsidy deals before 2009 and 14.59 bps after 2009. The difference between these two coefficients is statistically insignificant. This suggests that our results are not singularly driven by events belonging to either side of the financial crisis of 2009.

In Column (3), we report our results from more county-level economic considerations. There is some evidence that firms' decisions to locate in a region may increase house prices locally[3]. This may also be correlated with local household incomes. In this regard, we report our results in Columns (3) by additionally controlling for the values of log(household income) and log(house price index). We find that the main result is 10.92 bps, which is economically meaningful and statistically significant.

In the baseline specification, we do not include bond ratings so that we can analyze both rated and unrated bond transactions. Here, we check on the robustness of our main results using only those bonds for which the most recent bond ratings are available from S&P's credit ratings. We show this result in Column (4) of Table B2.8 by introducing the numeric equivalent of bond level ratings among the regressors. The magnitude goes up to over 14.02 bps, and the result is statistically significant.

*Subsidy Deals Sponsored by the State*

Our baseline sample includes subsidy deals sponsored by state and local governments. Due to limited information, it is difficult to precisely pin down the amount of subsidy borne by the state versus local governments. However, our main result is likely driven by the additional borrowing that local governments may have to undertake to finance the new economic activities after the subsidy. Nevertheless, we try to address any concerns due to state versus local apportionment of the subsidy package. Specifically, in Column (5) of Panel C, we drop deals involving subsidy package sponsored by the state and local governments. We rely on print media evidence to make this distinction, by searching for each deal individually. This leaves us with 157 deals in the sample. We find the coefficient estimate to be 16.29 bps. The effect is similar to the baseline magnitude and is statistically significant.

*Heterogeneity based on Bond Purpose*

Our analyses includes different types of bonds issued by the various county authorities. There may be a concern that some bonds that trade more frequently may be weighed disproportionately in our estimation. Different issuers and bonds in the same county may be associated with different underlying risks. To this end, we present our results in Panel D of Table B2.8. We categorize bonds into five groups based on their use of proceeds, namely: education, improvement and development, transportation and utilities, water-sewer, and other public services.

First, to control for unobserved heterogeneity based on the use of proceeds (bond's purpose), we add purpose fixed effects to the baseline. We report this coefficient in Column (1) as 14.78 bps. Thereafter, we also consider the time-varying unobserved heterogeneity based on the purpose of the bonds by incorporating bond purpose × year-month fixed effects. We show our results for this specification in Column (2) as 13.35 bps. In Column (3), we introduce county-pair × bond purpose fixed effects. This accounts for unobserved heterogeneity across bond purpose types within a given county-pair. We find that our estimate of 14.49 bps is similar to the baseline effect. Finally, in Column (4) we use a weighted regression approach. The weights correspond to the fraction of par volume traded in a given purpose of the bond (use of proceeds) relative to the county aggregate. For example, this could refer to education bonds, highways, healthcare, and other infrastructure. With this consideration, the winning counties' after-tax yield spreads increased by 15.12 bps after the subsidy announcement.

[3]https://www.wsj.com/articles/amazon-primed-to-boost-property-prices-in-winning-hq2-cities-1542715200

*Clustering*

Next, in Panel F, we consider alternative ways to cluster standard errors in our baseline specification of Equation (4.1). We report our results in Columns (1)-(3). In our baseline we double cluster standard errors at county specific bond issuer and year-month level to account for correlations in yields for a given issuer. Here, in Column (1), we show that our main result is not affected by single clustering standard errors at the county-specific bond issuer level. This would be relevant if there is a concern that yield spreads from bonds of the same municipal bond issuer may be correlated with another. We show that our baseline effect is statistically significant under this consideration. In Column (2), we single cluster standard errors by bond issue. Finally, in Column (3), we show our results by double clustering standard errors by county bond issue and year-month. This specification addresses concerns of correlation among standard errors in yield spreads within bond issues over time. The statistical significance in all of these considerations suggests that our results are robust to alternative strategies of clustering.

*Other Dependent Variables*

Finally, for the robustness of our main result to the choice of the dependent variable, we show our results in Panel F. First, in Column (1), we use the average yield as the dependent variable and estimate the coefficient of interest as 10.11 bps without using any controls in the regression framework. Next, we add the bond level and county level controls into the regression. Our results show an increase in yield of 10.75 bps in Column (2). The magnitudes are lower in this case as we do not adjust for tax differentials among states. Likewise, we repeat this scheme using after-tax yield as the dependent variable in Columns (3) and (4). Overall, we show that we find similar results by using dependent variables in which yields are not adjusted for taxes or only adjusted for taxes.

#### B2.3.3 Probability of Rating Downgrades

The results in the previous subsections highlight that winning counties with a lower debt capacity and a lower jobs multiplier observe an increase in yield spreads and thus a reduction in the value for municipal bondholders. Next, we test if rating agencies react to such an event.[4] Specifically, we evaluate the impact on bond ratings by considering the probability of rating downgrades. Our municipal bond ratings come from S&P ratings, as provided by FTSE Russell's municipal bonds database. We use a dummy variable that switches to one after a rating downgrade (and zero otherwise) as the dependent variable in Equation (4.1). Given the nature of our county-pair based cohort fixed effect, we rely on a simple ordinary least squares regression instead of a logistic regression [166]. Similar to our baseline regression, we expect that the coefficient of interest, $\beta_0$, would capture the differential probability of rating downgrades on winning counties after the deal. We report our results in Figure B2.9.

Rating downgrades in the municipal bond market are not very frequent. On average, the probability of rating downgrades in our sample is 18.57%. First, we show that the probability of downgrade across all bonds for winning counties in the sample is statistically indistinguishable from zero after the subsidy announcement. After that, we show the effect based on high and low values of the ex-ante debt capacity among winners using interest expenditure and net debt. Using the *Interest/Revenue*$_1$ ratio, we find that the probability of rating downgrade increases by 4.85% after the subsidy announcement for winners in the high group. This represents 26.1% (=4.85/18.57) of the average likelihood of rating downgrades. The differential impact between the high and low groups is 6.87 percentage points, which is statistically significant. Subsequently, we find that the differential increase of the probability of a rating downgrade is 4.65% for winning counties with a high *Interest/Revenue*$_2$ ratio.[5] Similarly, our results using the ratio of interest to debt suggest a higher probability of rating downgrades among winning counties with above median values of this metric.

#### B2.3.4 Relative Bargaining Power of the County versus the Firm

Previously, we considered the impact of the debt capacity of the county and the anticipated jobs multiplier effects of the subsidy deal on the municipal bond yield spreads. The amount of subsidy offered likely could depend on the relative bargaining power between the counties and the firm involved in the deal. We argue that while firms may hire site consultants to conduct their search through a bidding mechanism[6], local governments may not have access to such sophisticated resources. To assess the relative bargaining power between the county and the firm, we use the following: a) *Proposed Value*, b) ratio of investment to state revenue, c) intensity of bidding competition, and d) county's unemployment rate. We present our results in Table B2.9.

[4]Moody's downgraded Racine County's credit worthiness in Wisconsin after the announcement of Foxconn's incentives. For details, see: https://www.bondbuyer.com/news/foxconn-incentives-costly-for-a-wisconsin-countys-rating

[5]The increase in bond yield spreads after the subsidy announcement could be associated with agency costs whereby politicians seek short-term outcomes for future elections. While this is interesting by itself, it is difficult to measure the agency problems due to data limitations. In this paper, we focus on the underlying debt capacity as the mechanism.

First, we divide the winning counties based on the median value of *Proposed Value*. Our measure is obtained by taking the ratio of the differential between proposed investment and subsidy to the county's lagged revenue. We hypothesize that if the excess value proposed from the investment, beyond the subsidy, is small relative to the county (size represented by revenue), then the county's relative bargaining power is likely to be lower. Therefore, the impact on borrowing costs would be higher. In Column (1) of Table B2.9, we report our results for the baseline Equation (4.1) interacted with dummies of this measure. We also control for group-year month fixed effects. We find that yield spreads go up by 14.91 bps for deals with low proposed value. We find a similar impact by using our second measure of investment to state revenue ratio. We argue that when the value of the investment to be made by the firm is relatively larger than the state's revenue, the county has a lower bargaining power. As shown in Column (2), the cost increases by 16.51 bps for the high group with a relatively milder effect when the corresponding ratio is low.

Given that state-level governments often support the competition for firms' investments, we construct our next measure at the state level. We calculate the ratio of the state-level budget surplus to revenue and use the gap between the winning and losing states as a measure of the intensity of competition. As the gap between states widens, the competition is likely to be lower, and the county's bargaining power is expected to be higher. In Column (3), we show our results based on the interaction with the intensity of competition. We find that the secondary market yields go up by 20.30 bps when the intensity of competition is high (surplus to revenue gap is low). Finally, we show our results based on the county-level unemployment rate in Column (4). We expect counties with a high unemployment rate to have lower bargaining power in the bidding process, resulting in a greater impact on after-tax yield spreads (19.06 bps).

Overall, we argue that the four measures of bargaining power highlight the differential impact of relative bargaining power between the county and the firm. The municipal bond market reaction to the deal is linked to this bargaining power, which is also related to the amount of subsidy offered. A lower bargaining power causes a greater increase in yields.

---

[6]For instance, The Wall Street Journal reported on a cadre of consultants who help companies decide the location of their projects: https://www.wsj.com/articles/meet-the-fixers-pitting-states-against-each-other-to-win-tax-breaks-for-new-factories-11558152005

B2.3.5 Impact of Local Political Environment

The local political environment may also influence the competition for bidding with subsidies. Politicians may be motivated to aggressively bid for corporations and the promised jobs when they are contesting in close elections. We examine this in our analysis by interacting our baseline Equation (4.1) with dummies corresponding to close elections among bidding counties and otherwise. We modify the baseline specification to include group × month fixed effects to control for the average effect within a particular group of counties. We show these results in Table B2.10. Our analysis focuses on 5% (Columns (1)-(2)) and 10% (Columns (3)-(4)) of the victory margin, respectively. Moreover, we show results by separately considering the voting margin among winners only versus among bidding counties. We use the most recent victory margin from the top two candidates in a county before the subsidy from [167]. In Column (1), we show that the after-tax yield spread increases by 40.44 bps for the winning counties when there is a close election among winners. We find similar results in using other approaches to capture close elections in Columns (2)-(4). However, our analysis here is limited by the small number of counties that indeed fall in the "close" elections group. This may affect the precision of our estimates.

As an alternative approach to shed light on the importance of political motivation, we also consider the distance to upcoming county legislative elections. We use similar data as above to identify the number of months between the upcoming (next) election and the month of the subsidy announcement. Using our baseline Equation (4.1) interacted with dummies corresponding to upcoming elections, we show our results in Table B2.11. We divide the winning counties into two groups around the median and around the top tercile, respectively. First, we show results by assigning counties with a below median value of months to election as "upcoming" elections. The interacted version of the baseline model with group × month fixed effects in Column (2) shows that after-tax yields increase by 23.67 bps with an upcoming election. Following a similar approach with distance to elections in the bottom tercile designated as "upcoming", we find that yields increase by 37.24 bps in Column (4). Taken together, these results suggest that the impact of winning a subsidy deal is higher when there may be higher political motivation to win the bidding competition. We try to capture these effects through close elections and upcoming elections at the county level.

Overall, our results in this section point toward how the political environment before the subsidy announcement may affect municipal bond yields. These results suggest that the impact on municipal bond yields is higher when politicians may be more aggressive due to closely contested local elections or upcoming local elections.

#### B2.3.6 Heterogeneity based on Subsidy-to-Jobs Ratio

We also consider the heterogeneity in the subsidy-to-jobs ratio among deals. In Table B2.12, we use the ratio of the subsidy offered to the number of jobs promised at the deal level. We interact the baseline Equation (4.1) with dummies corresponding to above and below median levels of the subsidy-to-jobs-ratio. In Columns (1)-(2), we show results with after-tax yield spread from the secondary market as the dependent variable. We find that the impact on yields is higher (19.09 bps in Column (2)) when the deal is above the median value of the subsidy-to-jobs ratio. We find similar results when we use the offering yield of new municipal bonds from the primary market as the dependent variable. In fact, the primary offering yields actually reduce for subsidy deals with low subsidy-to-jobs ratio.

#### B2.3.7 Impact on Local House Prices and Property Tax Revenue

We analyze the impact on the local house price index and local property tax revenue using the baseline Equation (4.1). We also show results for the interaction effect based on the debt capacity of counties. When we sort the winning counties by $\textit{Interest/Revenue}_1$, we preserved the winning-losing county deal pairs. From our results in Figure B2.6, we show that there are no economic differences by dividing the winners based on their debt capacity using $\textit{Interest/Revenue}_1$ ratio. First, in Table B2.13, we show results using the logged value of the county-level house price index from the Federal Housing Finance Agency (FHFA) as the dependent variable. In Columns (1) and (2), we show results around three and five years of the subsidy announcement, respectively. We find that there is an aggregate increase in prices, as indicated by the positive coefficient on the *Post* dummy. However, the differential effect on winning counties seems to be negative. Further, in Columns (3) and (4), we show estimates from the interaction of the main coefficient with dummies corresponding to the debt capacity. We find that the decline in house prices of winning counties is primarily driven by low debt capacity (high interest expenditure) counties. In Table B2.14, we perform a similar analysis using the logged value of the property tax revenue at the county level as the dependent variable. We do not find any significant differences for the overall effect on winning counties around three and five years in Columns (1) and (2), respectively. Subsequently, the estimates in Columns (3) and (4) do not show much difference between low versus high debt capacity winning counties.

# Bid to woo biomed firm: $90 million: Leaders hope to lure a California research center to Orlando

Mark Schlueb and David Damron . Knight Ridder Tribune Business News ; Washington [Washington]11 Apr 2006: 1.

Local officials have refused to reveal details of the economic- development bid they've dubbed "Project Power," fearing that doing so would tip off as many as a dozen other suitors. Public records and those involved indicate the total value of the bid from Orange County, Orlando and the developer of Lake Nona -- the sprawling development where the institute's satellite operation would be built -- was still in flux Monday.

"This is a competition, and it is something of a long shot. But I believe we have assembled here a very strong team," Orange County Mayor Rich Crotty said. He cited estimates by local economic- development officials that the

**Figure B2.1:** Project Secrecy:
This figure provides an excerpt representing an instance of secrecy maintained by local officials in their process to bid for a project by offering incentives.

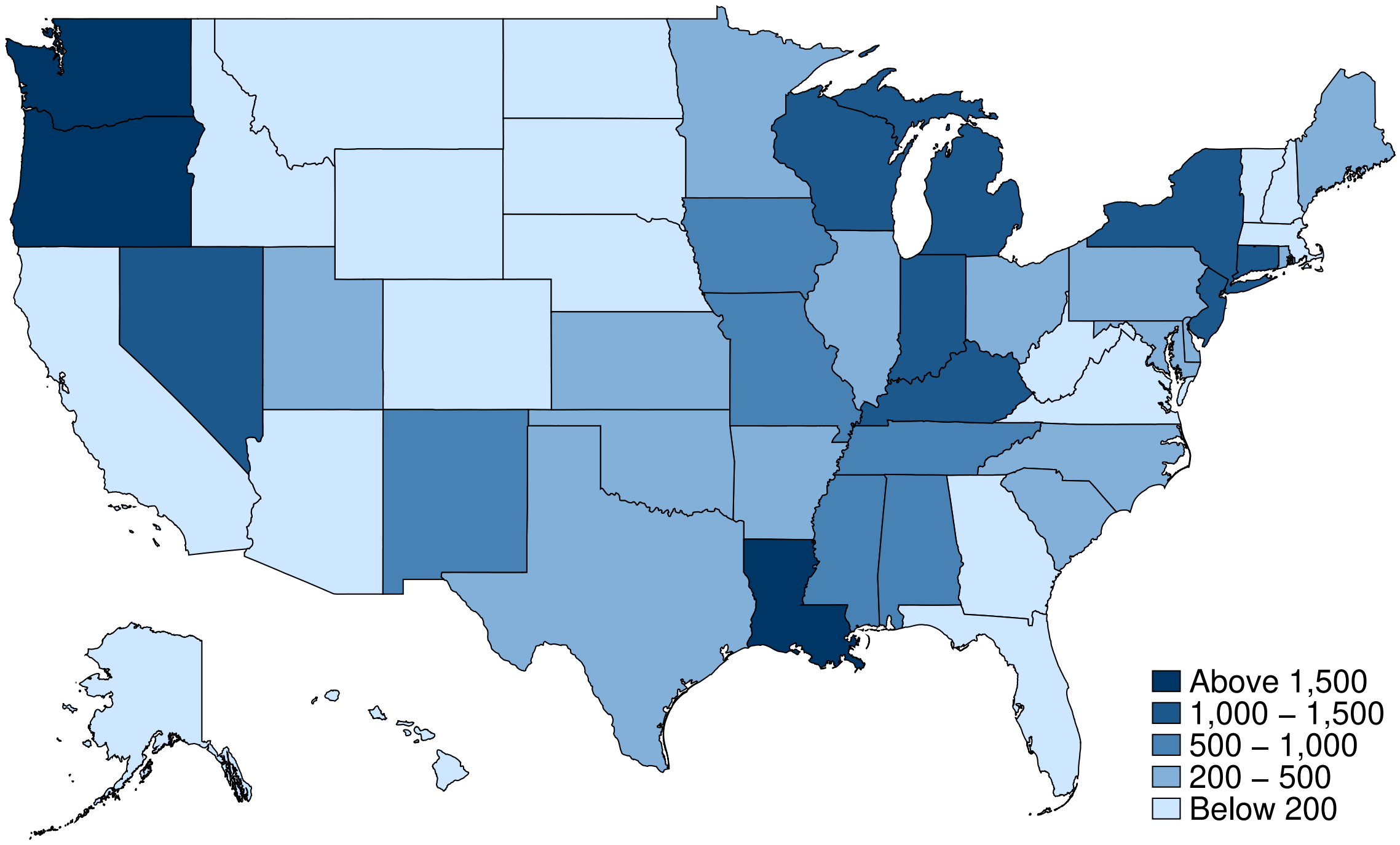

**Figure B2.2:** Subsidy per Capita:
The state-level distribution of subsidy per capita (in USD) is shown for the period 2005-2018. Calculated based on Source: *Good Jobs First, Subsidy Tracker*

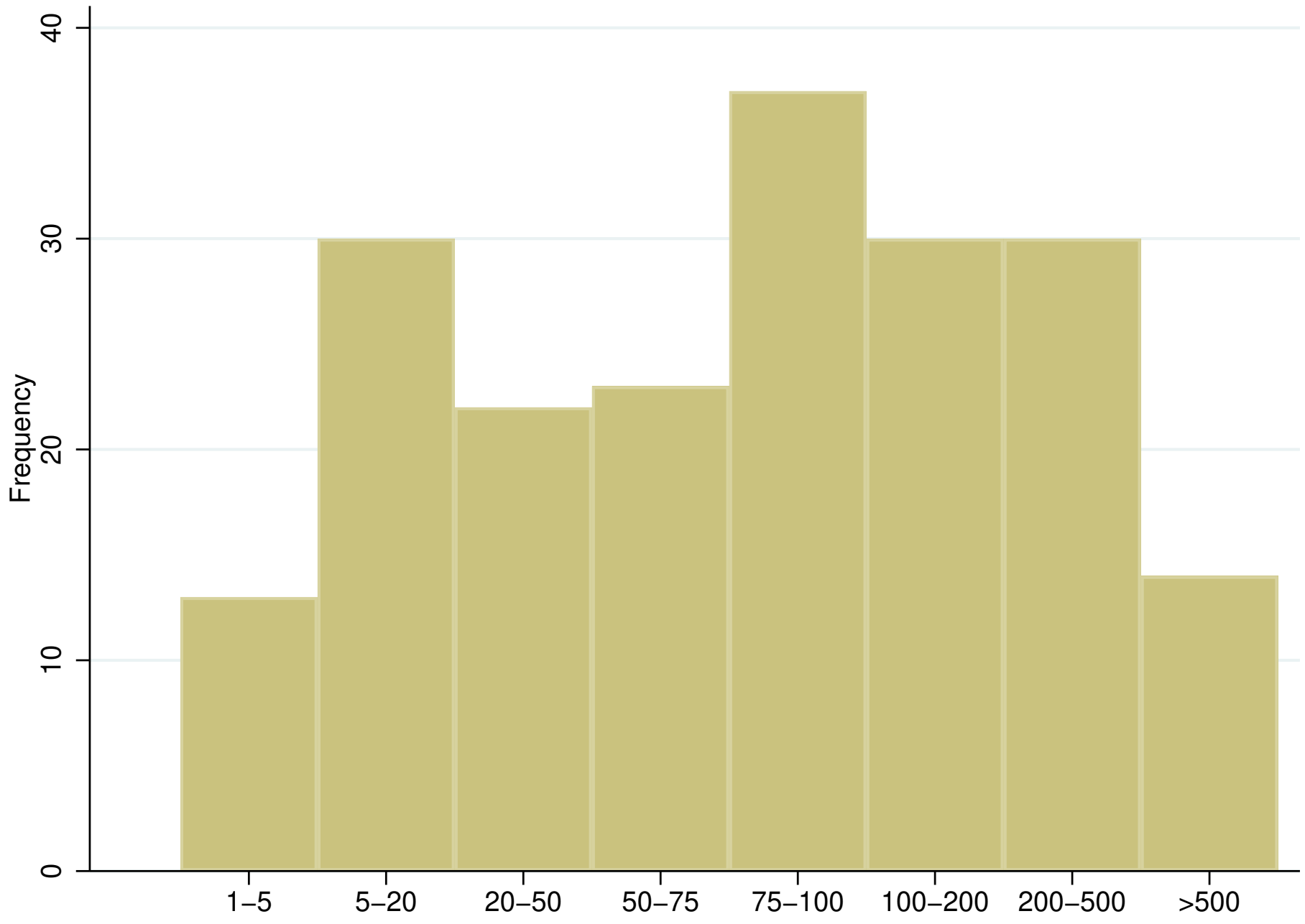

**Figure B2.3:** Distribution of Subsidy (USD million):
In this figure, we plot the number of deals in our sample of winner-loser pairs during 2005-2018 across different ranges of subsidy bins. The horizontal axis shows the subsidy bins (in USD million).

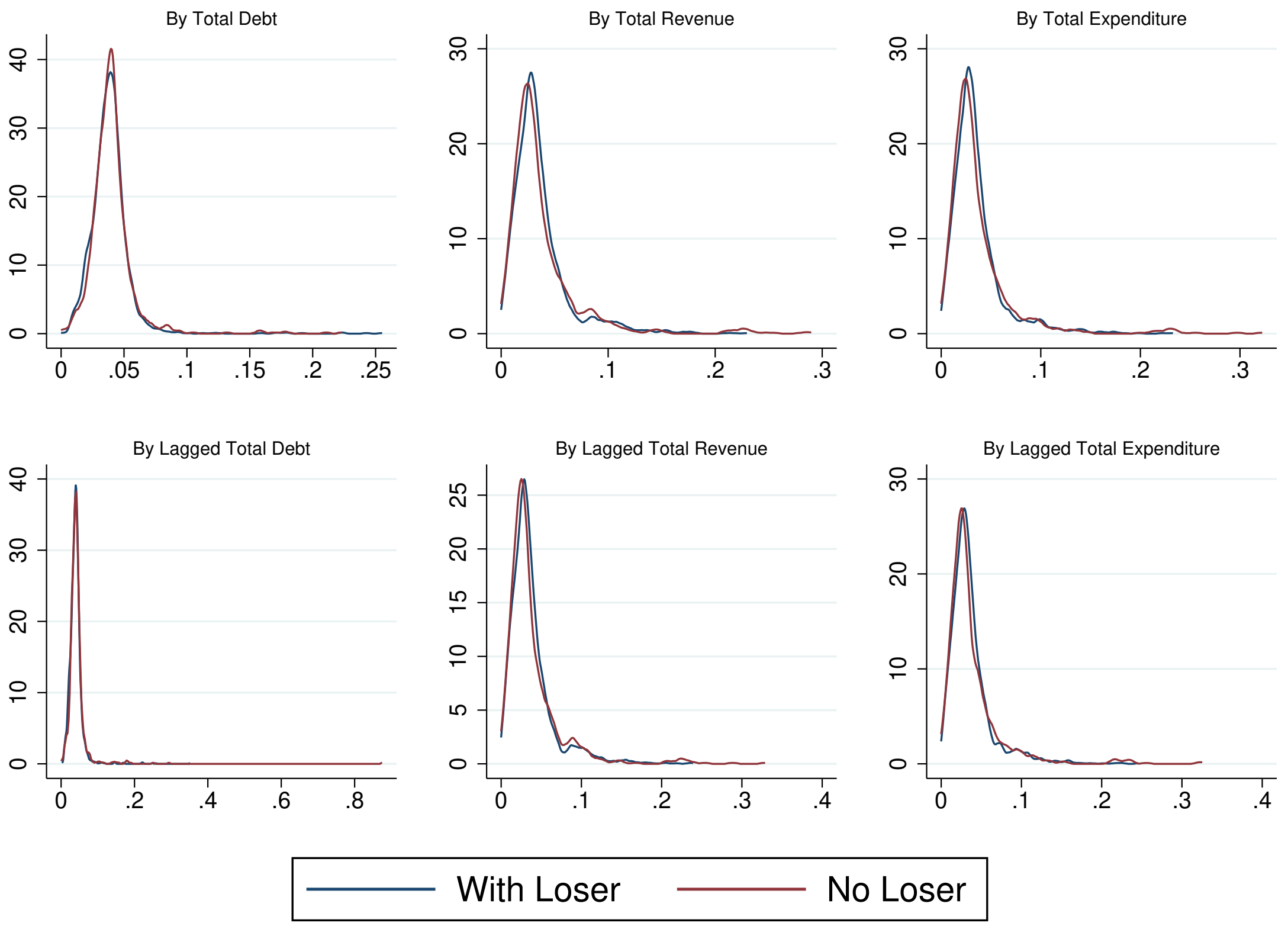

**Figure B2.4:** Comparing Counties: With Loser vs No Loser:
This figure shows the distribution of county-level interest on general debt, scaled by various fiscal metrics for the county. Specifically, we use county's total debt, total revenue and total expenditure, along with their corresponding lagged values. We compare the counties in the sample with loser versus others not in the sample during 2005-2018.

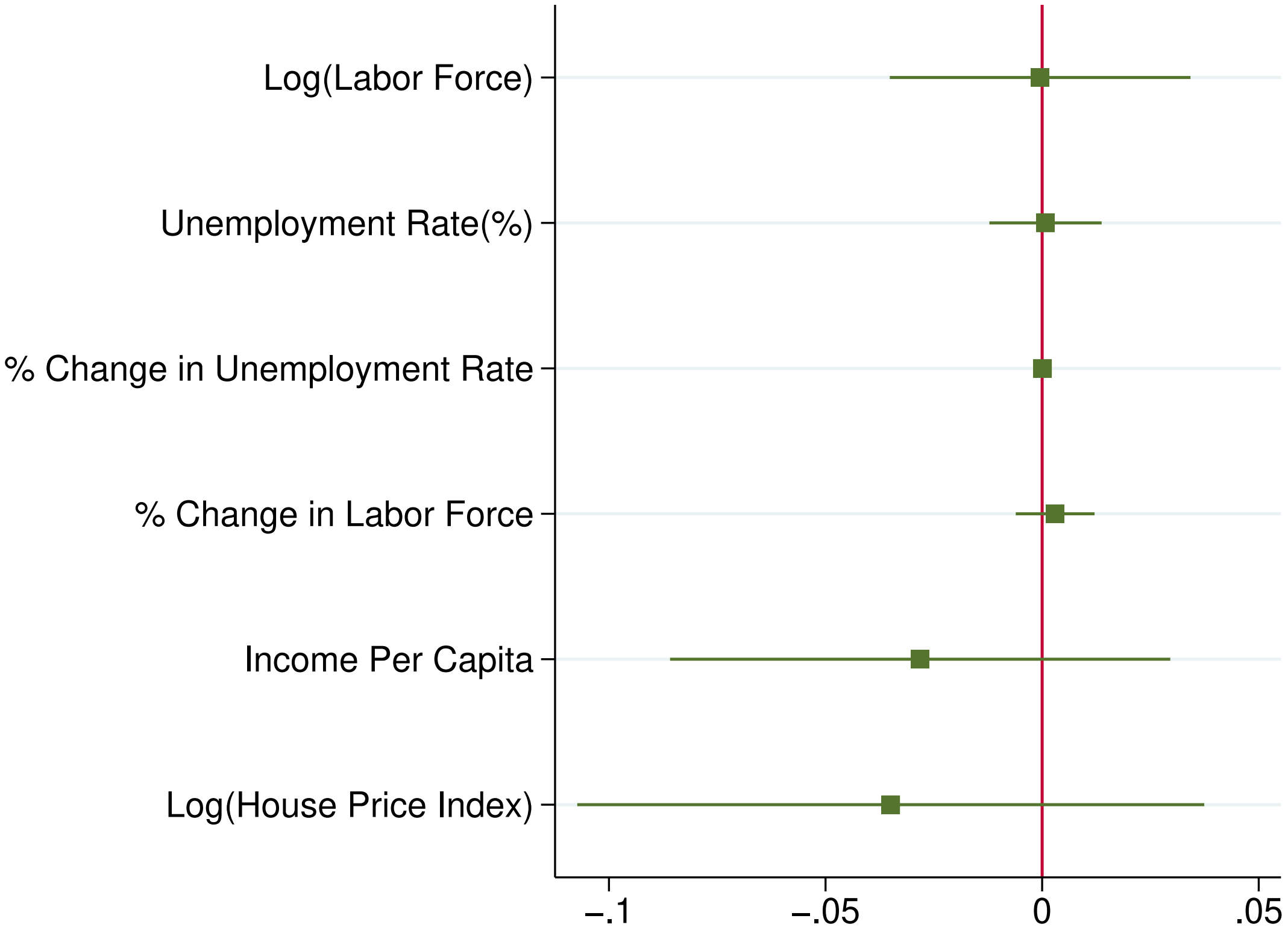

**Figure B2.5:** Predicting Winners:
This figure reports the regression coefficients of a linear probability model predicting the winners using local economic variables. We use local economic variables three years before the deal. Confidence intervals at the 95% level are plotted.

**Panel A: Log(Aggregate Employment) By Interest/Revenue$_1$:**

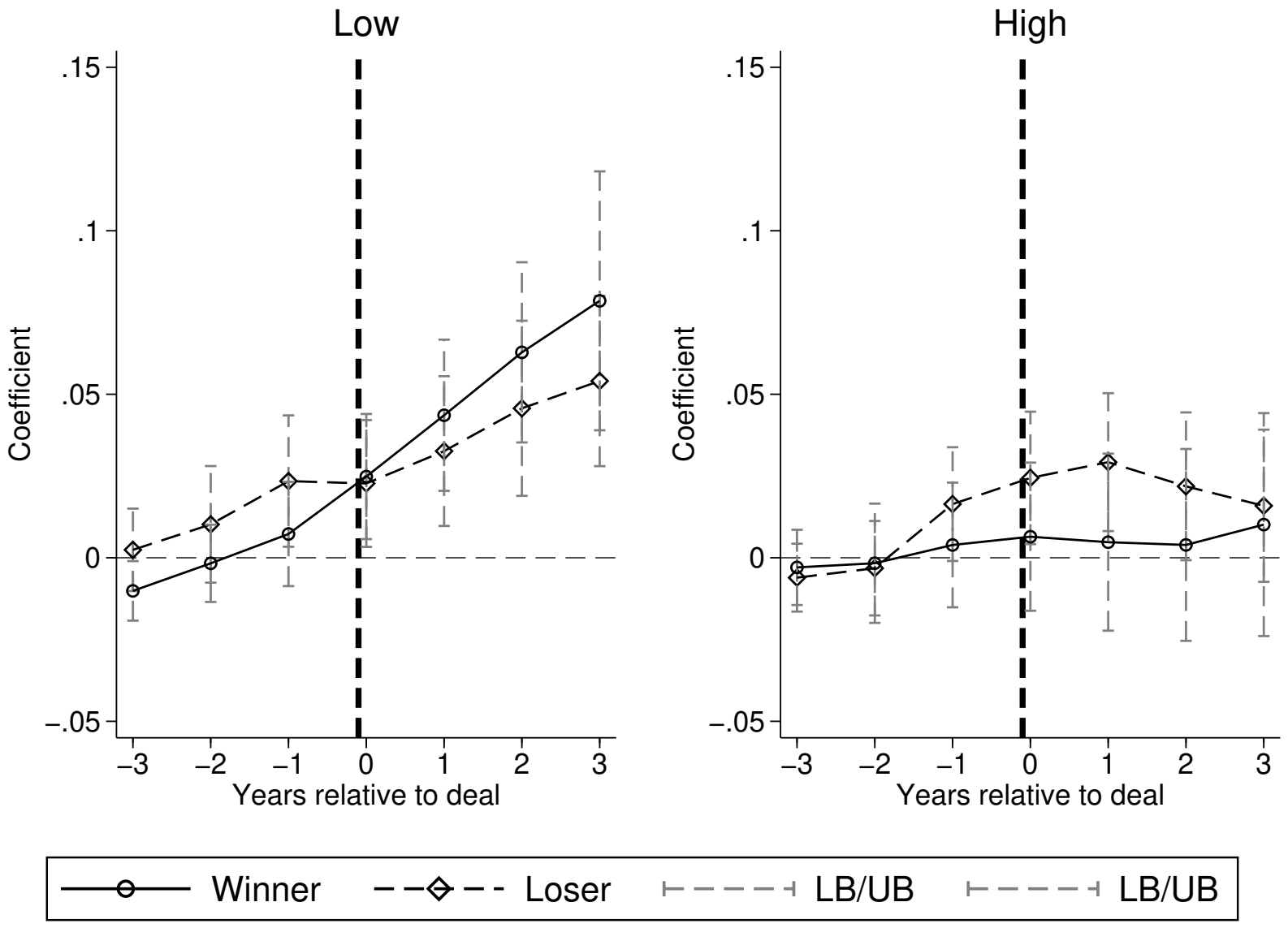

**Panel B: Unemployment Rate By Interest/Revenue$_1$:**

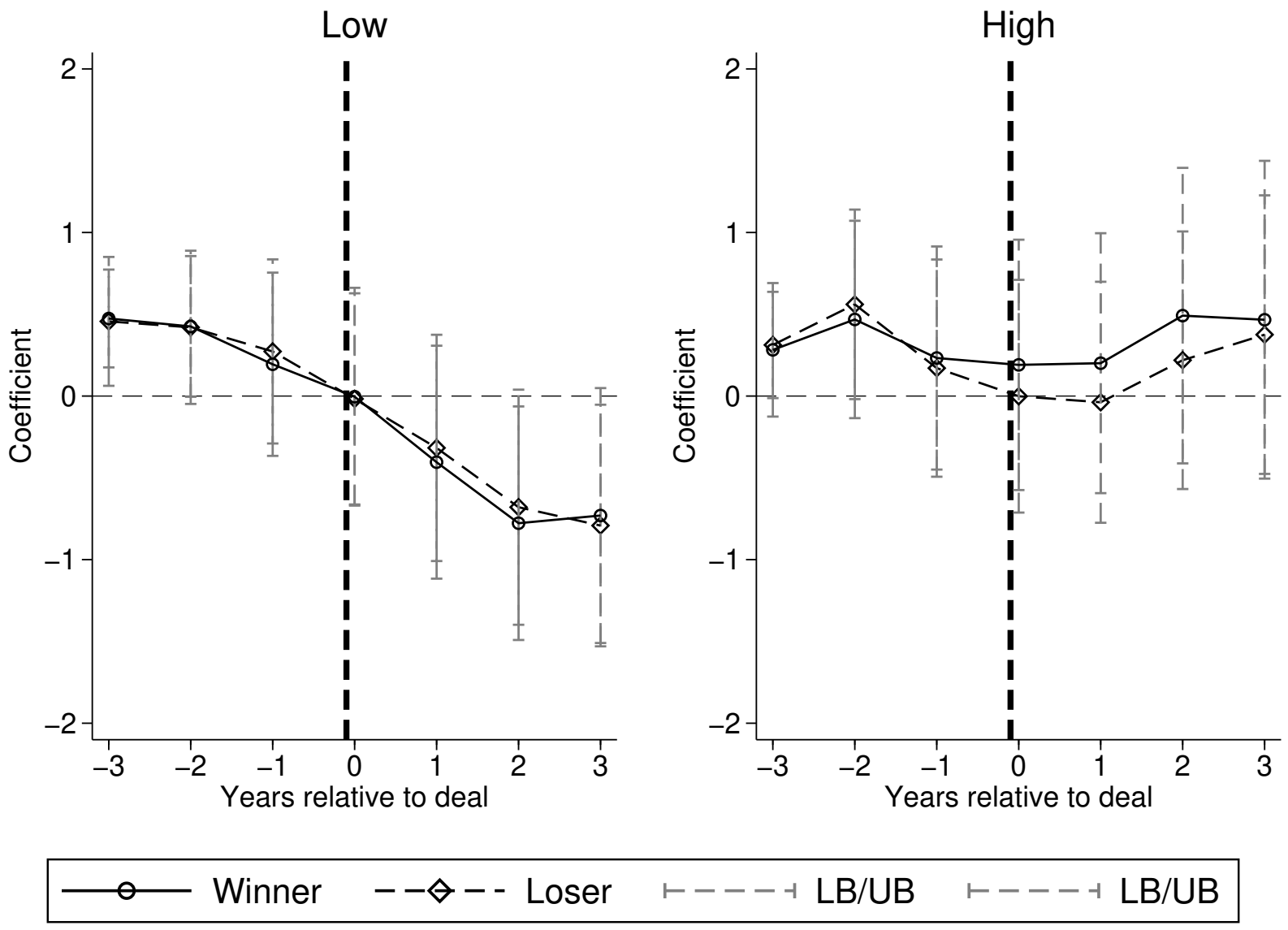

**Figure B2.6:** Identifying Assumption by Debt Capacity - Local Economy:
The figure shows the local economic conditions at the county level between the bidding counties, around the event of subsidy announcement by debt capacity. We split the sample based on the *Interest/Revenue*$_1$ ratio for the winning counties (defined as the ratio of interest on general debt to total revenue of the county). A low value of *Interest/Revenue*$_1$ ratio suggests a high debt capacity for the winning county. We use the annualized version of Equation (3.4). Here, we cluster standard errors at the deal level. Panel A shows aggregate employment. Panel B corresponds to unemployment rate. In Panel C, we focus on county-level credit ratings. Panel D shows local beta. The benchmark period is the year before the window (-3,3) years. The dashed lines represent 95% confidence intervals.

**Panel C: County Rating By Interest/Revenue$_1$:**

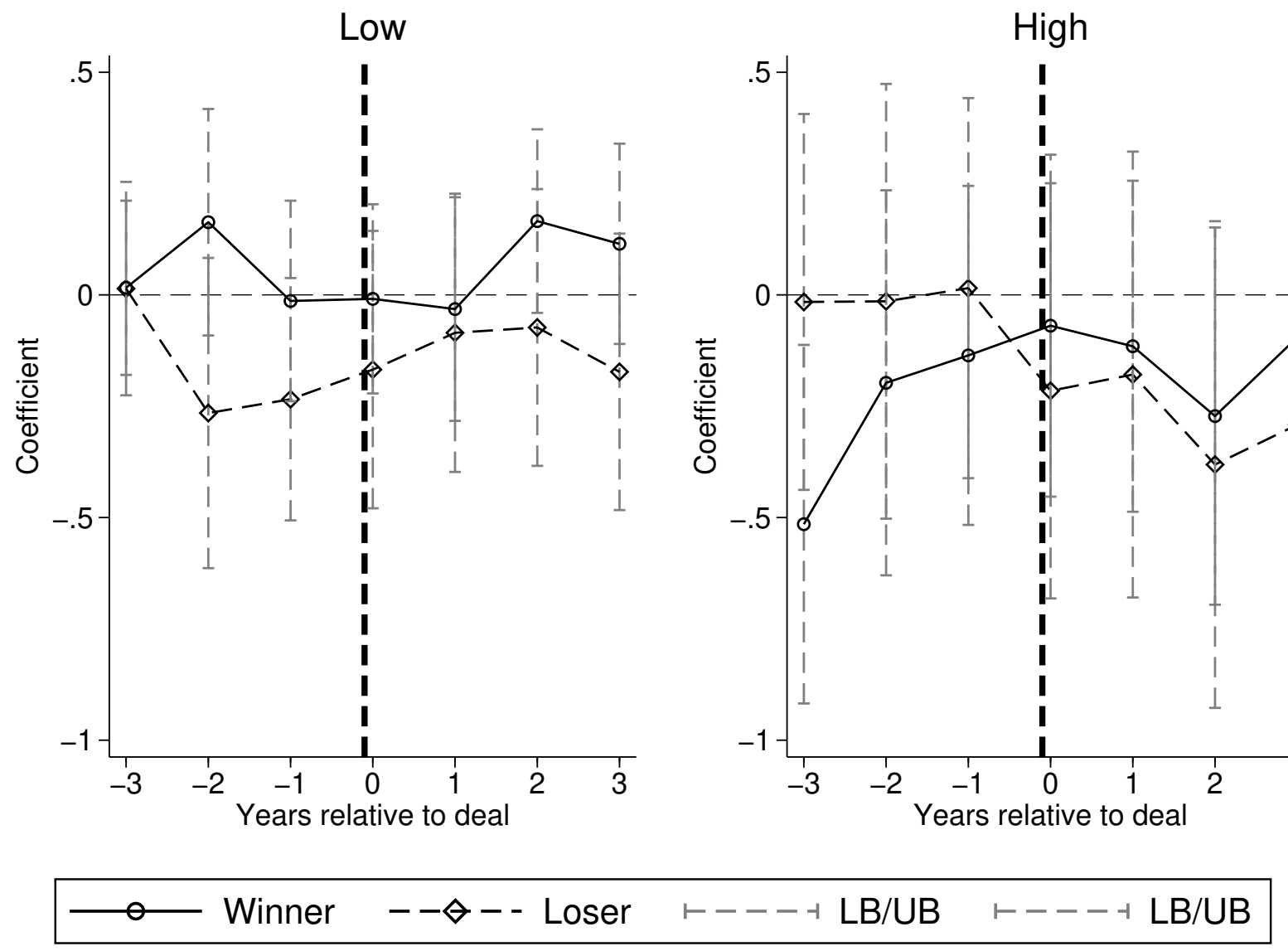

**Panel D: Local Beta By Interest/Revenue$_1$:**

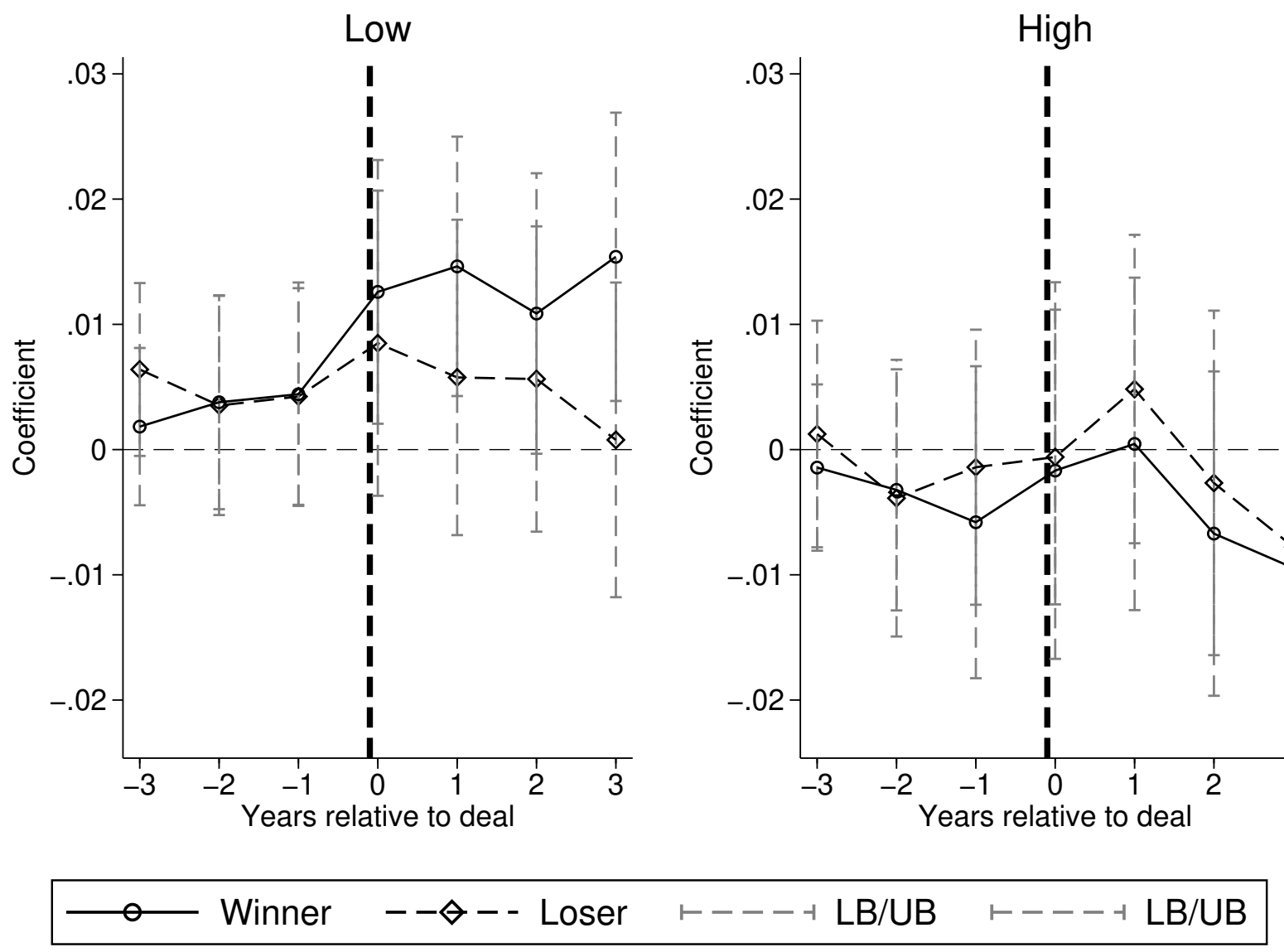

**Panel A: By Ex-ante Unemployment Rate**

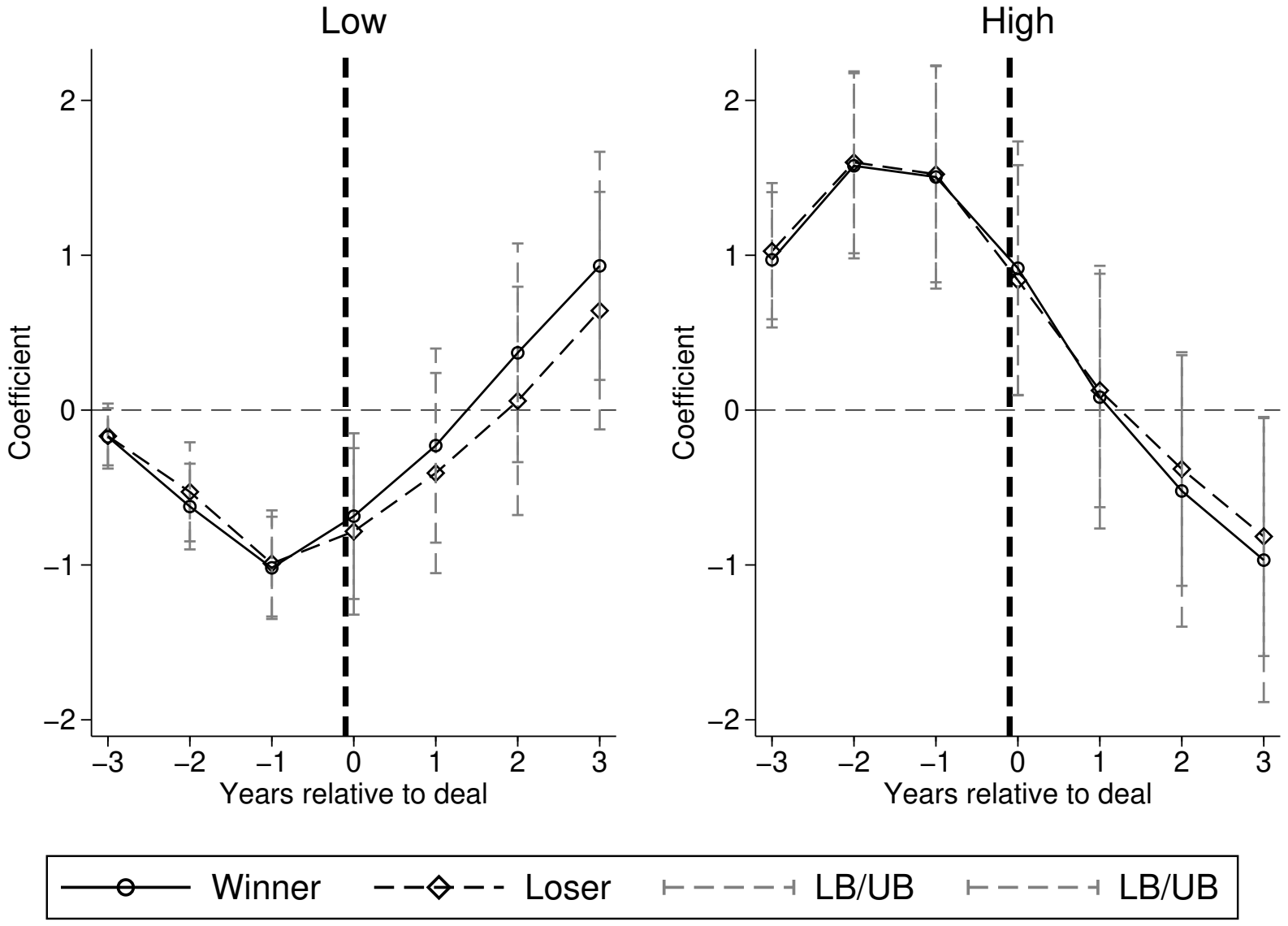

**Panel B: By Ex-ante Household Income**

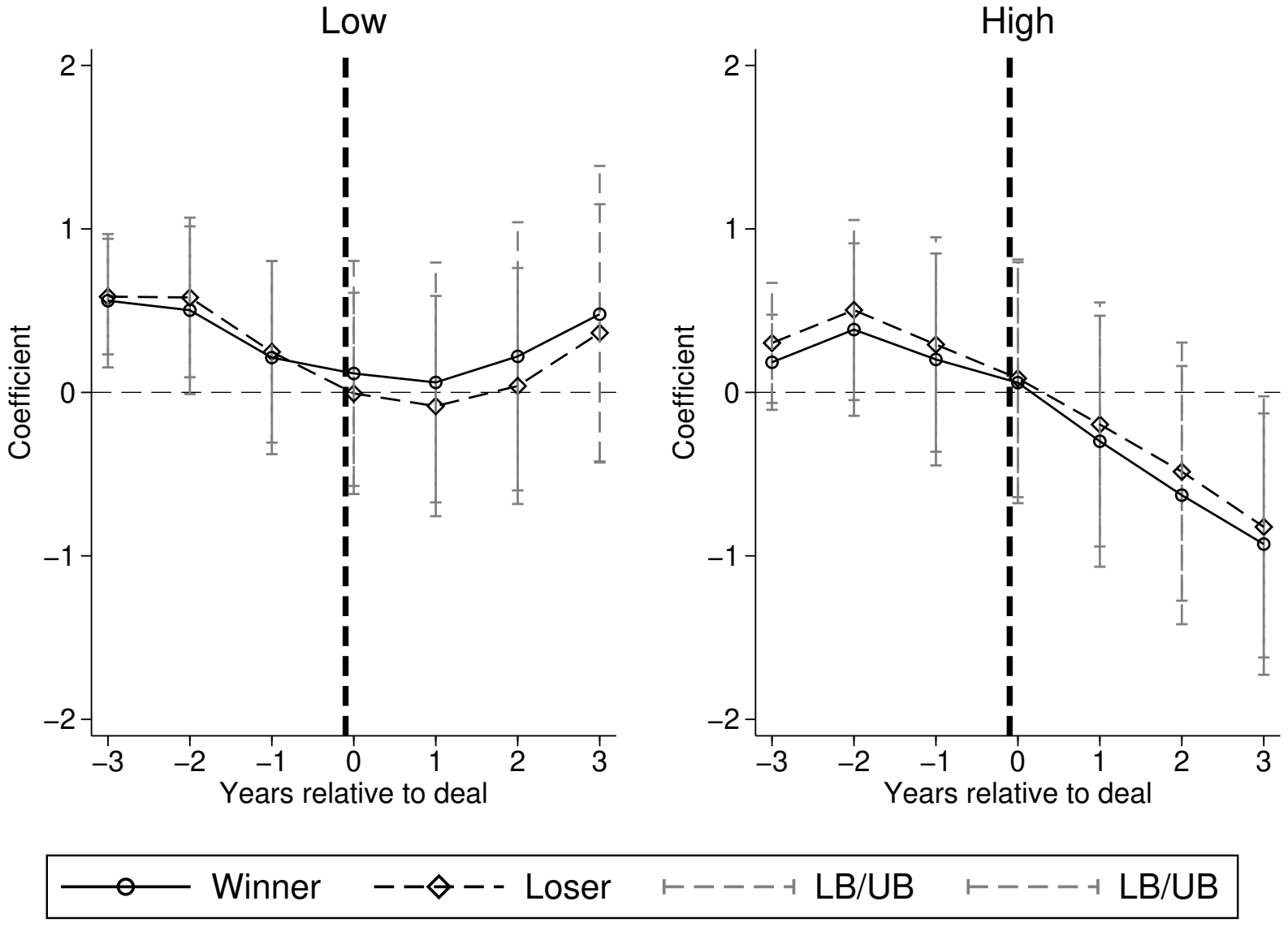

**Figure B2.7:** Winner vs Loser - Local Economy (Unemployment Rate):
The figure shows the unemployment rate at the county level for the bidding counties split into two groups (high and low), around the time of subsidy announcement. Panel A uses ex-ante unemployment rate among winning counties to divide the sample, while Panel B uses ex-ante median household income. We regress the unemployment rate against a set of interaction dummies for Winner $\times$ Post, split into half-yearly periods. We use deal and county fixed effects; standard errors are clustered by deal-pair. The benchmark period is from a year before the event window.

**Panel A: By Ex-ante County Rating**

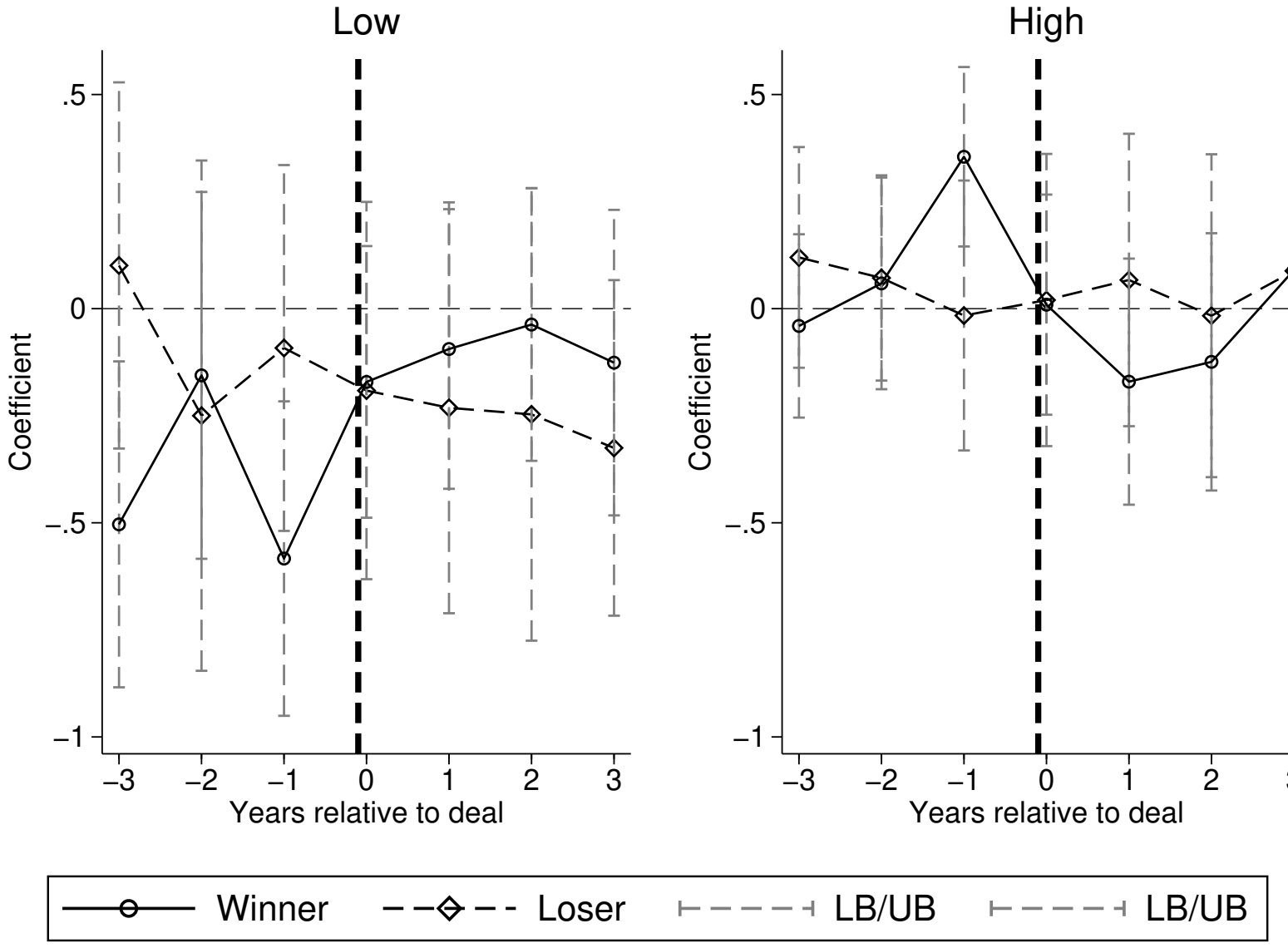

**Panel B: By Ex-ante Local Beta**

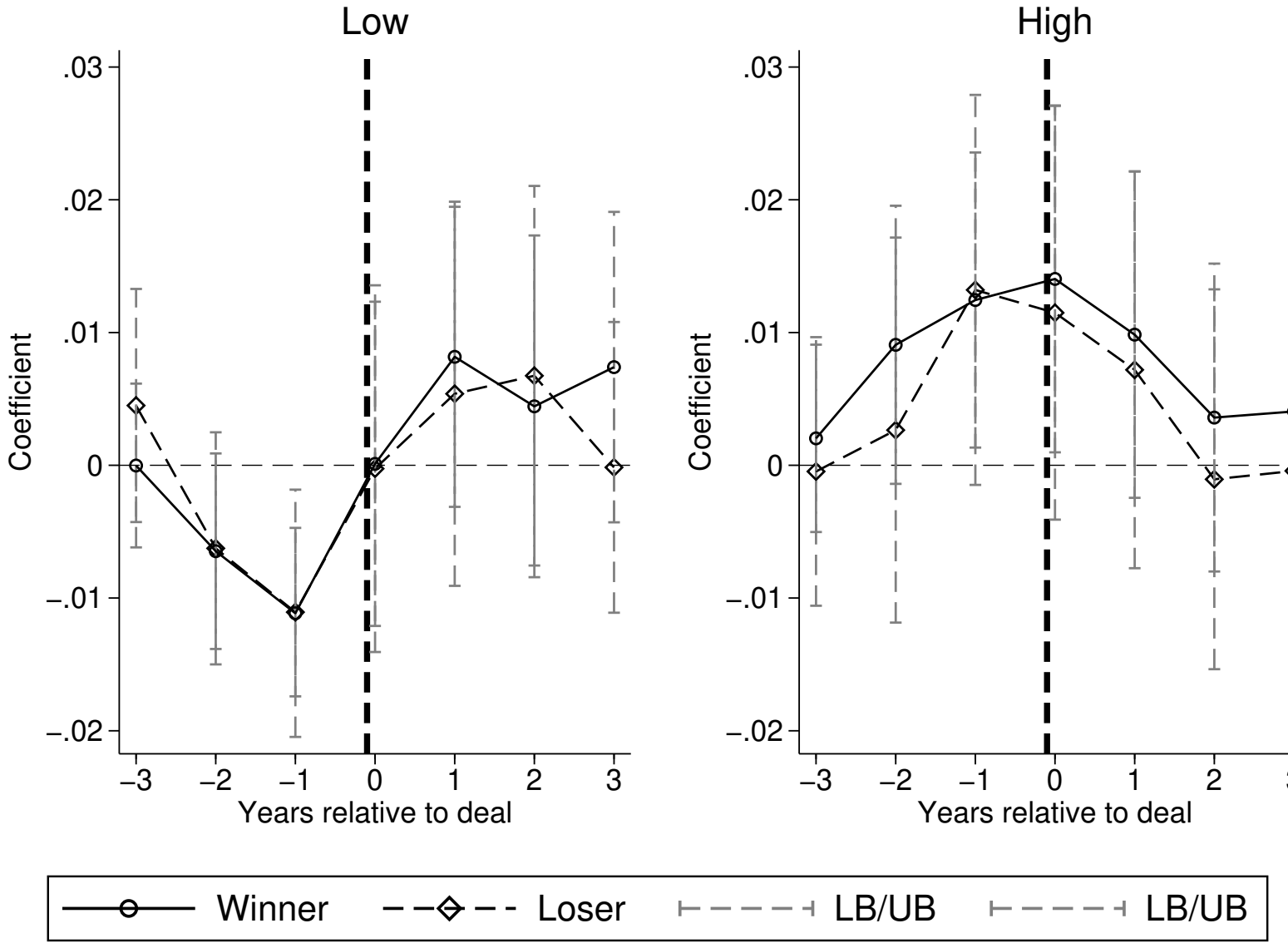

**Figure B2.8:** Winner vs Loser - Local Economy (Unemployment Rate):
The figure shows the rating at the county level for the bidding counties split into two groups (high and low), around the time of subsidy announcement. Panel A uses ex-ante county rating among winning counties to divide the sample, while Panel B uses ex-ante local beta. We regress the county rating in Panel A and county local beta in Panel B against a set of interaction dummies for Winner $\times$ Post, split into half-yearly periods. We use deal and county fixed effects; standard errors are clustered by deal-pair. The benchmark period is from a year before the event window.

**Probability of Bond Rating Downgrade by County Debt Capacity**

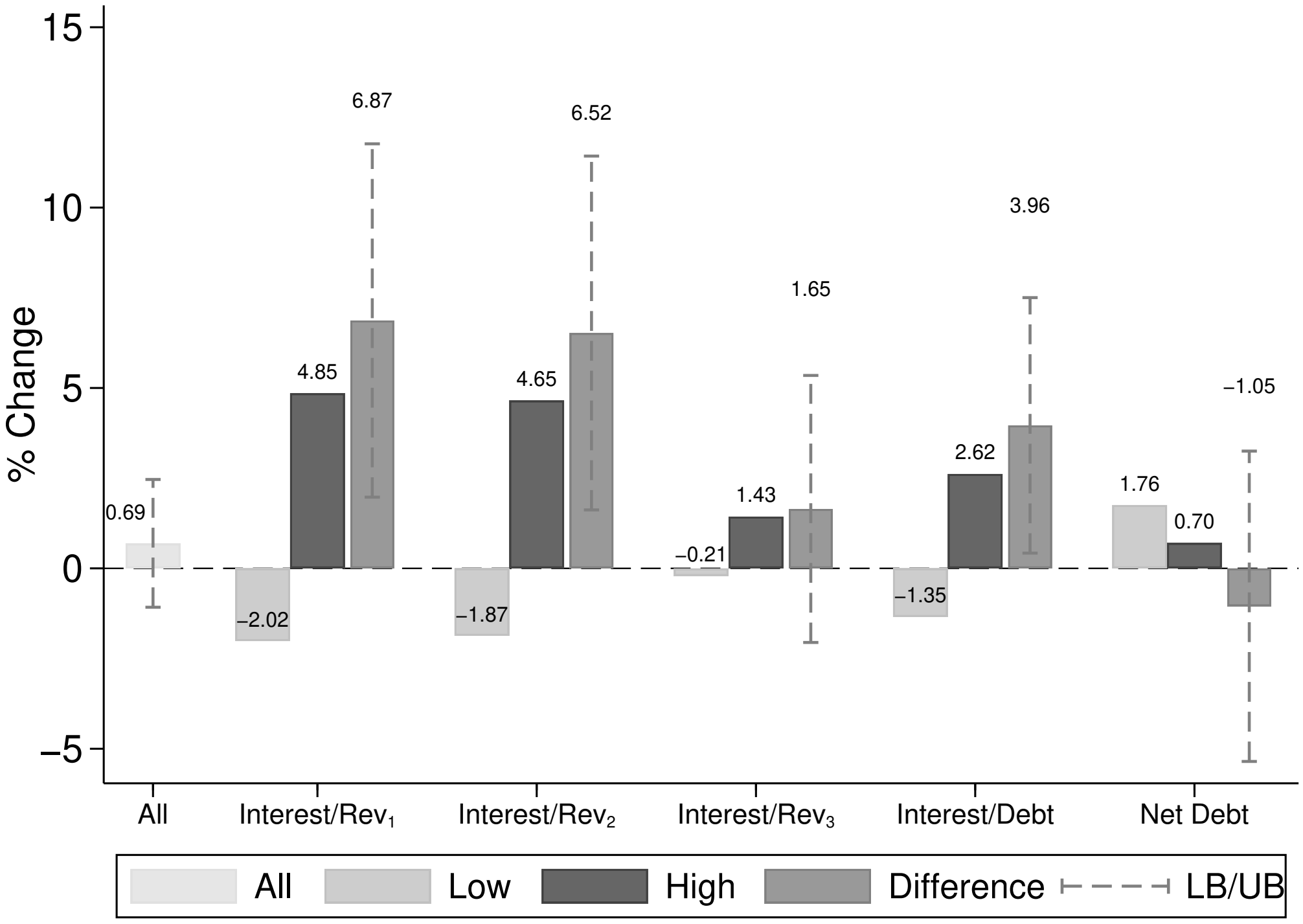

**Figure B2.9:** Probability of Bond Rating Downgrade:
The figure shows results for our main interaction term, $\beta_0$, from Equation 4.1. The dependent variable is a dummy variable equal to one indicating a bond rating downgrade (and zero otherwise). First, we show the baseline effect across all bonds in our sample. Next, we modify the baseline equation to interact with dummies for high and low values of ex-ante county level measures of debt capacity using interest expenditure and debt, namely: a) Interest/Revenue$_1$, b) Interest/Revenue$_2$, c) Interest/Revenue$_3$, d) Interest/Debt, e) Net Debt. See Table B1 for variables description. We additionally control for group-month fixed effects in these regressions. The corresponding bars for the low versus high groups and differences are indicated in the legend. Standard errors are double clustered by county bond issuer and year-month. The dashed lines represent 95% confidence intervals.

## **Table B2.1:** Comparison of Subsidy Datasets

This table provides a snapshot comparison of the information on subsidy deals between the original data from *Good Jobs First Subsidy Tracker* and the completed dataset prepared after hand-collection. Panel A shows a sample of data available from *Good Jobs First*. Panel B shows the information available in our completed dataset. "???" denotes that some information may be available, while "×" denotes that no information was available.

**Panel A: Good Jobs First**

| | | | | | Winner | | Loser | | | |
|---|---|---|---|---|---|---|---|---|---|---|
| Company | Year | Date | Subsidy ($ mil) | Investment ($ mil) | State | County | State | County | Jobs | Purpose |
| Baxter International | 2012 | × | 211 | ??? | GA | ??? | × | × | ??? | ??? |
| Foxconn | 2017 | × | 4,792 | 10,000 | WI | Racine | × | × | 13,000 | ??? |
| Vertex Pharmaceuticals | 2011 | × | 72 | ??? | MA | ??? | × | × | 500 | ??? |

**Panel B: Completed Dataset**

| | | | | | Winner | | Loser | | | |
|---|---|---|---|---|---|---|---|---|---|---|
| Company | Year | Date | Subsidy ($ mil) | Investment ($ mil) | State | County | State | County | Jobs | Purpose |
| Baxter International | 2012 | 4/19/2012 | 211 | 1,000 | GA | Newton | NC | Durham | 1,500 | New |
| Foxconn | 2017 | 7/26/2017 | 4,792 | 10,000 | WI | Racine | MI | Wayne | 13,000 | New |
| Vertex Pharmaceuticals | 2011 | 9/15/2011 | 72 | 2,500 | MA | Suffolk | MA | Middlesex | 500 | Relocation |

## **Table B2.2:** Names of Projects (Amounts in $ million)

This table shows some examples of project names under which the respective bidding processes were encoded by the winning local governments in order to maintain secrecy.

| Company | Year | State | Project Name | Investment | Subsidy |
|---|---|---|---|---|---|
| Burnham Institute for Medical Research | 2006 | FL | Power | 90 | 310 |
| Eastman Chemical | 2007 | TN | Reinvest | 1,300 | 100 |
| Freightquote | 2012 | MO | Apple | 44 | 64 |
| Airbus (EADS) | 2012 | AL | Hope | 600 | 158.5 |
| Benteler Steel/Tube | 2012 | LA | Delta | 900 | 81.75 |
| Northrop Grumman | 2014 | FL | Magellan | 500 | 471 |

## **Table B2.3:** Determinants of Subsidy

This table reports a linear regression of the amount of subsidy in our sample of deals from 2005-2018 on metrics potentially linked to the incentive. P-values are reported in brackets and standard errors are robust to heteroskedasticity. * $p < 0.10$, ** $p < 0.05$, *** $p < 0.01$

| | Subsidy (USD million) | | | | | |
|---|---|---|---|---|---|---|
| Jobs (1000) | 112.37*** | 67.49*** | 67.53*** | 74.35*** | 73.91*** | 72.55*** |
| | [9.89] | [5.64] | [5.60] | [4.80] | [4.73] | [4.68] |
| Investment (USD mil) | | 0.14*** | 0.14*** | 0.16*** | 0.16*** | 0.16*** |
| | | [7.30] | [7.26] | [6.88] | [6.78] | [6.69] |
| State Expenditure (USD mil) | | | 0.02 | -2.07 | | |
| | | | [0.04] | [-0.70] | | |
| HH-Income$_{t-1}(1000)$ | | | | | 1.59 | 1.34 |
| | | | | | [0.56] | [0.48] |
| State Surplus$_{t-1}(USDmil)$ | | | | | | -9.20** |
| | | | | | | [-2.01] |
| Constant | 34.82 | -8.90 | -10.26 | 138.23 | -98.56 | -64.97 |
| | [1.08] | [-0.30] | [-0.23] | [0.56] | [-0.82] | [-0.54] |
| State FE | | | | ✓ | ✓ | ✓ |
| Event-Year FE | | | | ✓ | ✓ | ✓ |
| Adj.-$R^2$ | 0.211 | 0.255 | 0.251 | 0.395 | 0.391 | 0.392 |
| Obs. | 199 | 194 | 194 | 186 | 181 | 181 |

**Table B2.4:** Sample Generation: Secondary Market

This table summarizes the construction of the municipal bond transactions sample. The steps involved in cleaning the transaction data include: removal of data errors such as dropping bonds with missing information in the MSRB data, coupons greater than 20%, maturities over 100 years, and fewer than 10 trades in the sample period; as well as dropping individual trades occurring at prices below 50 and above 150.

| | Number of CUSIPs | Number of Transactions |
|---|---|---|
| Customer Purchase trades (2005-2019) | 2,627,769 | 63,131,685 |
| Drop if maturity (days) $>$ 36,500 or $<$ 0 or missing | 2,622,312 | 63,111,155 |
| Drop if missing coupon or maturity | 2,560,586 | 59,503,049 |
| Drop if USD price $<$5 0 or $>$150 | 2,553,836 | 58,854,058 |
| Drop primary market trades | 1,798,931 | 46,563,129 |
| Drop trades within 15 days after issuance | 1,745,485 | 44,147,623 |
| Drop trades with less than 1 year to maturity | 1,632,037 | 42,387,878 |
| Drop if yield$<$0 or $>$50% | 1,618,440 | 41,489,989 |
| Drop if $<$ 10 transactions | 604,387 | 38,099,893 |
| Match CUSIPs from MSRB txns to MBSD features | 604,274 | 38,096,970 |
| Matching to FIPS using Bloomberg | 595,931 | 36,691,025 |
| Matching local bonds to corporate subsidy locations by FIPS | 228,389 | 14,231,762 |
| Aggregating to CUSIP-month txns. and plugging tax rates | 227,291 | 4,515,531 |
| Use tax-adjusted spread for event panel of 3 years using local bonds | 163,771 | 3,780,639 |
| - Winner | 85,834 | 1,433,430 |
| - Loser | 112,143 | 2,347,209 |

## **Table B2.5:** Trading Volume

This table reports the baseline results similar to Table 3.2 for our sample using Equation (4.1), with trading volume as the dependent variable. Columns (1)-(2) show the results for a sub-sample of customer buy trades. Columns (3)-(4) show the results for a sub-sample of customer sell trades. Finally, Columns (5)-(6) use the sum total of buy and sell trades (wherever both are available) as the dependent variable in the trading volume. T-statistics are reported in brackets and standard errors are double clustered at county bond issuer and year month level, unless otherwise specified. $^{*}$ $p < 0.10$, $^{**}$ $p < 0.05$, $^{***}$ $p < 0.01$

| *Dependent Variable*: | Trading volume (bond-month) | | | | | |
|---|---|---|---|---|---|---|
| | Customer Buy | | Customer Sell | | Total | |
| | (1) | (2) | (3) | (4) | (5) | (6) |
| Winner × Post | 27,004.26 | 29,588.63* | 34,596.39* | 37,751.27* | 66,609.67* | 73,005.50* |
| | [1.54] | [1.71] | [1.76] | [1.96] | [1.68] | [1.88] |
| Winner | 35,892.49 | 32,408.96 | 42,404.45* | 37,510.25 | 82,309.96* | 73489.76 |
| | [1.64] | [1.55] | [1.68] | [1.54] | [1.68] | [1.55] |
| Post ($t \geq 0$) | -16,059.37 | -17,288.61 | -74,75.86 | -86,95.33 | -22,056.45 | -24732.14 |
| | [-1.07] | [-1.15] | [-0.43] | [-0.50] | [-0.66] | [-0.73] |
| Deal FE | ✓ | ✓ | ✓ | ✓ | ✓ | ✓ |
| Year-Month FE | ✓ | ✓ | ✓ | ✓ | ✓ | ✓ |
| County FE | ✓ | ✓ | ✓ | ✓ | ✓ | ✓ |
| Bond Controls | ✓ | ✓ | ✓ | ✓ | ✓ | ✓ |
| County Controls | | ✓ | | ✓ | | ✓ |
| Adj.-$R^2$ | 0.050 | 0.050 | 0.050 | 0.050 | 0.056 | 0.056 |
| Obs. | 3,780,639 | 3,780,639 | 2,983,853 | 2,983,853 | 2,983,853 | 2,983,853 |

## **Table B2.6:** Predicting Winner

This table shows the results from a linear probability model using the 'winner' dummy as the dependent variable. We use the three-year event window, before the subsidy deal announcement. T-statistics are reported in brackets and standard errors are robust to heteroskedasticity. * $p < 0.10$, ** $p < 0.05$, *** $p < 0.01$

| | Winner | | | | | |
|---|---|---|---|---|---|---|
| | (1) | (2) | (3) | (4) | (5) | (6) |
| Log(Labor Force) | -0.005 | -0.004 | -0.004 | -0.004 | -0.014 | -0.000 |
| | [-0.43] | [-0.38] | [-0.34] | [-0.35] | [-0.92] | [-0.03] |
| Unemployment Rate(%) | | 0.002 | 0.003 | 0.003 | 0.004 | 0.001 |
| | | [0.39] | [0.44] | [0.46] | [0.73] | [0.12] |
| △ Unemployment Rate(%) | | | -0.000 | -0.000 | -0.000 | 0.000 |
| | | | [-0.24] | [-0.23] | [-0.30] | [0.15] |
| △ Labor Force | | | | 0.001 | 0.001 | 0.003 |
| | | | | [0.23] | [0.19] | [0.73] |
| Income Per Capita | | | | | -0.041 | -0.028 |
| | | | | | [-1.05] | [-0.73] |
| Log(House Price Index) | | | | | | -0.035 |
| | | | | | | [-0.96] |
| Constant | 0.554*** | 0.532*** | 0.524*** | 0.526*** | 0.641*** | 0.921** |
| | [4.32] | [3.79] | [3.65] | [3.65] | [3.33] | [2.31] |
| $R^2$ | 0.000 | 0.000 | 0.000 | 0.000 | 0.002 | 0.003 |
| Obs. | 1,155 | 1,155 | 1,155 | 1,155 | 1,152 | 1,058 |

**Table B2.7:** Falsification Tests: Pre-refunded bonds

This table shows a falsification test based on Equation (4.1) using transactions from insured bonds that have been pre-refunded. We provide detailed steps for creating this sample in Section B2.3.1. Column (1) shows the results using average yield as the dependent variable. In Column (2), we use yield spread as the outcome variable. Columns (3)-(5) report our results using only the subset of pre-refunded bonds with after-tax yield spread as the dependent variable. Specifically, Column (5) corresponds to our baseline specification. T-statistics are reported in brackets and standard errors are double clustered at county bond issuer and year month level, unless otherwise specified. * $p < 0.10$, ** $p < 0.05$, *** $p < 0.01$

| *Dependent Variable*: | Average Yield | Yield Spread | After-tax yield spread | | |
|---|---|---|---|---|---|
| | (1) | (2) | (3) | (4) | (5) |
| Winner x Post | 2.17 | 1.53 | 1.99 | 1.35 | 1.20 |
| | [0.50] | [0.37] | [0.29] | [0.21] | [0.18] |
| Winner | 6.42 | 5.94 | 10.59 | 11.81* | 9.34 |
| | [1.53] | [1.43] | [1.57] | [1.86] | [1.46] |
| Post ($t \geq 0$) | -3.84 | -1.70 | -4.08 | -2.83 | -2.65 |
| | [-1.59] | [-0.72] | [-1.07] | [-0.76] | [-0.72] |
| County-pair FE | ✓ | ✓ | ✓ | ✓ | ✓ |
| Year-Month FE | ✓ | ✓ | ✓ | ✓ | ✓ |
| County FE | ✓ | ✓ | ✓ | ✓ | ✓ |
| Bond Controls | | | | ✓ | ✓ |
| County Controls | | | | | ✓ |
| Adj.-$R^2$ | 0.529 | 0.794 | 0.644 | 0.692 | 0.692 |
| Obs. | 384,744 | 384,744 | 384,744 | 384,744 | 384,744 |

**Table B2.8:** More Robustness Tests

In this table we report results for additional robustness tests for our baseline specification, i.e., Column (6) of Table 3.2 (Panel A). In Panel A, we report results using only customer-buy trades with transaction size $\leq$ \$25,000, $\leq$ \$50,000, $\leq$ \$100,000, $>$ \$25,000, $>$ \$50,000, and $>$ \$100,000, respectively. Panel B show results by dropping bonds that were issued close to the announcement of the subsidy. We consider dropping bonds issued within 6, 12, 18, 24, and 36 months on either side of the announcement date, respectively. Similarly, in Columns (6)-(8), we show our results by only focusing on a sub-sample of bonds issued within 18, 24, and 36 months of the subsidy announcement date, respectively. In Panel C, we consider robustness to events based on the financial crisis of 2009 in Columns (1)-(2). Specifically, Column (1) reports the coefficient for the baseline equation from events before 2009. Column (2) corresponds to the main effect for events after 2009. There is no statistical difference between these coefficients. Columns (3)-(5) report results with other considerations, namely: controlling for log household income and log house price index (HPI) in Column (3). Column (4) shows the results for bonds with the most recent non-missing S&P credit ratings for a given CUSIP. In Column (5), we drop deals involving subsidy package sponsored by the state and local governments. We rely on print media evidence to make this distinction, by searching for each deal individually. Next, we report results controlling for unobserved factors corresponding to the purpose of the bond in Panel D. We categorize bonds into five groups based on their use of proceeds, namely: education, improvement and development, transportation and utilities, water-sewer, and other public services. First, in Column (1), we add bond purpose fixed effects to the baseline to control for the purpose for which the money was borrowed by the county. Column (2) shows the main result with bond purpose $\times$ year-month fixed effects added to the baseline. In Column (3), we introduce county-pair $\times$ bond purpose fixed effects. Finally, in Column (4) we use a weighted regression approach. The monthly weights correspond to the fraction of par volume traded in a given bond purpose relative to the county aggregate. In Panel E, we show robustness to the choice of clustering used in the baseline specification. Column (1) shows results by single clustering at the county-specific issuer level. In Column (2), we single cluster standard errors at the county-specific bond issue level. Finally, Column (3) uses double clustering at the county-specific bond issue and year-month level. Panel F shows results using other dependent variables. Columns (1)-(2) show the results for average yield as the dependent variable. In Column (1), we do not use any bond or county level controls, whereas Column (2) uses all controls as in the baseline specification. We follow the same strategy for Columns (3)-(4) using after-tax yield as the dependent variable. T-statistics are reported in brackets and standard errors are double clustered at county bond issuer and year-month level, unless otherwise specified. $^{*}$ $p < 0.10$, $^{**}$ $p < 0.05$, $^{***}$ $p < 0.01$

**Panel A: Size of Transactions**

| *Dependent Variable*: | After-tax Yield Spread | | | | | |
|---|---|---|---|---|---|---|
| | ≤25,000<br>(1) | ≤50,000<br>(2) | ≤100,000<br>(3) | >25,000<br>(4) | >50,000<br>(5) | >100,000<br>(6) |
| Winner × Post | 15.95*** | 15.74*** | 15.65*** | 17.22*** | 17.26*** | 16.75*** |
| | [4.44] | [4.49] | [4.52] | [4.43] | [4.09] | [3.59] |
| Adj.-$R^2$ | 0.613 | 0.607 | 0.602 | 0.599 | 0.588 | 0.577 |
| Obs. | 2,677,723 | 3,225,859 | 3,505,595 | 2,171,296 | 1,341,041 | 772,837 |

**Panel B: Drop/Keep Recently Dated Bonds**

| *Dependent Variable*: | After-tax Yield Spread | | | | | | | |
|---|---|---|---|---|---|---|---|---|
| | Drop Recently Dated Bonds Within | | | | | Keep Recently Dated Bonds Within | | |
| | 6 months<br>(1) | 12 months<br>(2) | 18 months<br>(3) | 24 months<br>(4) | 36 months<br>(5) | 18 months<br>(6) | 24 months<br>(7) | 36 months<br>(8) |
| Winner × Post | 14.36*** | 12.93*** | 12.77*** | 12.13*** | 10.61** | 12.50*** | 9.96*** | 9.18*** |
| | [4.06] | [3.57] | [3.36] | [3.05] | [2.43] | [3.85] | [3.13] | [3.17] |
| Adj.-$R^2$ | 0.589 | 0.583 | 0.577 | 0.571 | 0.554 | 0.778 | 0.773 | 0.769 |
| Obs. | 3,537,140 | 3,312,832 | 3,074,944 | 2,854,018 | 2,399,474 | 705,695 | 926,621 | 1,381,163 |

**Panel C: Additional Considerations**

| *Dependent Variable*: | After-tax Yield Spread | | | | |
|---|---|---|---|---|---|
| | Financial Crisis (2009) | | Other | | |
| | Before<br>(1) | After<br>(2) | More Controls<br>(3) | Add Rating<br>(4) | Drop State Subsidy Deals<br>(5) |
| Winner × Post | 17.61*** | 14.59*** | 10.92*** | 14.02*** | 16.29*** |
| | [2.82] | [3.63] | [3.19] | [3.55] | [4.84] |
| Adj.-$R^2$ | 0.595 | | 0.607 | 0.578 | 0.601 |
| Obs. | 3,780,639 | | 3,313,215 | 2,682,657 | 2,553,211 |

**Panel D: Heterogeneity based on Bond Purpose**

| *Dependent Variable*: | After-tax Yield Spread | | | |
|---|---|---|---|---|
| | Other Unobservables | | | |
| | Add Bond Purpose FE | Add Bond Purpose × YM FE | Add County-pair × Bond Purpose FE | Weighted by Bond Purpose |
| | (1) | (2) | (3) | (4) |
| Winner × Post | 14.78*** | 13.35*** | 14.49*** | 15.12*** |
| | [4.33] | [3.89] | [4.21] | [3.76] |
| Adj.-$R^2$ | 0.601 | 0.603 | 0.615 | 0.606 |
| Obs. | 3,780,195 | 3,780,195 | 3,780,194 | 3,780,639 |

**Panel E: Clustering**

| *Dependent Variable*: | After-tax Yield Spread | | |
|---|---|---|---|
| | By Issuer | By Issue | By Issue and Year-month |
| | (1) | (2) | (3) |
| Winner × Post | 15.25*** | 15.25*** | 15.25*** |
| | [4.56] | [7.31] | [6.39] |
| Adj.-$R^2$ | 0.595 | 0.595 | 0.595 |
| Obs. | 3,780,639 | 3,780,639 | 3,780,639 |

**Panel F: Other Dependent Variables**

| *Dependent Variable*: | Average yield | | After-tax Yield | |
|---|---|---|---|---|
| | No controls | All controls | No controls | All controls |
| | (1) | (2) | (3) | (4) |
| Winner × Post | 10.11*** | 10.75*** | 15.68*** | 16.84*** |
| | [4.25] | [5.23] | [3.95] | [4.94] |
| Adj.-$R^2$ | 0.311 | 0.579 | 0.288 | 0.567 |
| Obs. | 3,780,639 | 3,780,639 | 3,780,639 | 3,780,639 |

**Table B2.9:** Bargaining Power of Winning Counties

This table shows the heterogeneity in bargaining power across counties and states, using the baseline Equation (4.1). We interact the main equation with dummies corresponding to the economic variables in each column, as described hereafter. Group-month fixed effects are added to control for the average effect within a particular group for that month. Column (1) shows results based on *Proposed Value* which is obtained by taking the ratio of the differential between proposed investment and subsidy to the county's lagged revenue. A low value of the ratio indicates low bargaining power of the county. In Column (2), we use the $\frac{Investment}{StateRevenue}$ ratio to create two bins based on the median value among the winning counties. Counties which received a large investment from the firm compared to their state's revenue would represent higher desperation and lower bargaining power. Column (3) shows the interactions based on *Intensity of Competition*. We use the gap between winning and losing states in their budget surplus to state revenue ratio as a proxy for intensity of bidding competition. A low gap in ratios denotes high bidding competition, leading to low bargaining power for the winner. In Column (4), we provide evidence from the ex-ante *Unemployment Rate* of the winning counties in the year before the deal. A high value of the unemployment rate would signify a lower bargaining power for the winning county. T-statistics are reported in brackets and standard errors are double clustered at county bond issuer and year-month level. $^{*}$ $p < 0.10$, $^{**}$ $p < 0.05$, $^{***}$ $p < 0.01$

| *Dependent Variable*: | After-tax Yield Spread | | | |
|---|---|---|---|---|
| *Interaction Variable*: | Proposed Value | $\frac{\text{Investment}}{\text{State Revenue}}$ | Intensity of Competition | Unemployment Rate |
| Winner × Post | (1) | (2) | (3) | (4) |
| × Low | 14.91$^{***}$ | 12.46$^{***}$ | 13.18$^{***}$ | 7.27 |
| | [3.63] | [2.82] | [3.07] | [1.53] |
| × High | 13.58$^{**}$ | 16.51$^{***}$ | 20.30$^{***}$ | 19.06$^{***}$ |
| | [2.29] | [3.52] | [3.85] | [4.03] |
| Difference | -1.33 | 4.05 | 7.12 | 11.79 |
| P-value | 0.84 | 0.50 | 0.29 | 0.08 |
| County-pair FE | ✓ | ✓ | ✓ | ✓ |
| Year-Month FE | ✓ | ✓ | ✓ | ✓ |
| County FE | ✓ | ✓ | ✓ | ✓ |
| County Controls | ✓ | ✓ | ✓ | ✓ |
| Group-Month FE | ✓ | ✓ | ✓ | ✓ |
| Adj.-$R^2$ | 0.596 | 0.596 | 0.596 | 0.596 |
| Obs. | 3,633,616 | 3,633,616 | 3,762,008 | 3,780,639 |

**Table B2.10:** Close elections of the Bidding Counties

This table shows the effect of election contests across counties, using the baseline Equation (4.1). We interact the main equation with dummies corresponding to close elections (*Close*) among bidding counties and otherwise. In Columns (1)-(2), we use a 5% victory margin to define close elections. Columns (3)-(4) define close elections with a 10% victory margin between the top two candidates. We focus county level elections among winning counties only in Columns (1) and (3). In Columns (2) and (4), we use county level elections among both winning and losing counties. Group-month fixed effects are added to control for the average effect within a particular group for that month. T-statistics are reported in brackets and standard errors are double clustered at county bond issuer and year-month level. $^{*}$ $p < 0.10$, $^{**}$ $p < 0.05$, $^{***}$ $p < 0.01$

| *Dependent Variable*: | After-tax Yield Spread | | | |
|---|---|---|---|---|
| *Victory Margin*: | 5% | | 10% | |
| *Election for*: | Winner | Winner & Loser | Winner | Winner & Loser |
| Winner x Post | (1) | (2) | (3) | (4) |
| × Other | 14.21*** | 15.05*** | 14.82*** | 14.96*** |
| | [4.13] | [4.33] | [4.24] | [4.28] |
| × Close | 40.44*** | 22.11 | 26.19*** | 44.01* |
| | [2.70] | [0.98] | [2.98] | [1.67] |
| Difference | 26.23 | 7.059 | 11.37 | 29.06 |
| P-value | 0.09 | 0.76 | 0.21 | 0.28 |
| County-pair FE | ✓ | ✓ | ✓ | ✓ |
| Year-Month FE | ✓ | ✓ | ✓ | ✓ |
| County FE | ✓ | ✓ | ✓ | ✓ |
| County Controls | ✓ | ✓ | ✓ | ✓ |
| Group-Month FE | ✓ | ✓ | ✓ | ✓ |
| Adj.-$R^2$ | 0.595 | 0.595 | 0.595 | 0.595 |
| Obs. | 3,780,639 | 3,780,639 | 3,780,639 | 3,780,639 |

**Table B2.11:** Upcoming elections of the Bidding Counties

This table shows the effect of upcoming elections among counties, using the baseline Equation (4.1). We interact the main equation with dummies corresponding to upcoming elections (*Close*) for the winning counties. In Columns (1)-(2), we classify winning counties as *Upcoming* elections if the time (months) to next election is above the median. Columns (3)-(4) define winning counties as *Upcoming* elections if the time (months) to next election is in the top tercile. In Columns (2) and (4), we show results with group-month fixed effects to control for the average effect within a particular group for that month. T-statistics are reported in brackets and standard errors are double clustered at county bond issuer and year-month level. * $p < 0.10$, ** $p < 0.05$, *** $p < 0.01$

| *Dependent Variable*: | After-tax Yield Spread | | | |
|---|---|---|---|---|
| *Distance to election based on*: | Median Value | | Bottom Tercile | |
| Winner x Post | (1) | (2) | (3) | (4) |
| × Other | 11.14*** | 10.90*** | 12.57*** | 12.36*** |
| | [2.85] | [2.90] | [3.74] | [3.70] |
| × Upcoming | 22.19*** | 23.67*** | 33.61*** | 37.24*** |
| | [3.77] | [3.75] | [4.10] | [4.41] |
| Difference | 11.05 | 12.77 | 21.04 | 24.88 |
| p-val | 0.10 | 0.07 | 0.01 | 0.00 |
| County-pair FE | ✓ | ✓ | ✓ | ✓ |
| Year-Month FE | ✓ | ✓ | ✓ | ✓ |
| County FE | ✓ | ✓ | ✓ | ✓ |
| County Controls | ✓ | ✓ | ✓ | ✓ |
| Group-Month FE | | ✓ | | ✓ |
| Adj.-$R^2$ | 0.596 | 0.596 | 0.596 | 0.596 |
| Obs. | 3,780,639 | 3,780,639 | 3,780,639 | 3,780,639 |

**Table B2.12:** Heterogeneity: Based on Subsidy-to-jobs ratio

This table shows the evidence based on the ratio of subsidy offered to the number of jobs promised at the deal level, using the baseline Equation (4.1). We interact the main equation with dummies corresponding to above and below median levels of subsidy-to-jobs ratio. In Columns (1)-(2), we show results from the secondary market of municipal bonds with after-tax yield spread as the dependent variable. Columns (3)-(4) show results from the primary market of municipal bonds with offering yield as the dependent variable. This analysis also includes issuer fixed effects. We additionally control for the average impact within a particular group by adding group-month and group fixed effects in Columns (2) and (4), respectively. T-statistics are reported in brackets and standard errors are double clustered at county bond issuer and year month level. * $p < 0.10$, ** $p < 0.05$, *** $p < 0.01$

| *Dependent Variable*: | After-tax Yield Spread | | Offering yield | |
|---|---|---|---|---|
| *Municipal Market*: | Secondary | | Primary | |
| Winner × Post | (1) | (2) | (3) | (4) |
| × Below Median | 9.11** | 9.51*** | -5.81** | -5.81** |
| | [2.59] | [2.75] | [-2.40] | [-2.40] |
| × Above Median | 23.37*** | 22.07*** | 7.79*** | 7.79*** |
| | [4.06] | [3.94] | [2.81] | [2.81] |
| Difference | 14.26 | 12.56 | 13.59 | 13.59 |
| p-val | 0.02 | 0.04 | 0.00 | 0.00 |
| County-pair FE | ✓ | ✓ | ✓ | ✓ |
| County FE | ✓ | ✓ | ✓ | ✓ |
| County Controls | ✓ | ✓ | ✓ | ✓ |
| Year-month FE | ✓ | | | |
| Group-month FE | | ✓ | | |
| Issuer FE | | | ✓ | ✓ |
| Group FE | | | | ✓ |
| Adj.-$R^2$ | 0.595 | 0.596 | 0.822 | 0.822 |
| Obs. | 3,780,639 | 3,780,639 | 487,033 | 487,033 |

**Table B2.13:** Impact on House price index

This table shows the impact of subsidy on the logged value of house price index in the county. We use the annualized version of Equation (4.1) as the primary specification for this table. Columns (1)-(2) show the aggregate impact between winners and losers over three and five years, respectively. Columns (3)-(4) present the results based on sub-groups of the *Interest/Revenue*$_1$ ratio over three and five years, respectively. Specifically, we show the interacted form of the difference-in-differences estimate using a dummy variable (*Low Int./Rev.*$_1$ and *High Int./Rev.*$_1$) based on the median value of the *Interest/Revenue*$_1$ ratio among winning counties. In these interacted specifications, we replace the event-year fixed effects with group-event year fixed effects. T-statistics are reported in brackets and standard errors are clustered at the deal level. $^{*}$ $p < 0.10$, $^{**}$ $p < 0.05$, $^{***}$ $p < 0.01$

| *Dependent Variable*: | Log(House Price Index) | | | |
|---|---|---|---|---|
| *Event Window (years)*: | [-3,+3] | [-5,+5] | [-3,+3] | [-5,+5] |
| | (1) | (2) | (3) | (4) |
| Winner × Post | -0.02$^{**}$ | -0.02$^{**}$ | | |
| | [-2.31] | [-2.41] | | |
| Winner | -0.04 | -0.02 | | |
| | [-1.32] | [-1.02] | | |
| Post (t ≥ 0) | 0.05$^{***}$ | 0.07$^{***}$ | | |
| | [5.70] | [6.70] | | |
| Winner × Post | | | -0.02 | -0.02 |
| ×Low Int./Rev.$_1$ | | | [-1.48] | [-1.32] |
| Winner × Post | | | -0.02$^{*}$ | -0.03$^{**}$ |
| ×High Int./Rev.$_1$ | | | [-1.93] | [-2.22] |
| Difference | | | 0.00 | -0.01 |
| P-value | | | 0.68 | 0.46 |
| County-pair FE | ✓ | ✓ | ✓ | ✓ |
| County FE | ✓ | ✓ | ✓ | ✓ |
| County Controls | ✓ | ✓ | ✓ | ✓ |
| Event-year FE | ✓ | ✓ | | |
| Group Event-Year FE | | | ✓ | ✓ |
| Adj.-$R^2$ | 0.964 | 0.948 | 0.964 | 0.948 |
| Obs. | 2,711 | 4,124 | 2,683 | 4,080 |

**Table B2.14:** Impact on Local Property Tax Revenue

This table shows the impact of subsidy on the logged value of property tax revenue in the county. We use the annualized version of Equation (4.1) as the primary specification for this table. Columns (1)-(2) show the aggregate impact between winners and losers over three and five years, respectively. Columns (3)-(4) present the results based on sub-groups of the *Interest/Revenue*$_1$ ratio over three and five years, respectively. Specifically, we show the interacted form of the difference-in-differences estimate using a dummy variable (*Low Int./Rev.*$_1$ and *High Int./Rev.*$_1$) based on the median value of the *Interest/Revenue*$_1$ ratio among winning counties. In these interacted specifications, we replace the event-year fixed effects with group-event year fixed effects. T-statistics are reported in brackets and standard errors are clustered at the deal level. $^{*}$ $p < 0.10$, $^{**}$ $p < 0.05$, $^{***}$ $p < 0.01$

| *Dependent Variable*: | Log(Property Tax Revenue) | | | |
|---|---|---|---|---|
| *Event Window (years)*: | [-3,+3] | [-5,+5] | [-3,+3] | [-5,+5] |
| | (1) | (2) | (3) | (4) |
| Winner × Post | 0.01 | 0.01 | | |
| | [0.78] | [0.68] | | |
| Winner × Post | | | 0.00 | 0.01 |
| ×Low Int./Rev.$_1$ | | | [0.09] | [0.39] |
| Winner × Post | | | 0.02 | 0.01 |
| ×High Int./Rev.$_1$ | | | [1.02] | [0.61] |
| Difference | | | 0.02 | 0.00 |
| P-value | | | 0.48 | 0.86 |
| County-pair FE | ✓ | ✓ | ✓ | ✓ |
| County FE | ✓ | ✓ | ✓ | ✓ |
| County Controls | ✓ | ✓ | ✓ | ✓ |
| Event-year FE | ✓ | ✓ | | |
| Group Event-Year FE | | | ✓ | ✓ |
| Adj.-$R^2$ | 0.997 | 0.996 | 0.997 | 0.996 |
| Obs. | 2,630 | 3,994 | 2,616 | 3,972 |

# APPENDIX C
# MISCELLANEOUS SECTION FOR CHAPTER 3

## C3.1 Additional Results

In Section C3.1.1, we provide results for additional considerations of matching.

### C3.1.1 Additional Matching Strategies

We understand that identifying a suitable match for the treated counties using nearest-neighbor matching based on five key variables may not be the only possible approach. Since we lack a perfect counterfactual in our setting, we demonstrate robustness to the choice of our matching strategy by using eight additional approaches in Table C3.2. First, in Columns (1)-(3), we introduce a sixth variable to match counties based on their debt capacity. Our metrics for debt capacity are based on [32]. Using measures linked to interest expenditure as additional matching variables to identify control counties, we show that that the baseline effect ranges from 6.11 to 10.89 bps. The effect remains statistically significant and economically meaningful. We provide the kernel density plot between treated and control counties for our matching variables with debt capacity using interest expenditure in Figure C3.4, Figure C3.5 and Figure C3.6.

Next, we address the concern that the control groups identified in the baseline approach may be different based on the size of bonds issued or the maturity bucket of municipal bonds in Columns (4)-(5). By using the average amount issued and average maturity of bonds, respectively, we show that the increase in yield spreads amounts to 4.73 and 5.74 bps, respectively. We present a comparison of our treated and control county characteristics on these matching variables in Figure C3.7 and Figure C3.8. Finally, we show robustness to our matching with respect to the geographic region of the control groups in Columns (6)-(8). In Column (6), we require that the baseline choice of control group comes from within the same geographic region of the United States and find the magnitude to be 5.64 bps. Columns (7)-(8) consider three and five nearest neighbors matching (instead of one) among the control counties, respectively. Our results show an increase in bond yields of 5.12 and 4 bps, respectively. We argue that since the additional neighbors are all given an equal weight among the controls, this may introduce some noise to the estimates. We show the distribution of the matching variables between the treated and control counties under these additional geographic considerations in Figure C3.9 and Figure C3.10.

#### C3.1.2 Discussion on Budgetary Restrictions vs State Incentives

*State Imposed Budgetary Restrictions*

Tax and expenditure limits impose controls on the taxing and spending ability of local governments. Beyond taxes, local governments may also raise money by issuing public debt to finance their expenditures. In this regard, debt limits restrict the ability of local governments to access the public debt market through bond issuances. The primary motivation behind these restrictions is minimizing defaults and over-borrowing. Often, a super-majority of voters is required to exceed such debt limits. Some of these taxing and spending limitations on local governments can be traced back to Proposition 13 in California in 1978. Unfunded budget mandates from the state may also control some local government budget expenditures. [168] show that bond market participants consider fiscal institutions in assessing the risk characteristics of tax-exempt bonds. In this light, we hypothesize that local budgetary restrictions imposed by states reduce the ability of local governments to respond to large firm bankruptcies in their counties.

We present our results in Table C3.4 based on a modified version of the baseline Equation (4.1) using data from [169]. We additionally control for group-year month fixed effects to account for average differences across groups in a given calendar year month. In Columns (1)-(3), we divide the treated counties into two groups based on the presence or absence of local restrictions on debt issuance, overall property tax, and total expenditure, respectively. Column (1) shows that counties with debt issuance limits exhibit a bond yield spread increase of 11.65 bps, while there is no significant effect on counties without such limits in the three years after a firm files for bankruptcy. Similarly, Column (2) shows that counties with an overall limit on property tax exhibit an increase of 19.30 bps in the bond yield spreads after a bankruptcy filing. There is a similar effect when we use expenditure limits in Column (3), although the difference between the two groups is not statistically significant.

Finally, since these budgetary restrictions are not mutually exclusive among the counties, we consider a linear combination of these dummy indicators by summing them up for each treated county. The overall index used in this combination of restrictions (shown in Column (4)) ranges between zero and three. Our results in Column (4) suggest that multiple (two or more) budgetary restrictions result in an increase in bond yield spreads of 19.14 bps. The corresponding impact on counties with one or zero restrictions is much smaller but also statistically insignificant. Taken together, the evidence based on county budgetary restrictions suggests that the municipal bond market yield spreads are more adversely affected when a bankrupt firm is located in a county with greater state-imposed controls over the county's budget/debt issuance.

*State Level Business Incentives*

The business environment in a given county may differ based on various factors and policies by the corresponding state. One such crucial factor could be the state level business incentives at the time a firm files for bankruptcy. We use the Panel Database on Incentives and Taxes (PDIT) from the Upjohn Institute[1], which includes data on incentives from 33 states in the US from 1990 to 2015. The data provides information on the taxes paid and incentives received by a business for a new hypothetical facility opened in a given year across 45 industries. The resulting data are reported as a percentage of the value-added for that particular industry. For our purposes, we focus on the 19 export-based manufacturing industries. We manually match these to the corresponding 4-digit NAICS industry of the firm filing for bankruptcy. Since these data are produced at the state level, we weigh the percentage of value added incentives by the corresponding proportion of the industry's GDP at the county level relative to the manufacturing GDP of the county. We use the average value by discounting the present value of incentives over the 20-year simulation period using a 12% real discount rate. Such a high discount rate reflects the perspective of many corporate decision makers, who place a higher value on short-term factors than long-term factors.

We report our results in Table C3.5 by using the baseline Equation (4.1) and modify it suitably with group-year month fixed effects to control for the average trends in each group. First, in Column (1), we show the results by dividing the treated counties based on the median value of net tax (after incentives). We find that bond yield spreads increase by 14.04 bps among counties with high net tax. On the other hand, the increase in yield spreads is only 3.59 bps for counties with low net tax, but the coefficient is statistically indistinguishable from zero. That said, a higher tax incentive likely could be beneficial to businesses. We find consistent results in Columns (2)-(4), suggesting that the impact on bond yield spreads is higher when the state sponsors lower level of incentives before the bankruptcy event. For example, Column (3) compares treated counties using the investment tax credit. We find that counties with low incentives experience an increase in yields of 15.61 bps, while the effect on the high group is statistically insignificant. These results suggest that counties located in states where business incentives are less generous tend to experience a greater impact on their municipal bond yield spreads. Alternatively, state-level incentive provisions may help the counties to mitigate the broader economic impact after a large firm bankruptcy.

[1] https://www.upjohn.org/bied/home.php

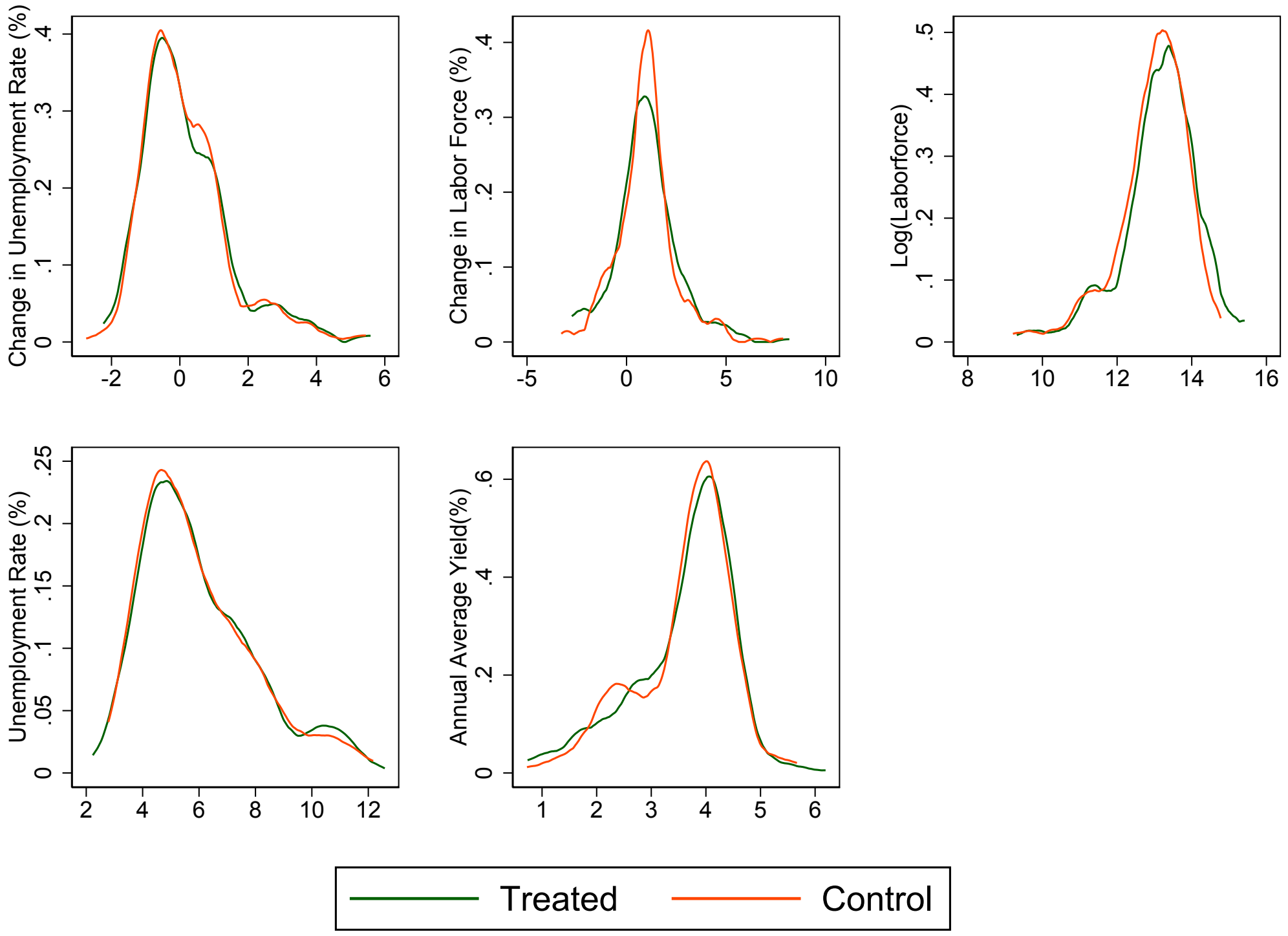

**Figure C3.1:** Matching County Characteristics:
The figure shows the kernel density plot based on matching the treatment counties using five variables in the year before firm bankruptcy: unemployment rate, change in unemployment rate, log(labor force), change in labor force, and average yield of the county in that year.

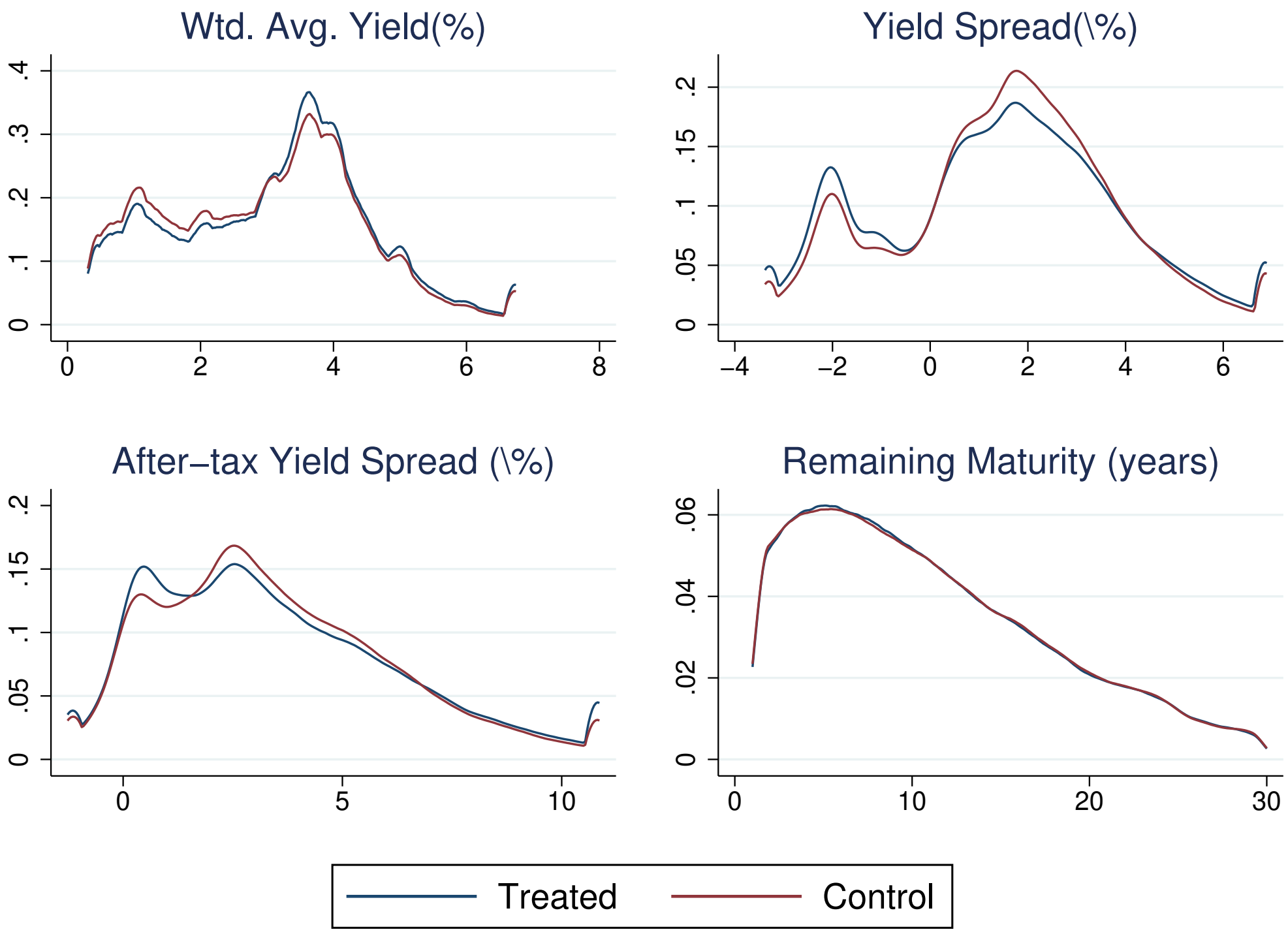

**Figure C3.2:** Secondary Market Bond Features:
The figure shows the kernel density plot for municipal bond aspects from the secondary market. The county-level matching is based on five variables in the year before firm bankruptcy: unemployment rate, change in unemployment rate, log(labor force), change in labor force, and average yield of the county in that year.

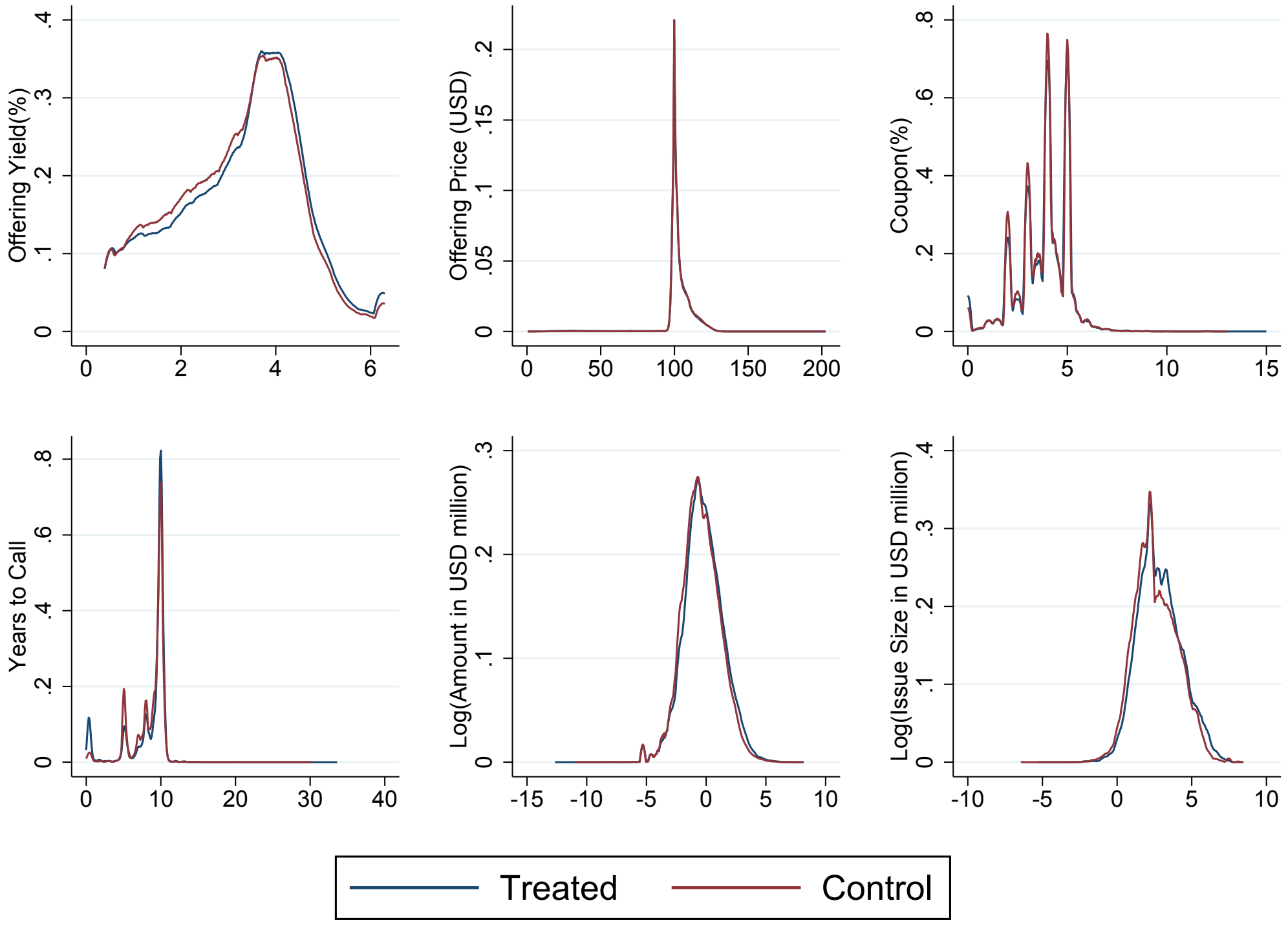

**Figure C3.3:** Primary Market Muni Bond Characteristics:
The figure shows the kernel density plot for municipal bond aspects from the primary market. The county-level matching is based on five variables in the year before firm bankruptcy: unemployment rate, change in unemployment rate, log(labor force), change in labor force, and average yield of the county in that year.

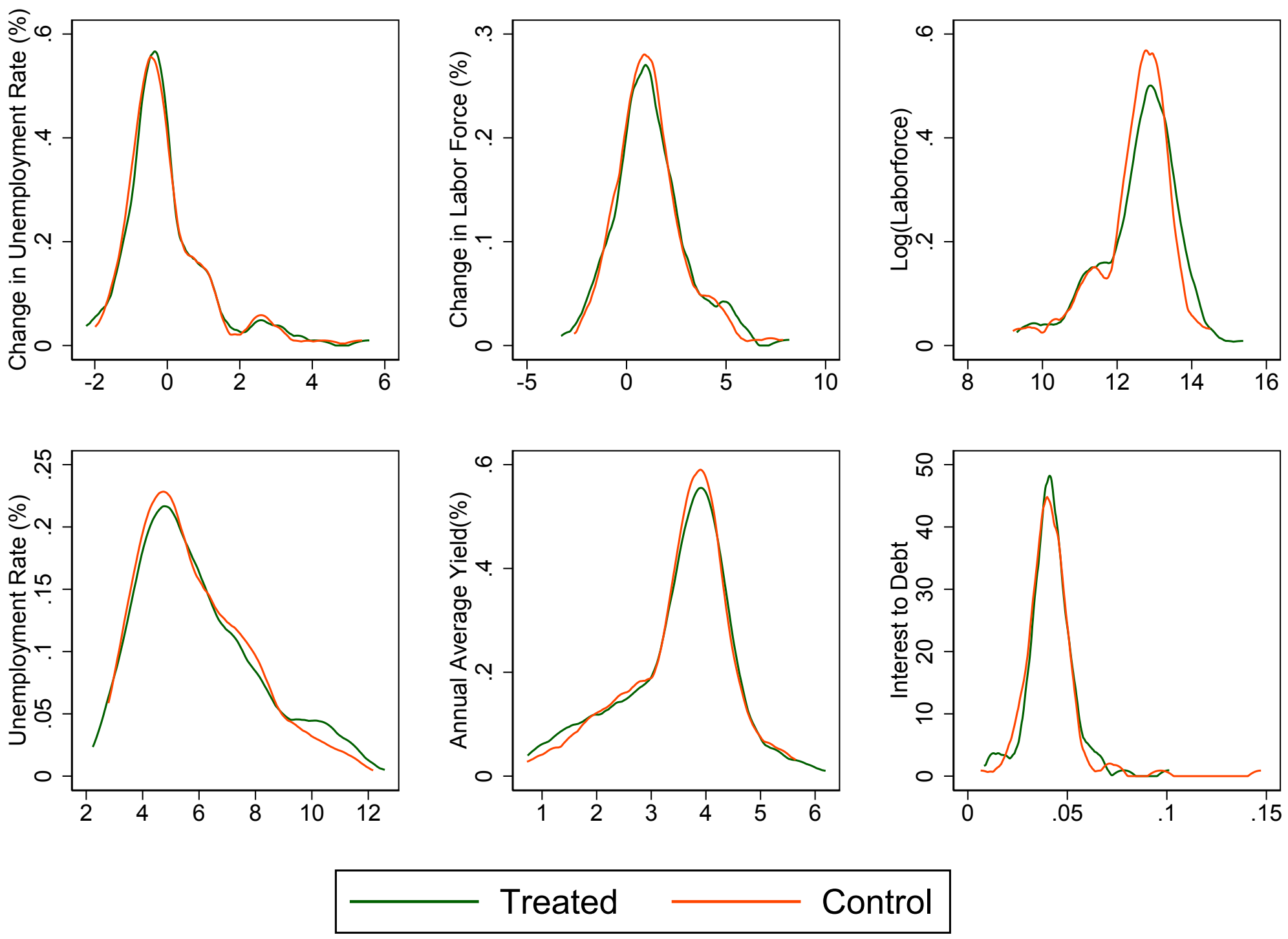

**Figure C3.4:** Matching County Characteristics - Interest to Debt:
The figure shows the kernel density plot based on matching the treatment counties using 5 variables in the year before firm bankruptcy: unemployment rate, change in unemployment rate, log(labor force), change in labor force, and average yield of the county in that year.

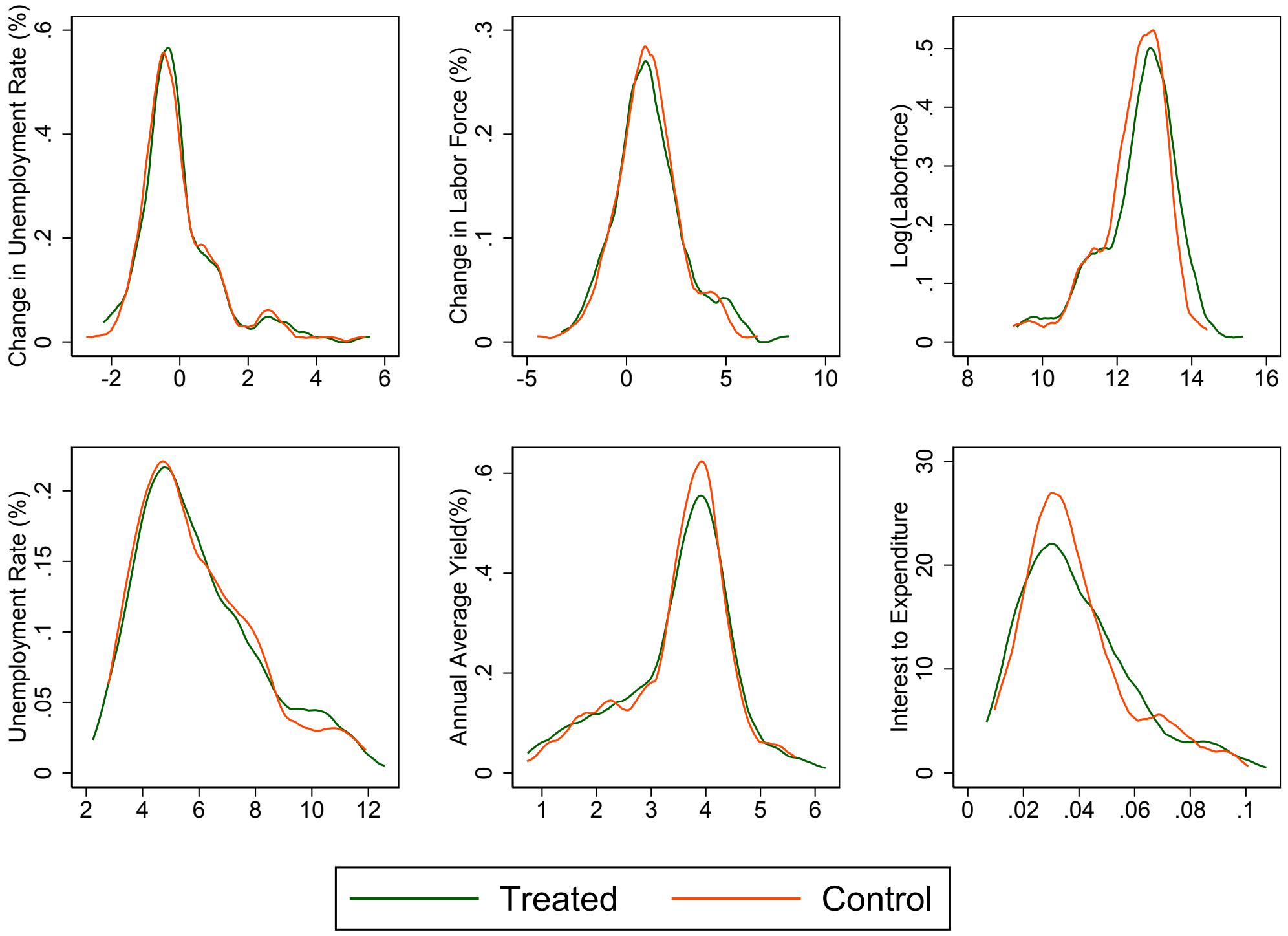

**Figure C3.5:** Matching County Characteristics - Interest to Expenditure:
The figure shows the kernel density plot based on matching the treatment counties using 5 variables in the year before firm bankruptcy: unemployment rate, change in unemployment rate, log(labor force), change in labor force, and average yield of the county in that year.

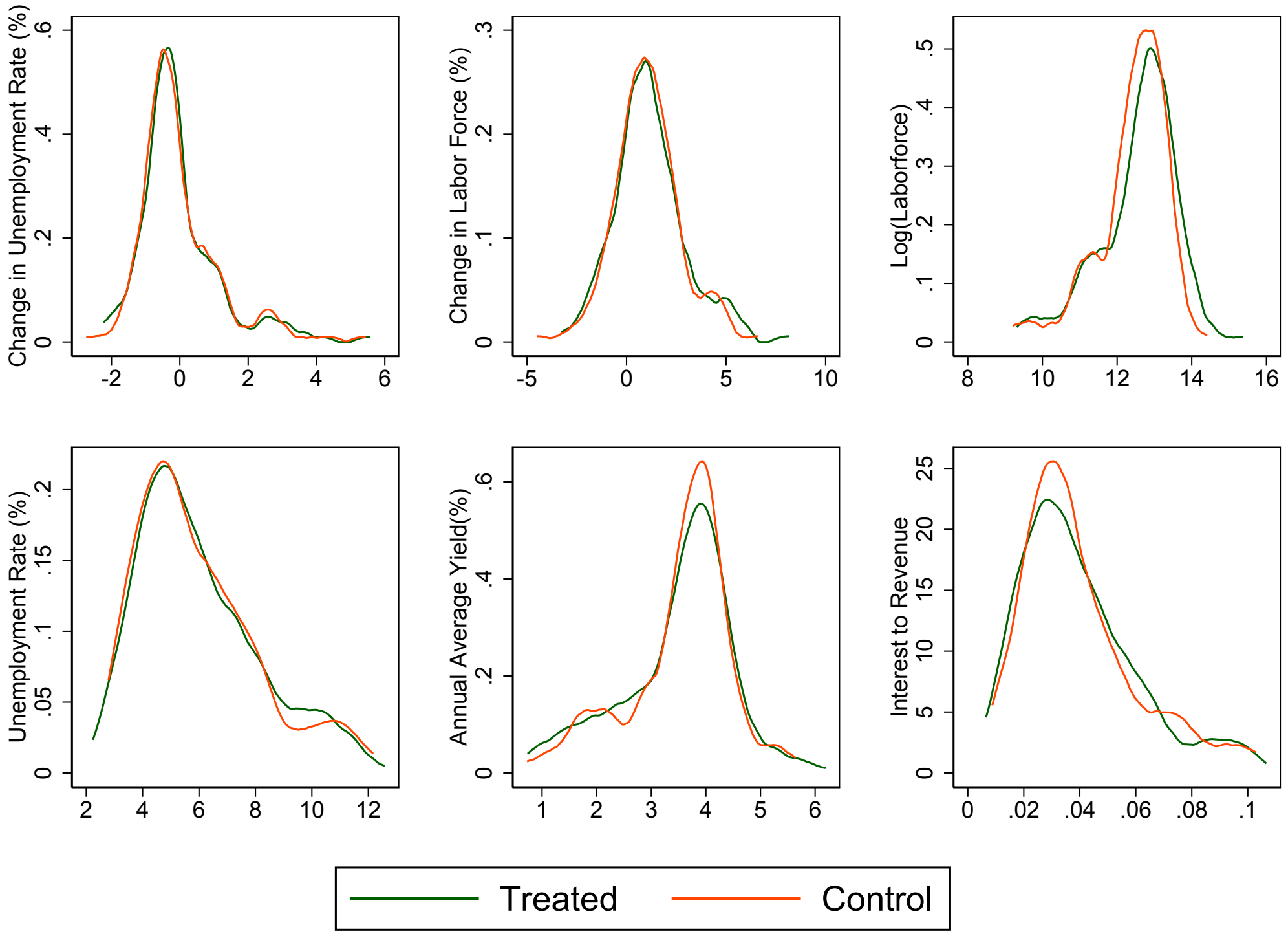

**Figure C3.6:** Matching County Characteristics - Interest to Revenue:
The figure shows the kernel density plot based on matching the treatment counties using 5 variables in the year before firm bankruptcy: unemployment rate, change in unemployment rate, log(labor force), change in labor force, and average yield of the county in that year.

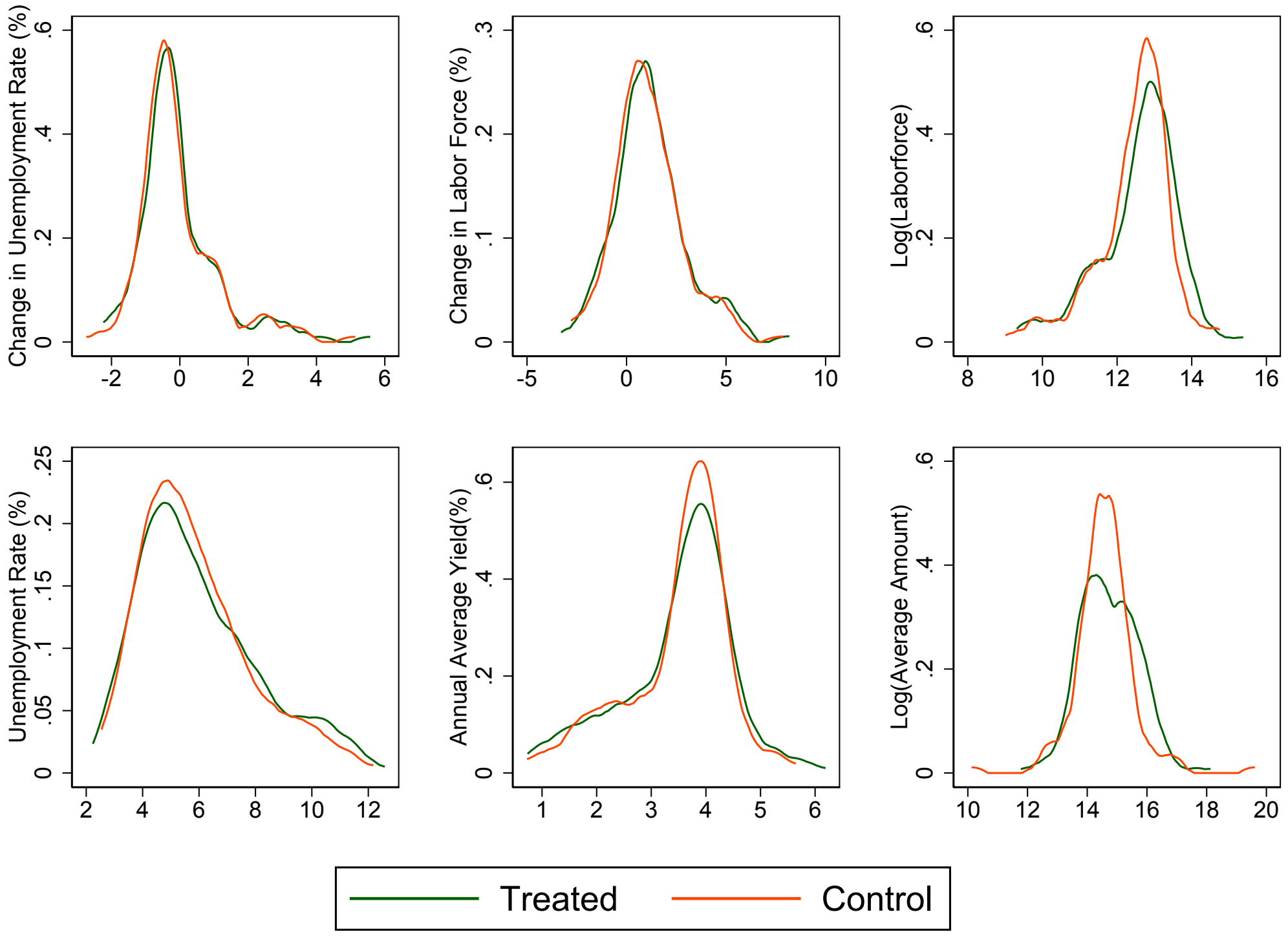

**Figure C3.7:** Matching County Characteristics - Log(Average Amount):
The figure shows the kernel density plot based on matching the treatment counties using 5 variables in the year before firm bankruptcy: unemployment rate, change in unemployment rate, log(labor force), change in labor force, and average yield of the county in that year.

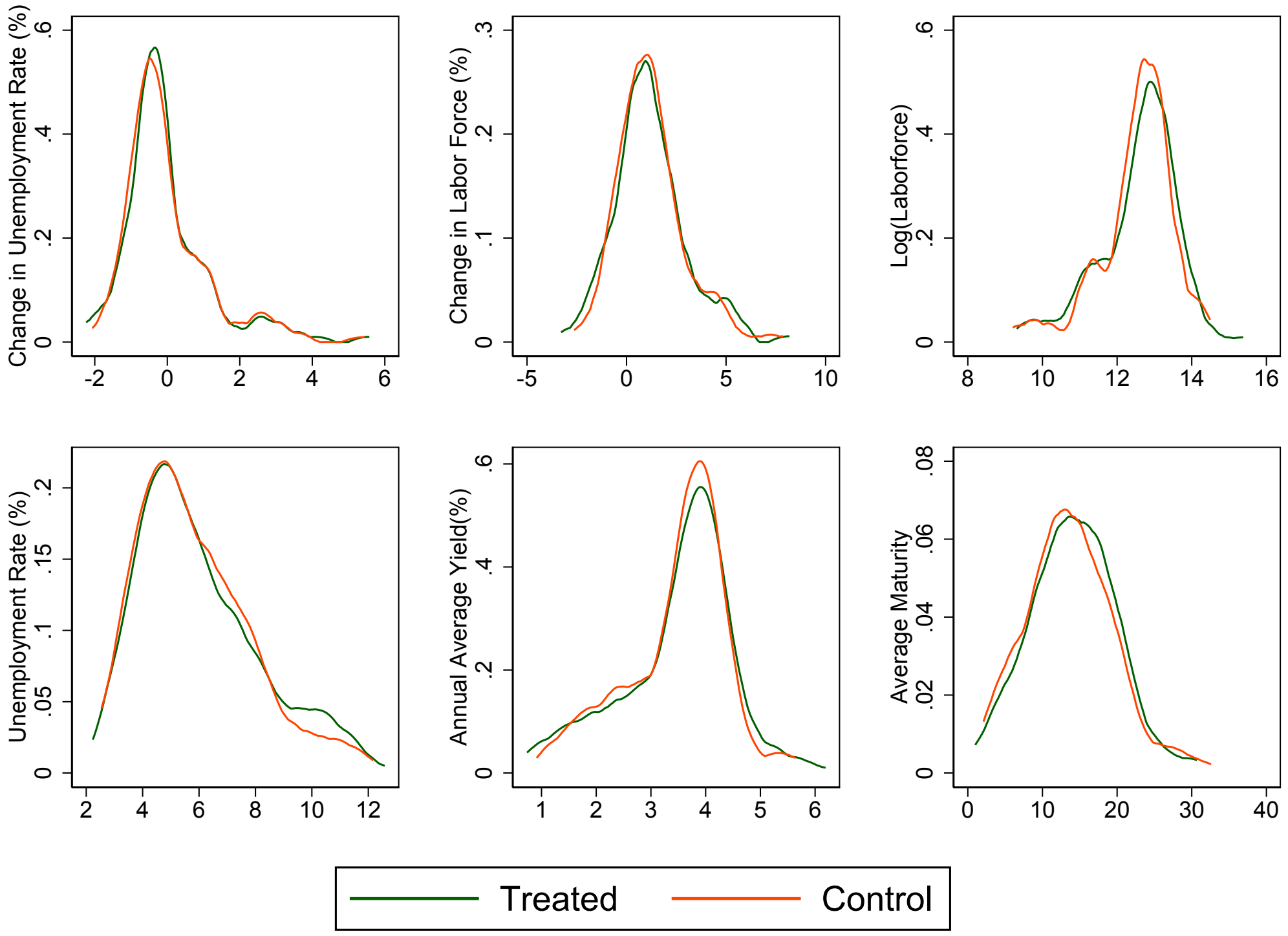

**Figure C3.8:** Matching County Characteristics - Average Maturity:
The figure shows the kernel density plot based on matching the treatment counties using 5 variables in the year before firm bankruptcy: unemployment rate, change in unemployment rate, log(labor force), change in labor force, and average yield of the county in that year.

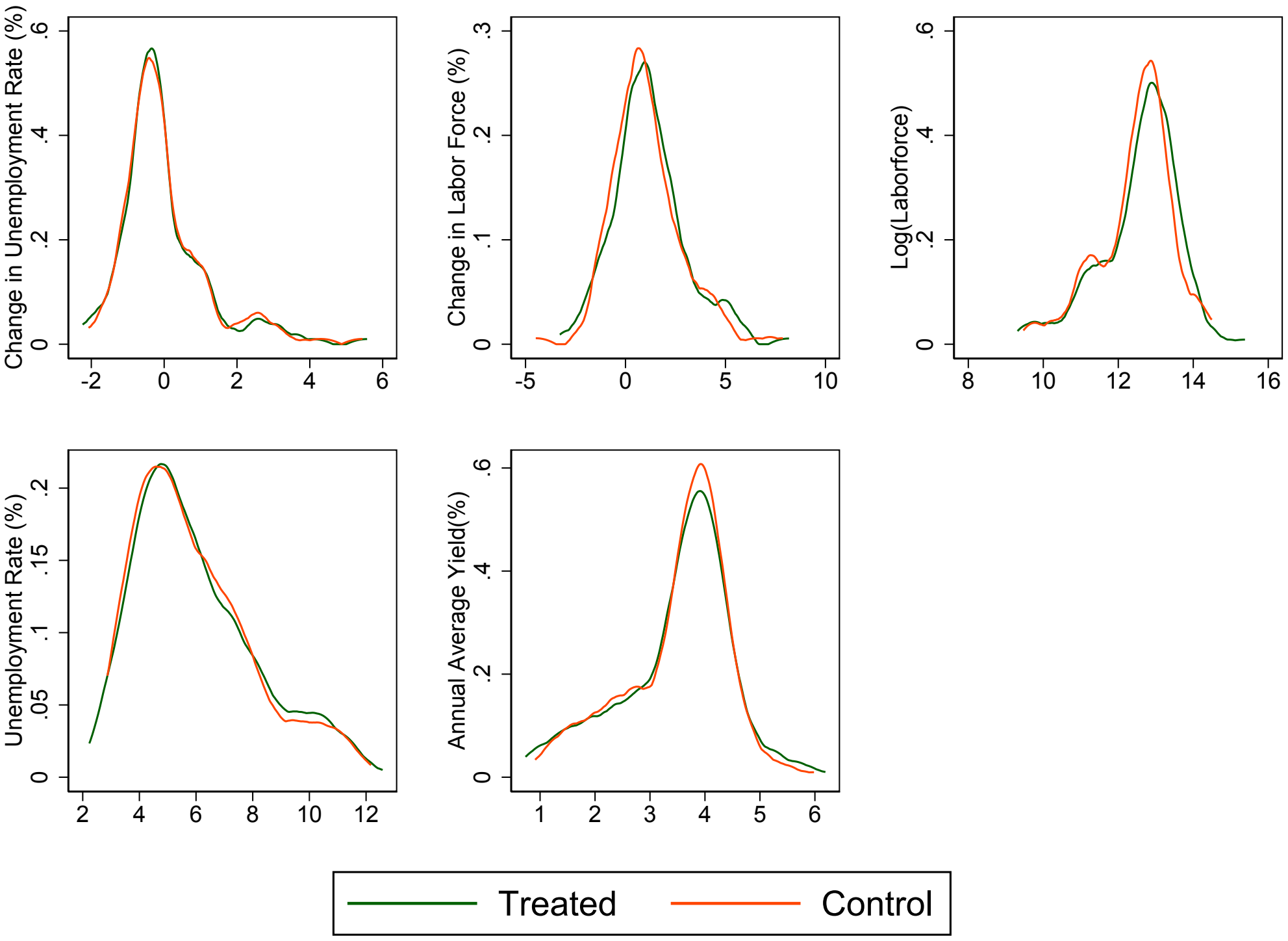

**Figure C3.9:** Matching County Characteristics - Same Region:
The figure shows the kernel density plot based on matching the treatment counties using 5 variables in the year before firm bankruptcy: unemployment rate, change in unemployment rate, log(labor force), change in labor force and average yield of the county in that year.

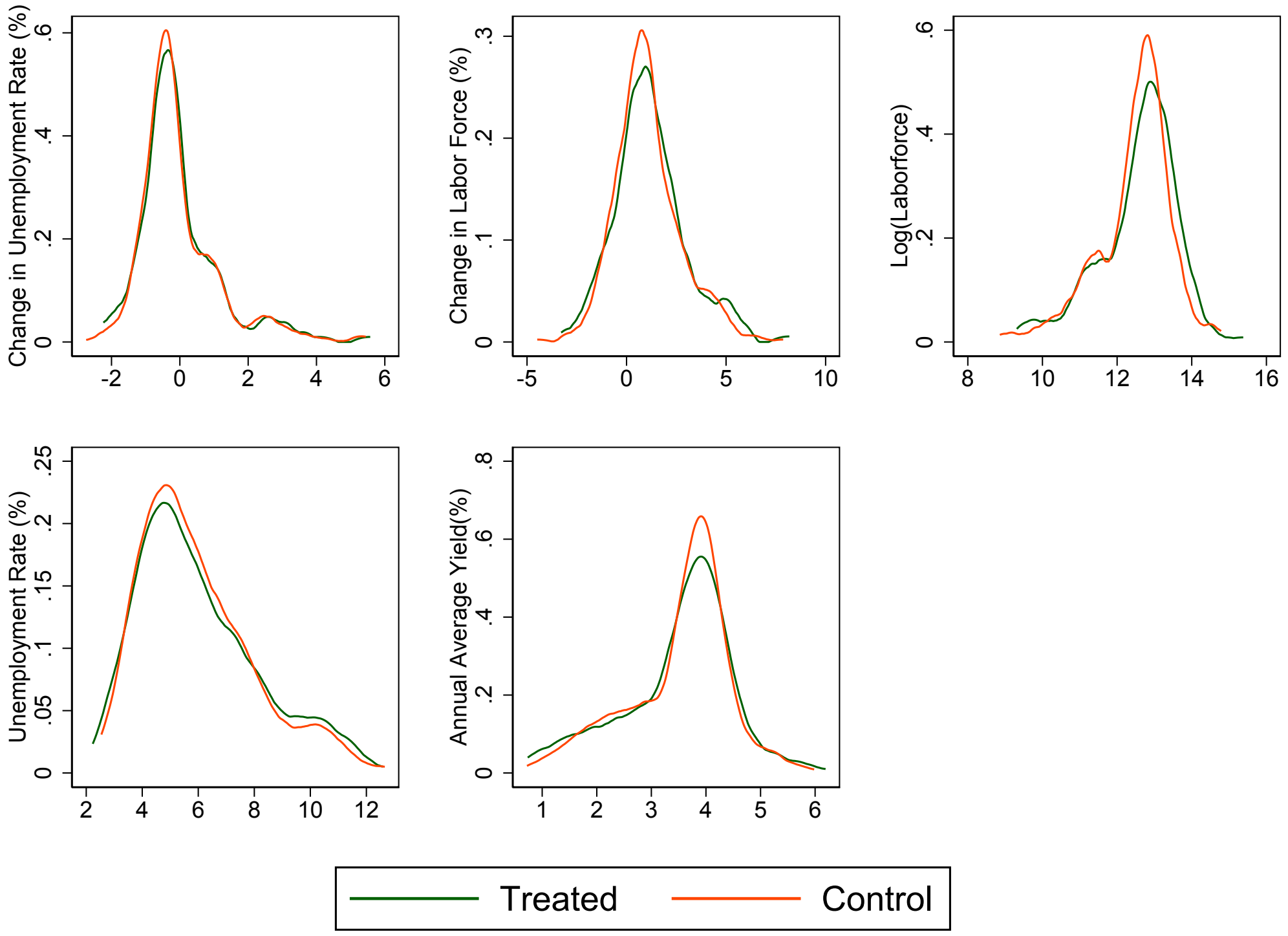

**Figure C3.10:** Matching County Characteristics - KNN(3):
The figure shows the kernel density plot based on matching the treatment counties using 5 variables in the year before firm bankruptcy: unemployment rate, change in unemployment rate, log(labor force), change in labor force and average yield of the county in that year.

**Based on County Dependence (Wages)**

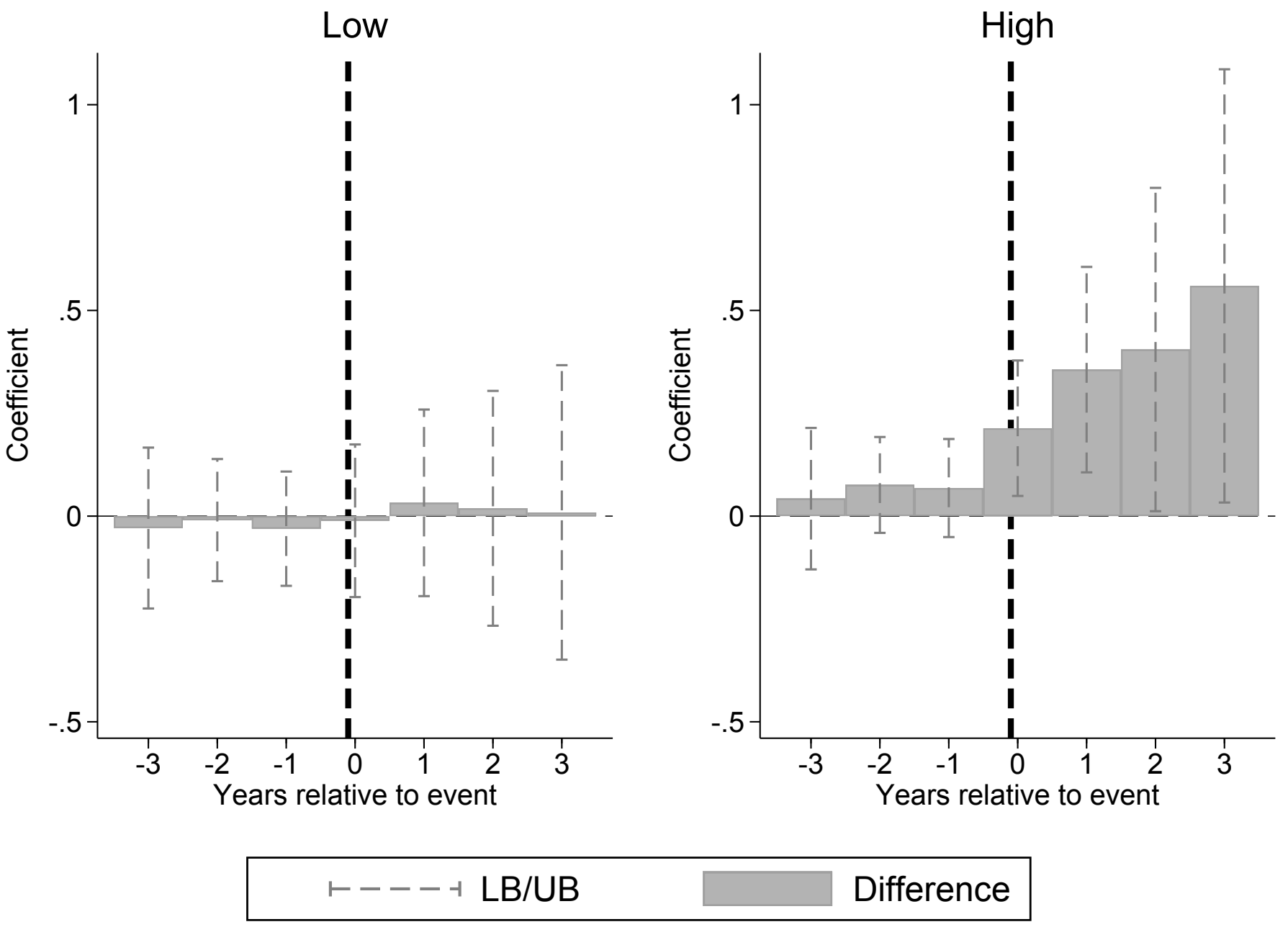

**Figure C3.11:** Unemployment Rate:
The figure shows the impact on unemployment rates between the treated and control counties. We report coefficients from Equation (3.6) using unemployment rate as the dependent variable. We cluster standard errors at the event pair level. We show the impact on sub-samples among the treated counties with low versus high shares of wages in the industry of the bankrupt firm. The control county was matched based on five variables in the year before firm bankruptcy: unemployment rate, change in unemployment rate, log(labor force), change in labor force, and average yield of the county in that year.

**Based on County Dependence (Wages)**

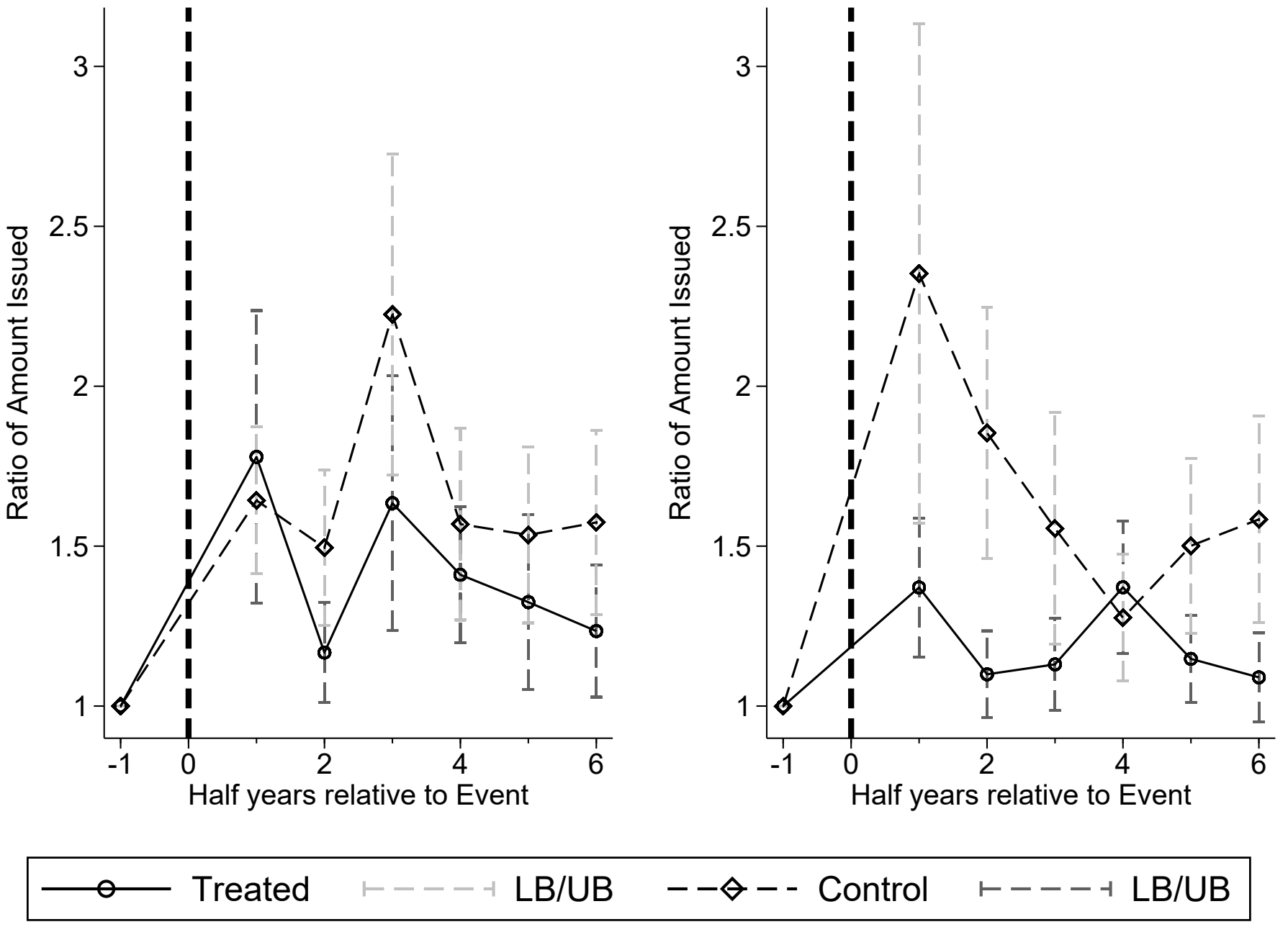

**Figure C3.12:** Primary Market Bond Issuance:

The figure shows the impact on unemployment rates between the treated and control counties. We report coefficients from Equation (3.6) using unemployment rate as the dependent variable. We cluster standard errors at the event pair level. We show the impact on sub-samples among the treated counties with low versus high shares of wages in the industry of the bankrupt firm. The control county was matched based on five variables in the year before firm bankruptcy: unemployment rate, change in unemployment rate, log(labor force), change in labor force, and average yield of the county in that year.

**Table C3.1:** Summary Statistics: Municipal Bonds

This table summarizes the bond level characteristics for our sample of bonds linked to bankruptcy counties during 2005-2019. Panel A reports the secondary market attributes. Panel B reports the primary market features. The key variables are described in Table C1.

**Panel A: Secondary market**

| | Count | Mean | Median | Std. Dev. |
|---|---|---|---|---|
| **Treated** | | | | |
| Wtd. Avg. Yield (%) | 1,520,693 | 3.1 | 3.3 | 1.5 |
| Yield Spread (%) | 1,520,693 | 1.4 | 1.5 | 2.4 |
| After-tax Yield Spread (%) | 1,520,693 | 3.4 | 3.0 | 2.8 |
| Remaining Maturity (years) | 1,520,693 | 11.1 | 9.8 | 7.2 |
| **Control** | | | | |
| Wtd. Avg. Yield (%) | 1,182,649 | 3.0 | 3.1 | 1.5 |
| Yield Spread (%) | 1,182,649 | 1.5 | 1.6 | 2.2 |
| After-tax Yield Spread (%) | 1,182,649 | 3.4 | 3.0 | 2.7 |
| Remaining Maturity (years) | 1,182,649 | 11.2 | 9.8 | 7.2 |

**Panel B: Primary market**

| | Count | Mean | Median | Std. Dev. |
|---|---|---|---|---|
| **Treated** | | | | |
| Offering Yield (%) | 217,814 | 3.2 | 3.5 | 1.3 |
| Offering Price (USD) | 217,811 | 102.3 | 101.4 | 11.3 |
| Coupon (%) | 217,804 | 3.8 | 4.0 | 1.3 |
| Years to Maturity | 217,814 | 10.2 | 9.1 | 6.8 |
| Years to Call | 99,464 | 8.5 | 9.8 | 2.8 |
| Amount (USD million) | 216,680 | 3.5 | 0.8 | 18.4 |
| Issue Size (USD million) | 217,814 | 53.1 | 15.0 | 124.7 |
| **Control** | | | | |
| Offering Yield (%) | 209,373 | 3.2 | 3.4 | 1.3 |
| Offering Price (USD) | 209,370 | 103.0 | 101.4 | 8.9 |
| Coupon (%) | 209,372 | 3.8 | 4.0 | 1.2 |
| Years to Maturity | 209,373 | 10.0 | 9.0 | 6.6 |
| Years to Call | 96,434 | 8.7 | 9.6 | 2.1 |
| Amount (USD million) | 208,310 | 2.8 | 0.6 | 20.1 |
| Issue Size (USD million) | 209,373 | 42.2 | 11.3 | 121.4 |

**Table C3.2:** Additional Matching Strategies

This table reports the results after incorporating additional matching strategies. We report results for our baseline specification from Equation 4.1. In Columns (1)-(3), we introduce additional variables for identifying the nearest neighbor matching county in the control group. These pertain to the county level debt capacity, similar to [32]. We describe these ratios in Table C1. Next, in Columns (4)-(5), we include ex-ante measures from the primary bond market to identify the nearest neighbor match. Specifically, we use average amount issued and average maturity of new bonds issued, respectively. In Columns (6)-(7), we consider geographical aspects in the matching strategy. First, in Column (6), we restrict the baseline matching to identifying a control county within the same geographic region. In Column (7), we consider three nearest neighbors using the baseline matching strategy. T-statistics are reported in brackets and standard errors are clustered at issue-year month level, unless otherwise specified. * $p < 0.10$, ** $p < 0.05$, *** $p < 0.01$

| *Dependent Variable*: | After-tax yield spread | | | | | | |
|---|---|---|---|---|---|---|---|
| *Additional Dimension*: | By Debt Capacity | | | By Primary Market | | By Geography | |
| | Interest to Debt | Interest to Expenditure | Interest to Revenue | Average Amount Issued | Average Maturity | Same Region | KNN(3) |
| | (1) | (2) | (3) | (4) | (5) | (6) | (7) |
| Post × Treated | 9.71*** | 6.96** | 6.64** | 5.34* | 5.81* | 6.00** | 4.54* |
| | [3.33] | [2.35] | [2.28] | [1.77] | [1.93] | [2.20] | [1.82] |
| Post | -4.24** | -3.39 | -2.18 | -0.20 | -0.41 | -3.22* | 1.36 |
| | [-2.00] | [-1.60] | [-1.01] | [-0.09] | [-0.20] | [-1.73] | [0.91] |
| Treated | -2.49 | -1.63 | -1.65 | -5.40 | -8.45** | 0.35 | 1.92 |
| | [-0.75] | [-0.46] | [-0.40] | [-1.60] | [-2.40] | [0.07] | [0.84] |
| Event FE | ✓ | ✓ | ✓ | ✓ | ✓ | ✓ | ✓ |
| County FE | ✓ | ✓ | ✓ | ✓ | ✓ | ✓ | ✓ |
| Year-month FE | ✓ | ✓ | ✓ | ✓ | ✓ | ✓ | ✓ |
| Bond Controls | ✓ | ✓ | ✓ | ✓ | ✓ | ✓ | ✓ |
| County Controls | ✓ | ✓ | ✓ | ✓ | ✓ | ✓ | ✓ |
| Adj.-$R^2$ | 0.654 | 0.659 | 0.655 | 0.656 | 0.660 | 0.663 | 0.642 |
| Obs. | 2,634,989 | 2,598,068 | 2,615,855 | 2,746,200 | 2,589,944 | 2,720,500 | 5,041,025 |

**Table C3.3:** Impact on Primary Market of Rated Municipal Bonds

This table shows the effect of bankruptcy filing on new bond issuances using a difference-in-differences estimate similar to the baseline specification. It is based on primary market bonds in Equation 3.5 for after-tax offering yields. In Column (1), we show the result by using only event-pair fixed effects and issuer fixed effects in the baseline equation. Next, in Column (2), we introduce bond-level controls. Column (3) shows the results with county controls and county fixed effects. We use S&P credit ratings at the time of issuance. P-values are reported in brackets and standard errors are clustered at issue level. $^{*}$ $p < 0.10$, $^{**}$ $p < 0.05$, $^{***}$ $p < 0.01$

| *Dependent Variable*: | After-tax Offering Yield | | |
|---|---|---|---|
| | (1) | (2) | (3) |
| Post × Treated | 10.52$^{***}$ | 8.92$^{***}$ | 6.34$^{***}$ |
| | [0.00] | [0.00] | [0.00] |
| Post | -8.21$^{***}$ | -9.77$^{***}$ | -8.91$^{***}$ |
| | [0.00] | [0.00] | [0.00] |
| Treated | -2.33 | -2.02 | -0.51 |
| | [0.34] | [0.25] | [0.76] |
| Event FE | ✓ | ✓ | ✓ |
| Issuer FE | ✓ | ✓ | ✓ |
| Bond Controls | | ✓ | ✓ |
| County Controls | | | ✓ |
| Adj.-$R^2$ | 0.466 | 0.851 | 0.853 |
| Obs. | 276,977 | 276,977 | 276,977 |

**Table C3.4:** Mechanism: State-Imposed Budgetary Restrictions

This table reports the results for the proposed mechanism at work based on county budgetary restrictions. We report results for our baseline specification from Equation (4.1), which is interacted with dummies corresponding to county budgetary restrictions. The dependent variable is after-tax yield spread. We additionally include group-month fixed effects in the modified baseline equation. We provide more details in Section C3.1.2. Column (1) shows the results by dividing the treated counties with firm bankruptcies into those which do or do not have *Debt Issuance* restrictions imposed by their respective states. In Column (2), we divide the treated counties based on whether they have *Overall Property Tax Restriction* or not. Column (3) shows the results using *Expenditure Limit* to distinguish between the types of treated counties. Finally, in Column (4), we show the impact based on a linear combination of these indicators (for Columns (1)-(3)) by summing them. We divide the treated counties into groups based on the total number of restrictions. T-statistics are reported in brackets and standard errors are double clustered at county bond issuer and year month level, unless otherwise specified. $^{*}$ $p < 0.10$, $^{**}$ $p < 0.05$, $^{***}$ $p < 0.01$

| *Dependent Variable*: | After-tax yield spread | | | |
|---|---|---|---|---|
| | Debt Issuance | Overall Property Tax Restriction | Expenditure Limit | Combination of Restrictions |
| Post x Treated | (1) | (2) | (3) | (4) |
| No | 5.15 | 1.93 | 8.25** | |
| | [0.83] | [0.53] | [2.22] | |
| Yes | 11.65*** | 19.30*** | 12.90** | |
| | [3.46] | [3.86] | [2.49] | |
| None | | | | 4.14 |
| | | | | [0.67] |
| One | | | | 3.65 |
| | | | | [0.92] |
| Multiple | | | | 19.14*** |
| | | | | [3.85] |
| Difference | 6.50 | 17.37 | 4.66 | 15.00 |
| P-value | 0.34 | 0.01 | 0.46 | 0.05 |
| Event FE | ✓ | ✓ | ✓ | ✓ |
| County FE | ✓ | ✓ | ✓ | ✓ |
| Month FE | ✓ | ✓ | ✓ | ✓ |
| Group Month FE | ✓ | ✓ | ✓ | ✓ |
| Controls | ✓ | ✓ | ✓ | ✓ |
| Adj.-$R^2$ | 0.649 | 0.649 | 0.650 | 0.650 |
| Obs. | 2,703,342 | 2,703,342 | 2,703,342 | 2,703,342 |

## **Table C3.5:** Mechanism: State-Level Business Incentives

This table reports the results for the proposed mechanism at work based on state-level business incentives. We report results for our baseline specification from Equation 4.1, which is interacted with dummies corresponding to the level of business incentives from the state. The dependent variable is after-tax yield spread. We additionally include group-month fixed effects in the modified baseline equation. We divide the treated counties into two groups based on the median value of the incentives. We provide more details in Section C3.1.2. Column (1) shows the results by dividing the treated counties into two groups using *Net Tax (after incentives)*. In Column (2), we divide the treated counties based on *Total Incentives*. Next, in Column (3), we show the impact based on the level of *Investment Tax Credit*. Finally, Column (4) shows the main result based on *R&D Credit*. T-statistics are reported in brackets and standard errors are double clustered at county bond issuer and year month level, unless otherwise specified. $^{*}$ $p < 0.10$, $^{**}$ $p < 0.05$, $^{***}$ $p < 0.01$

| *Dependent Variable*: | After-tax yield spread | | | |
|---|---|---|---|---|
| | Net Tax (after incentives) | Total Incentives | Investment Tax Credit | R&D Tax Credit |
| Post × Treated | (1) | (2) | (3) | (4) |
| Low | 3.59 | 10.35$^{**}$ | 15.61$^{***}$ | 13.72$^{***}$ |
| | [0.73] | [2.42] | [3.86] | [2.79] |
| High | 14.04$^{***}$ | 9.06$^{**}$ | -0.83 | 7.24$^{**}$ |
| | [3.01] | [1.99] | [-0.16] | [2.01] |
| Difference | 10.45 | 1.29 | 16.44 | 6.48 |
| P-value | 0.15 | 0.84 | 0.01 | 0.26 |
| Event FE | ✓ | ✓ | ✓ | ✓ |
| County FE | ✓ | ✓ | ✓ | ✓ |
| Month FE | ✓ | ✓ | ✓ | ✓ |
| Group-Month FE | ✓ | ✓ | ✓ | ✓ |
| Controls | ✓ | ✓ | ✓ | ✓ |
| Adj.-$R^2$ | 0.648 | 0.648 | 0.649 | 0.648 |
| Obs. | 2,625,214 | 2,625,214 | 2,625,214 | 2,625,214 |

## APPENDIX D

## MISCELLANEOUS SECTION FOR CHAPTER 4

**Table D4.1:** Labels in Training Data

This table shows the label names and the number of sequences under each label in the training data generated from the sustainability documents.

| Label | Number of Sequences |
|---|---|
| Environmental | 5,141 |
| Social | 6,921 |
| Governance | 2,291 |
| Business Model Innovation | 3,563 |
| Economic | 634 |

**Table D4.2:** Earnings Call Snippets of Top Firms with E&S Discussion

| Firm | Exchange | Ticker | Time | E&S Phrases | Snippet |
|---|---|---|---|---|---|
| Jaguar Animal Health | NASDAQ | JAGX | 2017Q3 | HIV; LGBT | In addition, we have many nonpersonal promotional efforts including a large advertising campaign, which was launched in late September. The campaign is called, ”Enough is Enough”. It is currently running in print and online digital advertising in nine HIV and gastroenterology medical journals and their websites, five national HIV magazines including Pause and HIV-Positive and Plus magazines along with their digital version and then 23 regional LGBT magazines and newspapers and their digital outlets, which run across 16 major metropolitan areas in the United States. |
| Tri-Tech Holding Inc | NASDAQ | TRIT | 2012Q2 | industrial waste; wastewater; water treatment; treatment plant | For that part, again industrial wastewater treatment project, we are going to complete pipeline placing, and the stable compressing of the wastewater treatment plant, equipment procurement and the signing of contract. |
| Aemetis Inc | NASDAQ | AMTX | 2019Q3 | zero carbon; renewable; low carbon | We plan to produce increasing amount of valuable below zero carbon renewable fuels through the use of patented and proprietary technologies to convert dairy waste, waste wood and other cellulosic feedstocks into low carbon and below zero carbon renewable fuels. |

| Firm | Exchange | Ticker | Time | E&S Phrases | Snippet |
|---|---|---|---|---|---|
| 22nd Century Group Inc | NYSEMKT | XXII | 2016Q2 | public health; nicotine | On an international level we are pleased to hear that public health agencies around the world are beginning to broadly announce positions and support of very low nicotine tobacco cigarettes. |
| China Ming Yang Wind Power Group Ltd | NYSE | MY | 2015Q4 | renewable; hydropower; generated energy; carbon emission; energy conservation | As for power grid connection and consumption on March the 3rd, NDRC released instructions on building a guidance system on development and utilization target of renewable energy, which state that, renewable energy excluding hydropower, should take up more than 9% of total generated energy by 2020. A renewable energy green certificates trading mechanism shall be established, allowing green certificate holders to participate in the trading of carbon emission reduction and energy conservation, while green profits will be generated for renewable energy. |
| Hudson Technologies Inc | NASDAQ | HDSN | 2014Q2 | environmentally; ozone; global warming; EPA | R-22 is very environmentally harmful substance from both an ozone depletion and global warming perspective, and EPA as heard from a broad cross-section of stakeholders, including members of Congress that a more aggressive phase out will have the best outcome. |
| FuelCell Energy Inc | NASDAQ | FCEL | 2010Q3 | fuel cell; low carbon; green energy; renewable | Our FuelCell technology helps South Korea achieve their low carbon green energy goals, allows agriculture and municipal customers to transform waste problems into renewable energy solutions and generates reliable secure power for commercial and government facilities. |

| Firm | Exchange | Ticker | Time | E&S Phrases | Snippet |
|---|---|---|---|---|---|
| China Recycling Energy Corp | NASDAQ | CREG | 2013Q4 | biomass; waste heat | Two biomass system to Shenqiu Phase I and II for 11 years and Shenqiu Phase II for 9.5 years; five power and steam generating systems from waste heat from metal refining to Erdos for 20 years; one waste heat system to Zhongbao for nine year; one waste heat system to Jilin Ferroalloys for 24 years; and two TRT systems for Datong for 30 years. |
| SJW Corp | NYSE | SJW | 2016Q1 | treatment plant; surface water | Recall that Montevina Treatment Plant project is being constructed using a progressive design build model that allows surface water production through the plant between construction phases. |
| Blue Earth Inc | NASDAQ | BBLU | 2015Q2 | renewable; solar project | Construction sales are derived from the installation of alternative renewable energy systems and installation and maintenance of back up energy systems. |

## **Table D4.3:** Additional Summary Statistics

This table summarizes the descriptive statistics for independent variables used in the paper. Panel A corresponds to the fraction of E&S phrases (reported as %) and control variables related to the market reaction analysis. Panel B shows the distribution of the proportion of environmental and social phrases in our sample, aggregated up to the firm–year level. Panel C presents the number of green patents granted in the 2-year and 3-year windows after the earnings calls in the samples of the Poisson regression.

| | Count | Mean | Median | Std. Dev. |
|---|---|---|---|---|
| **Panel A: Transcript-Level Variables** | | | | |
| Level(E+S) | 81,662 | 0.051 | 0.024 | 0.089 |
| Level(E) | 81,662 | 0.035 | 0.011 | 0.081 |
| Level(S) | 81,662 | 0.016 | 0.000 | 0.036 |
| Level(E-Material) | 81,662 | 0.012 | 0.000 | 0.037 |
| Level(E-Immaterial) | 81,662 | 0.023 | 0.000 | 0.062 |
| Level(S-Material) | 81,662 | 0.008 | 0.000 | 0.024 |
| Level(S-Immaterial) | 81,662 | 0.009 | 0.000 | 0.024 |
| Earnings surprise | 81,662 | 0.000 | 0.001 | 0.012 |
| Negative EPS | 81,662 | 0.228 | 0.000 | 0.420 |
| HighUE | 81,662 | 0.098 | 0.000 | 0.298 |
| LowUE | 81,662 | 0.098 | 0.000 | 0.298 |
| Pre-event return | 81,662 | 0.000 | 0.001 | 0.003 |
| Pre-event volume (Millions) | 81,662 | 2.060 | 0.718 | 3.930 |
| ROA | 81,662 | 0.004 | 0.009 | 0.037 |
| Size | 81,662 | 7.765 | 7.750 | 1.766 |
| Accruals | 81,662 | -0.014 | -0.011 | 0.036 |
| Earnings volatility | 81,662 | 0.024 | 0.012 | 0.033 |
| MTB | 81,662 | 1.894 | 1.461 | 1.271 |
| Leverage | 81,662 | 0.230 | 0.199 | 0.205 |
| Return volatility | 81,662 | 0.104 | 0.088 | 0.061 |
| # Analysts (log) | 81,662 | 1.611 | 1.609 | 0.918 |
| Firm age (log) | 81,662 | 2.924 | 2.962 | 0.781 |
| Uncertainty | 81,662 | 1.018 | 0.996 | 0.248 |
| Sentiment | 81,662 | 0.580 | 0.584 | 0.598 |
| **Panel B: Firm-Year Aggregation** | | | | |
| Level(E+S) | 39,388 | 0.056 | 0.029 | 0.093 |
| Level(E) | 39,388 | 0.037 | 0.013 | 0.085 |
| Level(S) | 39,388 | 0.018 | 0.008 | 0.038 |
| Level(E-Material) | 39,388 | 0.013 | 0.000 | 0.040 |
| Level(E-Immaterial) | 39,388 | 0.025 | 0.007 | 0.064 |
| Level(S-Material) | 39,388 | 0.008 | 0.000 | 0.024 |
| Level(S-Immaterial) | 39,388 | 0.010 | 0.003 | 0.026 |
| **Panel C: Number of Green Patents** | | | | |
| *2-Year Window* | | | | |
| Overall | 32,593 | 4.625 | 0.000 | 18.037 |
| Env&Water | 32,593 | 0.936 | 0.000 | 4.315 |
| Climate | 32,593 | 3.913 | 0.000 | 15.382 |
| *3-Year Window* | | | | |
| Overall | 32,686 | 7.132 | 0.000 | 27.796 |
| Env&Water | 32,686 | 1.453 | 0.000 | 6.764 |
| Climate | 32,686 | 6.034 | 0.000 | 23.623 |

**Table D4.4:** Robustness: Firm Pollution Using Alternative Scaling and E&S Discussion

This table reports the results from Equation (4.1) using yearly firm pollution as the dependent variable during 2007-2019. *Level(E+S)* corresponds to the fraction of environmental and social phrases discussed in earnings conference calls during the year. Similarly, *Level(E)* denotes the fraction of environmental phrases discussed in earnings conference calls during the year. Each column corresponds to a different variable used to scale the pollution emissions in the outcome variable. The T-statistics (in brackets) are based on double clustering standard errors at the firm and year levels, unless otherwise specified. * $p < 0.10$, ** $p < 0.05$, *** $p < 0.01$

**Panel A: Using Environmental & Social Discussion**

| *Dependent Variable*: | Log (1+Pollution/Scaling Variable$_{t-1}$) | | | |
|---|---|---|---|---|
| *Scaling Variable*: | COGS$_t$ | Revenue$_{t-1}$ | Sale$_{t-1}$ | Total Assets$_{t-1}$ |
| | (1) | (2) | (3) | (4) |
| Level(E+S) | -56.07*** | -68.82*** | -68.25*** | -43.31*** |
| | [-3.09] | [-3.85] | [-3.79] | [-3.21] |
| Firm FE | ✓ | ✓ | ✓ | ✓ |
| Industry-Year FE | ✓ | ✓ | ✓ | ✓ |
| Controls | ✓ | ✓ | ✓ | ✓ |
| Adj.-$R^2$ | 0.942 | 0.947 | 0.946 | 0.941 |
| Obs. | 8,971 | 9,046 | 8,968 | 9,054 |

**Panel B: Using Only Environmental Discussion**

| *Dependent Variable*: | Log (1+Pollution/Scaling Variable$_{t-1}$) | | | |
|---|---|---|---|---|
| *Scaling Variable*: | COGS$_t$ | Revenue$_{t-1}$ | Sale$_{t-1}$ | Total Assets$_{t-1}$ |
| | (1) | (2) | (3) | (4) |
| Level(E) | -52.48** | -67.43*** | -66.82*** | -40.80** |
| | [-2.54] | [-3.29] | [-3.24] | [-2.76] |
| Firm FE | ✓ | ✓ | ✓ | ✓ |
| Year FE | | | | |
| Industry-Year FE | ✓ | ✓ | ✓ | ✓ |
| Controls | ✓ | ✓ | ✓ | ✓ |
| Adj.-$R^2$ | 0.942 | 0.946 | 0.946 | 0.941 |
| Obs. | 8,971 | 9,046 | 8,968 | 9,054 |

## **Table D4.5:** Robustness: Firm Pollution and E&S Discussion Using Matched Panel

This table reports the results from Equation (4.1) using yearly firm pollution from the EPA Toxic Release Inventory (TRI) as the dependent variable during 2007-2019 using a matched panel. We do not impute zero levels of environmental or social discussion to firms for which we do not find earnings conference call transcripts. *Level(E+S)* corresponds to the fraction of words devoted to discussion of environmental and social topics in the earnings conference calls during the year. *Level(E)* denotes the fraction of words devoted to discussion of environmental topics in the earnings conference calls during the year. In Panel A, we show regression results using *Level(E+S)* as the explanatory variable. Panel B shows results using *Level(E)* as the independent variable. In Panel C, we show results using *Level(E)* with at least one environmental word in the sample. Industry definition is based on Fama-French 30 classification. The T-statistics (in brackets) are based on double clustering standard errors at the firm and year levels, unless otherwise specified. $^{*}$ $p < 0.10$, $^{**}$ $p < 0.05$, $^{***}$ $p < 0.01$

### Panel A: Using Environmental & Social Discussion

| *Dependent Variable*: | Log (1+Pollution/COGS$_{t-1}$) | | | | | |
|---|---|---|---|---|---|---|
| | Total | | | Air | Water | Ground |
| | (1) | (2) | (3) | (4) | (5) | (6) |
| Level(E+S) | -66.97$^{**}$ | -61.98$^{**}$ | -56.45$^{**}$ | -66.51$^{**}$ | -12.92 | 32.65 |
| | [-2.76] | [-2.81] | [-2.63] | [-2.59] | [-0.86] | [0.52] |
| Log(Total Revenue) $_{t-1}$ | | -0.71$^{***}$ | -0.72$^{***}$ | -0.71$^{***}$ | -0.32$^{***}$ | -0.23$^{**}$ |
| | | [-7.62] | [-7.57] | [-6.92] | [-3.87] | [-2.91] |
| Leverage $_{t-1}$ | | 0.19 | 0.12 | 0.03 | 0.04 | -0.10 |
| | | [1.10] | [0.69] | [0.18] | [0.54] | [-0.86] |
| Log(Assets) $_{t-1}$ | | 0.16$^{*}$ | 0.15$^{*}$ | 0.26$^{***}$ | 0.04 | 0.07 |
| | | [1.99] | [1.88] | [3.14] | [0.61] | [1.06] |
| NetPPEA $_{t-1}$ | | 0.11 | 0.13 | 0.16 | 0.02 | 0.05 |
| | | [0.47] | [0.55] | [0.53] | [0.08] | [0.19] |
| Firm FE | ✓ | ✓ | ✓ | ✓ | ✓ | ✓ |
| Year FE | ✓ | ✓ | | | | |
| Industry-Year FE | | | ✓ | ✓ | ✓ | ✓ |
| Controls | | ✓ | ✓ | ✓ | ✓ | ✓ |
| Adj.-$R^2$ | 0.942 | 0.945 | 0.947 | 0.933 | 0.953 | 0.959 |
| Obs. | 7,347 | 7,347 | 7,318 | 7,318 | 7,318 | 7,318 |

**Panel B: Using Only Environmental Discussion**

| *Dependent Variable*: | Log (1+Pollution/COGS$_{t-1}$) | | | | | |
|---|---|---|---|---|---|---|
| | Total | | | Air | Water | Ground |
| | (1) | (2) | (3) | (4) | (5) | (6) |
| Level(E) | -59.37** | -54.09** | -51.49** | -64.87** | -13.42 | 33.05 |
| | [-2.20] | [-2.23] | [-2.20] | [-2.36] | [-0.87] | [0.48] |
| Firm FE | ✓ | ✓ | ✓ | ✓ | ✓ | ✓ |
| Year FE | ✓ | ✓ | | | | |
| Industry-Year FE | | | ✓ | ✓ | ✓ | ✓ |
| Controls | | ✓ | ✓ | ✓ | ✓ | ✓ |
| Adj.-$R^2$ | 0.942 | 0.945 | 0.947 | 0.933 | 0.953 | 0.959 |
| Obs. | 7,347 | 7,347 | 7,318 | 7,318 | 7,318 | 7,318 |

**Panel C: At Least One Environmental Word**

| *Dependent Variable*: | Log (1+Pollution/COGS$_{t-1}$) | | | | | |
|---|---|---|---|---|---|---|
| | Total | | | Air | Water | Ground |
| | (1) | (2) | (3) | (4) | (5) | (6) |
| Level(E) | -60.00** | -56.16** | -53.09** | -66.73** | -14.24 | 32.37 |
| | [-2.21] | [-2.33] | [-2.27] | [-2.40] | [-0.93] | [0.46] |
| Log(Total Revenue) $_{t-1}$ | | -0.73*** | -0.76*** | -0.75*** | -0.36*** | -0.27** |
| | | [-8.08] | [-8.25] | [-7.30] | [-4.08] | [-2.98] |
| Leverage $_{t-1}$ | | 0.13 | 0.02 | -0.09 | 0.04 | -0.14 |
| | | [0.82] | [0.15] | [-0.50] | [0.48] | [-1.18] |
| Log(Assets) $_{t-1}$ | | 0.13* | 0.15* | 0.27*** | 0.04 | 0.08 |
| | | [1.80] | [1.98] | [3.28] | [0.57] | [1.04] |
| NetPPEA $_{t-1}$ | | 0.11 | 0.08 | 0.17 | -0.07 | 0.04 |
| | | [0.44] | [0.34] | [0.56] | [-0.26] | [0.12] |
| Firm FE | ✓ | ✓ | ✓ | ✓ | ✓ | ✓ |
| Year FE | ✓ | ✓ | | | | |
| Industry-Year FE | | | ✓ | ✓ | ✓ | ✓ |
| Controls | | ✓ | ✓ | ✓ | ✓ | ✓ |
| Adj.-$R^2$ | 0.945 | 0.948 | 0.949 | 0.935 | 0.953 | 0.960 |
| Obs. | 6,714 | 6,714 | 6,677 | 6,677 | 6,677 | 6,677 |

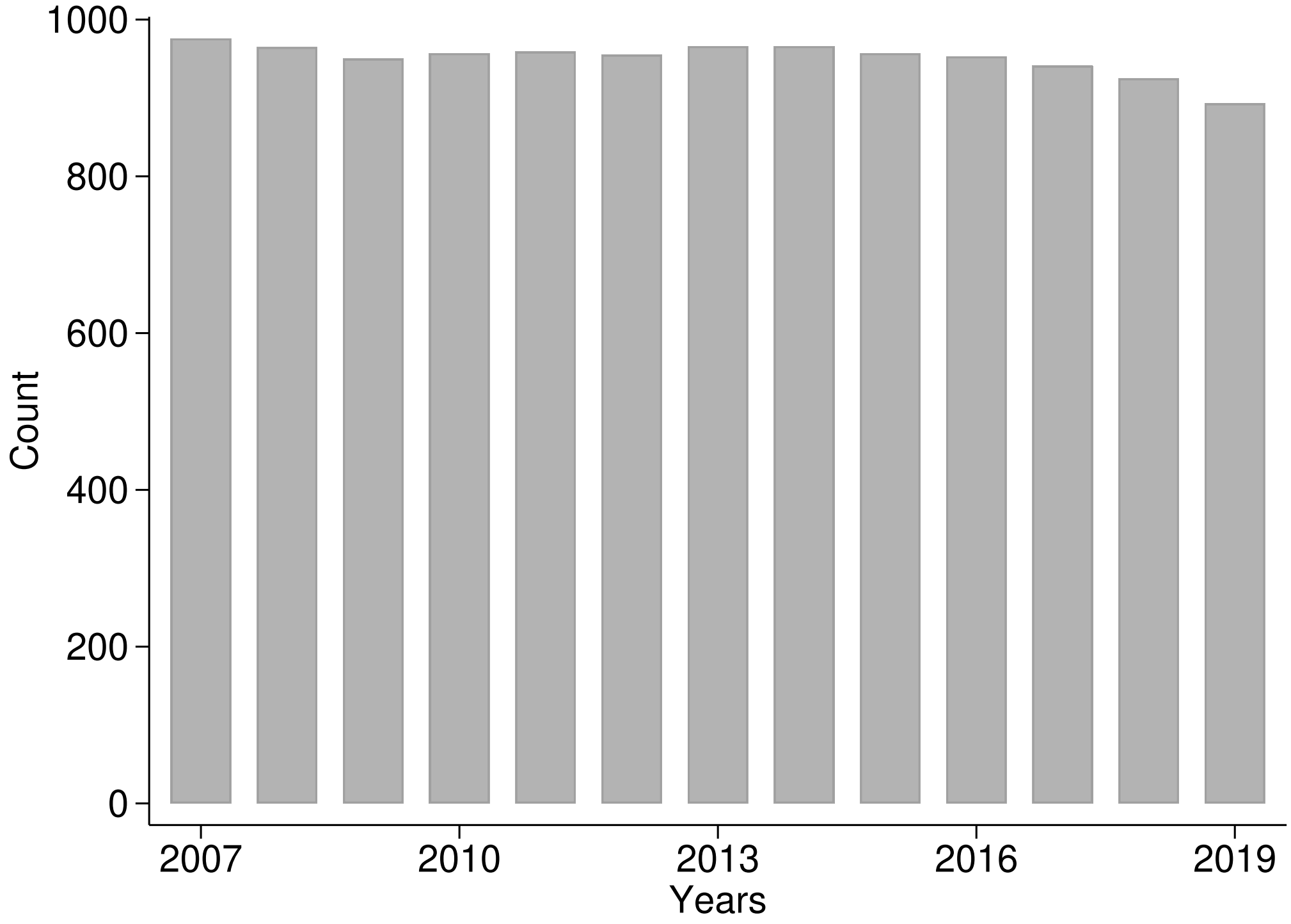

**Figure D4.1:** Toxic Releases Inventory (TRI) - Number of Firms:
The figure shows the the number of firms in our sample for which we obtain data on toxic releases from EPA's TRI.

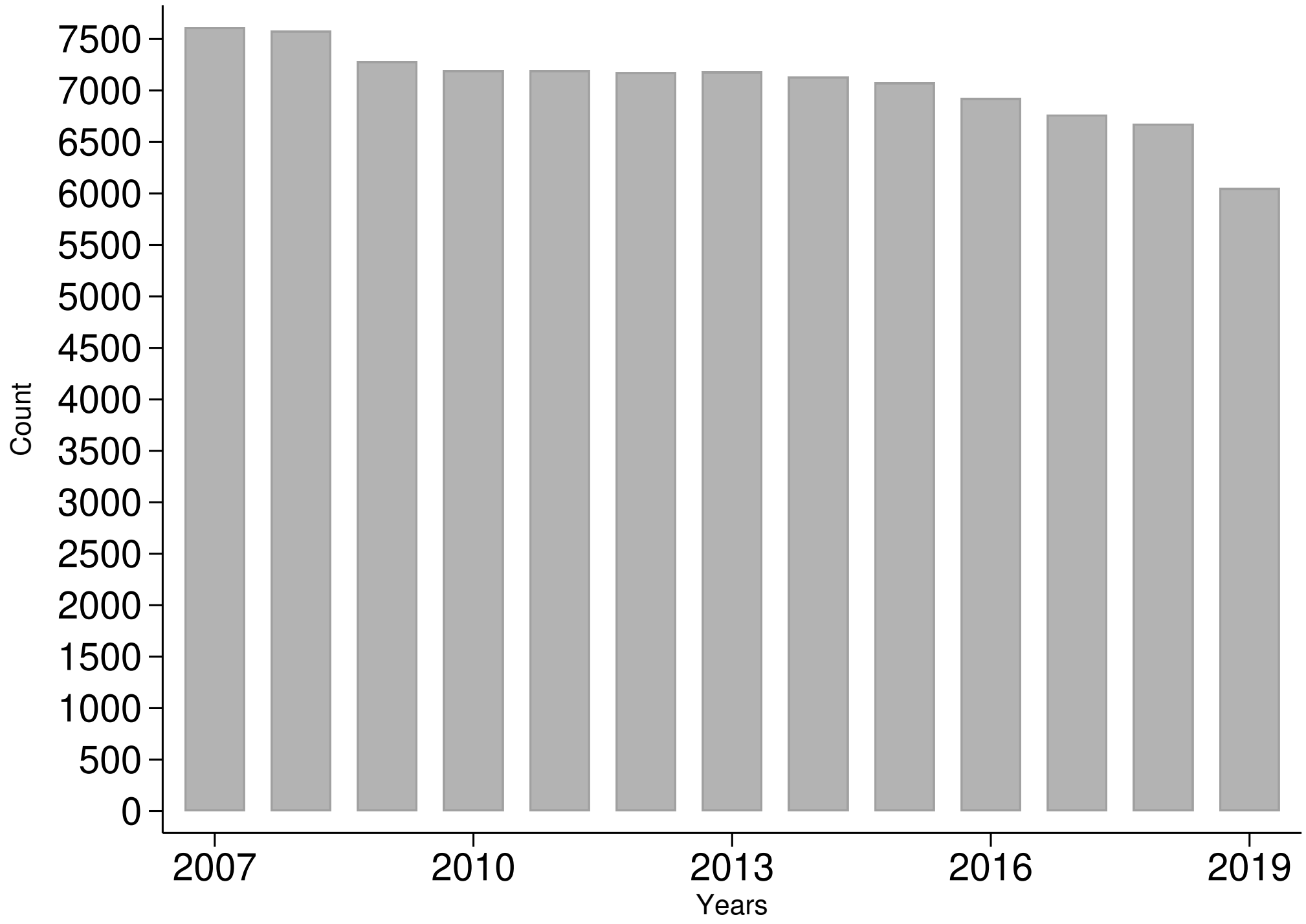

**Figure D4.2:** Toxic Releases Inventory (TRI) - Number of Facilities:
The figure shows the the number of facilities in our sample for which we obtain data on toxic releases from EPA's TRI.

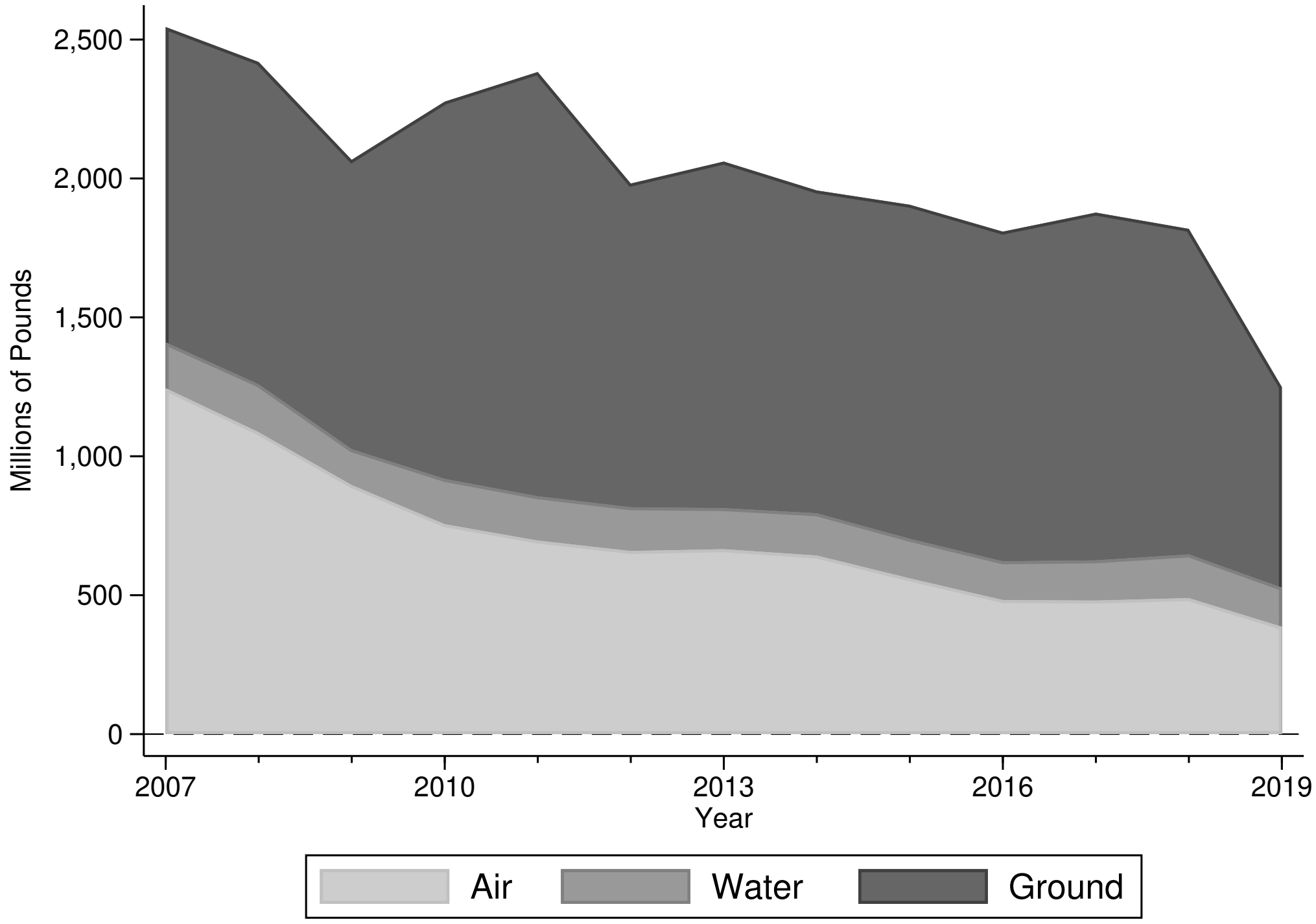

**Figure D4.3:** Total Toxic Releases:
This figure shows the aggregate pollution (in millions of pounds) across different media of the environment for the firms in our sample during 2007-2019, as reported to the EPA's Toxic Release Inventory. The sample excludes firms that were not required to report emissions during the entire period.

# VITA

Baridhi Malakar completed his Ph.D. in Finance at the Scheller College of Business, Georgia Institute of Technology. Before that, Baridhi received a Master's degree in Economics from the School of Economics, Georgia Institute of Technology and an MBA in Finance from XLRI Jamshedpur. He holds a Bachelor's degree in Civil Engineering from the Indian Institute of Technology Roorkee.

Baridhi's research interests include empirical corporate finance, responsible and sustainable finance, and municipal bonds. His research has won the Best Paper Award at Annual Meeting of AEFIN Finance Forum. His papers have been presented at numerous prestigious conferences such as AFA, SFS Cavalcade, MFA, and the European Finance Association, and accepted for publication at the Review of Finance. He has taught classes on the Management of Financial Institutions and on Finance and Investments.